\documentclass[]{aa}

\usepackage{txfonts}
\usepackage{natbib,graphicx,amssymb}
\usepackage[breaklinks=false]{hyperref}
\usepackage{xspace}
\usepackage{lscape}

\newcommand{\kms}{km\,s$^{-1}$\xspace}
\newcommand{\mum}{$\mu$m\xspace}
\newcommand{\av}{$A_V$\xspace}
\newcommand{\cii}{[C\,{\sc ii}]\xspace}
\newcommand{\nii}{[N\,{\sc ii}]\xspace}
\newcommand{\niii}{[N\,{\sc iii}]\xspace}
\newcommand{\oi}{[O\,{\sc i}]\xspace}
\newcommand{\oiii}{[O\,{\sc iii}]\xspace}
\newcommand{\ciil}{[C\,{\sc ii}]$_{157}$\xspace}
\newcommand{\niila}{[N\,{\sc ii}]$_{122}$\xspace}
\newcommand{\niilb}{[N\,{\sc ii}]$_{205}$\xspace}
\newcommand{\niiil}{[N\,{\sc iii}]$_{57}$\xspace}
\newcommand{\oila}{[O\,{\sc i}]$_{63}$\xspace}
\newcommand{\oilb}{[O\,{\sc i}]$_{145}$\xspace}
\newcommand{\oiiila}{[O\,{\sc iii}]$_{52}$\xspace}
\newcommand{\oiiilb}{[O\,{\sc iii}]$_{88}$\xspace}
\newcommand{\hers}{\textit{Herschel}\xspace}
\newcommand{\spit}{\textit{Spitzer}\xspace}
\newcommand{\oii}{[O\,{\sc ii}]\xspace}
\newcommand{\siii}{[S\,{\sc iii}]\xspace}
\newcommand{\siv}{[S\,{\sc iv}]\xspace}
\newcommand{\neii}{[Ne\,{\sc ii}]\xspace}
\newcommand{\neiii}{[Ne\,{\sc iii}]\xspace}
\newcommand{\silii}{[Si\,{\sc ii}]\xspace}
\newcommand{\oiv}{[O\,{\sc iv}]\xspace}
\newcommand{\ariii}{[Ar\,{\sc iii}]\xspace}
\newcommand{\arii}{[Ar\,{\sc ii}]\xspace}
\newcommand{\feiii}{[Fe\,{\sc iii}]\xspace}
\newcommand{\feii}{[Fe\,{\sc ii}]\xspace}
\newcommand{\siiila}{[S\,{\sc iii}]$_{18}$\xspace}
\newcommand{\siiilb}{[S\,{\sc iii}]$_{33}$\xspace}
\newcommand{\sivl}{[S\,{\sc iv}]$_{10}$\xspace}

\newcommand{\neiiil}{[Ne\,{\sc iii}]$_{15}$\xspace}

\newcommand{\ha}{H$\alpha$\xspace}
\newcommand{\hua}{Hu$\alpha$\xspace}
\newcommand{\htwo}{H$_{\rm 2}$\xspace}
\newcommand{\hi}{H\,{\sc i}\xspace}
\newcommand{\hii}{H\,{\sc ii}\xspace}
\newcommand{\zsun}{$Z_{\odot}$\xspace}
\newcommand{\msun}{$M_{\odot}$\xspace}
\newcommand{\lsun}{$L_{\odot}$\xspace}
\newcommand{\ltir}{${L_{\rm TIR}}$\xspace}
\newcommand{\cm}{cm$^{-3}$\xspace}

\begin{document}

\title{The \hers Dwarf Galaxy Survey: \\
II.~Physical conditions, origin of \cii emission, and \\
porosity of the multiphase low-metallicity ISM
\thanks{Tables of observed line fluxes and predicted line
luminosities from the models are only available in electronic form
at the CDS via anonymous ftp to cdsarc.u-strasbg.fr (130.79.128.5)
or via http://cdsweb.u-strasbg.fr/cgi-bin/qcat?J/A+A/}
}

\author{
  D.~Cormier\inst{1}
  \and N.~P.~Abel\inst{2}
  \and S.~Hony\inst{3}
  \and V.~Lebouteiller\inst{1}
  \and S.~C.~Madden\inst{1}
  \and F.~L.~Polles\inst{4}
  \and F.~Galliano\inst{1}
  \and I.~De~Looze\inst{5,6}
  \and M.~Galametz\inst{1}
  \and A.~Lambert-Huyghe\inst{1}
}

\institute{ 
  AIM, CEA, CNRS, Universit\'e Paris-Saclay,
  Universit\'e Paris Diderot, Sorbonne Paris Cit\'e,
  F-91191 Gif-sur-Yvette, France\\
\email{diane.cormier@cea.fr} 
\and
 University of Cincinnati, Clermont College,
 Batavia, OH, 45103, USA
\and
  Institut f\"ur theoretische Astrophysik, 
  Zentrum f\"ur Astronomie der Universit\"at Heidelberg, 
  Albert-Ueberle Str. 2, D-69120 Heidelberg, Germany
\and 
  LERMA, Observatoire de Paris, PSL Research University,
  CNRS, Sorbonne Universit\'es, 75014 Paris, France
\and
 Department of Physics and Astronomy,
 University College London, Gower Street,
 London WC1E 6BT, UK
\and
 Sterrenkundig Observatorium, Universiteit Gent,
 Krijgslaan 281 S9, B-9000 Gent, Belgium
}

\date{Received 19 October 2018 / Accepted 3 April 2019}

\abstract
{The sensitive infrared telescopes, \spit and \hers, have been
used to target low-metallicity star-forming galaxies, allowing us
to investigate the properties of their interstellar medium (ISM)
in unprecedented detail.
Interpretation of the observations in physical terms relies
on careful modeling of those properties. 
We have employed a multiphase approach to model the ISM
phases (\hii region and photodissociation region) with the
spectral synthesis code {Cloudy}. 
Our goal is to characterize the physical conditions (gas densities,
radiation fields, etc.) in the ISM of the galaxies from the \hers
Dwarf Galaxy Survey. 
We are particularly interested in correlations between those
physical conditions and metallicity or star-formation activity.
Other key issues we have addressed are the contribution
of different ISM phases to the total line emission, especially
of the \cii157\,\mum line, and the characterization of the porosity
of the ISM.
We find that the lower-metallicity galaxies of our sample
tend to have higher ionization parameters and galaxies with
higher specific star-formation rates have higher gas densities.
The \cii emission arises mainly from PDRs and the contribution
from the ionized gas phases is small, typically less than 30\%
of the observed emission.
We also find a correlation -- though with scatter -- between
metallicity and both the PDR covering factor and the fraction
of \cii from the ionized gas. Overall, the low metal abundances
appear to be driving most of the changes in the ISM structure
and conditions of these galaxies, and not the high specific
star-formation rates. These results demonstrate in a quantitative
way the increase of ISM porosity at low metallicity. Such
porosity may be typical of galaxies in the young Universe.
}

\titlerunning{Modeling the multiphase ISM of star-forming dwarf galaxies}
\authorrunning{Cormier et al.}
\maketitle

\section{Introduction}
\label{sect:intro}
The interstellar medium (ISM) of galaxies is a complex
environment composed of phases with inhomogeneous
structures and different physical conditions, evolving
through dynamical effects such as feedback from massive
stars or turbulence
\citep[e.g.,][]{hennebelle-2008,hopkins-2012,naab-2017}.
Characterizing the ISM in galaxies is necessary
to understand how dense structures are formed and how
the environmental properties (metal content, activity,
stellar mass) may affect the star-formation process
and thereby galaxy evolution.
Space and airborne missions in the infrared, with IRAS, ISO,
KAO, \spit, \hers, and SOFIA, have been fundamental for
unveiling a large part the energy budget of galaxies and
enabling for a better census of the cooling of the ISM.
Advances in sensitivity have allowed detection of emission
lines probing regions related to star formation in low-metallicity
galaxies for the first time \citep[e.g.,][]{hunter-2001,
madden-2006,wu-2006,hunt-2010,cormier-2015,cigan-2016}.
These galaxies stand out compared to more metal-rich galaxies.
In particular, the prominent \oiii line at 88\,\mum in
low-metallicity star-forming regions highlights the presence
of an extended ionized gas phase, while the lack of CO
emission suggests an extremely clumpy dense gas distribution
\citep[e.g.,][]{indebetouw-2013,cormier-2014,lebouteiller-2017,
turner-2017,vallini-2017,jameson-2018}.
Those observations provide evidence of a possibly
different, porous ISM structure in those galaxies as opposed
to more metal-rich galaxies.
Establishing whether there is such a metallicity-dependent
change in the ISM structure is very important for understanding
the characteristics of low-mass galaxies in the high-redshift
Universe. It is particularly relevant for galaxies at the epoch
of reionization because enhanced porosity could be a
precondition that facilitates the escape of ionizing photons
from those galaxies \citep{stark-2016}.
However, a quantitative and systematic characterization
of the topology and composition of the ISM in galaxies has
yet to be achieved.
Because of the complexity of the ISM and the fact that
current observations do not resolve individual ISM components
in external galaxies, such a characterization can currently only
be reliably achieved via multiphase modeling.

Emission lines in the mid-infrared (MIR) and far-infrared
(FIR) are powerful diagnostics of the ISM properties
\citep[e.g.,][]{diaz-santos-2013,cormier-2015,fernandez-ontiveros-2016,
lapham-2017,spinoglio-2017,ucci-2017,herrera-camus-2018a,
polles-2019,vandertak-2018} because, contrary to optical lines,
they probe a wider range of physical conditions (temperatures,
densities, pressure, etc.), including more embedded regions.
In addition, their observed intensities are less affected by extinction.
The various lines that are accessible in the MIR and FIR may arise
from different ionized and neutral gas phases that can be
heated by several mechanisms such as ultraviolet photoionization
from young stars, photoelectric effect on dust grains, X-rays,
and shocks \citep{tielens-book}. Among those lines is the \cii
at 157\,\mum, which is one of the brightest cooling lines in
galaxies, used as a star-formation rate and mass tracer
\citep[e.g.,][]{delooze-2011,delooze-2014,herrera-camus-2015,
vallini-2015,jameson-2018,zanella-2018}, but with uncertainties
linked in part to the possibility of \cii being emitted in diffuse
ionized gas as well as in photodissociation regions
(PDR) with moderate densities.

In a first paper, \citealt{cormier-2015}, we present
observations of the FIR fine-structure lines observed
with the PACS instrument onboard \hers
\citep{poglitsch-2010,pilbratt-2010} in the galaxies
of the ``Dwarf Galaxy Survey'' \citep[DGS;][]{madden-2013}.
We found strikingly higher FIR line ratios of \oiiilb/\niila,
\niiil/\niila, \ciil/\niila, and \oiiilb/\oila 
in dwarf galaxies compared to more metal-rich galaxies.
Along with observations from the \spit IRS spectrograph
\citep{werner-2004,houck-2004}, we interpret those
ratios in terms of radiation field hardness and
high filling factor of ionized gas relative to the neutral gas.
However, our analysis in that paper is focused on the average properties
of the dwarf galaxy sample and the model assumptions
are quite simplified (for example, we assume a single set of abundances).
The DGS sample is also biased toward dwarf galaxies
with high specific star-formation rate (sSFR) for detectability
reasons. This bias does not always make it easy to infer
whether differences with metal-rich galaxies are driven by
their higher sSFR or the lower metallicity. 
Our goal in this paper is to perform a more detailed
modeling of the ISM properties (physical conditions,
porosity, etc.) of the DGS survey by modeling each
galaxy of the sample and by tuning some of the model
parameters to more appropriate values (from previous
studies or ancillary data) whenever possible. We aim
to investigate correlations of parameters characteristic
of the ISM structure and conditions with
a galaxy metallicity ($Z$) and star-formation activity, and,
ultimately, to infer what drives changes in the ISM of galaxies, 
which is important to understand how we can expect galaxies
in the early Universe to be affected by their low metallicities.

The outline of the paper is as follows.
In Section~\ref{sect:data} we present the galaxy sample and
dataset used. Section~\ref{sect:modeling} lays out the
modeling strategy, grid assumptions, and the method to
compare models and observations. We present the model
results in Section~\ref{sect:results} and their sensitivity
to some of the choices in the modeling strategy in
Section~\ref{sect:discparams}. Finally, we discuss the
origin of \cii emission as well as the ISM porosity in
Section~\ref{sect:discussion}. Our results are summarized
in Section~\ref{sect:conclusion}.

\section{Data}
\label{sect:data}
\subsection{Sample selection}
Our sample is drawn from the \hers Dwarf Galaxy Survey
\citep{madden-2013} which consists of 48 nearby galaxies
at distances between 0.5 and 200\,Mpc. In the closest
galaxies, only specific regions were observed with the PACS
spectrometer while the more distant galaxies were observed
in their entirety.
We have focussed our study on the galaxies of the $compact$ subsample
(43 galaxies) that were observed in full (thus excluding
NGC\,2366, NGC\,4861, and UM\,311), resulting in $40$ galaxies.
Among those, $37$ have ancillary \spit IRS spectroscopic
data (all but HS\,0017, UGC\,4483, and UM\,133).
We further require that our galaxies are detected
in at least three spectral lines (IRS or PACS) to proceed
with the modeling (thus excluding HS\,1236 and HS\,2352).
This leaves us with $38$ galaxies.
Details on the data reduction can be found in \cite{cormier-2015}.
Most galaxies were observed in both high-resolution (HR)
and low-resolution (LR) modes with the IRS instrument.
Since many lines are better detected in HR, we used
mainly fluxes from this mode. In cases where the LR
observation is deeper and provides a detection in LR
while the line is undetected in HR, we use the LR flux.
We also verify that fluxes in LR and HR are consistent.
Several lines are close to each other in wavelength
and blended spectrally in the LR mode (\htwo12.28\,\mum
and \hua\,12.36\mum, \oiv\,25.89\mum (not used) and
\feii\,25.99\mum). For those, we consider only the HR fluxes.
For convenience, we have provided a table with all the
PACS and IRS line fluxes used for the modeling as online material. 

We note that NGC\,4214 ($D=3$\,Mpc) is quite extended
and has not been observed in full in spectroscopy.
The physical properties of its two main star-forming
regions have been studied in \cite{dimaratos-2015}
using both IRS and PACS data. Here we have included those
two regions in our sample because they have (rare)
observations of the \nii 122\,\mum line as well as of the
\nii 205\,\mum line from the SPIRE FTS instrument. Those
are important constraints for the origins of \cii emission
discussed in Section~\ref{sect:disccii}.
We refer the reader to \cite{dimaratos-2015} for fluxes
and details on the modeling of the ionized and PDR
gas in this object.

Continuum measurements for our sample have been
obtained with \spit and \hers. We make use of the total infrared
luminosities (\ltir), derived by modeling the dust spectral
energy distributions, from \cite{remy-2015}. 
We also use bolometric luminosities, $L_{\rm BOL}$, as
measured from the continuum SEDs presented in
\cite{remy-2015}, extrapolating the curve in the UV/optical
range with a slope of $-0.6$ (in $\nu L_{\nu}$ versus $\lambda$)
and integrating the SED from 0.1 to 1,000\,microns.

\subsection{Properties of the main observed lines}
The standard ionic lines observed by the IRS and PACS
are relatively bright in low-$Z$ galaxies.
Their properties are summarized in Table~\ref{table:general}.
The main ionic lines used in this study to constrain models
are: \arii6.99\,\mum, \ariii8.99\,\mum, \siv10.51\,\mum,
\hua12.36\,\mum, \neii12.81\,\mum, \neiii15.56\,\mum,
\siii18.71\,\mum, \ariii21.83\,\mum, \siii33.48\,\mum,
\niii57.32\,\mum, \oiii88.36\,\mum, and \nii121.9\,\mum.
These lines are found in the ionized gas only, and usually
trace rather dense \hii regions. 
The ratios of \siv/\siii, \neiii/\neii, \ariii/\arii, \niii/\nii are
indicators of the hardness and strength of the radiation field.
They are independent of elemental abundances and less
dependent on density than optical lines.
The ratios of the two \siii lines or \ariii lines measure the
electron density in the \hii regions for densities in the range
$100-10,000$\,\cm and $3,000-300,000$\,\cm respectively
as they are not very sensitive to the electron temperature
nor to extinction \citep{houck-1984,keenan-1993,osterbrock-book}.
The \hi recombination line \hua is very useful because
it is a good tracer of the total ionized gas mass. It can be
used to calibrate elemental abundances from MIR lines or
to measure the extinction in the optical by comparison to
\ha \citep[e.g.,][]{hummer-1987,lebouteiller-2008}.

Other ionic lines such as \feiii (22.93\,\mum) and \feii
(17.93\,\mum and 25.99\,\mum)
are present in the IRS spectral range but their detection
rate is lower compared to the main lines described above.
The abundance of iron in the gas phase is enhanced in
the presence of shocks. As a result, we have not used those iron
lines to constrain the models but we compared a posteriori
predictions from the models to observations in
Section~\ref{sect:mirsecond}.
Many ionic lines are also detected in the optical.
However, optical observations often do not match our
infrared spatial coverage. Optical lines are also sensitive
to extinction and generally do not probe embedded
regions as well as the infrared lines. Again, we have not
used those optical lines to constrain our models. Instead,
we compared optical line ratio predictions from the models
to observations a posteriori in Section~\ref{sect:optextinct}.

The main lines observed by the IRS and PACS tracing
the neutral ISM are: \silii34.82\,\mum, \oi63.18\,\mum,
\oi145.5\,\mum, and \cii157.7\,\mum.
The ionization potentials of Si$^0$ and C$^0$ being below
that of hydrogen, both \silii and \cii can be found outside of
\hii regions, in the neutral phase.  
\oi is only found in neutral gas and usually arises from warm
dense regions. \oi tracks the high-\av layer of PDRs
better than \cii. This is because the transition of C$^+$ into
C$^0$ ($\sim$1-3\,mag) occurs before the formation of CO
occurs ($\sim$5\,mag), and because at high-\av, 
almost all the carbon is converted into
molecules while some oxygen remains in atomic form.
The ratio of the two \oi lines is an indicator of the gas
temperature for T$\simeq$300\,K. It is a density tracer
for high temperatures and high densities \citep{hollenbach-1999}.
In our modeling, the main constraints for the PDR are the \cii,
\oi lines and the infrared continuum radiation.
We have not used \silii as a constraint since its abundance and
emission can be linked to shocks, but we compared \silii
observations to model predictions a posteriori.
Mid-IR \htwo rotational lines are also detected in nine
of our galaxies. However, models are very sensitive to the
treatment of \htwo and shocks can also contribute to this
emission. Hence, we also compared \htwo observations to
model predictions a posteriori (see Section~\ref{sect:mirsecond}).

\begin{center}
\begin{table}[!t]\small
  \caption{General properties of the main MIR and FIR used in this work.} 
  \hfill{}
\begin{tabular}{lcccc}
    \hline     \hline
     \vspace{-8pt}\\
     {Species~~~~~} & $\lambda$  & $E$   & $T_{\rm ex}$  & $n_{\rm crit}$ \\
        & ($\mu$m)      & (eV)  & (K)   & (cm$^{-3}$) \\
     \vspace{-8pt}\\
    \hline 
     \multicolumn{5}{c}{Used to constrain the models}\\
     \hline
     \vspace{-8pt}\\
     \neiii     & 15.56 & 40.96 & 925   & 3$\times$10$^5$\,[e] \\
     \oiii      & 88.36 & 35.12 & 163   & 2$\times$10$^3$\,[e] \\
     \siv       & 10.51 & 34.79 & 1\,369        & 4$\times$10$^4$\,[e] \\
     \niii      & 57.32 & 29.60 & 251   & 1$\times$10$^3$\,[e] \\
     \ariii     & 8.99  & 27.63 & 1\,600        & 3$\times$10$^5$\,[e] \\
     \ariii     & 21.83 & 27.63 & 2\,259        & 5$\times$10$^4$\,[e] \\
     \siii      & 18.71 & 23.34 & 1\,199        & 2$\times$10$^4$\,[e] \\
     \siii      & 33.48 & 23.34 & 430   & 5$\times$10$^3$\,[e] \\
     \neii      & 12.81 & 21.56 & 1\,123        & 6$\times$10$^5$\,[e] \\
     \arii      & 6.99  & 15.76 & 2\,060        & 4$\times$10$^5$\,[e] \\
     \nii       & 121.9 & 14.53 & 188   & 3$\times$10$^2$\,[e] \\
     \hua       & 12.36 & 13.60 & 1\,163        & ... \\
     \cii       & 157.7 & 11.26 & 91            & 50\,[e], 3$\times$10$^3$\,[H] \\
     \oi        & 63.18 & ...           & 228   & 2$\times$10$^5$\,[H] \\
     \oi        & 145.5 & ...           & 327   & 4$\times$10$^4$\,[H] \\
     \ltir       & 3-1,000 & ...                & ...           & ...   \\
     \vspace{-8pt}\\
     \hline
     \multicolumn{5}{c}{Compared a posteriori to models}\\
     \hline
     \vspace{-8pt}\\
     \feiii     & 22.93 & 16.19 & 628   & 1$\times$10$^5$\,[e] \\
     \silii     & 34.82 & 8.15  & 413   & 2$\times$10$^3$\,[e], 4$\times$10$^5$\,[H] \\
     \feii      & 25.99 & 7.90  & 554   & 3$\times$10$^4$\,[e] \\ 
     \feii      & 17.93 & 7.90  & 3\,496        & 7$\times$10$^4$\,[e] \\ 
     \htwo S(0)   & 28.21       & ...   & 510   & 1$\times$10$^3$\,[H] \\
     \htwo S(1)   & 17.03       & ...   & 1\,015        & 4$\times$10$^4$\,[H] \\
     \htwo S(2)   & 12.28       & ...   & 1\,682        & 1$\times$10$^6$\,[H] \\
     \htwo S(3)   & 9.66        & ...   & 2\,504        & 2$\times$10$^6$\,[H] \\
     $L_{\rm BOL}$ & 0.1-1,000 & ...    & ...   & ...    \\
    \hline     \hline
  \end{tabular}
  \hfill{}
  \tablefoot{
Columns: 
Wavelength of the transition, energy to remove electrons to reach this state, 
excitation temperature required to populate the transition level from the ground state, 
critical density with electrons [e] (for $T=10,000\,K$) or hydrogen atoms [H] (for $T=100\,K$)
used in Cloudy. We note that collisions of \feii with H atoms are not considered in this version of Cloudy.
}
\label{table:general}
\end{table}
\end{center}

\section{Methodology}
\label{sect:modeling}
\hers and \spit spectral data provide powerful diagnostics
on the ISM physical conditions. The observables that we
have in hand trace various conditions and phases of the ISM,
including diffuse ionized or neutral gas, compact \hii regions,
and dense PDRs. In order to interpret them correctly, a
self-consistent multiphase modeling is required. In particular,
we are interested in fitting the ionic and neutral gas lines
simultaneously to account for the contribution from all phases
modeled to the line emission and to better constrain conditions
at the ionized-neutral phase boundary when few spectral lines
are observed. Our goal is to identify a parameter space
(density, ionization parameter, neutral gas covering factor)
that fits all our tracers, thus requiring one or two model components
(see Section~\ref{sect:results}).

\subsection{Grid initial conditions}
\label{sect:cloudyin}
We used the spectral synthesis code {Cloudy} v.17.00 \citep{ferland-2017} 
to model the ionized and neutral gas phases of our galaxies. 
The geometry is 1D plane-parallel. The main model input
parameters are explained below and listed in Table~\ref{table:params}.

Since the modeling results are sensitive to the gas and dust
abundances, we build five grids of models with the following
metallicity values: $Z_{\rm bin}=[0.05,0.1,0.25,0.5,1]$. 
$Z_{\rm bin} = 1$ corresponds to the default set of ISM
abundances in Cloudy, for which $12+\log({\rm O/H})=8.5$.
For the other metallicities, we scale the gas and dust abundances
by $Z_{\rm bin}$.
The abundance of several elements (C, N, Ne, S, Si, Ar, Fe, Cl)
are further scaled to measured patterns in low-metallicity galaxies
according to \cite{izotov-2006}. Table~\ref{table:params} reports
the adopted abundances of those elements.
For the dust, we use grain properties similar to those of the SMC
and include polycyclic aromatic hydrocarbons (PAHs) which
are important for neutral gas heating by photoelectric effect
\citep{abel-2008}. The dust-to-gas ratio (DGR) is $5\times10^{-3}$
for $Z_{\rm bin}=1$ and decreases linearly with metallicity. 
The abundance of PAHs is further scaled with $Z_{\rm bin}$
to the power $1.3$, as determined by \cite{remy-2015}.
The implications of these assumptions on the DGR and
PAH abundance are further discussed in Section~\ref{sect:testdust}.

 \begin{figure}[t]
\centering
\includegraphics[clip,width=8.8cm]{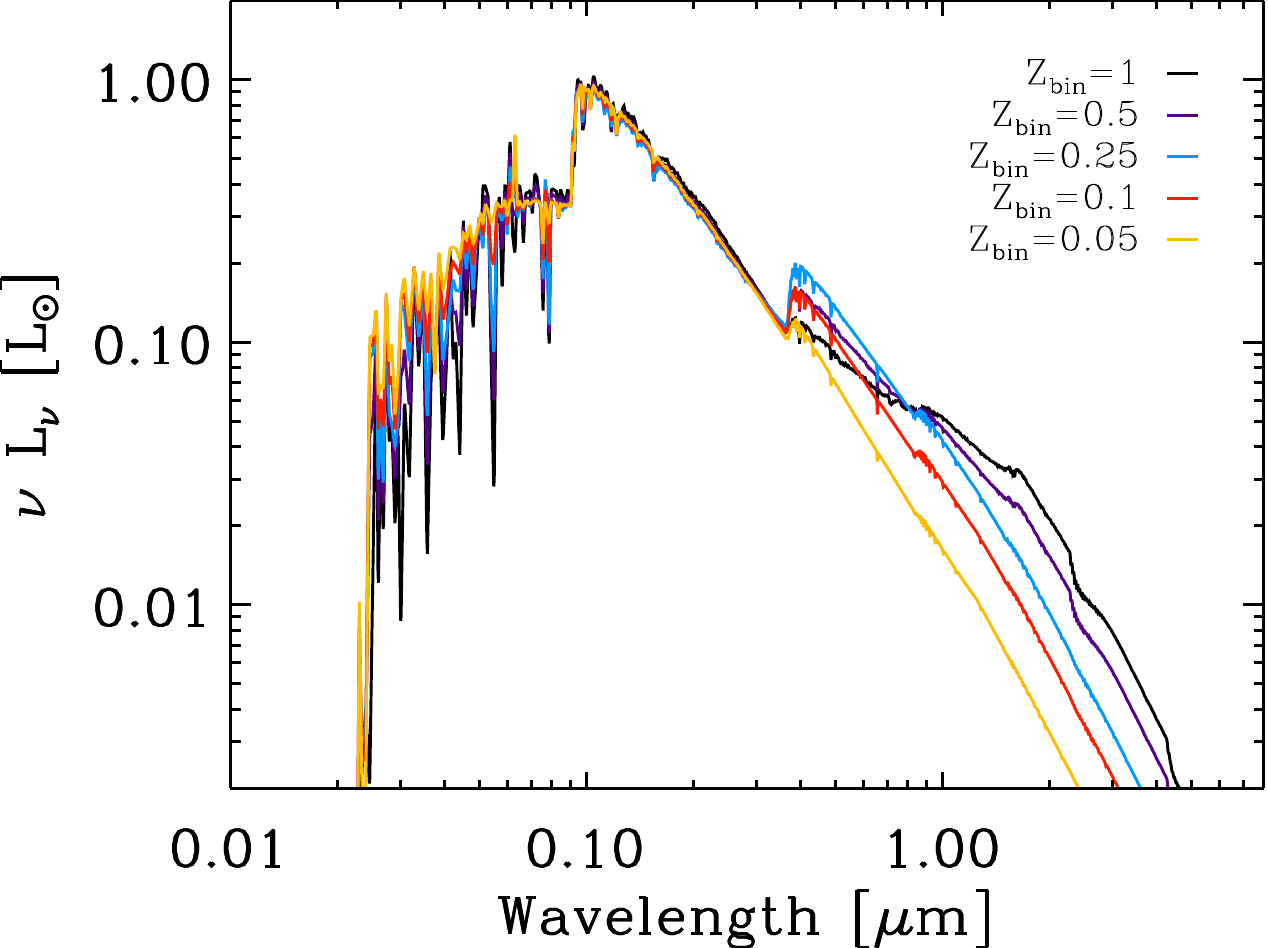}
\caption{
Spectra from Starburst99 used as input for Cloudy.
Spectra are normalized and shown for each metallicity bin
value of our grid. There are more hard photons at low metallicity.
}
\label{fig:gridsb99}
\end{figure}
 \begin{figure}[t]
\centering
\includegraphics[clip,width=8.8cm]{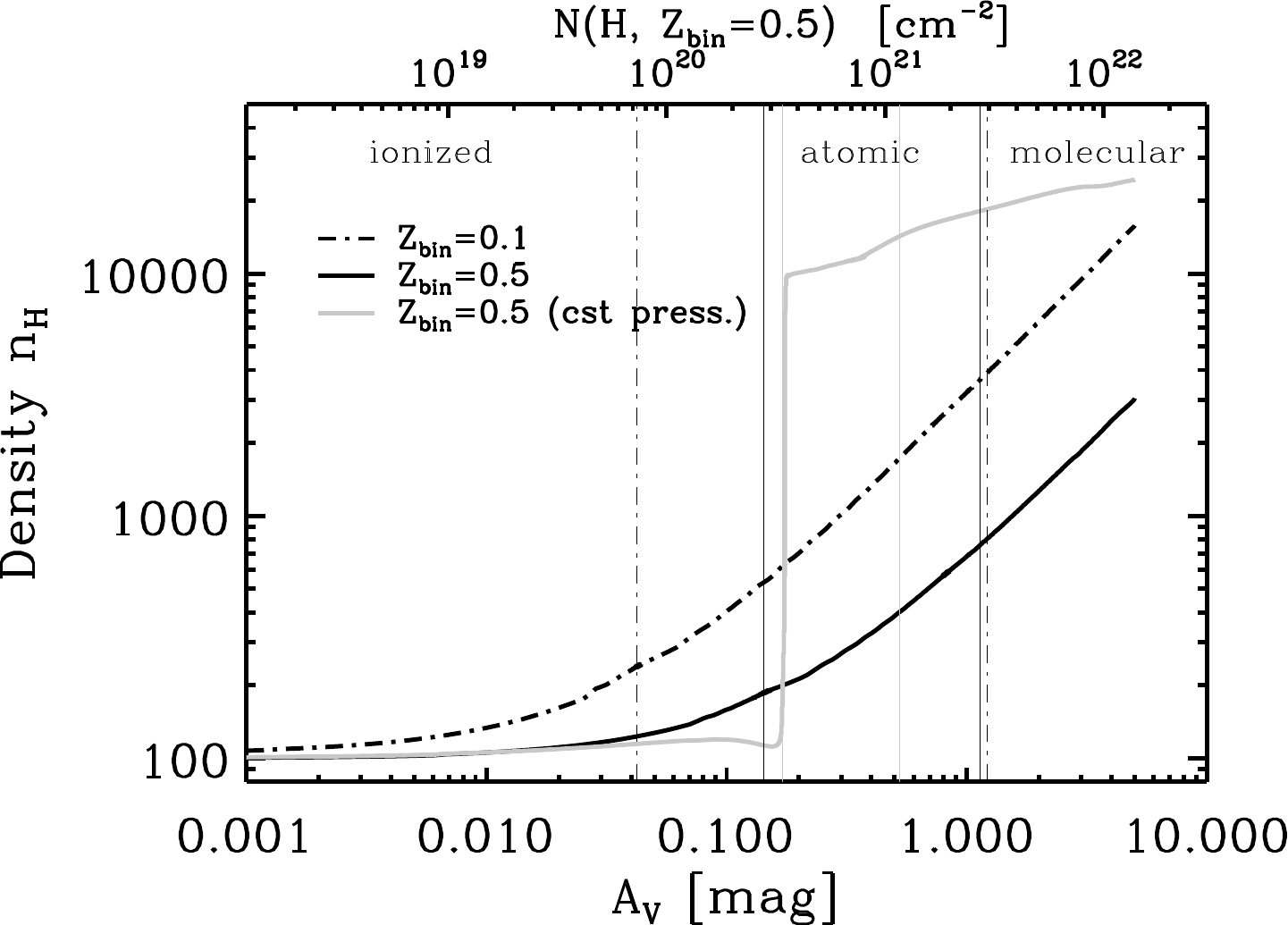}
\caption{
Adopted hydrogen density profile in the default grid of models
as a function of visual extinction, shown for $Z_{\rm bin}=0.5$
(solid line) and $Z_{\rm bin}=0.1$ (dot-dash line).
We also show the density profile for the constant pressure
test performed in section~\ref{sect:denslaw} (gray solid line).
Model parameters are set to
$n_{\rm H}=10^2$\,cm$^{-3}$ and $\log U = -2$.
The phase transitions from H$^+$-dominated to
\hi-dominated and from \hi-dominated to \htwo-dominated
are indicated with vertical lines. We note that \av
of 1\,mag corresponds to a hydrogen column density
that is $5 \times$ larger at $Z_{\rm bin}=0.1$ than at
$Z_{\rm bin}=0.5$.
}
\label{fig:griddens}
\end{figure}

The source of radiation is chosen as a young starburst that
we simulated with Starburst99 \citep{leitherer-2010}. 
We opted for a continuous star formation spectral energy
distribution (SED) computed out to an age of 10\,Myr.
The stars were selected from the standard Geneva tracks
(without rotation) and distributed on a Kroupa IMF (slope $-1.3$
between 0.1 and 0.5\,\msun and $-2.3$ between 0.5 and 100\,\msun;
\citealt{kroupa-2001}).
The Starburst99 spectra are shown in Figure~\ref{fig:gridsb99}
for the different $Z_{\rm bin}$ values.
We also added a small soft X-ray component (blackbody peaking
at $\simeq$1.5\,keV) to the input radiation field and adopt
a cosmic ray rate of $6\times10^{-16}$\,s$^{-1}$ (see Section
~\ref{sect:xrcr} for a justification of those parameters).
We adopted a density profile that is roughly constant in the
\hii region and increases linearly with the hydrogen column
density in the neutral gas (see Fig.~\ref{fig:griddens}, and
Table~\ref{table:params}).This density profile is an intermediate
choice between the more extreme, common assumptions
of a constant density or of constant pressure (that we test
in Section~\ref{sect:discparams}).
In the following, unless stated otherwise, $n_{\rm H}$
refers to the initial density. It is the density in the \hii region.
All models were stopped at an \av of 5\,mag to ensure
going deep enough in the cloud for complete PDR predictions.
We note that this \av corresponds to different $N({\rm H})$
values for different $Z$ values.
Stopping the models at a fixed $N({\rm H})$ value instead
of a fixed \av value would not change our results as long as
$N({\rm H})$ is chosen large enough so that the PDR
lines (\cii, \oi) are fully produced.
The transitions from the ionized to the atomic and from the
atomic to the molecular regions are defined where the fractions
of $e^-$ and H$^0$, and H$^0$ and H$_2$ respectively,
are equal to 0.5.
We note that in environments with high ionization parameters,
radiation pressure could become dynamically important if it
dominates the other pressure terms. This could be considered
an unstable model. For our models, we have verified that
this is not the case.

We generated a grid of models for each of the five
metallicity bins by varying the following parameters:
the initial density ($n_{\rm H}$) from $10^{0.5}$
to $10^{3.5}$\,\cm, and the inner radius such that we cover
an ionization parameter ($U$) from $10^{-4}$ to $10^{-1}$.
Values of $n_{\rm H}$ and $U$ are those at the start
of the calculation.
Since the PDR conditions are set by the \hii region model
(by continuity), there is little degree of freedom on the PDR
parameters. The radiation field is set by that in the \hii
region and the density is set by the adopted density profile.
As a free parameter we chose the covering of the PDR with
respect to the \hii region which always has a covering factor
of unity.
The PDR covering factor ($cov_{\rm PDR}$) is not varied
within the grid but a posteriori, from zero to unity, because
it just corresponds to a linear scaling of the PDR intensities.
We note that with this setup, the \hii region is always
radiation bounded, in other words no ionizing radiation escapes.
In the case of a low PDR covering factor, segments of
the \hii region that are not covered by neutral gas could be
stopped at lower \av, before the ionization front
is reached (making the model matter
bounded with ionizing radiation escaping). However,
this stopping parameter can only be constrained by specific
lines tracing the ionization front (e.g., \arii, \neii), and it
introduces degeneracies with other parameters
\citep[see][for a detailed analysis with this method]{polles-2019}.
Hence we prefer not to add an additional free parameter
and to keep all models radiation bounded. For the following
discussion, we use the term porosity when referring to a
reduced covering factor of the dense neutral gas (PDR)
relative to the ionized gas as determined by the modeling.
For convenience, we have provided line luminosities
from the grid of models predicted at the ionization front
and at an \av of 5\,mag as online material.

Most of the choices for the model parameters described
above are motivated by observations or set to default
values. However, uncertainties in those choices can remain.
In Section~\ref{sect:discparams}, we change some of the
input parameters (namely the radiation field, the presence
of cosmic rays and X-rays, the turbulent velocity, the density
law, and the dust abundance) and we discuss their influence
on the results.

\begin{center}
\begin{table}[!t]\small
  \caption{List of input parameters and abundances for the model grids.} 
  \hfill{}
\begin{tabular}{ll}
    \hline     \hline
       \vspace{-8pt}\\
      \multicolumn{2}{c}{Fixed parameters} \\
      \hline
       \vspace{-8pt}\\
     {Radiation field}                  & Starburst99 continuous SF of 10\,Myr,         \\
     {}                                         & ... Geneva stellar track, scaled to $Z$, \\
     {}                                         & ... Kroupa IMF, \\
     {}                                         & ... log($L/$\lsun) = 9        \\
     {}                                         & blackbody $T=5 \times 10^6$\,K (X-ray), \\
     {}                                         & ... log($L/$\lsun) = 5.2      \\
     {}                                         & cosmic rays background 0.5 log \\
     {}                                         & CMB   \\
     {}                                         & table ISM     \\
      {$v_{\rm turb}$}                  & 1.5\,\kms     \\
      {$D/G$ mass ratio}                & $(5\times10^{-3}) \times Z$   \\
      {metal, PAH abund.}               & set to values below   \\
     {density law}                              & $n_{\rm H,~PDR} = n_{\rm H,~HII}  \times (1+N({\rm H})/[10^{21}\,{\rm cm^{-2}}])$        \\
      {stopping criterion}              & \av = 5\,mag  \\
    \hline     \hline 
         \vspace{-8pt}\\
      \multicolumn{2}{c}{Varied parameters and grid values} \\
      \hline
       \vspace{-8pt}\\
      {$\log n_{\rm H}$}                & [0.5, 1.0, 1.5, 2.0, 2.5, 3.0, 3.5]\,\cm       \\
      {$\log U$}                                & [-4, -3.5, -3, -2.5, -2, -1.5, -1, -0.5]         \\
      {$cov_{\rm PDR}$}         & [0, 0.2, 0.4, 0.6, 0.8, 1]    \\
      {$Z_{\rm bin}$}                   & [~~~0.05, ~~0.10, ~~~0.25, ~~0.50, ~~~1.00]        \\
     {$... \log {\rm O/H}$}             & [$-4.80$, $-4.50$, $-4.10$, $-3.80$, $-3.50$]        \\
     {$... \log {\rm C/O}$}             & [$-0.99$, $-0.87$, $-0.69$, $-0.56$, $-0.44$]        \\
     {$... \log {\rm N/O}$}             & [$-1.60$, $-1.60$, $-1.60$, $-1.30$, $-1.00$]        \\
     {$... \log {\rm Ne/O}$}            & [$-0.82$, $-0.79$, $-0.75$, $-0.73$, $-0.70$]        \\
     {$... \log {\rm S/O}$}             & [$-1.70$, $-1.71$, $-1.72$, $-1.73$, $-1.74$]        \\
     {$... \log {\rm Si/O}$}            & [$-2.00$, $-2.00$, $-2.00$, $-2.00$, $-2.00$]        \\
     {$... \log {\rm Ar/O}$}            & [$-2.43$, $-2.42$, $-2.41$, $-2.39$, $-2.38$]        \\
     {$... \log {\rm Fe/O}$}            & [$-1.37$, $-1.55$, $-1.79$, $-1.98$, $-2.16$]        \\
     {$... \log {\rm Cl/O}$}            & [$-3.50$, $-3.47$, $-3.42$, $-3.38$, $-3.35$]        \\
     {$... \log {\rm PAH/H}$}   & [$-9.51$, $-8.82$, $-7.90$, $-7.22$, $-6.52$] \\
      \hline \hline
  \end{tabular}
  \hfill{}
  \tablefoot{
Other metals (not shown here) have abundances
that are based on the ISM abundance pattern, and
are then scaled to the metallicity ($Z_{\rm bin}$) of the grid. 
We note that the luminosity of the different radiation field
components are the reference values used for the model
computation, but models are scaled afterwards to the
observed luminosities of each source.
}
  \label{table:params}
\end{table}
\end{center}

\subsection{Comparison to observations}
\label{sect:compareobs}
\subsubsection{Observed and adopted model abundances}
Elemental abundances have been measured from optical
spectroscopy in several of our galaxies. Their values are
reported in Table~\ref{table:ab_obs} in the appendix.
Oxygen abundances are estimated with the \cite{pt05}
calibration. We refer the reader to \cite{remy-2014}
for further details and references on abundances and to
\cite{madden-2013,devis-2017} for a comparison of
metallicities obtained with other methods in the DGS galaxies.
Since abundances in the model grid are fixed, a correction
needs to be applied when comparing models to observations
of a given galaxy. For small abundance variations, abundances,
and intensities scale linearly. For each galaxy, we select the grid
of models with metallicity (oxygen abundance) closest to
the observed value (tabulated in Tables~\ref{table:modelres1}
and \ref{table:modelres2} as $Z_{\rm bin}$). Then, for elements
with measured abundances, we multiplied the predicted line
intensities by the offset between the observed abundances
(Table~\ref{table:ab_obs}) and the adopted model
abundances (Table~\ref{table:params}). For example,
for a galaxy with $\log({\rm O/H})=-4.40$, we select the grid
in the metallicity bin $Z_{\rm bin}=0.1$ and multiply the oxygen
line intensities by $10^{-4.40+4.50}=1.26$.
Oxygen abundances are measured for all of our galaxies.
For the other elements, if not measured, their abundances
are still assumed to follow the patterns in Table~\ref{table:params},
hence we interpolate values in that table considering
the observed $\log({\rm O/H})$ value.
The abundance values quoted here can be uncertain and
vary within galaxies or according to the datasets and
methods used to measure them. For our galaxies, literature
studies indicate that those variations are on the order of
$\sim$0.2\,dex. We see similar offsets between the measured
abundances and the abundance patterns adopted in
Table~\ref{table:params}. Hence we took into account an
uncertainty from abundances that
we set to 50\% of the observed line intensities.

\subsubsection{Observed and modeled luminosities}
\label{sect:comparelum}
To compare the absolute fluxes observed to model predictions,
we needed to scale the luminosity of the model to that of the source.
We calculated a nominal scaling factor as the ratio of the average
observed luminosity to the average predicted luminosity of the
main ionic lines to fit.
In addition to reproducing the line luminosities, the successful
scaled model should also reproduce the continuum luminosity
(\ltir) because this corresponds to the amount of stellar light
reprocessed by dust in the PDR.
Contrary to \ltir which is used as a formal constraint
on the models, the bolometric luminosity of the galaxies
($L_{\rm BOL}$) is not used as a direct constraint.
In principle, the input luminosity of the models should
be equal to the bolometric luminosity. However, the correct
value of the input luminosity in case of partial PDR
coverage is difficult to assess.
By energy conservation, the input luminosity of the models
must be equal to \ltir (if $cov_{\rm PDR}=1$) or greater
than \ltir (in case of escaping photons, $cov_{\rm PDR}<1$).
It should be, at most, the bolometric luminosity of the
galaxy (i.e., all energy is associated with the model stellar
cluster). 
$L_{\rm BOL}$ is compared a posteriori to the input
luminosity of the models to identify large mismatches
that could, for example, indicate sources of heating other
than stars. We identify a limited number of such
cases (see Section~\ref{sect:bolum}).

\subsubsection{Goodness of fit}
We searched for the best-fitting models by minimizing
the $\chi^2$ defined as 
\begin{align}
   \chi ^2=\sum_{i=1}^{n_{\rm lines}}{\frac{(M_i - O_i)^2}{({\rm min}\{O_i;M_i\} \times \sigma_i)^2}} 
\end{align}
\noindent where  
$M_i$ and $O_i$ are the modeled and observed intensities, 
and $\sigma_i$ is the fractional error on the observed flux 
(uncertainty/flux) with calibration and abundance uncertainties
added in quadrature to the measured uncertainties.
A line is considered not detected when its flux is
lower than three times its measured uncertainty.
For non-detections with a 1$\sigma$ upper limit $\sigma_{ul}$,
we set $O_i$ to $M_i$ if $M_i \le 3\sigma_{ul}$ (i.e., no contribution
of the upper limit to the $\chi^2$) and we set $O_i$ to
$3\sigma_{ul}$ and $\sigma_i$ to 1 if $M_i > 3\sigma_{ul}$.
The number of free parameters in the models is three
(density, ionization parameter, and PDR covering factor).
The reduced $\chi^2$ is given by $\chi^2_{\nu} = \chi^2 / \nu$
where $\nu$ is the difference between the number of constraints
and the number of free parameters.

\subsubsection{Reliability of the derived physical parameters}
The $\chi^2$ minimization gives an idea on the models
that are closest to the observations. However, it is possible
that those models do not converge to a single value of
the free parameters that we aim to constrain.
To assess how well the free physical parameters are constrained,
we compute their probability distribution functions (PDFs). 
The marginalized PDF of a given parameter is calculated
at each fixed value of this parameter. It is taken as the sum
of the probabilities ($e^{-\chi^2 /2}$) of all the models having
this parameter set to its fixed value and all possible values
for the other parameters. The PDFs are then normalized
such that the sum of probabilities of all models in the grid
is equal to one.
We estimated uncertainties on the best-fit parameters by
considering the average and standard deviation of the values
of the free parameters taken by models with $\chi^2$ values
within a certain range of $\chi^2_{min}$.
For instance, for a number of degrees of freedom of four, and
a 1$\sigma$ confidence level (probability of 68.3\%), this range is
$\Delta\chi^2 = \chi^2 - \chi^2_{min} = 4.72$ \citep{press-1992}.

\section{Results}
\label{sect:results}
We present general grid results in Section~4.1
and results from fitting the grid models to the observations
in the following subsections.

\subsection{Grid results}
\begin{figure*}[t]
\centering
\includegraphics[clip,width=5.2cm]{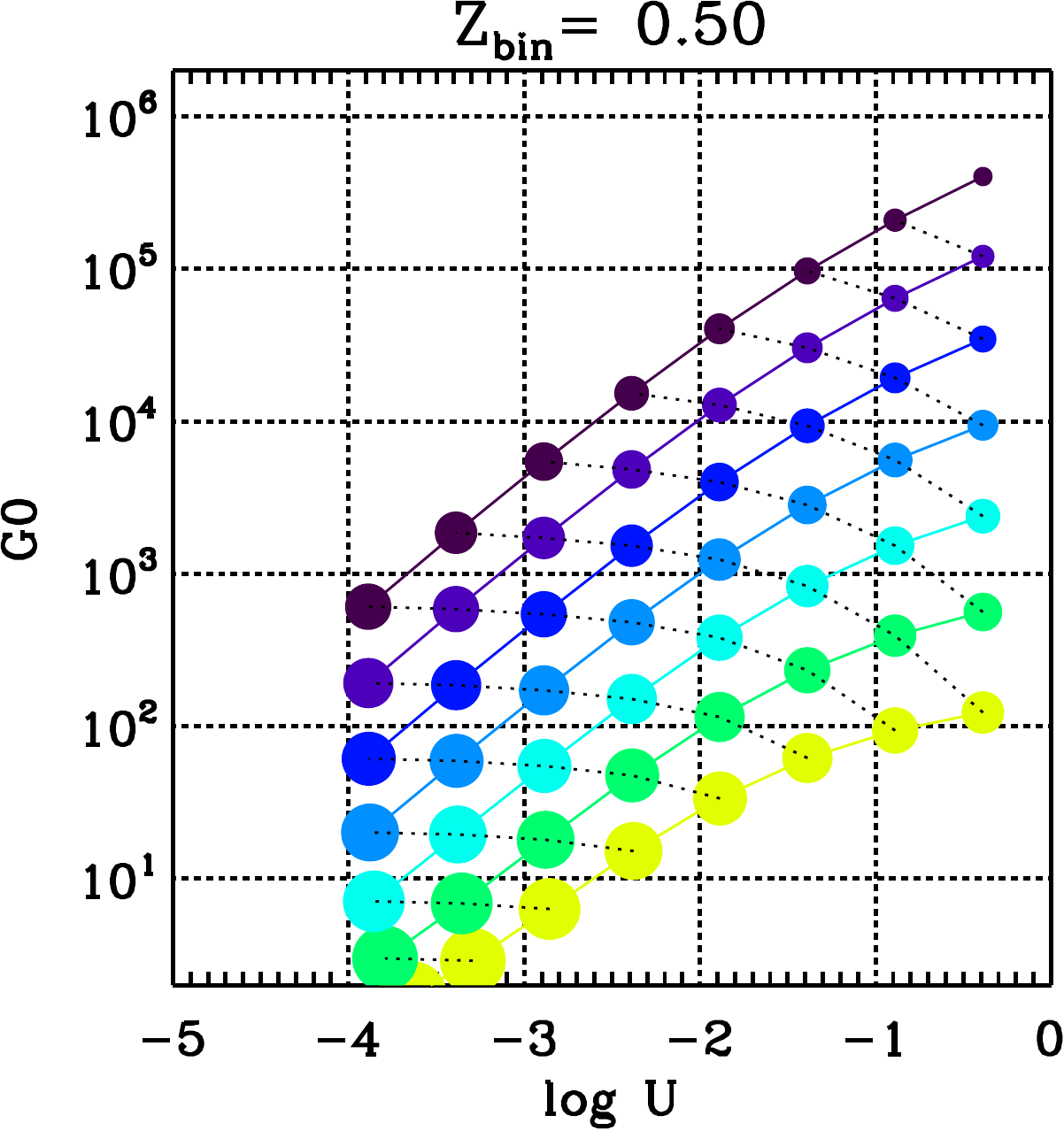}~
\includegraphics[clip,width=5.3cm]{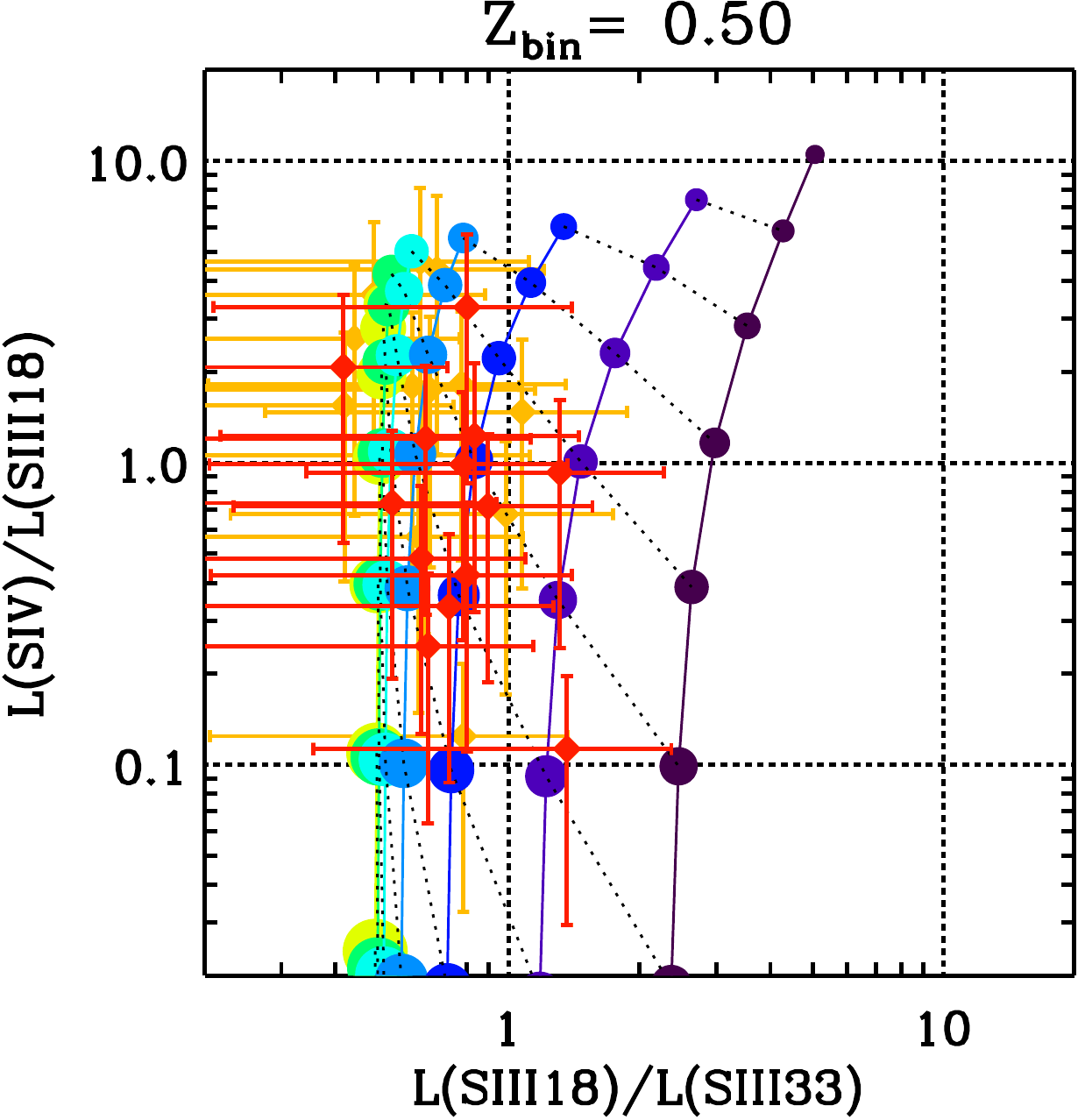}~
\includegraphics[clip,width=5.4cm]{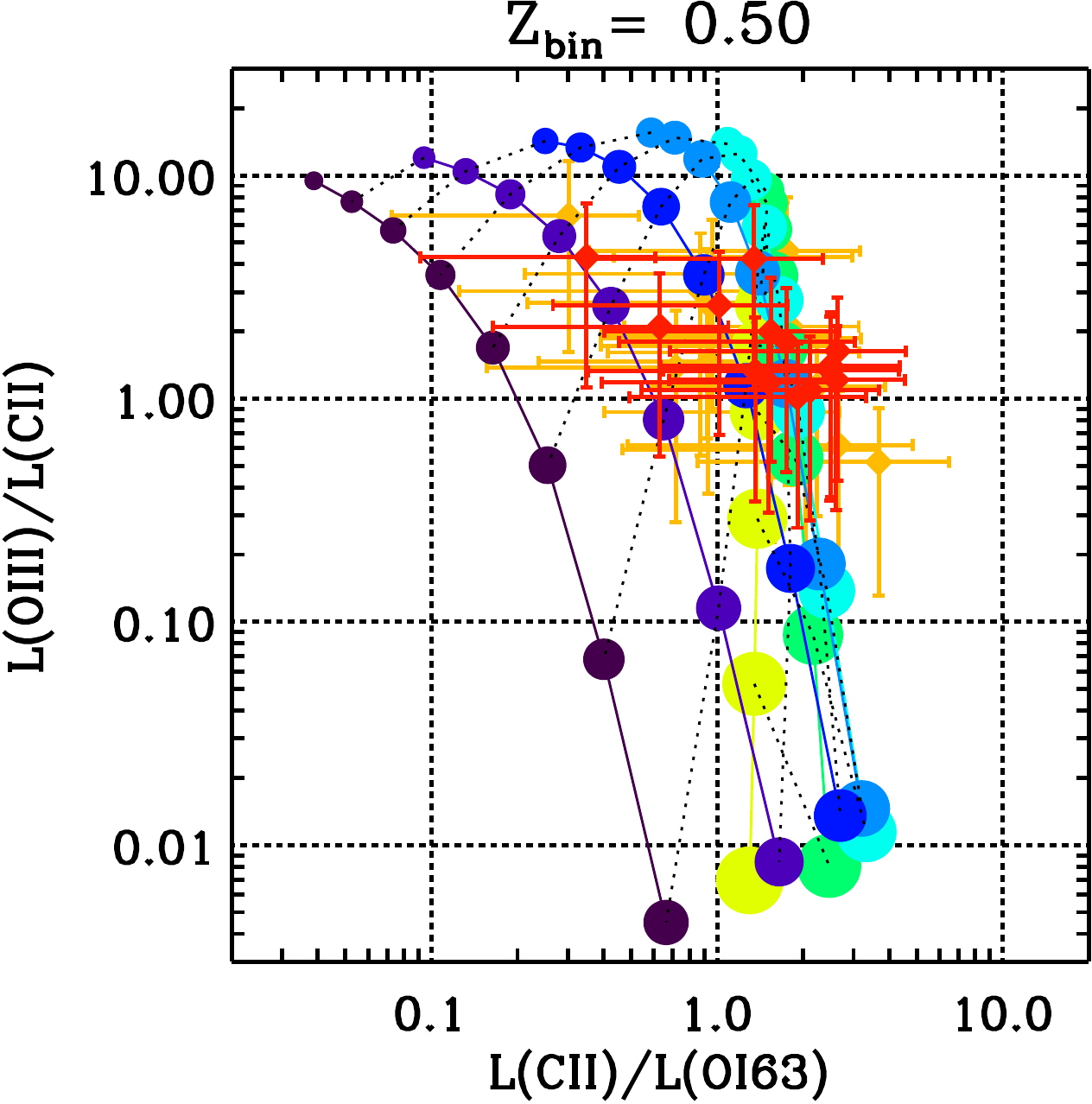}~
\vspace{3mm}
\includegraphics[clip,trim=0 -3cm 0 0,width=1.6cm]{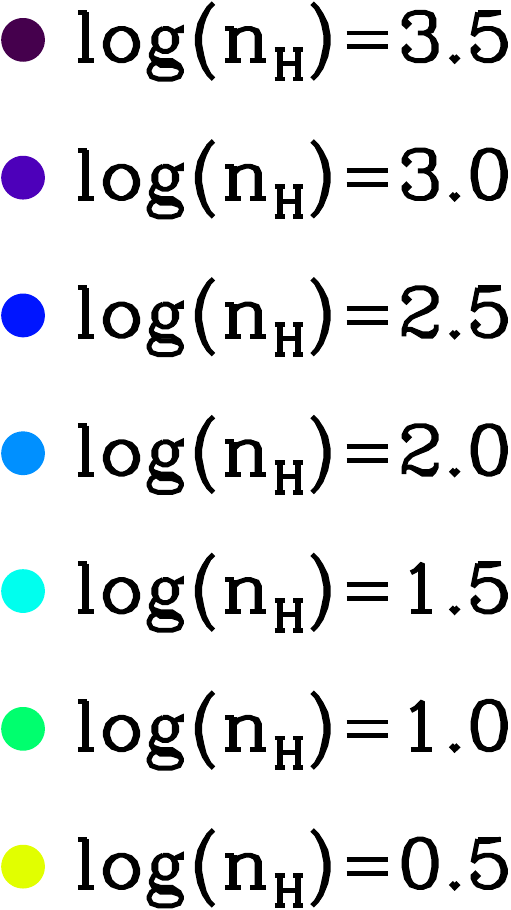}
\includegraphics[clip,width=5.2cm]{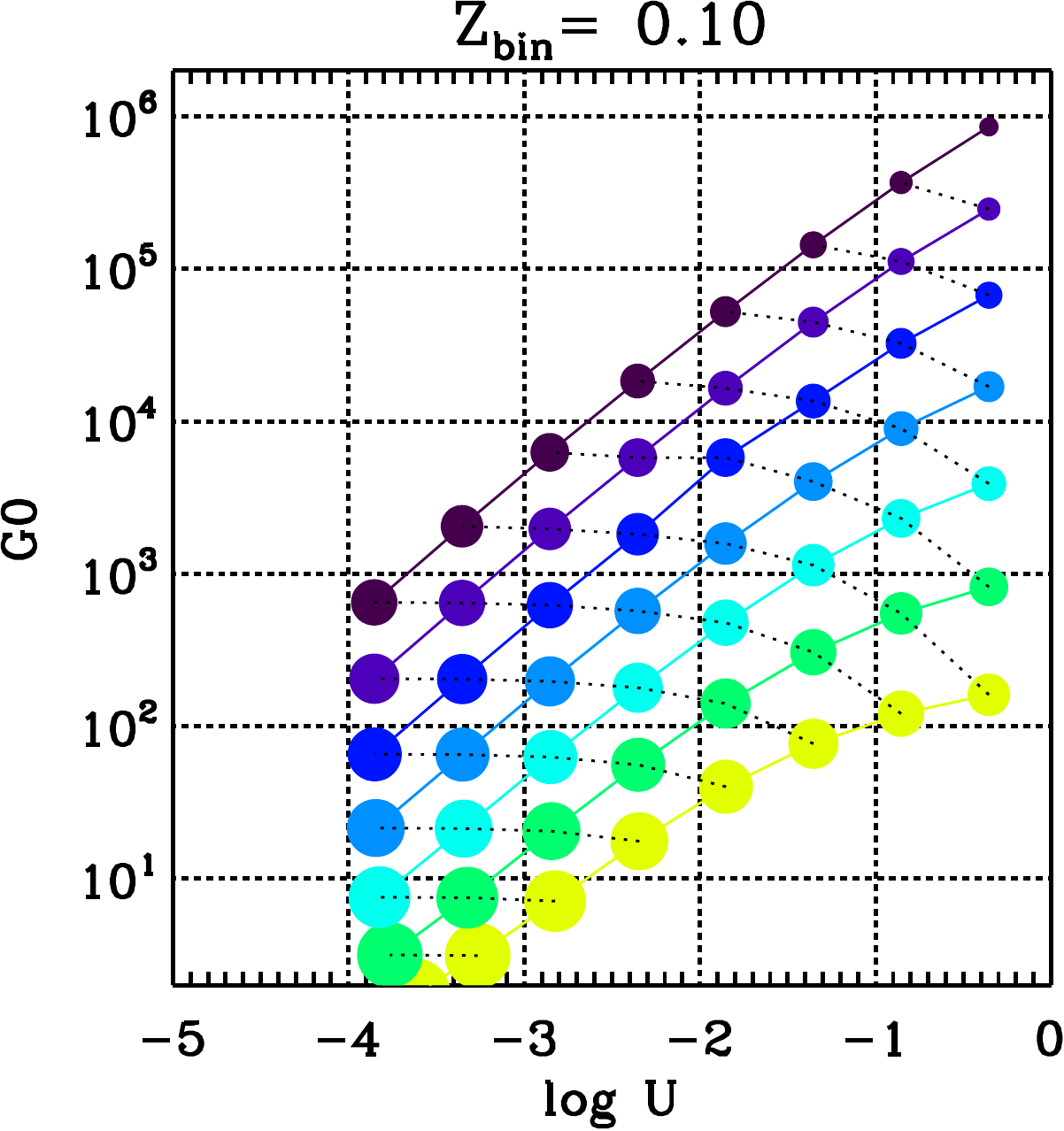}~
\includegraphics[clip,width=5.3cm]{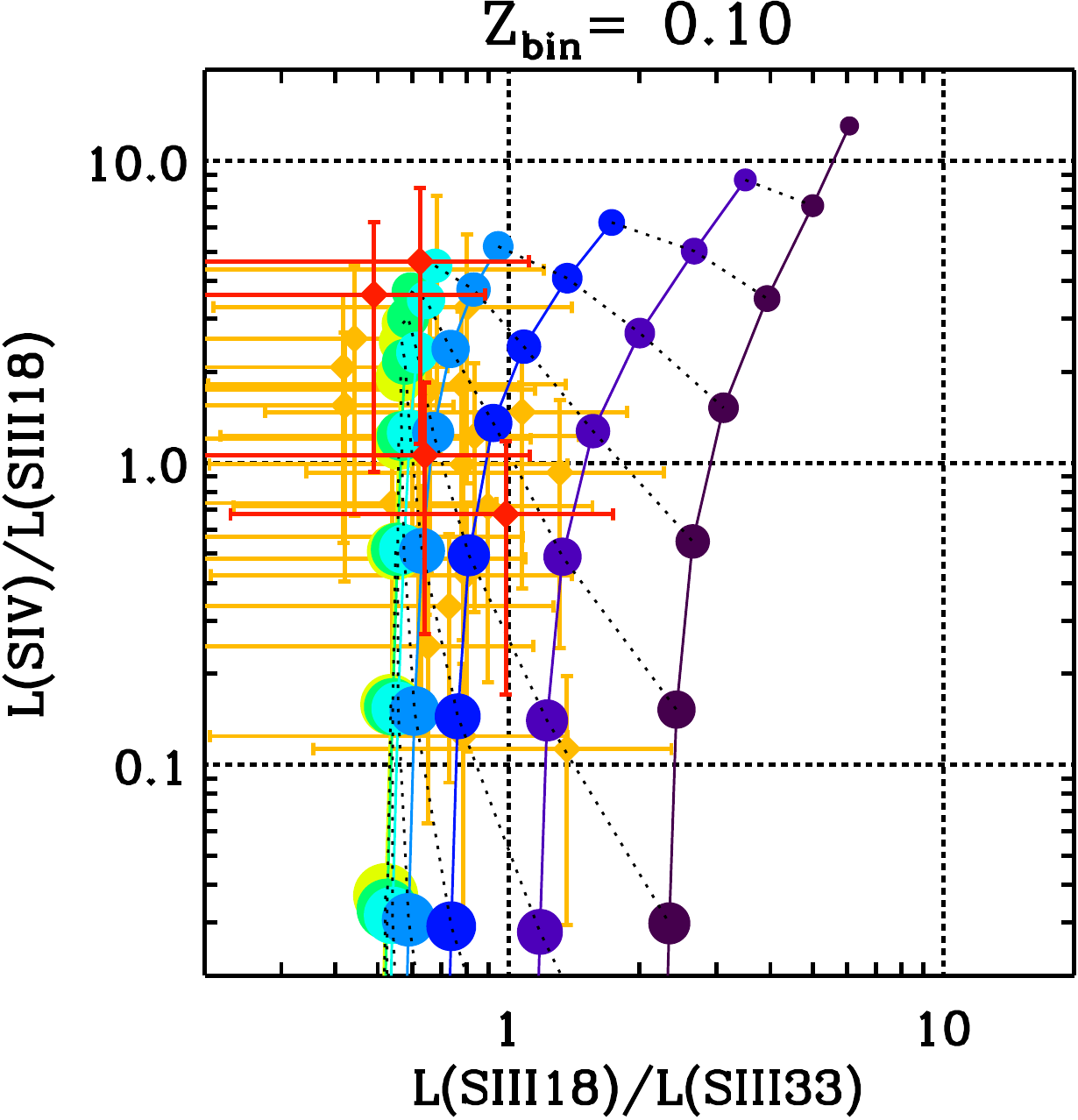}~
\includegraphics[clip,width=5.4cm]{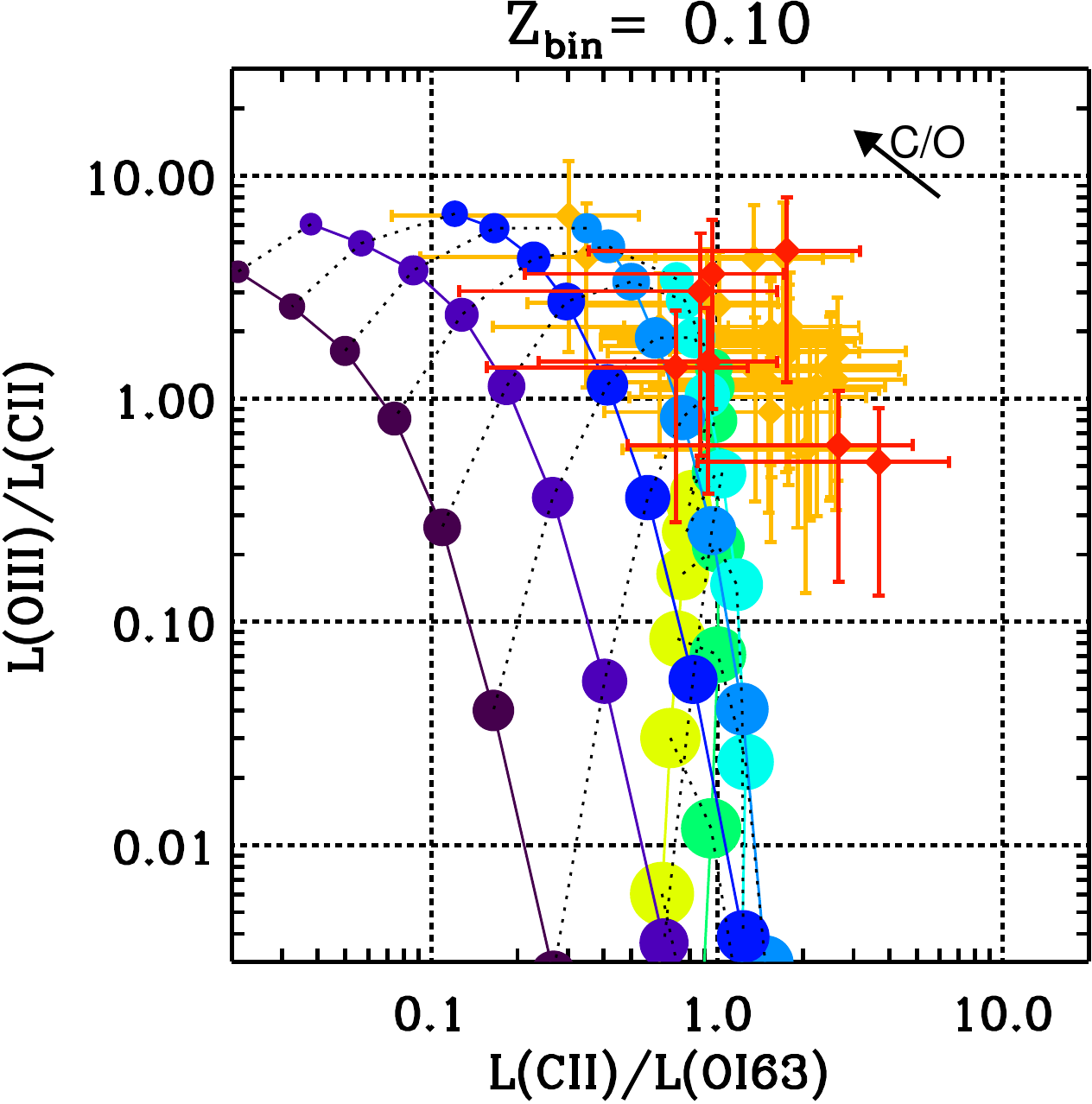}~
\hspace{1.6cm}
\caption{
Grid results: predictions of selected line intensities and model parameters,
for $Z_{\rm bin}=0.5$ (top panels) and $Z_{\rm bin}=0.1$ (bottom panels).
Color scheme: increasing initial density from $10^{0.5}$\,\cm (light green)
to $10^{3.5}$\,\cm (dark purple).
Symbol size: large symbols correspond to small inner radius (large $U$)
and small symbols to large inner radius (small $U$).
Dotted lines connect models with the same inner radius,
solid lines connect models with the same density.
All observations are overplotted in orange and galaxies
in the metallicity bin plotted are highlighted in red.
We note that the error bars are large because they include
abundance uncertainties (assumed to be 50\% of the flux).
The PDR covering factor is set to unity.
The top right panel gives a quantitative indication on how
the C/O abundance ratio affects the grid predictions from $Z_{\rm bin}=0.5$
to $Z_{\rm bin}=0.1$. Other effects due to metallicity (e.g.,
stellar spectra, column density, etc.) are discussed in the text.
}
\label{fig:gridres}
\end{figure*}

%
Figure~\ref{fig:gridres} shows the parameter space covered
by the models and the observations for $Z_{\rm bin}=0.1$
and $Z_{\rm bin}=0.5$.
In the left panels, we see the values of $G_0$ produced
when varying the ionization parameter and the density,
which are input parameters of the model grid. We remind
the reader that $G_0$ is the intensity of the ultraviolet
radiation field at the PDR front, given in units of the background
\cite{habing-1968} value, $1.6\times10^{-3}$\,erg\,cm$^{-2}$\,s$^{-1}$,
as defined by \cite{tielens-1985}. For a given density, $U$ and
$G_0$ scale linearly (except at the high-$U$, low-$n_{\rm H}$
end due to geometric dilution of the field; see e.g., \citealt{cormier-2015}).
The middle panels of Fig.~\ref{fig:gridres} show diagnostics
of the \hii region. The \siii line ratio traces intermediate
densities ($n_{\rm H}\simeq10^2-10^{3.5}$\,\cm), independently
of $Z$. Model predictions for the \siv/\siii line ratio are
slightly shifted up at $Z_{\rm bin}=0.1$ compared to
$Z_{\rm bin}=0.5$, especially for low-$U$ values, because
the starlight radiation field is harder. 
Observations span values between $0.1$ and $10$
for the \sivl/\siiila line ratio and between $0.4$ and $1.2$
for the \siiila/\siiilb line ratio. Both grids at $Z_{\rm bin}=0.1$
and $Z_{\rm bin}=0.5$ cover the observations quite well.
Observations are rather compatible with models of low/
intermediate-densities and high-$U$ values.
The right panels of Fig.~\ref{fig:gridres} show a diagnostic
linking the \hii region and the PDR. We see that predictions
of the \oiii/\cii and \cii/\oi ratios vary significantly with $Z$.
The main effects at work are the C/O abundance
ratio that decreases by a factor of two from
$Z_{\rm bin}=0.5$ to $Z_{\rm bin}=0.1$, the
harder stellar radiation field at $Z_{\rm bin}=0.1$
making more \oiii, and the fact that the PDR structure
is very sensitive to $Z$. In particular, with decreasing
$Z$, the PDR lines are formed up to a larger
column density and their emission is boosted
relative to the ionized gas metal lines
(i.e., predictions going down in 
Fig.~\ref{fig:gridres}, right panels).
Observations span values between $0.3$ and $10$
for the \oiiilb/\ciil line ratio and between $0.3$ and $5$
for the \ciil/\oila line ratio. At $Z_{\rm bin}=0.5$, models
cover quite well the observed ranges, while at $Z_{\rm bin}=0.1$,
we notice that model predictions are at the edge of the
observed ratio ranges. This suggests that a combination
of models and/or low covering factor of the PDR may
be needed to match better the observations.

We remind the reader that normal, star-forming, metal-rich
galaxies have similar \siiila/\siiilb and \ciil/\oila ratio values
than the DGS galaxies, but their \sivl/\siiila and \oiiilb/\ciil
values are respectively about seven times and four times
lower \citep{brauher-2008,dale-2009,cormier-2015}.
Those values are covered by our grid of models and
compatible with models of low or intermediate-densities
and intermediate-$U$ values (for $Z_{\rm bin}=1$).

\subsection{Best-fitting single \hii+PDR models}
Parameters of the best-fitting single models as well as
their corresponding $\chi^2$ and reduced $\chi^2$
($\chi^2_{\nu}$) are indicated in Tables~\ref{table:modelres1}
and \ref{table:modelres2}. The number of detected lines
in each galaxy varies from three (model under-constrained)
to $16$. 
Figure~\ref{fig:bestparams} shows how the observed
and predicted line intensities compare for the best-fitting
model of He\,2-10, as well as the PDFs of $n_{\rm H}$,
$\log(U)$, and $cov_{\rm PDR}$. Figures for the other
galaxies of the sample are shown in
Appendix~\ref{sect:append-d}.
The ionization parameters vary between $-2.9$ and $-0.3$
with median value $-2.4$, and the PDR covering factor
varies from 0.2 to one with a median value of 0.4.
Densities of the best-fitting models vary from $10^{0.5}$
to $10^{3.0}$\,\cm, with median at $10^{2.0}$\,\cm.
The medians are very close to the values $\log(U)=-2.5$
and $n_{\rm H}=10^{2.0}$\,\cm found in \cite{cormier-2015}
from average line ratios of the DGS sample.
The classical MIR density tracers (\siii and \ariii line ratios)
are not particularly sensitive to those low densities,
and although we can recover such densities by modeling
a suite of lines, future determinations of electron densities
in dwarf galaxies would benefit from FIR observations of
line doublets such as the \oiiila and \oiiilb or \niila and \niilb
lines.

About half of the galaxies (16/39) have minimum $\chi^2_{\nu}$
below two and their observed lines are satisfyingly
reproduced by a single model. 
Inspection of the results indicates that galaxies with larger
$\chi^2_{\nu}$ usually have one or more lines for which
predictions are off by a factor of more than two. 
The lines that are most difficult to reproduce simultaneously 
with the other lines are \siv and \neii.

\begin{center}
\begin{table*}[!t]\scriptsize
  \caption{Results of the \hii+PDR best-fitting models for the DGS galaxies where single models are preferred.} 
  \hfill{}
\begin{tabular}{lll|ccccc|cccccc}
    \hline\hline
     \vspace{-8pt}\\
     \multicolumn{3}{l|}{} &
     \multicolumn{5}{c|}{1-component model} &
     \multicolumn{6}{c}{2-component model} \\ \cline{4-8} \cline{9-14}
    \multicolumn{1}{l}{Galaxy\hspace{7mm}\hfill{}} & 
    \multicolumn{1}{l}{$Z_{\rm bin}$} & 
    \multicolumn{1}{l|}{$N_{\rm c}^{(a)}$} & 
    \multicolumn{1}{c}{log\,$n_{\rm H,HII}^{(b)}$} &
    \multicolumn{1}{c}{log\,$n_{\rm H,PDR}^{(c)}$} &
    \multicolumn{1}{c}{log\,$U$} &
    \multicolumn{1}{c}{$cov_{\rm PDR}$} &
    \multicolumn{1}{c|}{$\chi^2_{\rm min}~[\chi^2_{\nu, {\rm min}}]$} &
    \multicolumn{1}{c}{log\,$n_{\rm H,HII}$} &
    \multicolumn{1}{c}{log\,$n_{\rm H,PDR}$} &
    \multicolumn{1}{c}{log\,$U$} &
    \multicolumn{1}{c}{$cov_{\rm PDR}$} &
    \multicolumn{1}{c}{$f_c^{(d)}$} &
    \multicolumn{1}{c}{$\chi^2_{\rm min}~[\chi^2_{\nu, {\rm min}}]$} \\
        \vspace{-6pt}\\
    \hline
\vspace{-8pt} \\ NGC1705\dotfill{} & 0.50 & 8 & $\bf 1.00^{1.87}_{0.93}$ & $\bf 2.02^{2.59}_{2.00}$ & $\bf -2.39^{-2.33}_{-2.84}$ & $\bf 0.60^{0.84}_{0.36}$ & $\bf 8.41~[1.68]$ & $1.00$ & $2.02$ & $-2.39$ & $1.00$ & $0.50$ & $6.85~[2.28]$ \\  & & & & & & & & $1.00$ & $ - $ & $-2.88$ & $0.00$ & $0.50$ \\ 
\vspace{-8pt} \\ IIZw40\dotfill{} & 0.50 & 13 & $\bf 2.50^{2.62}_{2.04}$ & $\bf 3.41^{3.53}_{2.93}$ & $\bf -1.89^{-1.89}_{-1.89}$ & $\bf 0.60^{0.78}_{0.55}$ & $\bf 19.33~[1.93]$ & $2.50$ & $3.41$ & $-1.89$ & $0.80$ & $0.80$ & $18.41~[2.30]$ \\  & & & & & & & & $2.50$ & $ - $ & $-1.89$ & $0.00$ & $0.20$ \\ 
\vspace{-8pt} \\ Mrk930\dotfill{} & 0.25 & 12 & $\bf 2.50^{2.62}_{2.04}$ & $\bf 3.60^{3.73}_{3.13}$ & $\bf -2.38^{-2.38}_{-2.38}$ & $\bf 0.60^{0.78}_{0.55}$ & $\bf 17.41~[1.93]$ & $0.50$ & $2.28$ & $-0.38$ & $0.60$ & $0.60$ & $5.77~[0.82]$ \\  & & & & & & & & $0.50$ & $ - $ & $-3.29$ & $0.00$ & $0.40$ \\ 
\vspace{-8pt} \\ NGC1569\dotfill{} & 0.25 & 13 & $\bf 1.50^{1.97}_{0.92}$ & $\bf 2.63^{3.01}_{2.38}$ & $\bf -2.38^{-2.37}_{-2.38}$ & $\bf 0.40^{0.71}_{0.26}$ & $\bf 11.43~[1.14]$ & $0.50$ & $2.28$ & $-0.38$ & $0.60$ & $0.70$ & $3.35~[0.42]$ \\  & & & & & & & & $0.50$ & $ - $ & $-2.85$ & $0.00$ & $0.30$ \\ 
\vspace{-8pt} \\ HS1330\dotfill{} & 0.25 & 5 & $\bf 0.50$ & $\bf 2.28$ & $\bf -1.38^{-0.43}_{-1.33}$ & $\bf 0.60^{0.81}_{0.59}$ & $\bf 2.38~[1.19]$ & $0.50$ & $2.28$ & $-0.88$ & $0.80$ & $0.80$ & $1.97~[-]$ \\  & & & & & & & & $0.50$ & $ - $ & $-3.63$ & $0.00$ & $0.20$ \\ 
\vspace{-8pt} \\ HS1304\dotfill{} & 0.25 & 5 & $\bf 0.50^{1.61}$ & $\bf 2.28^{2.98}$ & $\bf -1.38^{-0.78}_{-2.12}$ & $\bf 0.40^{0.64}_{0.45}$ & $\bf 17.83~[8.91]$ & $0.50$ & $2.28$ & $-0.38$ & $0.60$ & $0.50$ & $9.72~[-]$ \\  & & & & & & & & $0.50$ & $ - $ & $-3.29$ & $0.00$ & $0.50$ \\ 
\vspace{-8pt} \\ Mrk153\dotfill{} & 0.25 & 9 & $\bf 1.50^{1.91}_{0.80}$ & $\bf 2.63^{2.97}_{2.33}$ & $\bf -2.38^{-2.12}_{-2.50}$ & $\bf 0.40^{0.56}_{0.24}$ & $\bf 8.74~[1.46]$ & $0.50$ & $2.28$ & $-0.88$ & $0.60$ & $0.60$ & $2.34~[0.59]$ \\  & & & & & & & & $0.50$ & $ - $ & $-2.85$ & $0.00$ & $0.40$ \\ 
\vspace{-8pt} \\ UM133\dotfill{} & 0.25 & 4 & $\bf 1.00^{1.60}_{0.90}$ & $\bf 2.42^{2.57}_{2.40}$ & $\bf -2.87^{-2.87}_{-2.88}$ & $\bf 0.40^{0.64}_{0.36}$ & $\bf 0.37~[0.37]$ & $3.00$ & $3.83$ & $-3.38$ & $0.40$ & $0.10$ & $0.03~[-]$ \\  & & & & & & & & $1.00$ & $ - $ & $-3.79$ & $0.00$ & $0.90$ \\ 
\vspace{-8pt} \\ HS1319\dotfill{} & 0.25 & 3 & $\bf 0.50^{3.50}$ & $\bf 2.28^{5.05}$ & $\bf -1.88^{-0.38}_{-3.88}$ & $\bf 0.20^{1.00}_{0.00}$ & $\bf 0.30~[-]$ & $2.00$ & $3.28$ & $-1.38$ & $0.20$ & $0.30$ & $<0.01~[-]$ \\  & & & & & & & & $1.00$ & $ - $ & $-3.79$ & $0.00$ & $0.70$ \\ 
\vspace{-8pt} \\ HS1222\dotfill{} & 0.25 & 4 & $\bf 0.50^{1.04}$ & $\bf 2.28^{2.44}$ & $\bf -0.38^{-0.47}_{-1.29}$ & $\bf 0.20^{0.42}_{0.18}$ & $\bf 1.32~[1.32]$ & $2.50$ & $3.72$ & $-1.88$ & $0.20$ & $0.80$ & $0.39~[-]$ \\  & & & & & & & & $0.50$ & $ - $ & $-3.63$ & $0.00$ & $0.20$ \\ 
\vspace{-8pt} \\ Mrk209\dotfill{} & 0.10 & 8 & $\bf 1.00^{1.48}_{0.95}$ & $\bf 2.86^{3.14}_{2.83}$ & $\bf -0.85^{-0.44}_{-1.55}$ & $\bf 0.20^{0.39}_{0.18}$ & $\bf 3.67~[0.73]$ & $1.00$ & $2.87$ & $-0.35$ & $1.00$ & $0.30$ & $2.55~[0.85]$ \\  & & & & & & & & $0.50$ & $ - $ & $-0.85$ & $0.00$ & $0.70$ \\ 
\vspace{-8pt} \\ Pox186\dotfill{} & 0.10 & 7 & $\bf 2.00^{2.45}_{1.55}$ & $\bf 3.66^{4.11}_{3.27}$ & $\bf -1.35^{-0.58}_{-1.63}$ & $\bf 0.20$ & $\bf 1.47~[0.37]$ & $2.00$ & $3.73$ & $-0.85$ & $0.20$ & $0.90$ & $0.93~[0.46]$ \\  & & & & & & & & $1.00$ & $ - $ & $-3.76$ & $0.00$ & $0.10$ \\ 
\vspace{-8pt} \\ SBS1249\dotfill{} & 0.10 & 4 & $\bf 1.00^{1.46}_{0.88}$ & $\bf 2.86^{3.13}_{2.79}$ & $\bf -0.85^{-0.35}_{-1.35}$ & $\bf 0.20$ & $\bf 0.52~[0.52]$ & $1.00$ & $2.87$ & $-0.35$ & $0.20$ & $0.60$ & $0.02~[-]$ \\  & & & & & & & & $0.50$ & $ - $ & $-3.26$ & $0.00$ & $0.40$ \\ 
\vspace{-8pt} \\ HS0017\dotfill{} & 0.10 & 3 & $\bf 2.00^{3.50}_{0.50}$ & $\bf 3.57^{5.44}_{2.67}$ & $\bf -1.85^{-0.35}_{-3.85}$ & $\bf 0.80^{1.00}_{0.00}$ & $\bf 0.03~[-]$ & $2.50$ & $3.98$ & $-2.35$ & $1.00$ & $0.30$ & $<0.01~[-]$ \\  & & & & & & & & $1.50$ & $ - $ & $-3.34$ & $0.00$ & $0.70$ \\ 
\vspace{-8pt} \\ HS1442\dotfill{} & 0.10 & 4 & $\bf 1.00^{1.52}_{0.98}$ & $\bf 2.87^{3.23}_{2.85}$ & $\bf -0.35^{-0.35}_{-0.88}$ & $\bf 0.40^{0.98}_{0.42}$ & $\bf 0.34~[0.34]$ & $1.00$ & $2.87$ & $-0.35$ & $0.80$ & $0.60$ & $0.19~[-]$ \\  & & & & & & & & $1.00$ & $ - $ & $-0.35$ & $0.00$ & $0.40$ \\ 
\vspace{-8pt} \\ SBS1211\dotfill{} & 0.10 & 5 & $\bf 2.00^{2.10}_{1.40}$ & $\bf 3.46^{3.54}_{3.01}$ & $\bf -2.35^{-1.75}_{-2.46}$ & $\bf 0.20$ & $\bf 0.87~[0.44]$ & $1.50$ & $3.09$ & $-1.85$ & $0.60$ & $0.30$ & $0.28~[-]$ \\  & & & & & & & & $0.50$ & $ - $ & $-2.34$ & $0.00$ & $0.70$ \\ 
\vspace{-8pt} \\ SBS1415\dotfill{} & 0.10 & 8 & $\bf 2.00^{1.76}_{0.97}$ & $\bf 3.46^{3.27}_{2.81}$ & $\bf -2.35^{-1.60}_{-2.38}$ & $\bf 0.40^{0.55}_{0.25}$ & $\bf 4.47~[0.89]$ & $1.00$ & $2.87$ & $-1.85$ & $0.60$ & $0.50$ & $2.76~[0.92]$ \\  & & & & & & & & $1.00$ & $ - $ & $-2.35$ & $0.00$ & $0.50$ \\ 
\vspace{-8pt} \\ Tol1214\dotfill{} & 0.10 & 6 & $\bf 1.00^{1.38}_{0.88}$ & $\bf 2.86^{3.08}_{2.79}$ & $\bf -0.85^{-0.50}_{-1.46}$ & $\bf 0.20$ & $\bf 9.01~[3.00]$ & $1.00$ & $2.86$ & $-0.85$ & $0.20$ & $0.60$ & $7.13~[7.13]$ \\  & & & & & & & & $1.00$ & $ - $ & $-3.76$ & $0.00$ & $0.40$ \\ 
\vspace{-8pt} \\ UGC4483\dotfill{} & 0.10 & 4 & $\bf 1.50^{1.62}_{1.04}$ & $\bf 2.95^{3.02}_{2.87}$ & $\bf -2.85^{-1.85}_{-2.85}$ & $\bf 0.20^{0.45}_{0.22}$ & $\bf 2.06~[2.06]$ & $1.50$ & $3.01$ & $-2.35$ & $0.80$ & $0.10$ & $0.40~[-]$ \\  & & & & & & & & $0.50$ & $ - $ & $-3.62$ & $0.00$ & $0.90$ \\ 
\vspace{-8pt} \\ SBS1159\dotfill{} & 0.10 & 6 & $\bf 2.00^{3.50}_{0.50}$ & $\bf 3.46^{5.44}_{2.67}$ & $\bf -2.35^{-0.35}_{-3.85}$ & $\bf 0.20^{1.00}_{0.00}$ & $\bf 14.05~[4.68]$ & $1.00$ & $2.86$ & $-1.35$ & $0.40$ & $0.20$ & $4.88~[4.88]$ \\  & & & & & & & & $1.00$ & $ - $ & $-3.76$ & $0.00$ & $0.80$ \\ 
\vspace{-8pt} \\ HS0822\dotfill{} & 0.05 & 5 & $\bf 1.50^{1.96}_{1.38}$ & $\bf 3.48^{3.90}_{3.34}$ & $\bf -0.83^{-0.88}_{-1.45}$ & $\bf 0.20$ & $\bf 4.63~[2.32]$ & $1.50$ & $3.48$ & $-0.83$ & $0.20$ & $0.70$ & $4.14~[-]$ \\  & & & & & & & & $1.00$ & $ - $ & $-3.75$ & $0.00$ & $0.30$ \\ 
\vspace{-8pt} \\ SBS0335\dotfill{} & 0.05 & 9 & $\bf 2.00^{2.02}_{1.48}$ & $\bf 4.01^{4.01}_{3.46}$ & $\bf -0.83^{-0.37}_{-1.29}$ & $\bf 0.40^{0.41}_{0.19}$ & $\bf 7.42~[1.24]$ & $2.00$ & $4.01$ & $-0.83$ & $0.60$ & $0.70$ & $5.98~[1.49]$ \\  & & & & & & & & $1.00$ & $ - $ & $-1.33$ & $0.00$ & $0.30$ \\ 
\vspace{-8pt} \\ IZw18\dotfill{} & 0.05 & 8 & $\bf 1.50^{1.82}_{0.98}$ & $\bf 3.43^{3.68}_{3.12}$ & $\bf -1.33^{-0.58}_{-1.88}$ & $\bf 0.20$ & $\bf 9.88~[1.98]$ & $1.00$ & $3.18$ & $-1.33$ & $0.80$ & $0.10$ & $0.62~[0.21]$ \\  & & & & & & & & $1.00$ & $ - $ & $-1.83$ & $0.00$ & $0.90$ \\ 
    \vspace{-6pt}\\
    \hline \hline
  \end{tabular}
  \tablefoot{
    This table reports results of galaxies for which single models are preferred,
    and Table~\ref{table:modelres2} reports results for galaxies where mixed
    models are preferred. The following notes apply to both tables.
  ~$(a)$~$N_{\rm c}$ is the number of observational constraints
  (number of detected lines plus $1$ for the luminosity constraint).
  ~$(b)$~$n_{\rm H,HII}$ corresponds to the initial density.
  ~$(c)$~$n_{\rm H,PDR}$ corresponds to the density of the PDR
  at the \hi-\htwo transition (i.e., \av$\sim1$\,mag).
  ~$(d)$~$f_c$ ($1-f_c$) is the contribution of the first (second) model
  to the mixed model. We note that the values of the covering factor
  are those of the initial grid (with full coverage of the \hii region).
  For 2-component models, $cov_{\rm PDR}$ needs to be further
  multiplied by $f_c$ to obtain the effective covering factor.
  Galaxies are sorted by decreasing metallicity.
  For one-component models, uncertainties on the free parameters are
  given in upper and lower scripts and are calculated as the average 
  $\pm$ the standard deviation of models within the 1$\sigma$ 
  confidence level (see Section~\ref{sect:compareobs} for details). 
  The number of free parameters $n_{\rm free}$ is three for
  the one-component model, and five for the two-component model.
  The degrees of freedom are given by $\nu = N_{\rm c}-n_{\rm free}$.
  The reduced $\chi^2$ is given by $\chi^2_{\nu} = \chi^2 / \nu$.
  }
  \label{table:modelres1}
\end{table*}
\end{center}
\begin{center}
\begin{table*}[!t]\scriptsize
  \caption{Results of the \hii+PDR best-fitting models for the DGS galaxies where mixed models are preferred.} 
  \hfill{}
\begin{tabular}{lll|ccccc|cccccc}
    \hline\hline
     \vspace{-8pt}\\
     \multicolumn{3}{l|}{} &
     \multicolumn{5}{c|}{1-component model} &
     \multicolumn{6}{c}{2-component model} \\ \cline{4-8} \cline{9-14}
    \multicolumn{1}{l}{Galaxy\hspace{7mm}\hfill{}} & 
    \multicolumn{1}{l}{$Z_{\rm bin}$} & 
    \multicolumn{1}{l|}{$N_{\rm c}$} & 
    \multicolumn{1}{c}{log\,$n_{\rm H,HII}$} &
    \multicolumn{1}{c}{log\,$n_{\rm H,PDR}$} &
    \multicolumn{1}{c}{log\,$U$} &
    \multicolumn{1}{c}{$cov_{\rm PDR}$} &
    \multicolumn{1}{c|}{$\chi^2_{\rm min}~[\chi^2_{\nu, {\rm min}}]$} &
    \multicolumn{1}{c}{log\,$n_{\rm H,HII}$} &
    \multicolumn{1}{c}{log\,$n_{\rm H,PDR}$} &
    \multicolumn{1}{c}{log\,$U$} &
    \multicolumn{1}{c}{$cov_{\rm PDR}$} &
    \multicolumn{1}{c}{$f_c$} &
    \multicolumn{1}{c}{$\chi^2_{\rm min}~[\chi^2_{\nu, {\rm min}}]$} \\
        \vspace{-6pt}\\
    \hline
\vspace{-8pt} \\ He2-10\dotfill{} & 1.00 & 16$^{**}$ & $2.50$ & $2.88$ & $-2.90$ & $0.60^{0.96}_{0.44}$ & $74.67~[5.74]$ & $\bf 2.50$ & $\bf 2.88$ & $\bf -2.90$ & $\bf 0.60$ & $\bf 0.80$ & $\bf 21.14~[1.92]$ \\  & & & & & & & & $\bf 1.50$ & $\bf - $ & $\bf -3.87$ & $\bf 0.00$ & $\bf 0.20$ \\ 
\vspace{-8pt} \\ NGC1140\dotfill{} & 1.00 & 13$^*$ & $0.50$ & $1.55$ & $-2.39$ & $0.60^{1.00}$ & $24.67~[2.47]$ & $\bf 0.50$ & $\bf 1.54$ & $\bf -1.90$ & $\bf 1.00$ & $\bf 0.80$ & $\bf 13.78~[1.72]$ \\  & & & & & & & & $\bf 0.50$ & $\bf - $ & $\bf -3.64$ & $\bf 0.00$ & $\bf 0.20$ \\ 
\vspace{-8pt} \\ NGC4214-s$^{(e)}$\dotfill{} & 0.50 & 11$^*$ & $3.00$ & $3.83$ & $-2.39$ & $0.60^{0.84}_{0.56}$ & $47.71~[5.96]$ & $\bf 3.00$ & $\bf 3.83$ & $\bf -2.39$ & $\bf 0.40$ & $\bf 0.90$ & $\bf 13.36~[2.23]$ \\  & & & & & & & & $\bf 0.50$ & $\bf - $ & $\bf -3.64$ & $\bf 0.00$ & $\bf 0.10$ \\ 
\vspace{-8pt} \\ Haro11\dotfill{} & 0.50 & 16$^{**}$ & $3.00$ & $3.71$ & $-2.89$ & $0.40^{0.64}_{0.36}$ & $1.59E+02~[12.22]$ & $\bf 3.00$ & $\bf 3.95$ & $\bf -1.89$ & $\bf 1.00$ & $\bf 0.90$ & $\bf 10.04~[0.91]$ \\  & & & & & & & & $\bf 1.50$ & $\bf - $ & $\bf -3.86$ & $\bf 0.00$ & $\bf 0.10$ \\ 
\vspace{-8pt} \\ UM448\dotfill{} & 0.50 & 14$^*$ & $3.00$ & $3.71$ & $-2.89^{-2.89}_{-2.89}$ & $0.60^{0.80}_{0.40}$ & $29.96~[2.72]$ & $\bf 2.50$ & $\bf 3.30$ & $\bf -2.39$ & $\bf 1.00$ & $\bf 0.80$ & $\bf 16.42~[1.82]$ \\  & & & & & & & & $\bf 2.50$ & $\bf - $ & $\bf -3.89$ & $\bf 0.00$ & $\bf 0.20$ \\ 
\vspace{-8pt} \\ Haro3\dotfill{} & 0.50 & 14$^*$ & $2.50^{2.88}_{2.38}$ & $3.30^{3.60}_{3.19}$ & $-2.39^{-2.26}_{-2.76}$ & $0.80^{0.96}_{0.44}$ & $28.94~[2.63]$ & $\bf 2.50$ & $\bf 3.30$ & $\bf -2.39$ & $\bf 0.80$ & $\bf 0.90$ & $\bf 6.15~[0.68]$ \\  & & & & & & & & $\bf 0.50$ & $\bf - $ & $\bf -3.64$ & $\bf 0.00$ & $\bf 0.10$ \\ 
\vspace{-8pt} \\ NGC5253\dotfill{} & 0.50 & 14$^*$ & $2.50$ & $3.30$ & $-2.39$ & $0.60^{0.84}_{0.56}$ & $34.76~[3.16]$ & $\bf 2.50$ & $\bf 3.41$ & $\bf -1.89$ & $\bf 0.80$ & $\bf 0.90$ & $\bf 5.68~[0.63]$ \\  & & & & & & & & $\bf 0.50$ & $\bf - $ & $\bf -3.64$ & $\bf 0.00$ & $\bf 0.10$ \\ 
\vspace{-8pt} \\ Haro2\dotfill{} & 0.50 & 10 & $3.00$ & $3.71$ & $-2.89$ & $0.40^{0.64}_{0.36}$ & $16.92~[2.42]$ & $\bf 2.50$ & $\bf 3.30$ & $\bf -2.39$ & $\bf 0.80$ & $\bf 0.50$ & $\bf 4.88~[0.98]$ \\  & & & & & & & & $\bf 1.00$ & $\bf - $ & $\bf -3.79$ & $\bf 0.00$ & $\bf 0.50$ \\ 
\vspace{-8pt} \\ NGC625\dotfill{} & 0.50 & 11 & $2.50$ & $3.30$ & $-2.39$ & $0.80^{1.00}_{0.76}$ & $25.14~[3.14]$ & $\bf 0.50$ & $\bf 1.98$ & $\bf -0.89$ & $\bf 0.80$ & $\bf 0.60$ & $\bf 9.65~[1.61]$ \\  & & & & & & & & $\bf 0.50$ & $\bf - $ & $\bf -3.29$ & $\bf 0.00$ & $\bf 0.40$ \\ 
\vspace{-8pt} \\ NGC4214-c$^{(e)}$\dotfill{} & 0.50 & 12$^{**}$ & $2.50^{2.62}_{2.04}$ & $3.17^{3.29}_{2.71}$ & $-2.89^{-2.89}_{-2.89}$ & $0.40^{0.58}_{0.35}$ & $24.02~[2.67]$ & $\bf 2.00$ & $\bf 2.78$ & $\bf -2.39$ & $\bf 0.80$ & $\bf 0.70$ & $\bf 12.57~[1.80]$ \\  & & & & & & & & $\bf 2.00$ & $\bf - $ & $\bf -3.88$ & $\bf 0.00$ & $\bf 0.30$ \\ 
\vspace{-8pt} \\ Mrk1089\dotfill{} & 0.50 & 13$^*$ & $2.50$ & $3.17$ & $-2.89$ & $0.60^{0.80}_{0.40}$ & $22.99~[2.30]$ & $\bf 2.50$ & $\bf 3.30$ & $\bf -2.39$ & $\bf 1.00$ & $\bf 0.80$ & $\bf 5.82~[0.73]$ \\  & & & & & & & & $\bf 0.50$ & $\bf - $ & $\bf -3.64$ & $\bf 0.00$ & $\bf 0.20$ \\ 
\vspace{-8pt} \\ HS0052\dotfill{} & 0.25 & 9 & $2.50$ & $3.60$ & $-2.38^{-2.34}_{-2.92}$ & $0.60^{0.76}_{0.24}$ & $24.09~[4.01]$ & $\bf 0.50$ & $\bf 2.28$ & $\bf -0.38$ & $\bf 0.80$ & $\bf 0.40$ & $\bf 3.01~[0.75]$ \\  & & & & & & & & $\bf 0.50$ & $\bf - $ & $\bf -3.29$ & $\bf 0.00$ & $\bf 0.60$ \\ 
\vspace{-8pt} \\ SBS1533\dotfill{} & 0.25 & 9 & $2.50^{2.52}_{1.68}$ & $3.60^{3.61}_{2.79}$ & $-2.38^{-2.38}_{-2.38}$ & $0.80^{0.81}_{0.47}$ & $18.80~[3.13]$ & $\bf 0.50$ & $\bf 2.28$ & $\bf -0.38$ & $\bf 0.60$ & $\bf 0.80$ & $\bf 5.56~[1.39]$ \\  & & & & & & & & $\bf 0.50$ & $\bf - $ & $\bf -3.29$ & $\bf 0.00$ & $\bf 0.20$ \\ 
\vspace{-8pt} \\ Mrk1450\dotfill{} & 0.25 & 11$^*$ & $1.00^{1.62}_{1.04}$ & $2.42^{2.68}_{2.44}$ & $-2.38^{-2.38}_{-2.38}$ & $0.20^{0.38}_{0.15}$ & $17.87~[2.23]$ & $\bf 1.00$ & $\bf 2.42$ & $\bf -1.38$ & $\bf 0.40$ & $\bf 0.70$ & $\bf 7.08~[1.18]$ \\  & & & & & & & & $\bf 1.00$ & $\bf - $ & $\bf -3.35$ & $\bf 0.00$ & $\bf 0.30$ \\ 
\vspace{-8pt} \\ UM461\dotfill{} & 0.10 & 9 & $2.00^{2.41}_{1.59}$ & $3.57^{4.01}_{3.17}$ & $-1.85^{-1.85}_{-1.85}$ & $0.40^{0.42}_{0.18}$ & $18.01~[3.00]$ & $\bf 1.50$ & $\bf 3.20$ & $\bf -0.85$ & $\bf 0.40$ & $\bf 0.90$ & $\bf 7.03~[1.76]$ \\  & & & & & & & & $\bf 1.50$ & $\bf - $ & $\bf -3.82$ & $\bf 0.00$ & $\bf 0.10$ \\ 
\vspace{-8pt} \\ VIIZw403\dotfill{} & 0.10 & 9 & $2.50^{2.62}_{2.12}$ & $3.83^{4.06}_{3.57}$ & $-2.85^{-2.23}_{-2.73}$ & $0.20^{0.56}_{0.24}$ & $29.86~[4.98]$ & $\bf 1.00$ & $\bf 2.86$ & $\bf -1.35$ & $\bf 0.60$ & $\bf 0.40$ & $\bf 7.15~[1.79]$ \\  & & & & & & & & $\bf 1.00$ & $\bf - $ & $\bf -3.32$ & $\bf 0.00$ & $\bf 0.60$ \\ 
    \vspace{-6pt}\\
    \hline \hline
  \end{tabular}
  \hfill{}
  \tablefoot{
  See caption of Table~\ref{table:modelres1} for details.
  The symbol (*) indicates if \niila is detected and (**) indicates
  if both \niila and \niilb are detected.
  ~$(e)$~NGC\,4214-c and NGC\,4214-s refer to the central and
  southern star-forming regions of NGC\,4214.
  }
  \label{table:modelres2}
\end{table*}
\end{center}

\subsection{Mixing two model components}
\subsubsection{Motivation}
\begin{figure}[t]
\centering
\includegraphics[clip,width=6cm]{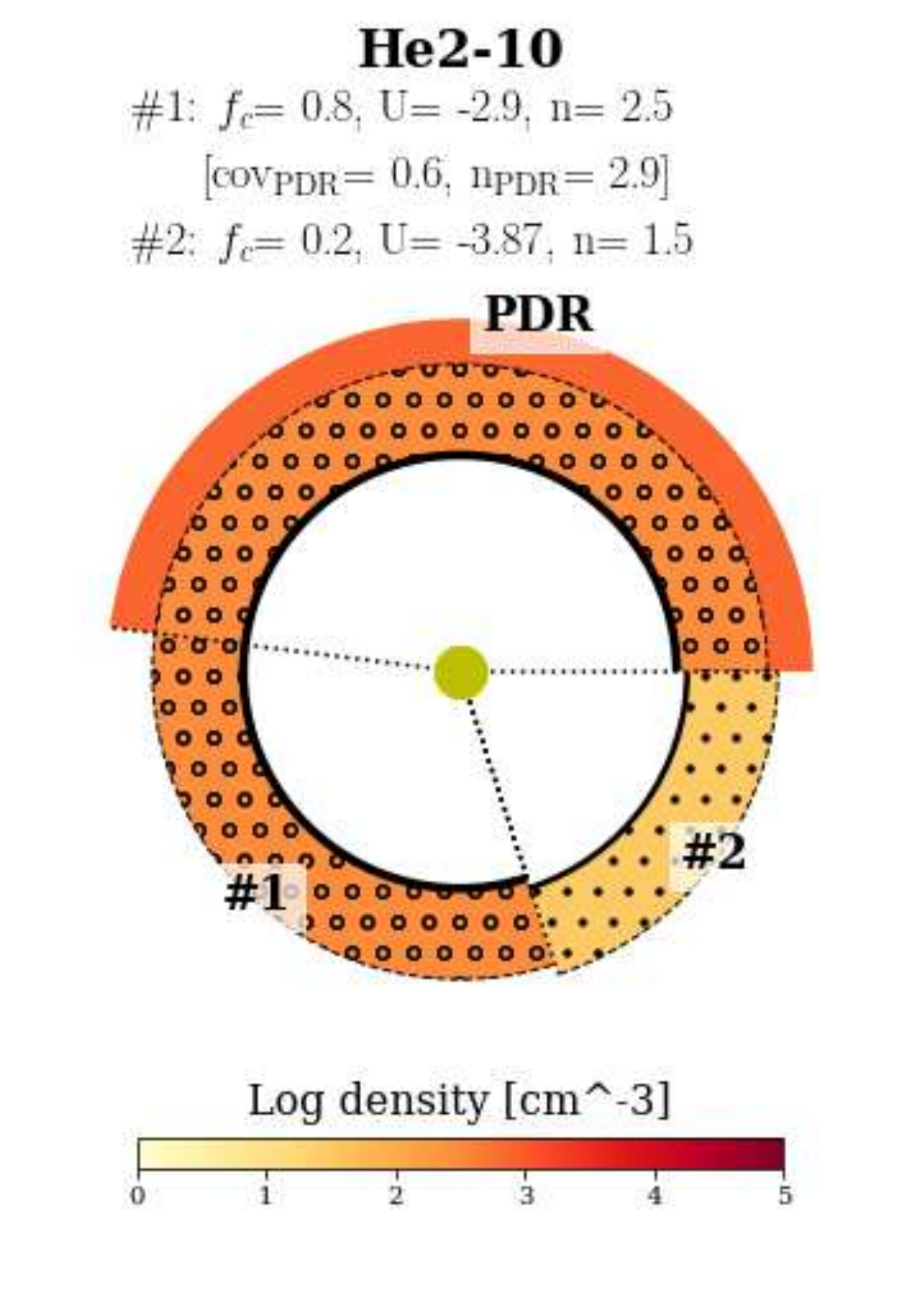}
\caption{
2D representation of the 1D best-fitting (mixed) model for
the example galaxy He\,2-10. Schematics for the other
galaxies of the sample are shown in Appendix~\ref{sect:append-b}.
The model parameters are the same as those reported in
Table~\ref{table:modelres2}.
Color coding corresponds to density. 
The central stellar cluster is represented in green.
The distance to the inner black shell is proportional to
the inner radius, and the thickness of the inner black
shell is proportional to the ionization parameter $U$, while
the thickness of the \hii and PDR regions are kept fixed.
$U$ is a function of the input stellar spectrum, density,
and inner radius. The angle shown by dotted
lines indicates the contribution from each component
(i.e., parameter $f_c$ in Table~\ref{table:modelres2}).
}
\label{fig:schematics}
\end{figure}

%
It is possible that the minimum $\chi^2$ for single models as
described above does not provide a satisfying fit to
the data because our model description is too simple. 
In addition to single models, we explored the possibility of
mixing two models of the grids: one dense high-$U$ model
(model \#1), which should represent the material that
gave birth to the starburst, and one more diffuse lower-$U$
model (model \#2), which should represent a hypothetical
more extended ionized phase. This two-component modeling
approach is supported by earlier work on those DGS galaxies
suggestive of a large filling factor, diffuse, ionized gas phase
\citep{cormier-2012,cormier-2015,dimaratos-2015}.
For this, we have generated a new set of models by combining
two models for which parameters are varied within the
parameter space of our original grid, and their relative 
contribution $f_c$ (in terms of luminosity) varies from 10\%
to 90\% in steps of 10\%. 
That is, for a given line, the intensity prediction of the combined 
model is the sum of the prediction of the first model multiplied
by $f_c$ and of the prediction of the second model times $1-f_c$.
Given the small number of constraints for the PDR, we
consider that only the dense \hii model produces a PDR.
In that case, we have as free parameters for the mixed models:
$n_{\rm H}$ (but $n_{\rm H,1} \ge n_{\rm H,2}$),
$U$ (but $U_1 \ge U_2$),
and $cov_{\rm PDR,1}$ (with $cov_{\rm PDR,2} =0$).
The number of free parameters is five.
Figure~\ref{fig:schematics} shows an illustration
of the best-fitting model for one example galaxy and
figures for each galaxy of the sample can be found in
Appendix~\ref{sect:append-b}.
We stress that for mixed models, $f_c$ and $cov_{\rm PDR}$
are different quantities. $f_c$ corresponds to the contribution
of a given model to the final (2-component) model, it is
therefore translated into a covering factor of the \hii region
in Fig.~\ref{fig:schematics}. $cov_{\rm PDR}$ is the covering
factor of the PDR relative to that of the \hii region, hence
the covering factor of the final model is
$cov_{\rm PDR} \times f_c$.

For galaxies with fewer than six lines detected (for which a
mixed model would be under-constrained) or with $\chi^2_{\nu}$
from the single models lower than  two (or lower than with the mixed
models), we consider the single best-fit model to be the best
representation. There are $23$ galaxies that satisfy this condition
(results reported in Table~\ref{table:modelres1}). For the other
$16$ galaxies, we consider the mixed best-fit model to be better
(results reported in Table~\ref{table:modelres2}).

\begin{figure*}[t]
\centering
(a)
\includegraphics[clip,width=11cm]{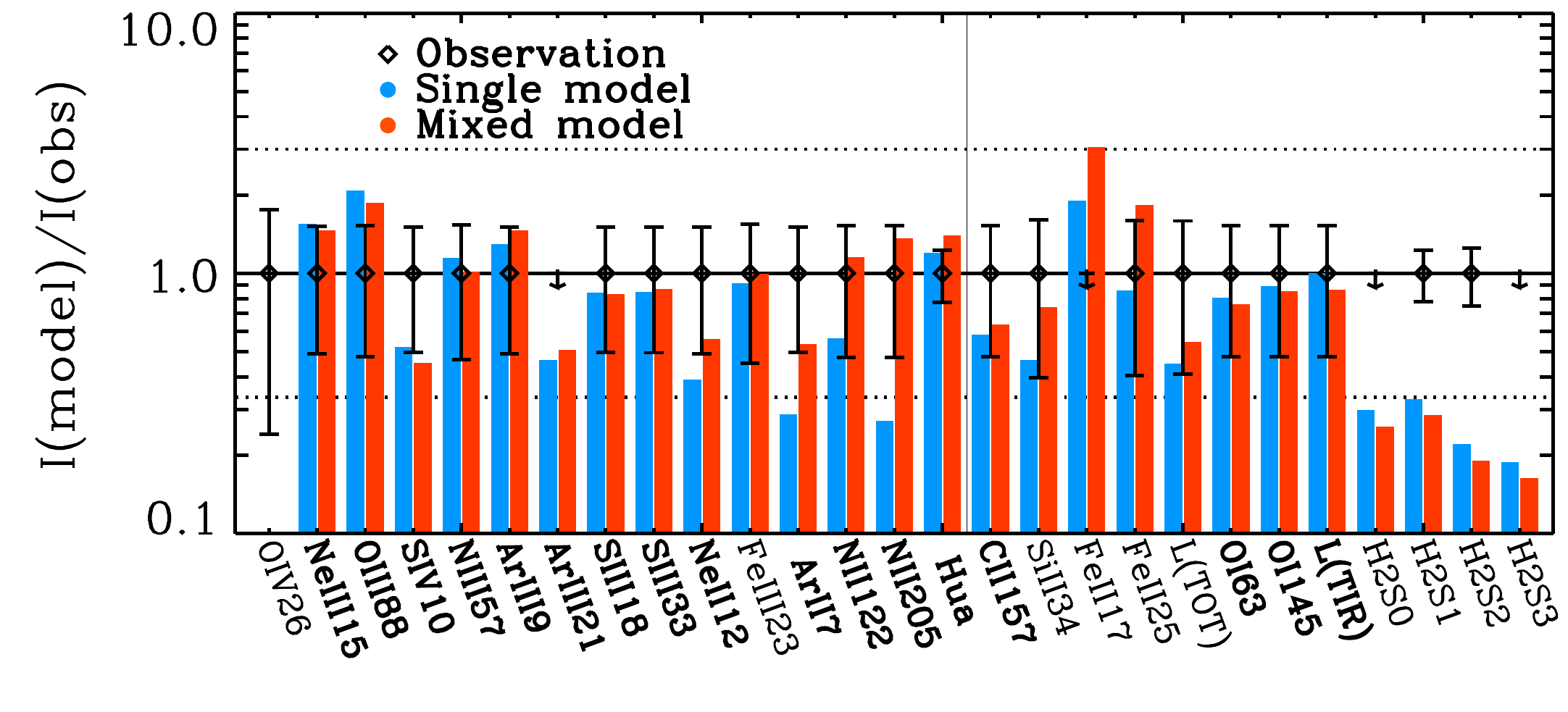} \hfill{}
(b)
\includegraphics[clip,width=6.2cm]{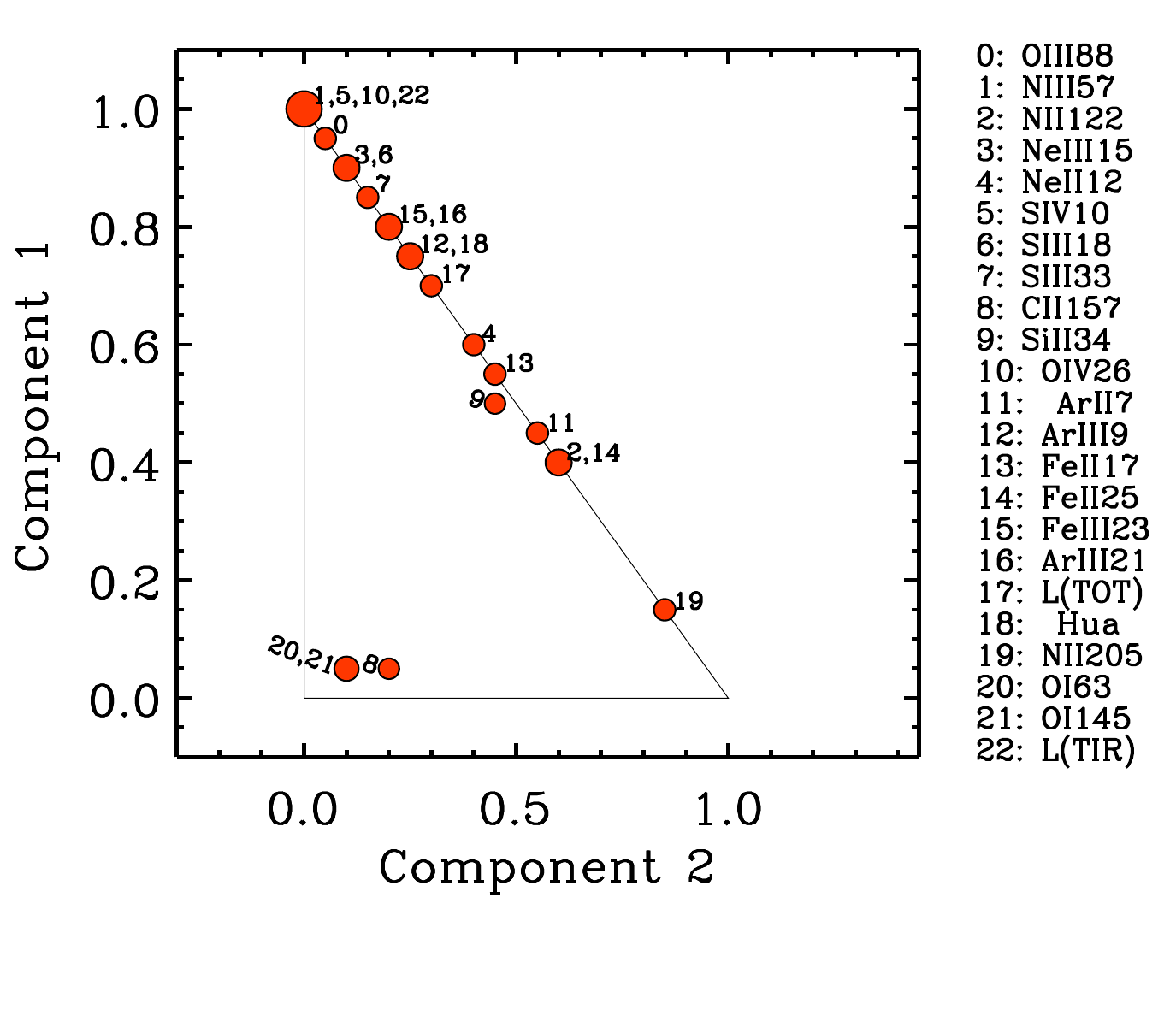}\\
(c)
\includegraphics[clip,width=8.8cm]{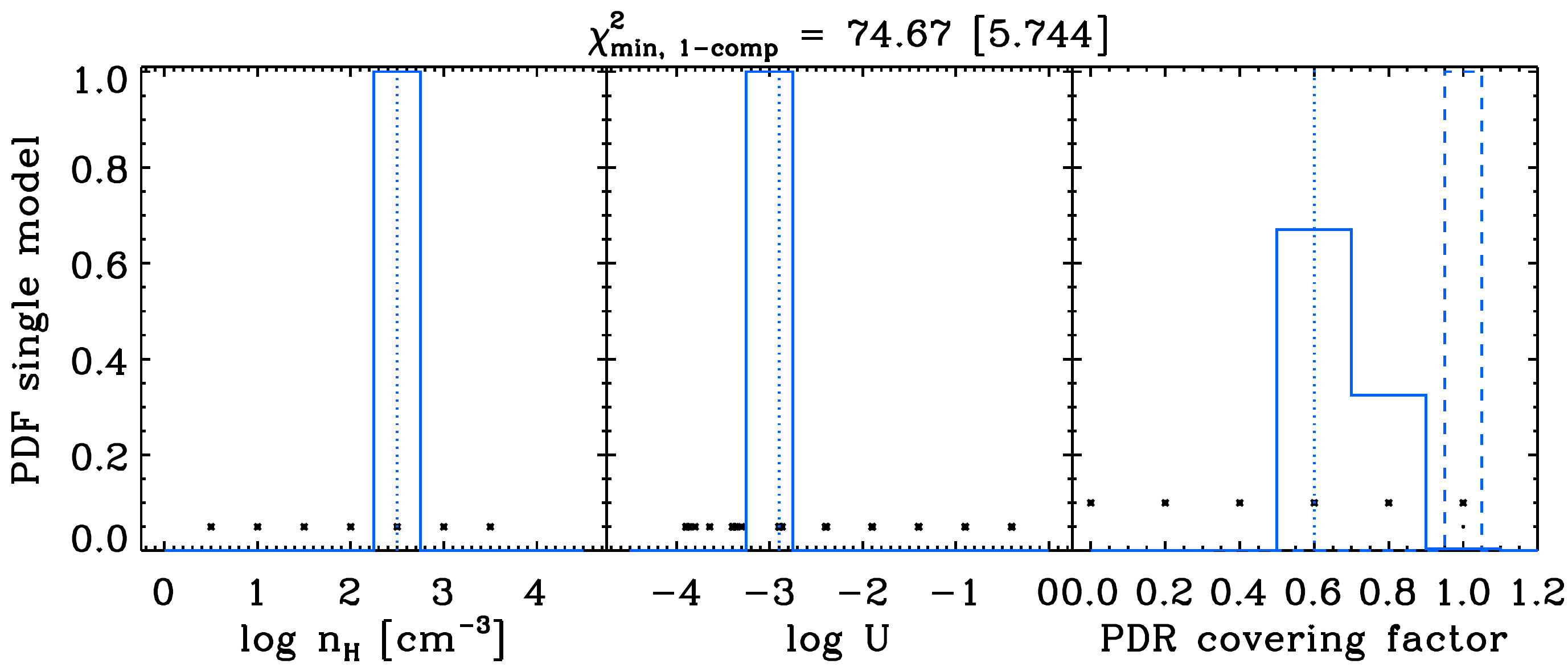} \hfill{}
\includegraphics[clip,width=8.8cm]{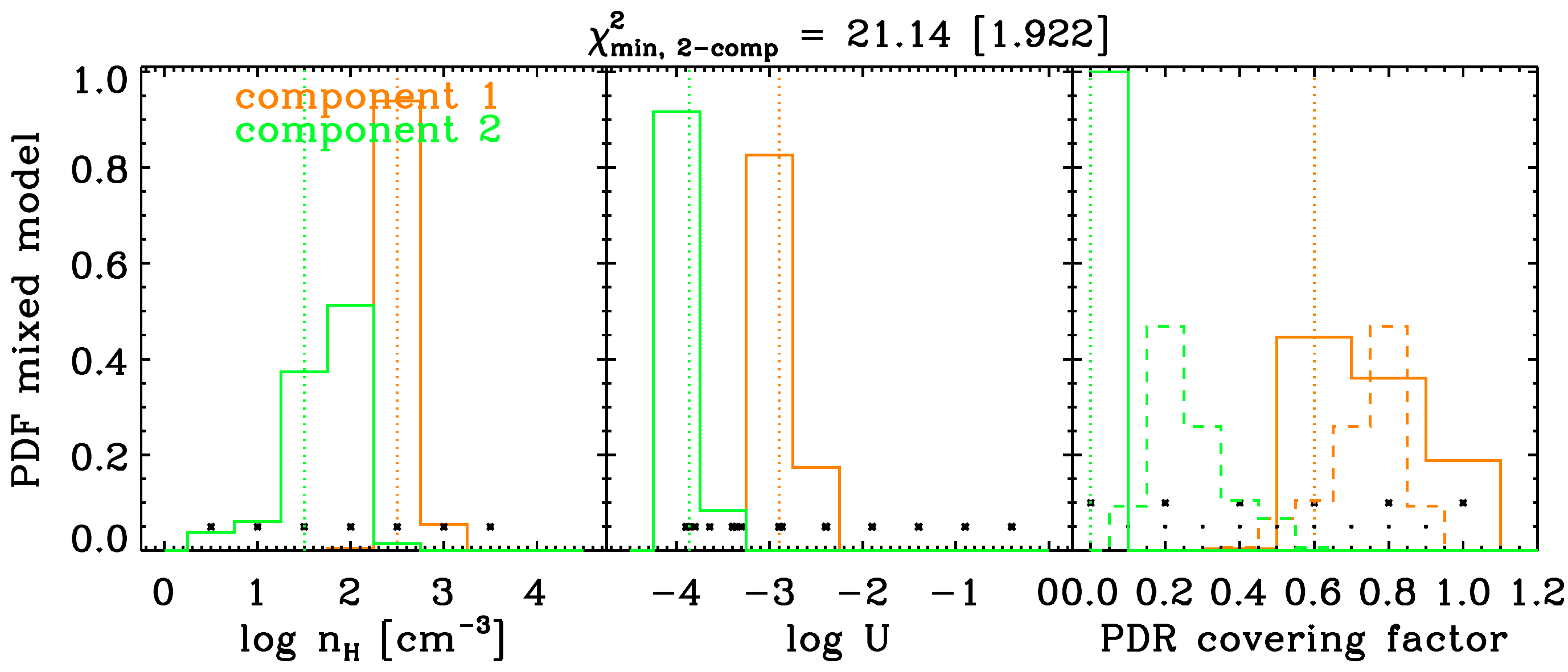}
\caption{
Model results, shown only for the example galaxy
He\,2-10 here and for all galaxies in Appendix~\ref{sect:append-d}.
$(a)$~Comparison of the observed (losange) and predicted 
intensities for our best-fitting single (blue bar) and mixed (red bar)
\hii region+PDR models. Lines are sorted by decreasing energy.
The fitted lines have labels in boldface.
$(b)$~For mixed models: respective contributions from
component \#1 and component \#2 to the line prediction.
For PDR lines (\cii, \oi, and \silii), only the contribution
from the ionized gas is shown; the PDR contribution is
one minus the sum of the contributions from the plot.
$(c)$~Probability density functions of the model parameters
($n_{\rm H}$, $U$, $cov_{\rm PDR}$/scaling factor)
for single (left panel) and mixed (right panel) models.
The scaling factor corresponds to the proportion in which
we combine two models in the mixed model case. It is
equal to unity otherwise. It is shown with dashed lines in
the same panels as the PDR covering factor. Vertical dotted lines
show values of the best-fitting model. Black asterisks
show the range and step of values of the grid parameters.
}
\label{fig:bestparams}
\end{figure*}

\subsubsection{Results for mixed models}
In order to reproduce all observed line intensities in galaxies 
where a single model fails ($\chi^2_{\nu} > 2$), we search
for a combination of two model components. We stress that
the second component is not added to the single model but
a new set of two model components is found (with some
constraints on those parameters as explained in the previous
subsection).
Parameters of the best-fitting mixed models as well as their 
corresponding $\chi^2$ are indicated in Tables~\ref{table:modelres1}
and \ref{table:modelres2}.
Similar figures as for the single model case are shown in
Fig.~\ref{fig:bestparams} and in Appendix~\ref{sect:append-d}
for all galaxies.

In general, the lines that were not satisfyingly reproduced
with single models are well fit by mixed models.
The low-$U$ component contributes mostly to the
\nii, \neii, \arii, and \siii lines. We note that the two
$U$-values of the mixed models always bracket the
$U$-value of the single models, while the $n_{\rm H}$-values
of the mixed models are sometimes lower than the
$n_{\rm H}$-value of the single models (for example,
NGC\,625, SBS\,1533). For those galaxies, the single
models tend to under-predict the \ciil/\oila ratio by a factor
of two to three while the lower $n_{\rm H}$-values
of the mixed models reproduce this ratio better.
Overall, our observations seem to be more sensitive to
variations in the ionization parameter, although there
might be also some degeneracy at play between density
and covering factor for the PDR lines.
The PDFs of the parameters of the mixed models are
sometimes broader than for the single models, especially
for the lower-density component, indicating that 
it is not very well constrained by the observations. 
In most cases, the denser component contributes more
than the lower-density component to the final mixed
model ($f_c>0.5$).

For most galaxies for which single models do not provide
a good fit, the mixed models perform better (lower
$\chi^2_{\nu, {\rm min}}$). All galaxies for which mixed
models are preferred have $\chi^2_{\nu, {\rm min}} \le 2$.
For several galaxies though, mixed models do not
improve the results over the single models but provide similar
$\chi^2_{\nu, {\rm min}}$ (II\,Zw\,40, Mrk\,209, Pox\,186, SBS\,0335,
SBS\,1159, SBS\,1415) or worse $\chi^2_{\nu, {\rm min}}$ (Tol\,1214).
For two of them (SBS\,1159, Tol\,1214), $\chi^2_{\nu, {\rm min}}$
remains above three. For those two galaxies, we find
that the combination of a low-$U$, high-$n_{\rm H}$ component
going in the PDR and a high-$U$, low-$n_{\rm H}$ component
would actually fit the data better (with $\chi^2_{\nu, {\rm min}}$
close to unity).
For the galaxies where a single model was already giving
a good fit, the mixed models provide lower $\chi^2$ but are
often under-constrained. The parameters of the two
model components are sometimes very similar to the
parameters of the single model or they bracket them
(especially for $U$ and less so for $n_{\rm H}$, as noted
above).

\noindent{\it Comparison to previous, similar studies:}\\
At this stage, it is worth comparing our findings to
two previous studies that we carried out to model
MIR and FIR lines using a similar approach. 
Results for Haro\,11 are in global agreement with previous 
modeling done in \cite{cormier-2012}. For the \hii region, 
we find similar densities ($10^{2.8}$\,\cm and $10^{1.0}$\,\cm
versus $10^{3.0}$\,\cm and $10^{1.5}$\,\cm here). The
$U$-values are somewhat different ($-2.5$ and $-1.7$
versus $-1.9$ and $-3.9$). This is mainly driven by
the different stellar spectra used in input (instantaneous
burst and single star versus continuous star-formation
scenario here). For the PDR, we find a larger PDR
covering factor because the PDR density is lower
than in \cite{cormier-2012} where the density was set by
pressure equilibrium between the \hii region and the PDR.
For the two star-forming regions of NGC\,4214, we also find
very similar results for single models to \cite{dimaratos-2015},
with lower density, ionization parameter, and PDR covering
factor in the central region compared to the southern region.
Mixed models (not included in \citealt{dimaratos-2015}) are
however better suited to reproduce the \nii emission in NGC\,4214.

\subsection{A posteriori checks}

\subsubsection{Mid-infrared lines}
\label{sect:mirsecond}
%
We have not used the silicon, iron, and \htwo lines from
the \spit IRS instrument to constrain the models due to
the ambiguous origin (possibly shocks) of those lines
and uncertain gas-phase abundances. However, we have
compared predictions from the best models a posteriori.
We report on the performance of the single and mixed models
in predicting those secondary lines in Table~\ref{table:secondperf}.

\begin{center}
\begin{table}[!t]\small
  \caption{Performance of the best-fitting models for the secondary lines and bolometric luminosity.} 
  \hfill{}
\begin{tabular}{lrcc}
    \hline     \hline
     \vspace{-8pt}\\
     {Line\hspace{1.3cm}} & $n_{\rm gal, det}$ & Single models & Mixed models \\
      &  & low | ok | high & low | ok | high \\ 
    \hline 
     \vspace{-8pt}\\
     \silii$_{34.82}$                   & $21$  & $21\,|\,0\,|\,0~~$    & $16\,|\,5\,|\,0~~$      \\
     \feii$_{17.94}$                    & $1$   & $0\,|\,1\,|\,0$               & $0\,|\,1\,|\,0$         \\
     \feii$_{25.99}$                    & $10$  & $0\,|\,10\,|\,0$      & $0\,|\,9\,|\,1$         \\
     \feiii$_{22.93}$                   & $11$  & $0\,|\,7\,|\,4$               & $0\,|\,8\,|\,3$         \\
     \htwo~S(0)$_{\,28.21}$             & $1$   & $1\,|\,0\,|\,0$               & $1\,|\,0\,|\,0$         \\
     \htwo~S(1)$_{\,17.03}$             & $8$   & $5\,|\,1\,|\,2$               & $5\,|\,0\,|\,3$         \\
     \htwo~S(2)$_{\,12.28}$             & $7$   & $4\,|\,3\,|\,0$               & $5\,|\,2\,|\,0$         \\
     \htwo~S(3)$_{\,9.66}$              & $0$   & $0\,|\,0\,|\,0$               & $0\,|\,0\,|\,0$         \\
     $L_{\rm BOL}$                      & $37$  & $0\,|\,37\,|\,0$      & $~~0\,|\,25\,|\,12$     \\
    \hline  \hline
  \end{tabular}
  \hfill{}
  \tablefoot{
  Columns\,1-2: line label and number of galaxies in which the line is detected.
  Columns\,3-4: number of galaxies in which the ratio of predicted over
  observed intensity is low, ok, and high, where the threshold is set to a factor of two.
  We report numbers for single and mixed models.
  }
\label{table:secondperf}
\end{table}
\end{center}

%
The \silii line at 34\,\mum can originate in the ionized gas
and in the neutral ISM. It is quite bright and detected in
many of the galaxies of our sample.   
We find that the \silii line is systematically under-predicted
by a factor of two to six by the best-fitting models.
Mixed models perform slightly better than single models
although \silii remains under-predicted in most cases.
Since our models include the \hii region and a diffuse,
low-excitation ionized phase in the case of mixed models,
any contribution of ionized gas to \silii emission is largely
taken into account by our models. Thus, reasons for
the under-prediction of \silii could be additional excitation
of \silii by shocks or uncertainty in the adopted gaseous
abundances. Given the difficulty in modeling iron
(see below), we cannot really compare our results
for \silii with those for \feii to rule out shock excitation.

The mid-infrared iron lines are faint and detected in only
a few galaxies of our sample. The \feii line at 25.99\,\mum
is generally well reproduced while the \feiii line at 22.93\,\mum
is over-predicted by a factor of two to three for half of
the sources with detection, even when adding those
lines in the list of lines to fit for.
We note that collisional excitation of \feii by H
and \htwo is not included in the version of Cloudy used
in this work, hence an additional contribution from PDRs
to the \feii emission (not included in our models) would
then over-predict the observations.
However, the main effect affecting the line prediction
is again the adopted abundances. Although measured
for many of our galaxies (Table~\ref{table:ab_obs}),
abundances can be uncertain because of unknown
ionization corrections, depletion of iron in the cold dense
medium \citep{savage-1996}, or enhancement of iron from
supernovae-driven shocks \citep{ohalloran-2008}.
In our case, adopted abundances may be systematically
too high. At solar metallicity, \cite{kaufman-2006} consider
an iron abundance ten times lower than our adopted
abundance.

The \htwo rotational lines are also faint and detected in
few galaxies in our sample. The S(1) transition at 17.03\,\mum
is sometimes over-predicted and sometimes under-predicted
by a factor of two to fifteen, while the S(2) transition at 12.28\,\mum
is usually under-predicted by a factor of three to ten.
The ratio of those two transitions is mostly sensitive to
density for $G_0>10^2$ \citep{kaufman-2006}.
When fitting for the \htwo lines, the densities required
for the PDR are indeed larger (typically $\ge10^2$\,\cm
at the PDR front) but $\chi^2_{\nu}$ increases a lot
(models performing worse for some \hii lines, \cii, \oi).
More \htwo detections are required to really investigate
the presence and importance of shocks.

\subsubsection{Bolometric luminosity}
\label{sect:bolum}
As explained in Section~\ref{sect:comparelum}, the total
luminosity of the models is not an easy parameter to constrain. 
Here we just verify that the total luminosity of the best-fitting
models does not exceed the observed bolometric luminosities.
Table~\ref{table:secondperf} shows that this condition
is satisfied for all galaxies in the case of single models.
In the case of mixed models, for $12/37$ galaxies, the
luminosity required by the models is larger than the
observed $L_{\rm BOL}$ by a factor of three to four
in most cases and up to $30$ for one galaxy. For all those
$12$ galaxies, the diffuse ionized component contributes
to the majority of the total luminosity, and for ten of those
galaxies, the single models are actually preferred. For the
two galaxies where mixed models are preferred (HS\,0052
and Mrk\,1089), $L_{\rm BOL}$ is over-predicted by
a factor of three. 
When interpreting this over-prediction, one should not
forget that there is a range of acceptable values for each
model parameter (PDFs in Fig.~\ref{fig:bestparams} and
in Appendix~\ref{sect:append-d}).
Indeed, a density of $10$\,\cm (value often within the
acceptable range) instead of $3$\,\cm for the
diffuse component of those two galaxies would reconcile
better the total observed and modeled luminosities.
We also note that we calculated the observed bolometric
luminosity relatively coarsely.

\subsubsection{Optical lines}
\label{sect:optextinct}
Since the optical lines are very sensitive to extinction and
to the distribution of dust in the models, as opposed to
the infrared lines, and their observations often cover a
smaller field-of-view, we do not include them in finding
the best-fitting models.
A posteriori comparisons indicate that the intensity ratios
of the optical lines relative to H$\beta$ are generally
well reproduced by the models (within a factor of three).
To assess how the best-fitting models would perform
for the total fluxes (i.e., not normalized to H$\beta$), we also
compare the ratio of total observed H$\alpha$ luminosity
(references in \cite{remy-2015}, mainly from \cite{gildepaz-2003})
to the \oiiilb luminosity. We find that the observed H$\alpha$/\oiiilb
ratio is matched by the models (within a factor of two) for
more than half of the galaxies of the sample, and it is
over-predicted for the other galaxies by a factor of three to eight.
The discrepancies with optical observations are likely
caused by extinction/geometry considerations and
reproducing them, simultaneously with IR lines, is not
trivial (and neither is it the goal of this study; see also
\citealt{polles-2019}).

\subsection{Contributions to the \cii and \silii emission}
The use of the \cii line at 157\,\mum as a reliable neutral gas/PDR
tracer has been questioned since it can also originate from diffuse
($n_{\rm crit, e^-, [CII]}\sim 50$\,\cm), low-excitation ionized gas.
To quantify such contribution to the observed \cii intensity,
it is often compared to the \nii lines at 122\,\mum or 205\,\mum,
which have low ionization potentials and low critical densities. 
In metal-rich environments, the fraction of \cii emission arising
in the \hii region ranges from 5\% to 60\% \citep[e.g.,][]{malhotra-2001,
pineda-2013,parkin-2014,hughes-2015,abdullah-2017,croxall-2017,
diaz-santos-2017,lapham-2017}. The low \niila/\ciil values
measured in the dwarfs \citep{cormier-2015} suggest little
contribution of the \hii region to the observed C$^+$ emission. 
Similarly, \silii can arise from low-excitation ionized gas but
it has higher critical density with electrons than \cii
($n_{\rm crit, e^-, [SiII]} \sim 1,000$\,\cm).
The \hii region produces significant \silii emission, more
than the PDR, at moderate densities and low-$U$
\citep[see also][]{abel-2005}.
With the physical conditions constrained in most of the DGS
galaxies, we can quantify more precisely how much \cii and
\silii emission is predicted to arise from the ionized gas. 
The definition that we use to measure this is to take the intensity
predictions at the phase transition from ionized to atomic gas
(i.e., where the fractions of $e^-$ and H$^0$ are equal to 0.5).

In the case of single models, we find that the fractions
of \cii and \silii predicted in the \hii region are small, typically
less than 10\% for \cii and less than 40\% for \silii.
In the case of mixed models, those fractions generally
increase slightly but remain below 40\% for \cii and 50\%
for \silii. Regardless of density, it is the low-$U$ component
which contributes most to those fractions. 
The fractions of \cii and \silii emission from the ionized gas
are generally low in the single models because most
ionic lines that we observed or detected require high-$U$
values. A possible additional low-$U$ component will
increase these fractions but the parameters of this
component can only be constrained when \nii and,
to a lesser extent, \neii and \arii are available.
We further discuss the fractions of \cii from the ionized
gas with other galaxy parameters and line diagnostics in
Section~\ref{sect:disccii}.

\section{Influence of some modeling choices}
\label{sect:discparams}
In the methodology described in Section~\ref{sect:modeling},
we have made choices regarding the set of spectral lines
to fit as well as the set of model parameters to fix or vary.
Not all of the input parameters can be varied and fit
by our observations. We have taken care to use the
best input fixed parameters and the motivation of their
choices for several of them is given below.
In this section, we further explore:
1)~how the inclusion or exclusion of specific lines
can change the best-fitting results (Section~\ref{sect:linetests});
and 2)~the influence of varying one of the fixed parameters
at a time (Sections~\ref{sect:rfchoice} to \ref{sect:testdust}).
For 2), we focus on the metallicity bins $Z_{\rm bin}=0.5$
and $Z_{\rm bin}=0.1$, and we make comparison to galaxies
in those bins (including Haro\,11 which we study
in detail in \citealt{cormier-2012}).

\subsection{Set of spectral lines to fit}
\label{sect:linetests}
As discussed in the previous section, spectral lines that are
known to be problematic for the adopted modeling strategy
were excluded from the fit (for instance if shock heating or
geometry effects can be important).
Now, within the default set of fitted lines, we aim to determine
the effect of excluding some specific lines on the best-fitting
results and to identify possible biases. For this, we have considered
excluding, one at a time, the \ciil line which can arise from
more diffuse gas than \oi, the \oila line which can be optically
thick, the \neiiil line which can be problematic to model
from population synthesis models (see Section~\ref{sect:rfchoice}),
and the \sivl line which is usually well detected but one
of the lines that is least well reproduced by our models.
We also consider including the \htwo lines in the fit and
fitting a small, fixed set of lines (\ciil, \oiiilb, \neiiil, and \ltir)
since several of the low-$Z$ galaxies are only
detected in those lines.

\begin{figure*}[t]
\centering
\includegraphics[clip,width=4.5cm]{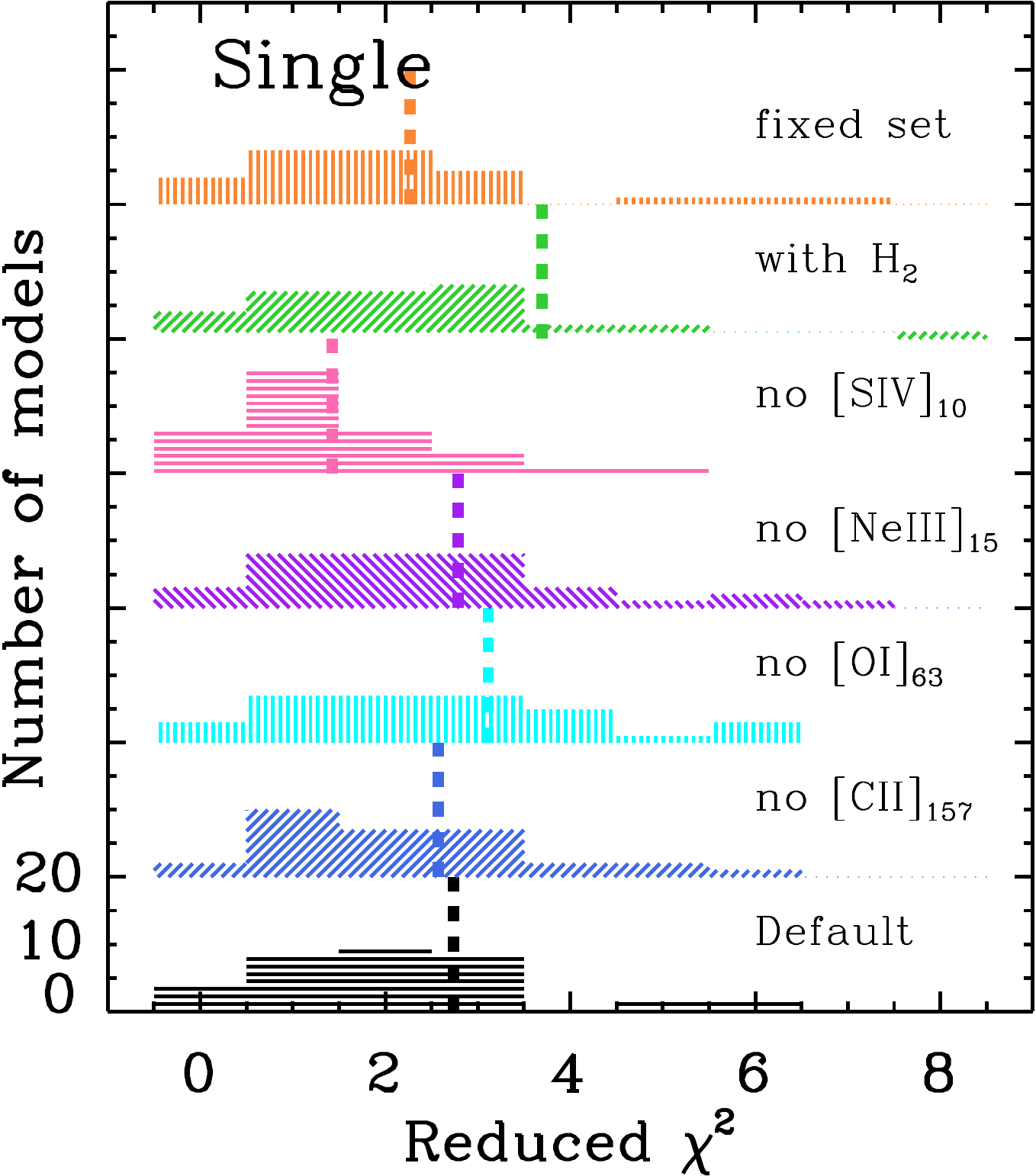}
\includegraphics[clip,width=4.5cm]{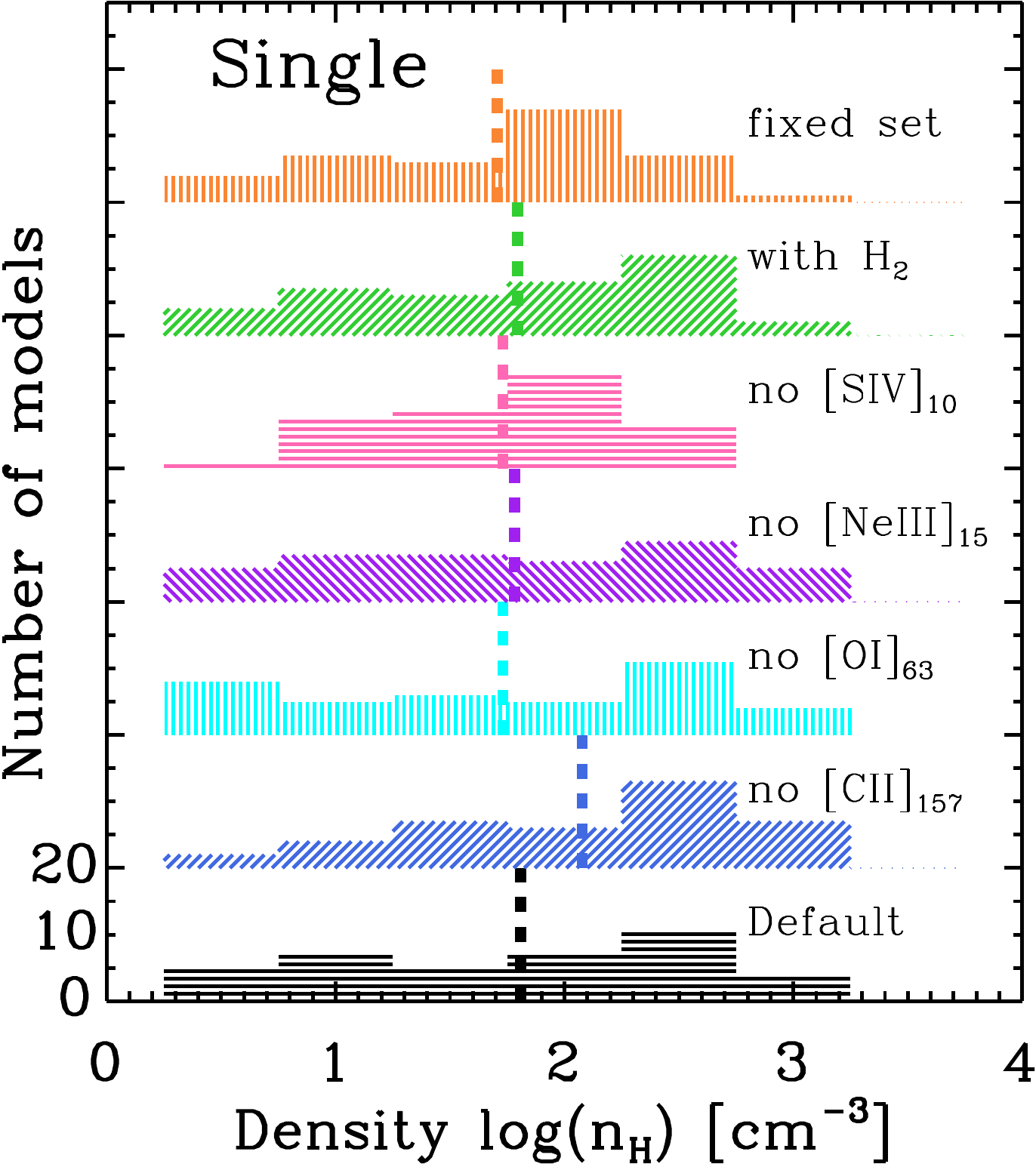}
\includegraphics[clip,width=4.5cm]{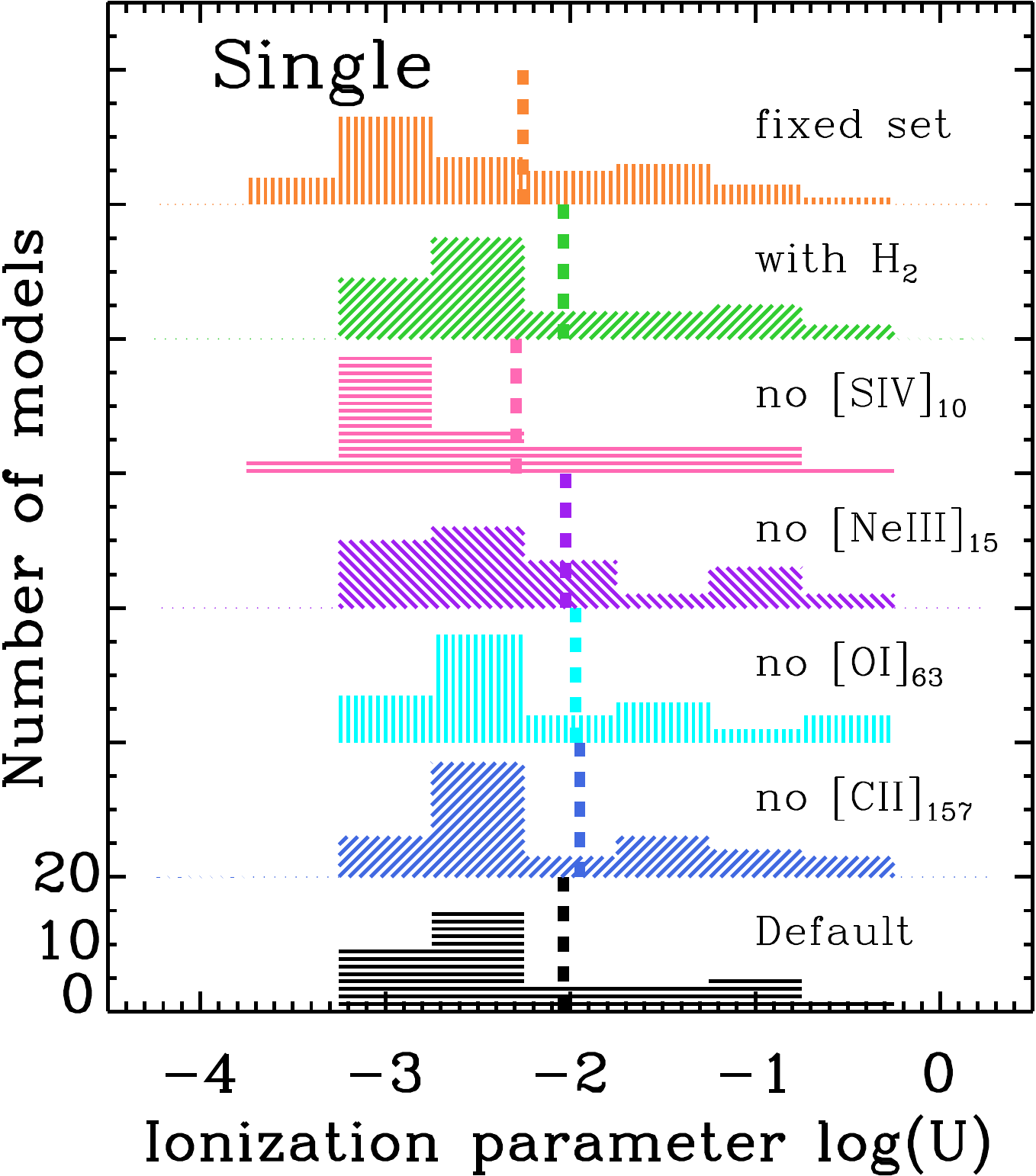}
\includegraphics[clip,width=4.5cm]{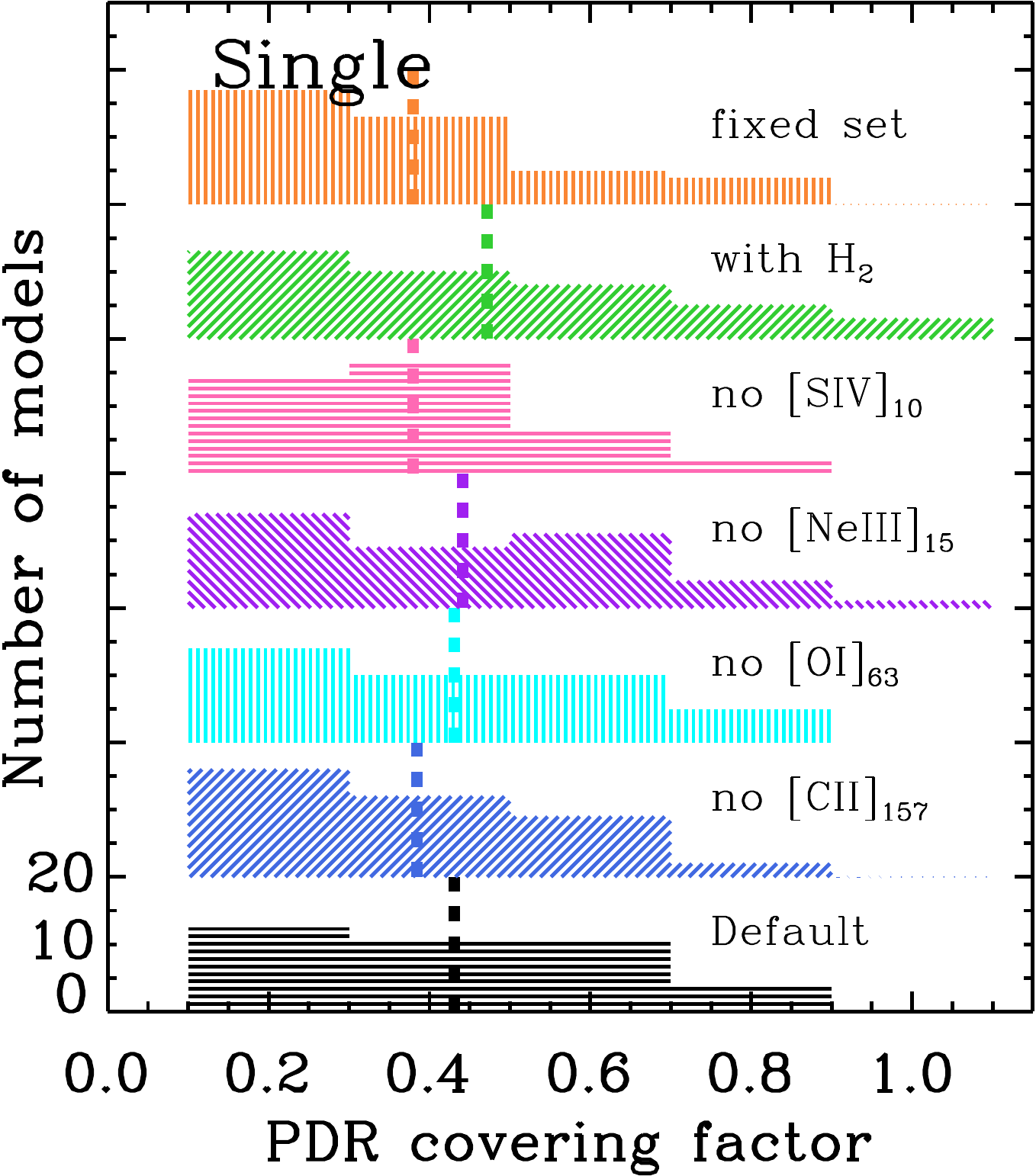} 
\includegraphics[clip,width=4.5cm]{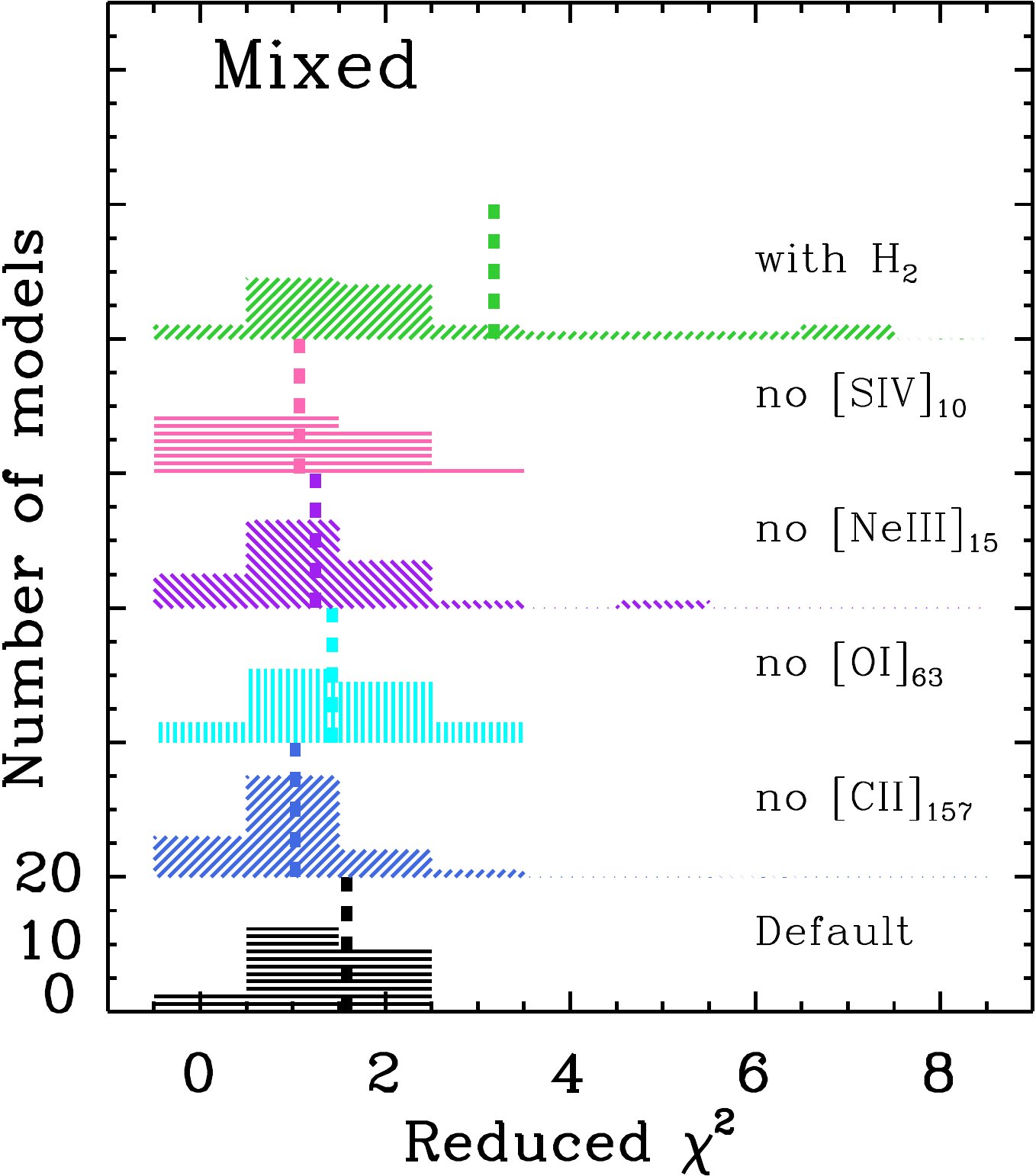}
\includegraphics[clip,width=4.5cm]{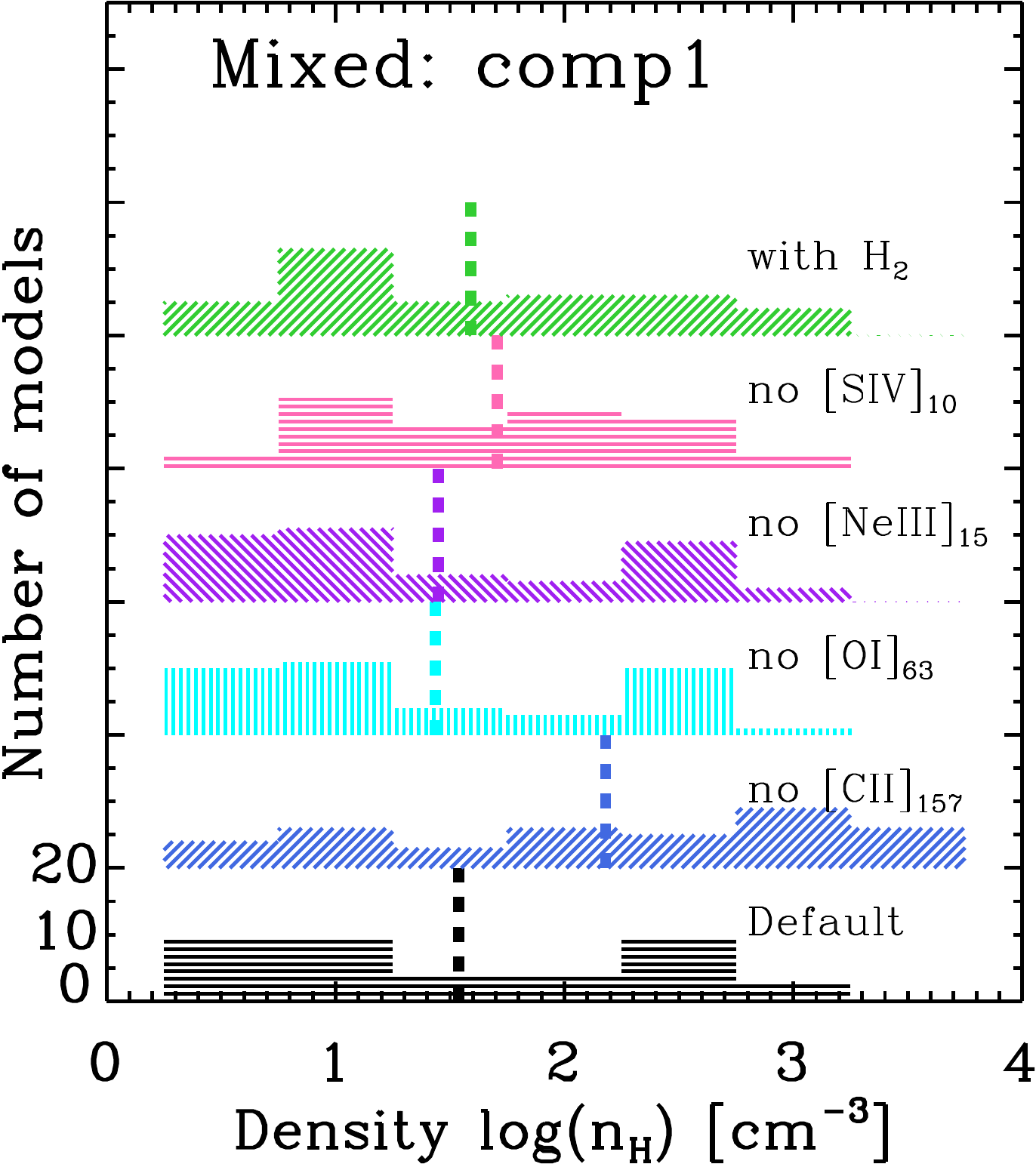}
\includegraphics[clip,width=4.5cm]{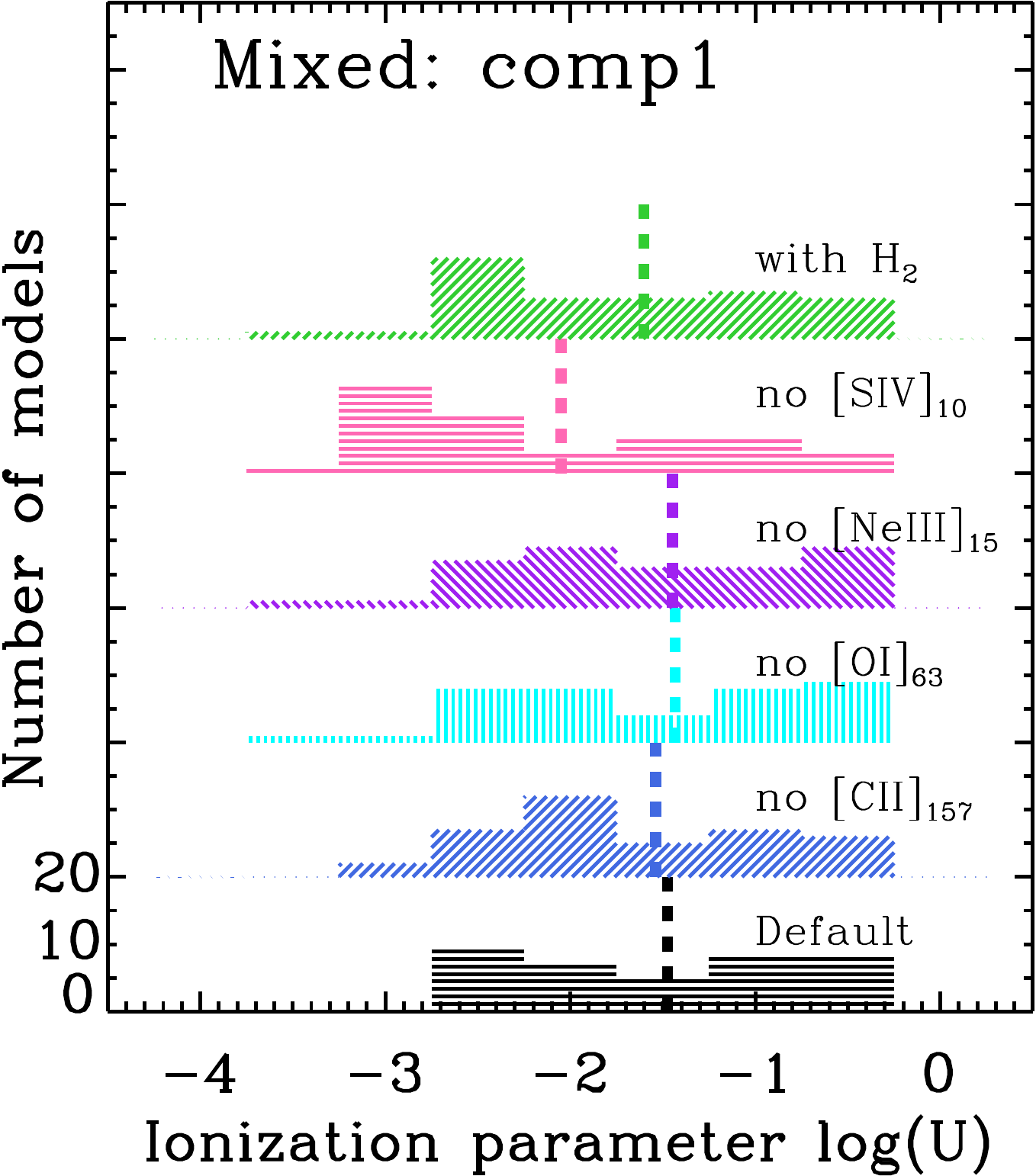}
\includegraphics[clip,width=4.5cm]{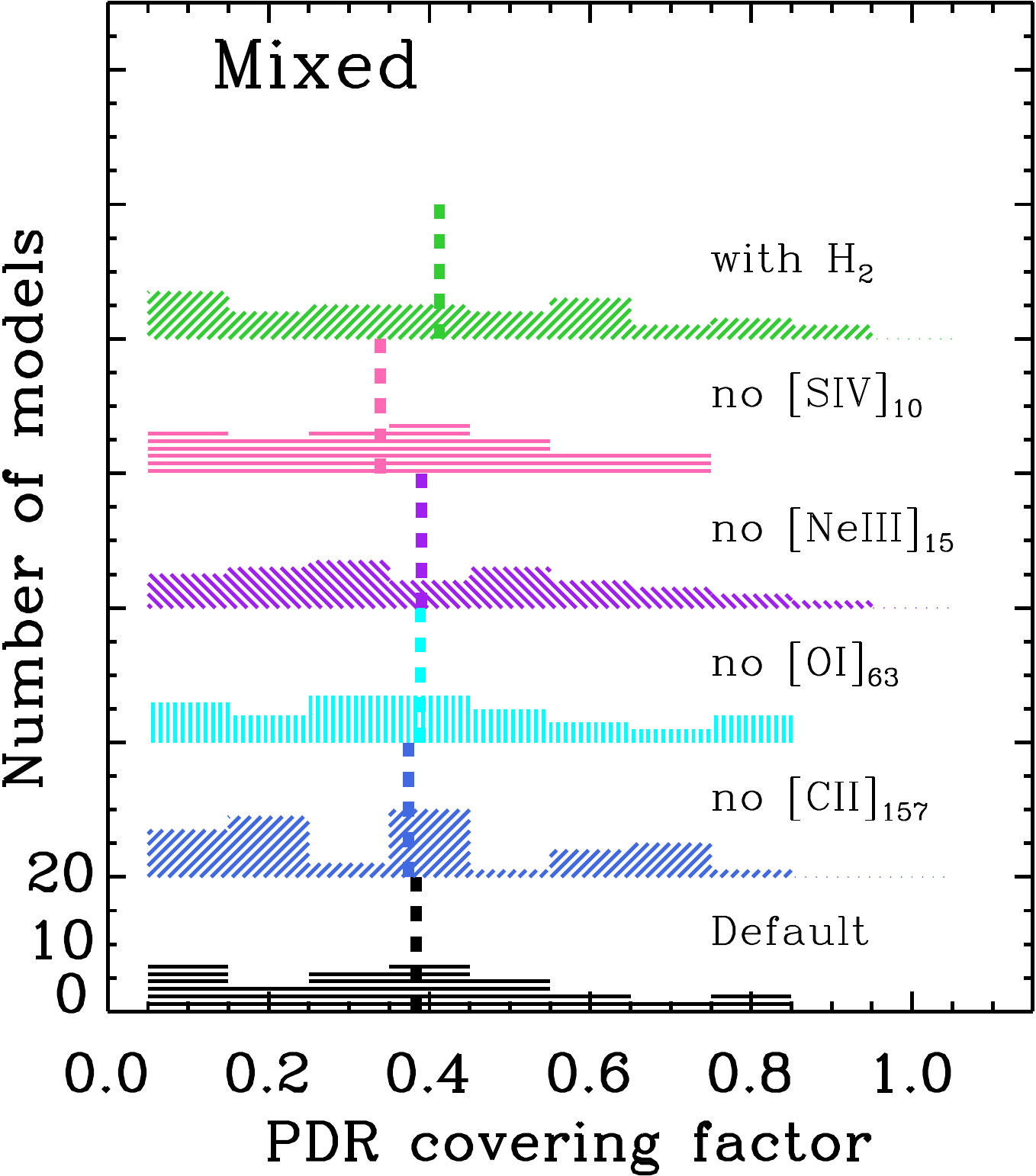}
\includegraphics[clip,width=4.5cm]{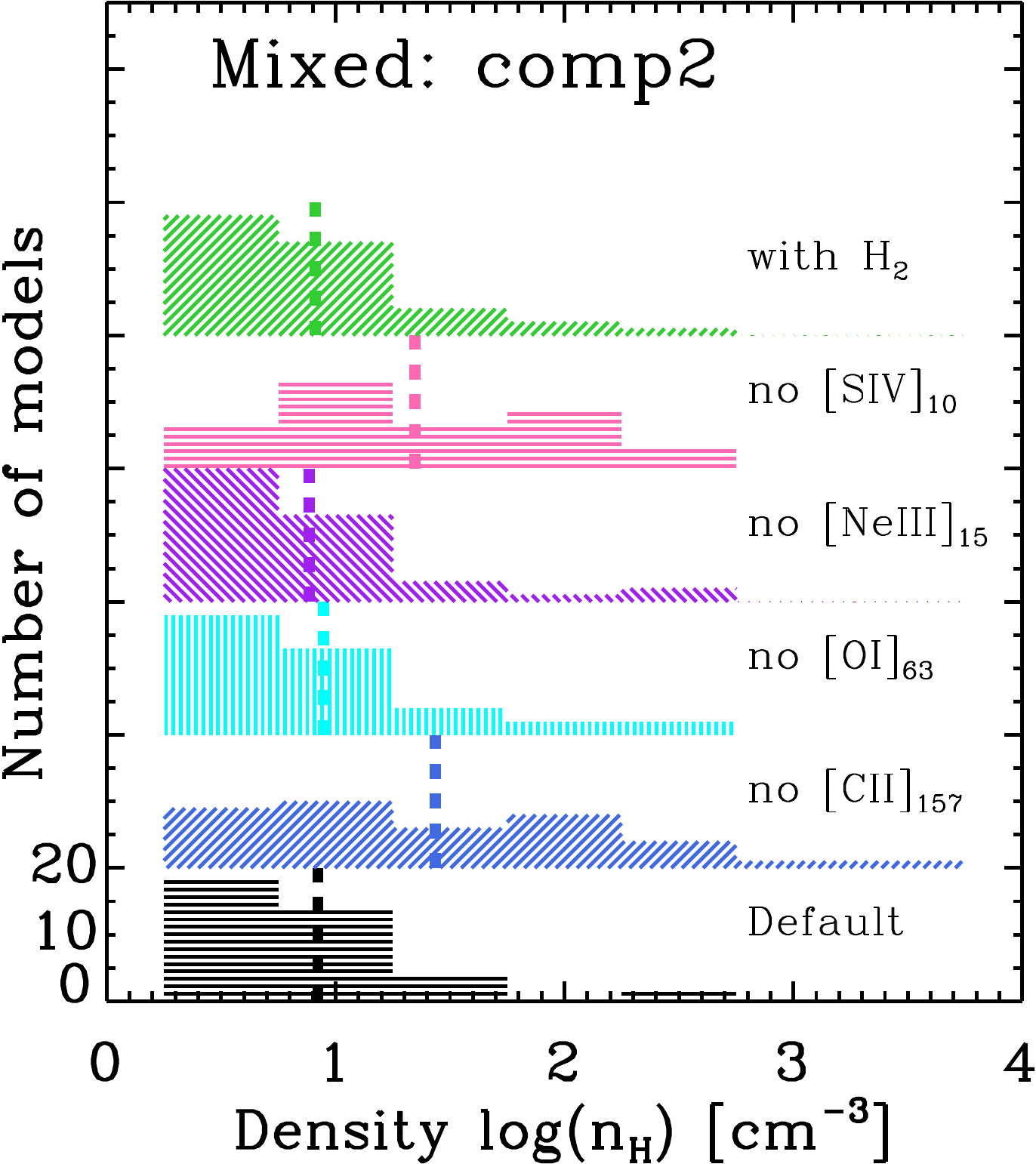}
\includegraphics[clip,width=4.5cm]{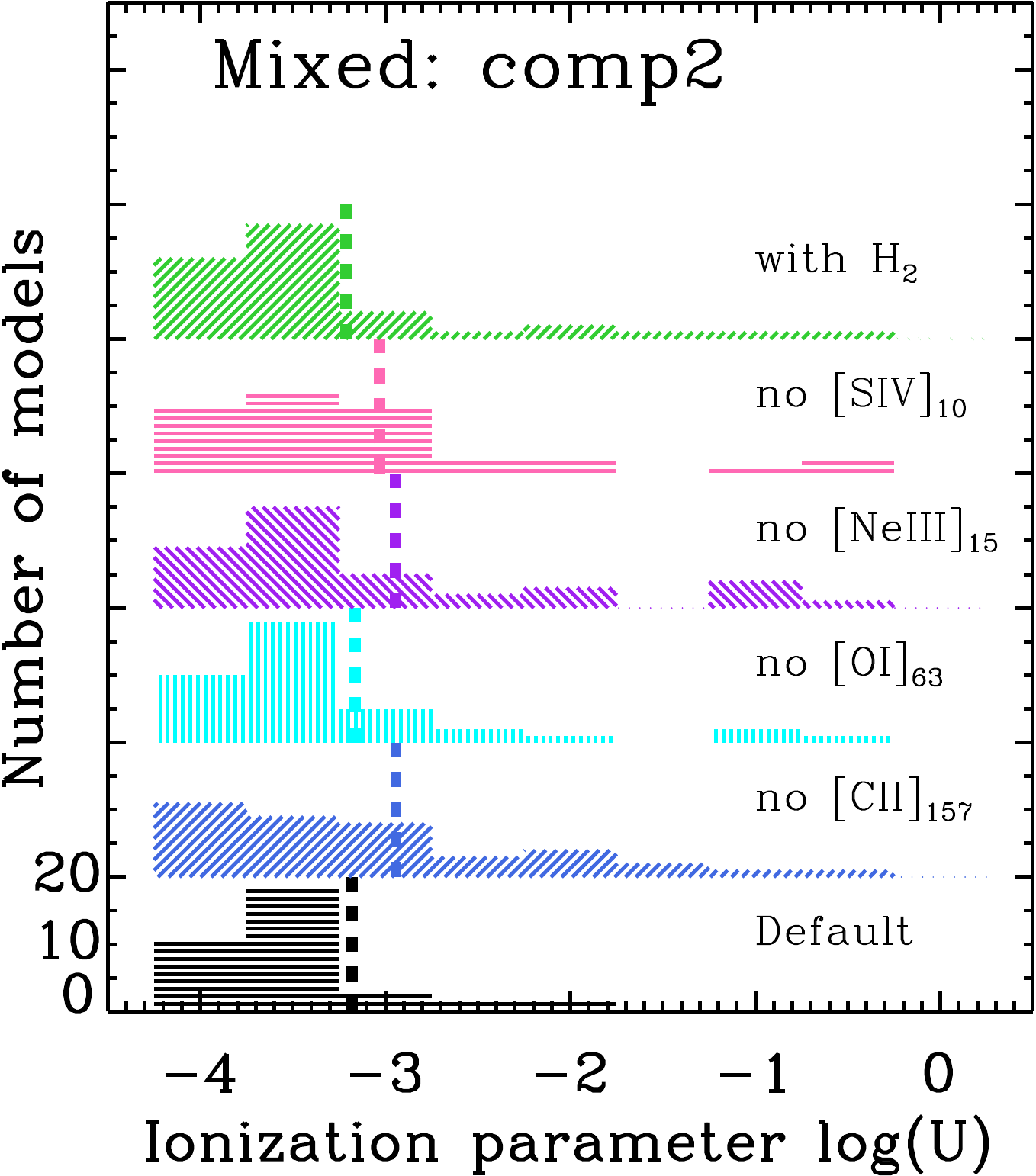}
\caption{
Sensitivity of the model results to the fitted lines.
We show histograms of the reduced $\chi^2$
($\chi^2_{\nu}$) and best-fitting parameters
($n_{\rm H}$, $\log U$, $cov_{\rm PDR}$) for
the entire sample of galaxies.
The results from the default grid of models and
the default set of lines are in black, and the results
are in color if the set of fitted lines is changed, that is,
when one spectral line is excluded or when the \htwo
lines included or when we consider a small, fixed
set of lines to fit (\neiiil, \oiiilb, \ciil, and \ltir). 
The vertical dotted lines indicate the mean values.
Each histogram is offset vertically by a value of $20$.
See Section~\ref{sect:linetests} for details.
}
\label{fig:setlines}
\end{figure*}

The results are shown in Figure~\ref{fig:setlines}.
We plot histograms of the best-fitting parameters and
$\chi^2_{\nu}$ values for both single and mixed models.
We note that all galaxies are included and uncertainties on
the best-fitting parameters as well as quality of the fits are not
conveyed on the plots.
For single models, $\chi^2_{\nu}$ seems to improve
quite significantly when excluding \siv.
For mixed models, the distributions of $\chi^2_{\nu}$
remain similar and the mean $\chi^2_{\nu}$ values
are only slightly lower when excluding a specific line.
Compared to the default case, the distribution of model
parameters for the tests performed have relatively similar
shapes, and individual galaxies typically move to
the neighboring bin at most.
Excluding \oila or \neiiil does not have a significant impact
on the best-fitting parameters, while the distribution of those
parameters change noticeably when excluding \siv or \cii.
Indeed, by excluding \siv, there are fewer galaxies with
best-fitting models that have very low density ($n_{\rm H}=10^{0.5}$\,\cm),
hence the mean density of the mixed, low-density component is larger.
The mean ionization parameters of the single and mixed,
high-density components are also slightly lower, probably
because \siv has one of the highest ionization potentials.
Similarly, when considering the small, fixed set of fitted
lines (which only allows for single models, not mixed models),
the distribution of the best-fitting solutions does not change
significantly but the uncertainties on the model parameters
are larger. There are fewer galaxies at the highest densities,
probably because the \siii lines are not available to constrain
further the density and the mean ionization parameter is
also slightly lower because \siv is excluded from the fit.
We want to stress that the trends of best-fitting parameters
with global galaxy properties (SFR, $Z$) discussed
in Section~\ref{sect:discussion} persist with this test.
By excluding \cii, the distributions of densities shift to larger
(factor of three on average) values and the distribution
of PDR covering factors also shifts to slightly lower values.
This can be understood as the PDR solution is driven by
\ltir and the \oi lines which generally trace the denser,
lower filling factor regions of PDRs than \cii.
Including the warm \htwo lines in the fit does not affect
the overall distributions and mean values of the model
parameters because those lines are detected in only
eight galaxies of our sample. Those lines are difficult to
reproduce though, yielding larger $\chi^2_{\nu}$ values
on average.

\subsection{Choice of input radiation field}
\label{sect:rfchoice}
\begin{figure}[t]
\centering
\includegraphics[clip,width=8.8cm]{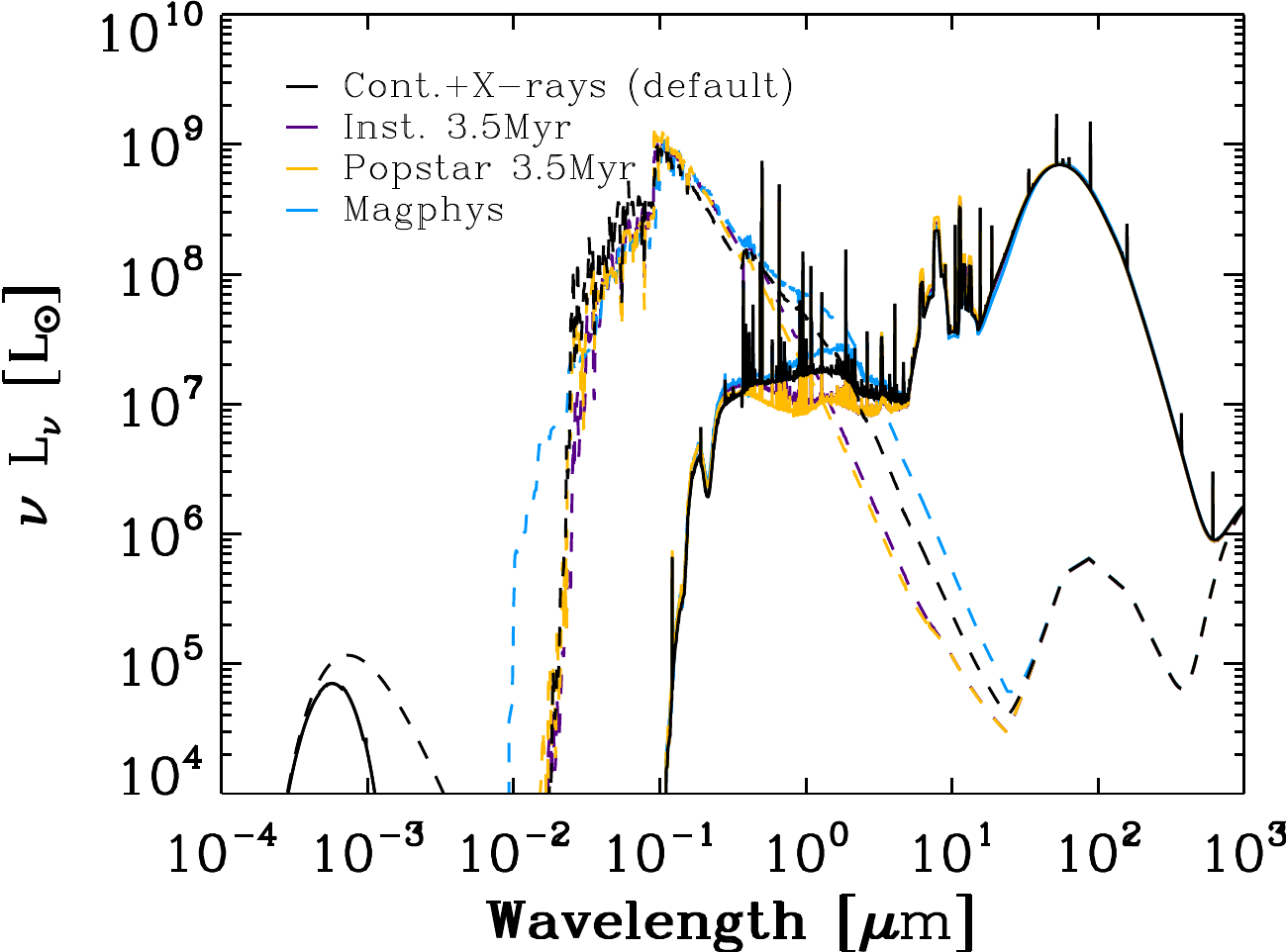}
\caption{
Cloudy input radiation fields tested. 
We show the input (dashed) and output (solid)
spectral energy distributions for:
a continuous star-formation scenario
of 10\,Myr (with an X-ray component),
a Starburst99 instantaneous burst of
3.5\,Myr, a Popstar burst of 3.5\,Myr,
a Magphys fit to the optical-IR photometry.
All have model parameters: $n_{\rm H}=10^2$\,\cm,
$r_{\rm in}=10^{20.4}$\,cm, and $Z_{\rm bin}=0.5$.
}
\label{fig:seds}
\end{figure}

The sources of radiation giving rise to the observed line
emission are multiple and mainly clustered in a few
star-forming regions. Given the limited geometry of
Cloudy, we consider only one central source per model
and need to choose its radiation field. Several input SEDs
were tested (Figure~\ref{fig:seds}). In this section, we
discuss the effects of different starburst spectra and
heating by X-rays and cosmic rays on the line predictions
and best-fitting models.

\subsubsection{Starburst spectrum}
%
We tried as input the Starburst99 stellar spectrum both
from an instantaneous burst with an age between 3 and
6\,Myr, and from a continuous (up to an age of 10\,Myr)
star-formation scenario. In the instantaneous case,
we find that the emission of lines tracing the radiation
field hardness is too sensitive to the step size for
sampling the burst age, making it easy to miss the
``best'' age (see also \citealt{polles-2019}). It is also
probably not the best representation of the cluster
age distribution in galaxies.
Moreover, the PDR lines are better fit with a continuous
star-formation scenario, in part because $G_0$ is too high
with an instantaneous burst (fewer lower-energy photons).
We also tried an instantaneous burst using the Popstar
evolutionary synthesis models \citep{molla-2009}. A different
treatment of the stellar atmosphere can produce significant
differences in the stellar spectrum for photons with energy
$>40$\,eV, hence affecting predictions of lines such as \neiii
\citep{morisset-2009}. However, similar problems persist,
as noted above for the Starburst99 instantaneous case.

For a more complex star-formation scenario, we also tried
to constrain the shape of the incident spectrum by taking
the optical photometry into account. Focusing on Haro\,11,
we used the SED code Magphys \citep{dacunha-2008} to
obtain an unattenuated spectrum and used that spectrum
as input in Cloudy.
However, for the lines that we fit (\hii region and PDR) the
results are not especially improved ($\chi^2_{\nu, {\rm min}}$
was not lower) than with the continuous scenario of Starburst99. 
Moreover, since the available optical photometry is limited for
some of our galaxies, we cannot use Magphys for all galaxies
of our sample. It is important to have a consistent method
in order to study trends. Therefore, we opt for a continuous
star-formation scenario for our grid of models.

\subsubsection{Heating of the PDR and \oi line ratio}
\label{sect:xrcr}
\begin{figure}[t]
\centering
\includegraphics[clip,width=8.8cm]{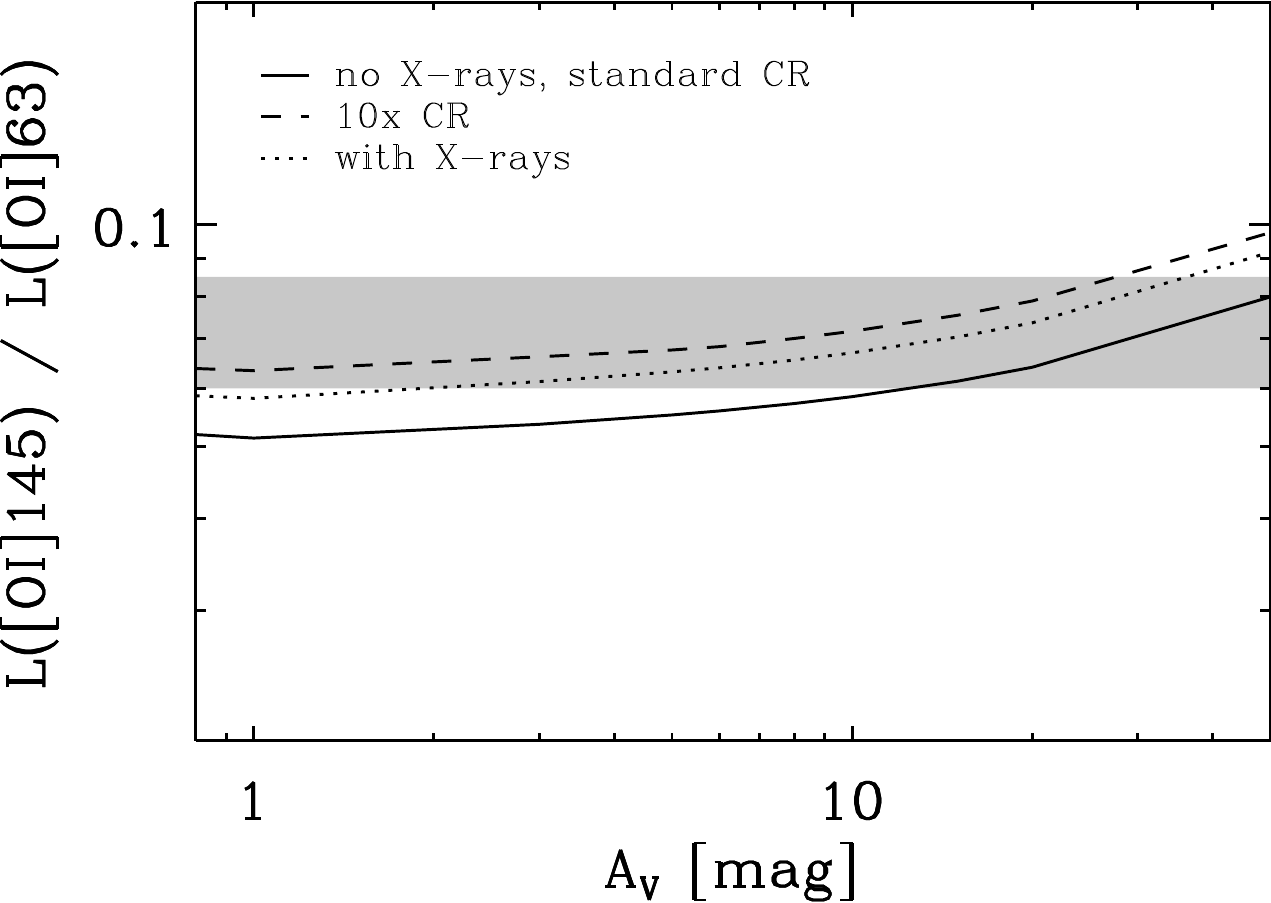}
\caption{
Model predictions for the \oilb/\oila luminosity ratio
as a function of visual extinction in the cloud.
The default model (solid curve) has parameters:
$n_{\rm H} = 10^3$\,\cm, $G_0 = 10^3$, $Z_{\rm bin}=0.5$.
Increasing the cosmic ray rate by a factor of $10$
(dashed curve) or adding a soft X-ray component to
the input radiation field (dotted curve) both increase
the PDR heating and predicted \oi line ratios.
The gray shading indicates the observed range of ratios.
}
\label{fig:oiratios}
\end{figure}

The standard model contains a soft X-ray component of
luminosity $\sim10^{-4}$ times the bolometric luminosity
and cosmic rays. This is primarily needed for understanding
the \oi line emission.
The \oi145\,\mum line is clearly detected in $10$ galaxies
and the \oi63\,\mum line in $30$ galaxies of the sample,
with \oilb/\oila luminosity ratios ranging from $0.06$ to $0.09$
\citep{cormier-2015}.
The \oi63\,\mum line is known to suffer from optical depth
effects or self-absorption in regions of the Milky Way and
dusty galaxies \citep[e.g.,][]{liseau-2006,abel-2007,rosenberg-2015},
giving rise to \oi line ratios above $0.1$. The emergent
line ratio is somewhat sensitive to the depth (\av) of
the models \citep{abel-2007}. 
Other heating sources than the stellar radiation field
can also affect the PDR chemistry and line emission,
such as cosmic rays or X-ray or shock heating.
We explored the effect of high column density (high \av),
cosmic rays, and soft X-ray heating in Figure~\ref{fig:oiratios}.

With no additional heating source than the stellar spectrum,
the observed range of \oi line ratios requires systematically
high \av, at which the CO optical depths are also very large,
and this seems unlikely. Indeed, the mean \av is expected to
decrease with decreasing mean metallicity, and CO emission
is known to be clumpy at low $Z$, suggesting that the \oi emission
integrated over entire galaxies is probably little affected by
high-\av ($>10$\,mag) sightlines.

Soft X-ray emission (0.2-2\,keV) is detected in several galaxies
of our sample, with an observed luminosity about $10^4$ times
lower than \ltir (except in I\,Zw\,18 which is strong in X-rays;
see \citealt{lebouteiller-2017}). The intrinsic (unabsorbed)
emission should be higher, however we do not know how much
of that emission goes into the heating of the dense ISM
(as opposed to diffuse isotropic emission). Hence we decide
to add a soft X-ray component to the input radiation field
of our models that we take as a blackbody spectrum
of temperature $5\times10^6$\,K (to peak around 1\,keV)
and luminosity in the soft X-ray range $10^{-4}\,L_{\rm BOL}$.
Adding the X-rays creates a hybrid XDR/PDR model
in Cloudy.
X-rays generally do not affect the predictions of the
inner ionized gas cloud but increases the heating
at the ionization front and in the PDR \citep[e.g.,][]{
meijerink-2005,lebouteiller-2017}. In the presence of
strong X-ray flux, the deeper penetrating X-rays produce
more Ne$^+$ and Ar$^+$, while the increased ionization
has an effect on CO, suppressing its formation due to reactions
with He$^+$ and H$_3^+$ \citep[e.g.,][]{meijerink-2005,abel-2009}.
This produces more free oxygen and a higher gas temperature.
In our case, the X-ray component has low luminosity.
The main effect is that the PDR temperature increases
roughly from 150\,K to 200\,K. The emission of the PDR
lines increases slightly, as well as the \neii and \arii emission
for models with low-density or high-$U$.
In the case tested in Fig.~\ref{fig:oiratios}, the predicted
\oi line ratios go up by a factor of $1.2$.

Regarding the cosmic rays, we show in Fig.~\ref{fig:oiratios}
the effect of increasing the standard cosmic ray rate
($2\times10^{-16}$\,s$^{-1}$; \citealt{indriolo-2007})
by an order of magnitude. Similarly to the case with soft
X-rays, the PDR heating and line emission are enhanced.
The cosmic ray rate varies significantly from sightline to sightline
in the Galactic ISM \citep[e.g.,][]{dalgarno-2006,neufeld-2017},
and the dwarfs of our survey have undergone recent star
formation, hence it seems reasonable to adopt a different
value than the standard rate. To match the observed \oi line
ratios, we decided to increase the cosmic ray rate by a factor
of three.

We note that cosmic rays and X-rays have similar
effects (ionization and heating) and distinguishing between
the two with no other constraints is not possible.
One way to constrain the cosmic ray flux would be to
study its effect on the ionization and coupling of the magnetic
field and the gas in the PDR \citep{padovani-2011}.

\subsection{Turbulent velocity}
The micro-turbulent velocity affects the shielding and
pumping of lines. It is included as a pressure term in
the equation of state. Hence, we expect that it impacts
mainly predictions of PDR and high optical depth lines.
We set the turbulent velocity $v_{\rm turb}$ to 1.5\,\kms
in the default grid of models, as in \cite{kaufman-1999},
and here, we explored the cases $v_{\rm turb}=0$\,\kms
and $v_{\rm turb}=5$\,\kms for the grid of models in the
metallicity bin $Z_{\rm bin}=0.5$.
We find that, by increasing the turbulent velocity from 0\,\kms
to 5\,\kms, the PDR temperature reduces marginally by about
10\%, the \oi line intensities are down by 15\%, the \cii intensity
is up by 20\%, and the \htwo lines are down by $\sim$30\%.
Overall, the turbulent velocity does not have a significant
impact on our resulting best-fitting models and therefore
on our trend analysis.

\subsection{Density law}
\label{sect:denslaw}
One of the model choices with the largest consequence
is the density profile to adopt. As default, we adopted
a density profile that increases smoothly with depth
\citep[see also][]{hosokawa-2005,wolfire-2010}.
Here we have tested, for the grid of models in the metallicity
bin $Z_{\rm bin}=0.5$, two other often-used options:
a constant density throughout the \hii region and the PDR,
and hydrostatic equilibrium. Hydrostatic equilibrium
assumes that the total pressure (including gas, turbulent,
radiation pressure terms) is constant and implies a jump
in density at the transition between the \hii region and
the PDR (of almost two orders of magnitude when
gas pressure dominates, see Fig.~\ref{fig:griddens}).
The adopted density law affects mostly the PDR
predictions with the constant density models predicting
less \oi emission and the constant pressure models
predicting more \oi emission than the default grid.
For Haro\,11, the dense component of the mixed model
prefers higher initial densities in constant density and lower
initial densities in constant pressure. However, we have
tested the constant density and constant pressure hypotheses
on eight other galaxies at $Z_{\rm bin}=0.5$ (Haro\,2, Haro\,3,
II\,Zw\,40, Mrk\,1089, NGC\,625, NGC\,1705, NGC\,5253,
UM\,448), and those trends are not systematic.
In general, we find that the constant density/pressure single
models do not provide better fits to the observations. With
mixed models, the constant pressure case provides (very
marginally) lower $\chi^2_{\nu, {\rm min}}$ than the default
grid for eight of those nine galaxies.
Overall, the smoothly increasing density law (adopted
as default) is a reasonable choice for single models.
It also provides satisfying fits for mixed models, although
constant pressure models could also be considered
but there is no clear evidence for the need to switch to
constant pressure.

\subsection{Abundance of PAHs and grains}
\label{sect:testdust}
Grains and PAHs are important for the energy balance, in
particular for the cooling of the neutral ISM and heating via
the photoelectric effect. The adopted abundances of grains
and PAHs are motivated by recent analysis of the \spit and
\hers data \citep{remy-2014,remy-2015}. However, there is
both large scatter at all metallicities investigated ($0.03 \le
Z \le 3$) and a limited number of observations at $Z\le0.2$.
For example, the relation linking the DGR with metallicity
seems to become steeper at $Z<0.2$ than the relation
calibrated on $Z\ge 0.2$ galaxies \citep[e.g.,][]{remy-2014}.
Concerning the PAHs, their abundance in the PDRs 
of dwarf galaxies is very uncertain as they are scarcely
detected for $Z<0.2$ with the \spit IRS instrument and
we expect a large variation of the PAH-to-dust ratio between
\hii regions and PDRs. We note that, if dust is present in
the ionized gas but PAHs are weak there, a lower PDR
covering factor at low $Z$ (see Section~\ref{sect:disc_cov})
would significantly influence the globally observed
PAH-to-dust emission ratios. Constraints on PAHs from
the {\it James Webb Space Telescope} (JWST) will be
very valuable.
Departure from the default values and thus the effect on
the results will be larger at low $Z$. Hence we carried out
two tests for $Z_{\rm bin}=0.1$ (and not $Z_{\rm bin}=0.5$
as in the previous subsections): $(a)$~we adopt a PAH
abundance $10\times$ lower; $(b)$~we adopt a DGR
$10\times$ lower than in Table~\ref{table:params}.

Results for the galaxies Mrk\,209, Pox\,186, SBS\,1159,
SBS\,1415, Tol\,1214, UM\,461, VII\,Zw\,403 (all at
$Z_{\rm bin}=0.1$) are inspected.
Test~$(a)$ does not change any of the results, probably
because the PAH abundance was already low by default
hence the PAHs contribute very little to the energy
balance of those galaxies. Dust grains are responsible
for the heating. 
Test~$(b)$ affects the energy balance noticeably.
By reducing the DGR by a factor of ten and therefore
the dust grain abundance, the photoelectric effect
remains the main heating source in the PDR but
it is less important, and the PDR temperature goes
down from $\simeq150$\,K to 80\,K.
The cloud is also more transparent and the stopping
criterion of \av of 5\,mag corresponds to a larger
depth. The emission of the \cii and \oi lines increases
by a factor of $\sim2.5$ and $1.4$, respectively.
This grid with lower DGR at $Z_{\rm bin}=0.1$ still
does not cover the highest observed \cii/\oila ratios
(lower, right panel of Fig.~\ref{fig:gridres}.
We find that the $\chi^2_{\nu, {\rm min}}$
values are lower, especially for SBS\,1159 and
Tol\,1214, and marginally for three other galaxies.
The best-fitting models have slightly higher densities
for four of the seven galaxies in this metallicity bin,
but the best-fitting solutions do not change much
and there are no clear systematic trends for all
galaxies (in particular for the PDR covering factor
which is discussed later).

In conclusion, we find that additional heating sources
are not required to reproduce the PDR emission, as
opposed to what was found by \cite{lebouteiller-2017}
for I\,Zw\,18, where X-rays are mainly responsible
for the PDR heating and not grains. For this particular
galaxy, our adopted DGR of $\simeq 2 \times 10^{-4}$
might be too high since \cite{lebouteiller-2017} adopt
a value of $\simeq 7 \times 10^{-6}$.
We also conclude that the abundance of PAHs is
not an important parameter for the models (as long
as the dust grain abundance is larger). The dust
grain abundance (and DGR) does, however,
change the temperature structure of the cloud.
For our study, this does not have a noticeable
impact on the results, but additional PDR observations
(such as MIR \htwo lines for which line ratios are
expected to change) and better constraints on the
DGR would be needed for a more thorough study
of the PDR properties.

\section{Discussion}
\label{sect:discussion}
\subsection{Origin of \cii emission}
\label{sect:disccii}
All of the ionic lines used in this study have critical densities
for collisions with electrons larger than that of C$^+$, making
it difficult to constrain a low-density, low-excitation ionized
phase from which \cii emission could arise.
Optical lines such as [S\,{\sc ii}] are often used
as density tracers, but they probe a similar range of
densities as \siii, i.e., 100 to 10,000\,\cm, as well
as shocks \citep{osterbrock-book}.
\niila is the best line that we have in hand but, in principle,
\niilb is better suited (see Table~\ref{table:general} for
critical densities).
\cite{croxall-2017} and \cite{diaz-santos-2017} carried out
studies on the origin of \cii emission in nearby late-type spiral
and luminous galaxies based on \niila and \niilb. In those
galaxies, \nii line ratios that are indicative of electron densities
between 1 and 300\,\cm with median values at 30 and 41\,\cm,
(\citealt{herrera-camus-2016} and \citealt{diaz-santos-2017},
respectively; see also \citealt{herrera-camus-2018a}).

 \begin{figure}[t]
\centering
\includegraphics[clip,width=8.8cm]{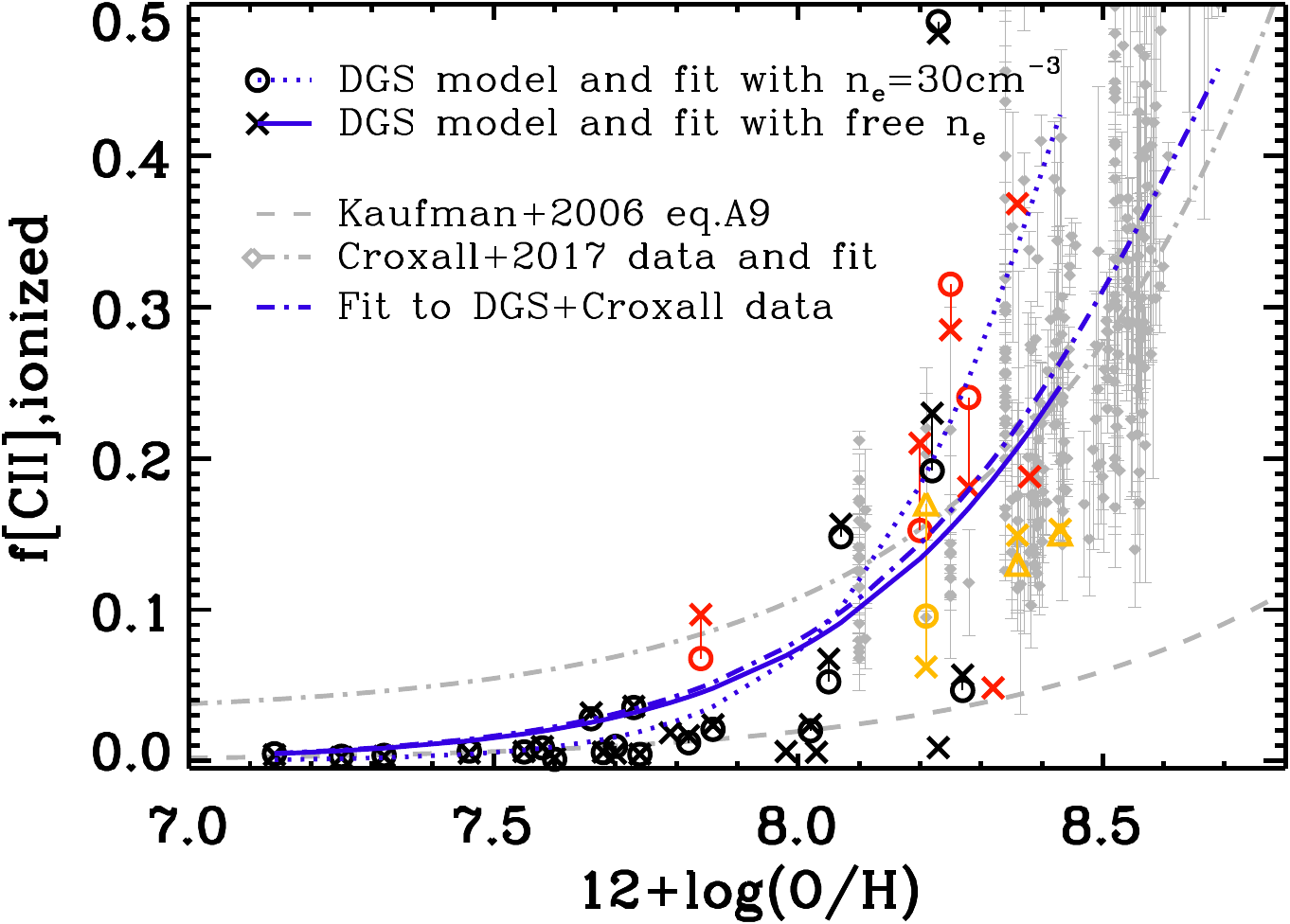}
\caption{
Fraction of \cii emission coming from the ionized gas phase.
We show predictions for each DGS galaxy based on:
(1)~the best single or mixed models reported in
Tables~\ref{table:modelres1} and \ref{table:modelres2} (crosses), and
(2)~the best single or mixed models with one model having
a fixed low density of 30\,\cm (circles); predictions
yielding $\chi^2_{\nu} > 3$ (all models are bad)
are not plotted.
Red symbols indicate that \niila is detected and orange
symbols indicate that both \niila and \niilb are detected.
For the three galaxies where \niilb is detected, we also
show the predicted values from simple theoretical
calculations, as done in \protect\cite{croxall-2017}.
We overlaid the data (diamonds) and fit from \protect\cite{croxall-2017}
as well as the prediction from PDR models from
\protect\cite{kaufman-2006} in gray.
Fits to the DGS data and to the DGS (free density) and
\protect\cite{croxall-2017} data are shown in blue.
}
\label{fig:ciifraction}
\end{figure}
%
%
 \begin{figure*}[t]
\centering
\includegraphics[clip,width=4.8cm]{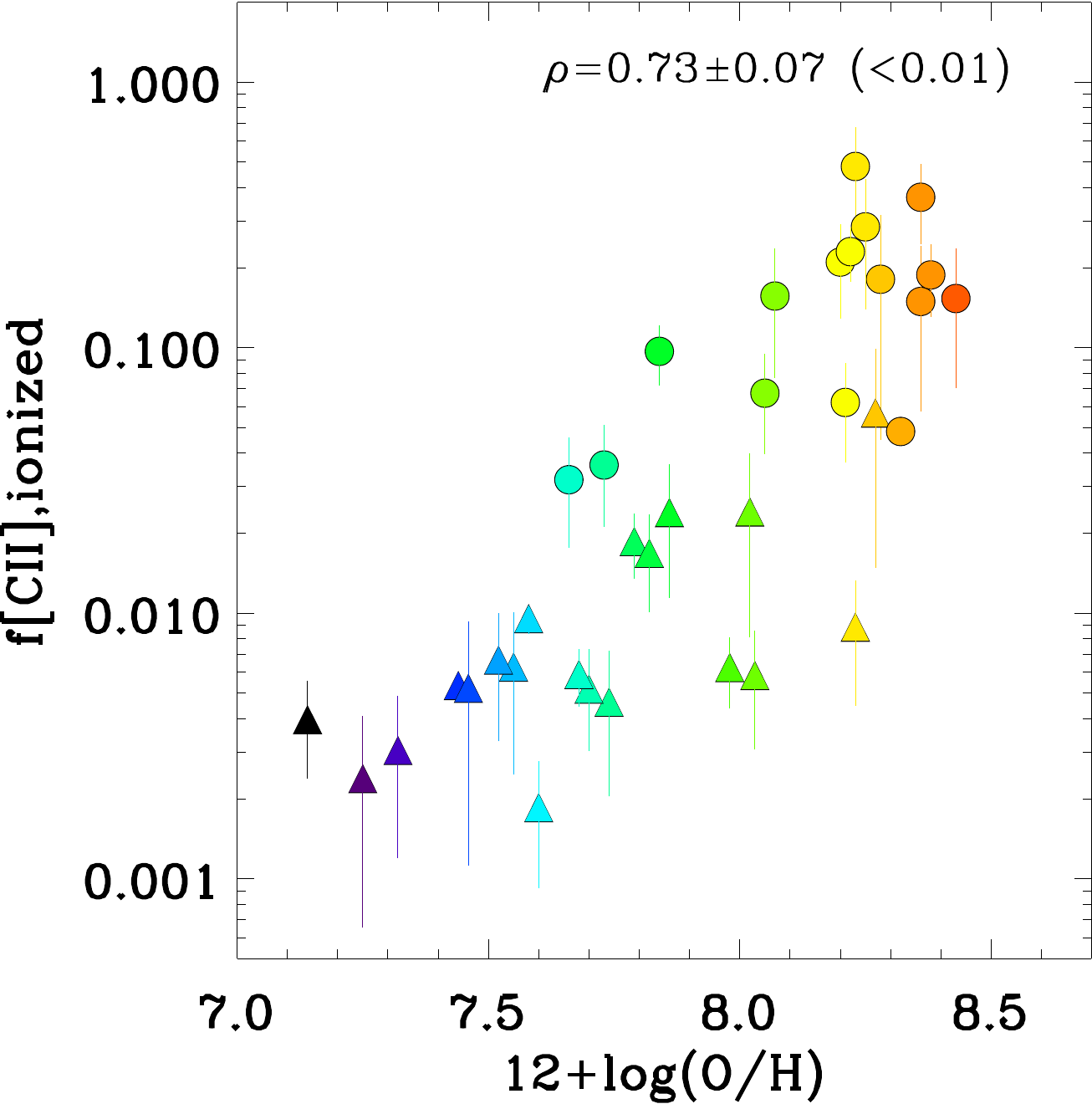} \hspace{3pt} \vspace{3mm}
\includegraphics[clip,width=4.8cm]{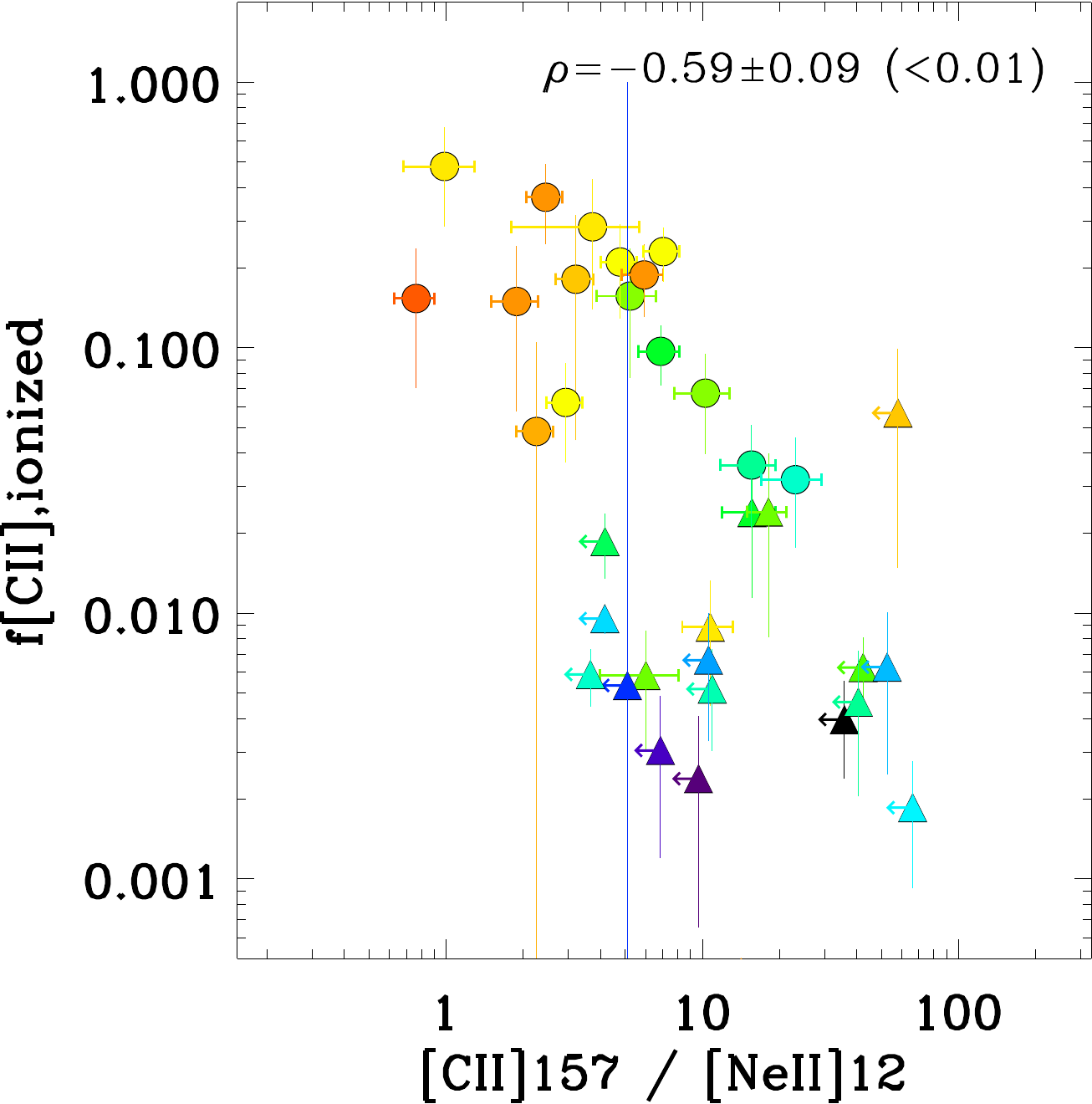}
\includegraphics[clip,trim=28mm 0 6mm 0,width=3.8cm]{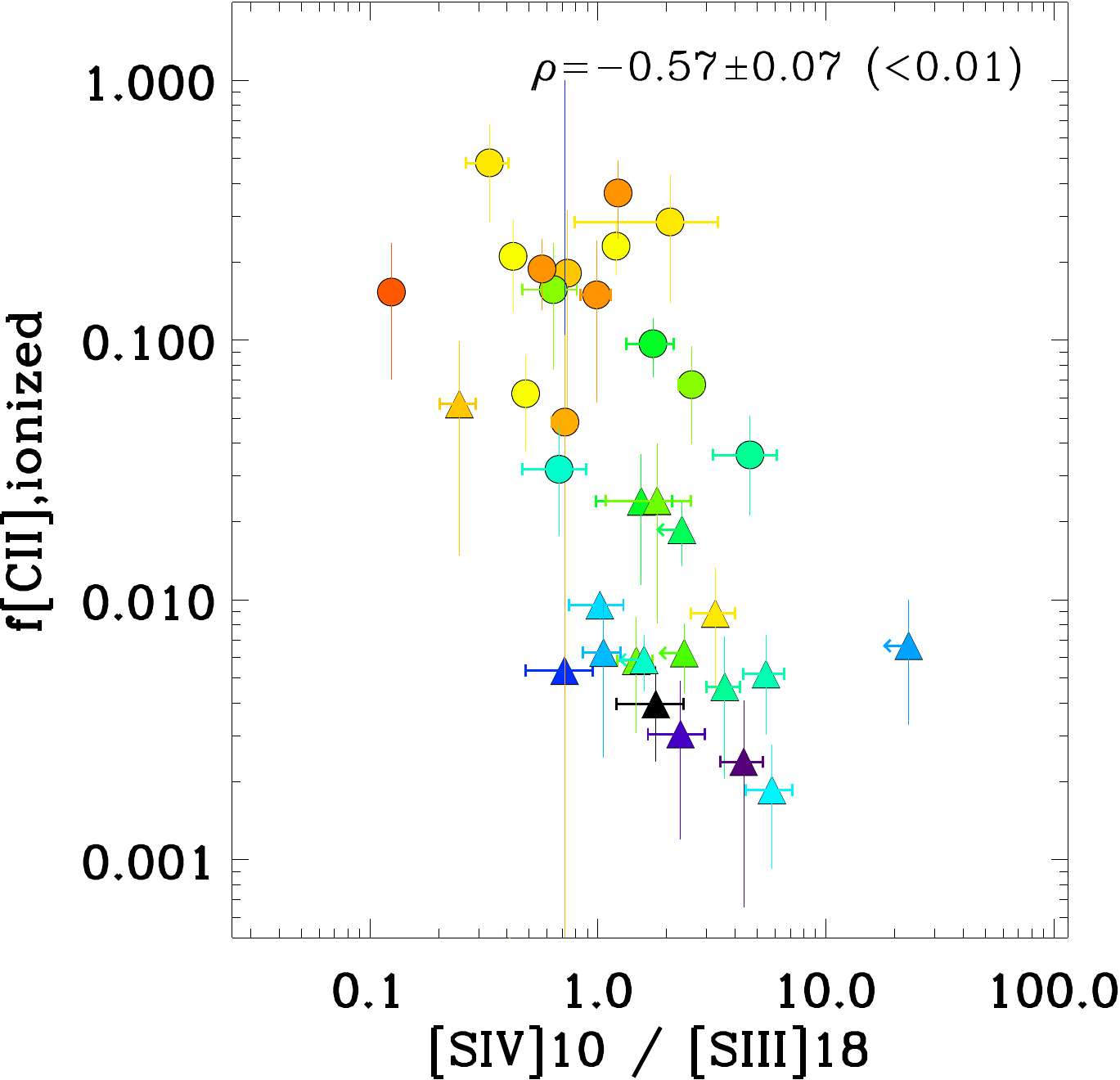} 
\includegraphics[clip,trim=28mm 0 0 0,width=3.8cm]{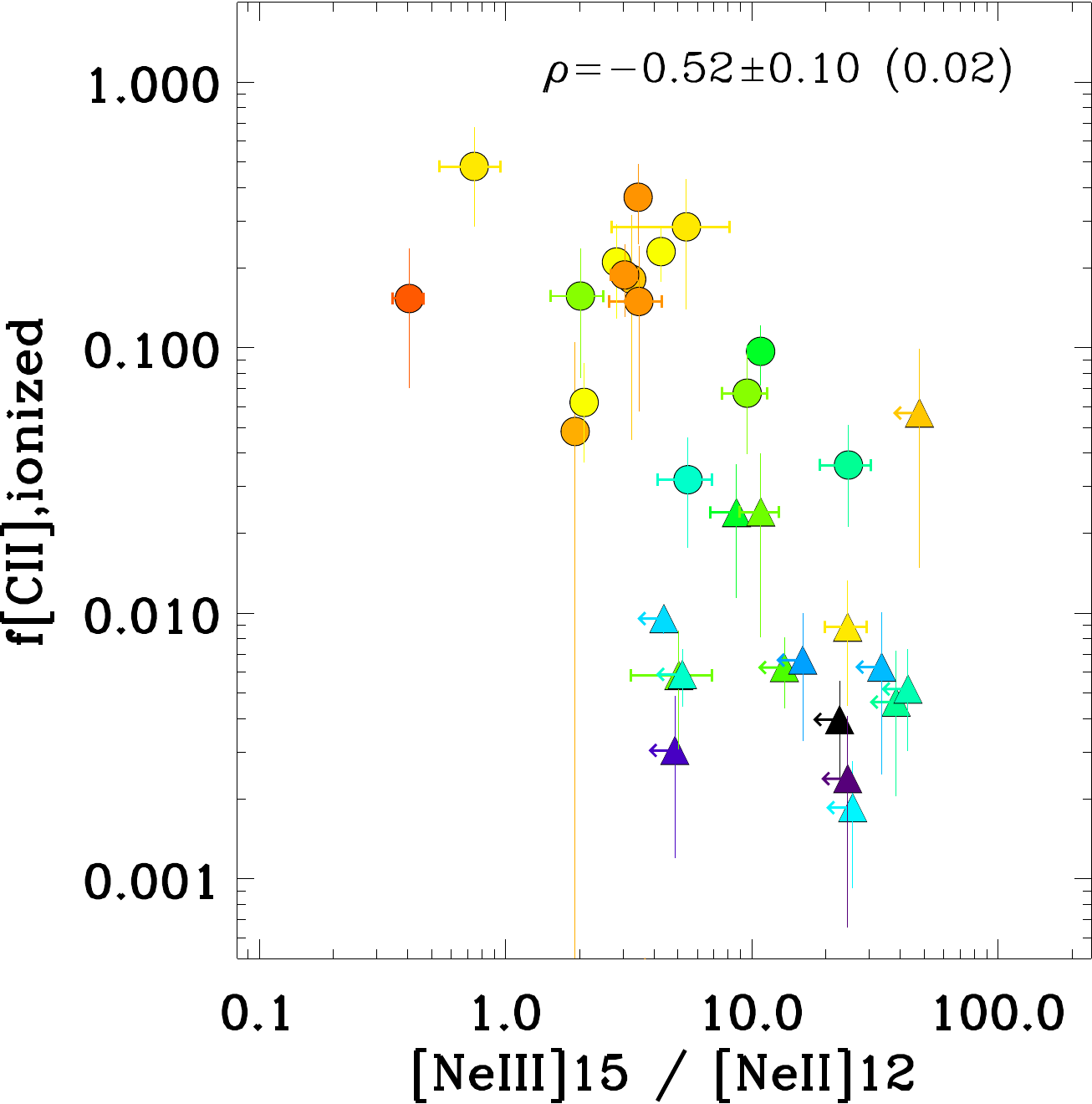}
\includegraphics[clip,width=4.8cm]{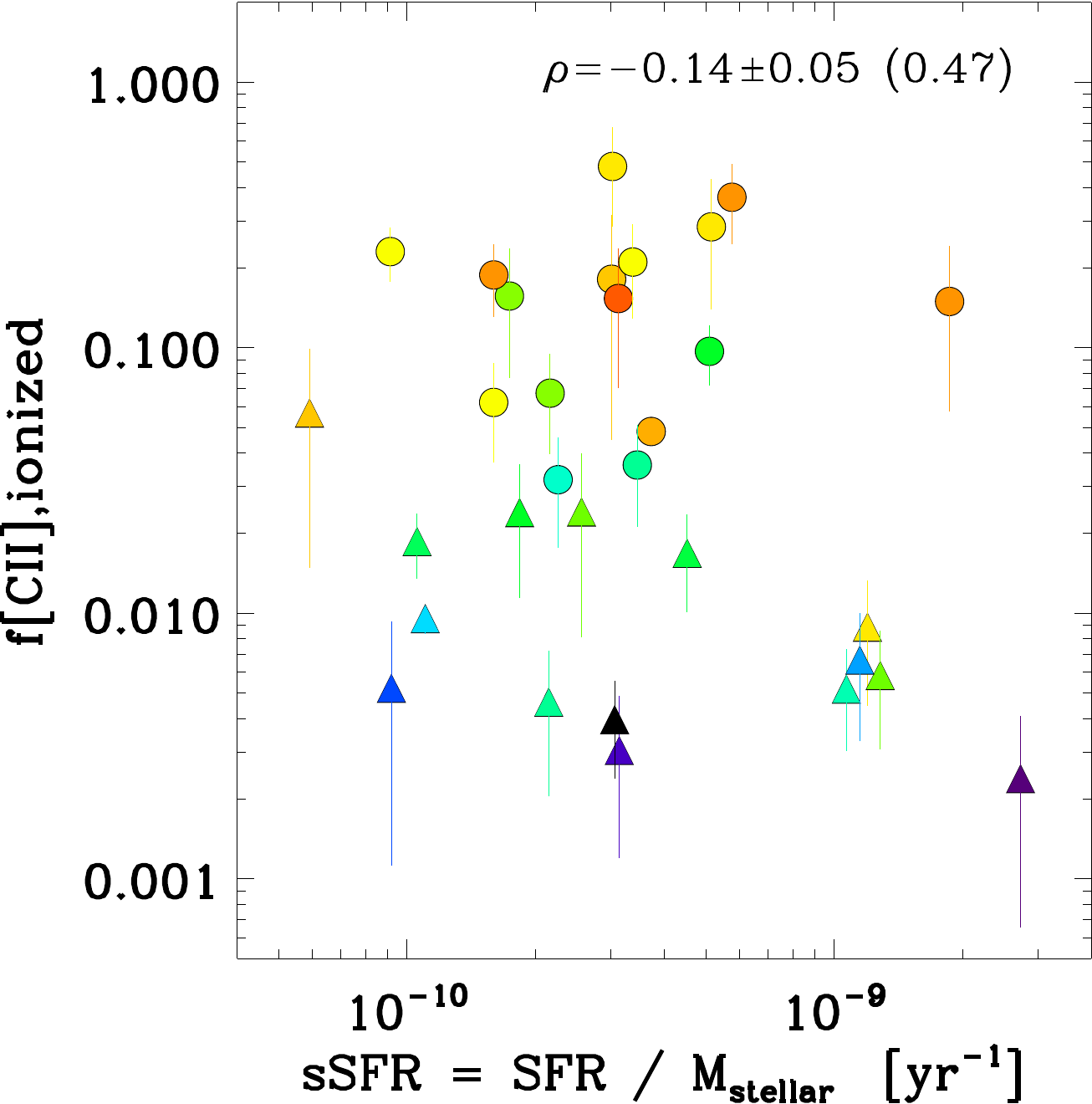} \hspace{3pt}
\includegraphics[clip,width=4.8cm]{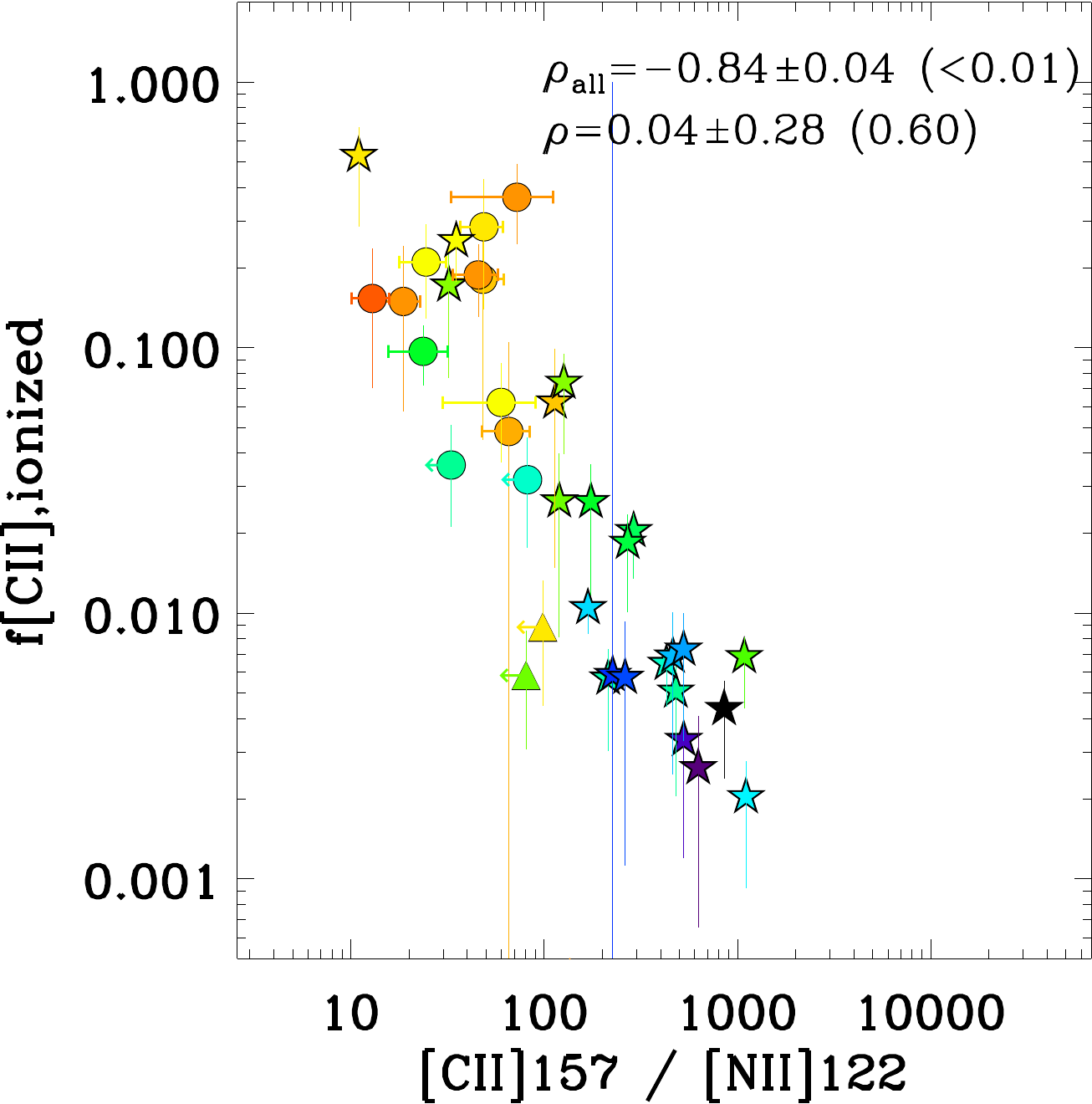}
\includegraphics[clip,trim=28mm 0 2mm 0,width=3.82cm]{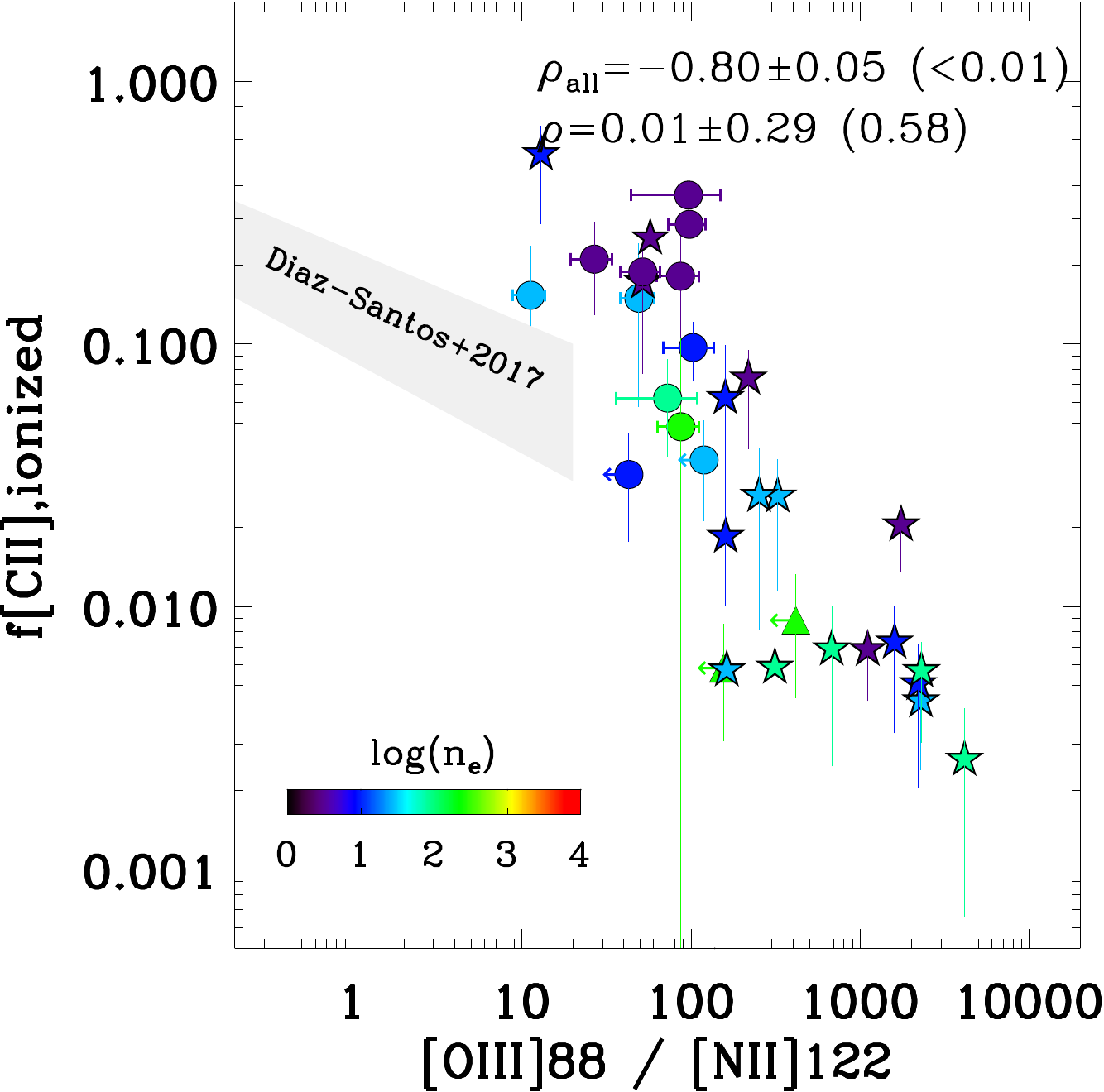}
\includegraphics[clip,trim=28mm 0 0 0,width=3.89cm]{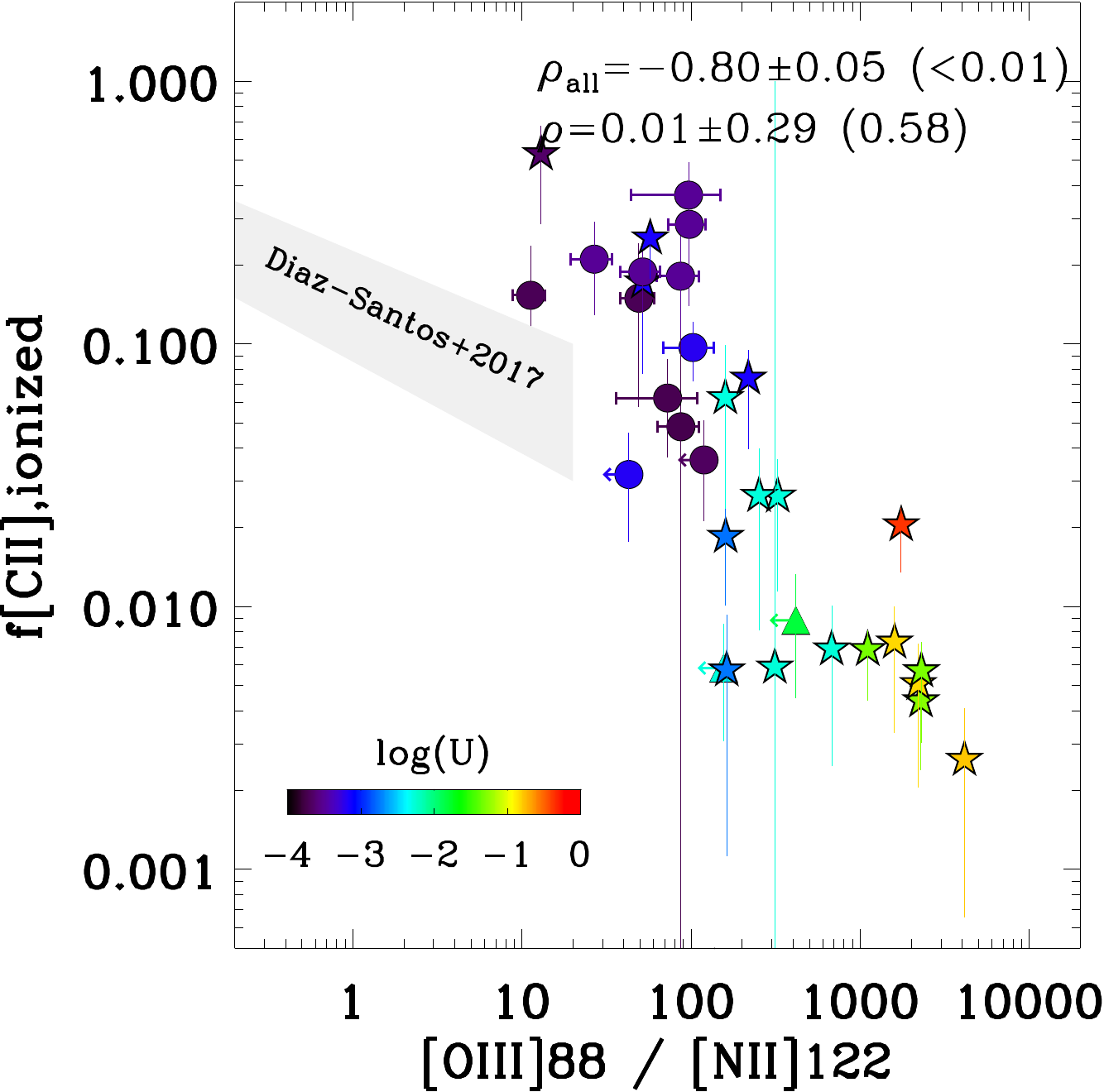}
 \caption{
Correlation of the fraction of \cii emission from the ionized gas with:
metallicity, specific star-formation rate, and line ratios for which
the strongest trends are found.
Values for the y-axis are taken from the models in boldface 
in Tables~\ref{table:modelres1} and \ref{table:modelres2}
(triangles for single models, filled circles for mixed models). 
Color coding corresponds to metallicity, except for the last
two panels which are color coded by electron density and
ionization parameter (of component \#2 in case of mixed models).
Spearman's rank correlation coefficient and significance of its deviation
from zero in parenthesis are indicated in the top-right corner ($\rho$).
In the last three panels, star symbols are the ratios of observed
\ciil or \oiiilb intensity divided by the predicted \niila intensity.
The correlation coefficients $\rho_{\rm all}$ include those
star symbols.
}
\label{fig:fcii_corr}
\end{figure*}

%
Because of the issue of constraining the low-density
ionized gas, it is interesting to compare our results
on densities and fractions of \cii emission from the ionized
gas with those obtained from theoretical calculations based
on the \nii lines. We have also explored the sensitivity of those
parameters to a fixed-density ionized gas component
in the models.
\begin{itemize}
\item
{\it Inferred electron densities from the \nii line ratio only:}\\
\niila is observed in seven of our galaxies and \niilb in only
three of our galaxies: Haro\,11, He\,2-10, and NGC\,4214-c.
The \niilb/\niila ratios are observed to be $0.67$, $0.35$,
and $0.4$, respectively and are matched by models
with densities between $30$ and $80$\,\cm.
\item
{\it Comparison of observed \nii emission with models:}\\
The single models often tend to under-predict \niila
while the mixed models match \niila observations
better. In Haro\,11, He\,2-10, and NGC\,4214-c, we also
fit for the \niilb line and the best mixed models reproduce
successfully both \nii lines, which emission arises from
the low-$U$ ($\log(U)\simeq-3.9$), low-density component
($n_{\rm H}$ of $30$, $30$ and $100$\,\cm, respectively,
values close to those inferred from the \nii line ratio only).
\item
{\it Forcing a low-density component in the models:}\\
For galaxies without \nii observations, we still would like
to estimate how much \cii emission could arise from
such low-density, low-excitation ionized phase.
Here, we assume $n_e \simeq 30$\,\cm.
We derived the fraction of ionized \cii by forcing the
single models or one component in our mixed models
to have a density of 30\,\cm and an ionization
parameter lower than that of the other component.
\end{itemize}
The results on the \cii emission arising from the
ionized gas are presented in Figure~\ref{fig:ciifraction}.
For the three galaxies with \niilb detected, we first estimated
the fraction of \cii emission from the ionized gas from
theoretical calculations, based on statistical equilibrium
of collisionally excited gas (see, e.g., \citealt{oberst-2006}
and \citealt{croxall-2017}).
Those fractions range from $13$\% to $17$\% (triangles in
Fig.~\ref{fig:ciifraction}) and are close to those derived from 
our models (circles and crosses in Fig.~\ref{fig:ciifraction}),
though on the high side for NGC\,4214-c.
For all galaxies, we then compared estimates from our
models with free $n_{\rm H}$ and with fixed, low $n_{\rm H}$.
In all cases, we find that the fraction of \cii emission
coming from the ionized phase is at most 50\%.
In galaxies where \nii is observed, the models predict
lower fractions of \cii in the ionized phase than
predictions from \cite{croxall-2017} because our observed
\nii/\cii ratios are on the low side of most of their observations.
Predictions based on assuming a fixed low-density component
($n_{\rm H}=30$\,\cm) can be unreliable because
they sometimes yield much higher $\chi^2$ than when
$n_{\rm H}$ is let free. Those predictions are not plotted
in Fig.~\ref{fig:ciifraction} if they yield $\chi^2_{\nu} > 3$.
Based on our models, we also see a trend of decreasing
fraction of \cii emission arising in the ionized gas with
decreasing $Z$. The trend is similar to that reported
by \cite{croxall-2017} and extends to lower $Z$ values.
For $Z \le 0.2$, our observations agree better with
predictions from \cite{kaufman-2006} (their equation
A9 for a low-density medium) than with predictions
from \cite{croxall-2017} (extrapolated to lower $Z$).
Following \cite{kaufman-2006}, we fit the DGS data
only as well as the DGS and \cite{croxall-2017} data
with a function of the form
\begin{equation}
f[{\rm CII}]_{\rm ionized~gas} = \frac{1}{1+a\times Z^b},
\end{equation}
where $Z$ is the metallicity normalized to the solar value
(i.e., $12+\log([{\rm O/H}]_{\odot}$) = 8.7). The metallicities
of both datasets are based on the \cite{pt05} calibration.
For the DGS galaxies and with $n_e \simeq 30$\,\cm,
we find coefficients $(a,b)\simeq(0.3,-2.3)$ (blue dotted
curve in Fig.~\ref{fig:ciifraction}). For the DGS and
DGS+\cite{croxall-2017} galaxies with free $n_e$, we find
coefficients $(a,b)\simeq(1.2,-1.4)$ (solid blue and
dash-dotted blue curves in Fig.~\ref{fig:ciifraction}).

In Figure~\ref{fig:fcii_corr}, we present correlations
of the fractions of \cii emission from the ionized gas,
$f[{\rm CII}]_{\rm ionized~gas}$, obtained from our
models with $n_{\rm H}$ free, with metallicity, specific
star-formation rate (sSFR) and several combinations
of observed line ratios to see if it is possible to predict
the \cii fractions more easily (from observables rather
than modeling).
We quantified correlations using the IDL procedure
\texttt{r\_correlate.pro} which outputs the Spearman's rank
correlation coefficient and the two-sided significance of its
deviation from zero. Here, we vary quantities within their
uncertainties a 1,000 times to obtain a distribution of correlation
coefficients. The plots report the median value $\pm$ the
standard deviation of the distribution, as well as the median
value of the significance distribution in parenthesis.
As seen in Fig.~\ref{fig:fcii_corr} (and already noted in
Fig.~\ref{fig:ciifraction}), $f[{\rm CII}]_{\rm ionized~gas}$ is
correlated with metallicity. $f[{\rm CII}]_{\rm ionized~gas}$
is not correlated with sSFR.
Regarding observed line ratios, we find the strongest trends
with \cii/\neii, \neiii/\neii, and \siv/\siiila. We also find strong
trends of $f[{\rm CII}]_{\rm ionized~gas}$ with \cii/\niila and
\oiii/\niila when taking into account predictions of \niila in
addition to observations. The trends with \neiii/\neii, \siv/\siiila,
and \oiii/\niila can be understood as those ratios trace
the average radiation field strength and, indirectly,
metallicity. Regarding the \oiii/\niila ratio, we overlay in
Fig.~\ref{fig:fcii_corr} the fit of \cite{diaz-santos-2017} for
local luminous galaxies. We find a similar but offset trend,
the dwarfs probing higher ratios of \oiii/\niila. 
We do not find any trend of $f[{\rm CII}]_{\rm ionized~gas}$
with the \siii line ratio, which is a density tracer, though
sensitive to higher densities than the critical density of \cii
with electrons.
From our model results, we find a weak anticorrelation
between electron densities and $f[{\rm CII}]_{\rm ionized~gas}$,
but with a lot of scatter. Hence, as in \cite{diaz-santos-2017},
$f[{\rm CII}]_{\rm ionized~gas}$ correlates more strongly
with the radiation field intensity than with the electron density
(conveyed in the color coding of the last two panels of
Fig.~\ref{fig:fcii_corr}, and see also Fig.~\ref{fig:append-c}
in the appendix).

\cite{accurso-2017a} and \cite{olsen-2017} investigate
the \cii emission based on simulations of galaxies.
In general, they find larger fractions of \cii in the ionized
gas than we find here. Equation\,16 of \cite{accurso-2017a}
provides a prescription of the fraction of \cii associated
with PDRs that depends (only) on a galaxy's sSFR.
This prescription predicts that 55-75\% of the \cii emission
in our galaxies should arise from the PDR, which is
somewhat coherent with our results. However, their
Equation\,12, which also depends on the dust mass
fraction, $Z$, and electron density that we set to 30\,\cm,
brings those fractions below 50\%. \cite{olsen-2017} also
find that the \cii emission from PDRs is below 50\% in
simulations of high-redshift star-forming galaxies.
Those fractions of \cii emission arising from the PDR are
low compared to our results and to \cite{croxall-2017}.
This discrepancy might be related to the fact that a
two-phase model is a simplistic view of a galaxy, or
alternatively that prescriptions or simulations are not
representative of our population of galaxies, each galaxy
being dominated by a few localized star-forming events.

Finally, our results may have implications for the use of
\cii as a SFR indicator. Most of the \cii emission arises in
PDRs at low metallicities, which indicates that the \ciil line
could still be a good tracer of the SFR in the low-$Z$ galaxies
at high redshift \citep[see also][]{stacey-2010,vallini-2015,
lagache-2018}. \cite{delooze-2014} found an offset
in the \cii-SFR relation between nearby dwarf galaxies, and
normal star-forming spirals and IR-luminous galaxies. This
offset might be linked to contamination of the ionized gas to
the observed \cii emission, resulting in higher \cii level of
emission for a given level of SFR in metal-rich galaxies.
Given our results and those of \cite{croxall-2017} on
higher-$Z$ galaxies, the contamination of the ionized gas
to the observed \cii emission remains however moderate.
Another important factor to consider in the \cii-SFR
relation is the \cii line deficit (with respect to \ltir) which
varies by orders of magnitude in galaxies. This line deficit
is closely related to the surface density of star formation
and therefore probably to local variations of the dense gas
conditions \citep{smith-2017}.

\subsection{Correlation of modeled physical conditions, metallicity, and SFR}

 \begin{figure}[t]
\centering
\includegraphics[clip,width=4.75cm]{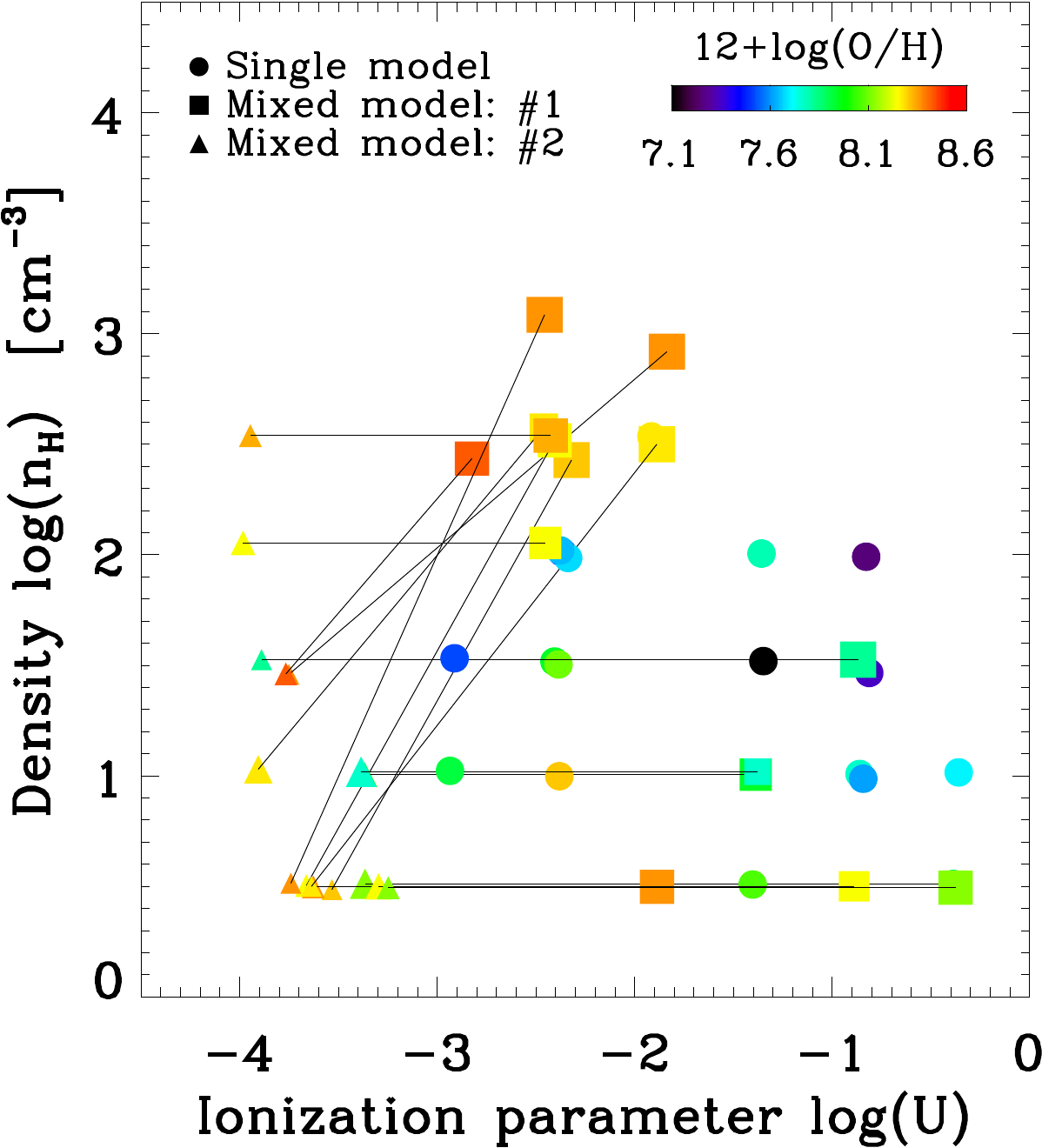} \vspace{3mm}
\includegraphics[clip,trim=16.5mm 0 0 0,width=4.11cm]{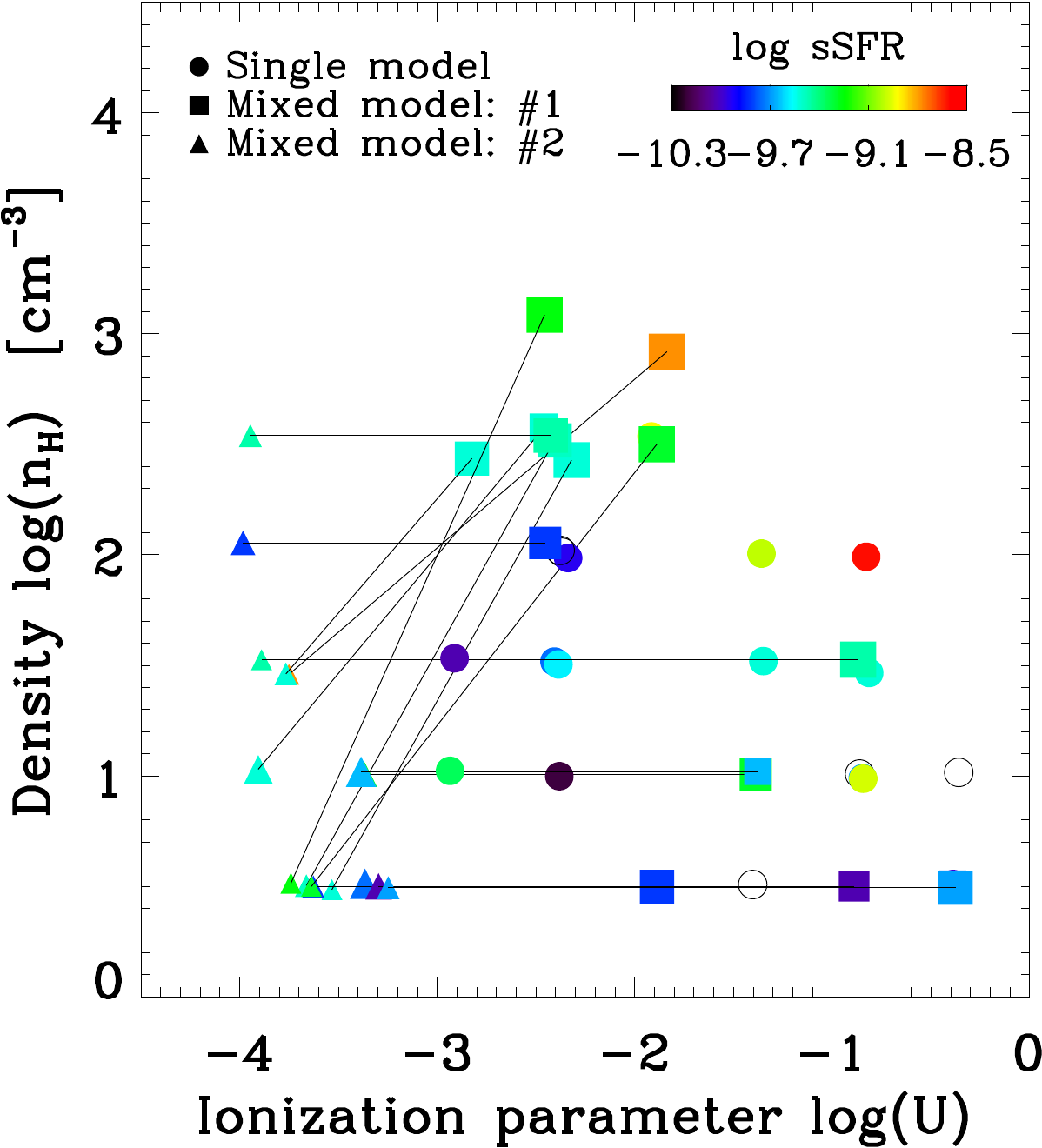}
\includegraphics[clip,width=4.75cm]{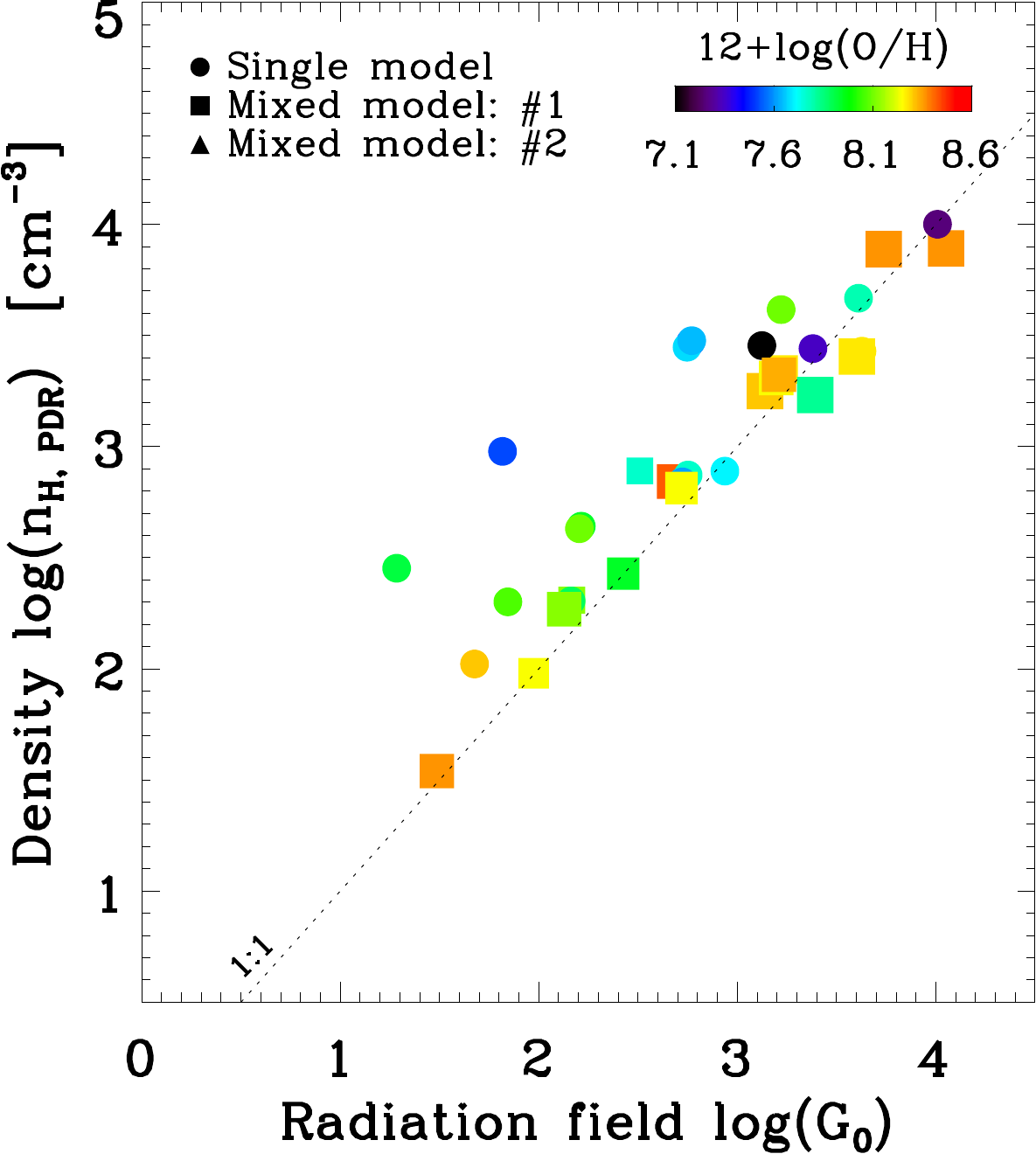}
\includegraphics[clip,trim=15.5mm 0 0 0,width=4.14cm]{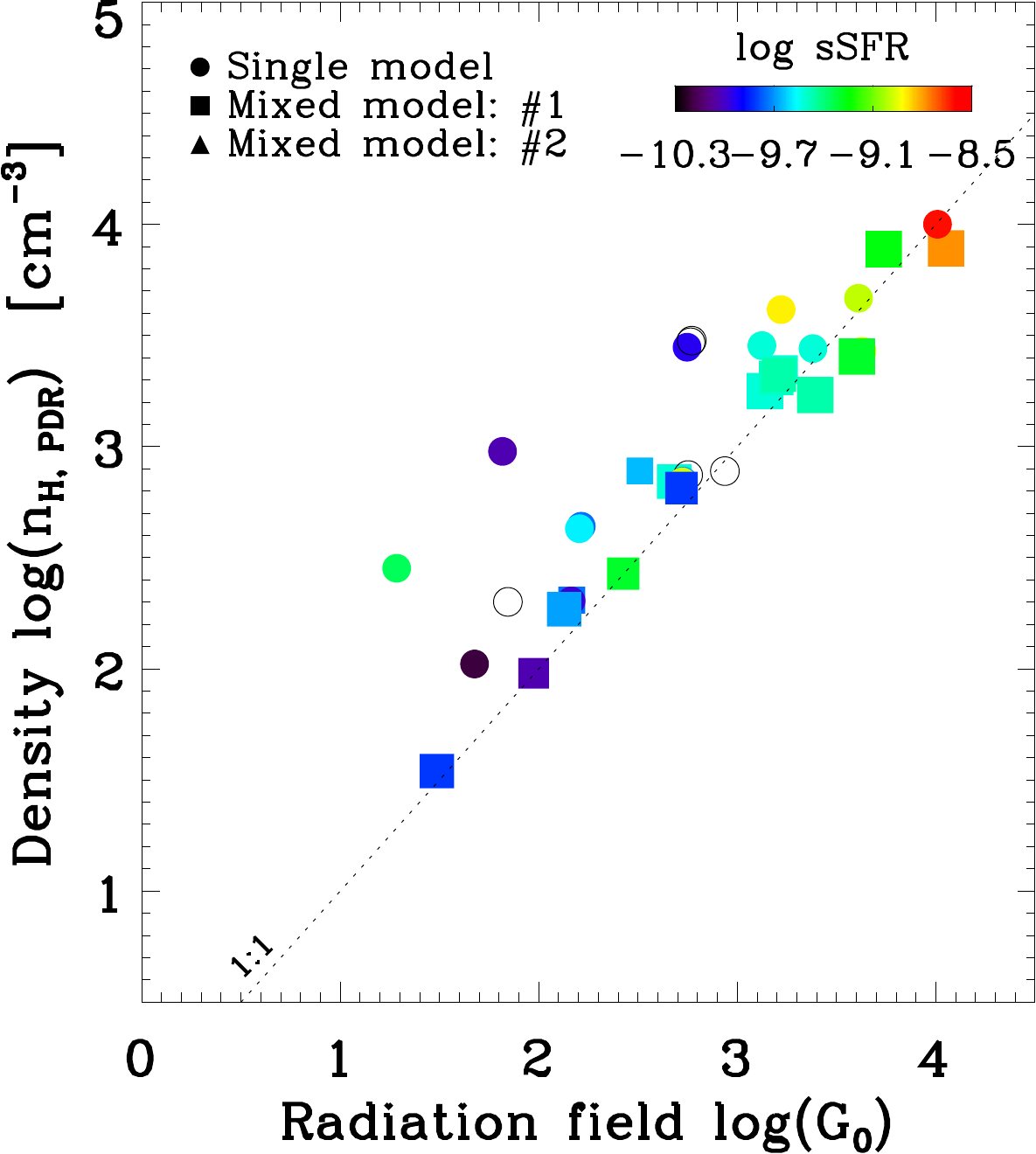}
\caption{
Distribution of model parameters.
\textit{Top row:} Ionization parameter $U$ and density
$n_{\rm H}$ for the best-fitting single models (circles) or
mixed models (component \#1: squares and component \#2:
triangles). For mixed models, the black lines link the two
model components and the symbol size is proportional
to the scaling factor $f_c$ of each component.
\textit{Bottom row:} Same for the radiation field $G_0$
impinging the PDR and the average PDR density $n_{\rm H, PDR}$.
Color coding corresponds to $Z$ (left panels) and sSFR (right panels).
Galaxies with no sSFR value are displayed with open symbols.
}
\label{fig:param_corr}
\end{figure}

We aim to see if we can link the physical conditions, namely
the density and radiation field in the \hii regions and PDRs,
to more global parameters such as $Z$ and sSFR.
Figure~\ref{fig:param_corr} shows the model parameters of
each galaxy with color codes corresponding to $Z$ and sSFR.

Regarding the model parameters of the grid (top panels in
Fig.~\ref{fig:param_corr}), we find that there are no obvious
systematics in the distribution of mixed models. For several
galaxies, the mixed models highlight a need to have components
with different $U$ rather than different $n_{\rm H}$. There is
a tendency for the low-$U$ models to also have smaller
scaling factors.
Galaxies with higher $Z$ tend to have lower ionization
parameters than galaxies with lower $Z$ (see also
Fig.~\ref{fig:append-c} in the appendix). Low-to-moderate
ionization parameters ($\log(U)<-2$) are indeed found in more
metal-rich nearby galaxies \citep[e.g.,][]{herrera-camus-2018b}.
It is in the dwarf galaxies with lower $Z$ that we find
the highest $U$ values ($\log(U)>-2$).
Finally, galaxies with higher sSFR tend to have higher
$n_{\rm H}$ than galaxies with lower sSFR.

Regarding the PDR parameters (bottom panels in Fig.~
\ref{fig:param_corr}), we find a clear correlation between
$G_0$ and $n_{\rm H, PDR}$.
We also find a tendency for lower-$Z$ galaxies
to have higher $G_0$ and $n_{\rm H, PDR}$ values,
but that tendency is clearer for high sSFR galaxies.
\cite{malhotra-2001} conducted a study on $60$ nearby
normal star-forming galaxies. They also found a relation
between $G_0$ and $n$, with $G_0 \propto n^{1.4}$,
and explain that such scaling is expected from \hii regions
(Str\"omgren spheres) surrounded by PDRs.
The ratio of the two, $G_0/n_{\rm H, PDR}$, drives the
heating of the PDR \citep[e.g.,][]{tielens-1985} and,
interestingly, it is roughly constant in the galaxies of our
sample. We find a median value of $0.5$ and a standard
deviation of 0.4\,dex. 
This is in the same range of values found by
\cite{diaz-santos-2017} for IR-luminous galaxies.
High sSFR galaxies also have slightly higher 
$G_0/n_{\rm H, PDR}$ values than the median,
indicating more compact/intense star-forming regions,
consistent with the increase of $G_0/n_{\rm H, PDR}$
at high IR surface densities (though for more moderate
values of $G_0/n_{\rm H, PDR}$ and $\Sigma_{\rm IR}$
in our sample) reported by \cite{diaz-santos-2017}.
Analytically, given our adopted density law, $G_0/n_{\rm H, PDR}$
is proportional to $U \times Z$. Since we find that lower-$Z$
galaxies tend to have higher-$U$ values, this naturally
leads to a roughly constant $G_0/n_{\rm H, PDR}$ ratio.
However, a constant ratio is not purely a consequence of model
assumptions because the PDR parameters have to fit
the (PDR) observations. A different density law would lead to
a similar relation.

\subsection{Phase filling factors}
\label{sect:disc_cov}

 \begin{figure}[t]
\centering
\includegraphics[clip,width=4.8cm]{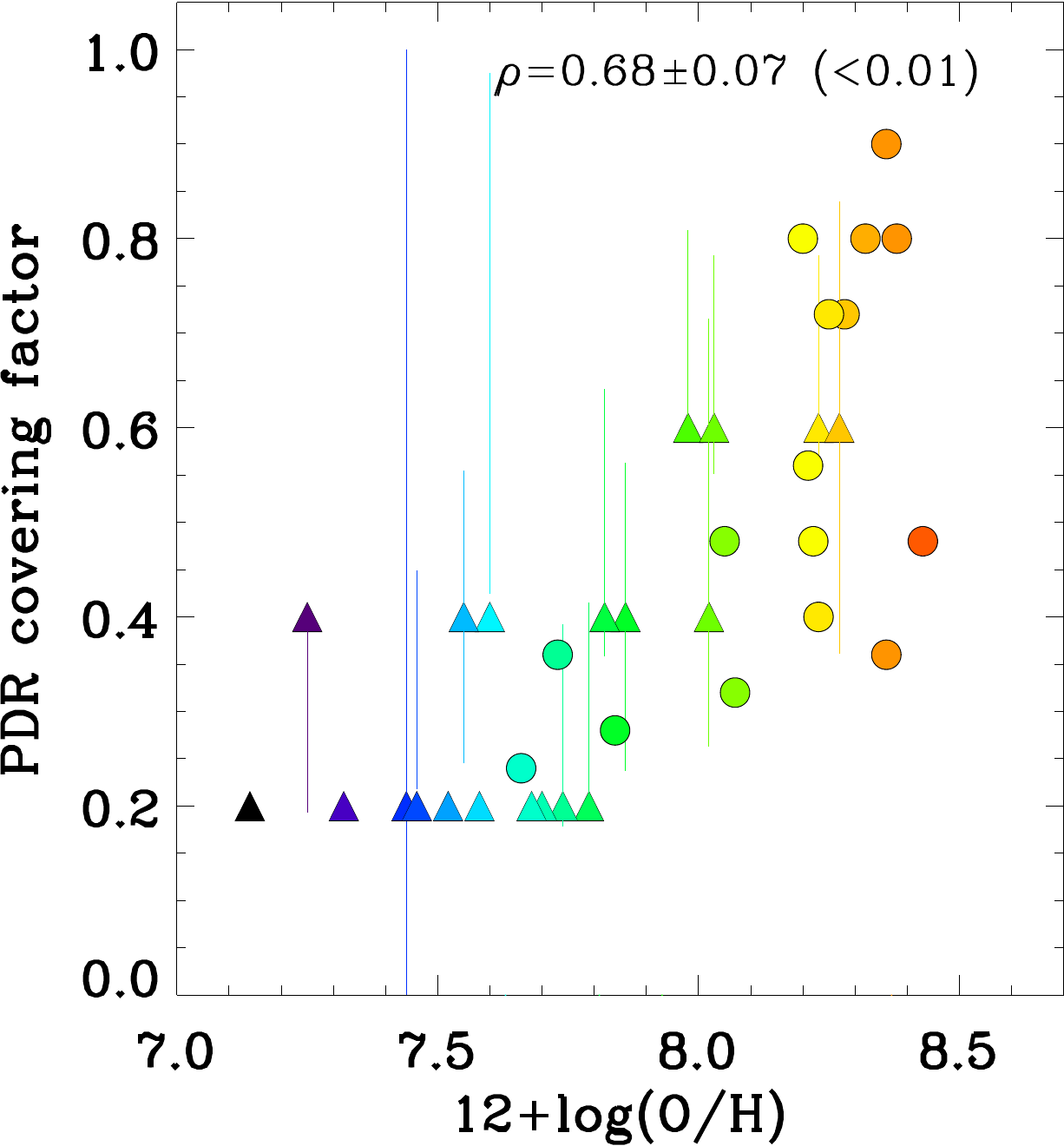} \vspace{3mm}
\includegraphics[clip,trim=20mm 0 0 0,width=4.03cm]{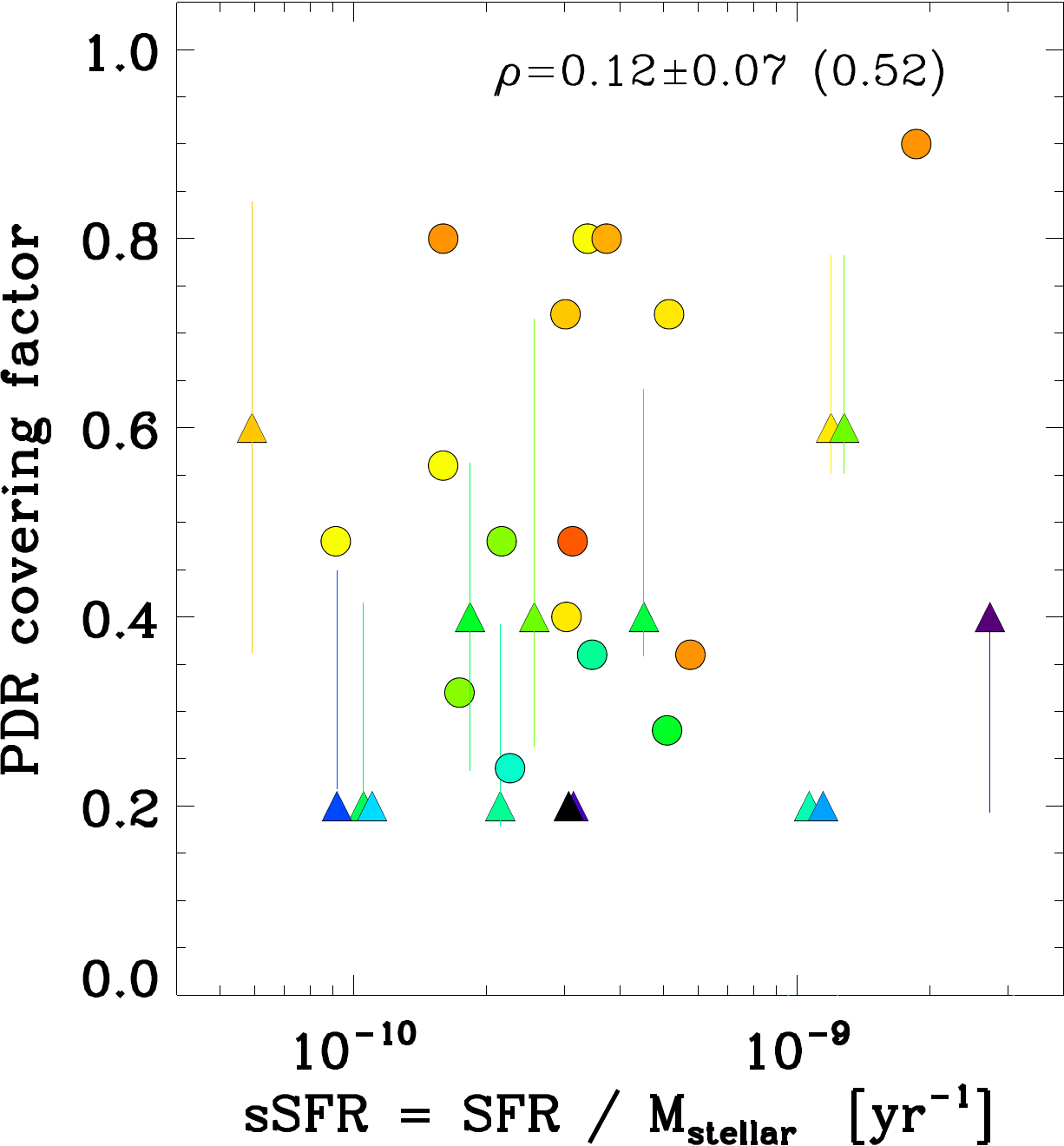}
\includegraphics[clip,width=4.8cm]{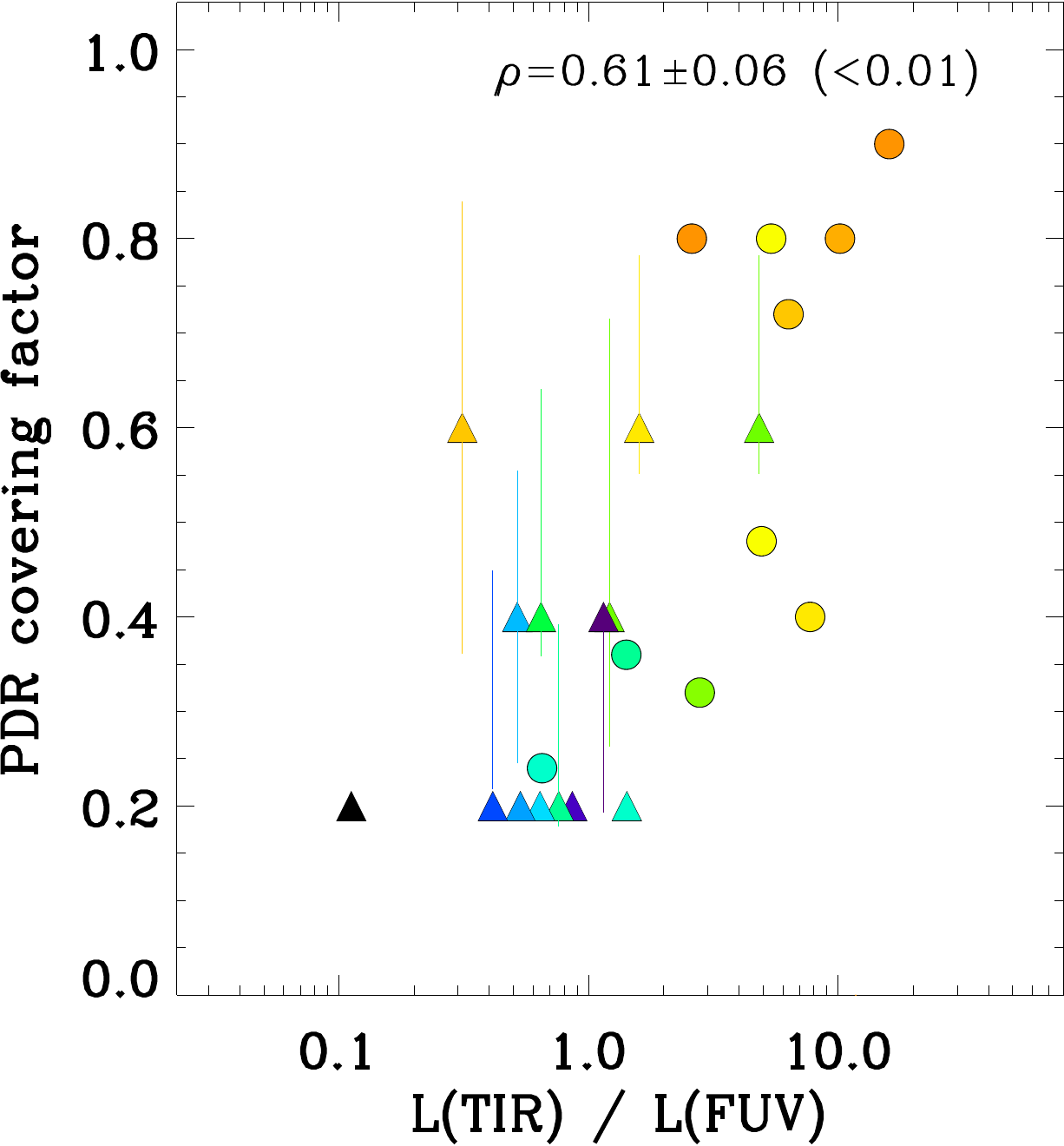}
\includegraphics[clip,trim=20mm 0 0 0,width=4.03cm]{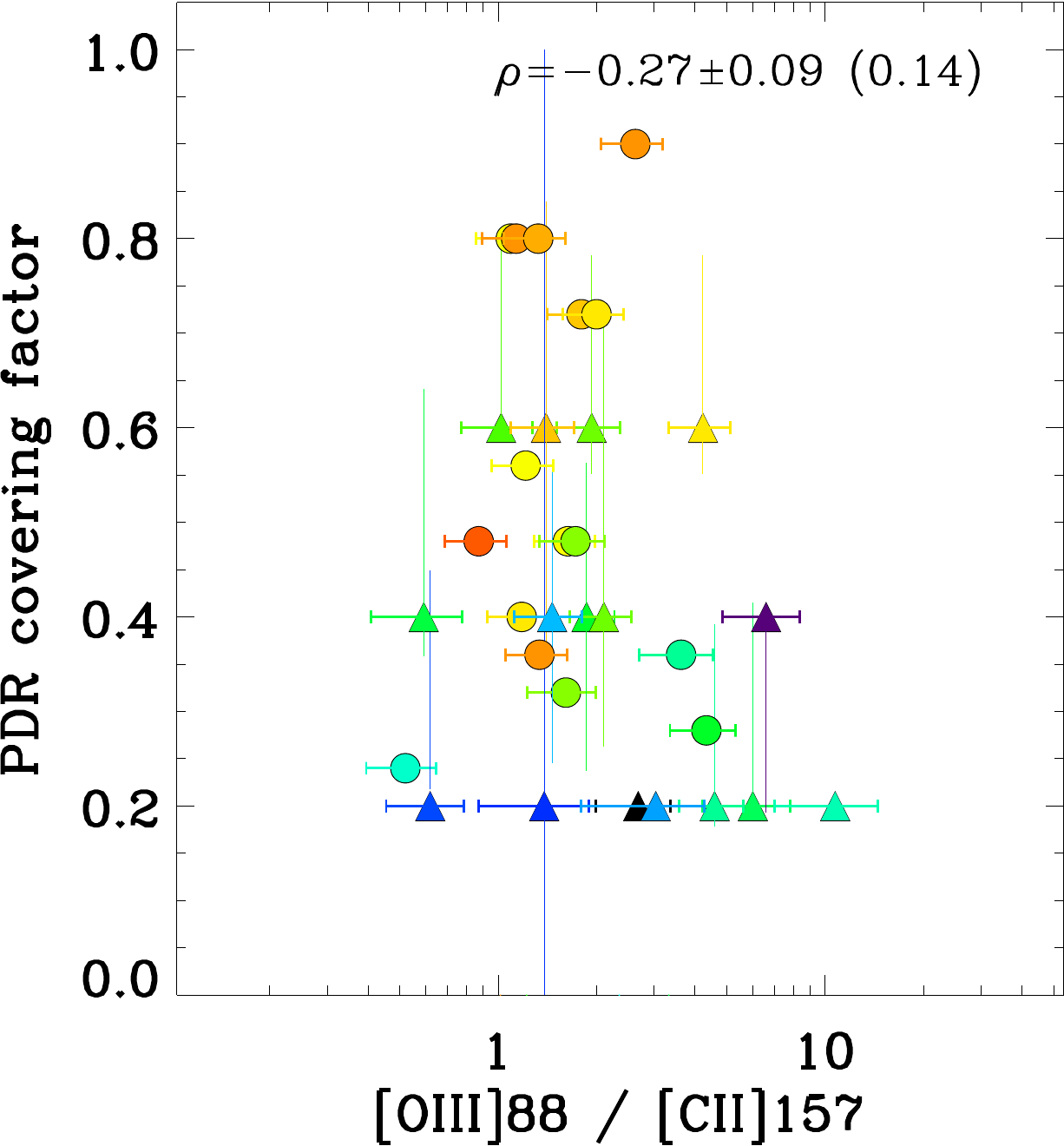}
\caption{
Correlations of the PDR covering factor with metallicity
(top left panel), specific star-formation rate (top right panel),
\ltir/$L_{\rm FUV}$ ratio (bottom left panel), and \oiiilb/\cii
ratio (bottom right panel). Galaxies are color coded by $Z$.
Spearman's rank correlation coefficient and significance
of its deviation from zero in parenthesis are indicated in
the top-right corner. 
Lower-$Z$ galaxies tend to have smaller covering factors
but we do not find strong trends with observed line ratios.
The apparent floor of $0.2$ in the PDR covering factor
is due to the chosen steps for this parameter in the grid.
}
\label{fig:cov_corr}
\end{figure}

One main result from analyzing \spit and \hers observations
of both resolved regions and entire galaxies is that the
ionized phase is extended and fills a large volume of
the ISM of star-forming dwarf galaxies
\citep{cormier-2012,chevance-2016,polles-2019}.
With the modeling that we have performed for each
individual galaxy of the \hers DGS, we can quantify further
the relative filling factor of the ionized and PDR phases. 
First, we investigate the relative volume filling factor
of the two ionized gas components in the case of mixed
models. We could expect a diffuse, low-excitation ionized
medium to fill a larger volume than denser \hii regions.
This is indeed the case for galaxies where the two components
have very different densities (for example, Haro\,11 and
NGC\,5253 for which the diffuse phase fills a volume
that is about ten and one hundred times larger than
the dense phase, respectively).
However, as discussed in the previous section (Fig~
\ref{fig:param_corr}, top panels), we do not find a
systematic contrast in density for the mixed models,
even in galaxies where \niila is detected. In particular,
the \oiii emission, often found extended in studies that
have sufficient spatial resolution
\citep{chevance-2016,polles-2019}, is well reproduced
by the high-$U$ (not necessarily low density) models.

Second, we investigate the covering factor of the PDR
phase relative to that of the ionized phase.
Figure~\ref{fig:cov_corr} shows the PDR covering factors
of the models as a function of metallicity, sSFR, and
selected line ratios. The covering factor is simply the
fraction of solid angle covered by PDR gas as seen from
the center of the ionized region. For the mixed models,
it is the product of the covering factor of the dense model
multiplied by the scaling factor (the covering factor of the
low-density model being set to zero by default).
Correlating the PDR covering factor with several line ratios,
such as \oiiilb/\ciil, \oiiilb/\ltir, or the optical line ratio of 
\oiii5007\,\AA\ -to- \oii3727\,\AA, does not yield strong trends.
Although the dwarf galaxies are offset in the observed parameter
space of those ratios compared to metal-rich galaxies
\citep{cormier-2015}, we do not find a robust line ratio
predictor of the PDR covering factor (i.e., ISM porosity).
It seems that the emission of the lines under study is too
dependent on the other physical conditions and only
the combination of many constraints provided by various
lines can predict such parameter. 
However, we find that there is a correlation between
the PDR covering factor and both the broadband
luminosity ratio of \ltir/$L_{\rm FUV}$ and $Z$, with large
scatter. The trend indicates that the porosity of the ISM
(relative covering factors of ionized and PDR gas)
increases at low $Z$.

A decrease of the PDR covering factor may facilitate
the escape of ionizing photons from the star-forming regions.
Although our models by design cannot be used to
quantify this escape, it is still a parameter worth discussing.
In the Magellanic Clouds (with $Z=1/2$ and $1/5$\,\zsun, 
where $12+\log(\rm{O/H})_{\odot}=8.69$; \citealt{asplund-2009}),
the fraction of ionizing photons escaping \hii regions is
of the order of $45$\% \citep{pellegrini-2012}.
Moreover, \cite{gazagnes-2018} show for galaxies with
known Lyman continuum leakage that the escape of
ionizing photons is related to the covering factor of the
\hi gas (rather than low \hi column densities).
Recent modeling of the Local Group dwarf galaxy IC\,10
also shows that on scales of individual star-forming regions,
clouds tend to be matter-bounded, allowing for the possibility
of photons escaping the regions, while on larger spatial
scales, clouds tend to be more radiation-bounded
\citep{polles-2019}. At large enough scales the escape
fraction is expected to be near zero.
However, given that our models are radiation-bounded
(i.e., we stopped them further than the ionization front)
and that several nearby dwarf galaxies have extended
\hi halos that we have not modeled, the PDR covering
factors derived here cannot be translated into galaxy
escape fractions. Such escape fractions have been
measured to be up to a few percent in nearby dwarf
galaxies (e.g., \citealt{leitet-2011,leitet-2013}; see also
\citealt{zastrow-2013}), but could be larger in dwarfs
at higher redshift \citep{razoumov-2010,yajima-2011}.
\cite{leitet-2013} find that the escape fraction is somewhat
larger in galaxies with lower metallicity, lower stellar mass
and higher sSFR for the energetics of the star-forming
regions to affect the whole galaxy \citep{clarke-2002}.
Here, we find correlations between the PDR covering
factor and metallicity (as well as with stellar mass and
SFR) but not with sSFR. A lower PDR covering factor
in nearby low-$Z$ galaxies might indicate that the escape
fraction is likely to be higher for low-$Z$ galaxies in
the high-redshift Universe.

\section{Conclusions}
\label{sect:conclusion}
We have modeled the MIR and FIR emission of the ISM
of galaxies from the \hers Dwarf Galaxy Survey with the
spectral synthesis code Cloudy. The goal of this study is to
characterize the gas physical conditions in the \hii region
and PDR of those galaxies (namely density, ionization parameter,
covering factor of the PDR), and to understand how or if they
correlate with a galaxy's metallicity or star-formation activity.
We focus our discussion on the phase origin of \cii emission
and on the ISM porosity. Our results are summarized as follows.

\begin{itemize}
\item Several galaxies are well reproduced by single models,
but for other galaxies we need mixed models, especially
models with a different ionization parameter (rather than
a different gas density) to reproduce satisfyingly the observed
MIR-FIR emission. We find values of ionization parameters
in the range $\log(U) \simeq -3.0$ to $-0.3$, and electron
densities in the range $n_e \simeq 10^{0.5}-10^{3.0}$\,\cm,
corresponding to PDR densities roughly ten times larger
(given the adopted density profile). The highest $U$
values are higher than in metal-rich galaxies and are found
in the lower-$Z$ galaxies of our sample. Radiation fields
and densities in the PDR are also found higher for the
lowest-$Z$ galaxies. Densities are higher in galaxies with
higher specific star-formation rates.
\item We evaluated the effect of the choice of spectral lines to fit
and of individual input parameters of the models on the results.
The ionization parameter is mainly sensitive to the choice
of input radiation field library, while choices on the density profile
and dust-to-gas ratio can affect the density in the PDR.
We find the need for a low-luminosity soft X-ray component
and/or increased cosmic ray rate to reproduce the observed
\oilb/\oila line ratios.
\item The contribution of the ionized phase to the \cii emission
is low, typically $<$30\%, making \cii a good tracer of PDRs
and of the SFR at low metallicities (and potentially at high redshift).
As found in \cite{croxall-2017}, there is a trend of increasing
\cii fractions from the ionized gas with metallicity, though with
scatter. This fraction seems more driven by the ionization
parameter (and indirectly metallicity) than by density. 
\item We find that the covering factor of the neutral gas relative
to that of the ionized gas decreases with decreasing metallicity
and TIR-to-FUV luminosity ratio.
This provides evidence for a change in the ISM porosity, which
we conjecture may facilitate the propagation and escape
of ionizing photon  in some systems. If such ISM porosity is present
in low-metallicity galaxies at high redshift, this might have implications
for the fraction of escaping photons and cosmic reionization.
\end{itemize}

The DGS galaxies span more than an order of magnitude
in metallicity and sSFR with a bias toward high sSFR values.
Metallicity (rather than sSFR) seems to be the main parameter
driving variations in the ISM properties of the DGS galaxies.
However, extending the modeling approach of this paper
to galaxies that cover a larger parameter space in terms
of metallicity, SFR, stellar mass, amongst others, will be important
to map the evolution of ISM properties through cosmic times.

Our suite of models makes predictions for many more lines
than observed and discussed in this paper. These predictions
will be used to guide follow-up observations that will allow us
to exploit unique PDR coolants, such as [C\,{\sc i}] (available
with ALMA), MIR \htwo and PAH features (available with the
JWST), and to test the importance of specific heating
mechanisms, such as shocks.
Spatially resolving the main ISM phases would also be
extremely useful for a more direct confirmation of the phase origin
of the emission lines, respective filling factors, ISM porosity
and potentially, the escape of ionizing photons.

\vspace{5mm}
\begin{acknowledgements}
The authors thank P.\,De\,Vis for providing the Magphys
SED of Haro\,11 and M.\,Krumholz for useful discussion.
We also thank the referee for the valuable comments
which helped to improve the manuscript.
DC is supported by the European Union's Horizon 2020 research
and innovation program under the Marie Sk\l{}odowska-Curie
grant agreement No 702622.
NPA acknowledges funding from NASA and SOFIA
through grant program SOF 05-0084.
SH is supported by the DFG program HO 5475/2-1.
IDL gratefully acknowledges the support of the 
Research Foundation -- Flanders (FWO).
We also acknowledge support from the DAAD/PROCOPE projects
57210883/35265PE and from the Programme National
``Physique et Chimie du Milieu Interstellaire'' (PCMI) of
CNRS/INSU with INC/INP co-funded by CEA and CNES.
\end{acknowledgements}

\bibliographystyle{aa}
\bibliography{../../BIB/references}

\appendix
\section{Observed abundances for the DGS galaxies}
\label{sect:append-a}
In Table~\ref{table:ab_obs}, we report the observed
abundances from the literature for the galaxies
of the \hers Dwarf Galaxy Survey \citep{madden-2013}
studied here.

\begin{center}
\begin{table}[!t]\small
  \caption{Observed elemental abundances in our galaxies.} 
  \hfill{}
\begin{tabular}{lcccc}
    \hline     \hline
     \vspace{-8pt}\\
     {Galaxy}           & {$\log {\rm O/H}$}     & {$\log {\rm C/O}$}   & {$\log {\rm N/O}$}      & {$\log {\rm Ne/O}$} \\
      \hline
     \vspace{-8pt}\\
     {Haro\,11}                 & $-3.64$       & ...   & $-0.92^{(1)}$         & $-0.66^{(1)}$   \\
     {Haro\,2}          & $-3.77$       & ...   & ...    & ...  \\
     {Haro\,3}          & $-3.72$       & ...   & $-1.35$        & $-0.67$ \\
     {He\,2-10}                 & $-3.57$       & ...   & ...    & ...  \\
     {HS\,0052+2536}    & $-3.93$       & ...   & ...    & ...  \\
     {HS\,0822+3542} & $-4.68$  & ...   & ...    & $-0.80$ \\
     {HS\,1222+3741}    & $-4.21$       & ...   & ...    & ...  \\
     {HS\,1304+3529}    & $-4.07$       & ...   & ...    & ...  \\
     {HS\,1330+3651}    & $-4.02$       & ...   & ...    & ...  \\
     {II\,Zw\,40}               & $-3.77$       & ...   & $-1.44$        & $-0.76$ \\
     {I\,Zw\,18}                & $-4.86$       & $-0.74$       & $-1.60$        & $-0.80$ \\
     {Mrk\,153}                 & $-4.14$       & ...   & ...    & ...  \\
     {Mrk\,209}                 & $-4.26$       & ...   & $-1.49$        & $-0.75$ \\
     {Mrk\,930}                 & $-3.97$       & ...   & $-1.39$        & $-0.71$ \\
     {Mrk\,1089}                & $-3.80$       & ...   & $-1.05$        & $-0.75$ \\
     {Mrk\,1450}                & $-4.16$       & ...   & $-1.45$        & $-0.64$ \\
     {NGC\,1140}        & $-3.62$       & ...   & $-1.20$        & $-0.67$ \\
     {NGC\,1569}        & $-3.98$       & ...   & $-1.39^{(2)}$  & $-0.82^{(2)}$ \\
     {NGC\,1705}        & $-3.73$       & ...   & $-1.34^{(3)}$  & $-0.70^{(3)}$ \\
     {NGC\,5253}        & $-3.75$       & $-0.56$       & $-1.45^{(4)}$  & ...   \\
     {NGC\,625}                 & $-3.78$       & ...   & ...   & ...   \\
     {Pox\,186}                 & $-4.30$       & ...   & ...   & ...   \\
     {SBS\,0335-052}    & $-4.75$       & $-0.83$       & $-1.58$        & $-0.80$ \\
     {SBS\,1159+545}    & $-4.56$       & ...   & $-1.58$       & $-0.73$         \\
     {SBS\,1211+540}    & $-4.42$       & ...   & $-1.59$       & $-0.75$         \\
     {SBS\,1249+493}    & $-4.32$       & ...   & $-1.59$       & $-0.68$         \\
     {SBS\,1415+437} & $-4.45$  & $-0.78$       & $-1.58$       & $-0.73$         \\
     {SBS\,1533+574} & $-3.95$  & ...   & $-1.54$        & $-0.71$ \\
     {Tol\,1214-277}    & $-4.48$       & $-0.80^{(5)}$ & $-1.64$        & ...   \\
     {UM\,448}          & $-3.68$       & ...   & $-1.01$        & $-0.76$ \\
     {UM\,461}          & $-4.27$       & ...   & $-1.50$        & $-0.86$ \\
     {VII\,Zw\,403}     & $-4.34$       & ...   & $-1.53$        & $-0.81$ \\
     {NGC\,4214-c}      & $-3.80$       & $-0.50^{(6)}$ & $-1.30^{(6)}$  & $-0.74^{(6)}$ \\
     {NGC\,4214-s}      & $-3.64$       & $-0.50^{(6)}$ & $-1.38^{(6)}$  & $-0.87^{(6)}$ \\
    \hline     \hline
     \vspace{-8pt}\\
     {Galaxy}           & {$\log {\rm S/O}$}     & {$\log {\rm Si/O}$}  & {$\log {\rm Ar/O}$}     & {$\log {\rm Fe/O}$} \\
      \hline
     \vspace{-8pt}\\
     {Haro\,11}                 & $-1.83^{(1)}$  & ...  & $-2.50^{(1)}$  & $-2.20^{(1)}$ \\
     {Haro\,3}          & ...   & ...   & $-2.30$       & $-2.09$       \\
     {HS\,0822+3542}    & ...   & ...   & ...    & $-1.50$ \\
     {II\,Zw\,40}               & ...   & ...   & ...    & $-1.80$ \\
     {I\,Zw\,18}                & $-1.65$       & $-1.48$       & ...    & $-1.47$ \\
     {Mrk\,209}                 & $-1.47$       & ...   & ...   & $-2.01$         \\
     {Mrk\,930}                 & $-1.65$       & ...   & ...   & $-1.76$         \\
     {Mrk\,1089}                & $-1.66$       & ...   & ...   & $-1.89$         \\
     {Mrk\,1450}                & $-1.60$       & ...   & ...   & $-1.86$         \\
     {NGC\,1140}        & ...   & ...   & $-2.35$       & $-1.77$       \\
     {NGC\,1569}        & $-1.70^{(2)}$ & ...   & ...    & ... \\
     {NGC\,1705}        & $-1.60^{(3)}$         & ...   & $-2.30^{(3)}$  & ...   \\
     {NGC\,5253}        & $-1.68^{(4)}$         & $-1.53$       & ...    & ...   \\
     {SBS\,0335-052}    & $-1.59$       & $-1.60$       & ...    & $-1.34$ \\
     {SBS\,1159+545}    & $-1.51$       & ...   & ...    & ...  \\
     {SBS\,1211+540}    & $-1.48$       & ...   & ...    & ...  \\
     {SBS\,1249+493}    & $-1.66$       & ...   & ...    & ...  \\
     {SBS\,1415+437} & $-1.59$  & $-1.46$       & ...   & ...   \\
     {SBS\,1533+574} & $-1.62$  & ...   & ...   & $-1.82$       \\
     {UM\,448}          & $-1.67$       & ...   & ...   & $-2.01$       \\
     {UM\,461}          & $-1.51$       & ...   & ...   & ...   \\
     {VII\,Zw\,403}     & $-1.58$       & ...   & ...   & ...   \\
     {NGC\,4214-c}      & $-1.59^{(6)}$         & ...   & ...   & ...   \\
     {NGC\,4214-s}      & $-1.57^{(6)}$         & ...   & ...   & ...   \\
    \hline     \hline 
  \end{tabular}
  \hfill{}
  \tablefoot{
Abundances from \cite{izotov-1999,izotov-2004},
and \cite{thuan-2005b} except for:
$(1)$~\cite{guseva-2012};
$(2)$~\cite{kobulnicky-1997};
$(3)$~\cite{annibali-2015};
$(4)$~\cite{westmoquette-2013};
$(5)$~\cite{garnett-1995};
$(6)$~\cite{kobulnicky-1996}.
}
  \label{table:ab_obs}
\end{table}
\end{center}

\section{Atlas of the \hii region+PDR best-fitting model schematics}
\label{sect:append-b}
For each galaxy, we show a 2D schematic of the best-fitting
single or mixed model, based on model results presented
in Tables~\ref{table:modelres1} and \ref{table:modelres2}.
Galaxies are sorted by metallicity (from highest to lowest).
Color coding corresponds to density. 
The central stellar cluster is represented in green. The
distance to the inner black shell varies with the inner radius,
and the thickness of the inner black shell is proportional
to the ionization parameter $U$, while the thickness of the
\hii and PDR regions are kept fixed. $U$ is a function of the
input stellar spectrum, density, and inner radius. In the case
of mixed models, the angle shown by dotted lines indicates
the contribution from each component (i.e., parameter $f_c$
in Tables~\ref{table:modelres1} and \ref{table:modelres2}).

 \begin{figure*}[thp]
\centering
\includegraphics[clip,trim=0 3cm 0 -1cm,width=4.5cm]{./figures_appendix/He2-10_mixed_a}
\includegraphics[clip,trim=0 3cm 0 -1cm,width=4.5cm]{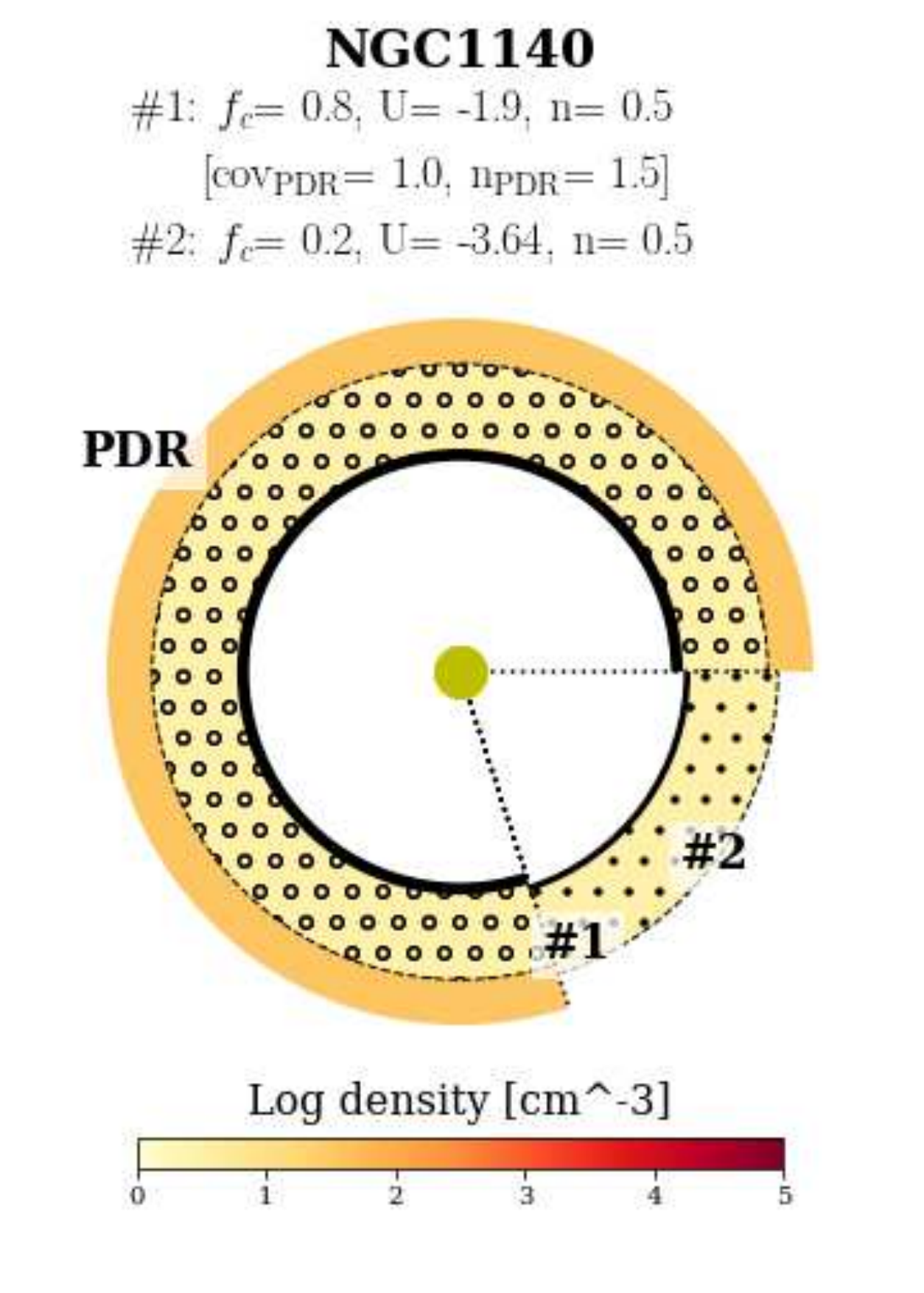}
\includegraphics[clip,trim=0 3cm 0 -1cm,width=4.5cm]{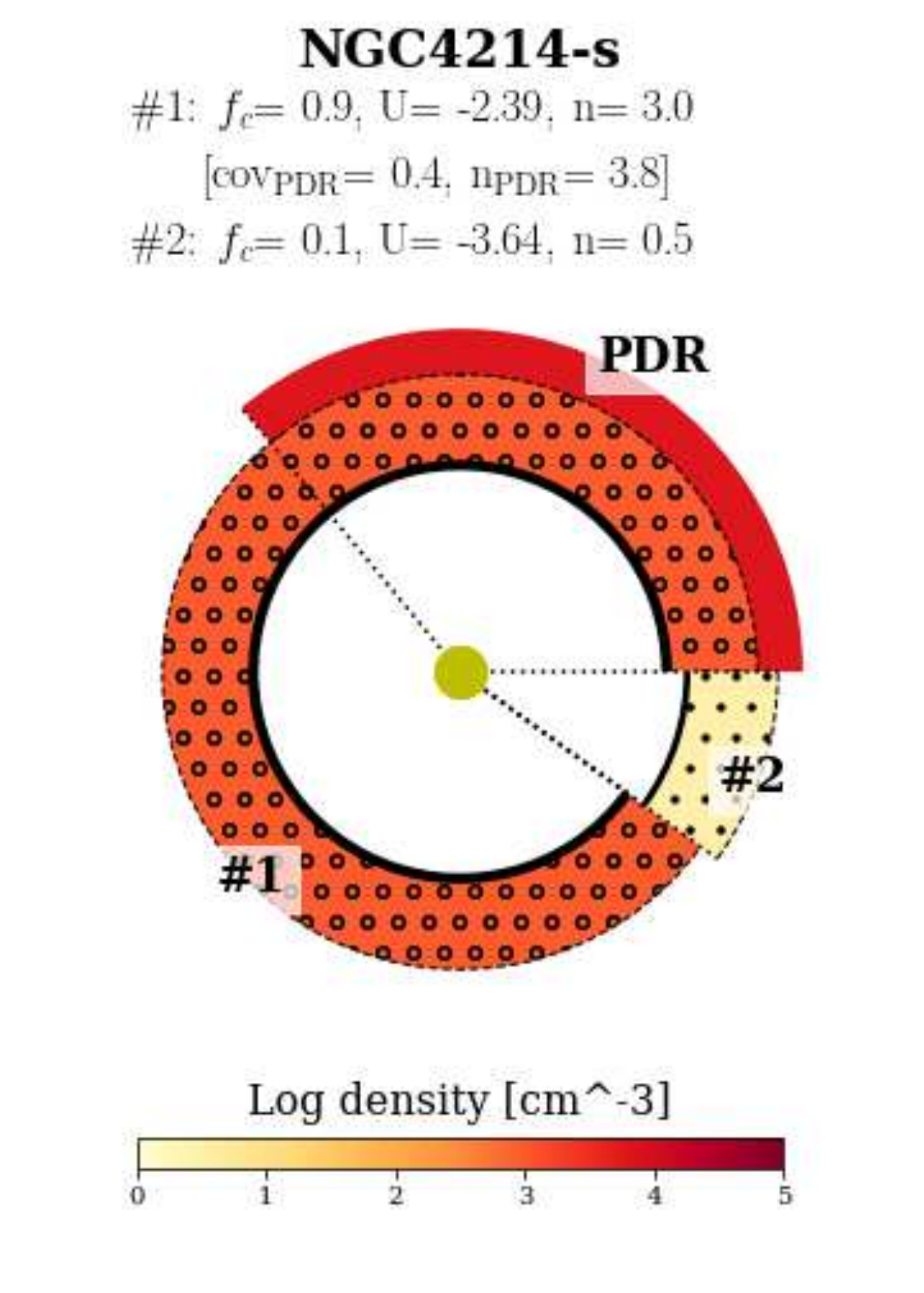}
\includegraphics[clip,trim=0 3cm 0 -1cm,width=4.5cm]{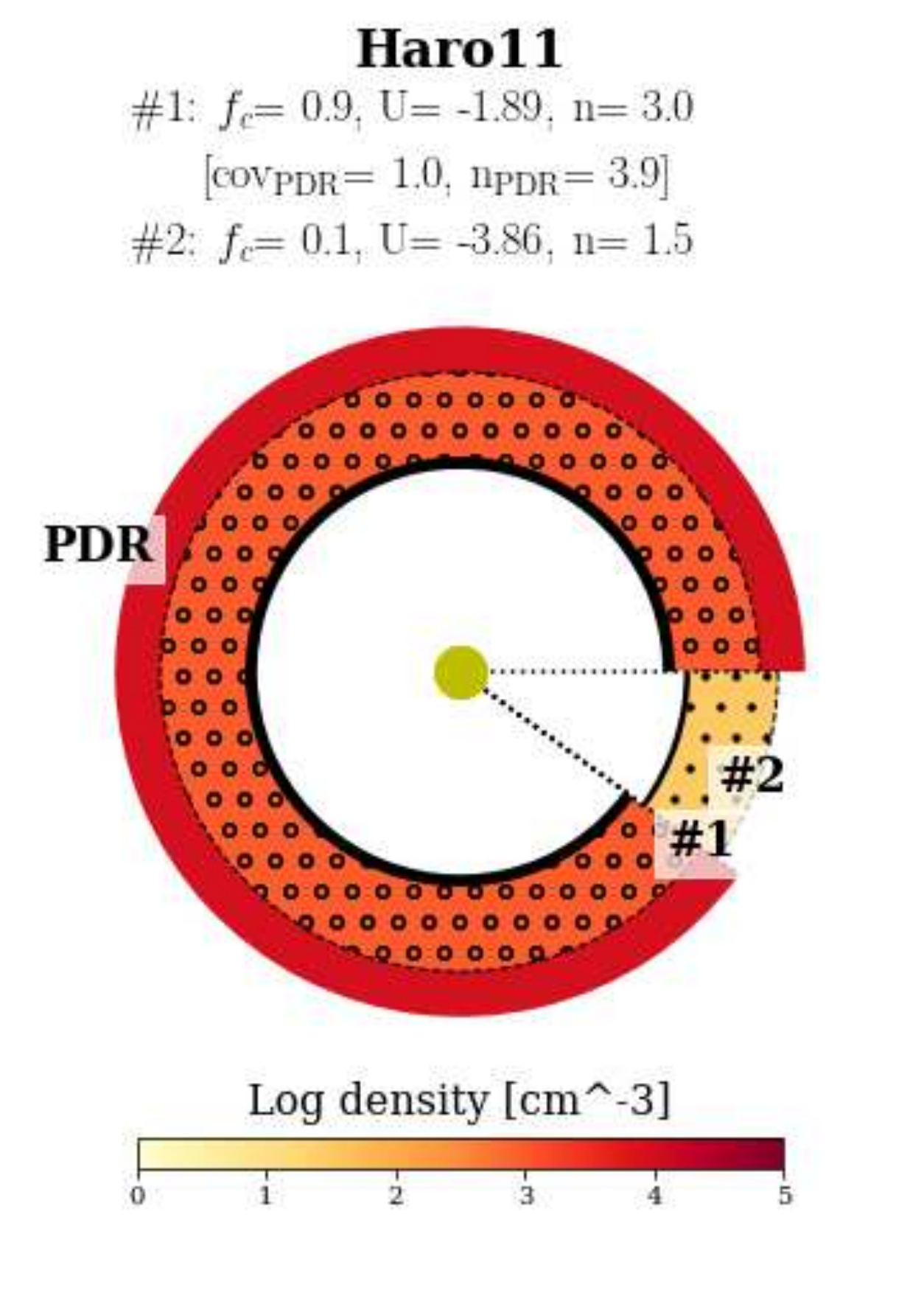}
\includegraphics[clip,trim=0 3cm 0 -1cm,width=4.5cm]{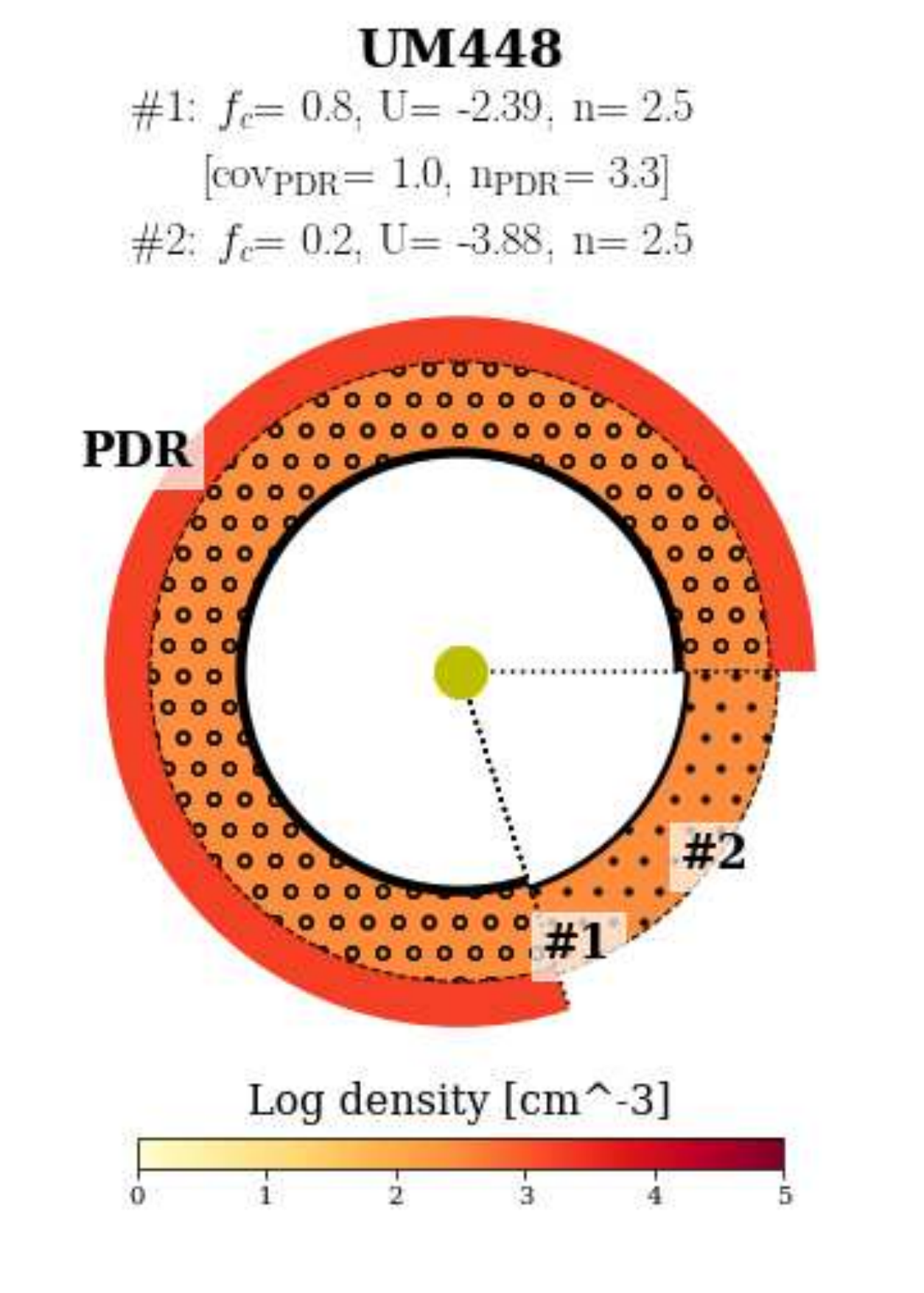}
\includegraphics[clip,trim=0 3cm 0 -1cm,width=4.5cm]{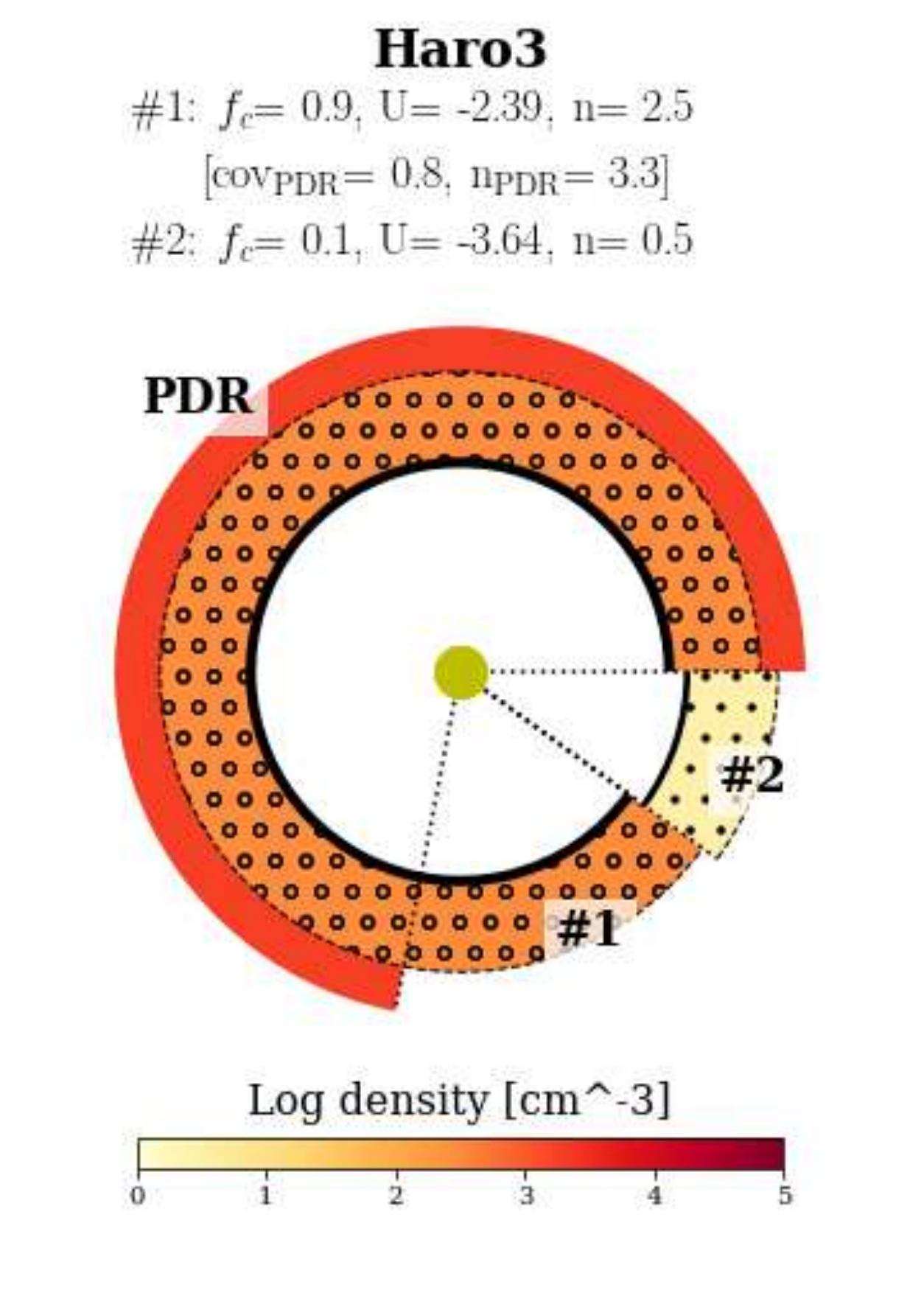}
\includegraphics[clip,trim=0 3cm 0 -1cm,width=4.5cm]{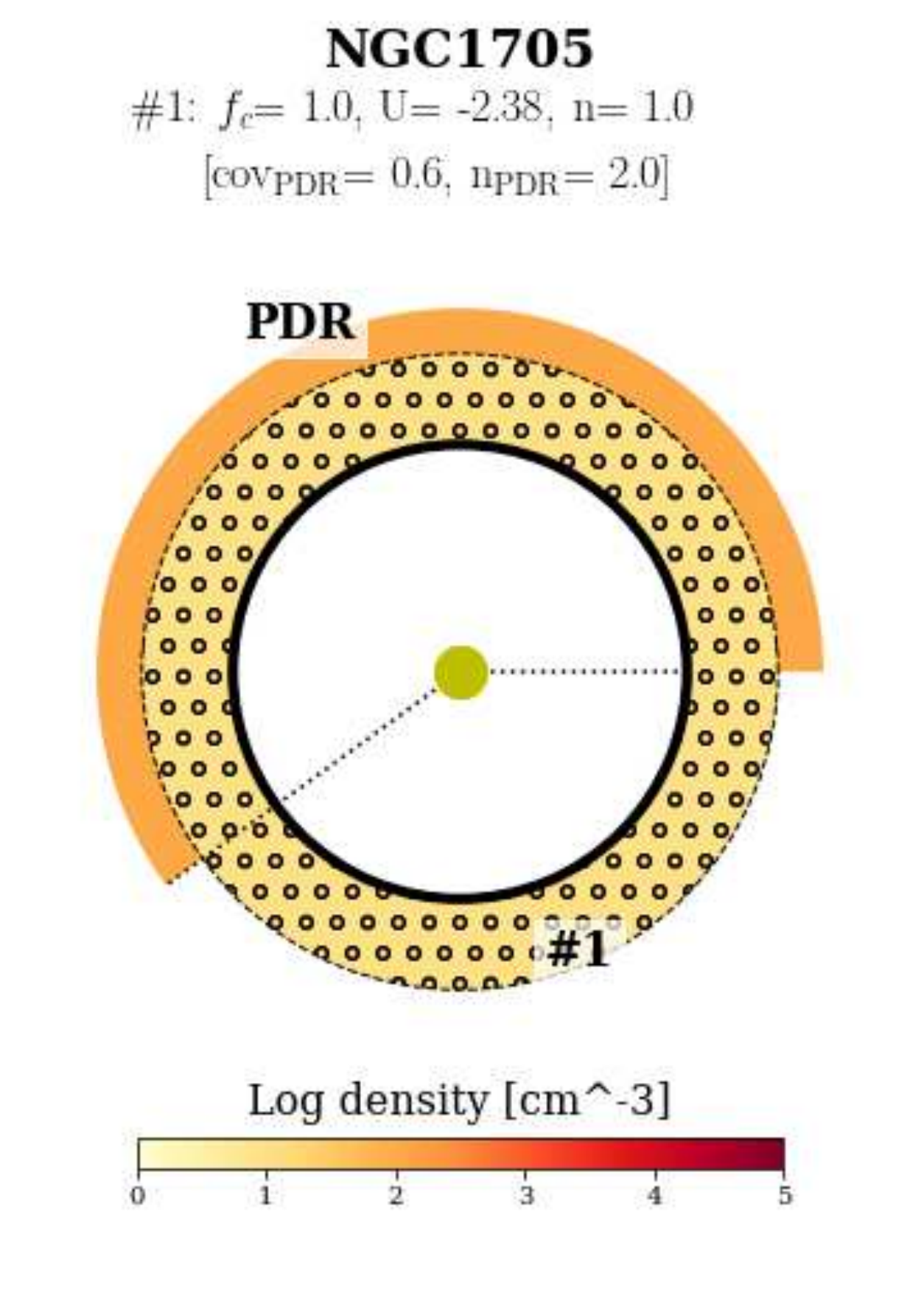}
\includegraphics[clip,trim=0 3cm 0 -1cm,width=4.5cm]{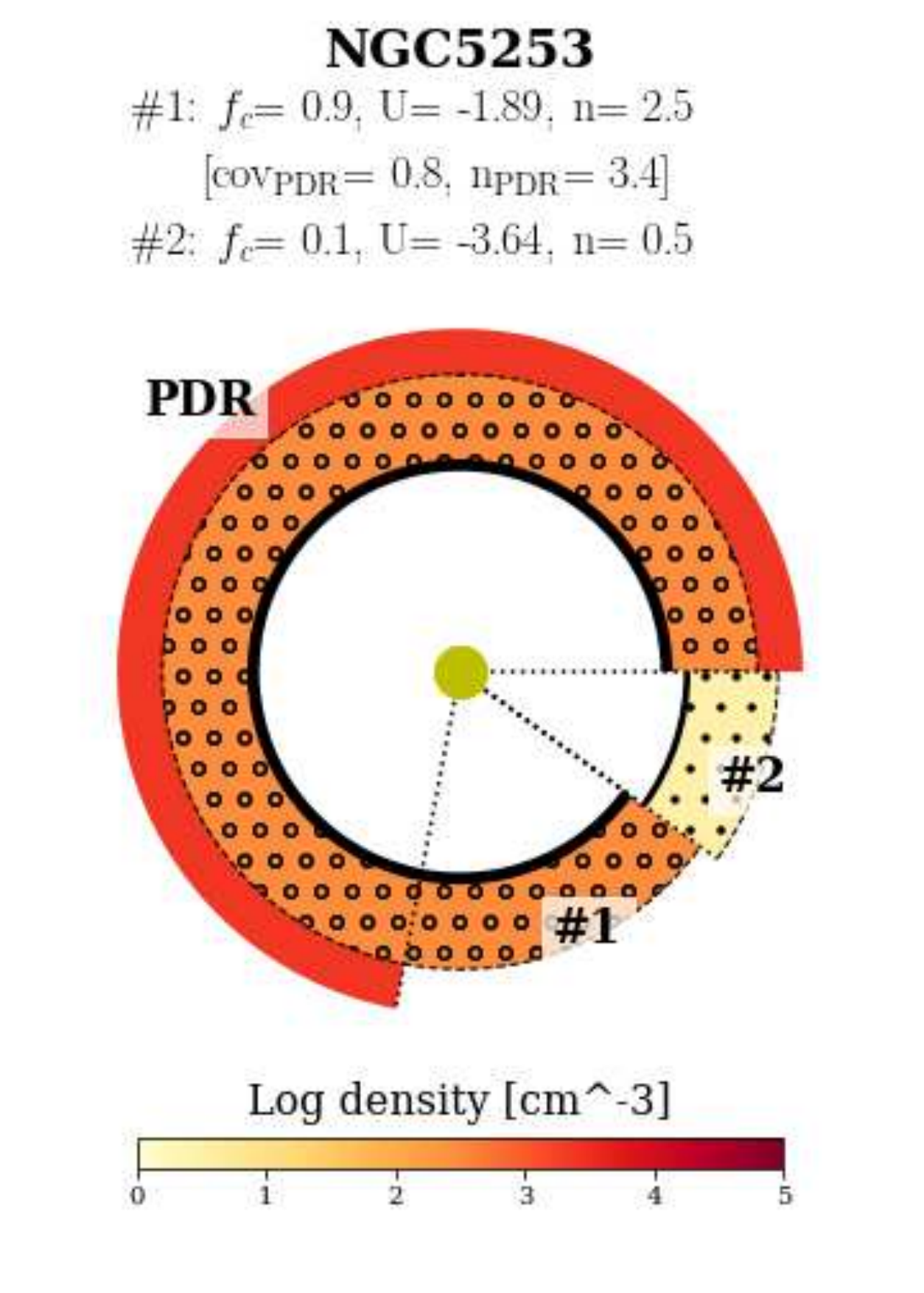}
\includegraphics[clip,trim=0 3cm 0 -1cm,width=4.5cm]{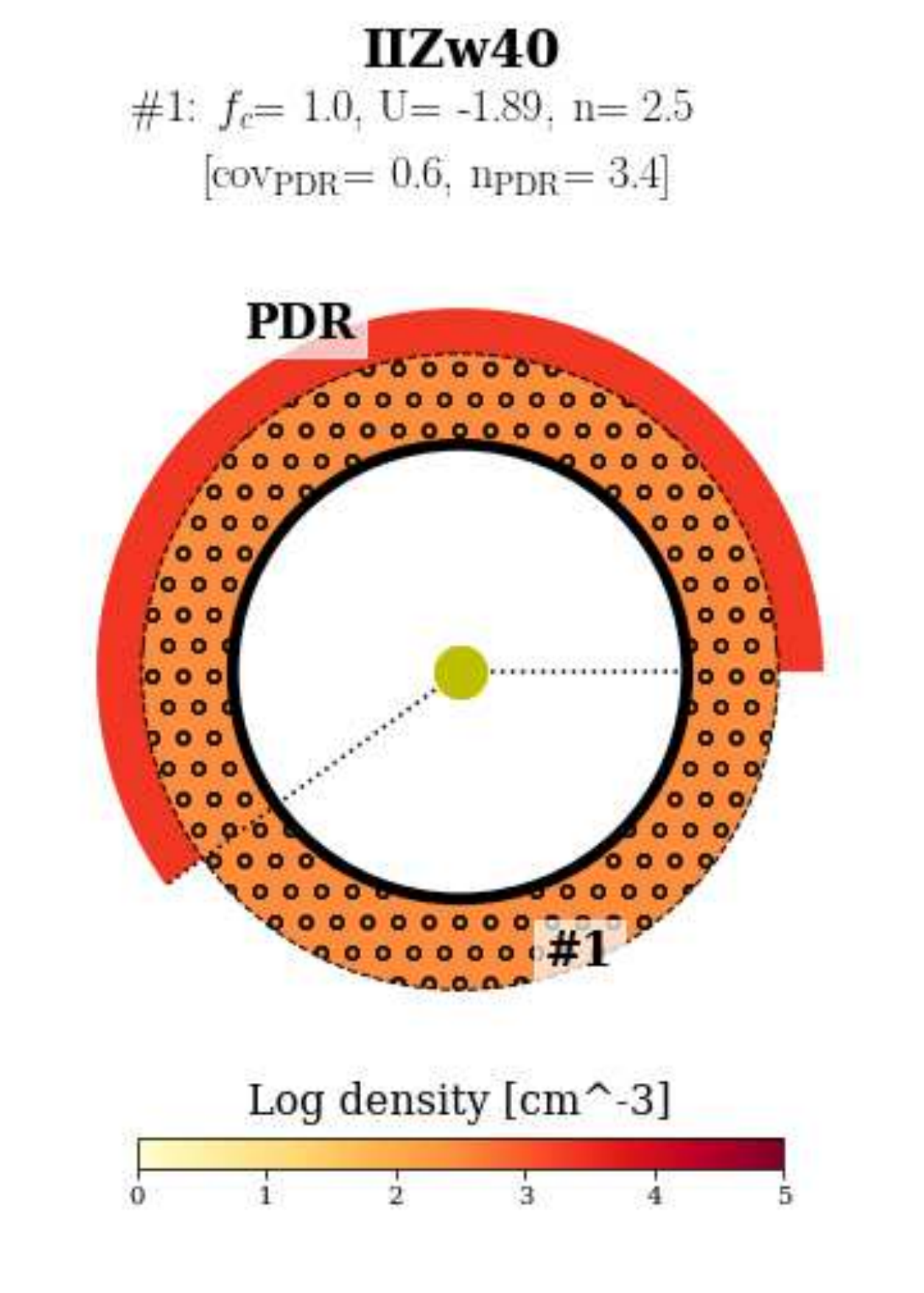}
\includegraphics[clip,trim=0 3cm 0 -1cm,width=4.5cm]{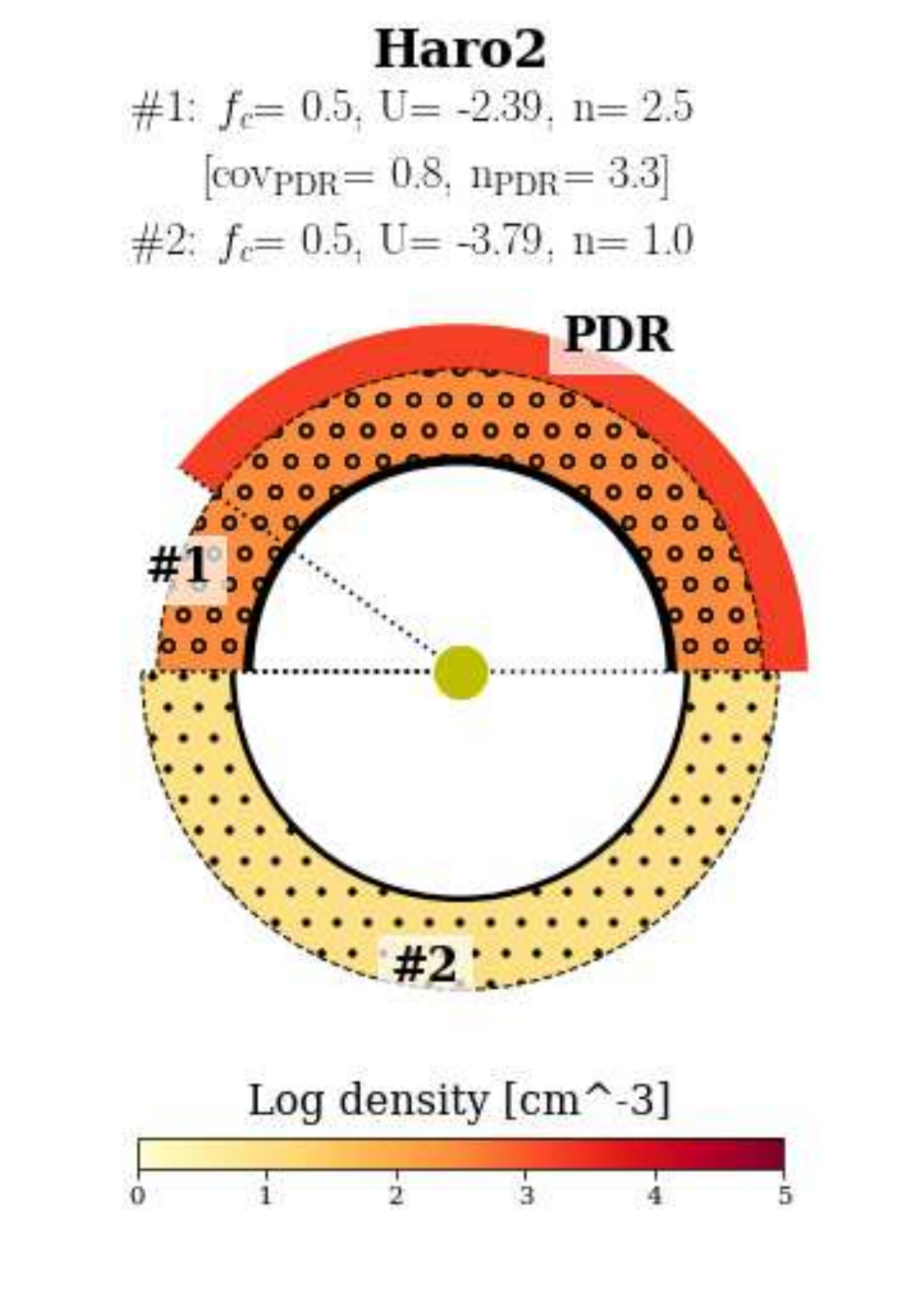}
\includegraphics[clip,trim=0 3cm 0 -1cm,width=4.5cm]{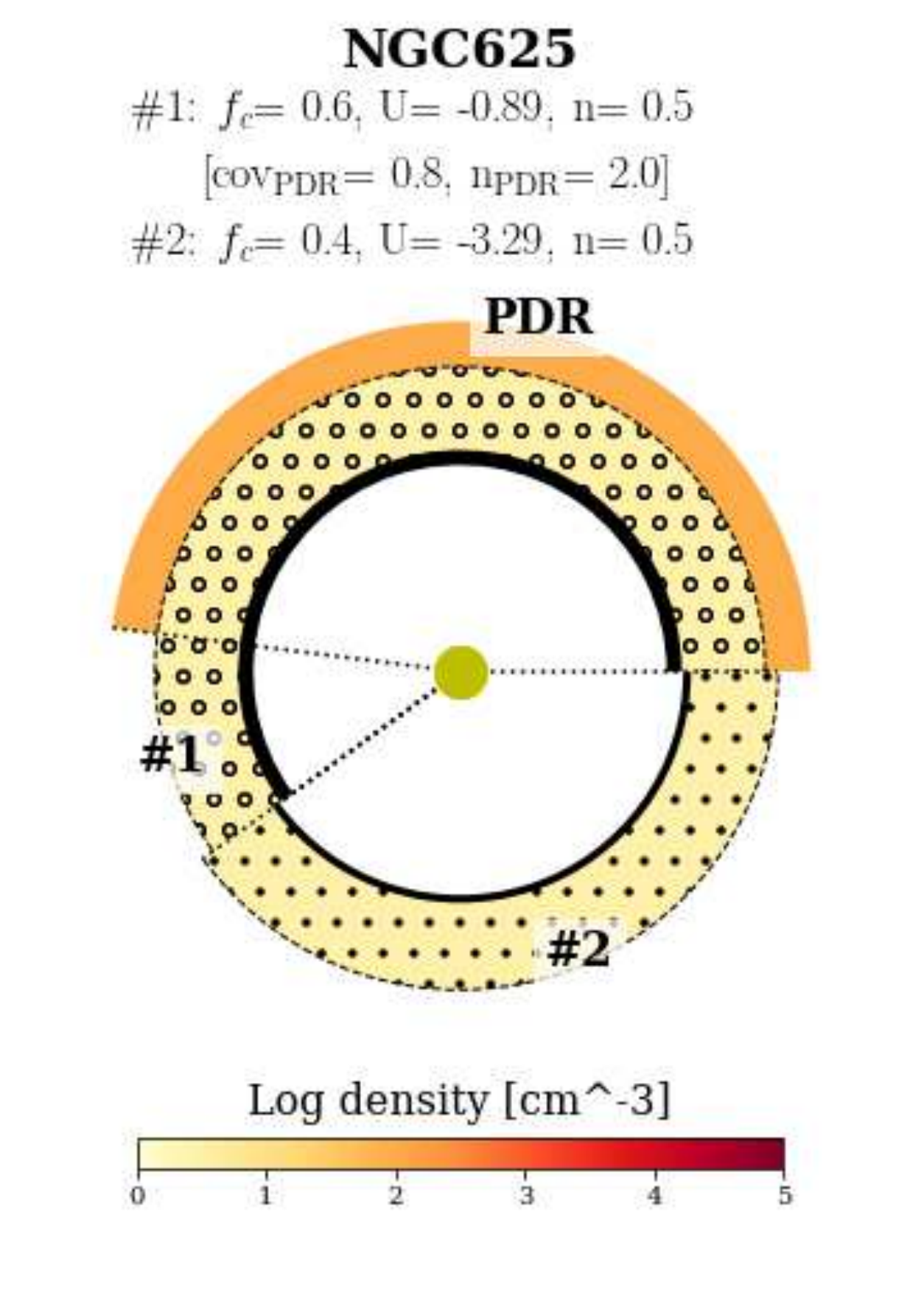}
\includegraphics[clip,trim=0 3cm 0 -1cm,width=4.5cm]{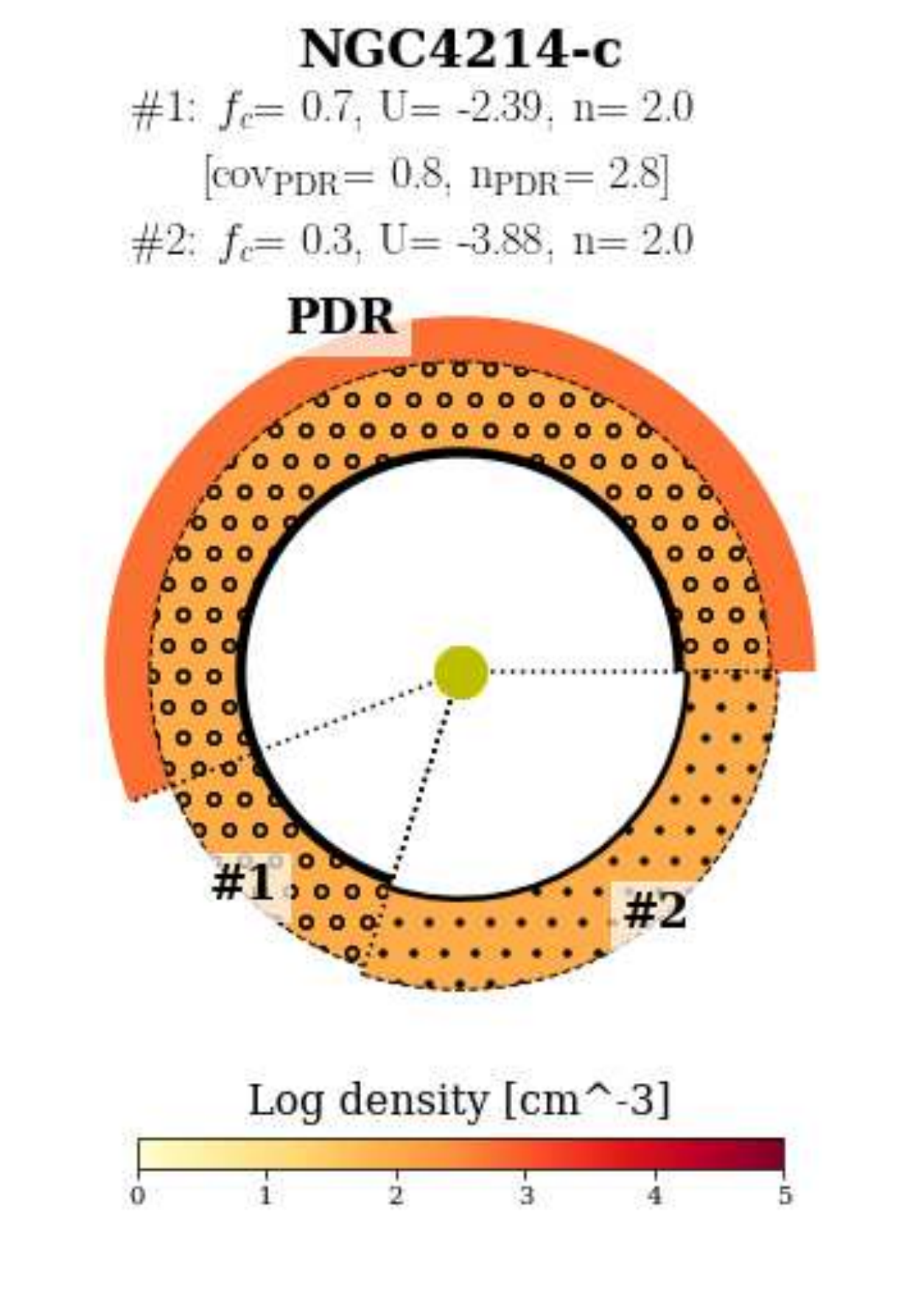}
\includegraphics[clip,trim=0 3cm 0 -1cm,width=4.5cm]{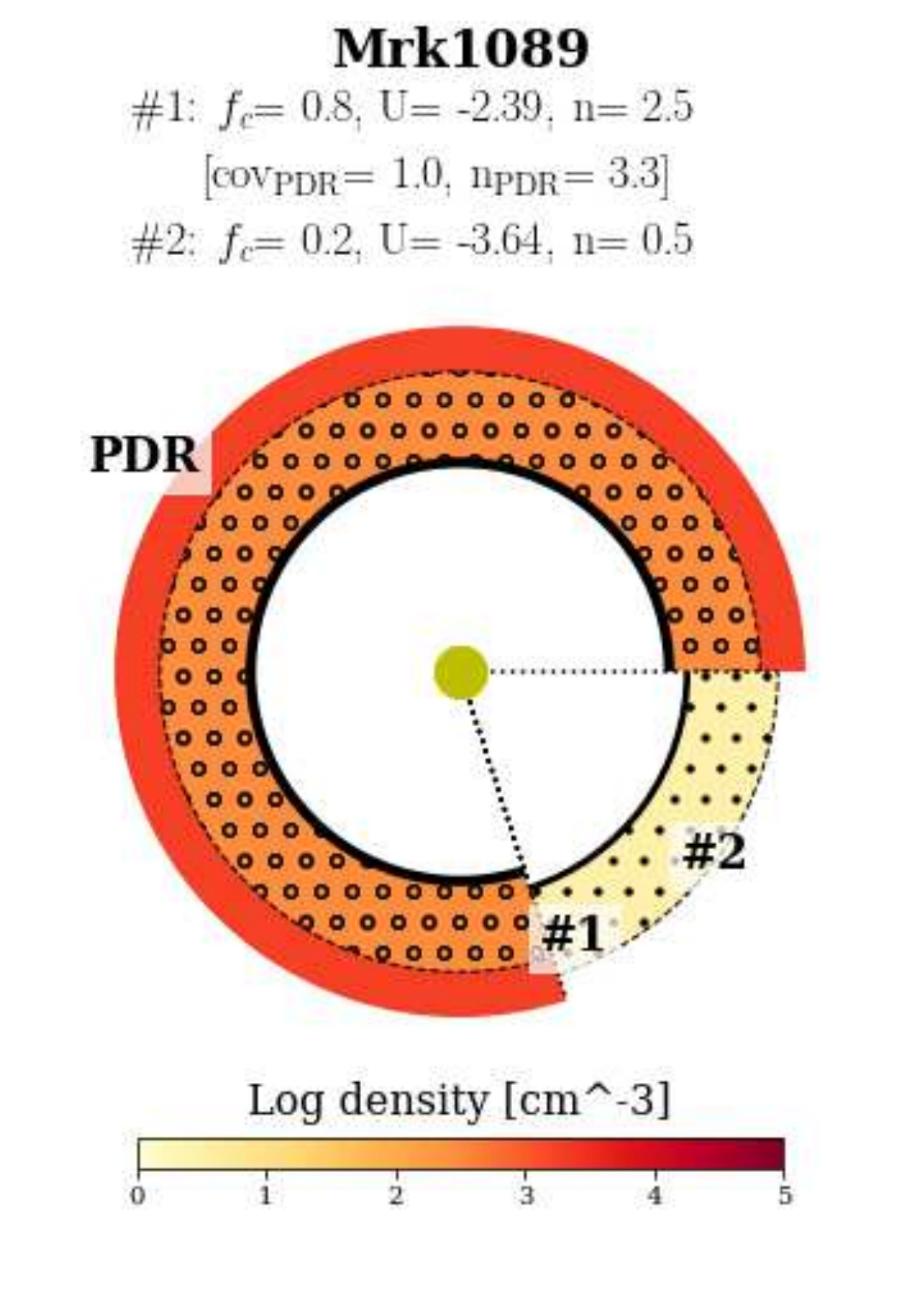}
\includegraphics[clip,trim=0 3cm 0 -1cm,width=4.5cm]{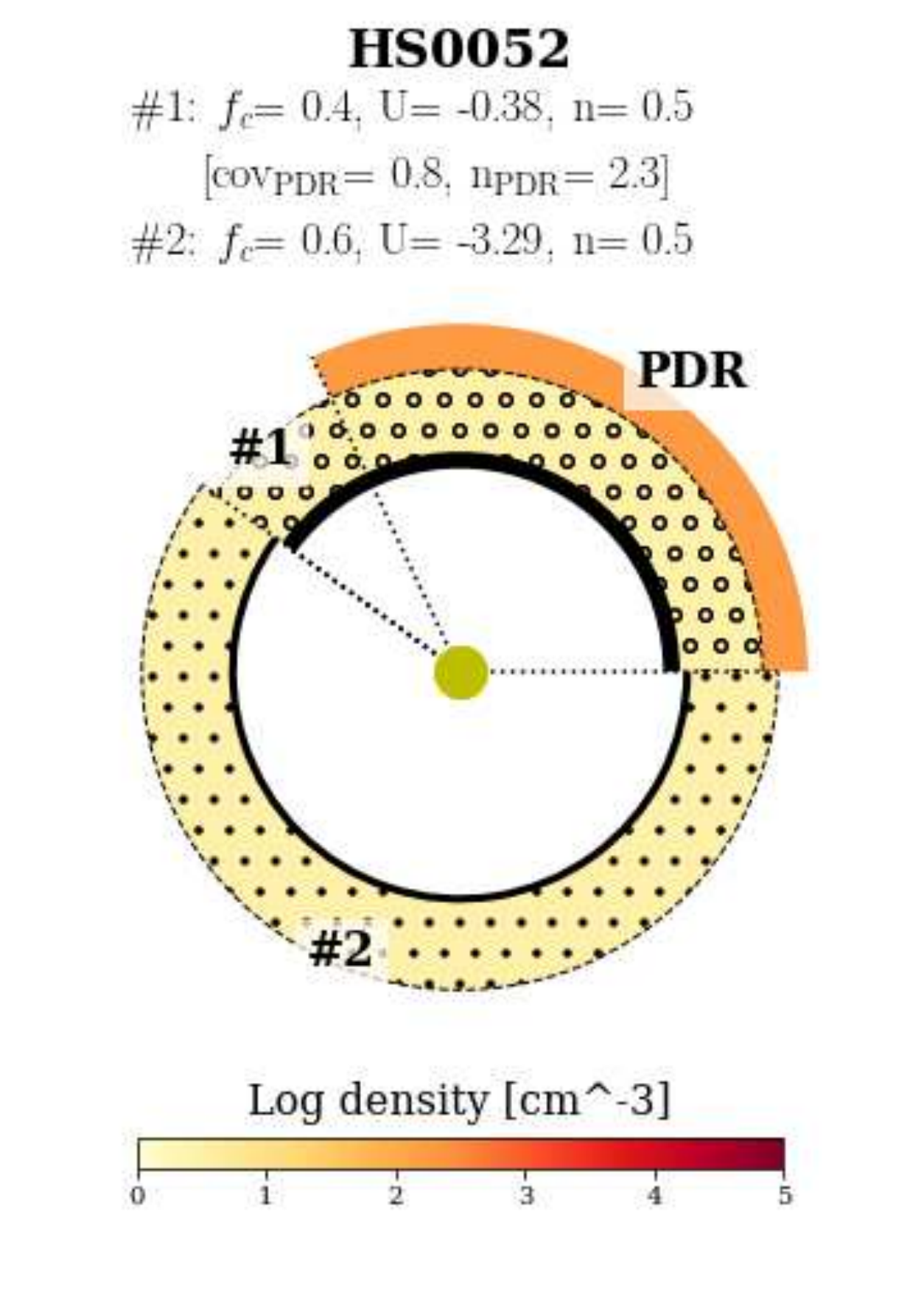}
\includegraphics[clip,trim=0 3cm 0 -1cm,width=4.5cm]{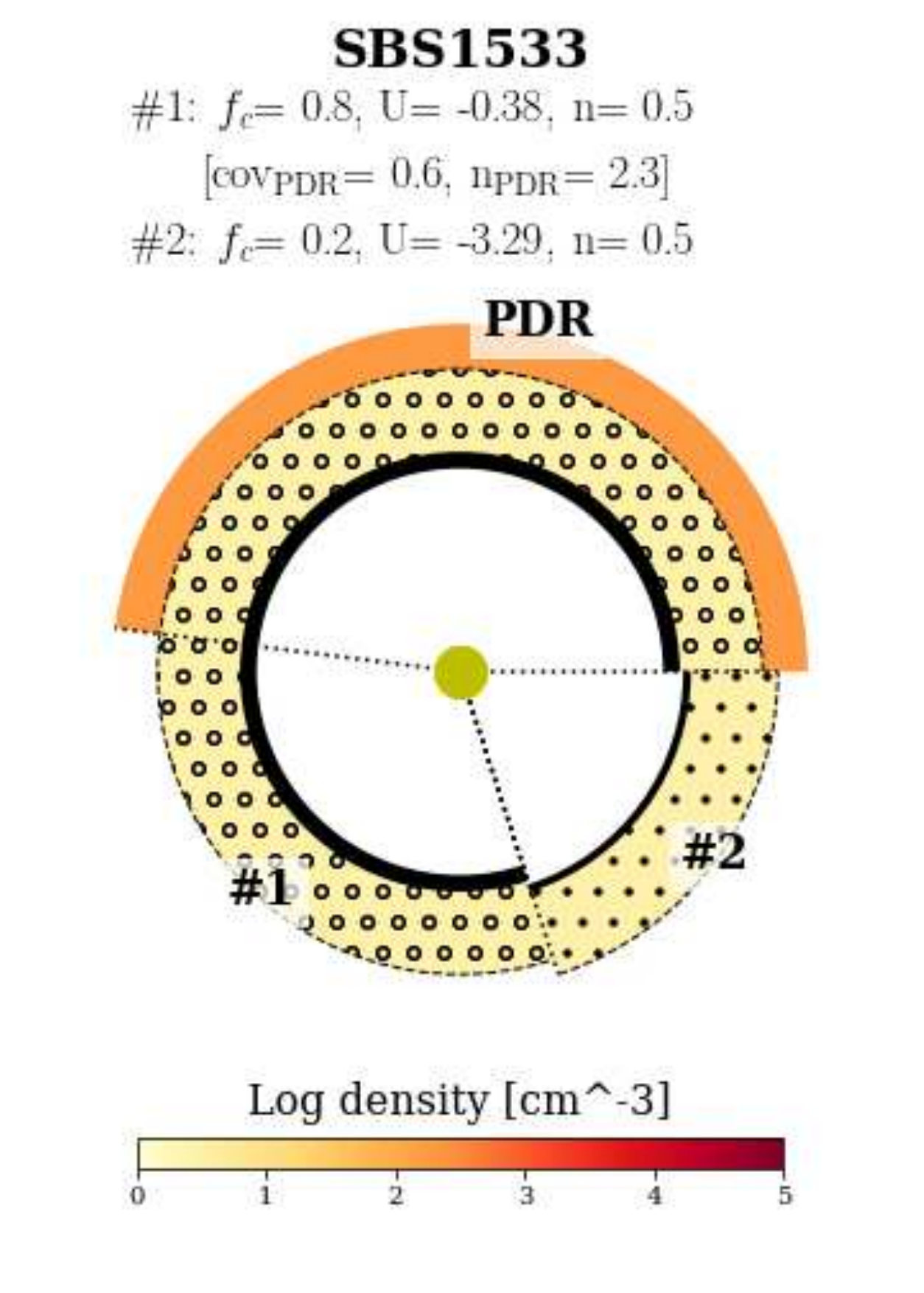}
\includegraphics[clip,trim=0 3cm 0 -1cm,width=4.5cm]{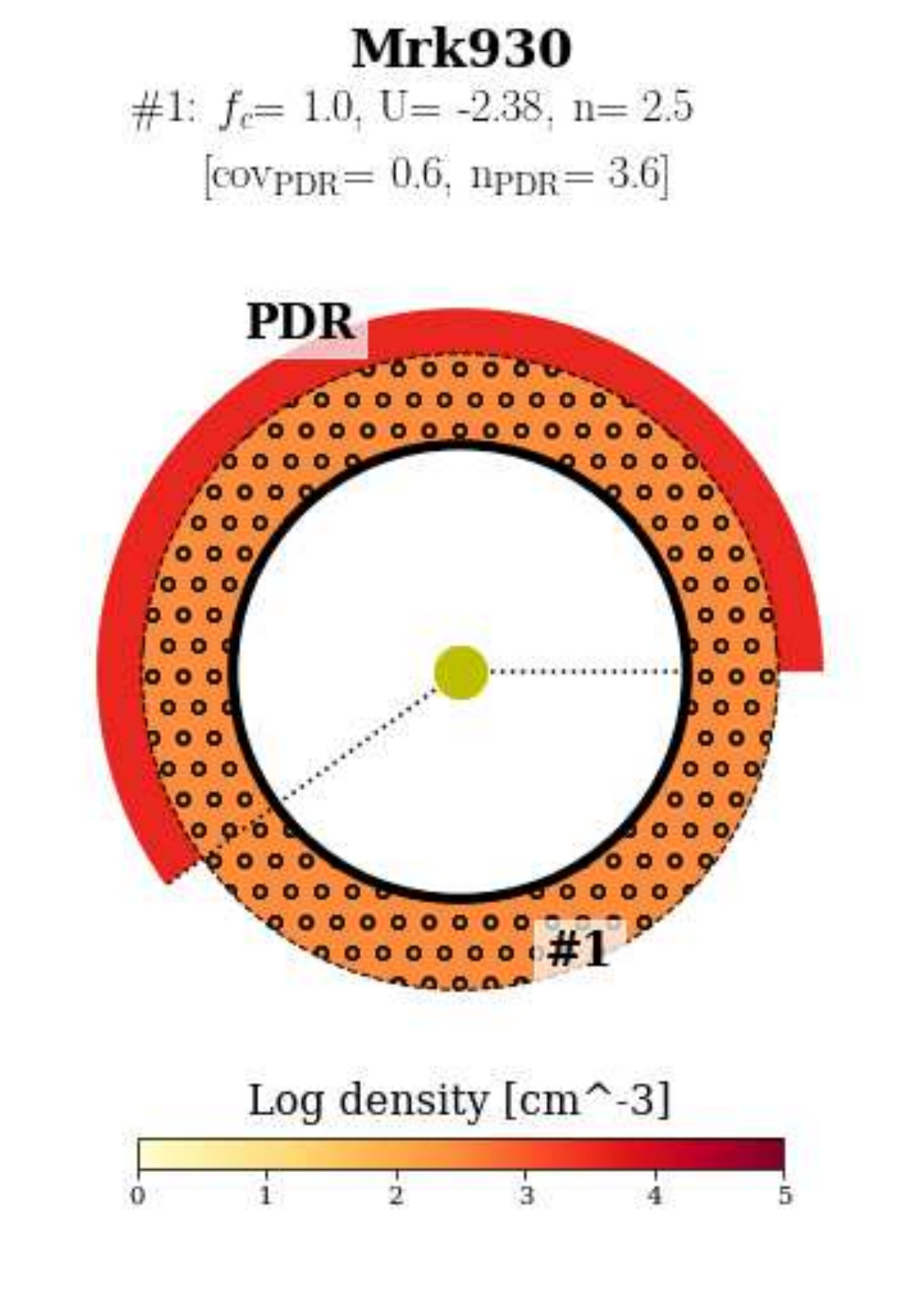}
\includegraphics[clip,trim=0 0 0 14cm,width=4.5cm]{./figures_appendix/NGC4214-c_mixed_a}
\caption{Best-fitting single or mixed model,
based on model results presented in
Tables~\ref{table:modelres1} and \ref{table:modelres2}.}
\end{figure*}

 \begin{figure*}[thp]
\centering
\includegraphics[clip,trim=0 3cm 0 -1cm,width=4.5cm]{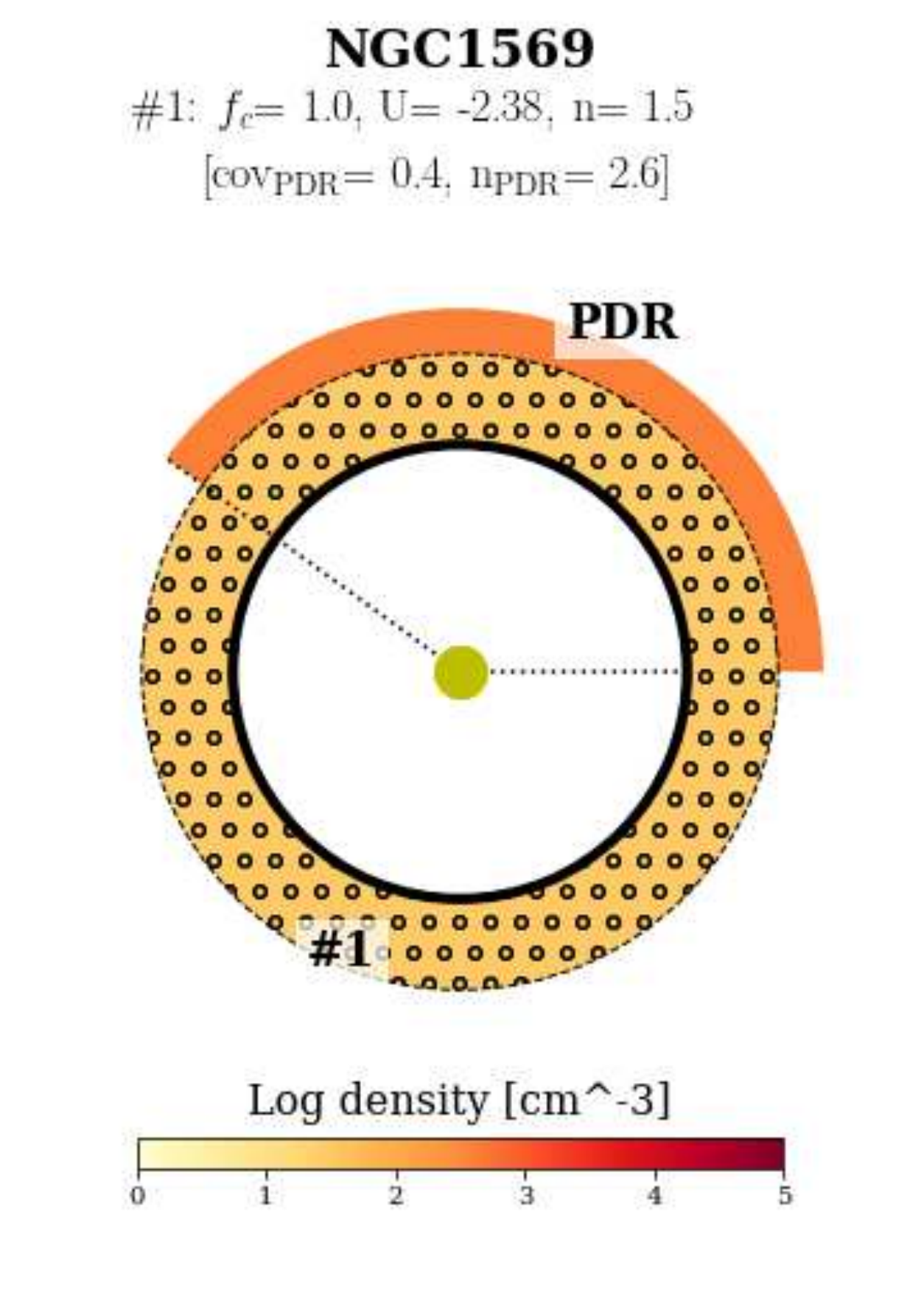}
\includegraphics[clip,trim=0 3cm 0 -1cm,width=4.5cm]{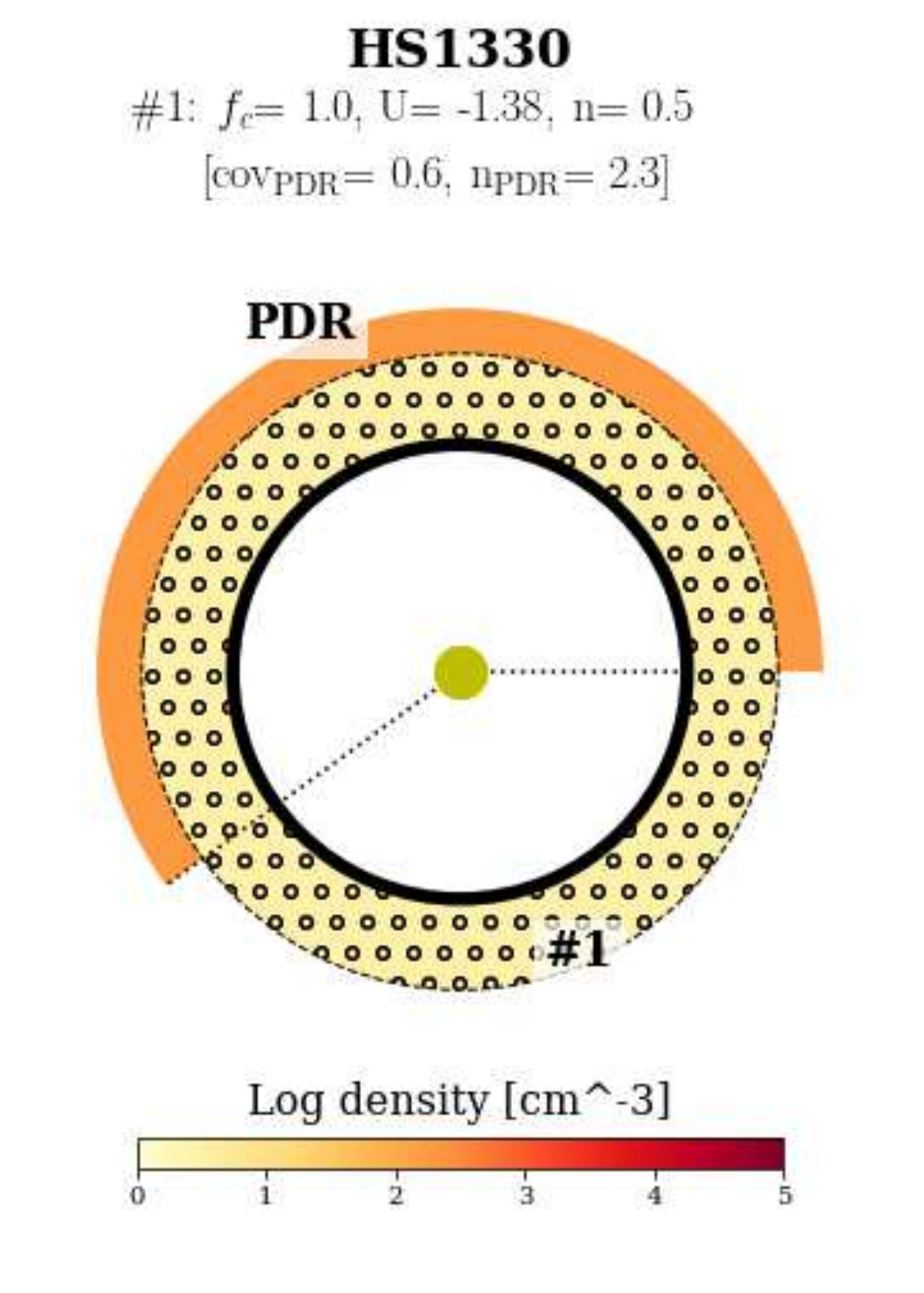}
\includegraphics[clip,trim=0 3cm 0 -1cm,width=4.5cm]{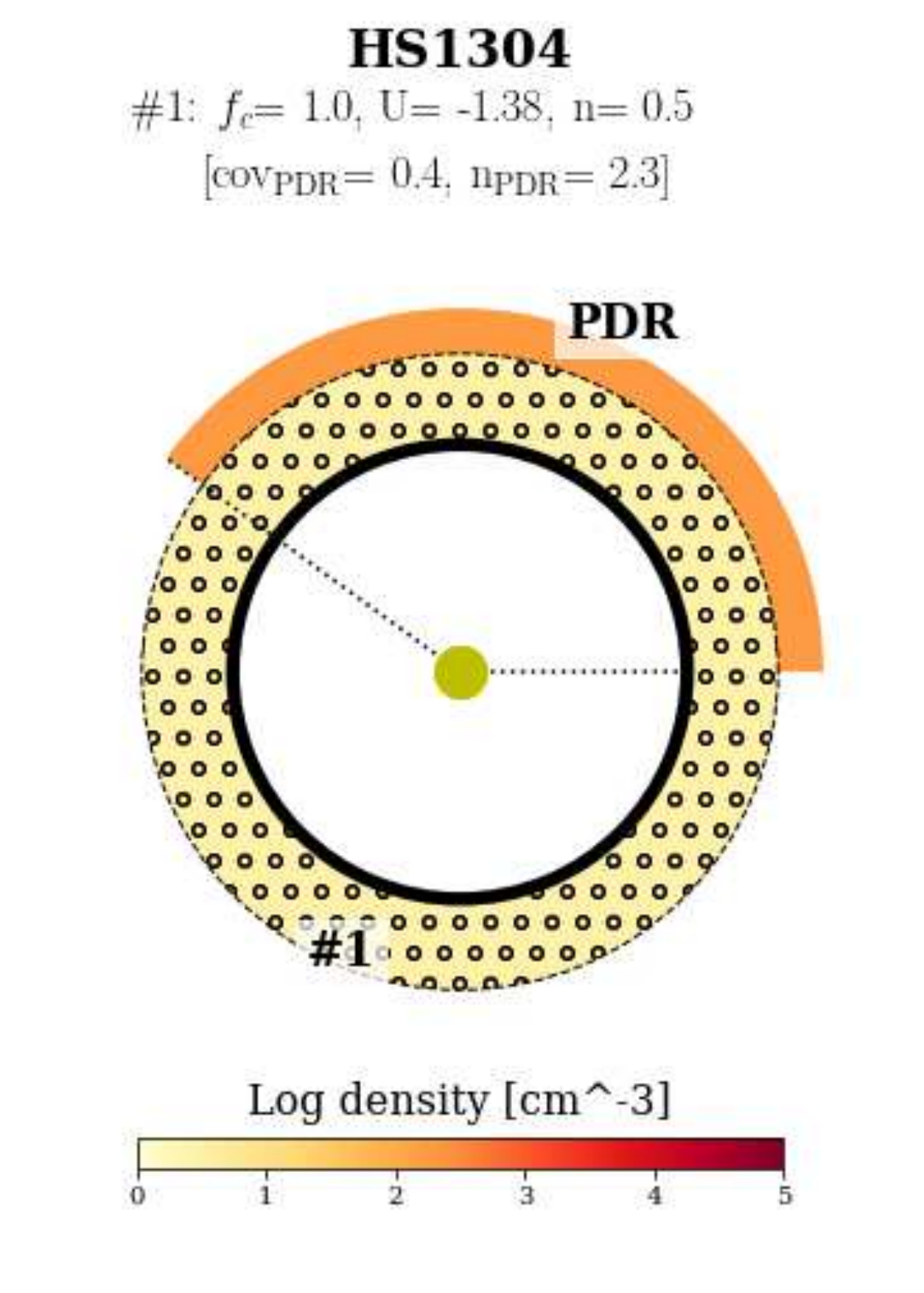}
\includegraphics[clip,trim=0 3cm 0 -1cm,width=4.5cm]{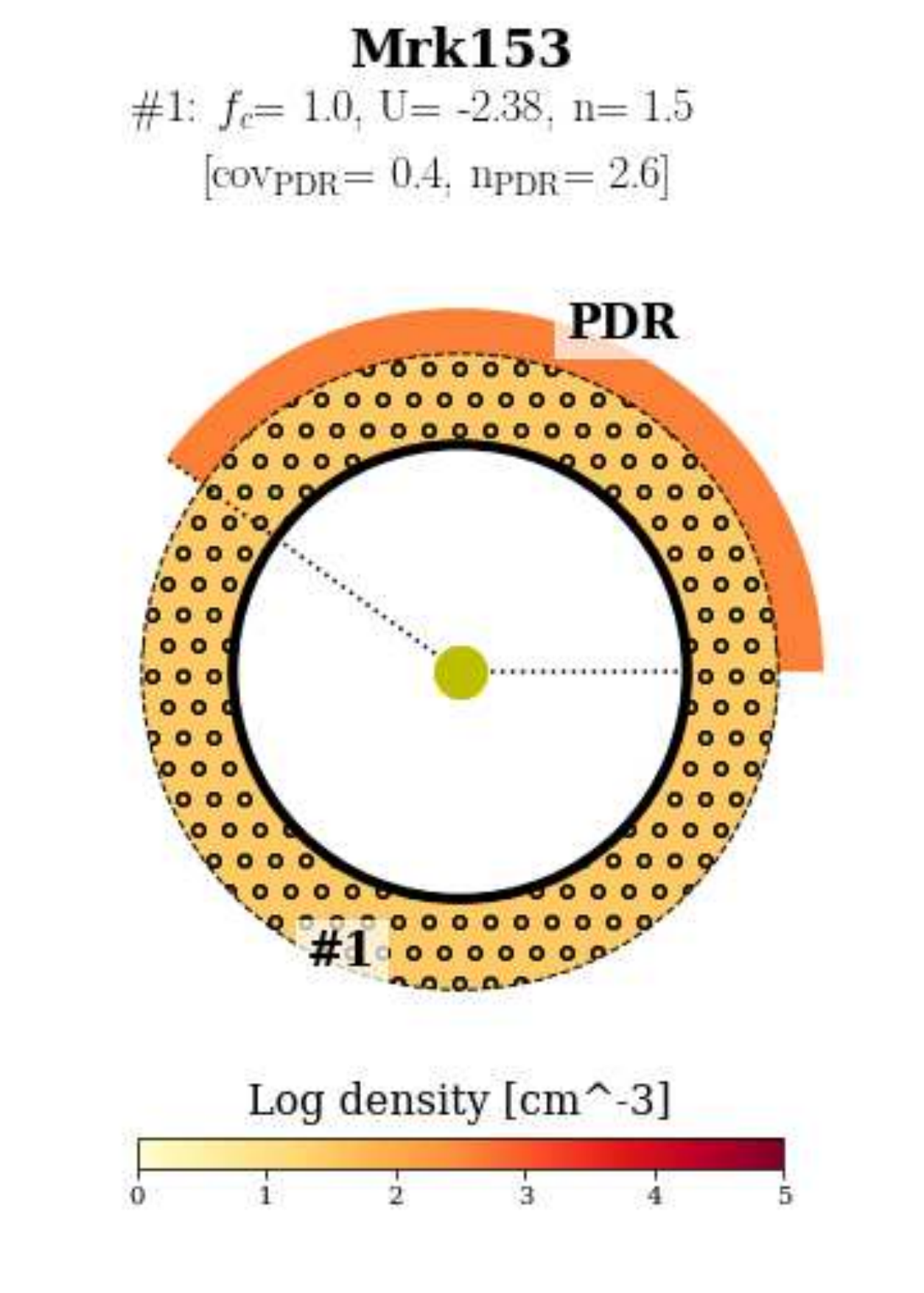}
\includegraphics[clip,trim=0 3cm 0 -1cm,width=4.5cm]{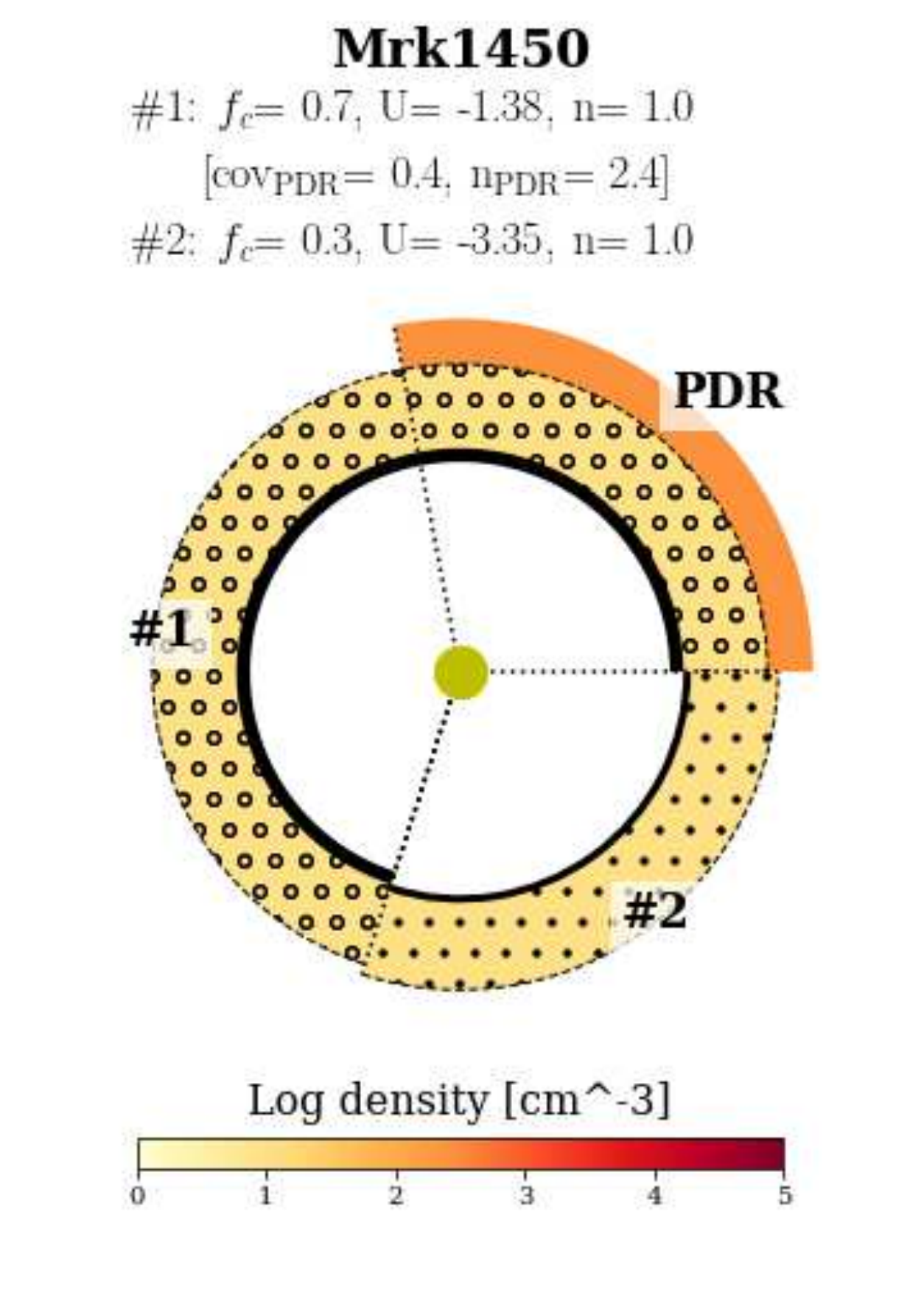}
\includegraphics[clip,trim=0 3cm 0 -1cm,width=4.5cm]{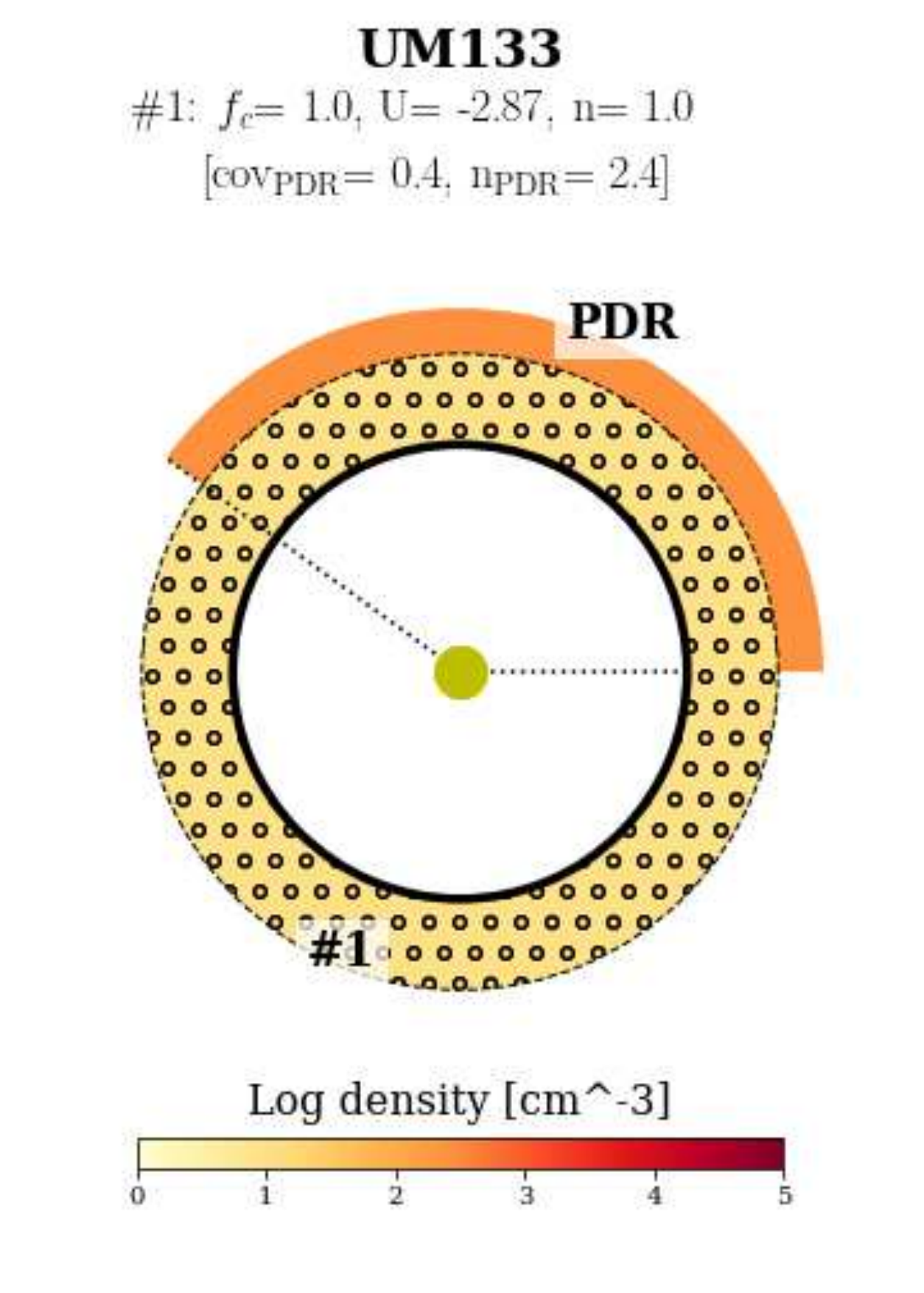}
\includegraphics[clip,trim=0 3cm 0 -1cm,width=4.5cm]{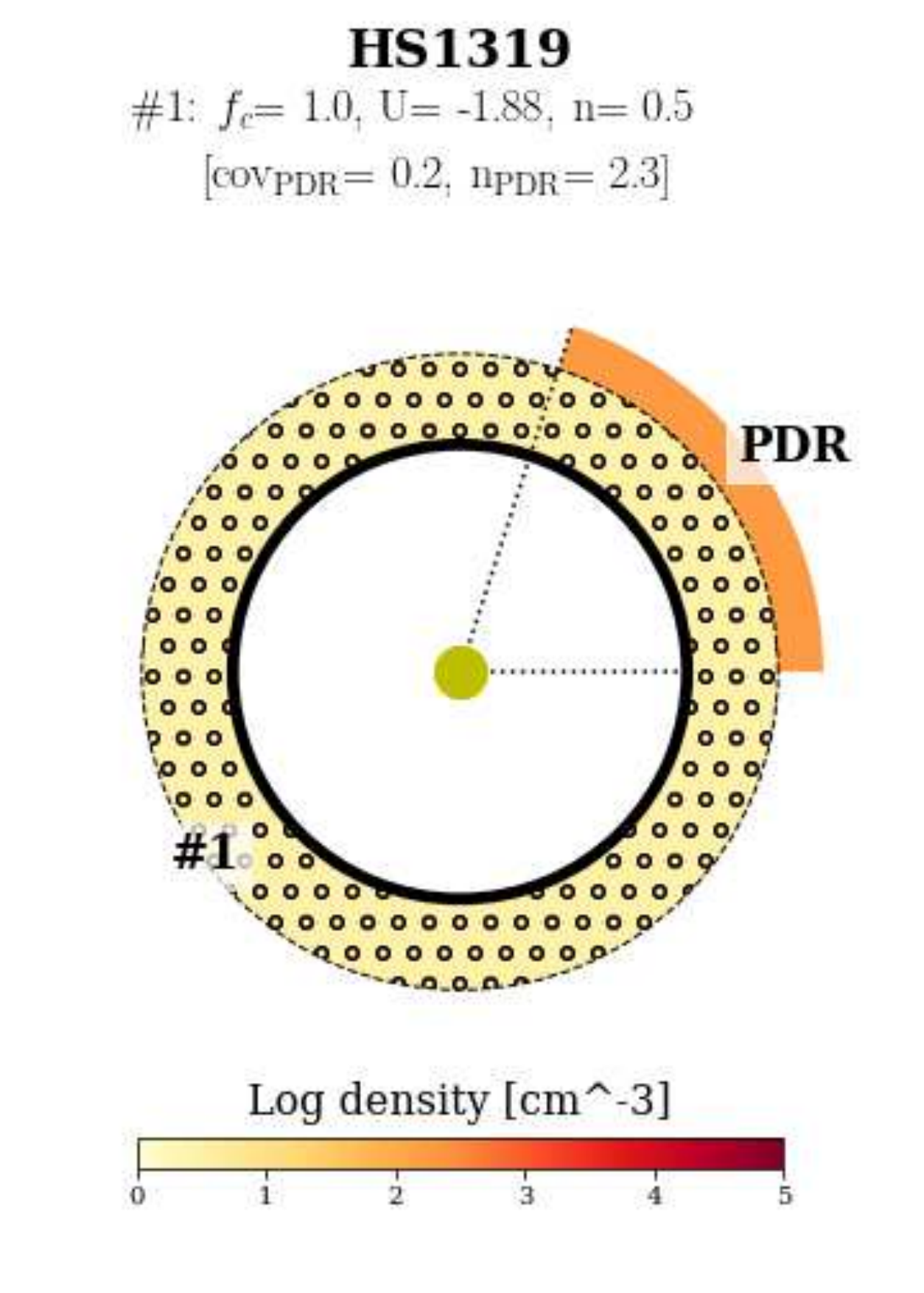}
\includegraphics[clip,trim=0 3cm 0 -1cm,width=4.5cm]{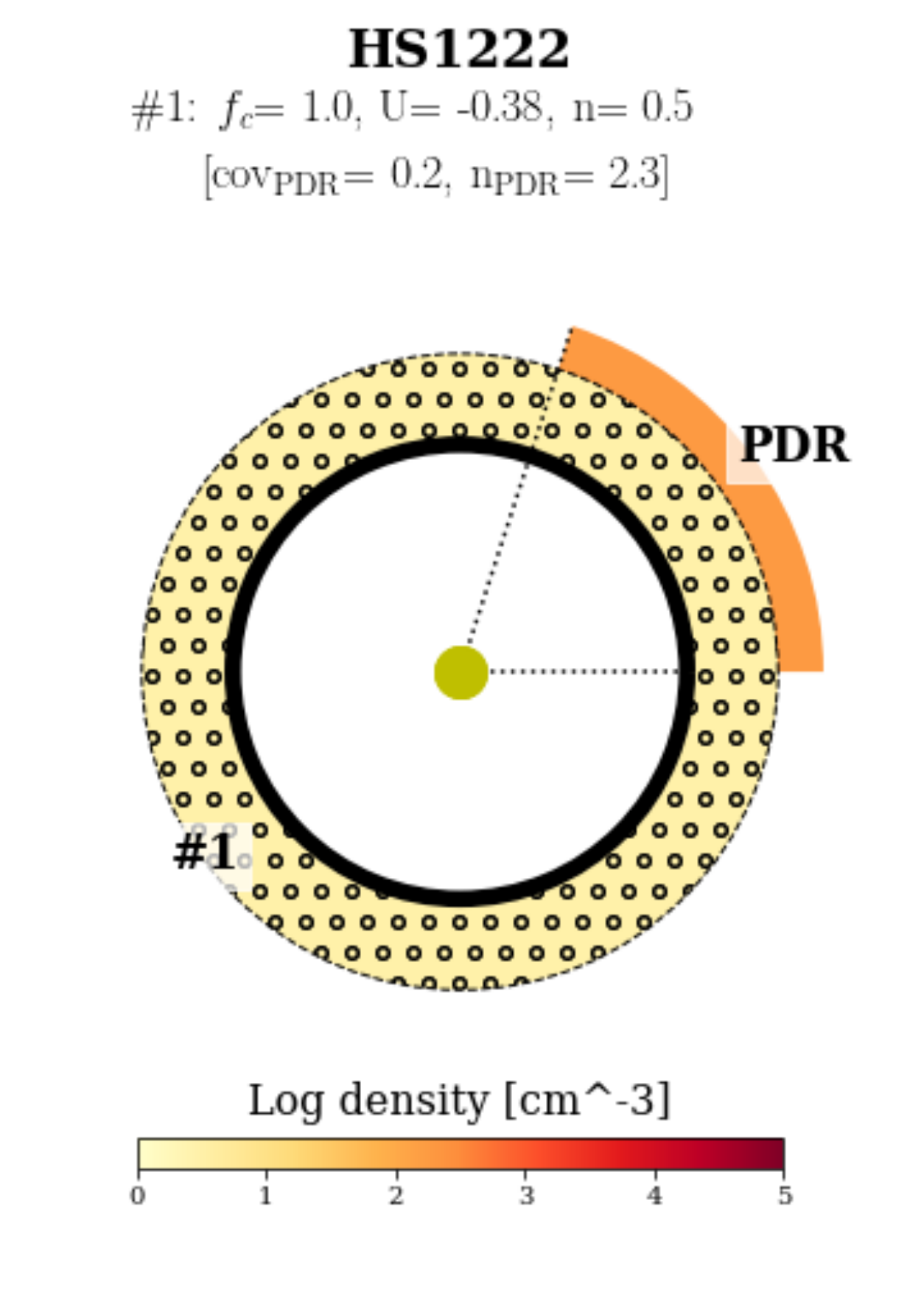}
\includegraphics[clip,trim=0 3cm 0 -1cm,width=4.5cm]{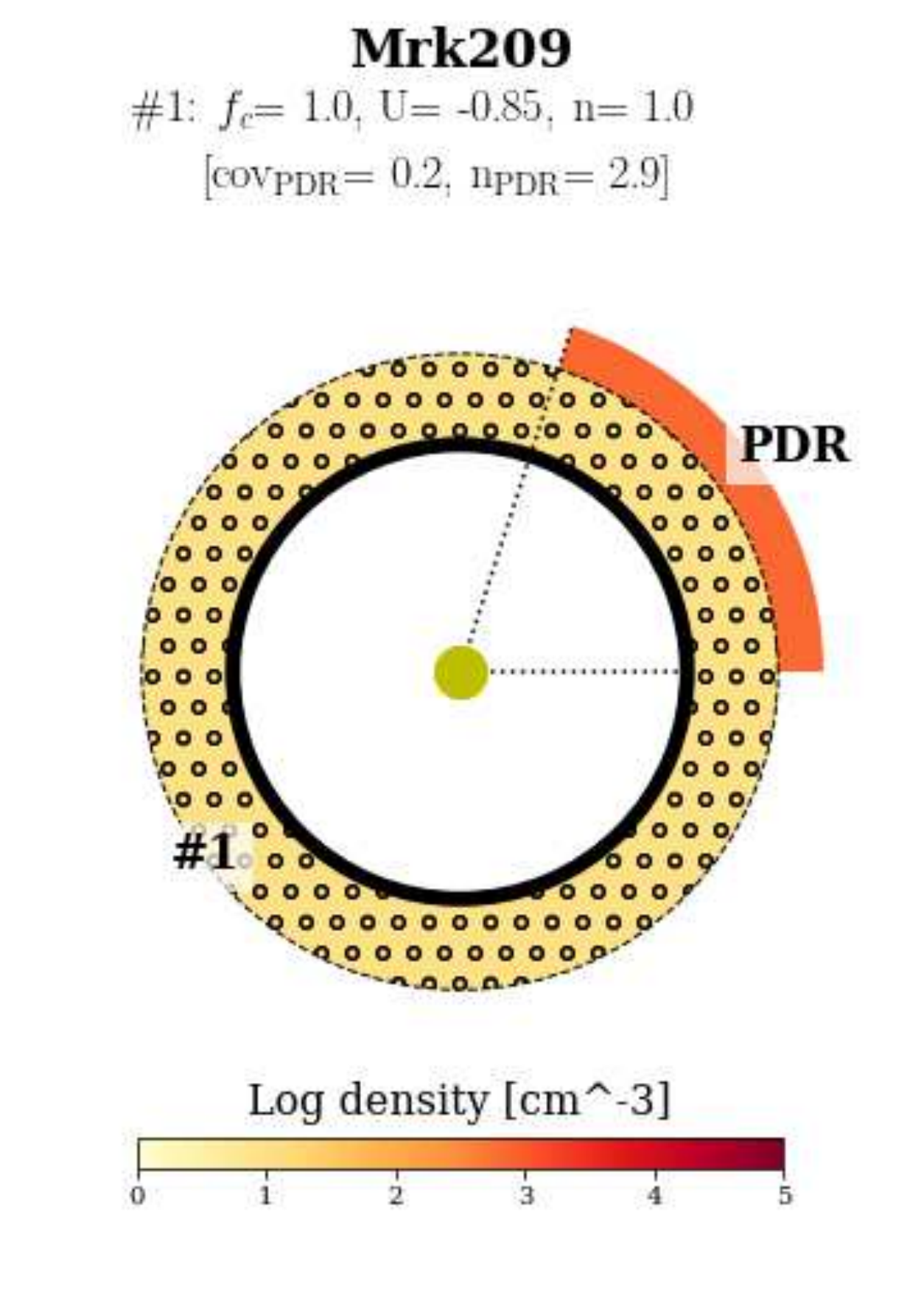}
\includegraphics[clip,trim=0 3cm 0 -1cm,width=4.5cm]{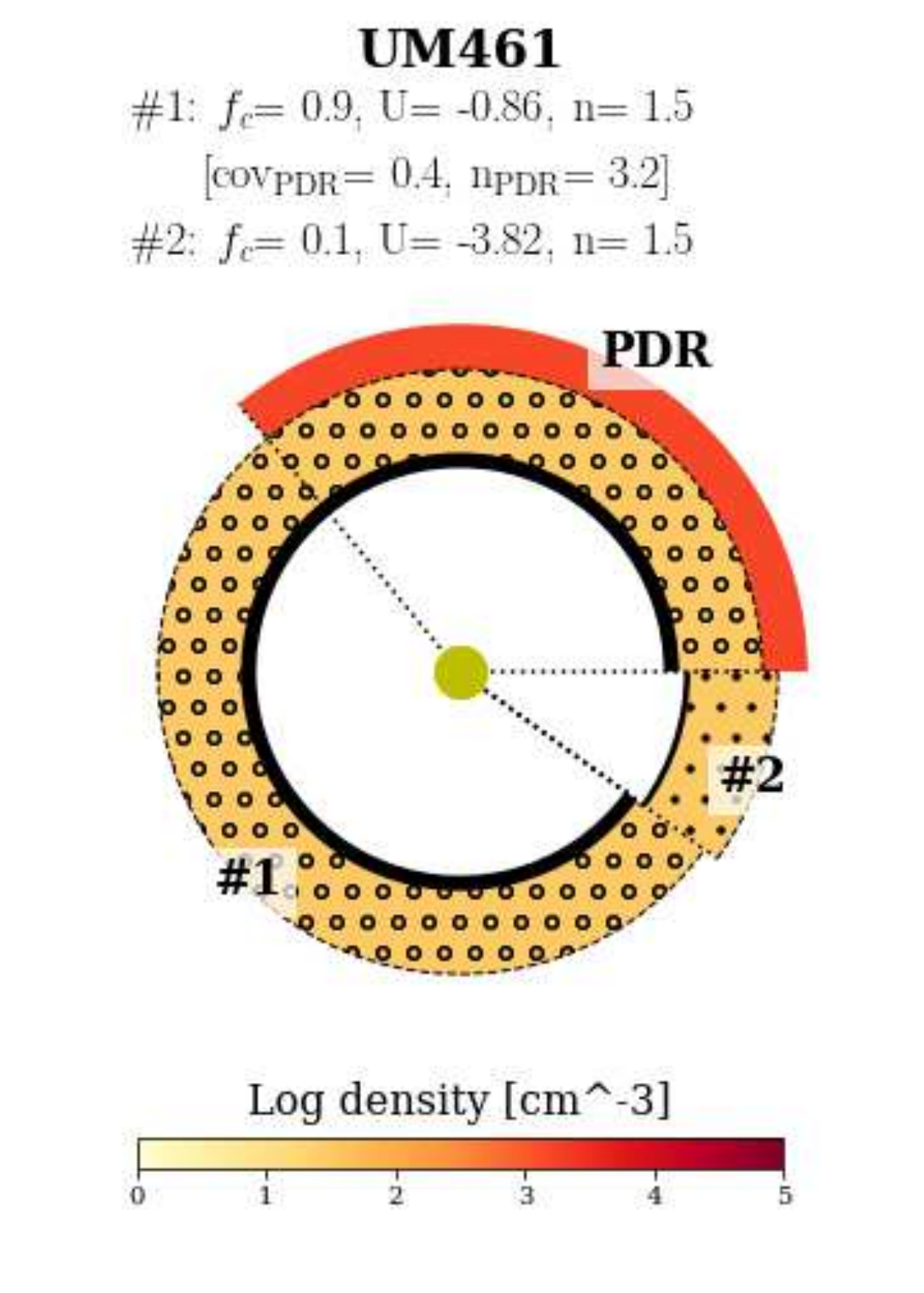}
\includegraphics[clip,trim=0 3cm 0 -1cm,width=4.5cm]{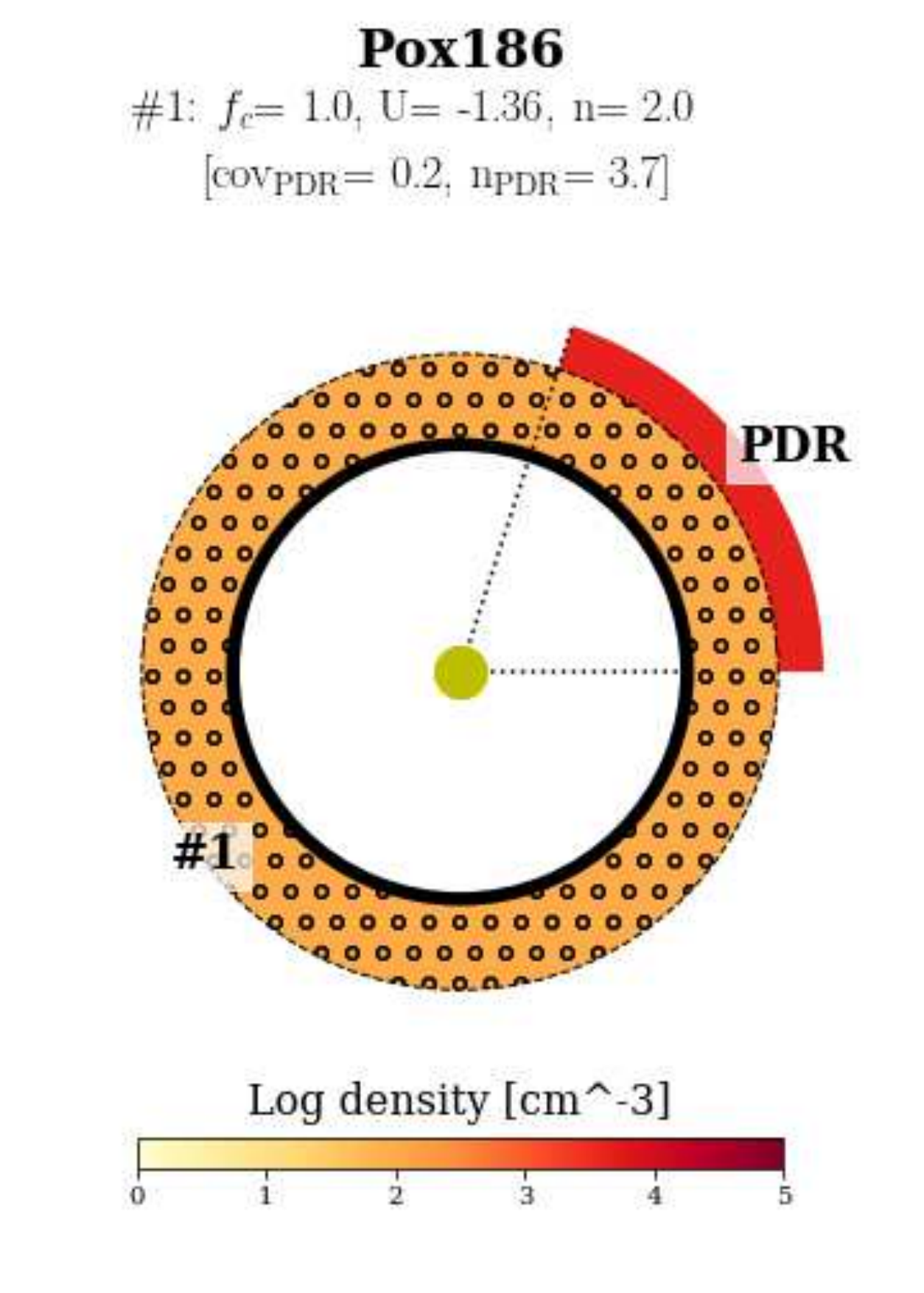}
\includegraphics[clip,trim=0 3cm 0 -1cm,width=4.5cm]{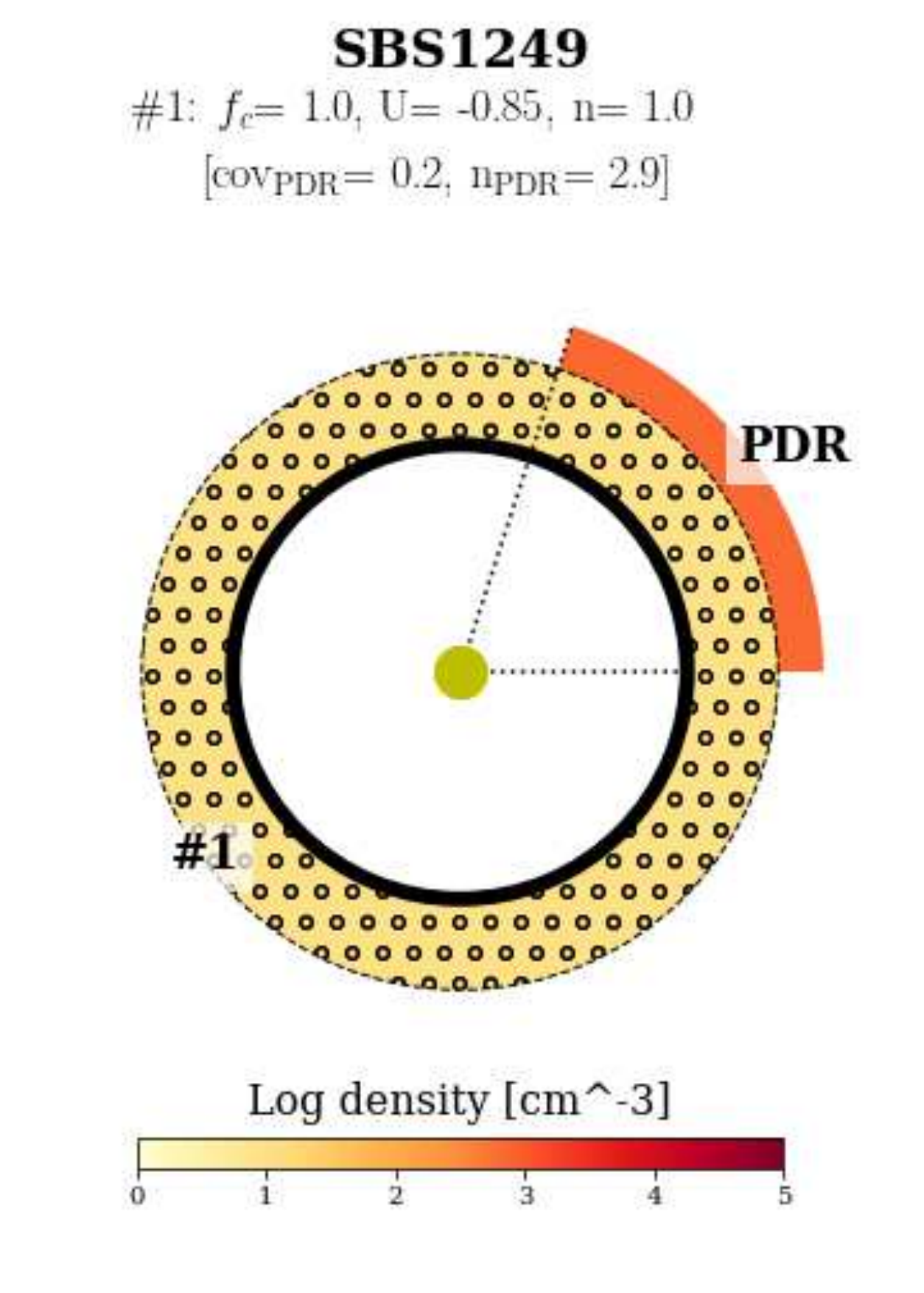}
\includegraphics[clip,trim=0 3cm 0 -1cm,width=4.5cm]{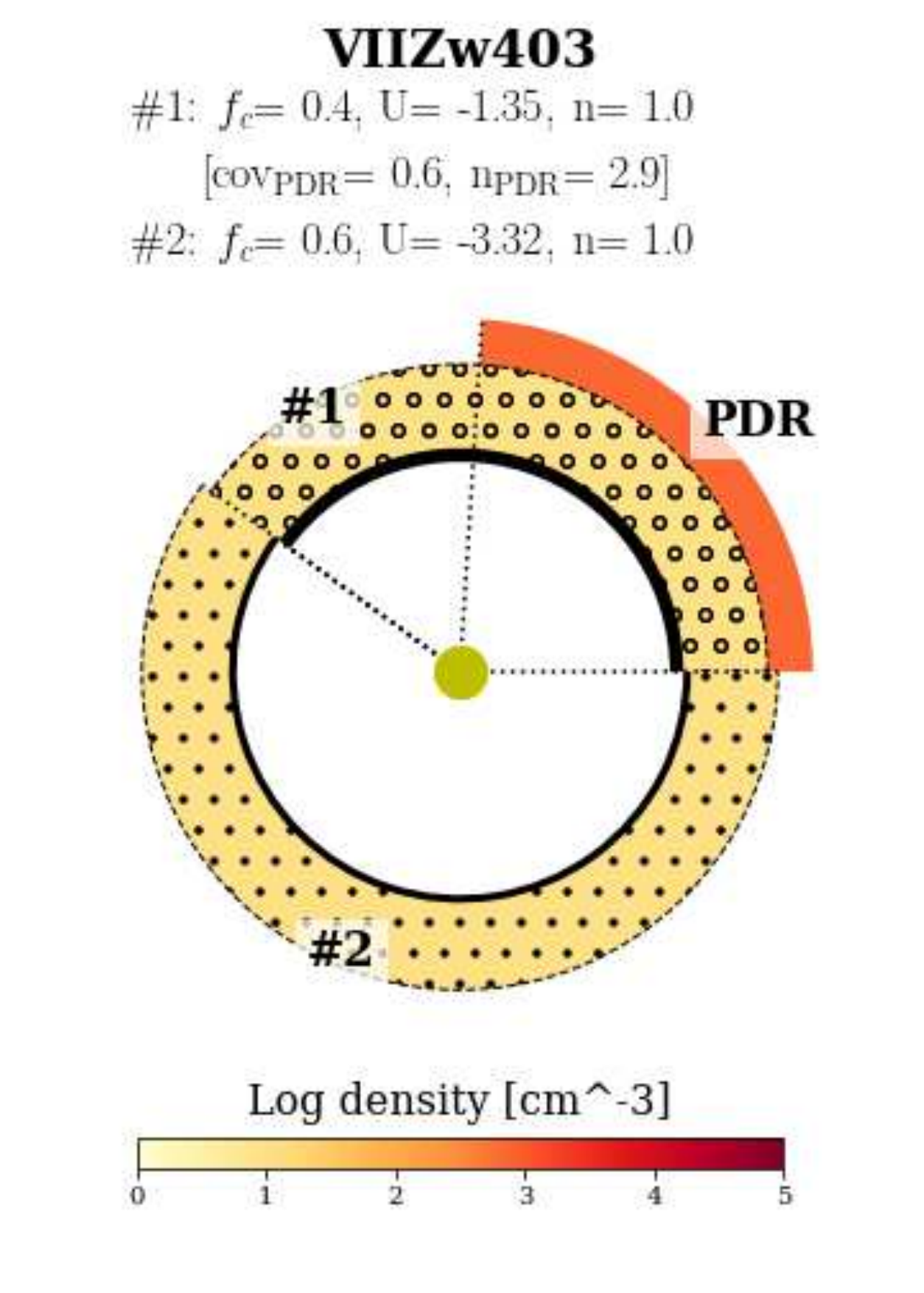}
\includegraphics[clip,trim=0 3cm 0 -1cm,width=4.5cm]{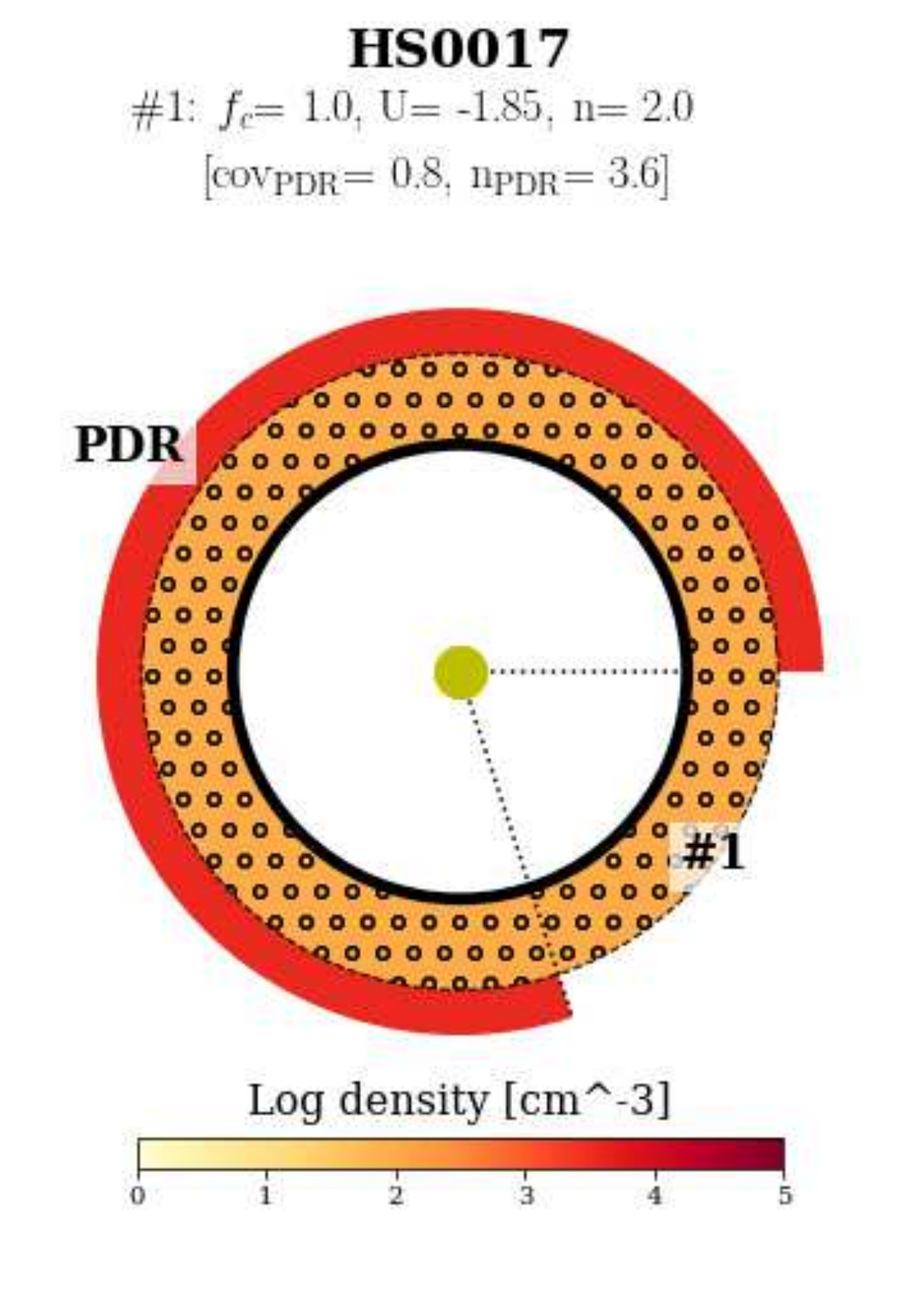}
\includegraphics[clip,trim=0 3cm 0 -1cm,width=4.5cm]{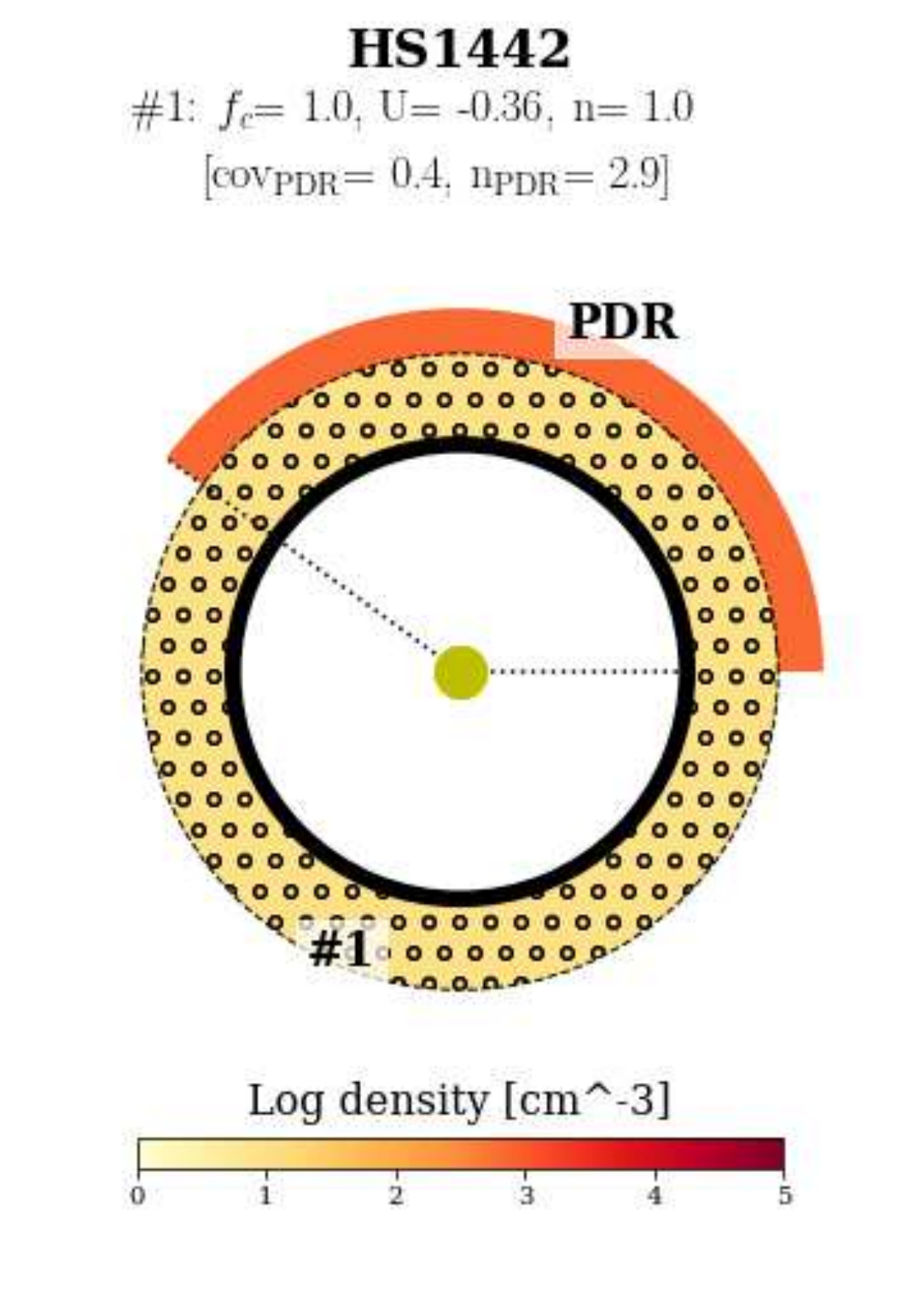}
\includegraphics[clip,trim=0 3cm 0 -1cm,width=4.5cm]{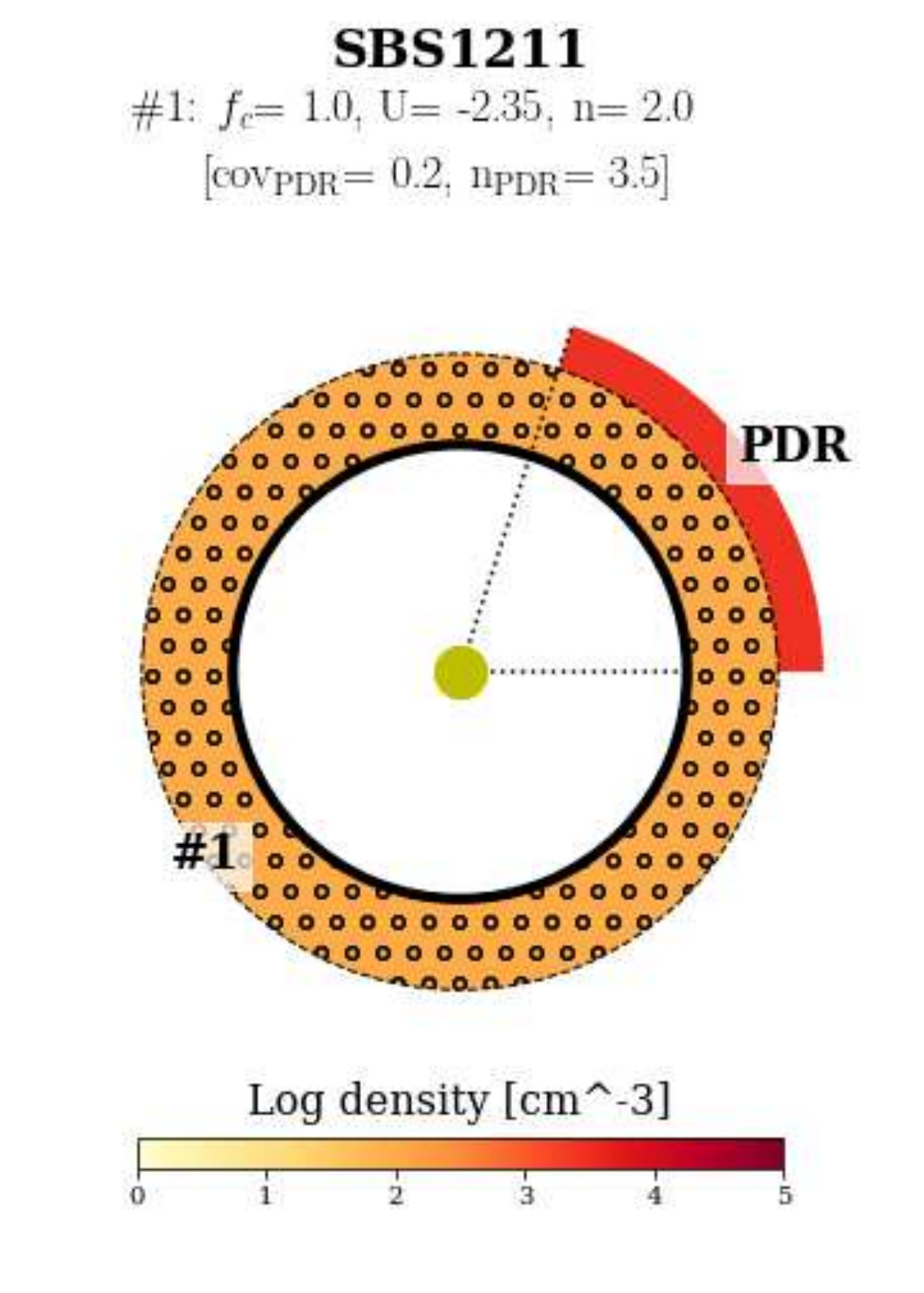}
\includegraphics[clip,trim=0 0 0 14cm,width=4.5cm]{./figures_appendix/NGC4214-c_mixed_a}
\caption{continued.}
\end{figure*}

 \begin{figure*}[thp]
\centering
\includegraphics[clip,trim=0 3cm 0 -1cm,width=4.5cm]{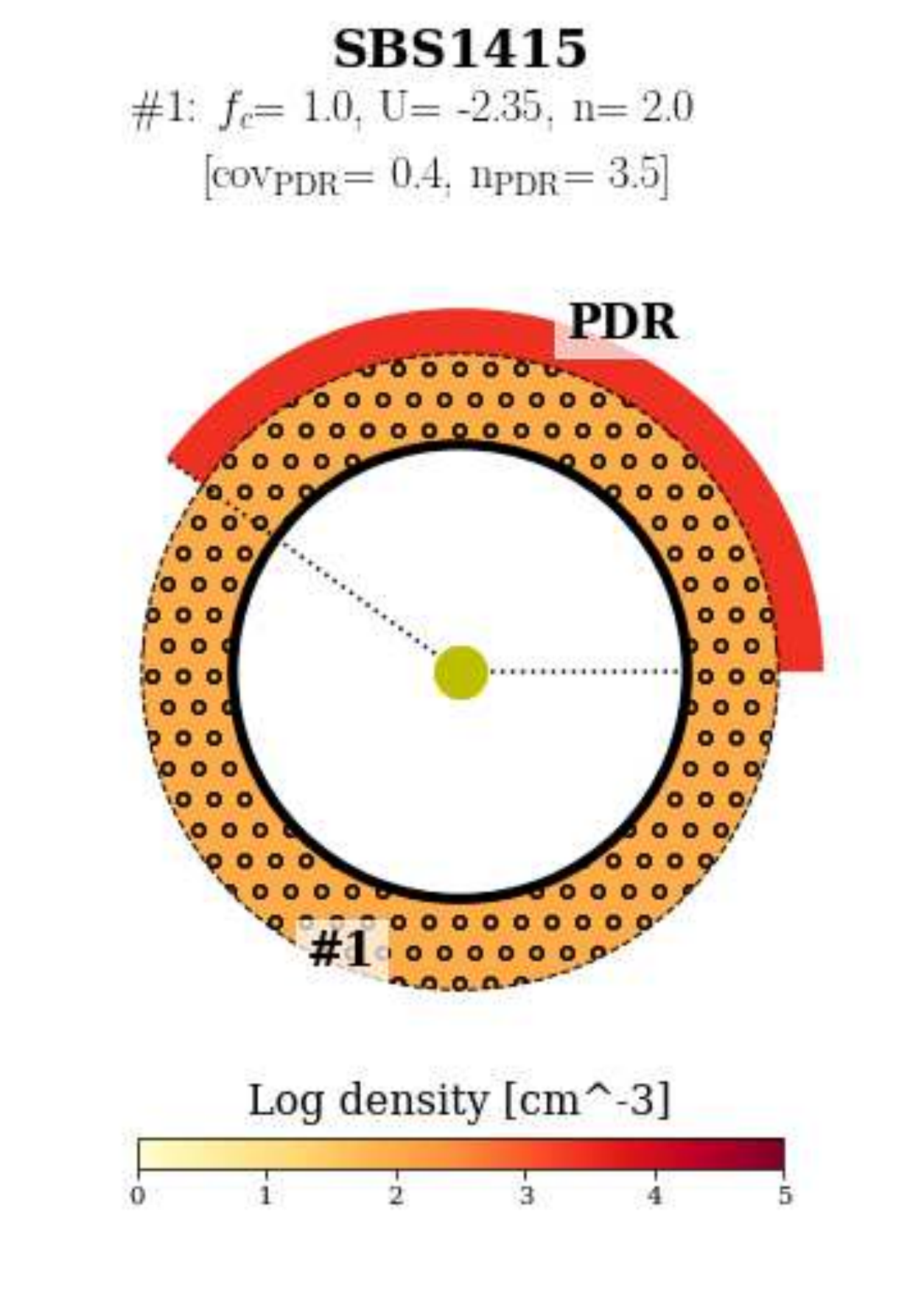}
\includegraphics[clip,trim=0 3cm 0 -1cm,width=4.5cm]{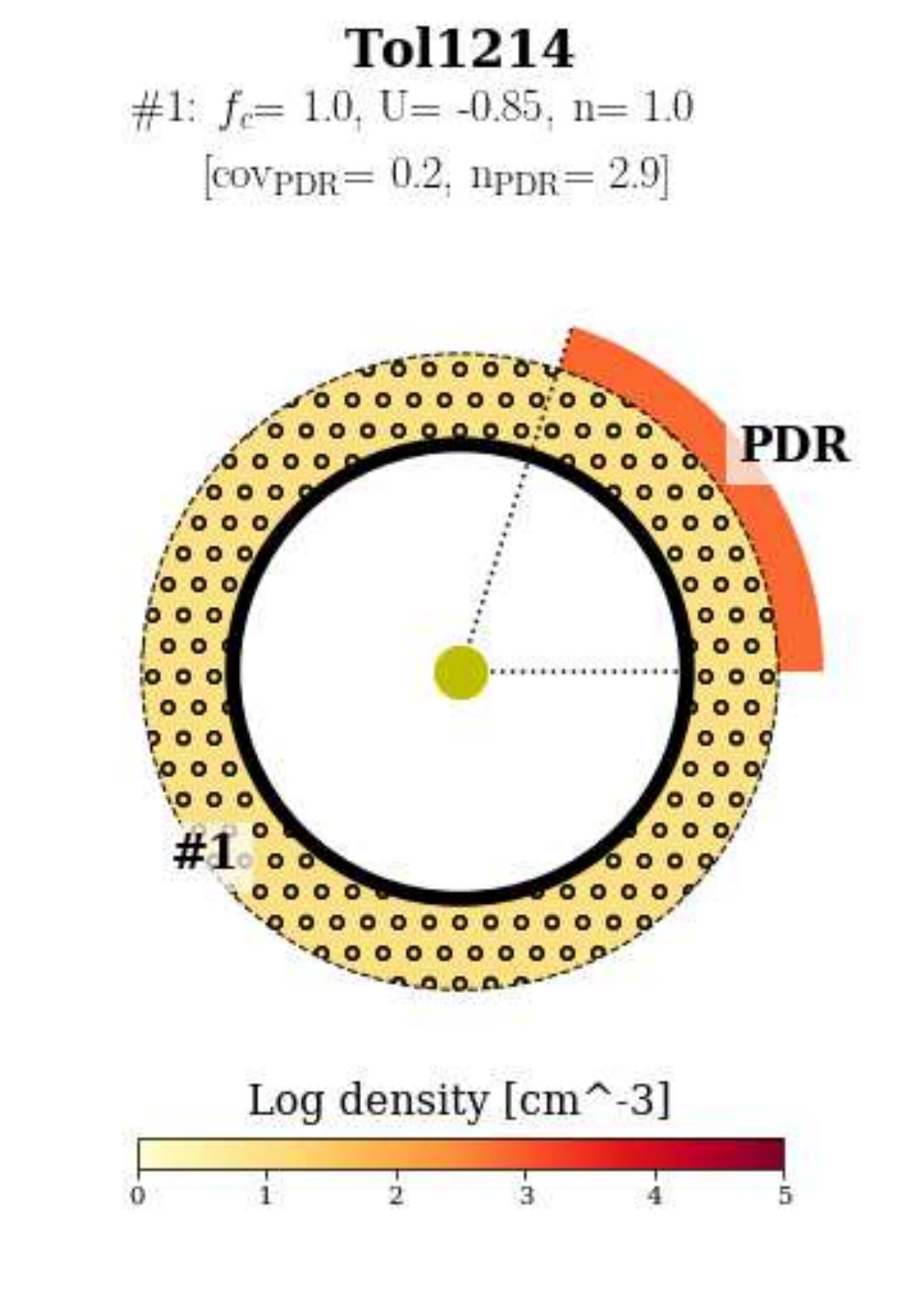}
\includegraphics[clip,trim=0 3cm 0 -1cm,width=4.5cm]{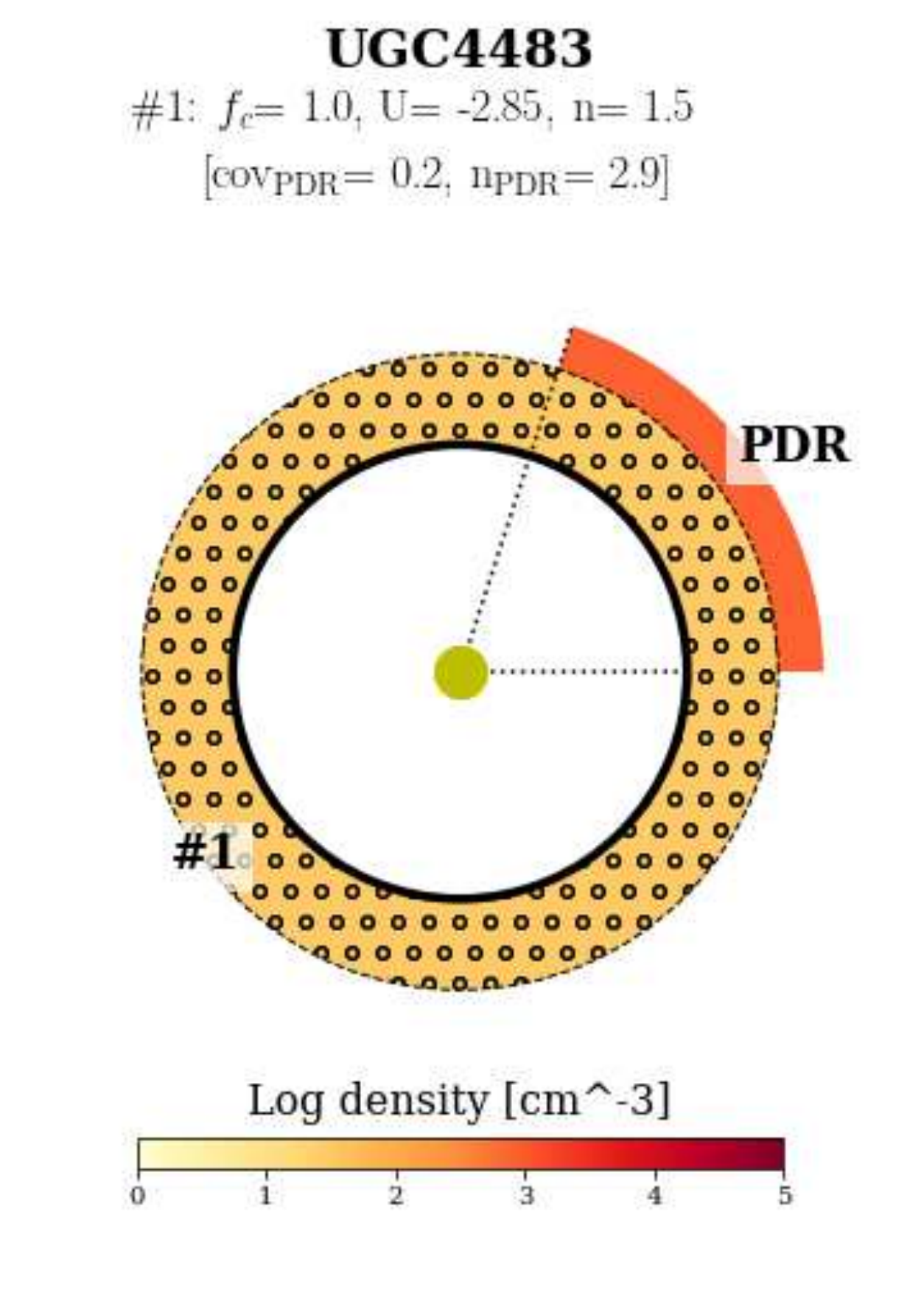}
\includegraphics[clip,trim=0 3cm 0 -1cm,width=4.5cm]{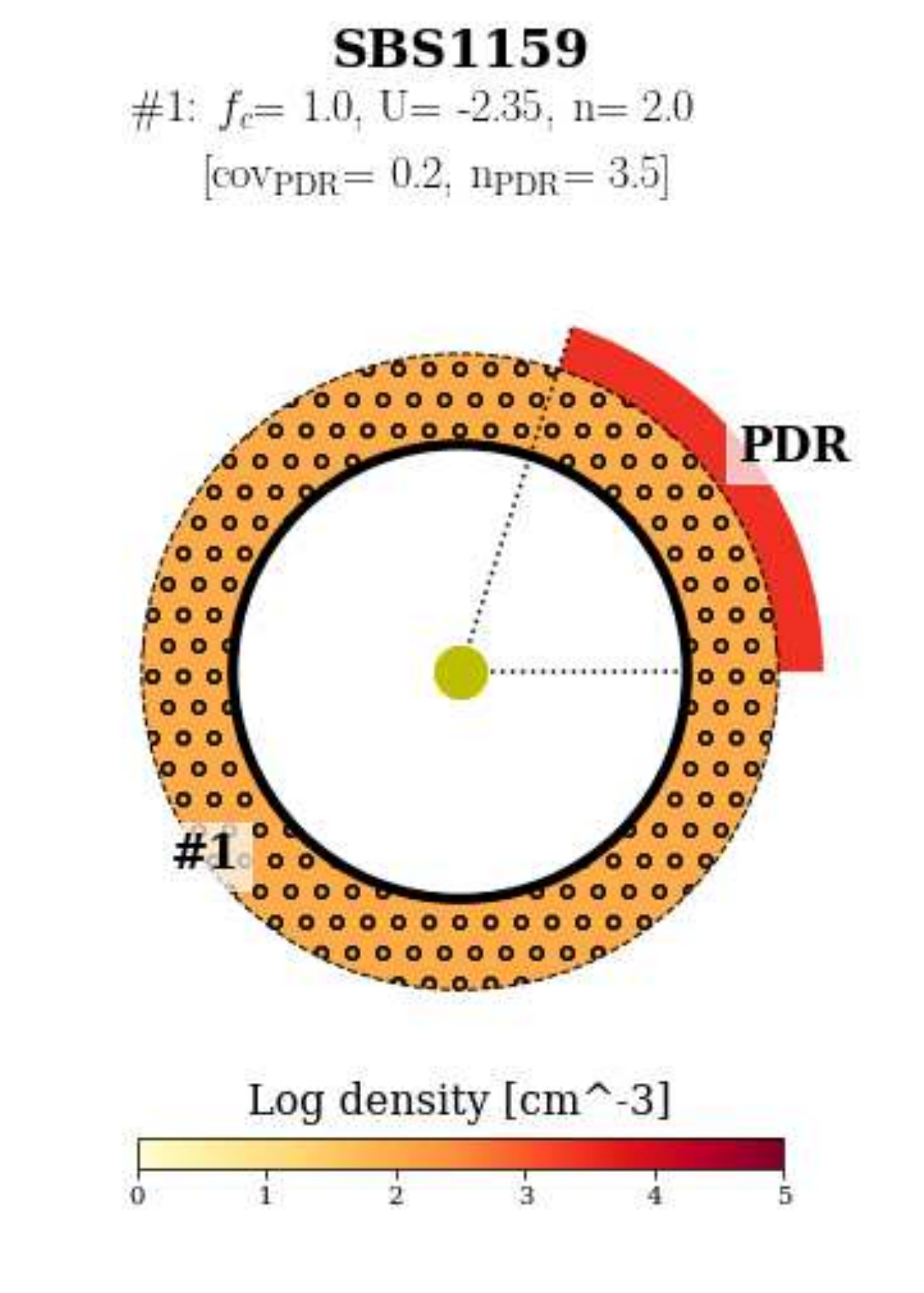}
\includegraphics[clip,trim=0 3cm 0 -1cm,width=4.5cm]{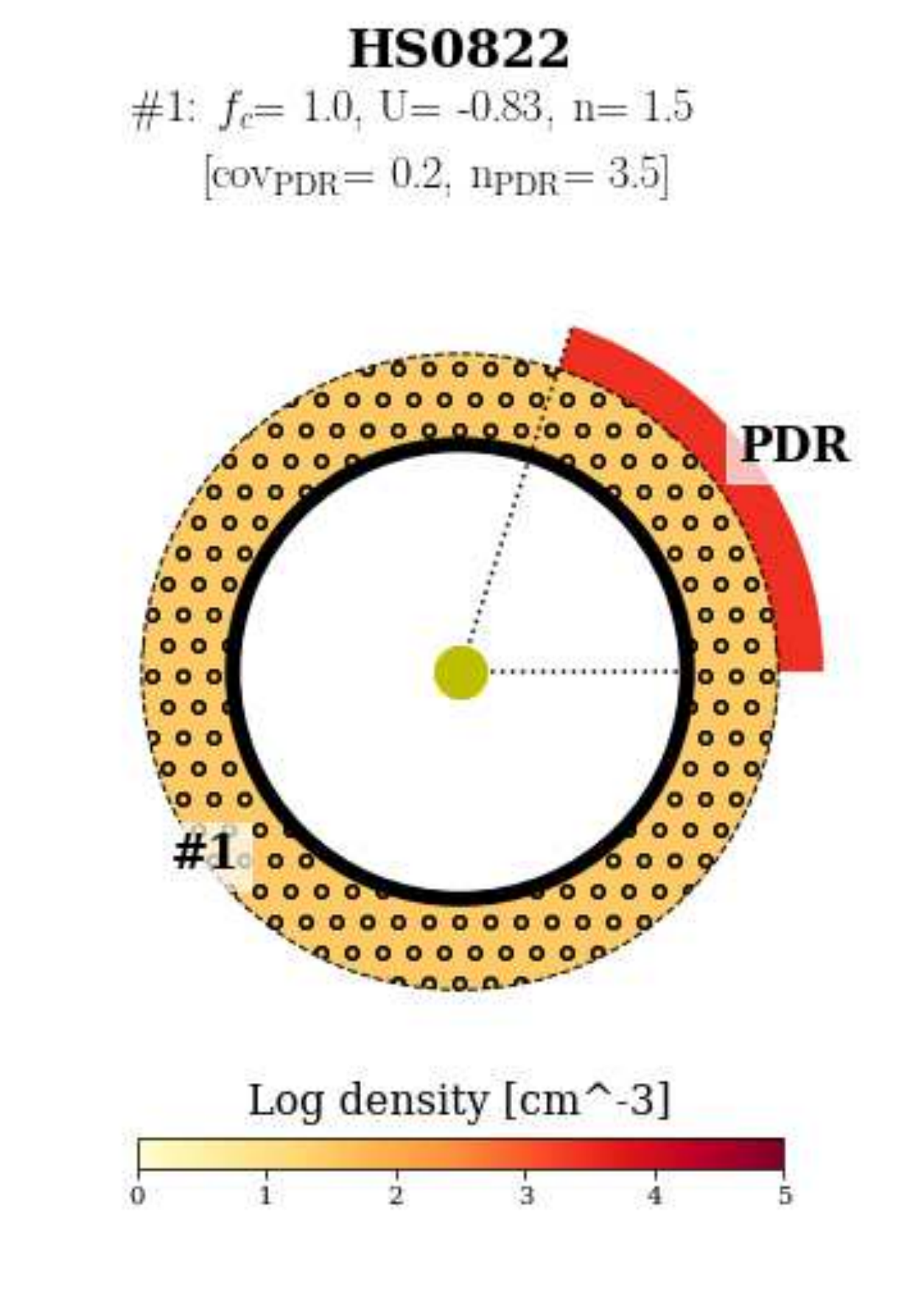}
\includegraphics[clip,trim=0 3cm 0 -1cm,width=4.5cm]{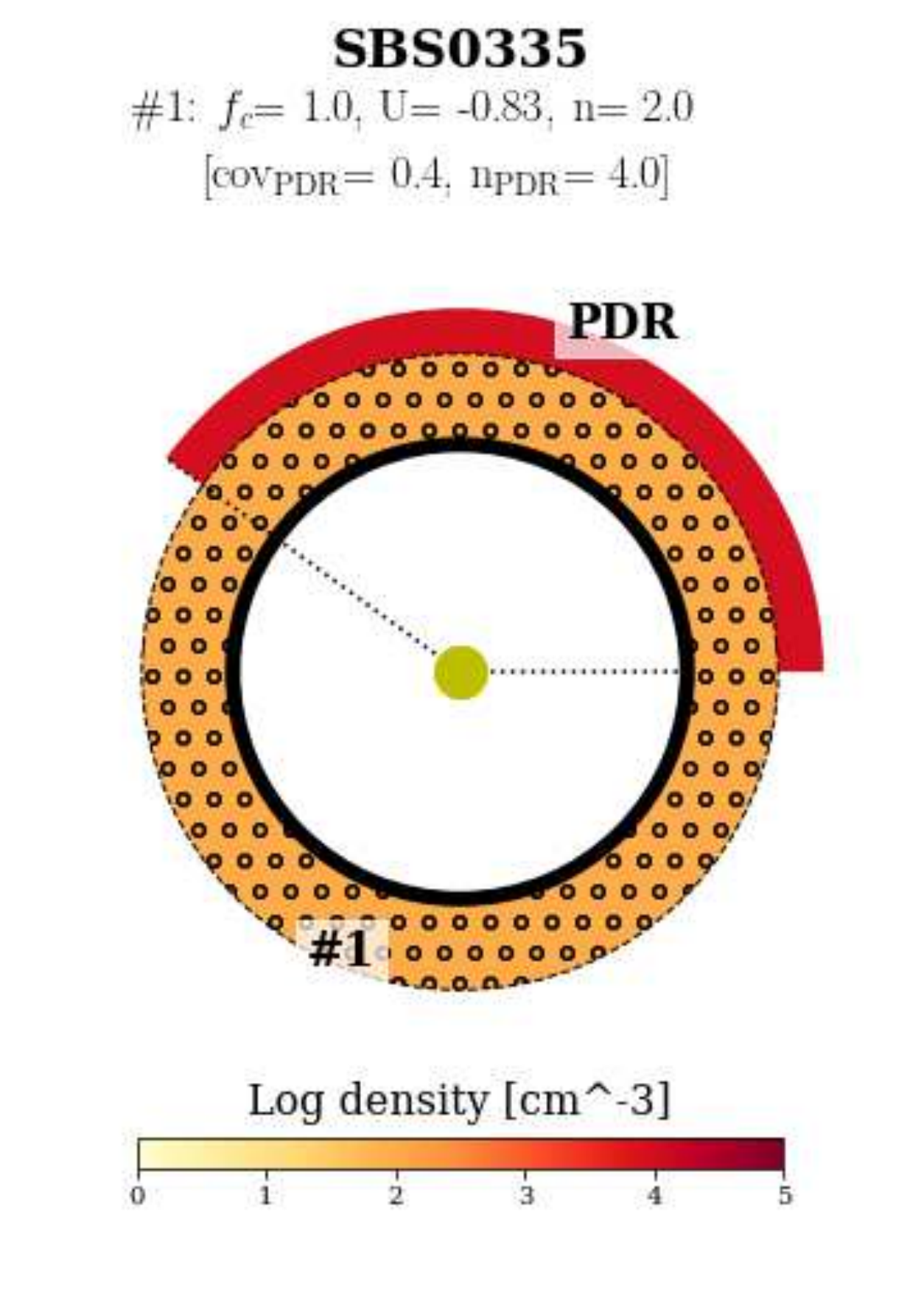}
\includegraphics[clip,trim=0 3cm 0 -1cm,width=4.5cm]{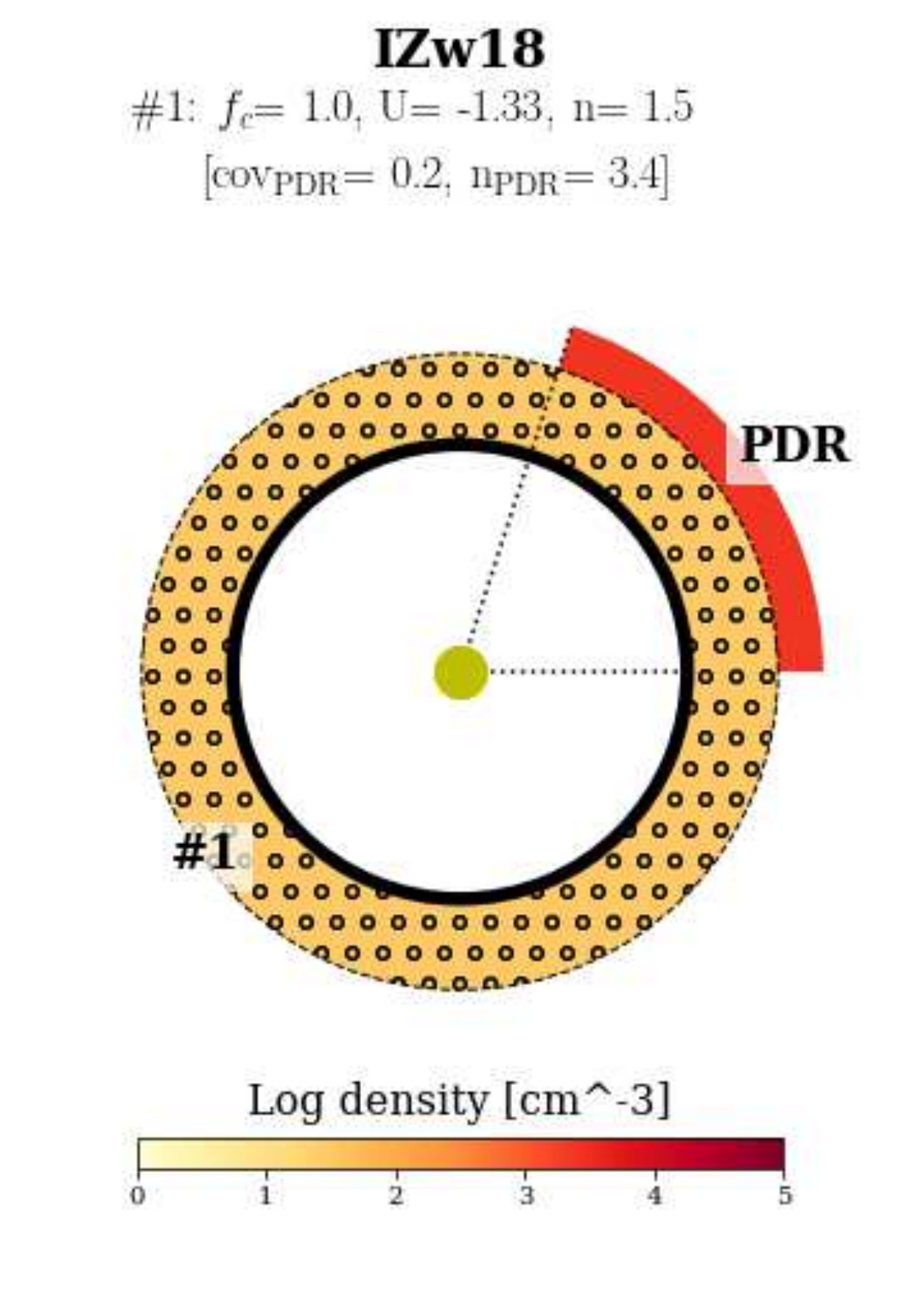}
\includegraphics[clip,trim=0 0 0 14cm,width=4.5cm]{./figures_appendix/NGC4214-c_mixed_a}
\caption{continued.}
\end{figure*}


\section{Additional correlations for the galaxy sample}
\begin{figure}
\centering
\includegraphics[clip,width=4.95cm]{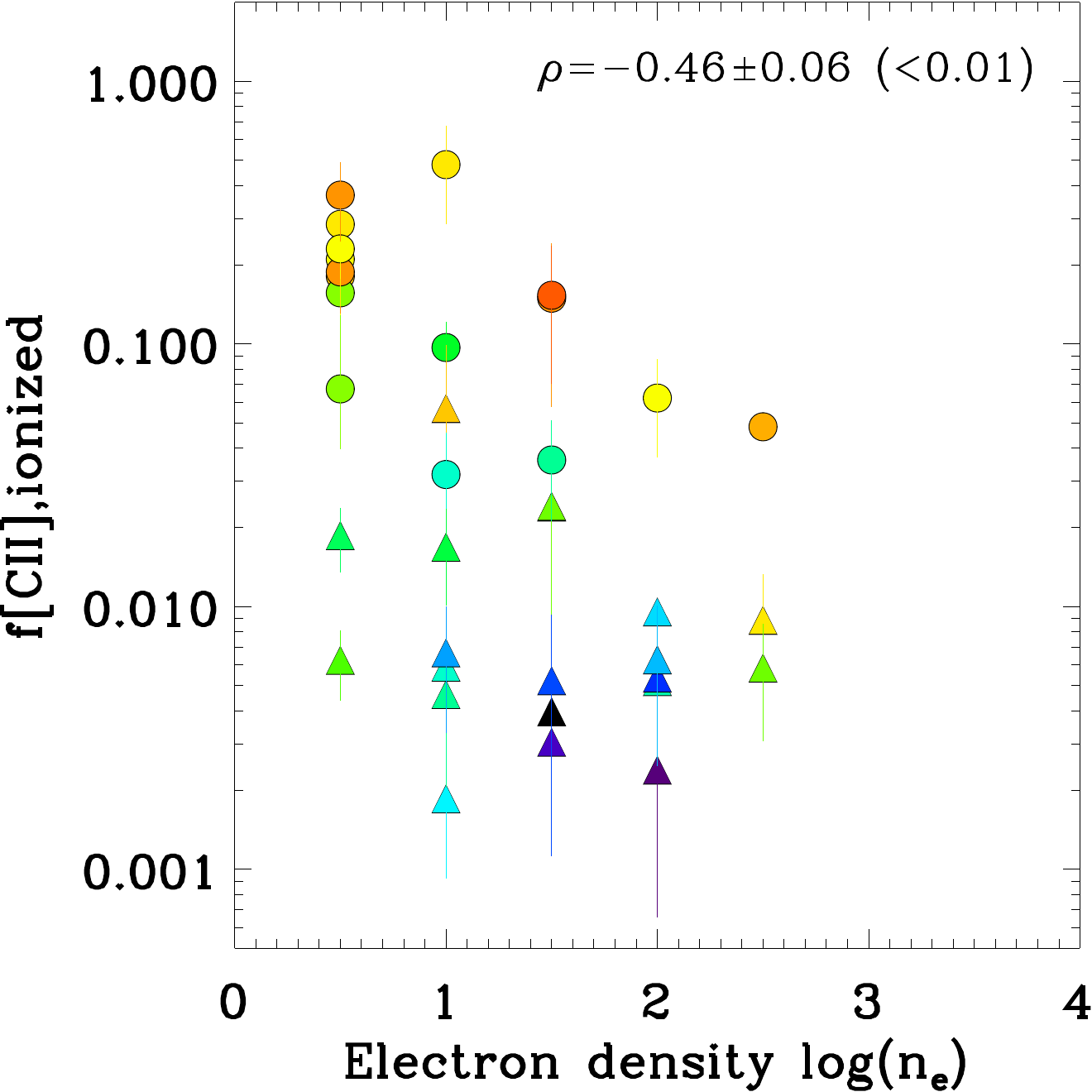}  \vspace{3mm}
\includegraphics[clip,trim=27mm 0 0 0,width=3.95cm]{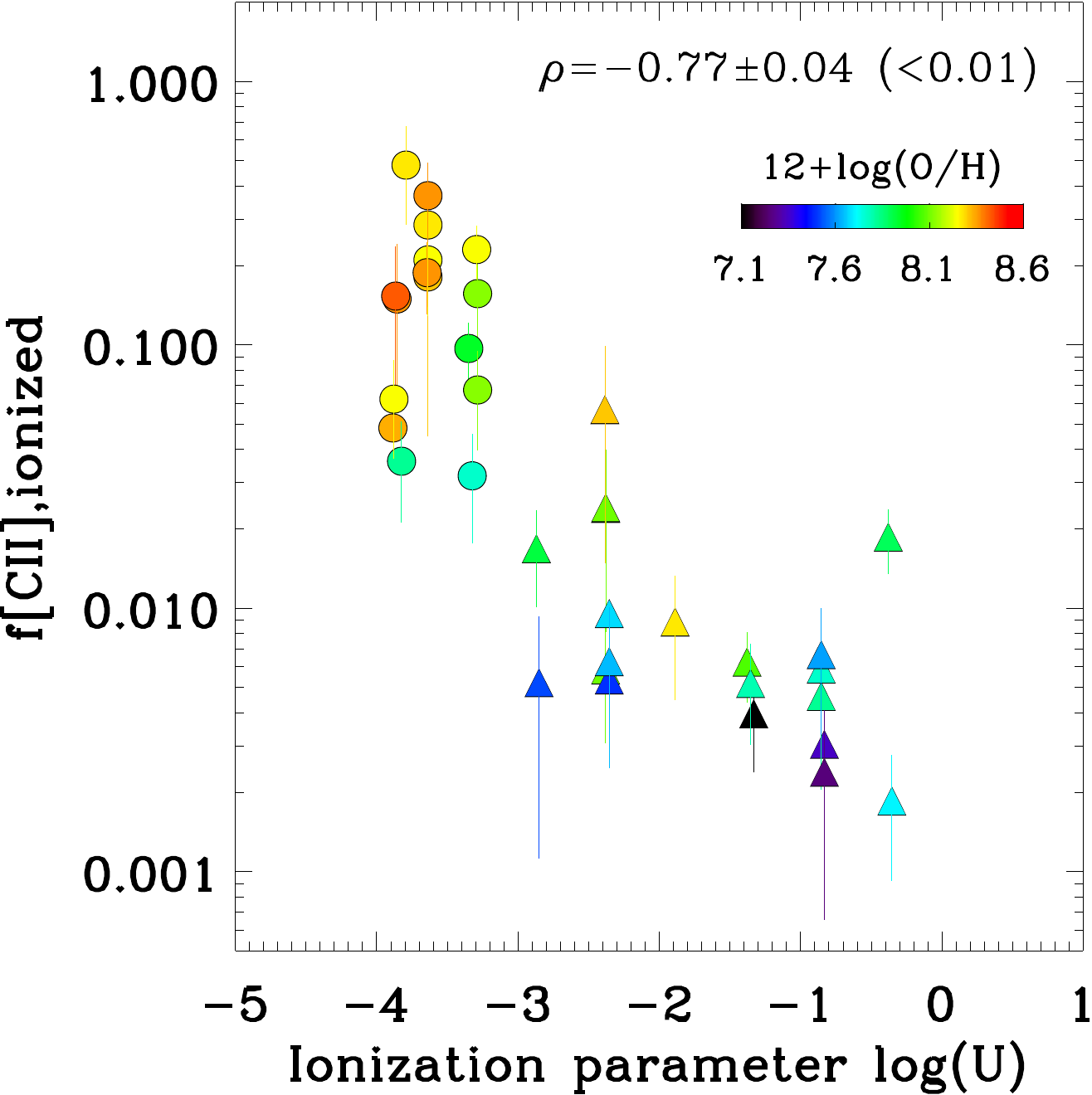}
\includegraphics[clip,width=5.2cm]{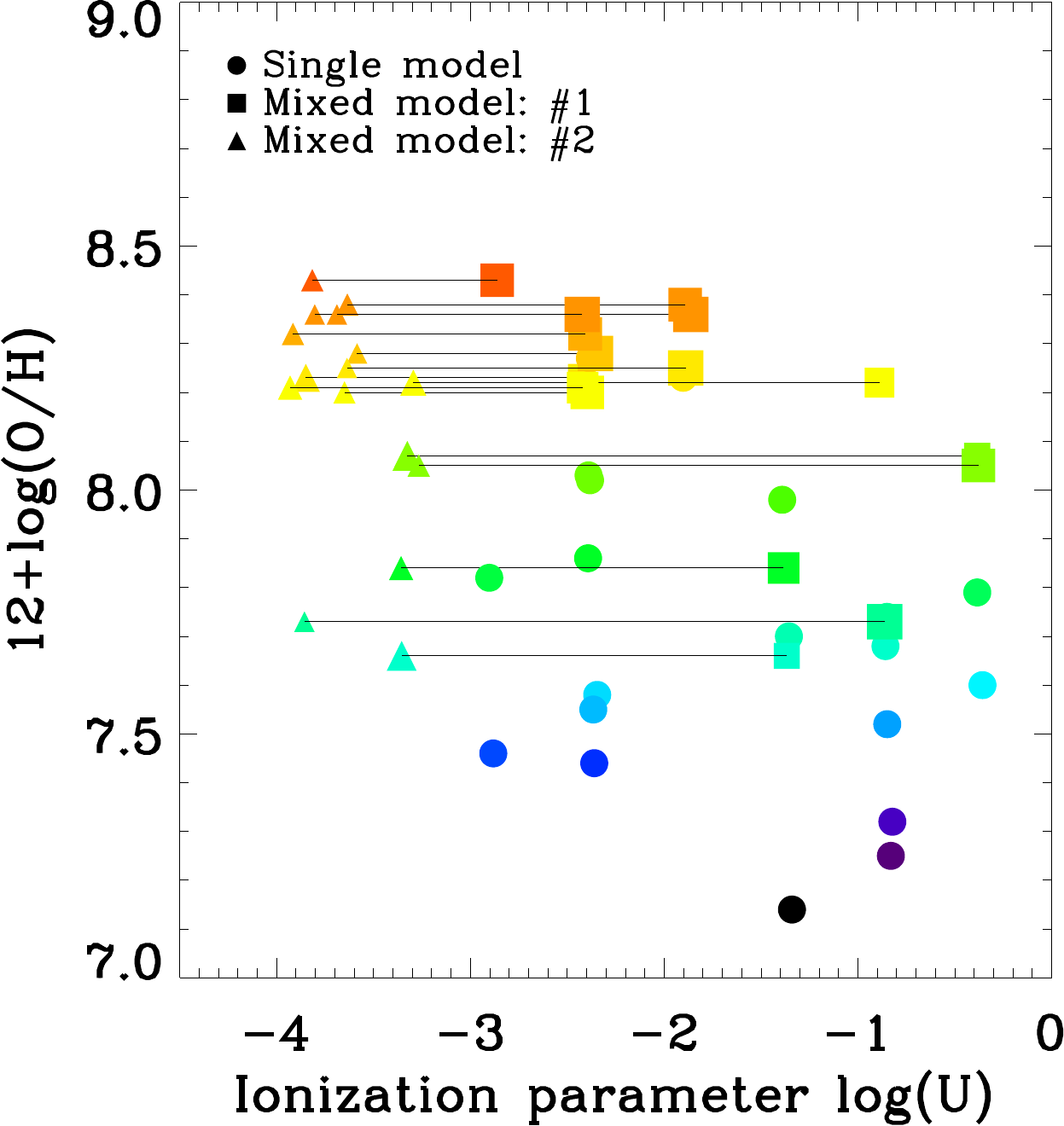} 
 \caption{
\textit{Top:} Correlation of the fraction of \cii emission
from the ionized gas with the model parameters:
the electron density (left panel) and the ionization
parameter (right panel). Values of component \#2
are plotted in case of mixed models.
Spearman's rank correlation coefficient and significance
of its deviation from zero in parenthesis are indicated in
the top-right corner.
\textit{Bottom:} Distribution of ionization parameter values
from the models with the observed galaxy metallicity.
A horizontal line links the two U-values in the case of mixed models.
In all panels, the data points are color coded by metallicity.
}
\label{fig:append-c}
\end{figure}

In Figure~\ref{fig:append-c}, we show correlations
of the fraction of \cii emission from the ionized gas
deduced from the models as a function of the model
density and ionization parameter, as well as the
correlation between the galaxy metallicity and the
model ionization parameter.

\section{Results of the \hii region+PDR models for individual galaxies}
\label{sect:append-d}
We show the results of the \hii region+PDR modeling
of individual galaxies with three main types of plots:
$(a)$~Comparison of the observed and predicted line intensities;
$(b)$~Contribution of each gas component to the total
line prediction in case of mixed models;
$(c)$~Probability density functions (PDFs) of the fitted parameters. 
Only the results of the galaxy He\,2-10 are shown in
the main text of the paper (Fig.~\ref{fig:bestparams}).

 \begin{figure*}[thp]
\centering
(a)
\includegraphics[clip,width=11cm]{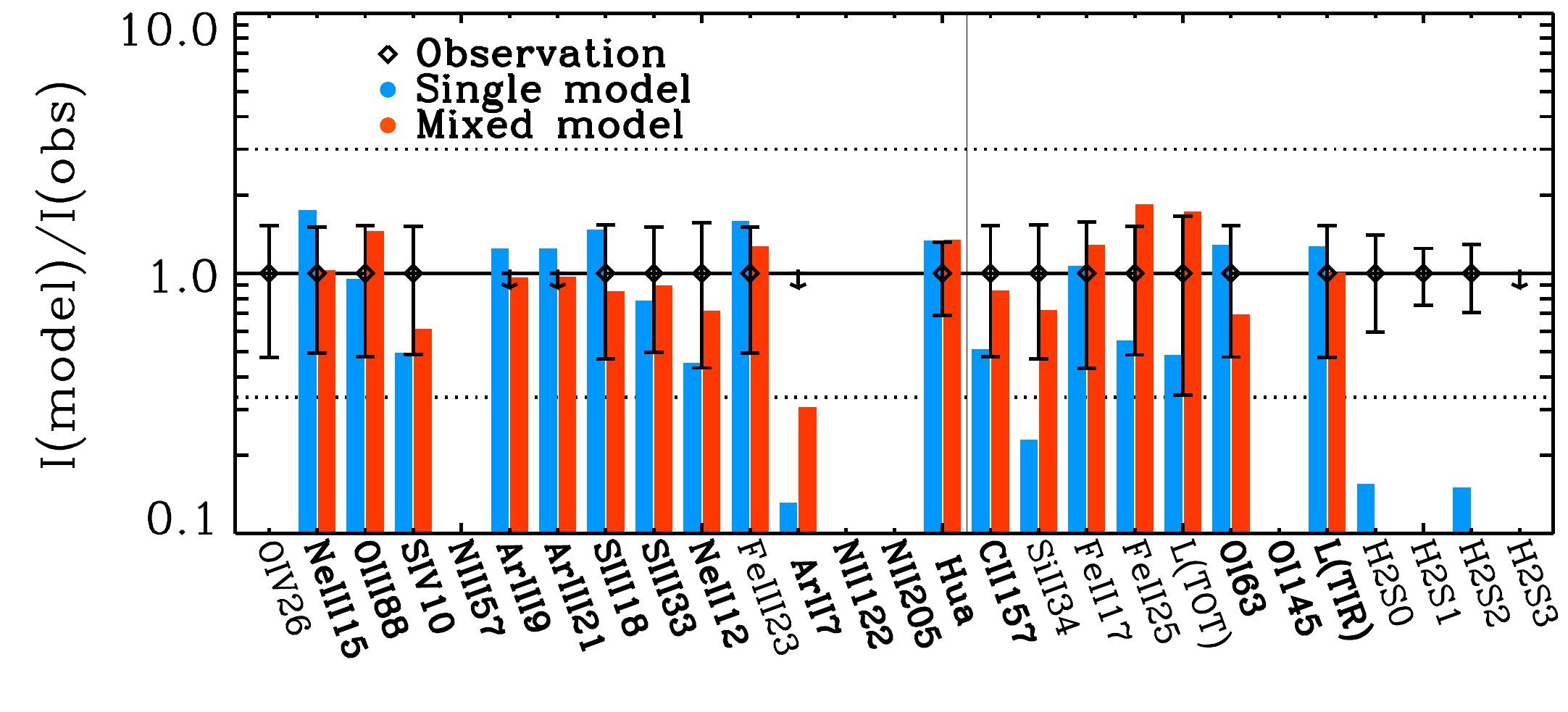} \hfill{}
(b)
\includegraphics[clip,width=6.2cm]{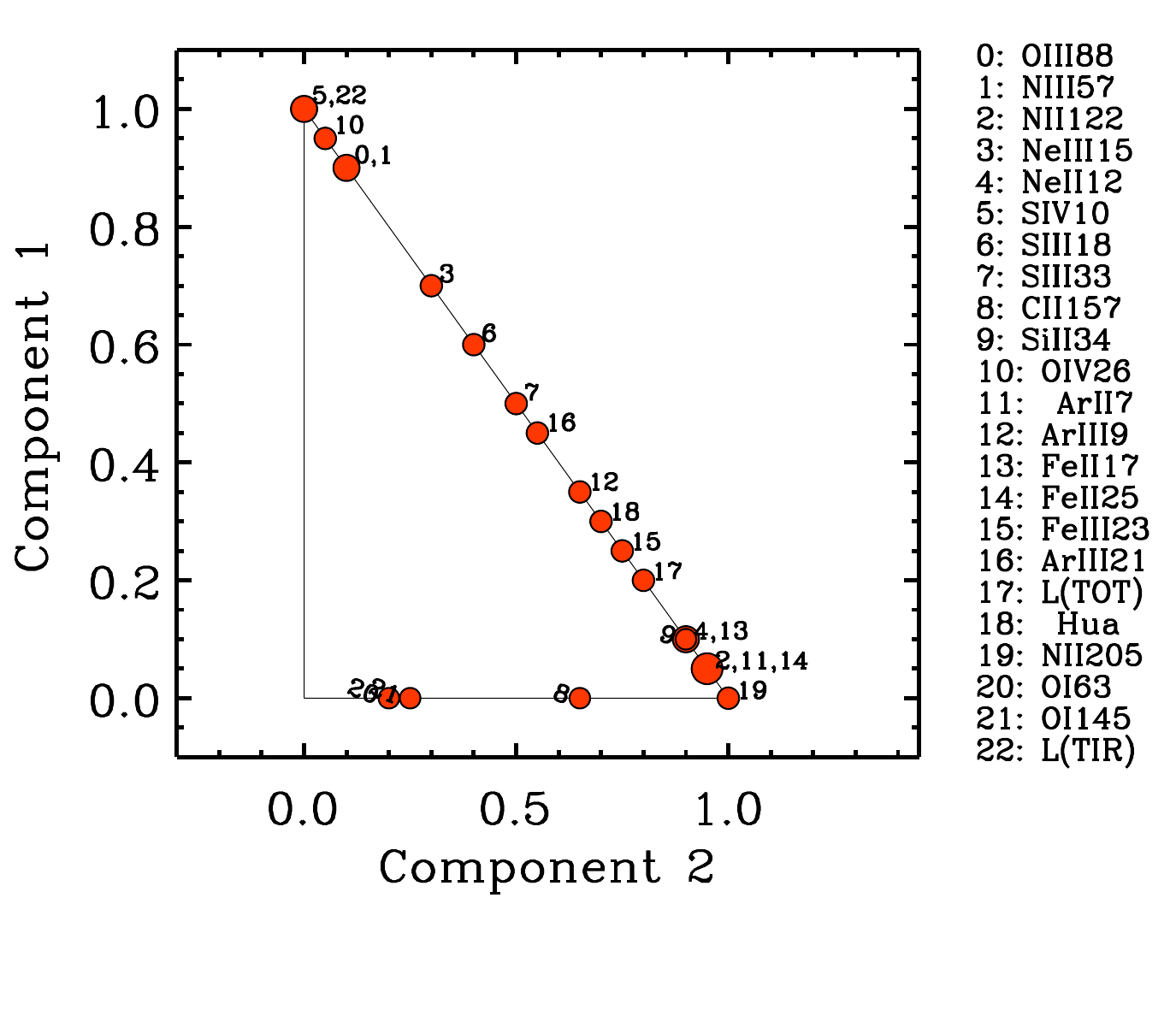}\\
(c)
\includegraphics[clip,width=8.8cm]{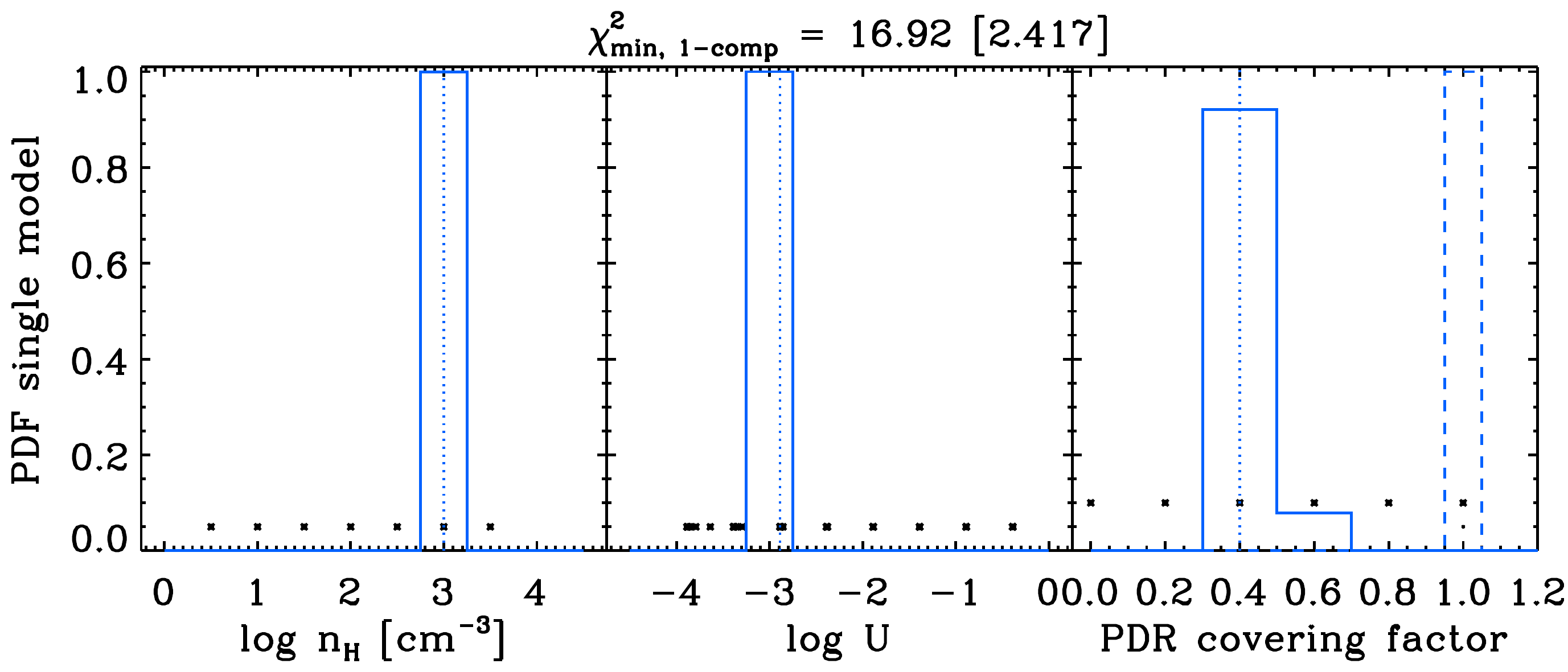} \hfill{}
\includegraphics[clip,width=8.8cm]{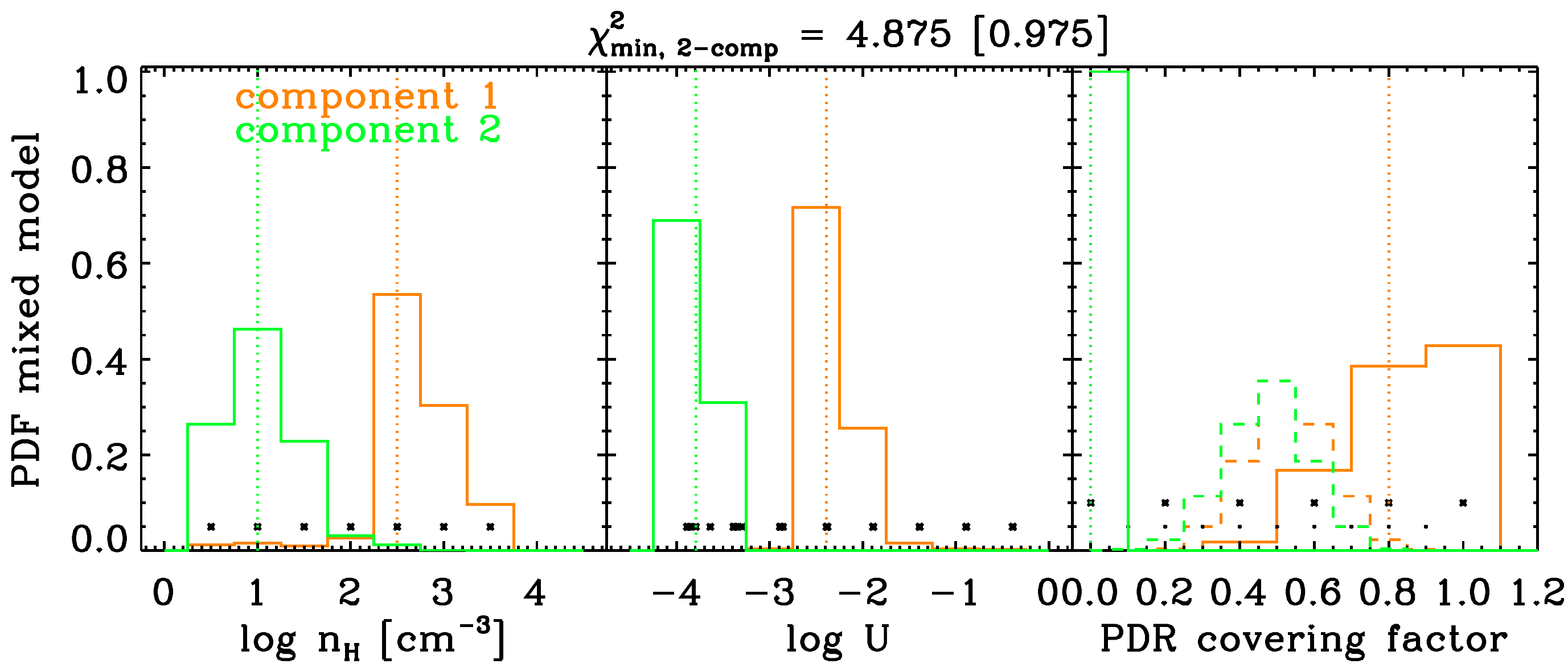}
\caption{ \textbf{Results for Haro\,2.}
$(a)$~Comparison of the observed (losange) and predicted 
intensities for our best-fitting single (blue bar) and mixed (red bar)
\hii region+PDR models. Lines are sorted by decreasing energy.
The fitted lines have labels in boldface.
$(b)$~For mixed models: respective contributions from
component \#1 and component \#2 to the line prediction.
For PDR lines (\cii, \oi, and \silii), only the contribution from
the ionized gas is shown; the PDR contribution is
one minus the sum of the contributions from the plot.
Contributions are expected to be dominated by one
or the other component if their model parameters
are noticeably different.
$(c)$~Probability density functions of the model parameters
($n_{\rm H}$, $U$, $cov_{\rm PDR}$/scaling factor)
for single (left panel) and mixed (right panel) models.
The scaling factor corresponds to the proportion in which
we combine two models in the mixed model case. It is
equal to unity otherwise. It is shown with dashed lines in
the same panels as the PDR covering factor. Vertical dotted lines
show values of the best-fitting model. Black asterisks
show the range and step of values of the grid parameters.
}
\label{fig:bests-haro2}
\end{figure*}
 \begin{figure*}[thp]
\centering
(a)
\includegraphics[clip,width=11cm]{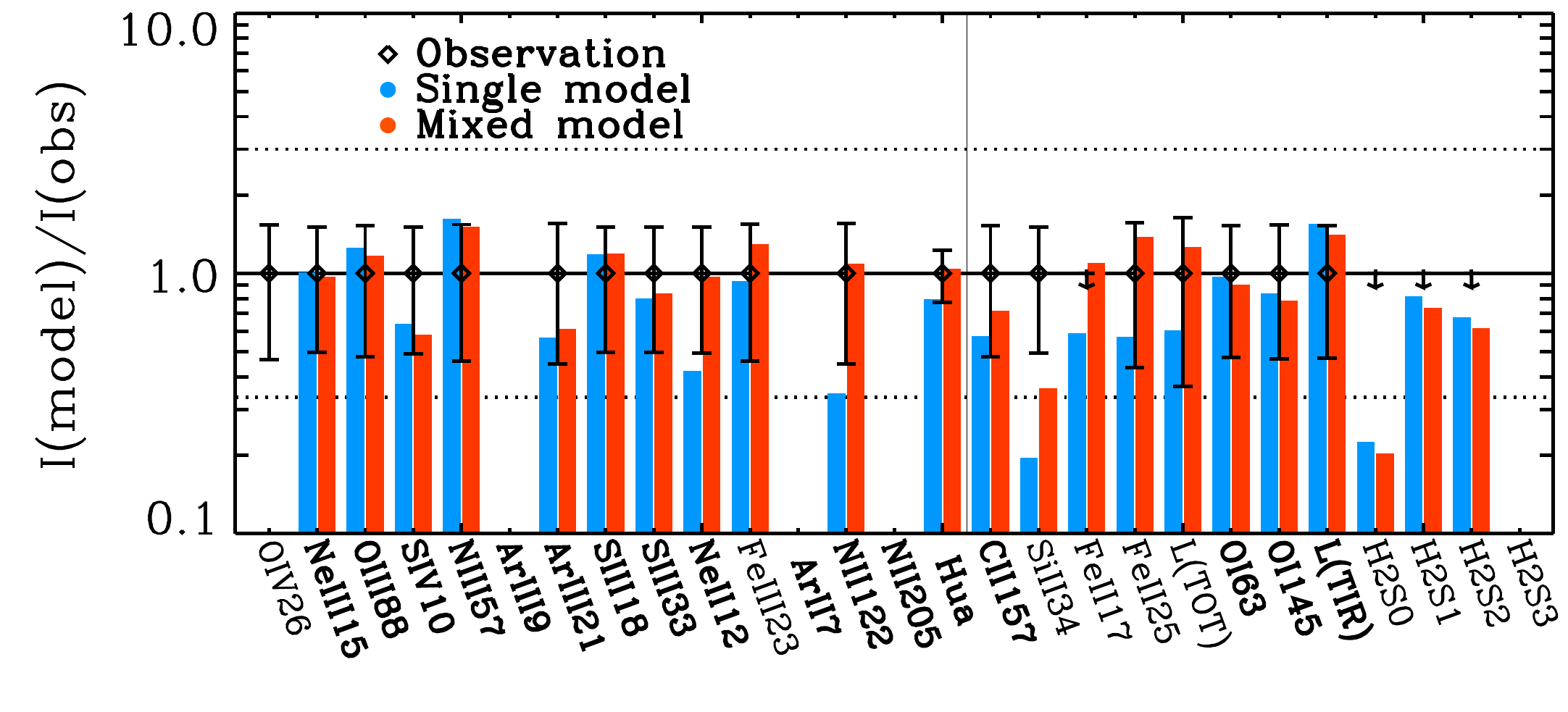} \hfill{}
(b)
\includegraphics[clip,width=6.2cm]{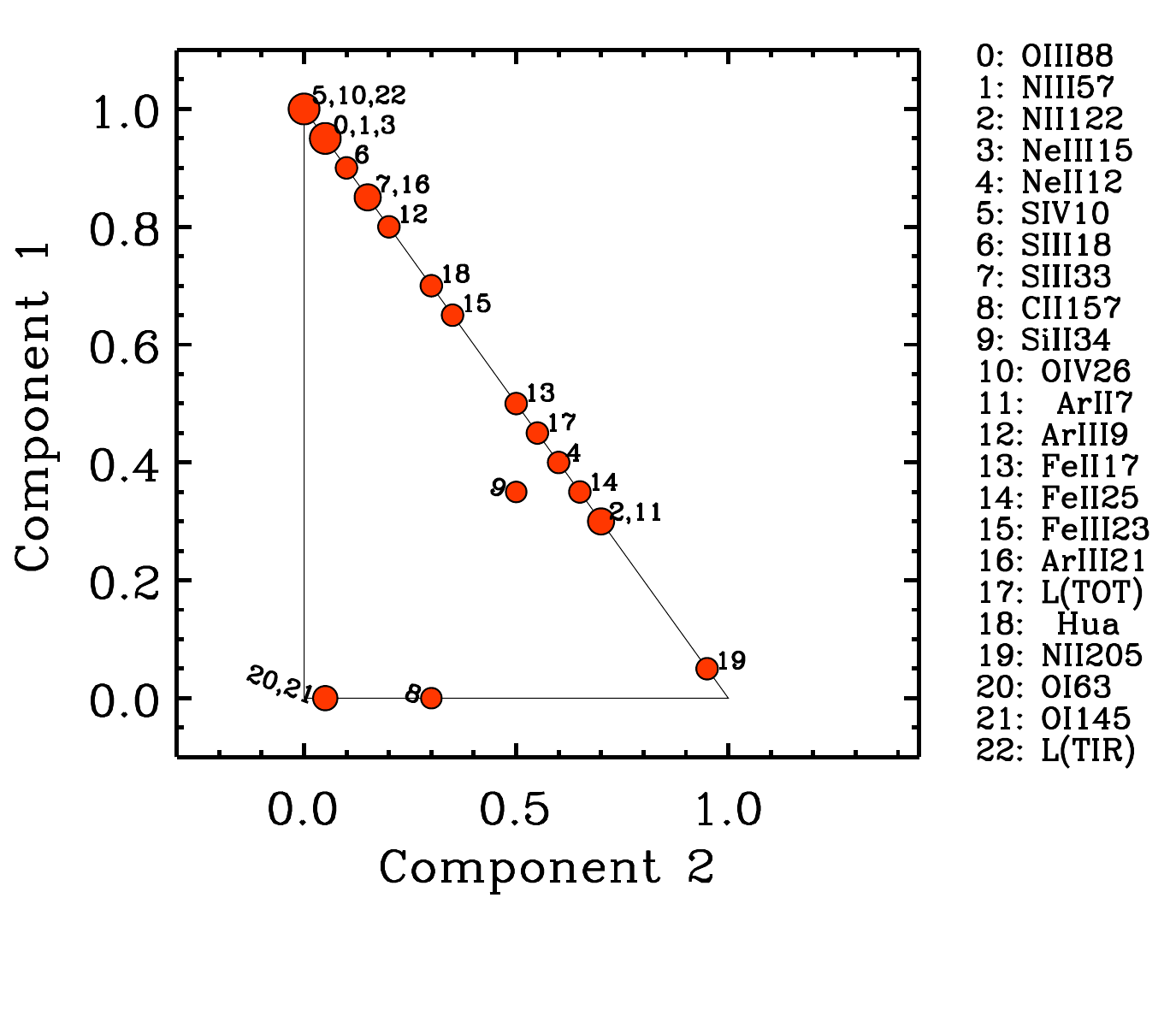}\\
(c)
\includegraphics[clip,width=8.8cm]{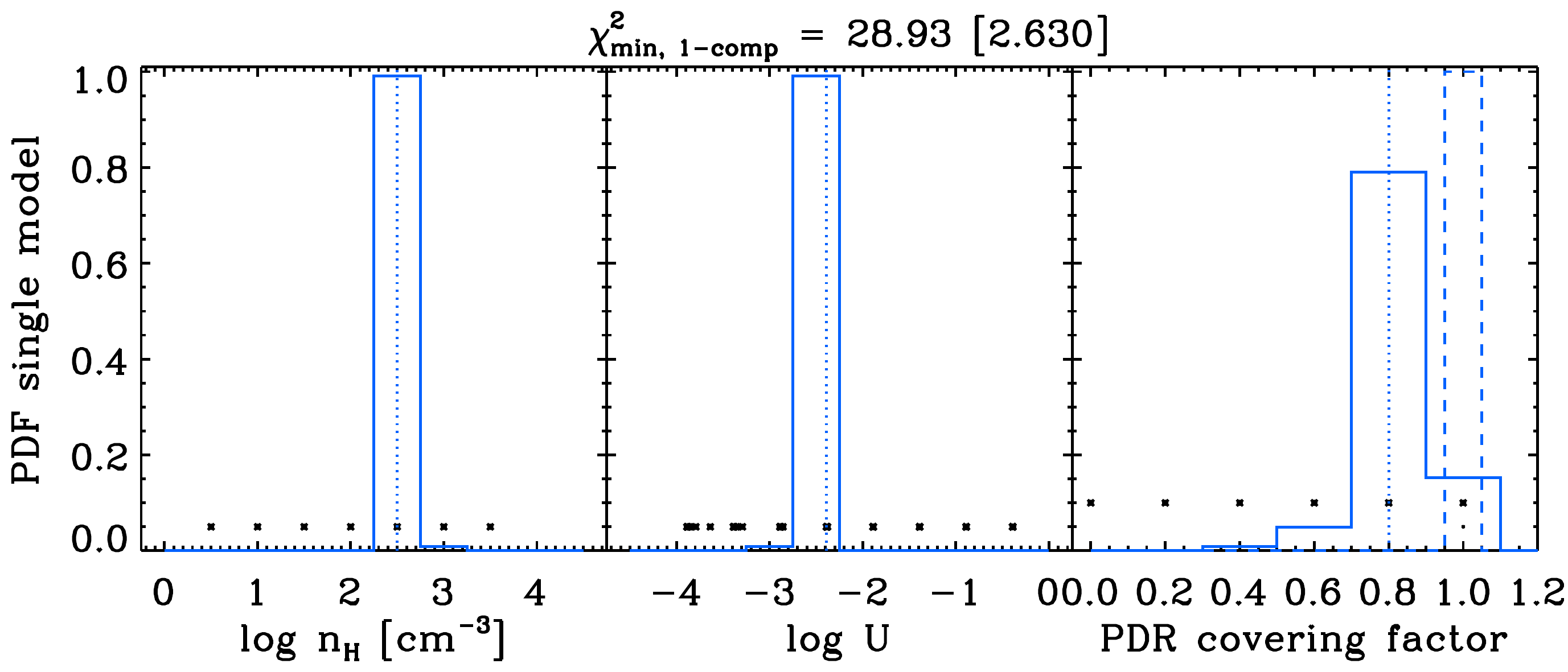} \hfill{}
\includegraphics[clip,width=8.8cm]{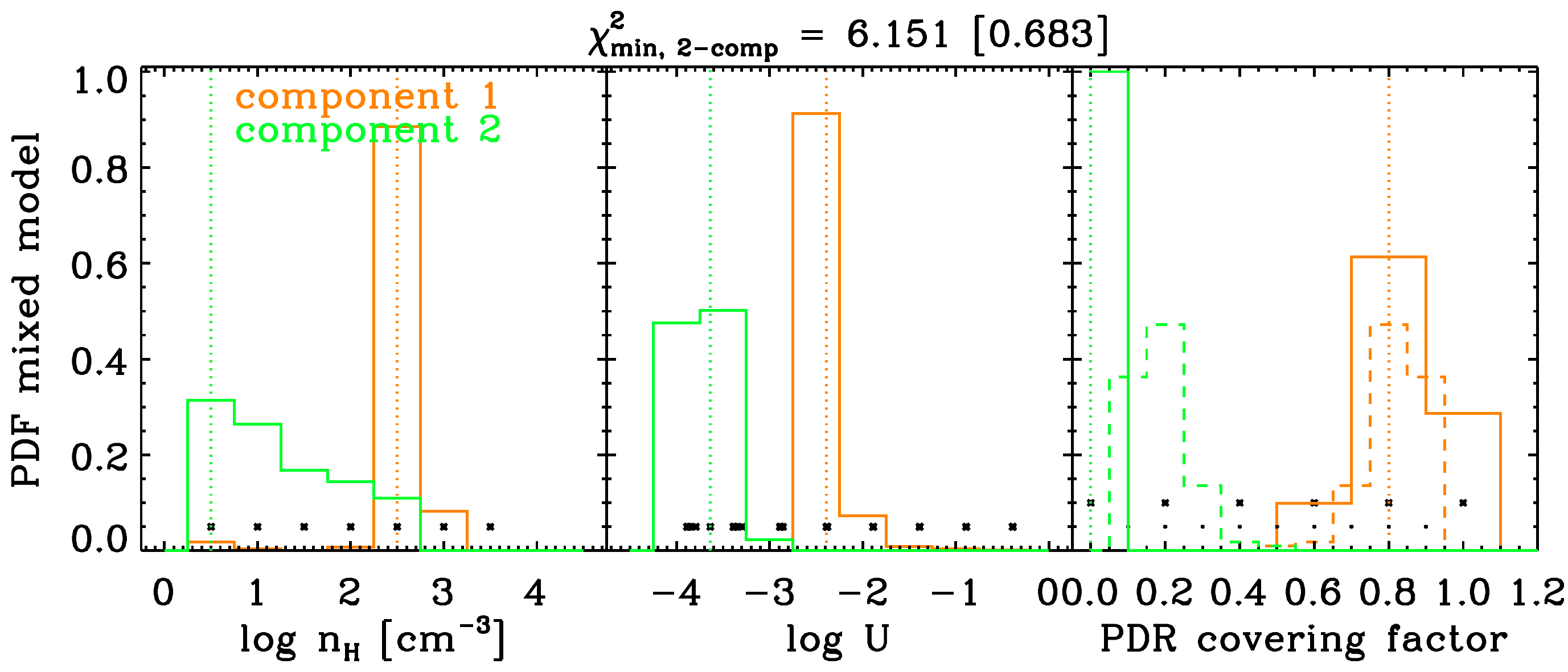}
\caption{ \textbf{Results for Haro\,3.}
See above for caption description.
}
\label{fig:bests-haro3}
\end{figure*}
\clearpage
 \begin{figure*}[thp]
\centering
(a)
\includegraphics[clip,width=11cm]{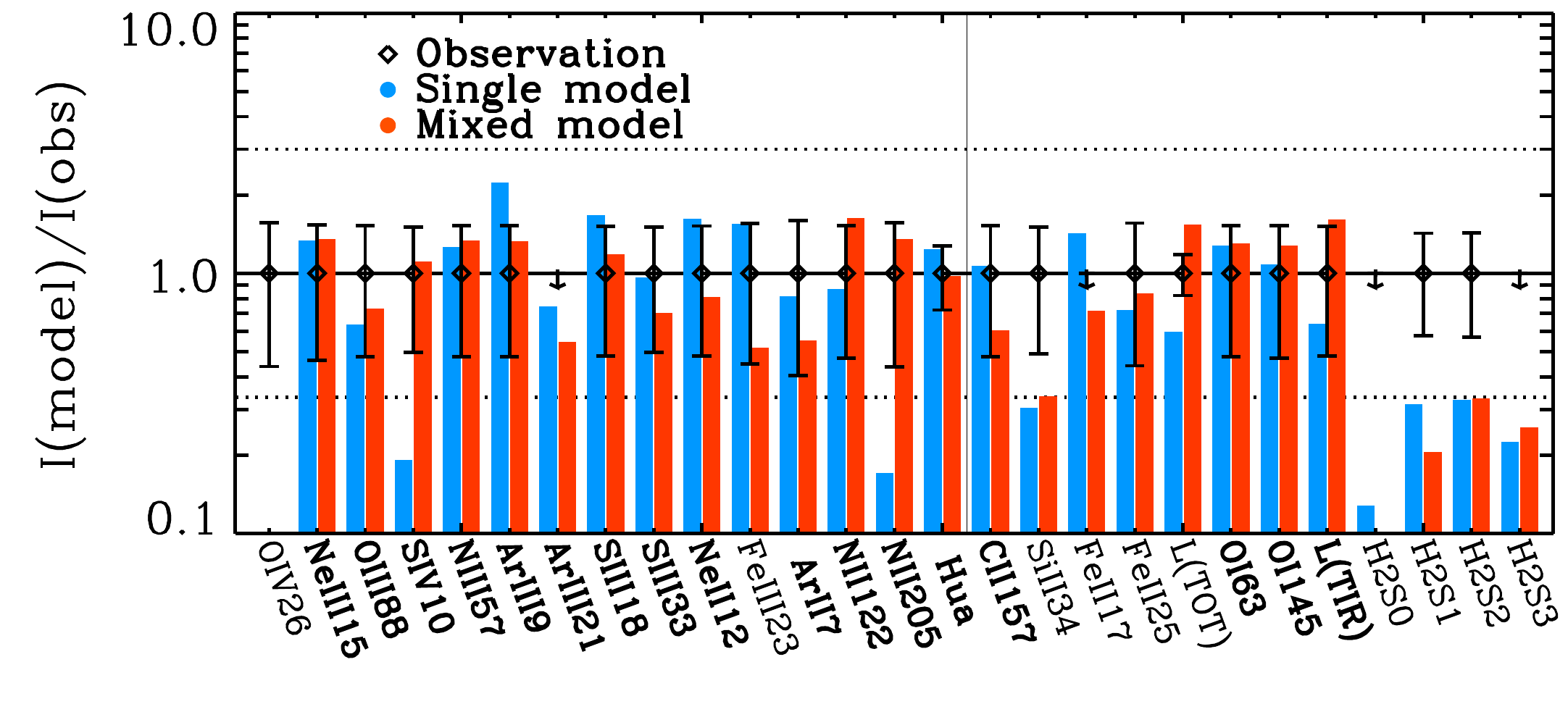} \hfill{}
(b)
\includegraphics[clip,width=6.2cm]{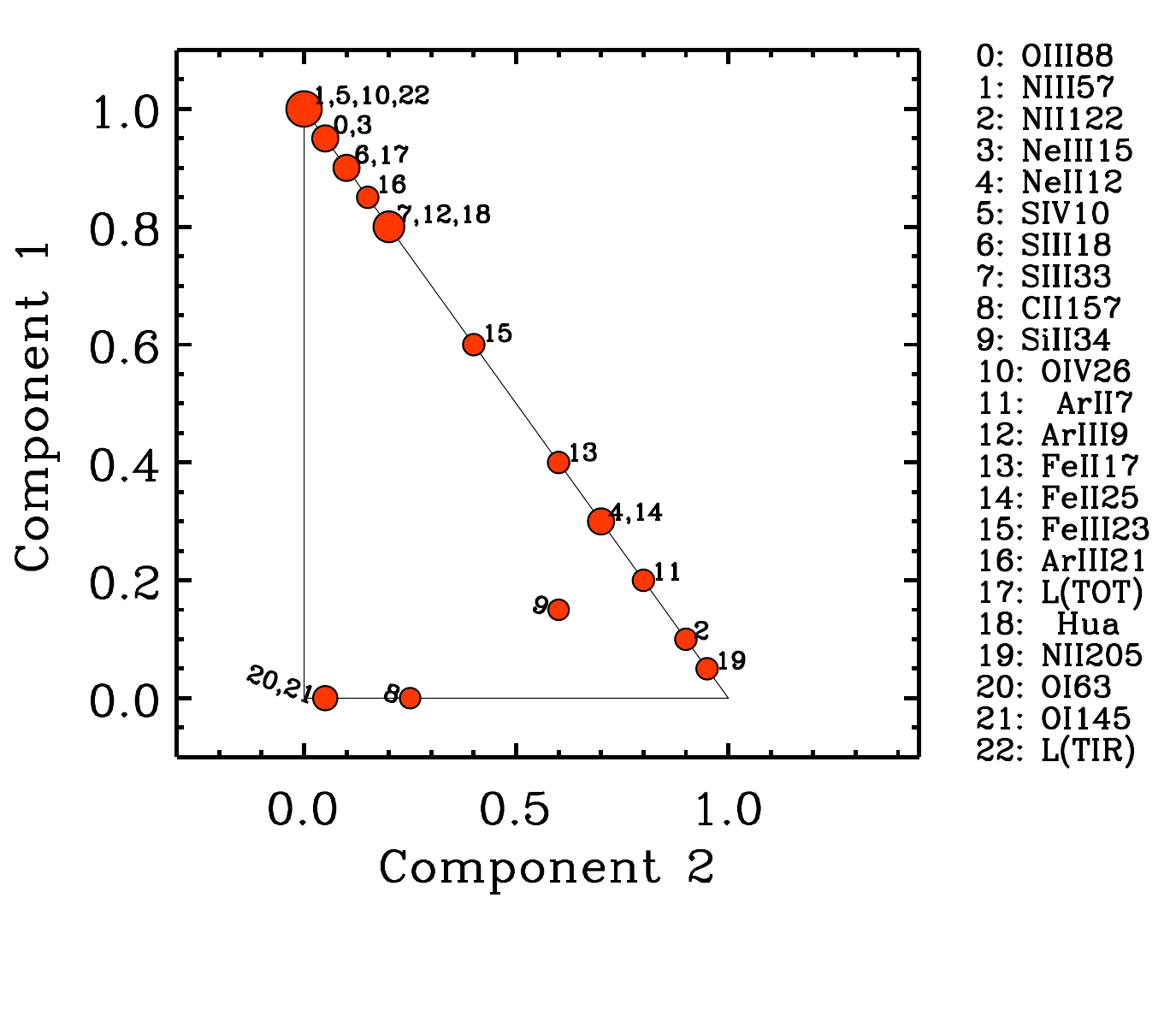}\\
(c)
\includegraphics[clip,width=8.8cm]{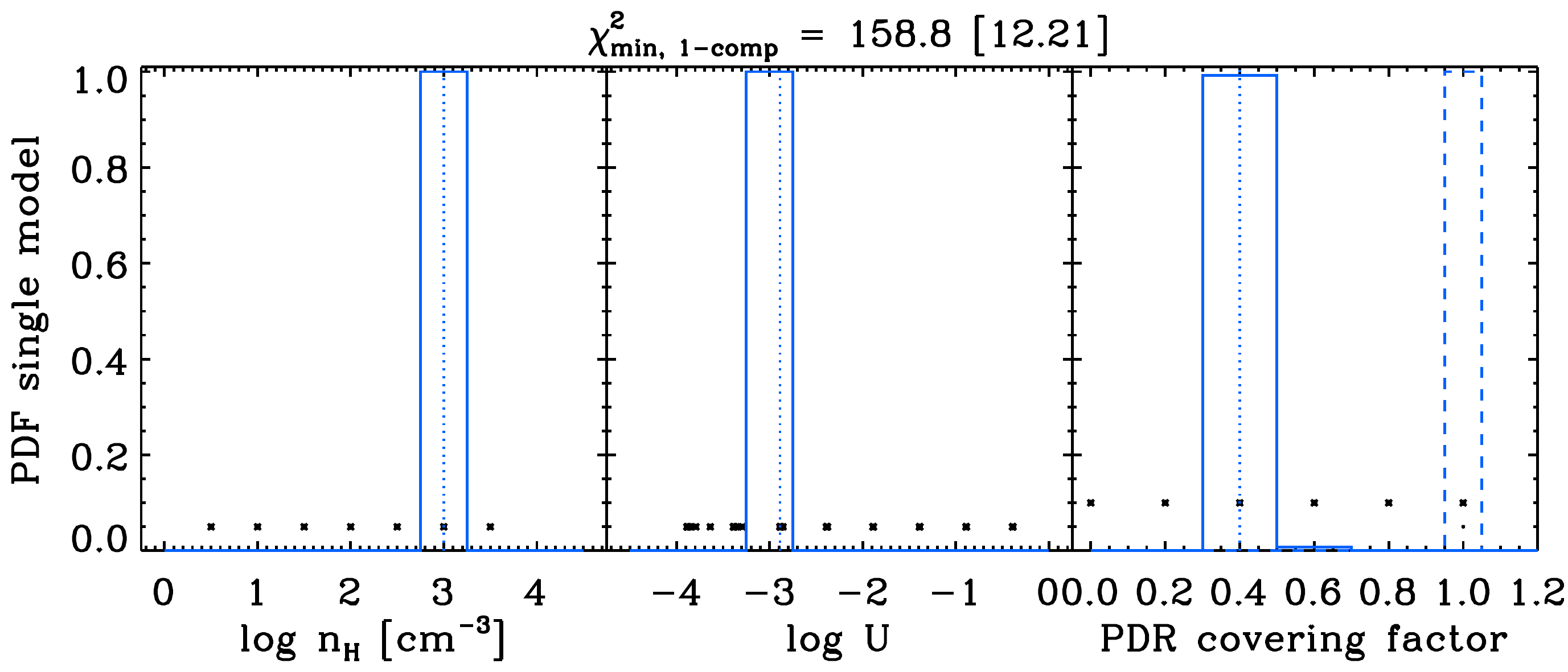} \hfill{}
\includegraphics[clip,width=8.8cm]{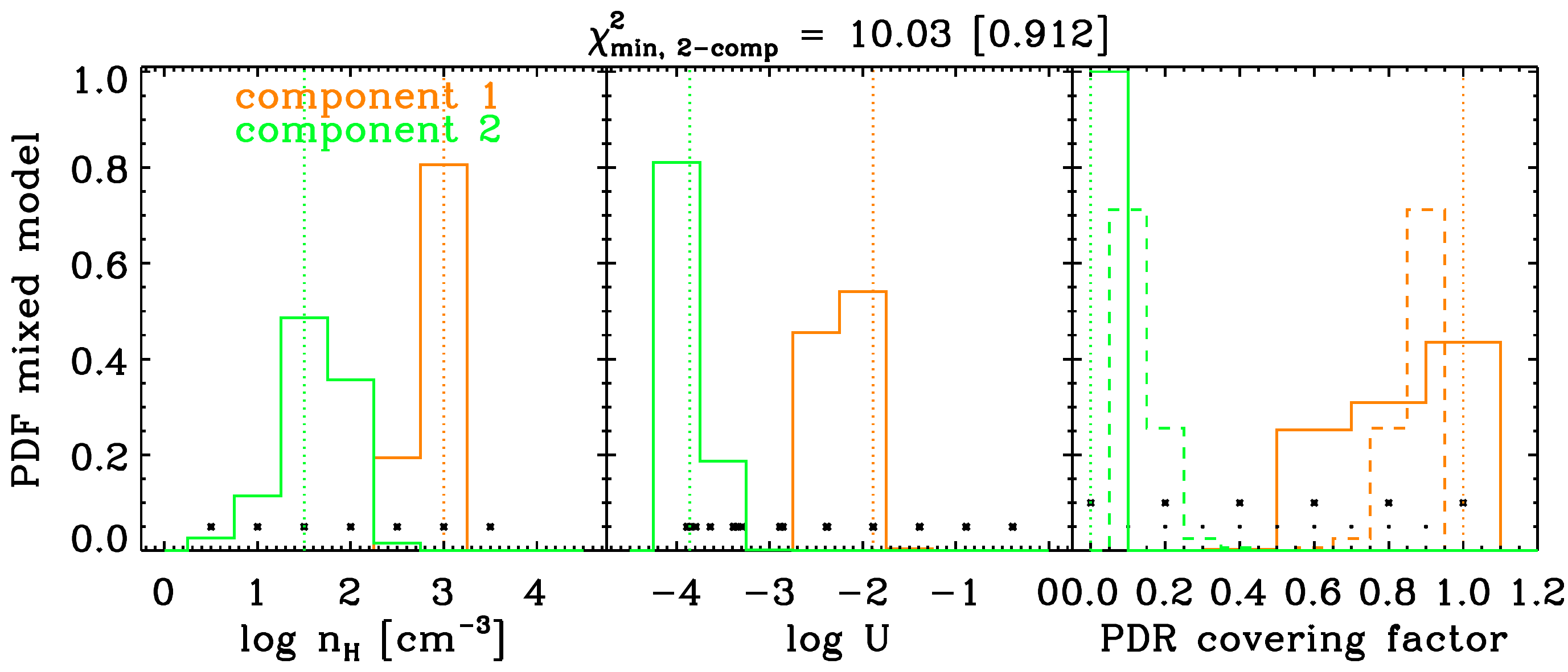}
\caption{ \textbf{Results for Haro\,11.} 
$(a)$~Comparison of the observed (losange) and predicted 
intensities for our best-fitting single (blue bar) and mixed (red bar)
\hii region+PDR models. Lines are sorted by decreasing energy.
The fitted lines have labels in boldface.
$(b)$~For mixed models: respective contributions from
component \#1 and component \#2 to the line prediction.
For PDR lines (\cii, \oi, and \silii), only the contribution from
the ionized gas is shown; the PDR contribution is
one minus the sum of the contributions from the plot.
Contributions are expected to be dominated by one
or the other component if their model parameters
are noticeably different.
$(c)$~Probability density functions of the model parameters
($n_{\rm H}$, $U$, $cov_{\rm PDR}$/scaling factor)
for single (left panel) and mixed (right panel) models.
The scaling factor corresponds to the proportion in which
we combine two models in the mixed model case. It is
equal to unity otherwise. It is shown with dashed lines in
the same panels as the PDR covering factor. Vertical dotted lines
show values of the best-fitting model. Black asterisks
show the range and step of values of the grid parameters.
}
\label{fig:bests-haro11}
\end{figure*}
\begin{figure*}[thp]
\centering
(a)
\includegraphics[clip,width=11cm]{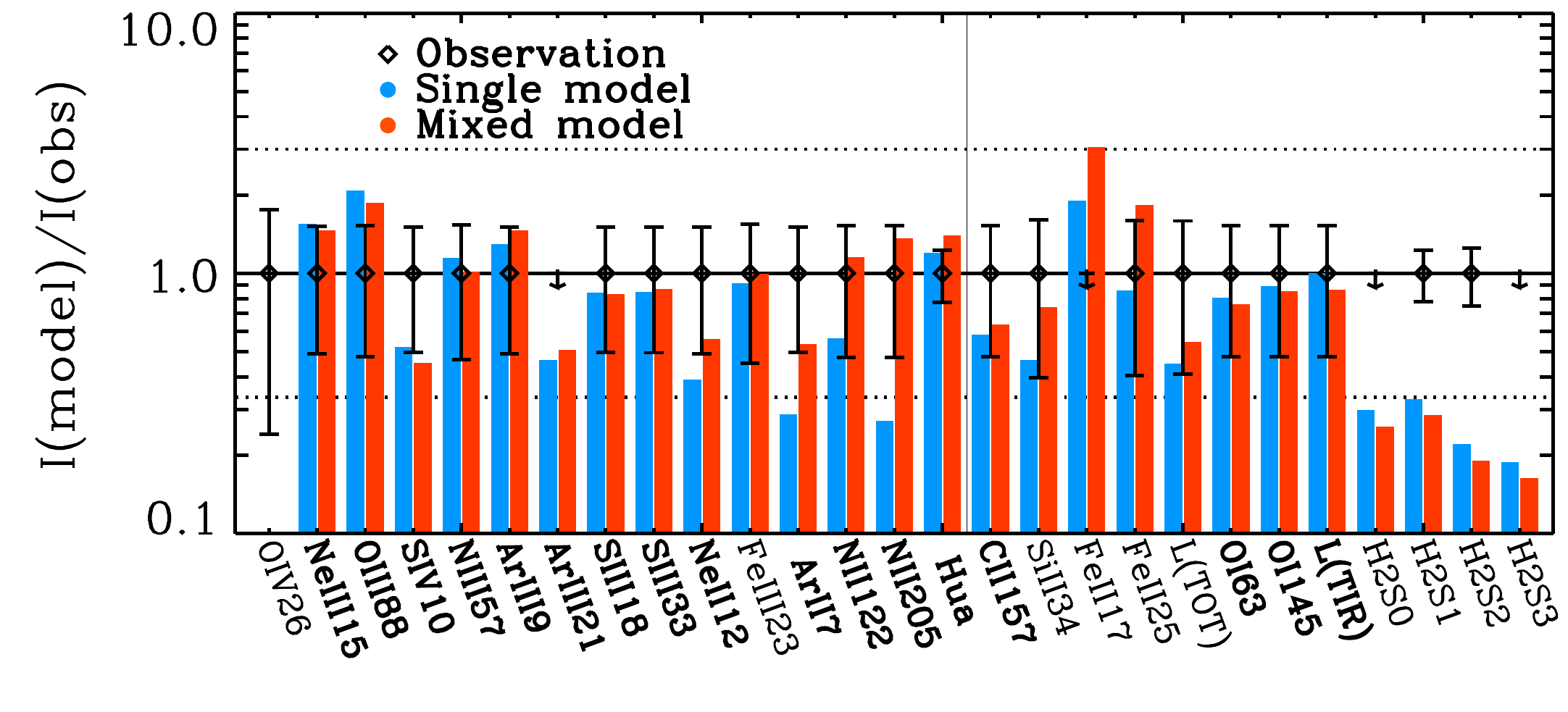} \hfill{}
(b)
\includegraphics[clip,width=6.2cm]{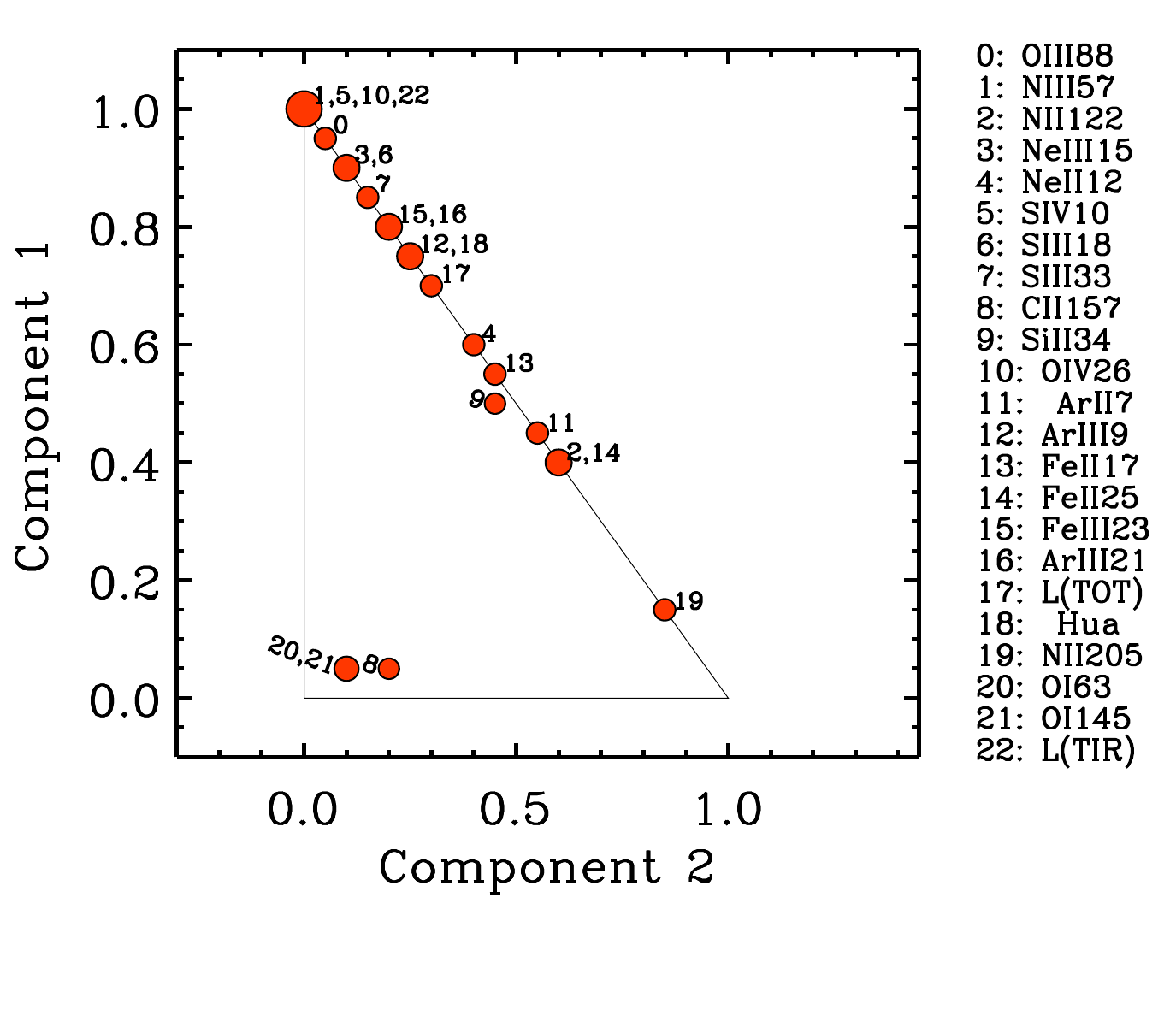}\\
(c)
\includegraphics[clip,width=8.8cm]{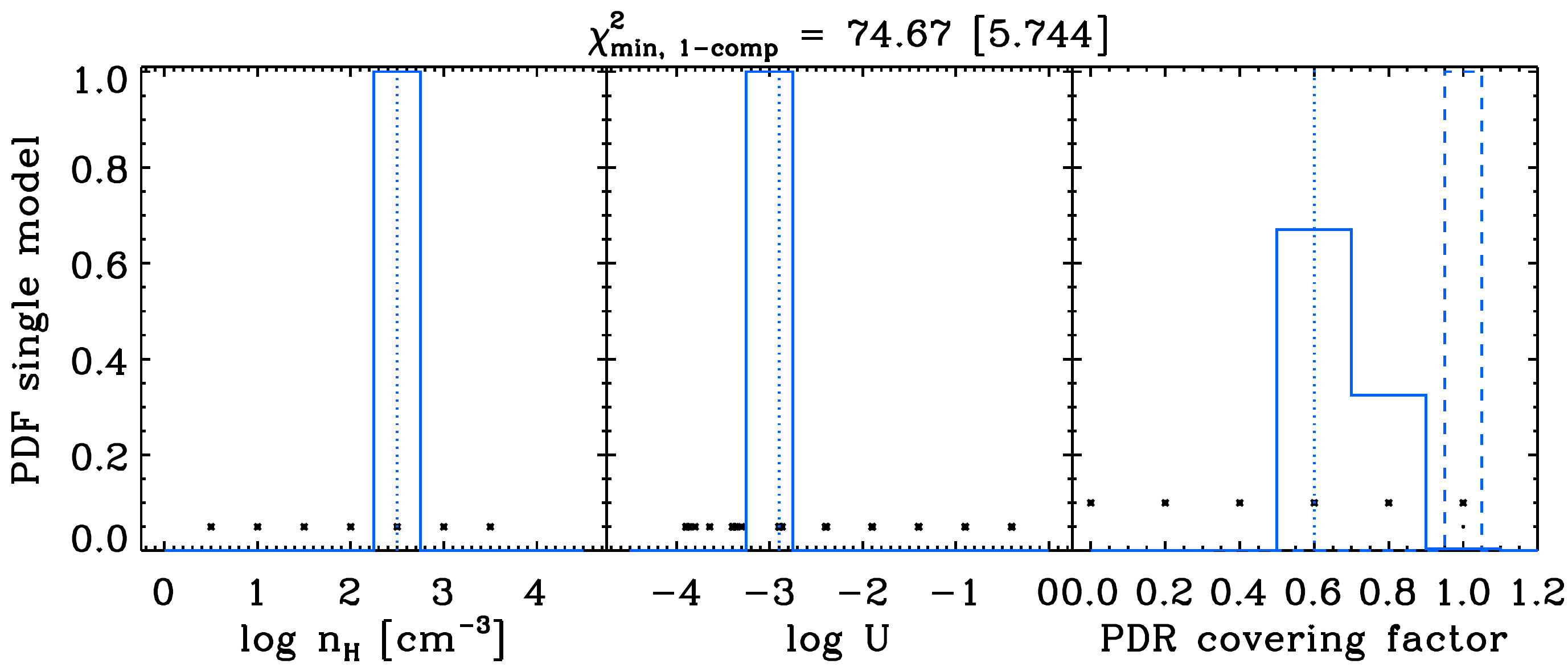} \hfill{}
\includegraphics[clip,width=8.8cm]{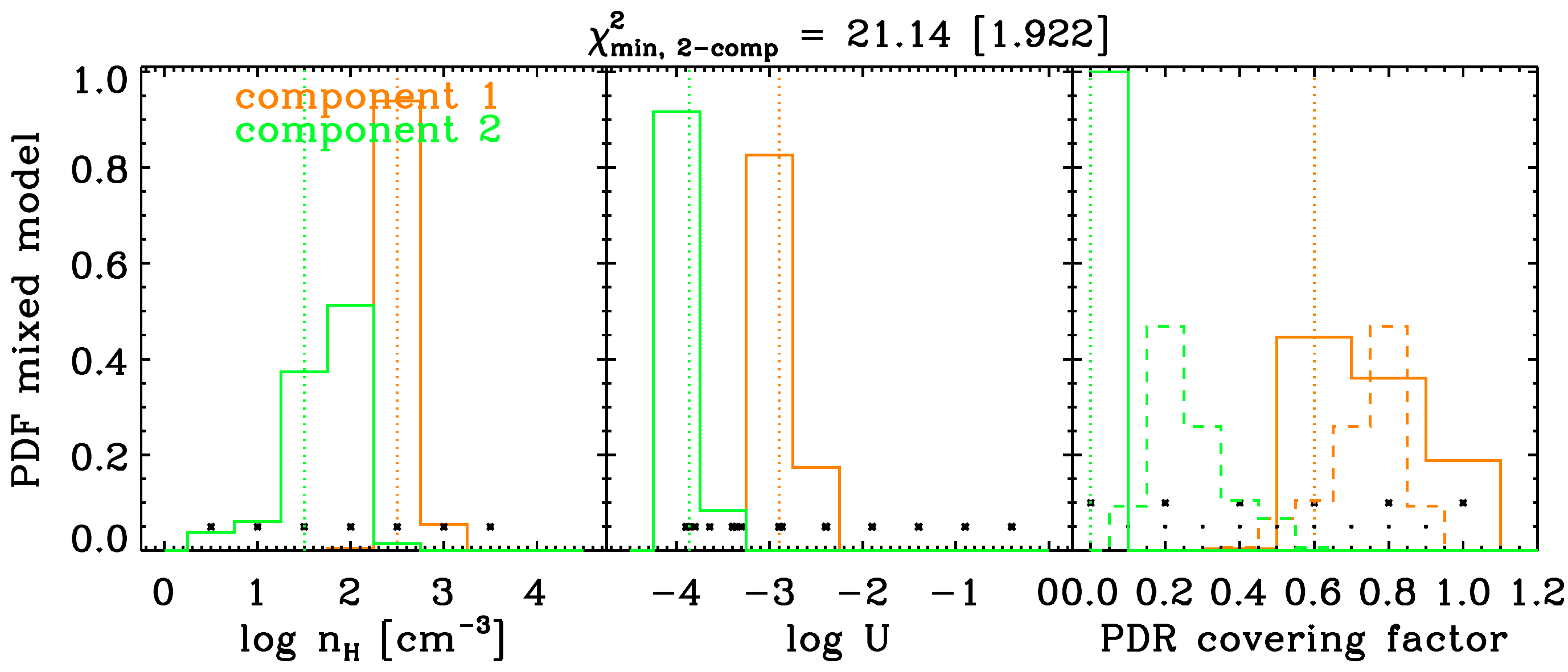}
\caption{ \textbf{Results for He\,2-10.}
See above for caption description.
}
\label{fig:bests-he2-10}
\end{figure*}
\clearpage
 \begin{figure*}[thp]
\centering
(a)
\includegraphics[clip,width=11cm]{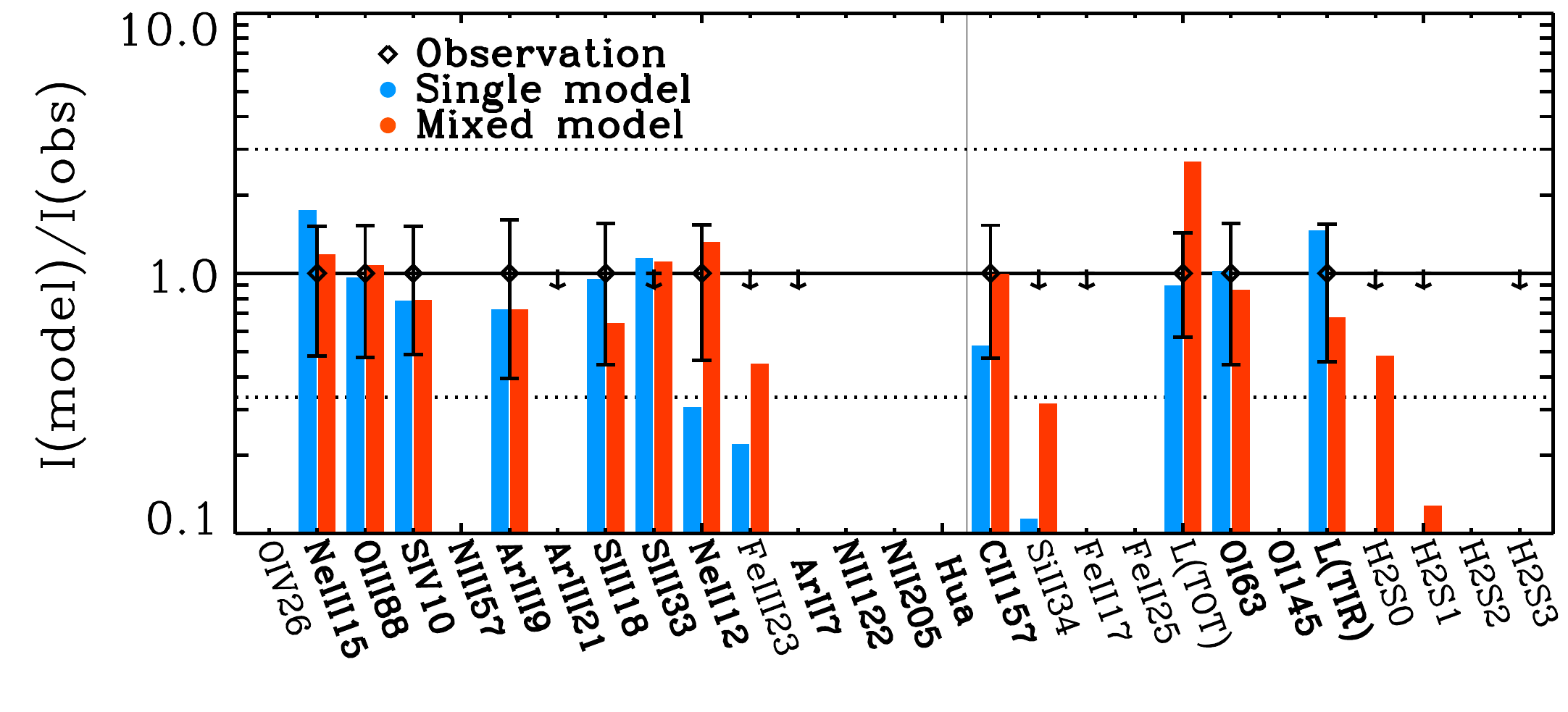} \hfill{}
(b)
\includegraphics[clip,width=6.2cm]{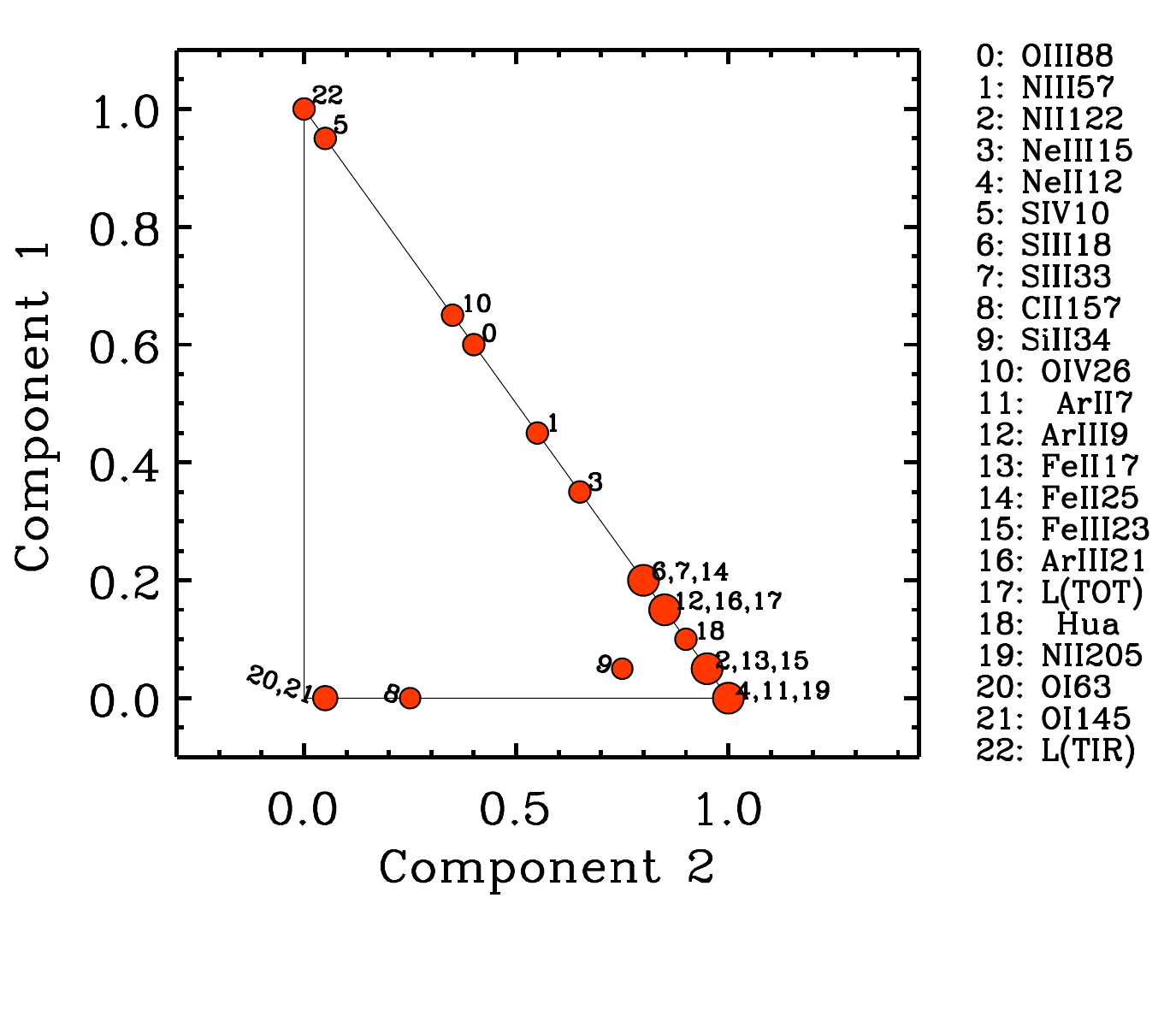}\\
(c)
\includegraphics[clip,width=8.8cm]{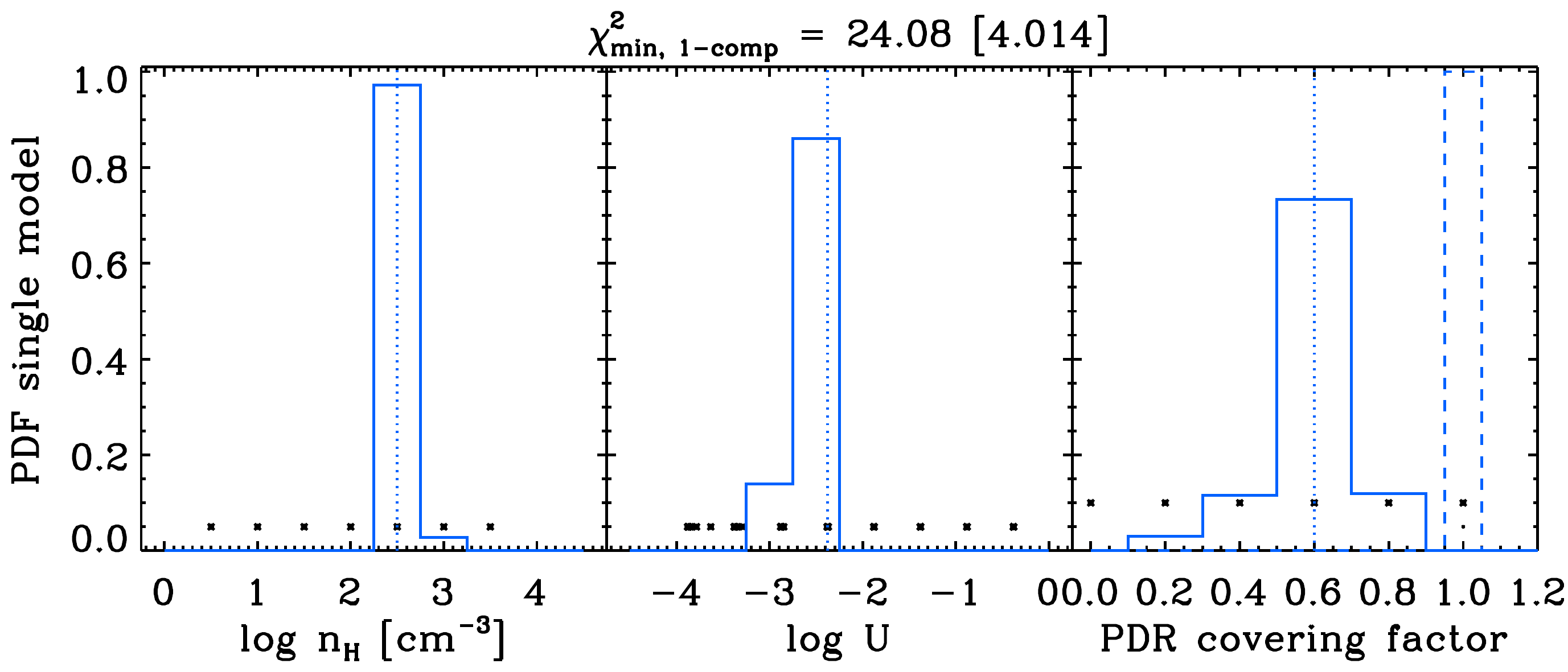} \hfill{}
\includegraphics[clip,width=8.8cm]{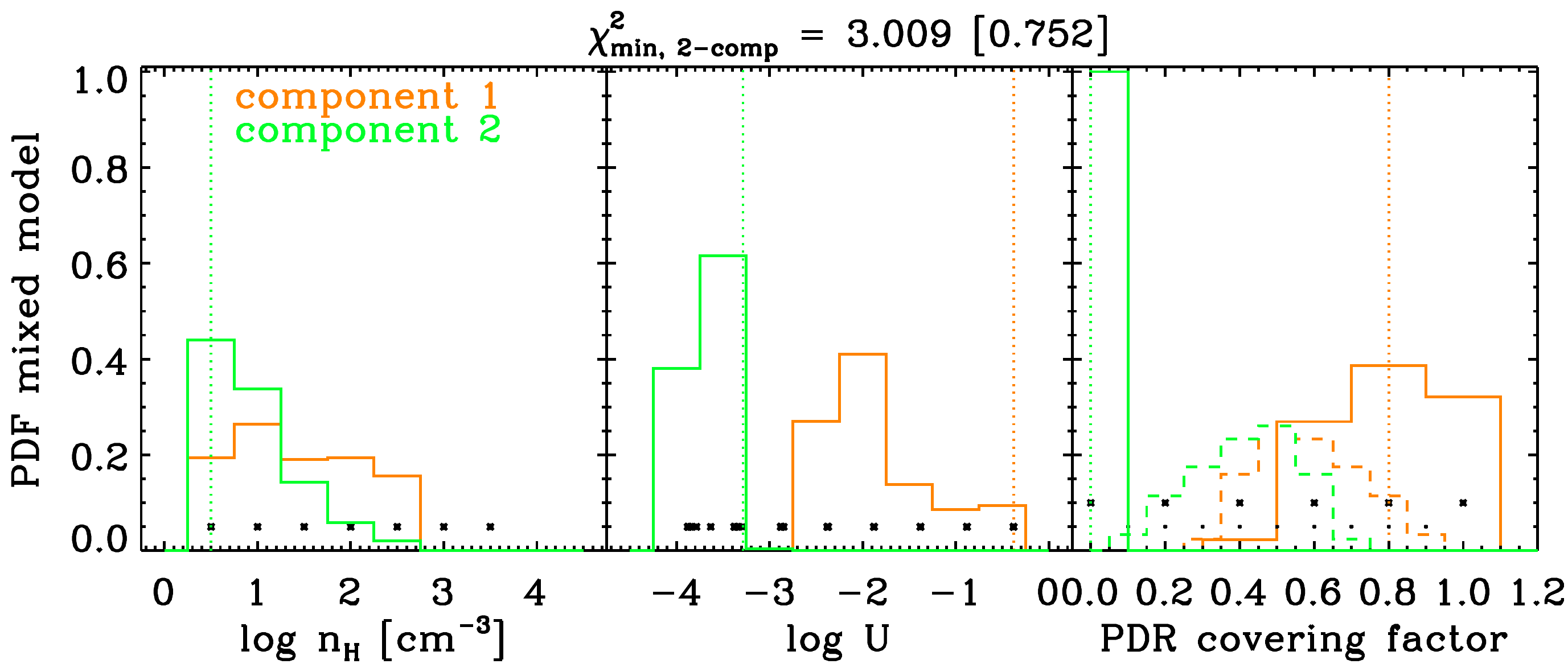}
\caption{ \textbf{Results for HS\,0052.} 
$(a)$~Comparison of the observed (losange) and predicted 
intensities for our best-fitting single (blue bar) and mixed (red bar)
\hii region+PDR models. Lines are sorted by decreasing energy.
The fitted lines have labels in boldface.
$(b)$~For mixed models: respective contributions from
component \#1 and component \#2 to the line prediction.
For PDR lines (\cii, \oi, and \silii), only the contribution from
the ionized gas is shown; the PDR contribution is
one minus the sum of the contributions from the plot.
Contributions are expected to be dominated by one
or the other component if their model parameters
are noticeably different.
$(c)$~Probability density functions of the model parameters
($n_{\rm H}$, $U$, $cov_{\rm PDR}$/scaling factor)
for single (left panel) and mixed (right panel) models.
The scaling factor corresponds to the proportion in which
we combine two models in the mixed model case. It is
equal to unity otherwise. It is shown with dashed lines in
the same panels as the PDR covering factor. Vertical dotted lines
show values of the best-fitting model. Black asterisks
show the range and step of values of the grid parameters.
}
\label{fig:bests-hs0052}
\end{figure*}
 \begin{figure*}[thp]
\centering
(a)
\includegraphics[clip,width=11cm]{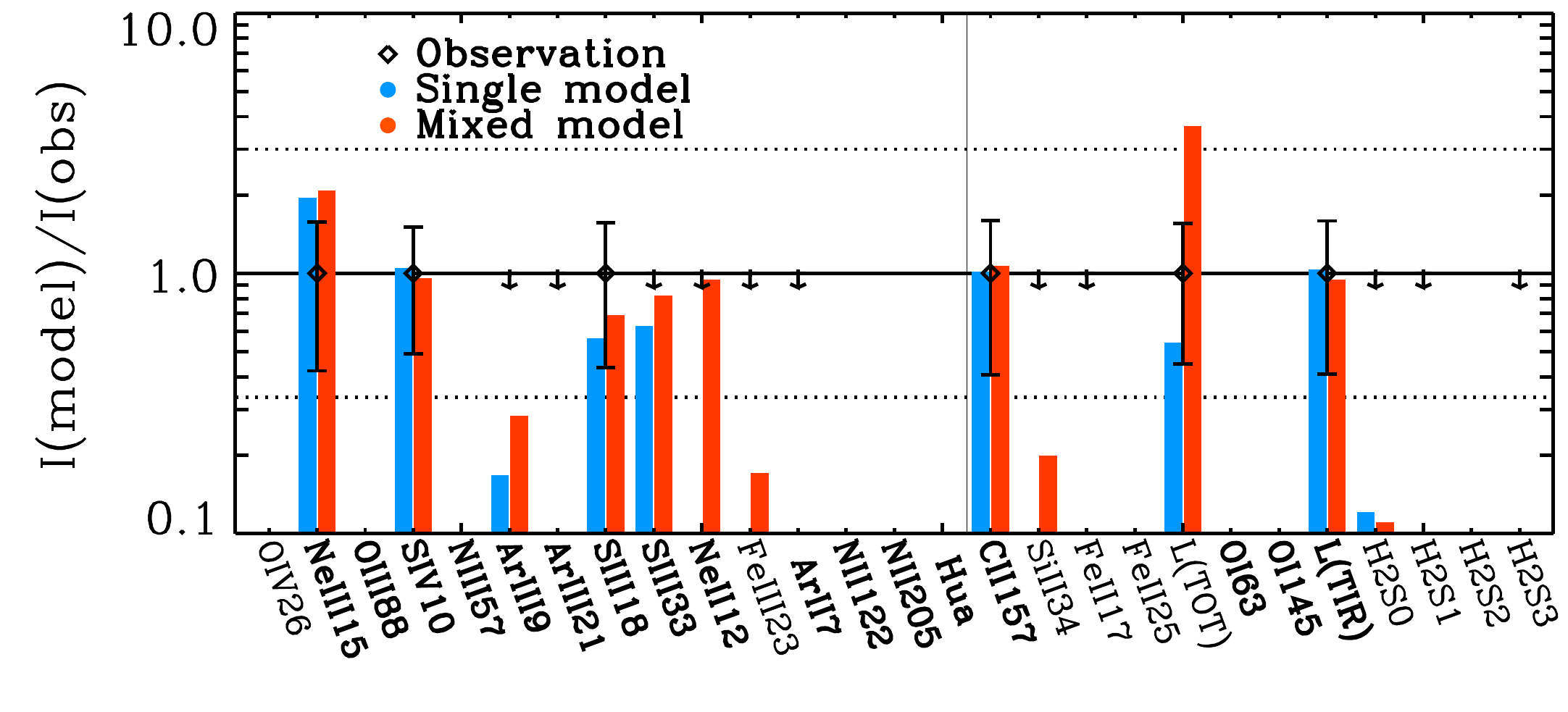} \hfill{}
(b)
\includegraphics[clip,width=6.2cm]{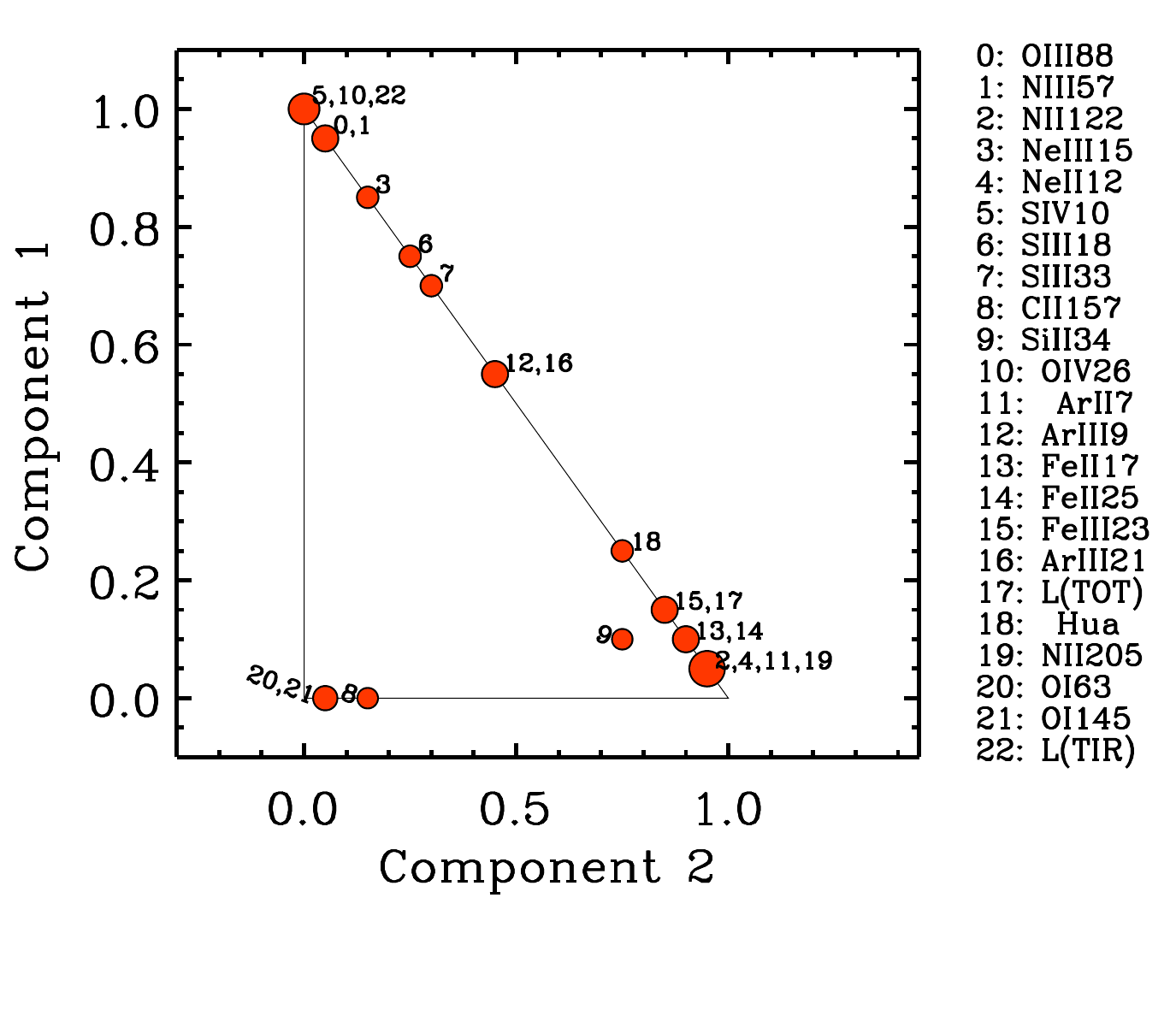}\\
(c)
\includegraphics[clip,width=8.8cm]{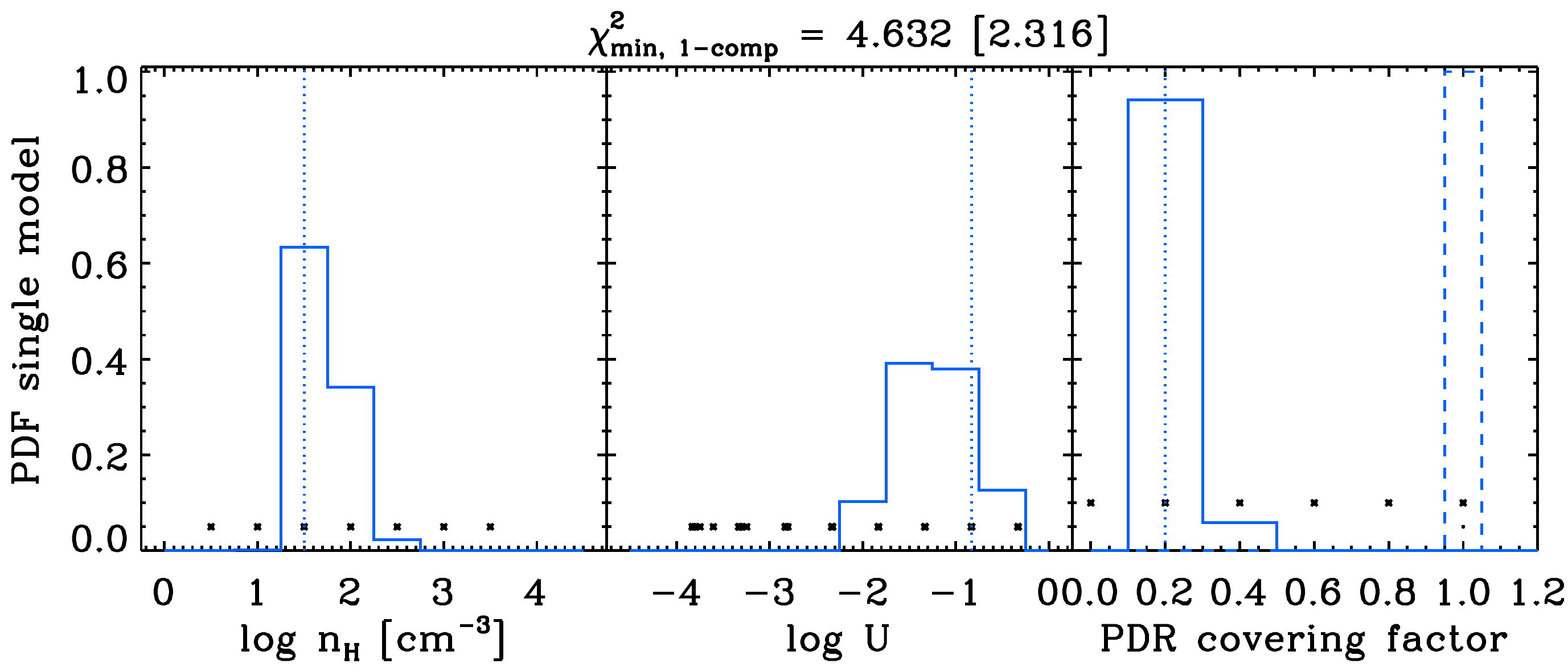} \hfill{}
\includegraphics[clip,width=8.8cm]{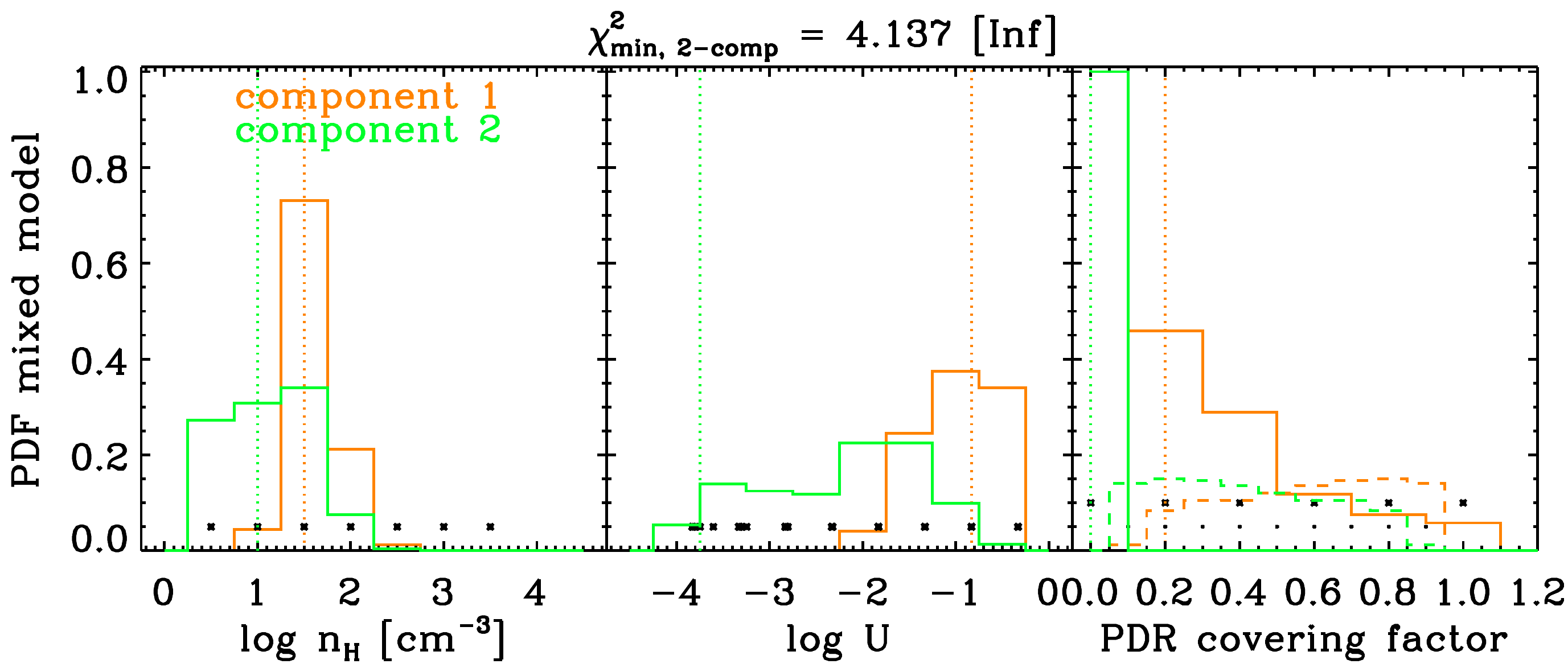}
\caption{ \textbf{Results for HS\,0822.}
See above for caption description.
}
\label{fig:bests-hs0822}
\end{figure*}
\clearpage
 \begin{figure*}[thp]
\centering
(a)
\includegraphics[clip,width=11cm]{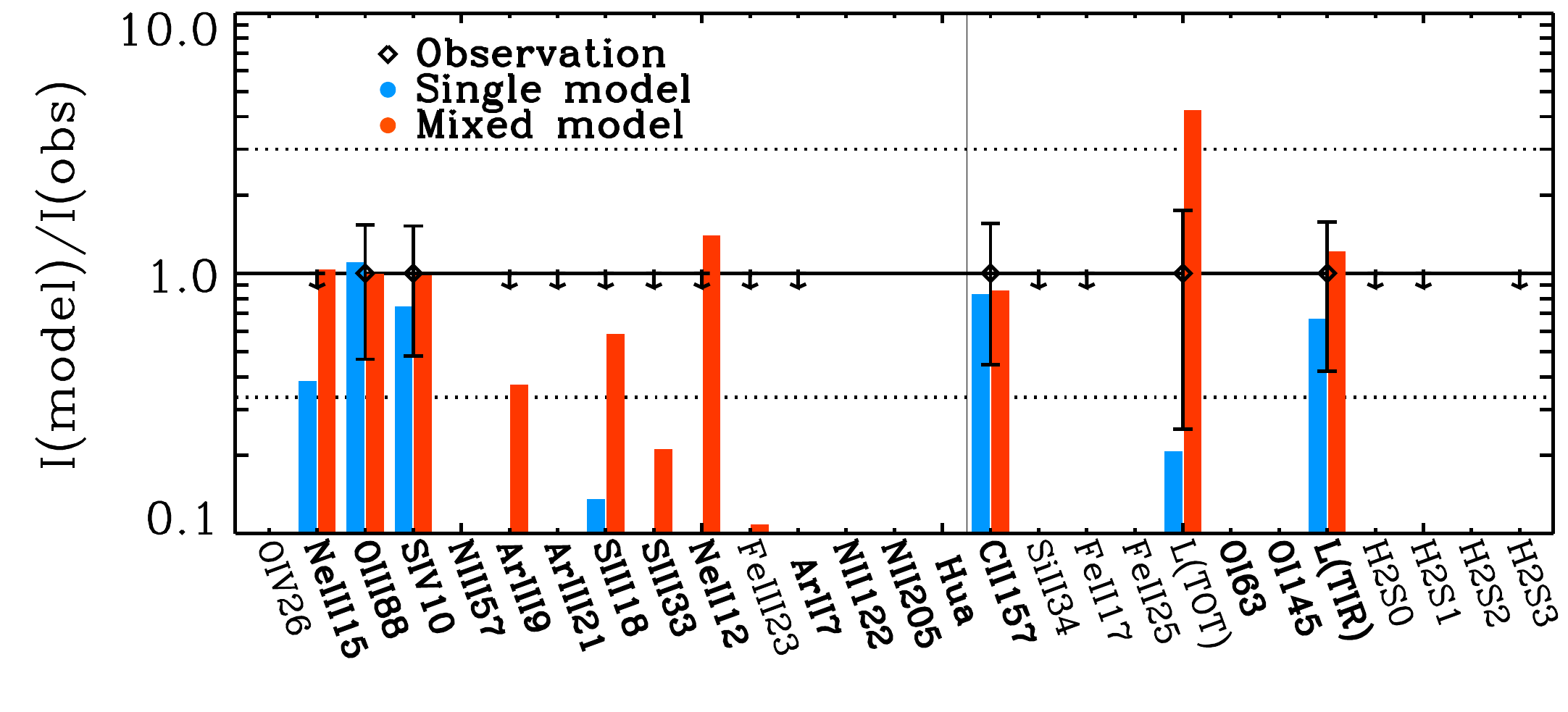} \hfill{}
(b)
\includegraphics[clip,width=6.2cm]{./figures_appendix/Haro11_HII_PDR_TIR_one_best_mixing_contrib}\\
(c)
\includegraphics[clip,width=8.8cm]{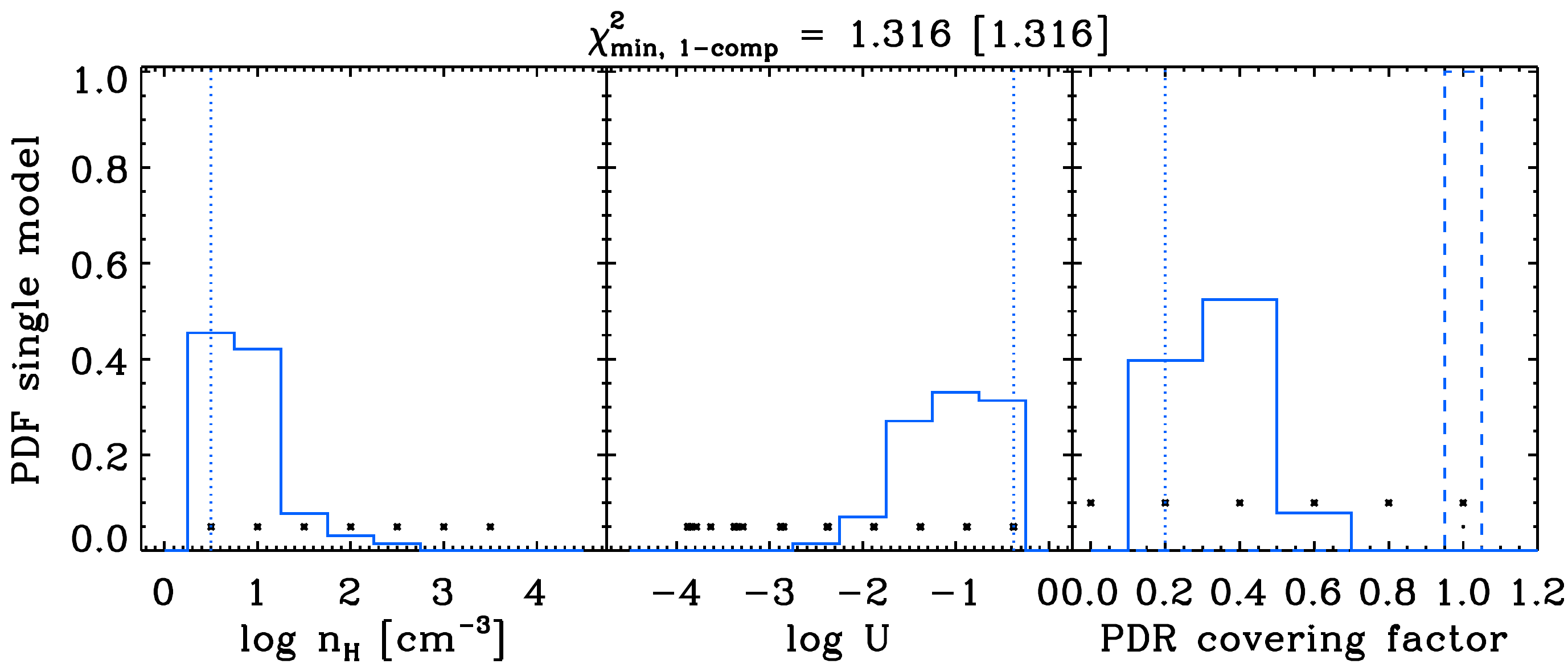} \hfill{}
\includegraphics[clip,width=8.8cm]{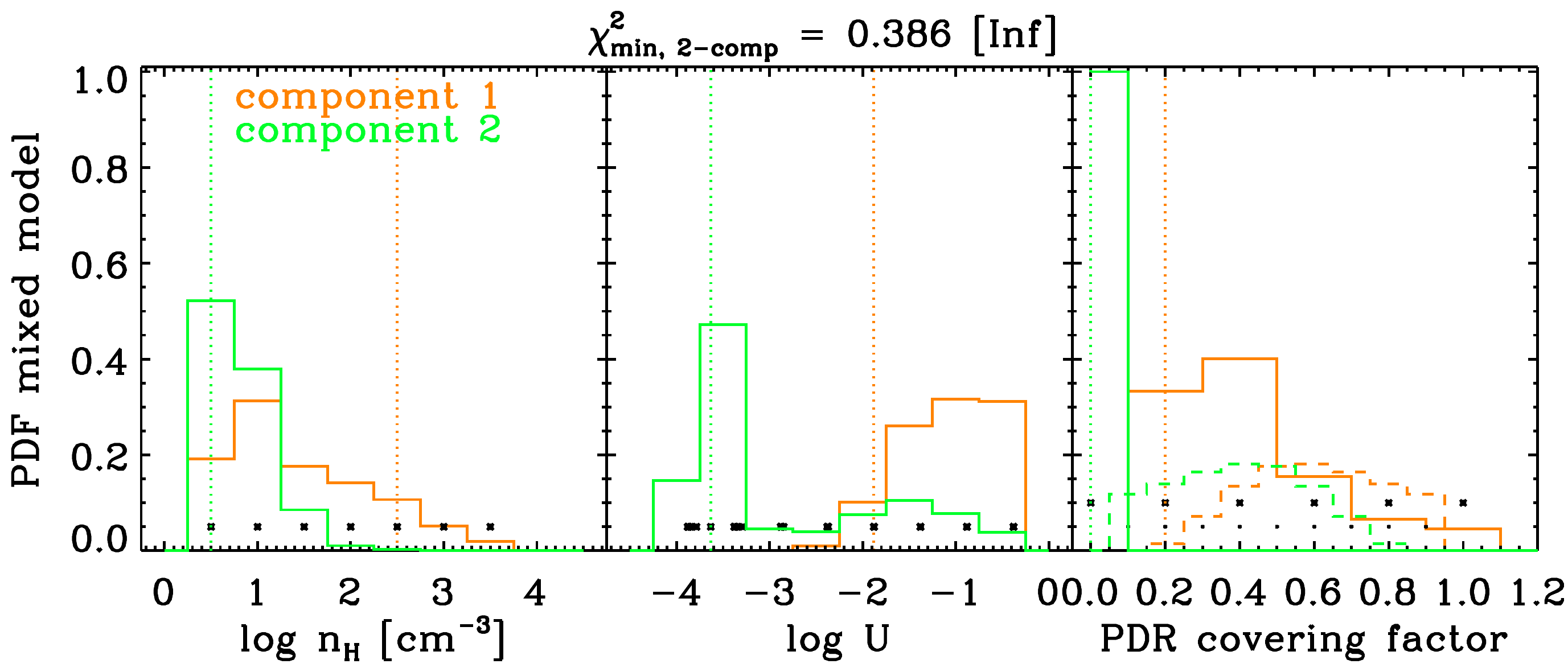}
\caption{ \textbf{Results for HS\,1222.} 
$(a)$~Comparison of the observed (losange) and predicted 
intensities for our best-fitting single (blue bar) and mixed (red bar)
\hii region+PDR models. Lines are sorted by decreasing energy.
The fitted lines have labels in boldface.
$(b)$~For mixed models: respective contributions from
component \#1 and component \#2 to the line prediction.
For PDR lines (\cii, \oi, and \silii), only the contribution from
the ionized gas is shown; the PDR contribution is
one minus the sum of the contributions from the plot.
Contributions are expected to be dominated by one
or the other component if their model parameters
are noticeably different.
$(c)$~Probability density functions of the model parameters
($n_{\rm H}$, $U$, $cov_{\rm PDR}$/scaling factor)
for single (left panel) and mixed (right panel) models.
The scaling factor corresponds to the proportion in which
we combine two models in the mixed model case. It is
equal to unity otherwise. It is shown with dashed lines in
the same panels as the PDR covering factor. Vertical dotted lines
show values of the best-fitting model. Black asterisks
show the range and step of values of the grid parameters.
}
\label{fig:bests-hs1222}
\end{figure*}
\begin{figure*}[thp]
\centering
(a)
\includegraphics[clip,width=11cm]{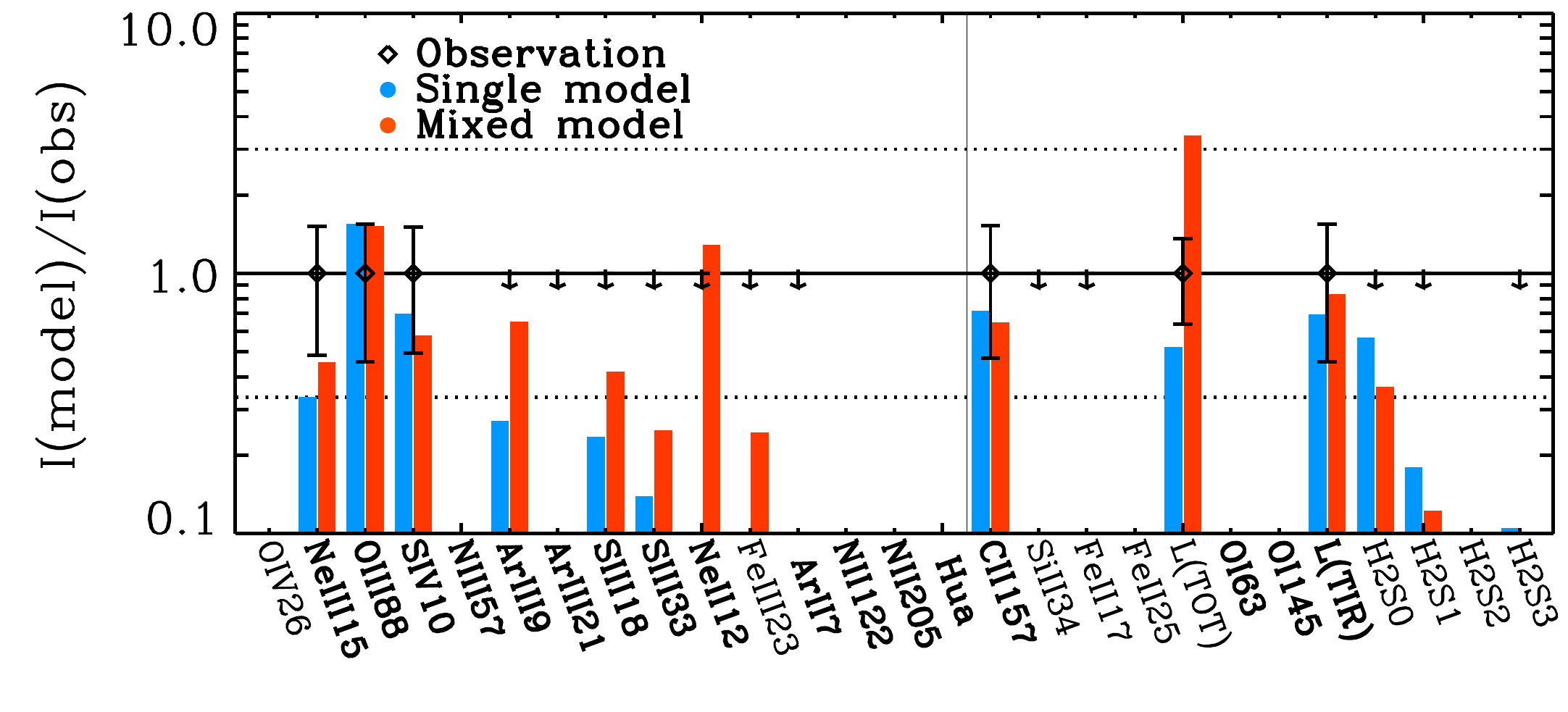} \hfill{}
(b)
\includegraphics[clip,width=6.2cm]{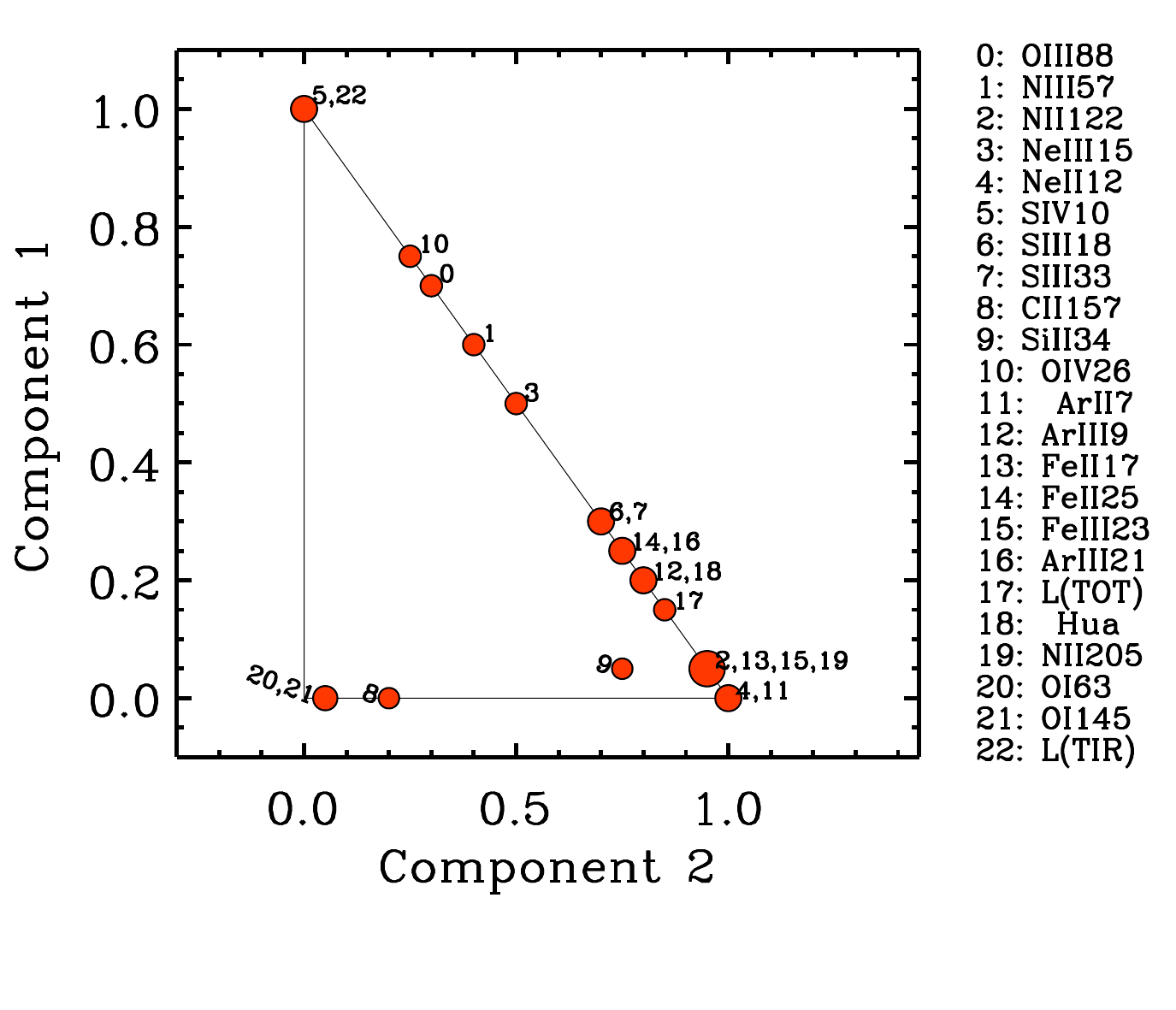}\\
(c)
\includegraphics[clip,width=8.8cm]{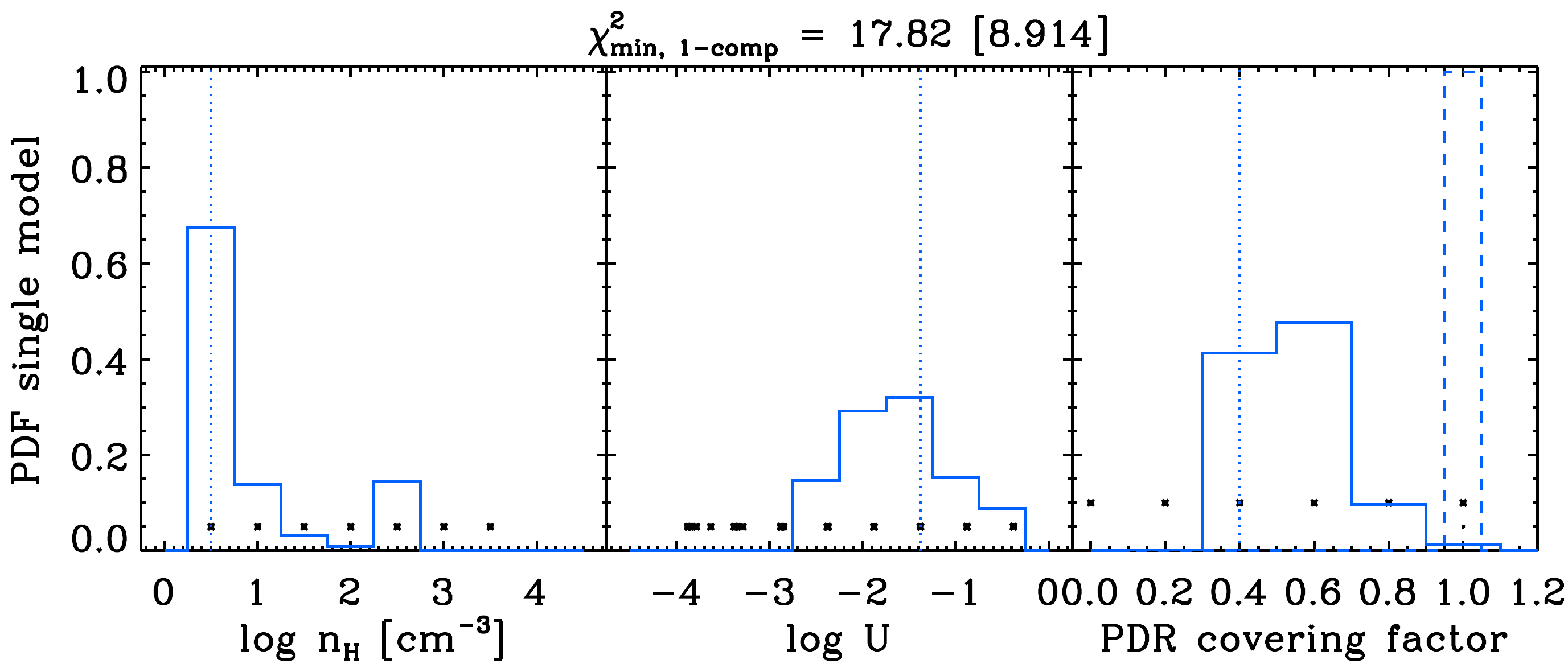} \hfill{}
\includegraphics[clip,width=8.8cm]{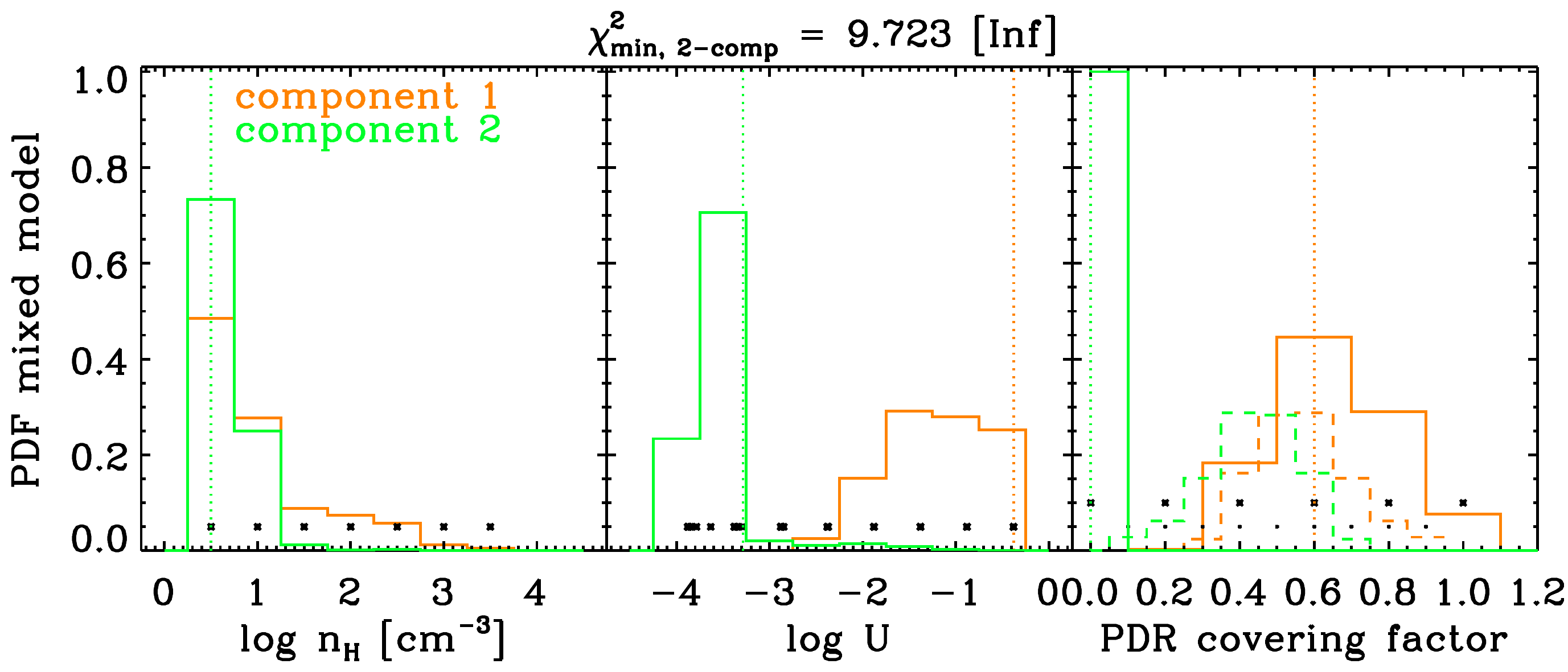}
\caption{ \textbf{Results for HS\,1304. } 
See above for caption description.
}
\label{fig:bests-hs1304}
\end{figure*}
 \clearpage
\begin{figure*}[thp]
\centering
(a)
\includegraphics[clip,width=11cm]{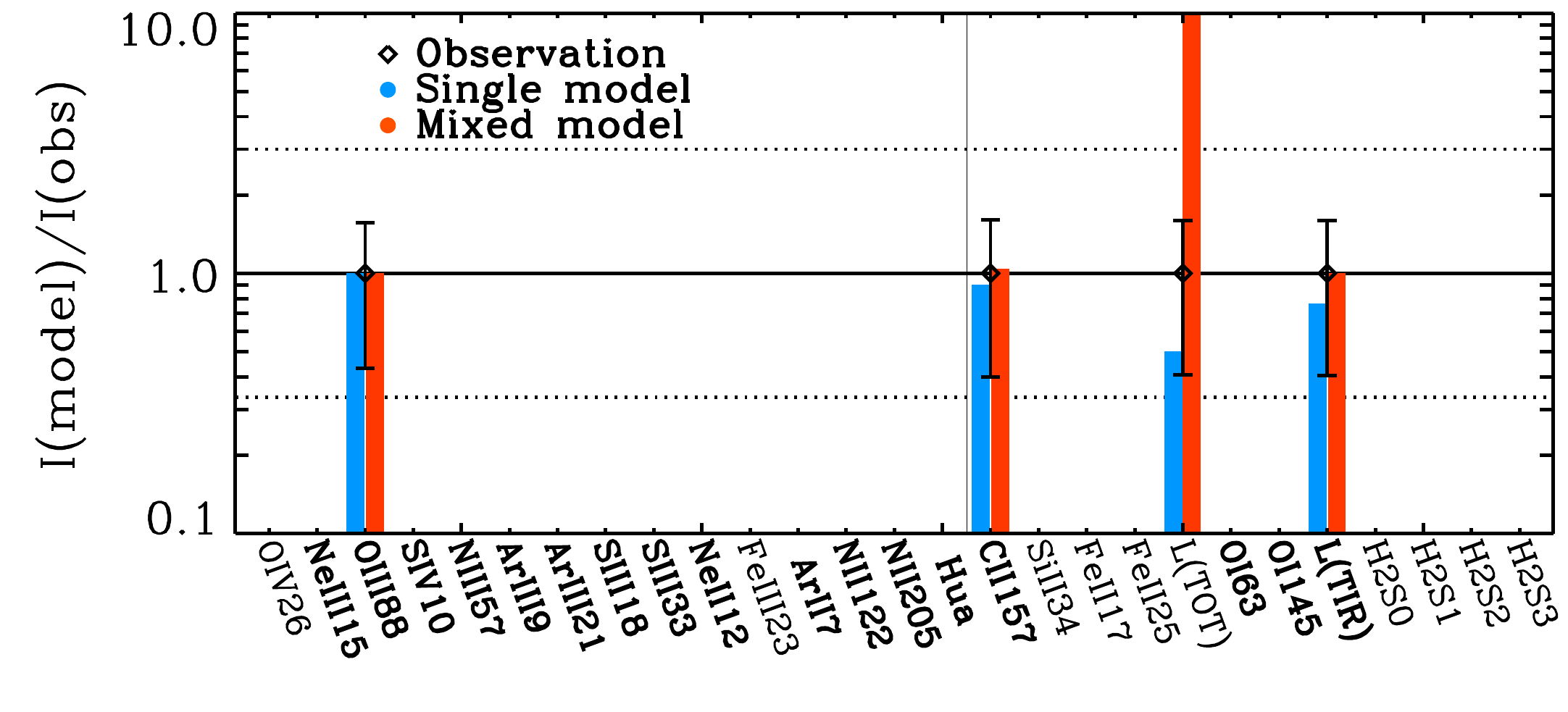} \hfill{}
(b)
\includegraphics[clip,width=6.2cm]{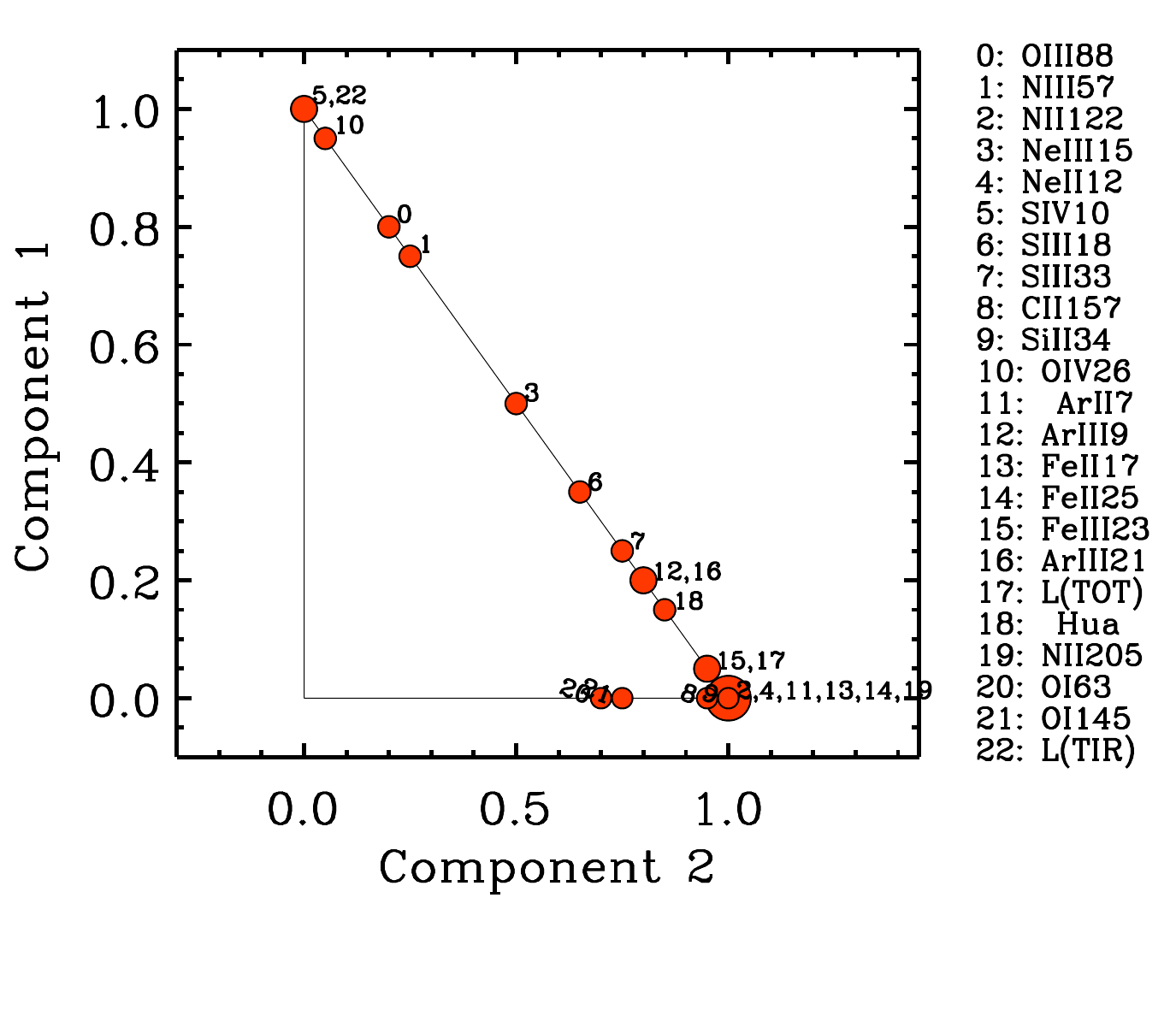}\\
(c)
\includegraphics[clip,width=8.8cm]{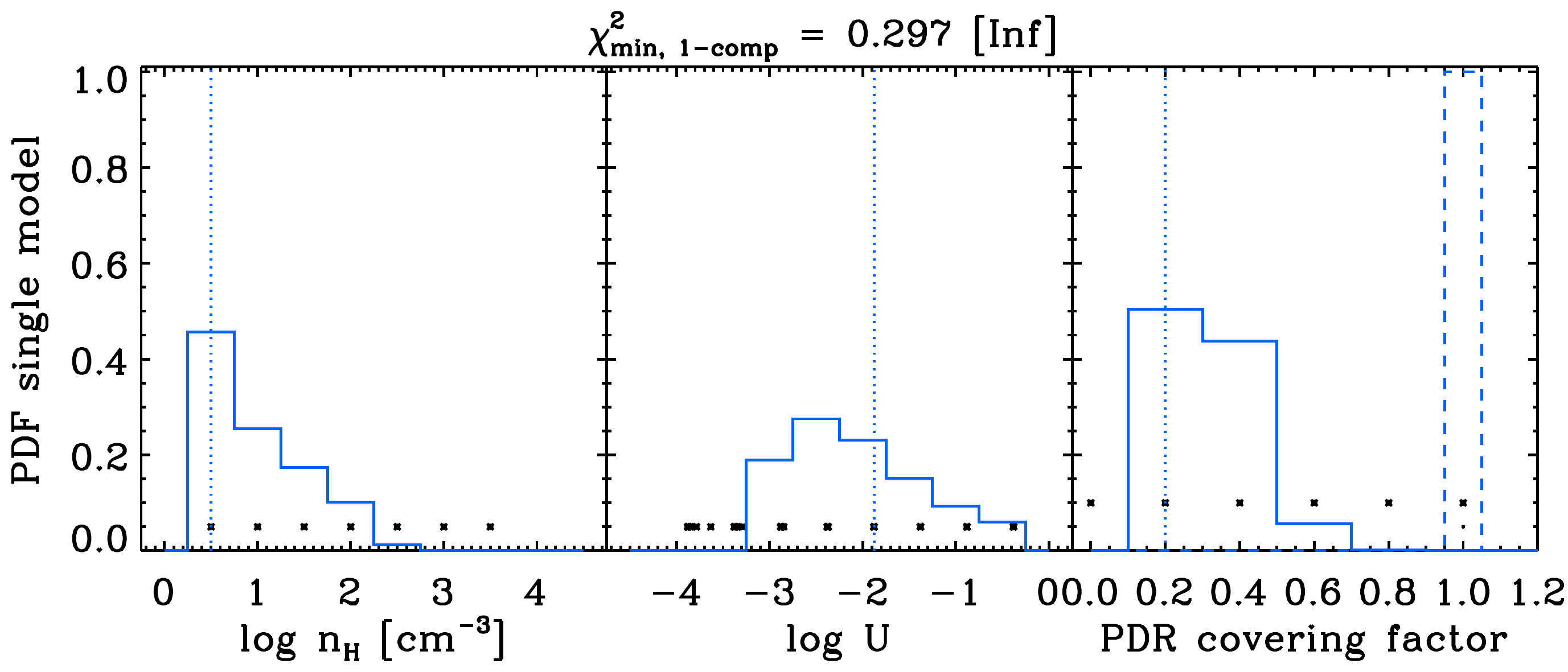} \hfill{}
\includegraphics[clip,width=8.8cm]{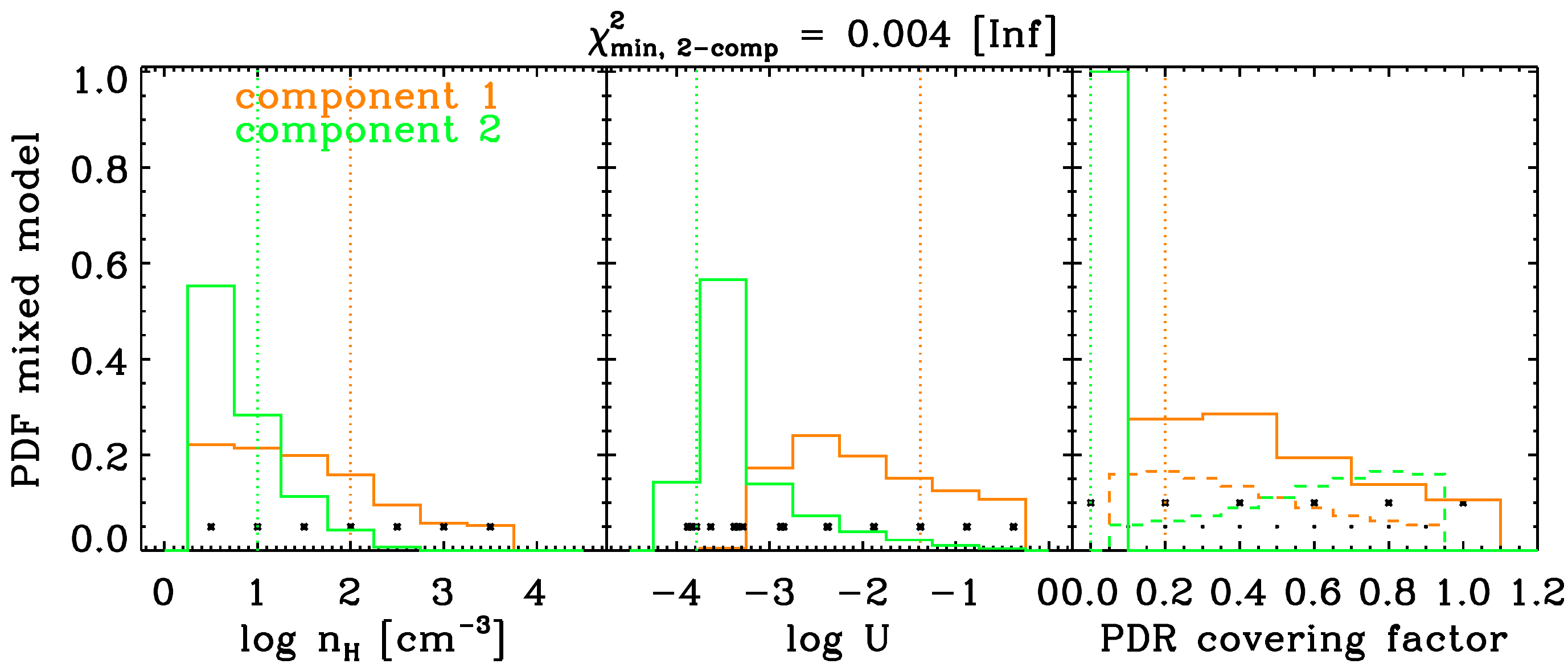}
\caption{ \textbf{Results for HS\,1319. }
$(a)$~Comparison of the observed (losange) and predicted 
intensities for our best-fitting single (blue bar) and mixed (red bar)
\hii region+PDR models. Lines are sorted by decreasing energy.
The fitted lines have labels in boldface.
$(b)$~For mixed models: respective contributions from
component \#1 and component \#2 to the line prediction.
For PDR lines (\cii, \oi, and \silii), only the contribution from
the ionized gas is shown; the PDR contribution is
one minus the sum of the contributions from the plot.
Contributions are expected to be dominated by one
or the other component if their model parameters
are noticeably different.
$(c)$~Probability density functions of the model parameters
($n_{\rm H}$, $U$, $cov_{\rm PDR}$/scaling factor)
for single (left panel) and mixed (right panel) models.
The scaling factor corresponds to the proportion in which
we combine two models in the mixed model case. It is
equal to unity otherwise. It is shown with dashed lines in
the same panels as the PDR covering factor. Vertical dotted lines
show values of the best-fitting model. Black asterisks
show the range and step of values of the grid parameters.
}
\label{fig:bests-hs1319}
\end{figure*}
 \begin{figure*}[thp]
\centering
(a)
\includegraphics[clip,width=11cm]{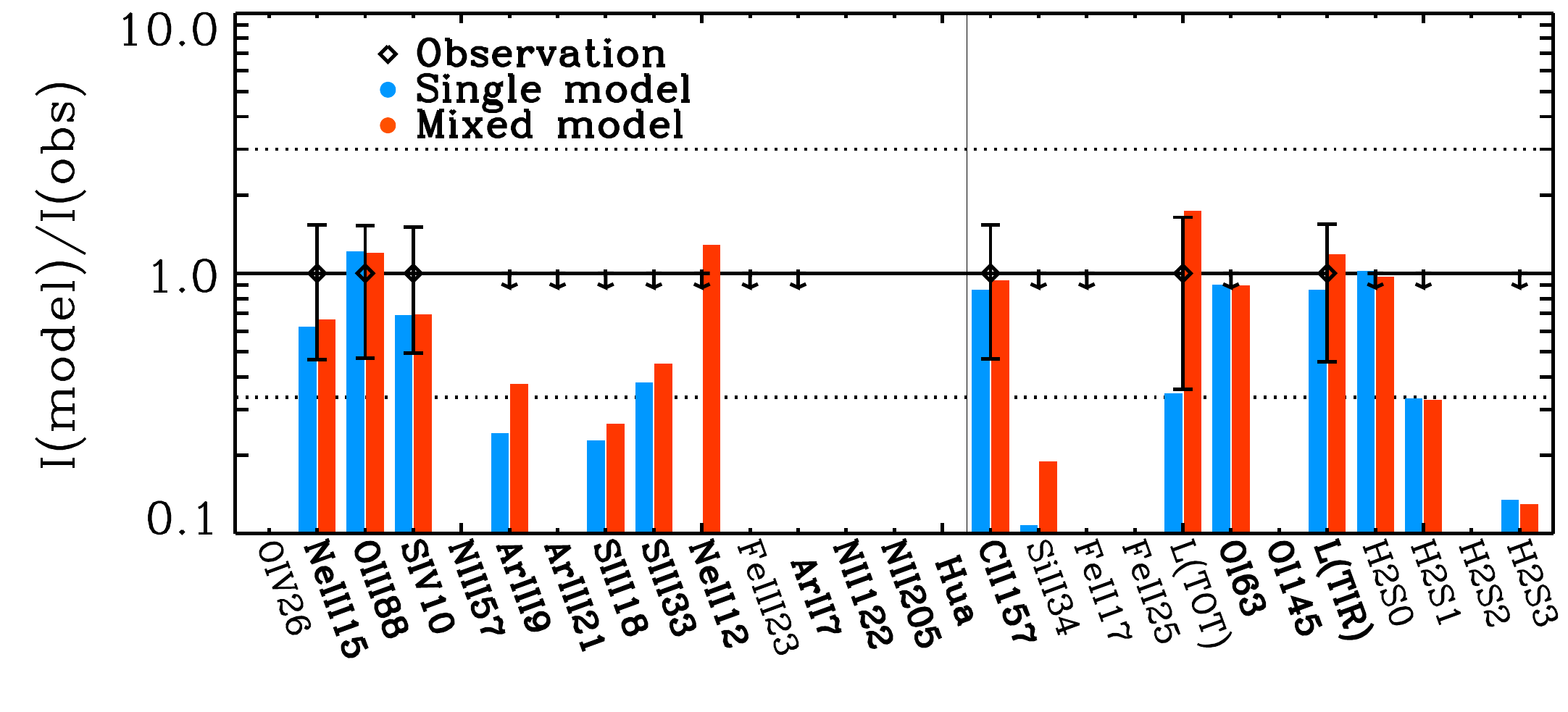} \hfill{}
(b)
\includegraphics[clip,width=6.2cm]{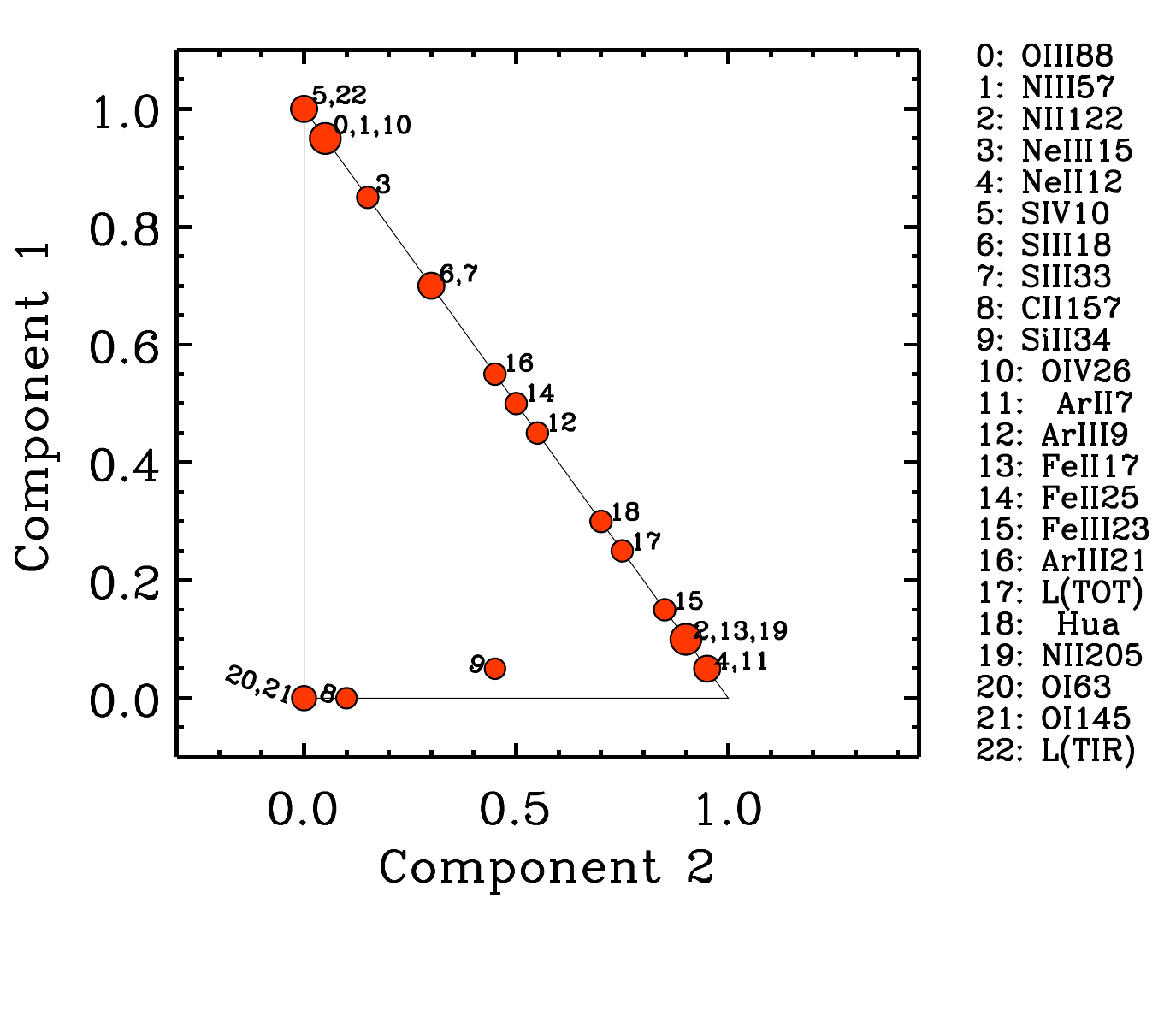}\\
(c)
\includegraphics[clip,width=8.8cm]{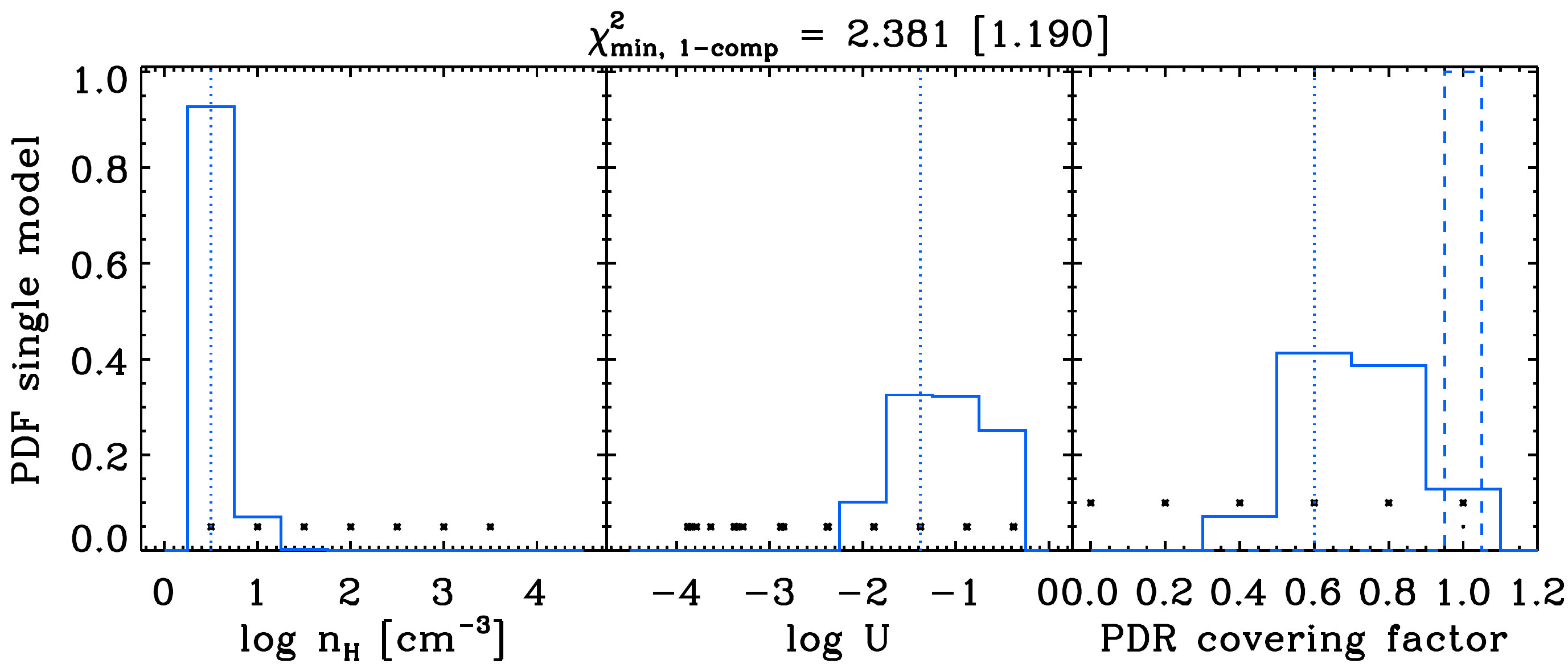} \hfill{}
\includegraphics[clip,width=8.8cm]{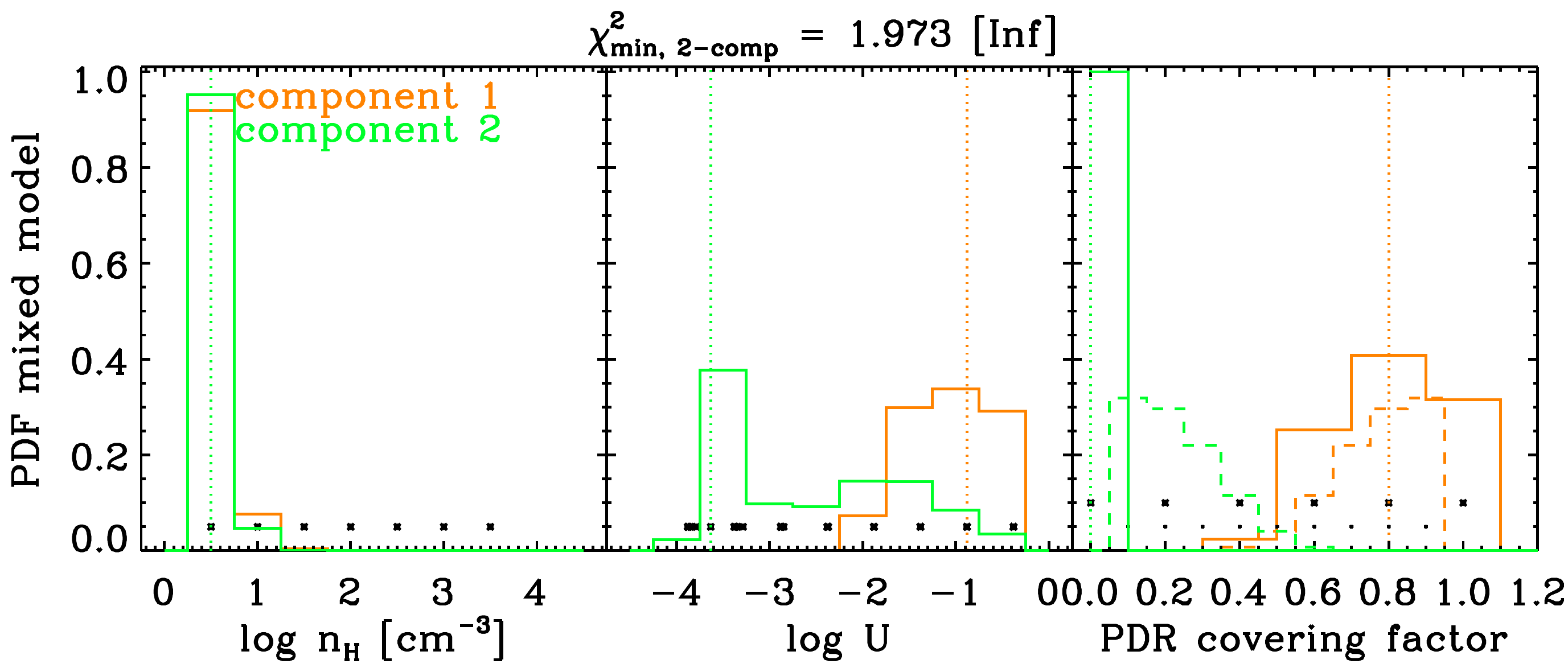}
\caption{ \textbf{Results for HS\,1330.} 
See above for caption description.
}
\label{fig:bests-hs1330}
\end{figure*}
\clearpage
 \begin{figure*}[thp]
\centering
(a)
\includegraphics[clip,width=11cm]{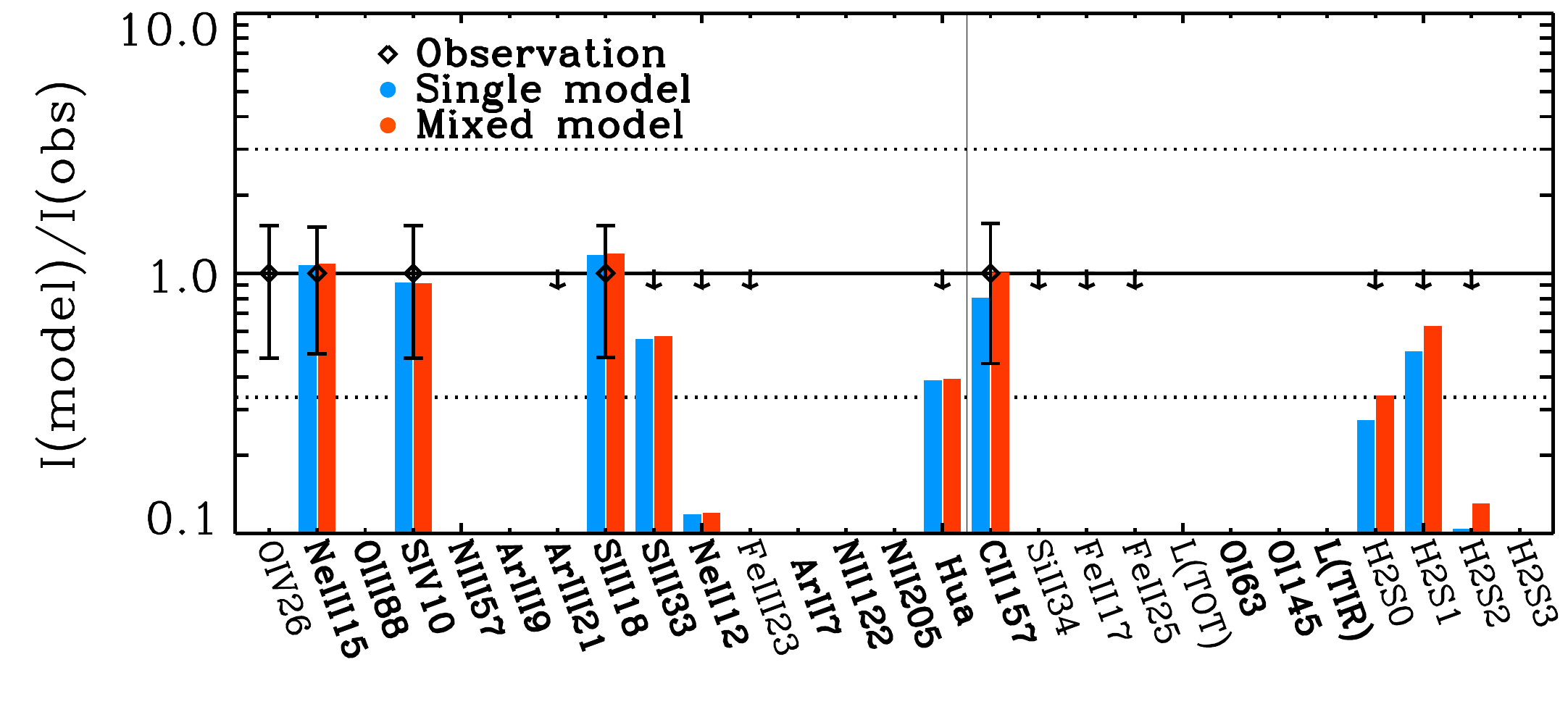} \hfill{}
(b)
\includegraphics[clip,width=6.2cm]{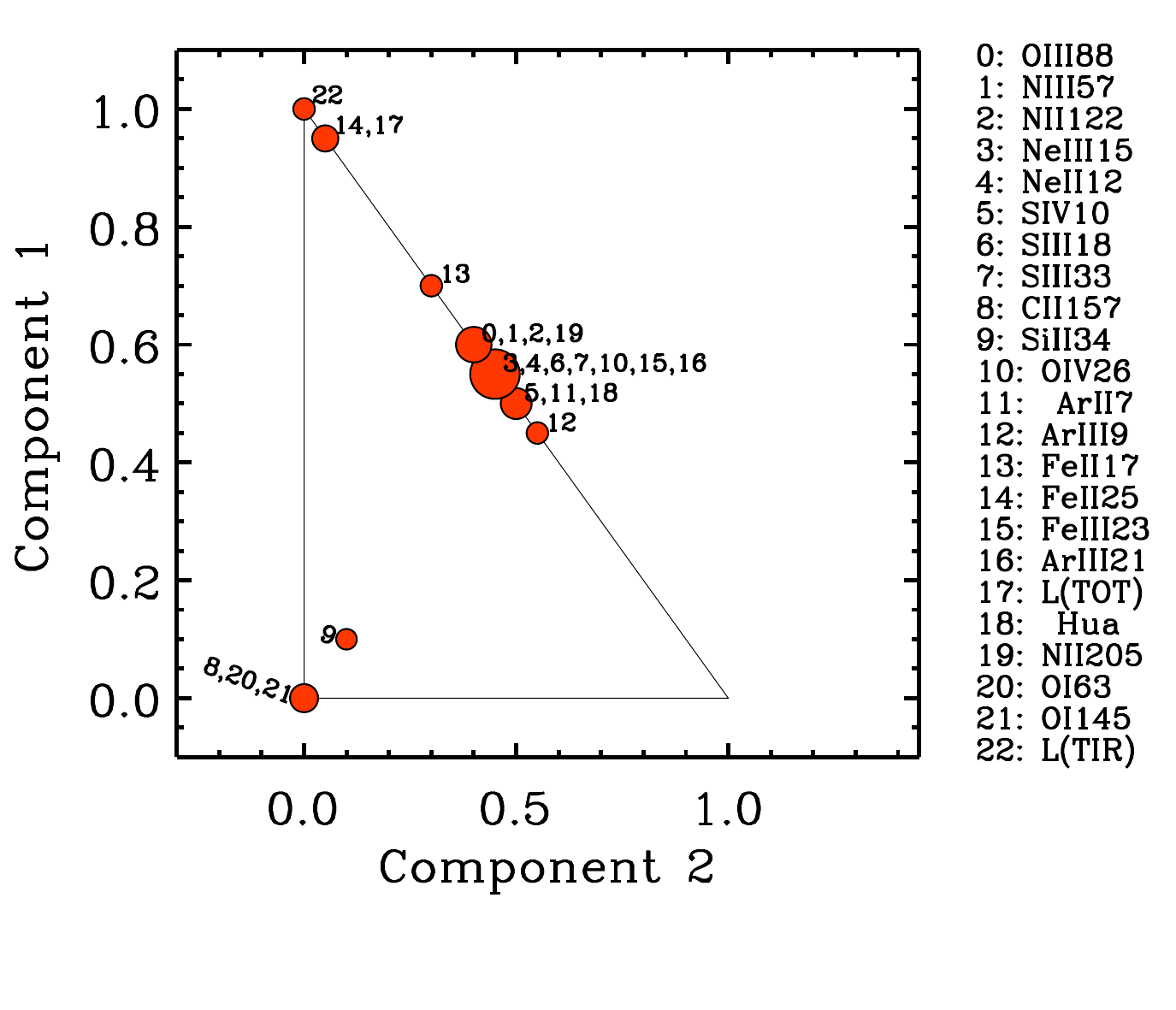}\\
(c)
\includegraphics[clip,width=8.8cm]{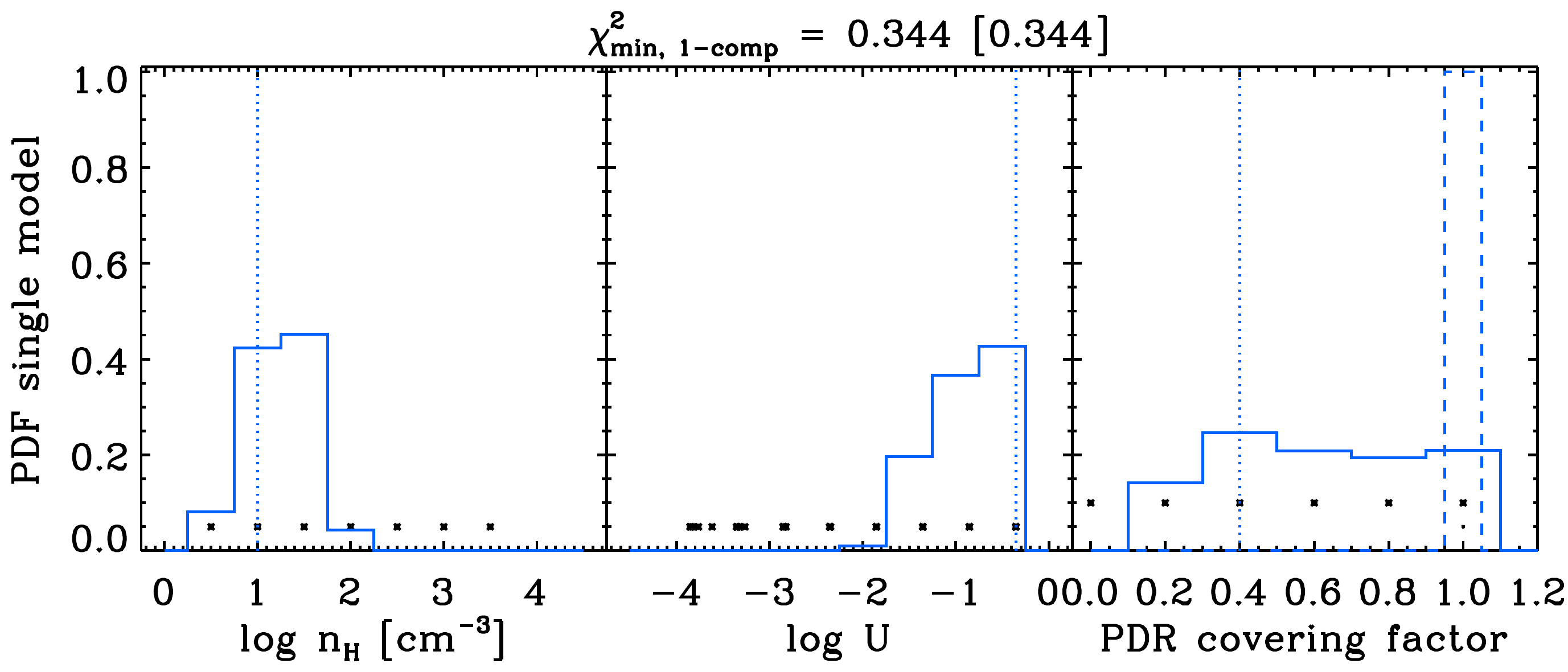} \hfill{}
\includegraphics[clip,width=8.8cm]{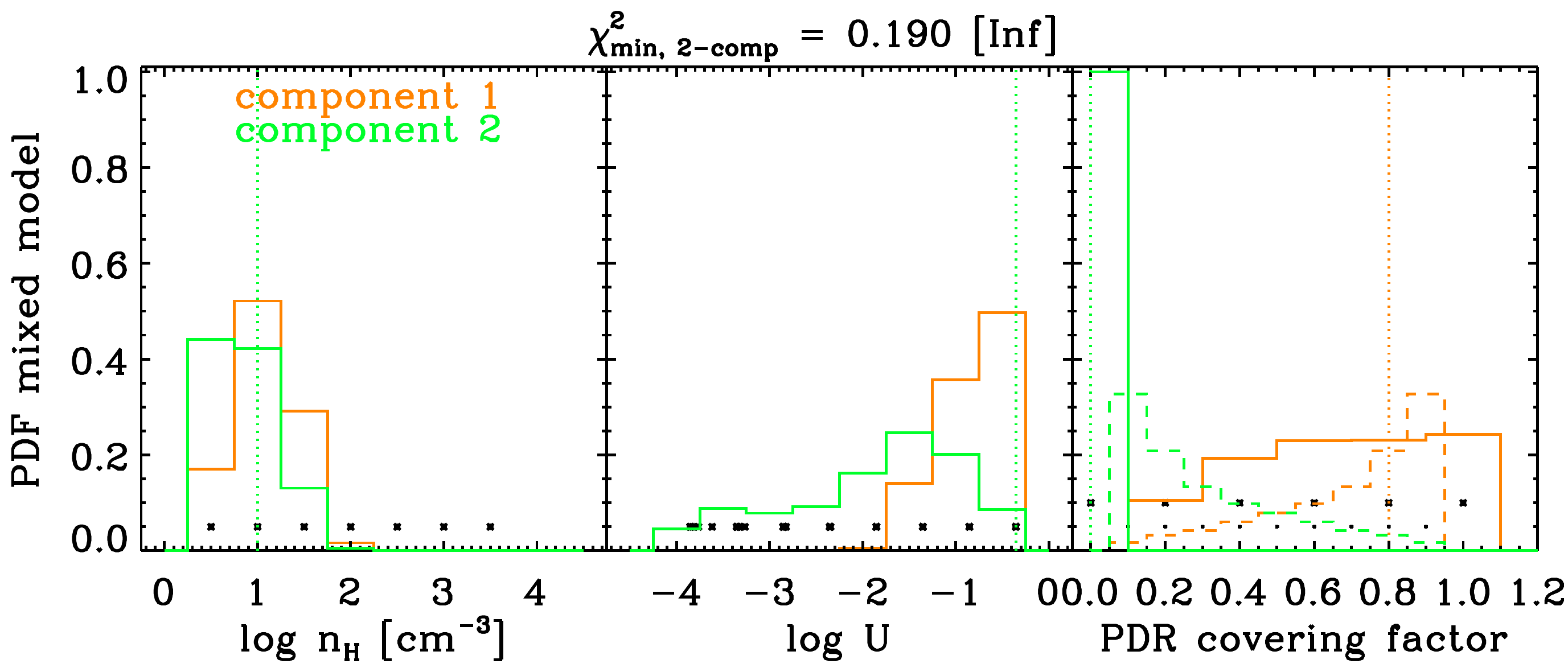}
\caption{ \textbf{Results for HS\,1442.}
$(a)$~Comparison of the observed (losange) and predicted 
intensities for our best-fitting single (blue bar) and mixed (red bar)
\hii region+PDR models. Lines are sorted by decreasing energy.
The fitted lines have labels in boldface.
$(b)$~For mixed models: respective contributions from
component \#1 and component \#2 to the line prediction.
For PDR lines (\cii, \oi, and \silii), only the contribution from
the ionized gas is shown; the PDR contribution is
one minus the sum of the contributions from the plot.
Contributions are expected to be dominated by one
or the other component if their model parameters
are noticeably different.
$(c)$~Probability density functions of the model parameters
($n_{\rm H}$, $U$, $cov_{\rm PDR}$/scaling factor)
for single (left panel) and mixed (right panel) models.
The scaling factor corresponds to the proportion in which
we combine two models in the mixed model case. It is
equal to unity otherwise. It is shown with dashed lines in
the same panels as the PDR covering factor. Vertical dotted lines
show values of the best-fitting model. Black asterisks
show the range and step of values of the grid parameters.
}
\label{fig:bests-hs1442}
\end{figure*}
 \begin{figure*}[thp]
\centering
(a)
\includegraphics[clip,width=11cm]{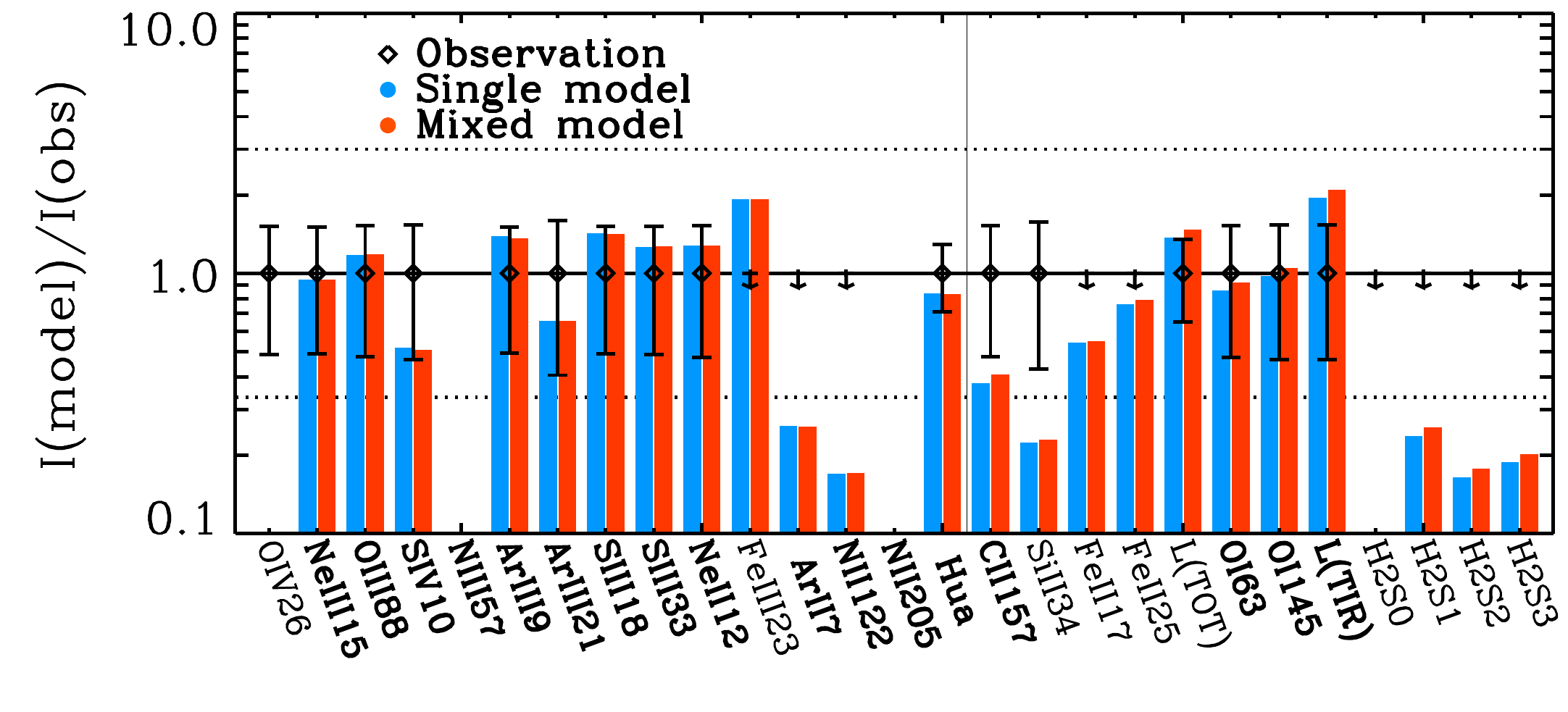} \hfill{}
(b)
\includegraphics[clip,width=6.2cm]{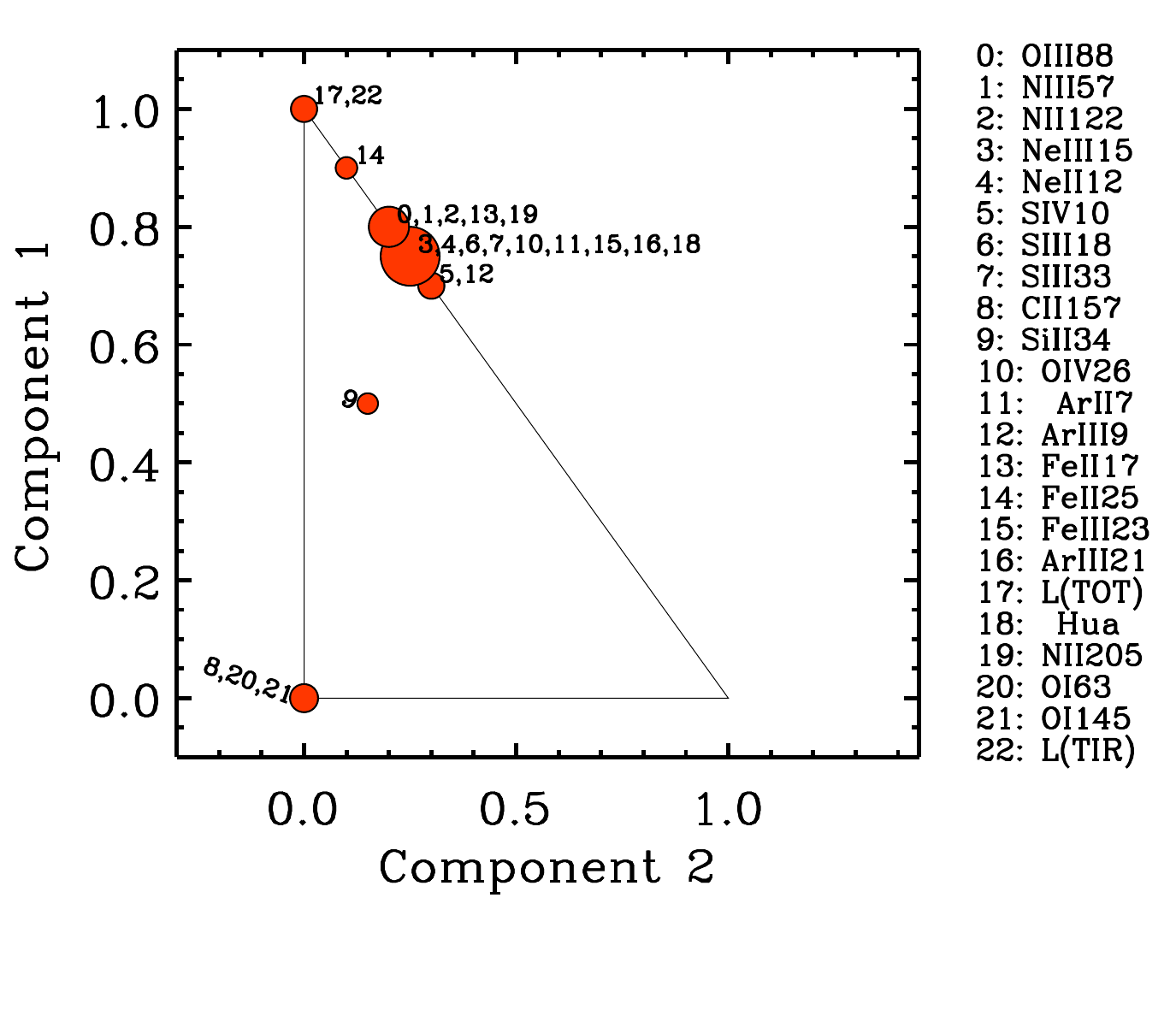}\\
(c)
\includegraphics[clip,width=8.8cm]{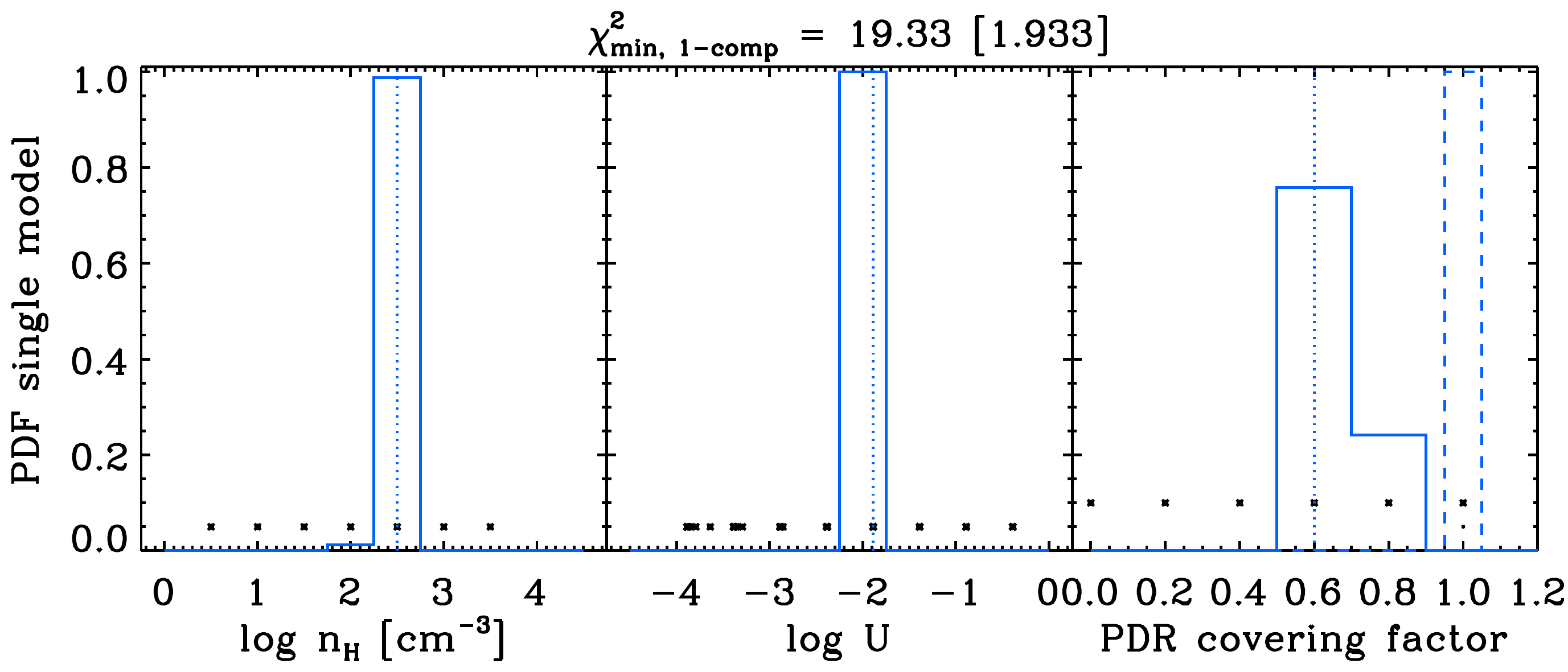} \hfill{}
\includegraphics[clip,width=8.8cm]{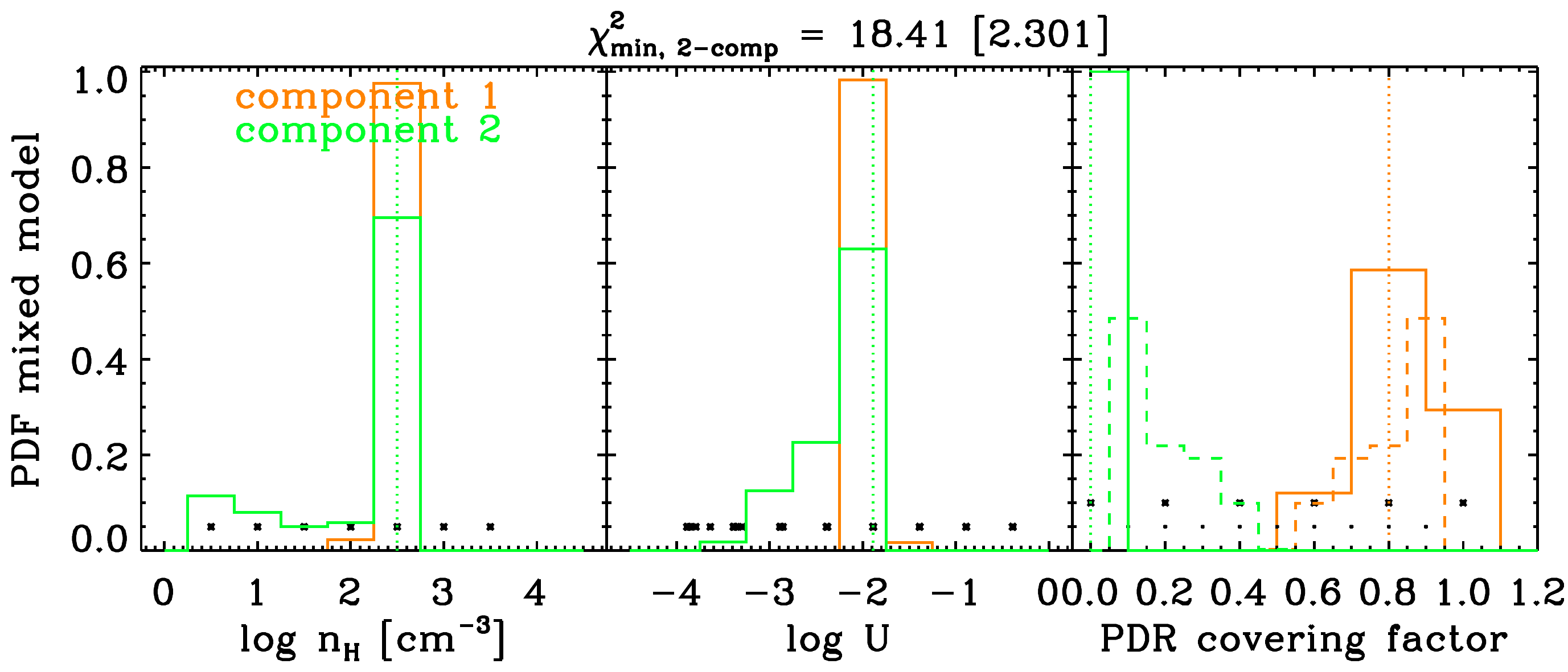}
\caption{ \textbf{Results for II\,Zw\,40.} 
See above for caption description.
}
\label{fig:bests-iizw40}
\end{figure*}
\clearpage
 \begin{figure*}[thp]
\centering
(a)
\includegraphics[clip,width=11cm]{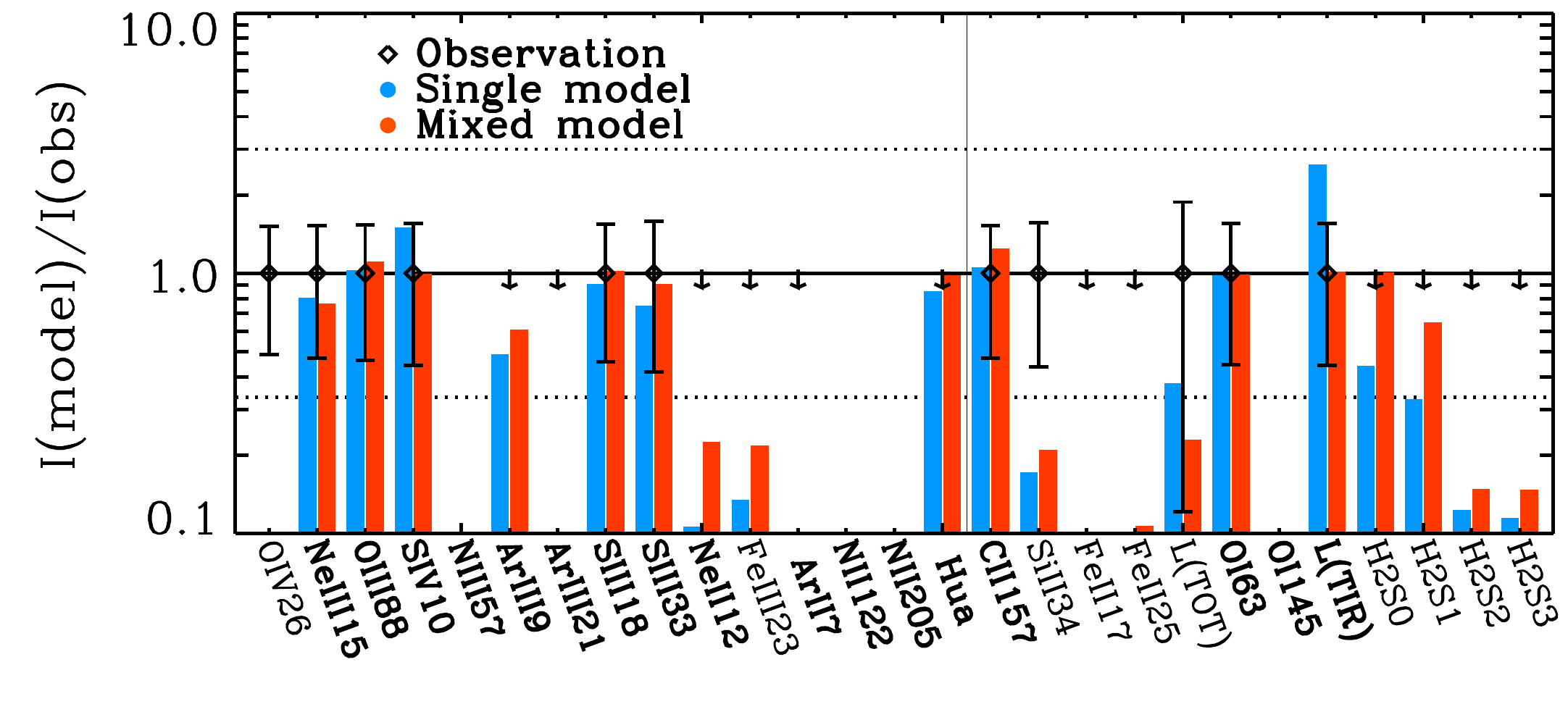} \hfill{}
(b)
\includegraphics[clip,width=6.2cm]{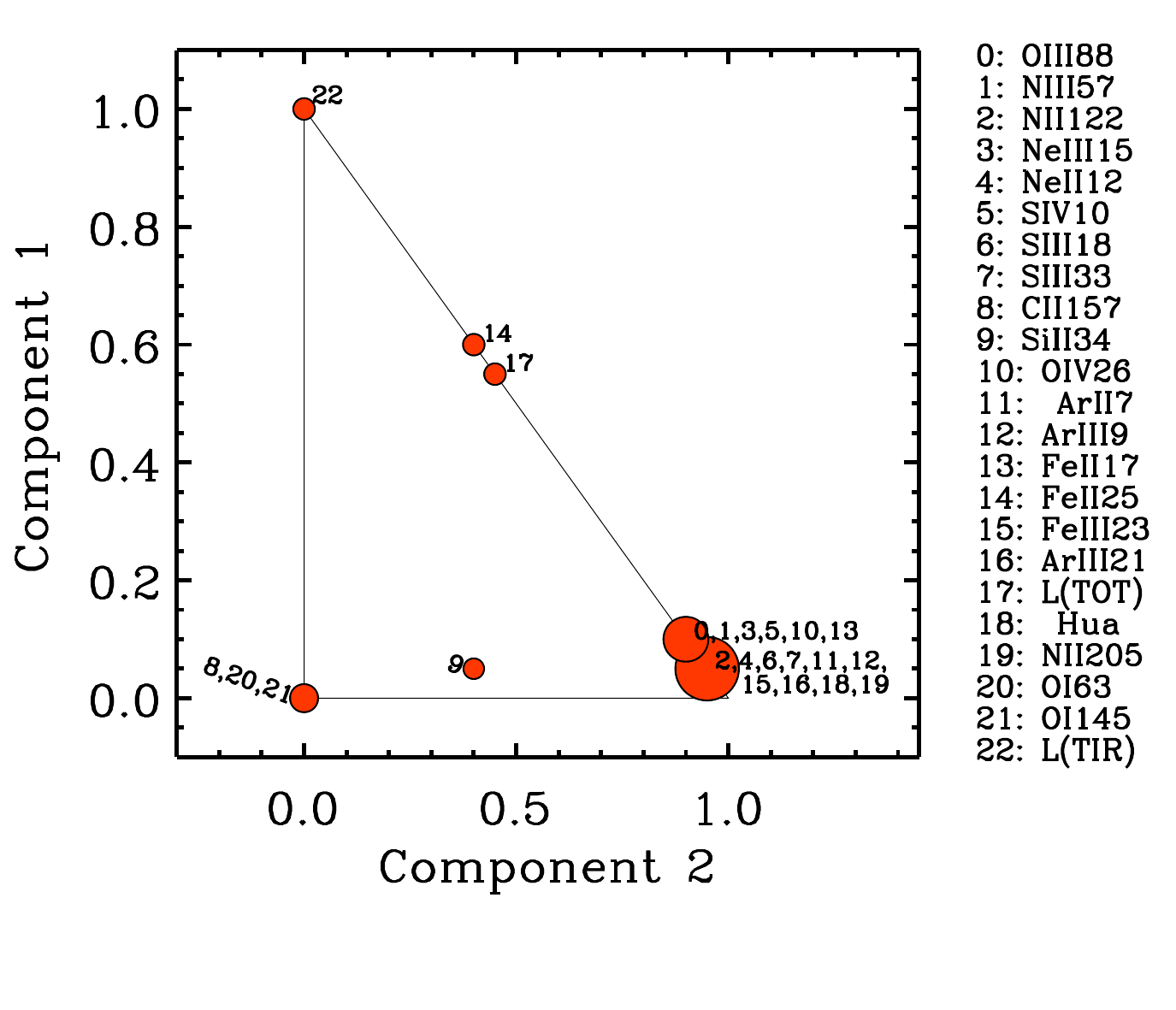}\\
(c)
\includegraphics[clip,width=8.8cm]{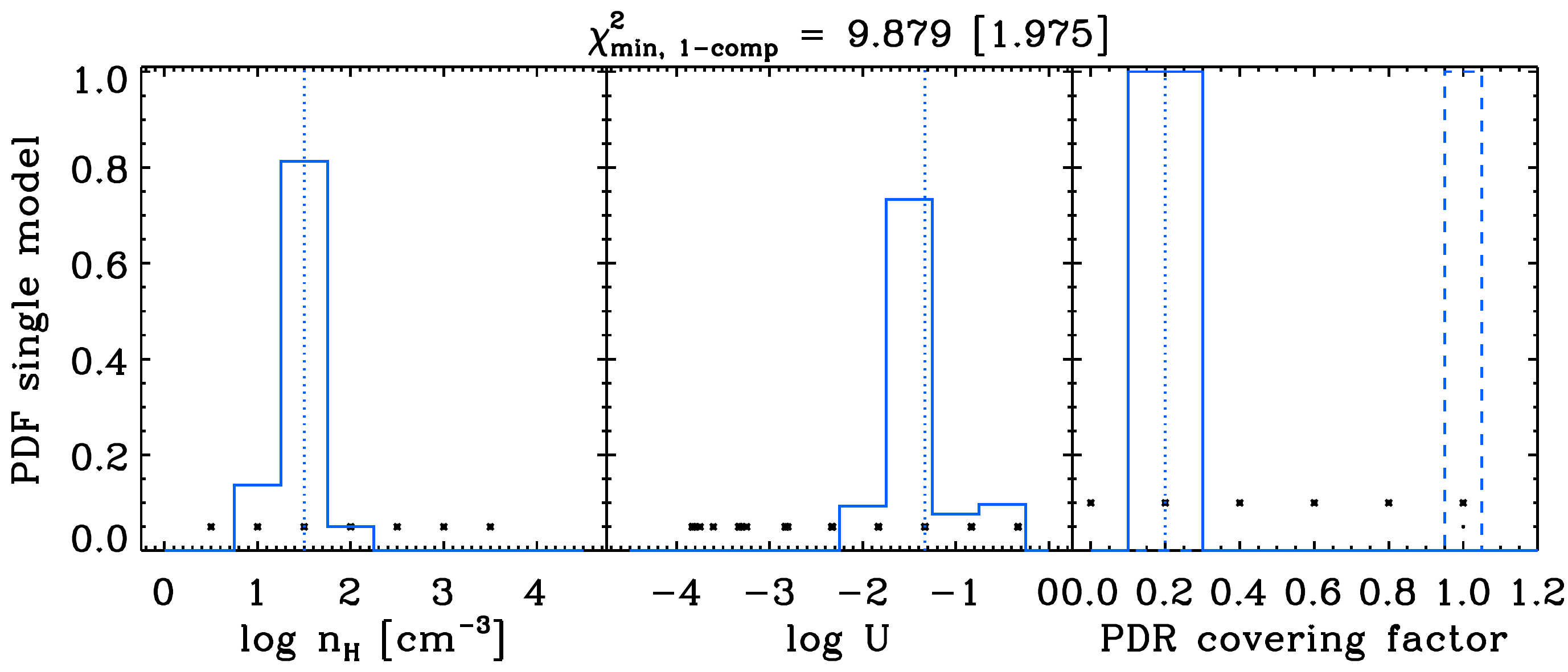} \hfill{}
\includegraphics[clip,width=8.8cm]{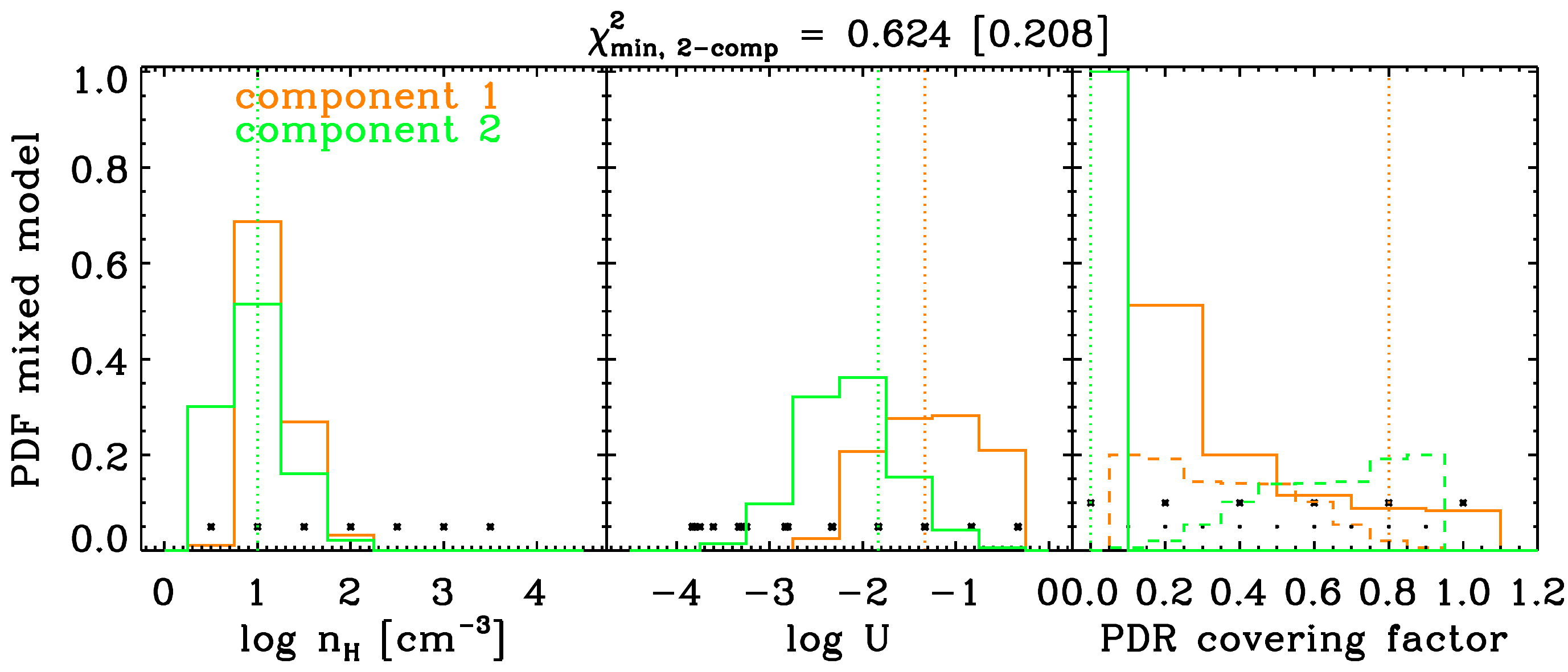}
\caption{ \textbf{Results for I\,Zw\,18.}
$(a)$~Comparison of the observed (losange) and predicted 
intensities for our best-fitting single (blue bar) and mixed (red bar)
\hii region+PDR models. Lines are sorted by decreasing energy.
The fitted lines have labels in boldface.
$(b)$~For mixed models: respective contributions from
component \#1 and component \#2 to the line prediction.
For PDR lines (\cii, \oi, and \silii), only the contribution from
the ionized gas is shown; the PDR contribution is
one minus the sum of the contributions from the plot.
Contributions are expected to be dominated by one
or the other component if their model parameters
are noticeably different.
$(c)$~Probability density functions of the model parameters
($n_{\rm H}$, $U$, $cov_{\rm PDR}$/scaling factor)
for single (left panel) and mixed (right panel) models.
The scaling factor corresponds to the proportion in which
we combine two models in the mixed model case. It is
equal to unity otherwise. It is shown with dashed lines in
the same panels as the PDR covering factor. Vertical dotted lines
show values of the best-fitting model. Black asterisks
show the range and step of values of the grid parameters.
}
\label{fig:bests-izw18}
\end{figure*}
 \begin{figure*}[thp]
\centering
(a)
\includegraphics[clip,width=11cm]{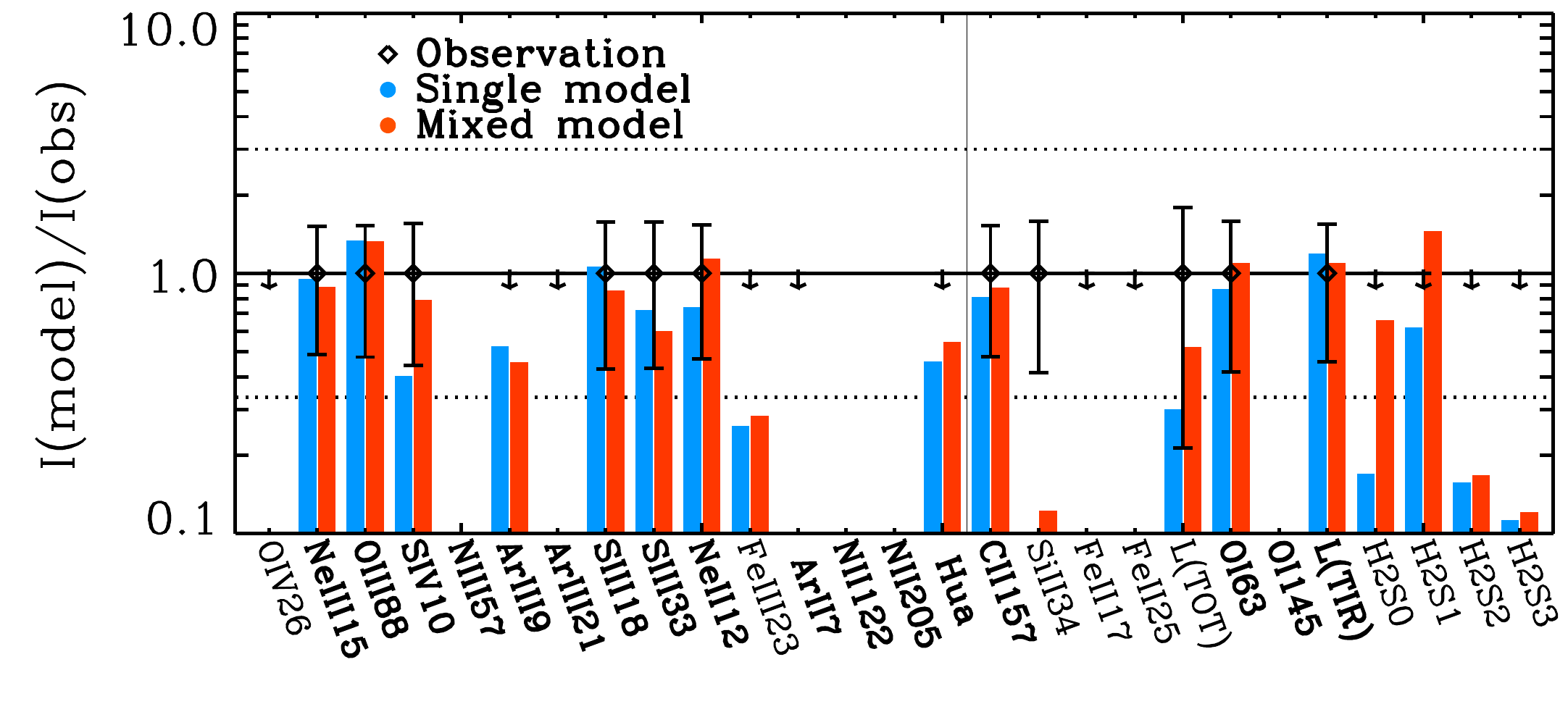} \hfill{}
(b)
\includegraphics[clip,width=6.2cm]{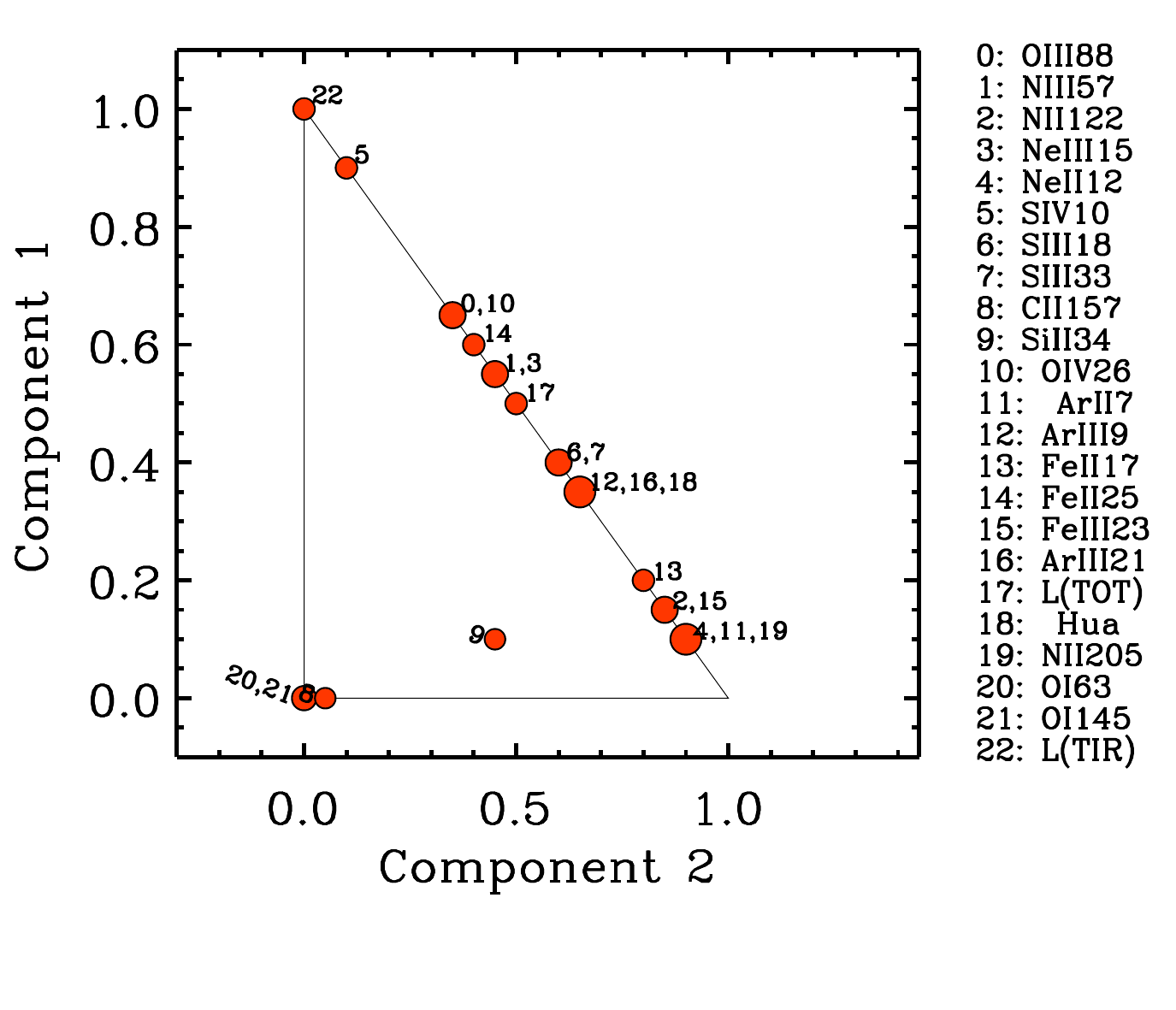}\\
(c)
\includegraphics[clip,width=8.8cm]{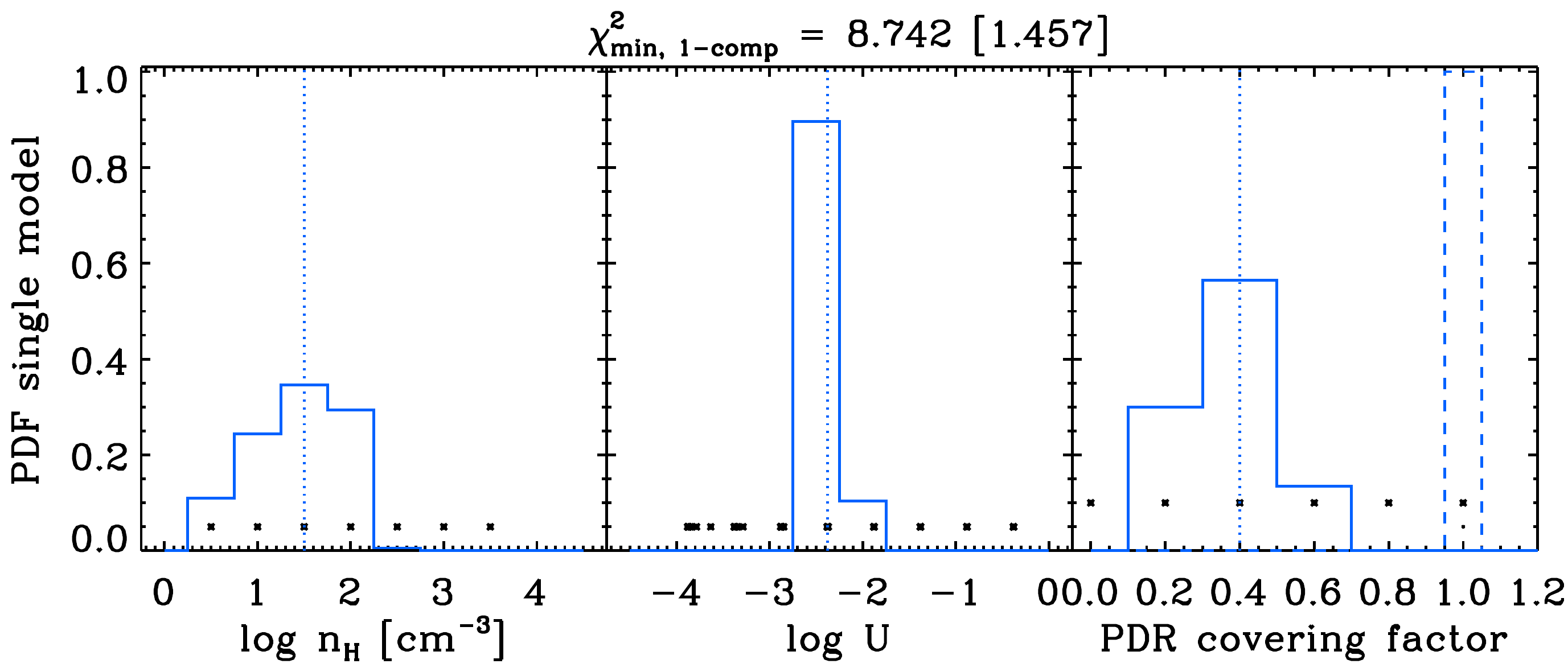} \hfill{}
\includegraphics[clip,width=8.8cm]{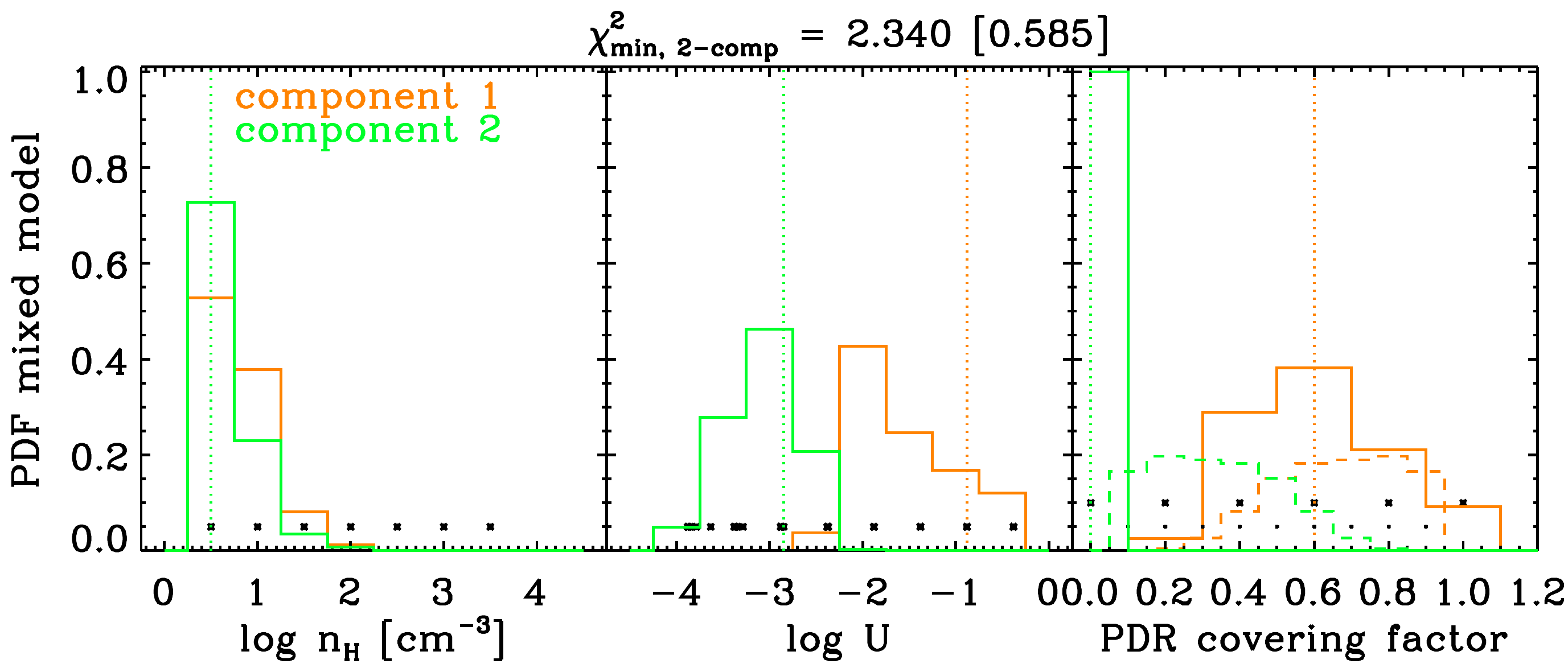}
\caption{ \textbf{Results for Mrk\,153.}
See above for caption description.
}
\label{fig:bests-mrk153}
\end{figure*}
\clearpage
 \begin{figure*}[thp]
\centering
(a)
\includegraphics[clip,width=11cm]{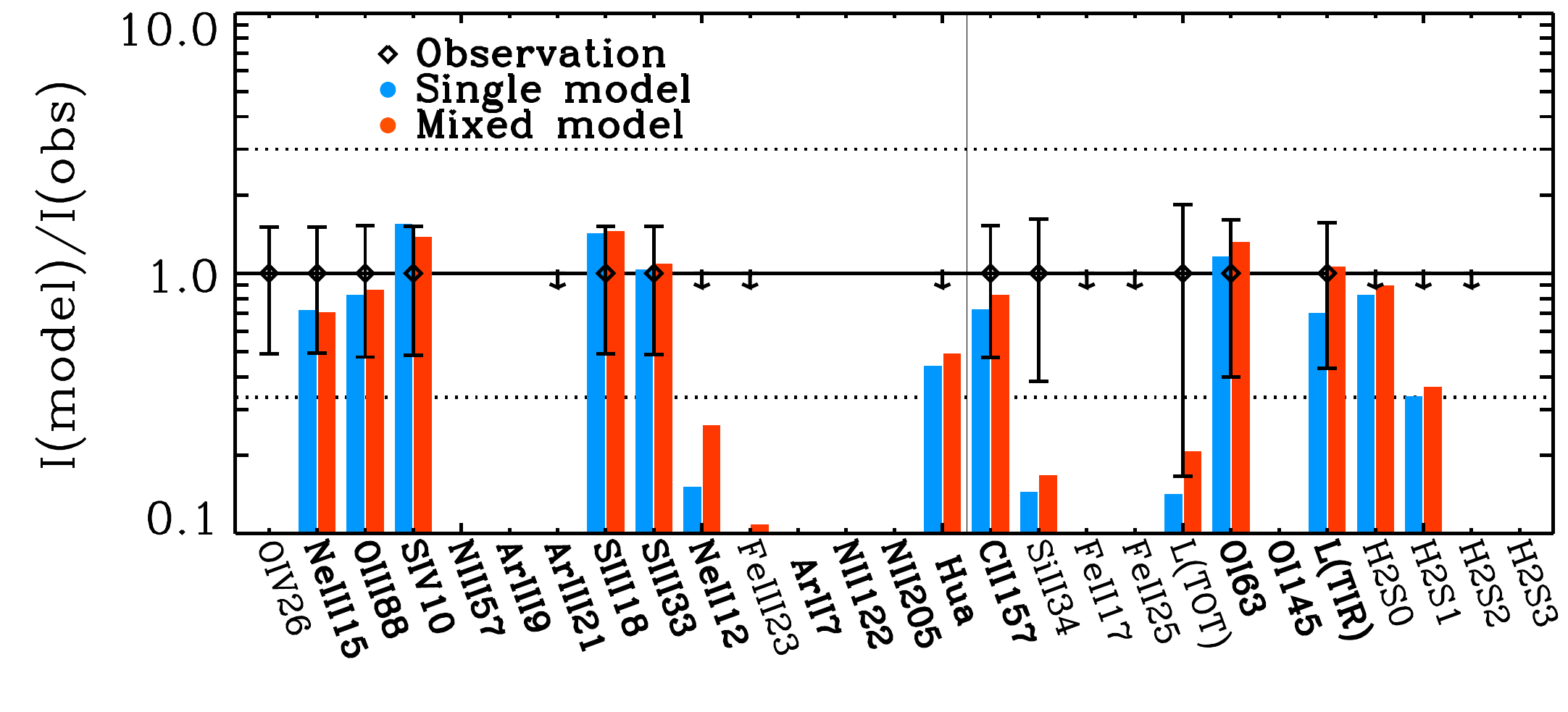} \hfill{}
(b)
\includegraphics[clip,width=6.2cm]{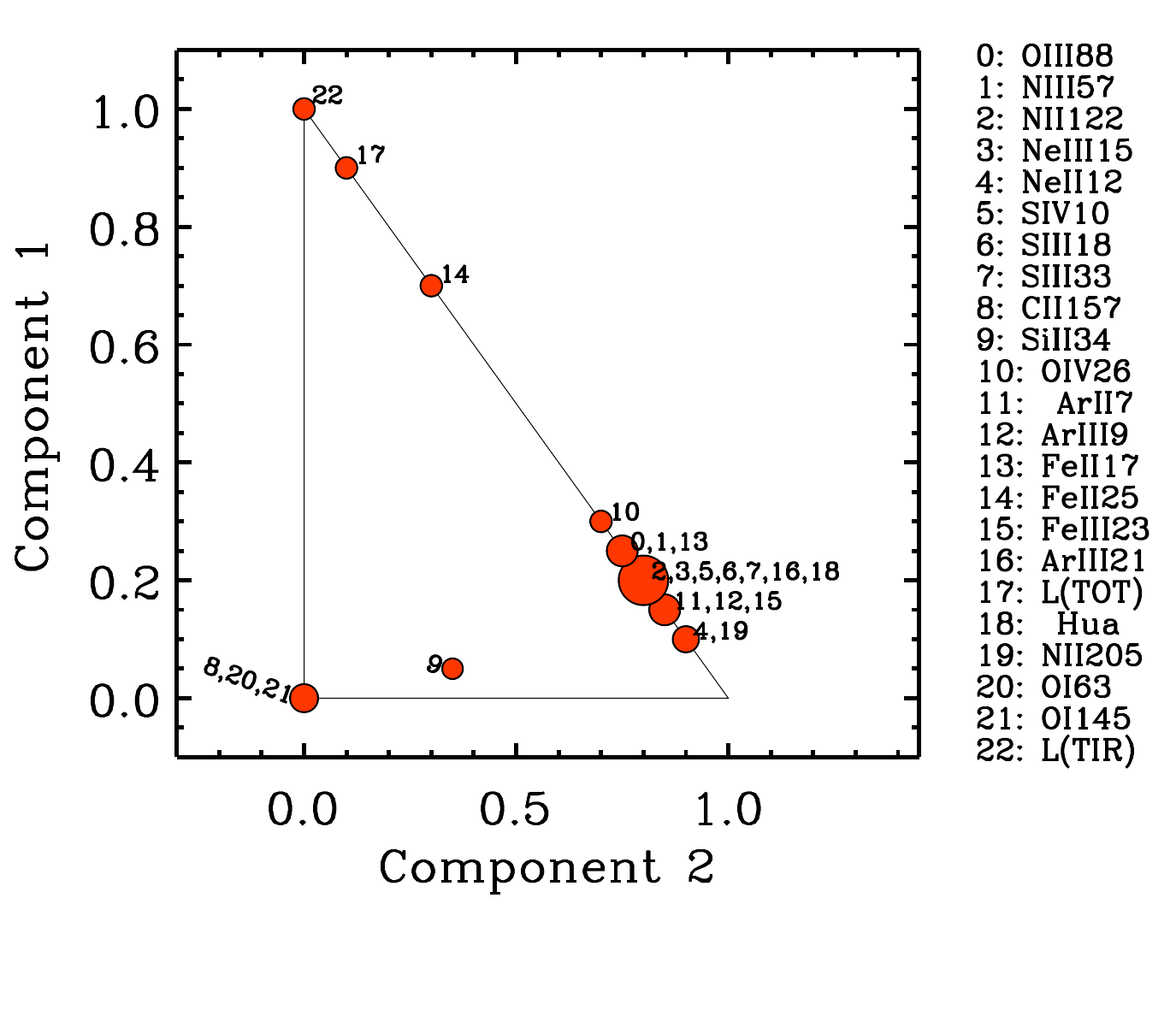}\\
(c)
\includegraphics[clip,width=8.8cm]{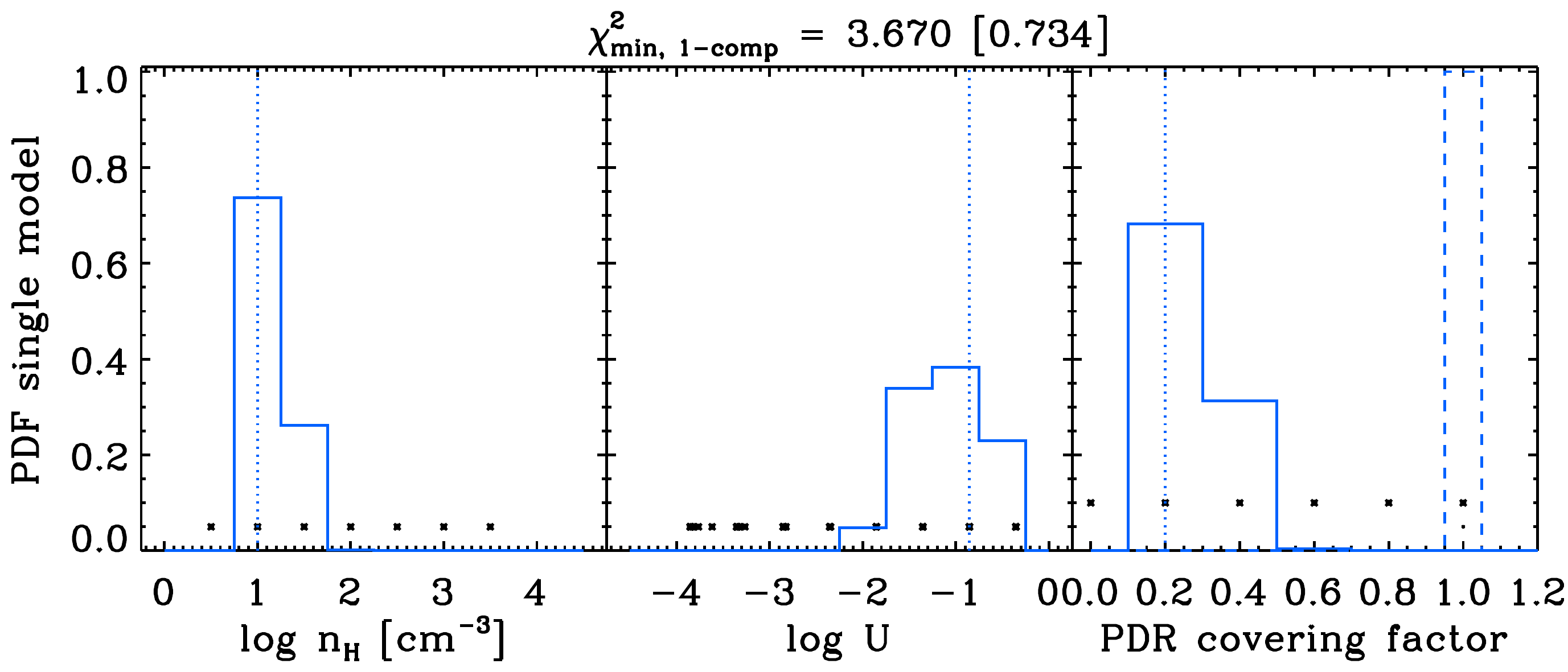} \hfill{}
\includegraphics[clip,width=8.8cm]{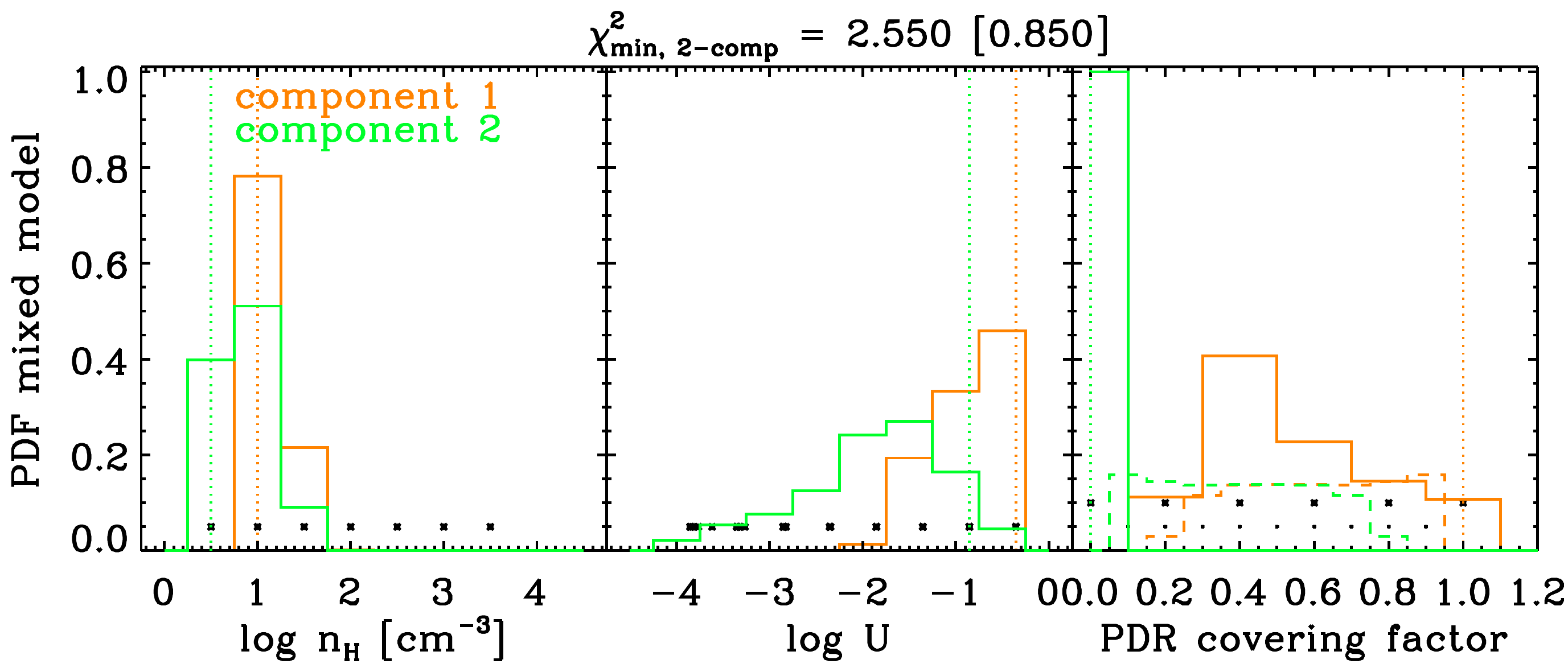}
\caption{ \textbf{Results for Mrk\,209.}
$(a)$~Comparison of the observed (losange) and predicted 
intensities for our best-fitting single (blue bar) and mixed (red bar)
\hii region+PDR models. Lines are sorted by decreasing energy.
The fitted lines have labels in boldface.
$(b)$~For mixed models: respective contributions from
component \#1 and component \#2 to the line prediction.
For PDR lines (\cii, \oi, and \silii), only the contribution from
the ionized gas is shown; the PDR contribution is
one minus the sum of the contributions from the plot.
Contributions are expected to be dominated by one
or the other component if their model parameters
are noticeably different.
$(c)$~Probability density functions of the model parameters
($n_{\rm H}$, $U$, $cov_{\rm PDR}$/scaling factor)
for single (left panel) and mixed (right panel) models.
The scaling factor corresponds to the proportion in which
we combine two models in the mixed model case. It is
equal to unity otherwise. It is shown with dashed lines in
the same panels as the PDR covering factor. Vertical dotted lines
show values of the best-fitting model. Black asterisks
show the range and step of values of the grid parameters.
}
\label{fig:bests-mrk209}
\end{figure*}
 \begin{figure*}[thp]
\centering
(a)
\includegraphics[clip,width=11cm]{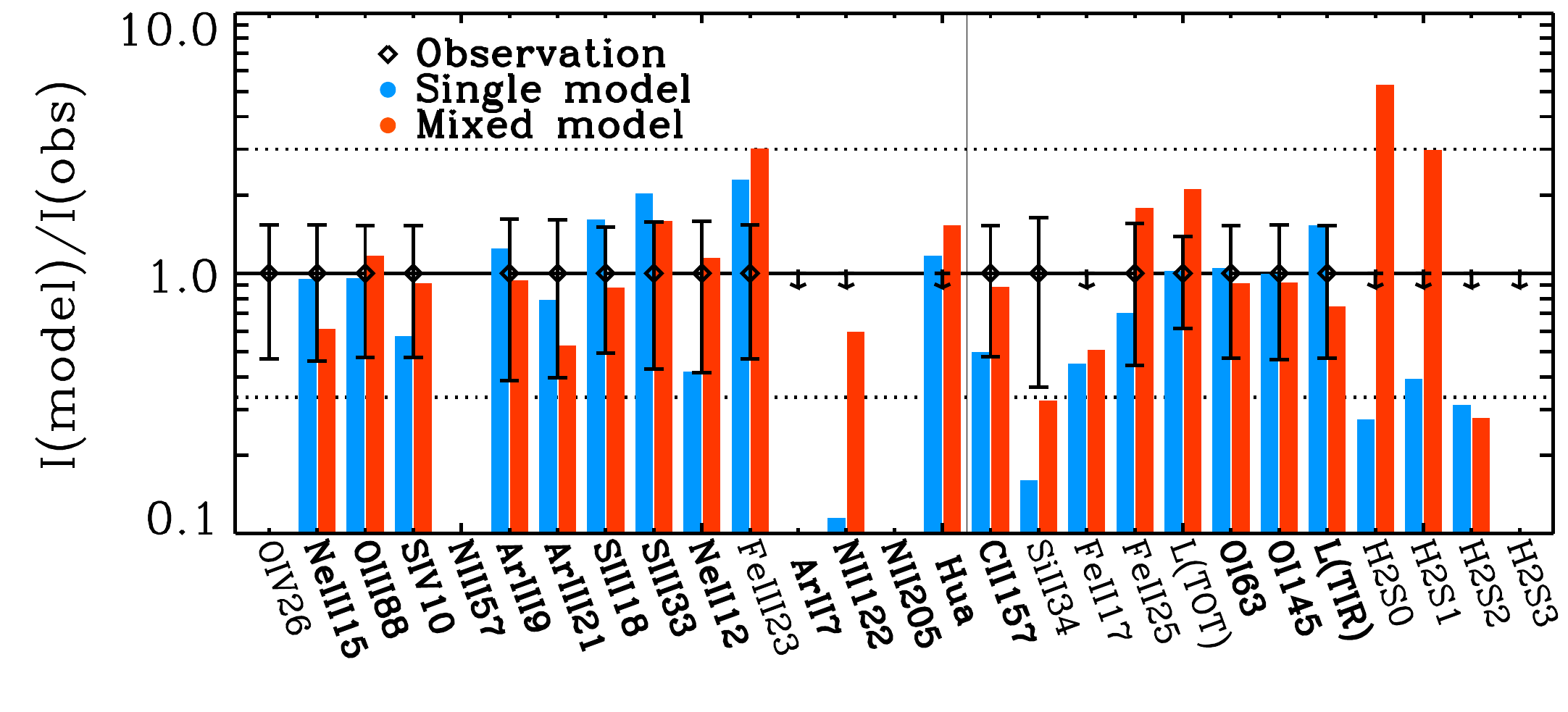} \hfill{}
(b)
\includegraphics[clip,width=6.2cm]{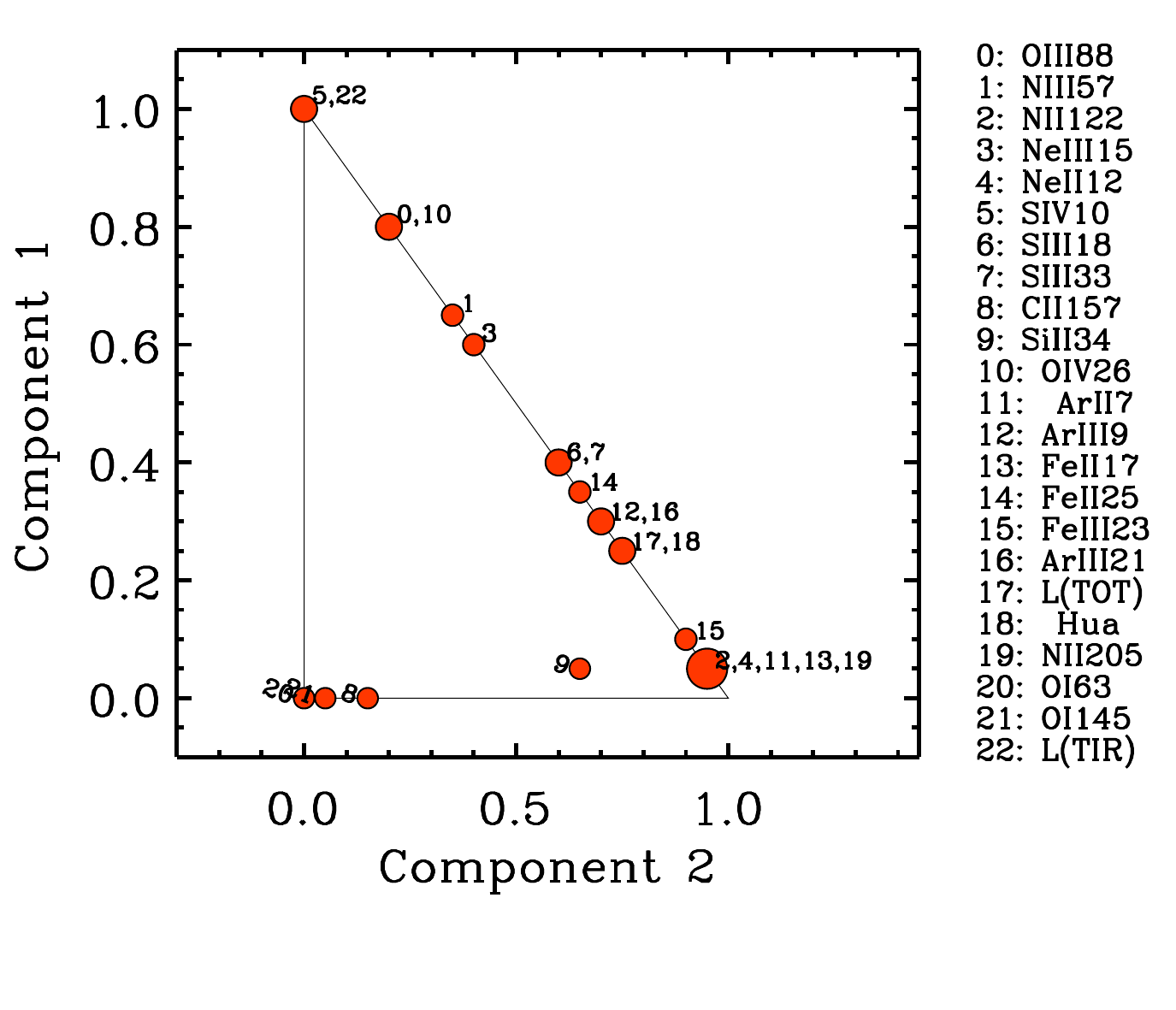}\\
(c)
\includegraphics[clip,width=8.8cm]{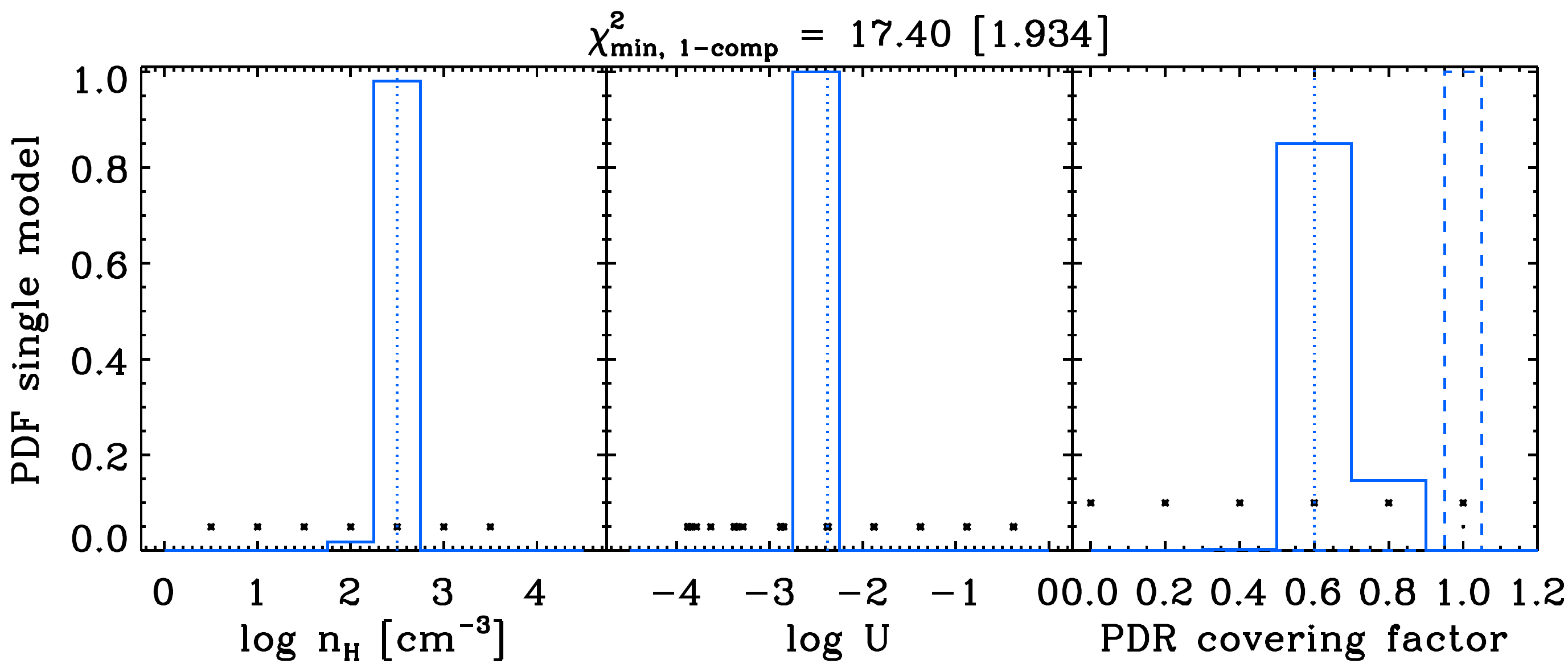} \hfill{}
\includegraphics[clip,width=8.8cm]{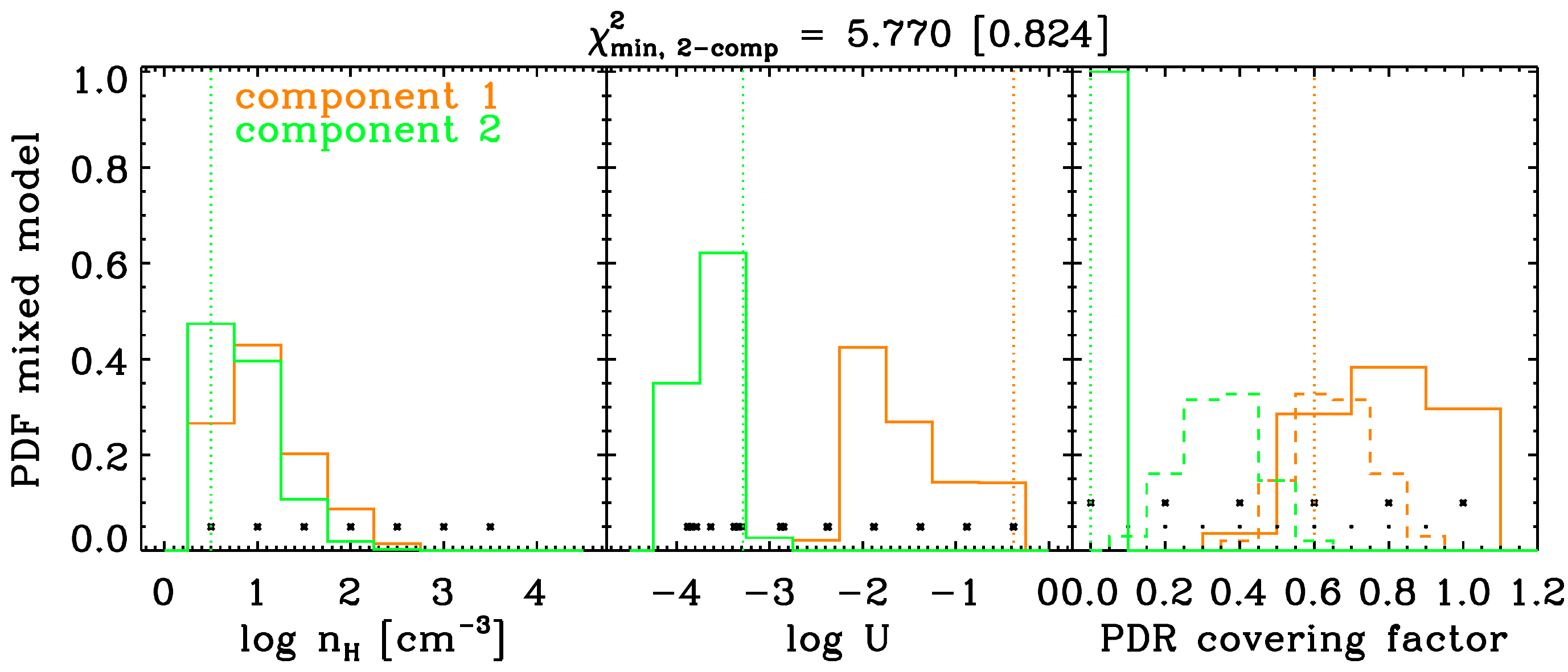}
\caption{ \textbf{Results for Mrk\,930.}
See above for caption description.
}
\label{fig:bests-mrk930}
\end{figure*}
 \begin{figure*}[thp]
\centering
(a)
\includegraphics[clip,width=11cm]{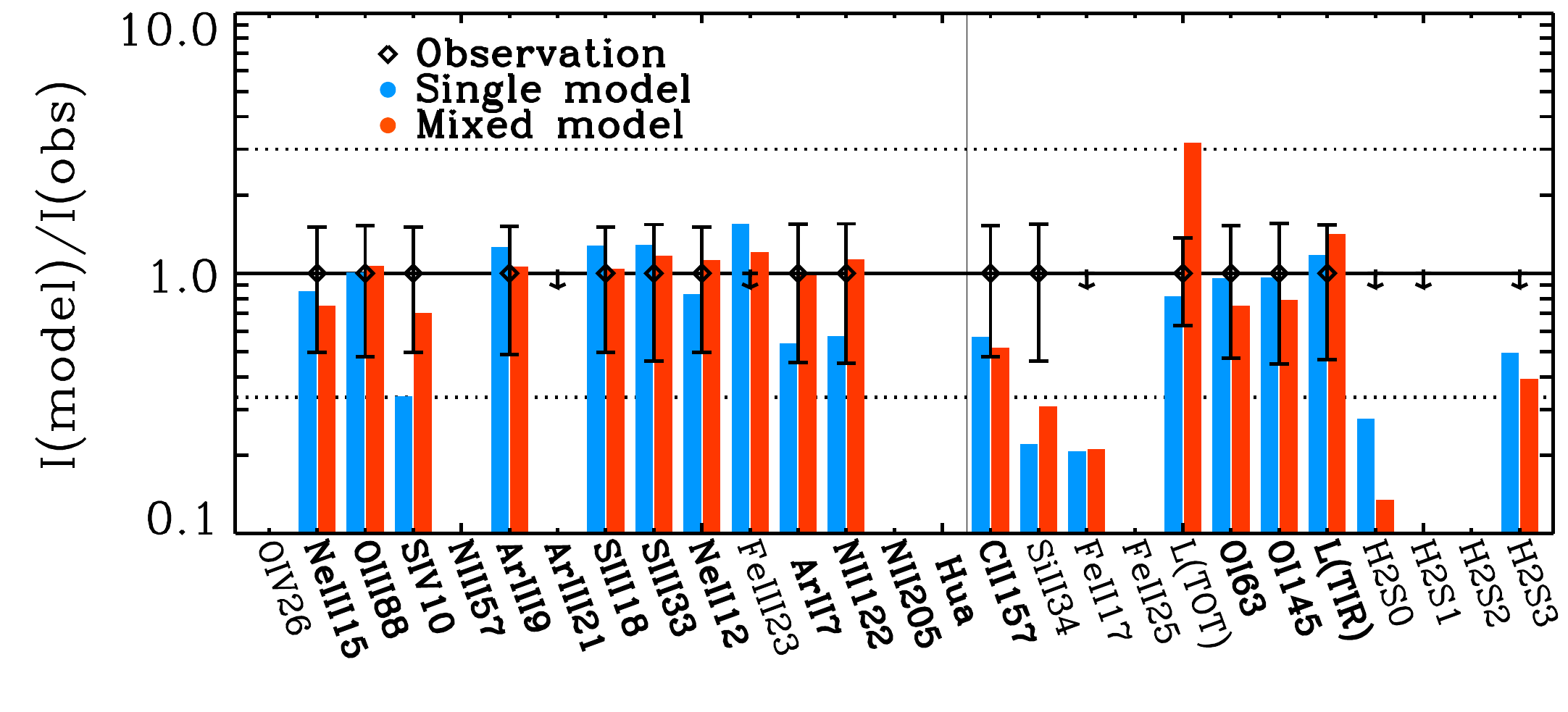} \hfill{}
(b)
\includegraphics[clip,width=6.2cm]{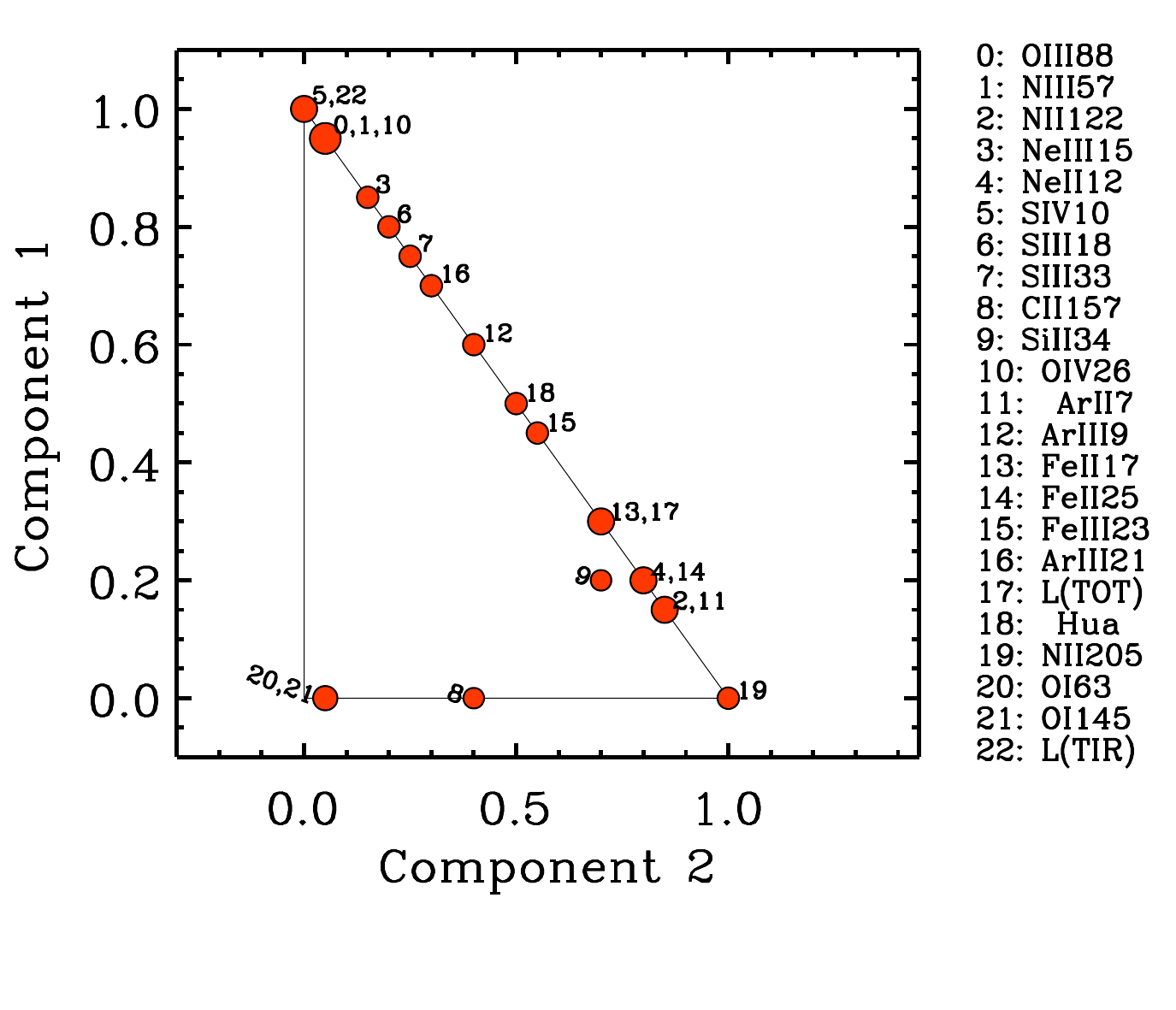}\\
(c)
\includegraphics[clip,width=8.8cm]{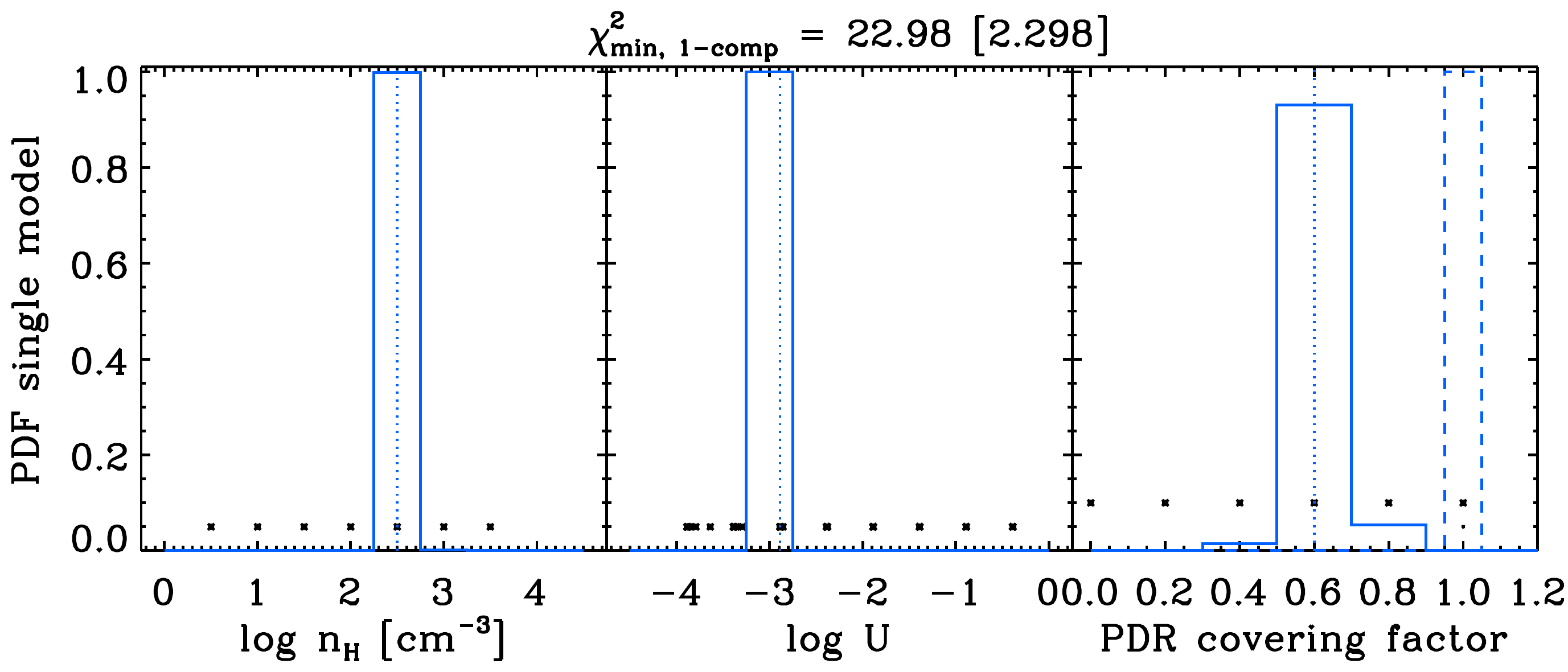} \hfill{}
\includegraphics[clip,width=8.8cm]{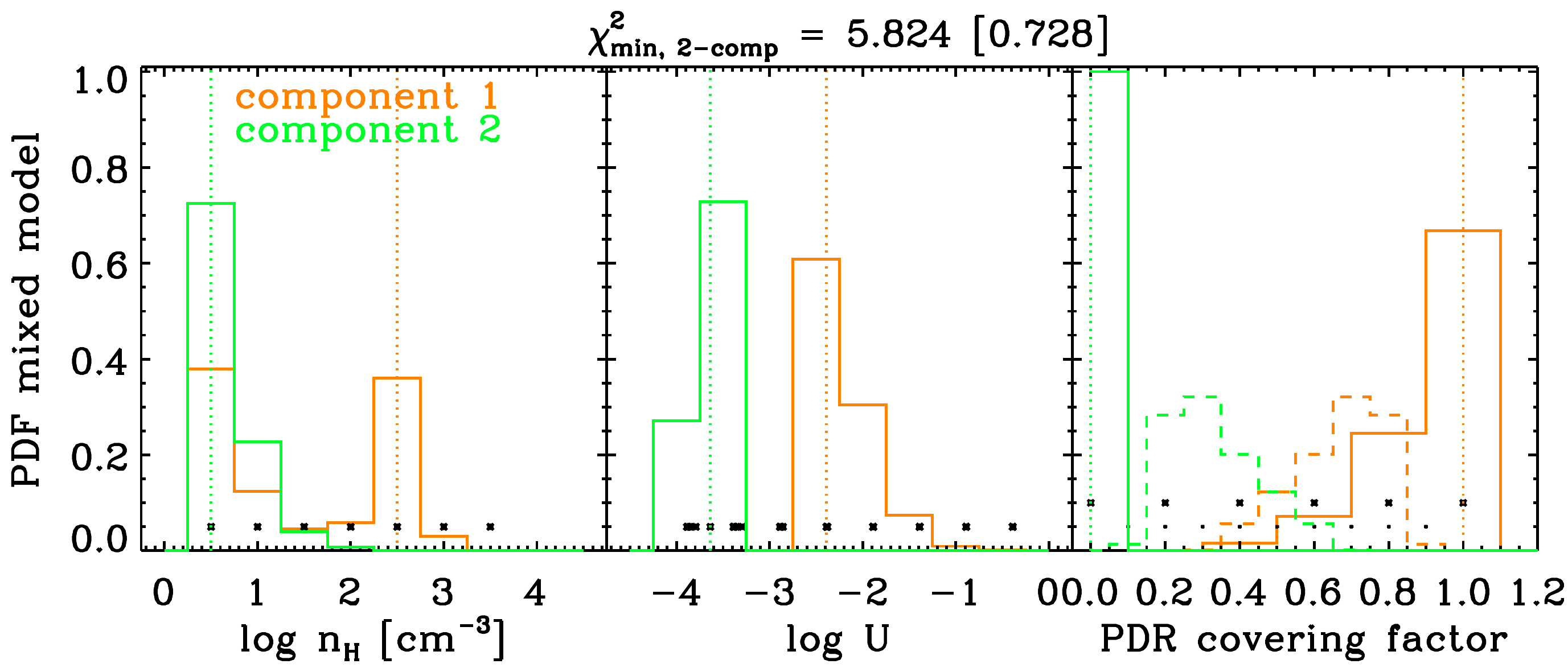}
\caption{ \textbf{Results for Mrk\,1089.}
$(a)$~Comparison of the observed (losange) and predicted 
intensities for our best-fitting single (blue bar) and mixed (red bar)
\hii region+PDR models. Lines are sorted by decreasing energy.
The fitted lines have labels in boldface.
$(b)$~For mixed models: respective contributions from
component \#1 and component \#2 to the line prediction.
For PDR lines (\cii, \oi, and \silii), only the contribution from
the ionized gas is shown; the PDR contribution is
one minus the sum of the contributions from the plot.
Contributions are expected to be dominated by one
or the other component if their model parameters
are noticeably different.
$(c)$~Probability density functions of the model parameters
($n_{\rm H}$, $U$, $cov_{\rm PDR}$/scaling factor)
for single (left panel) and mixed (right panel) models.
The scaling factor corresponds to the proportion in which
we combine two models in the mixed model case. It is
equal to unity otherwise. It is shown with dashed lines in
the same panels as the PDR covering factor. Vertical dotted lines
show values of the best-fitting model. Black asterisks
show the range and step of values of the grid parameters.
}
\label{fig:bests-mrk1089}
\end{figure*}
 \begin{figure*}[thp]
\centering
(a)
\includegraphics[clip,width=11cm]{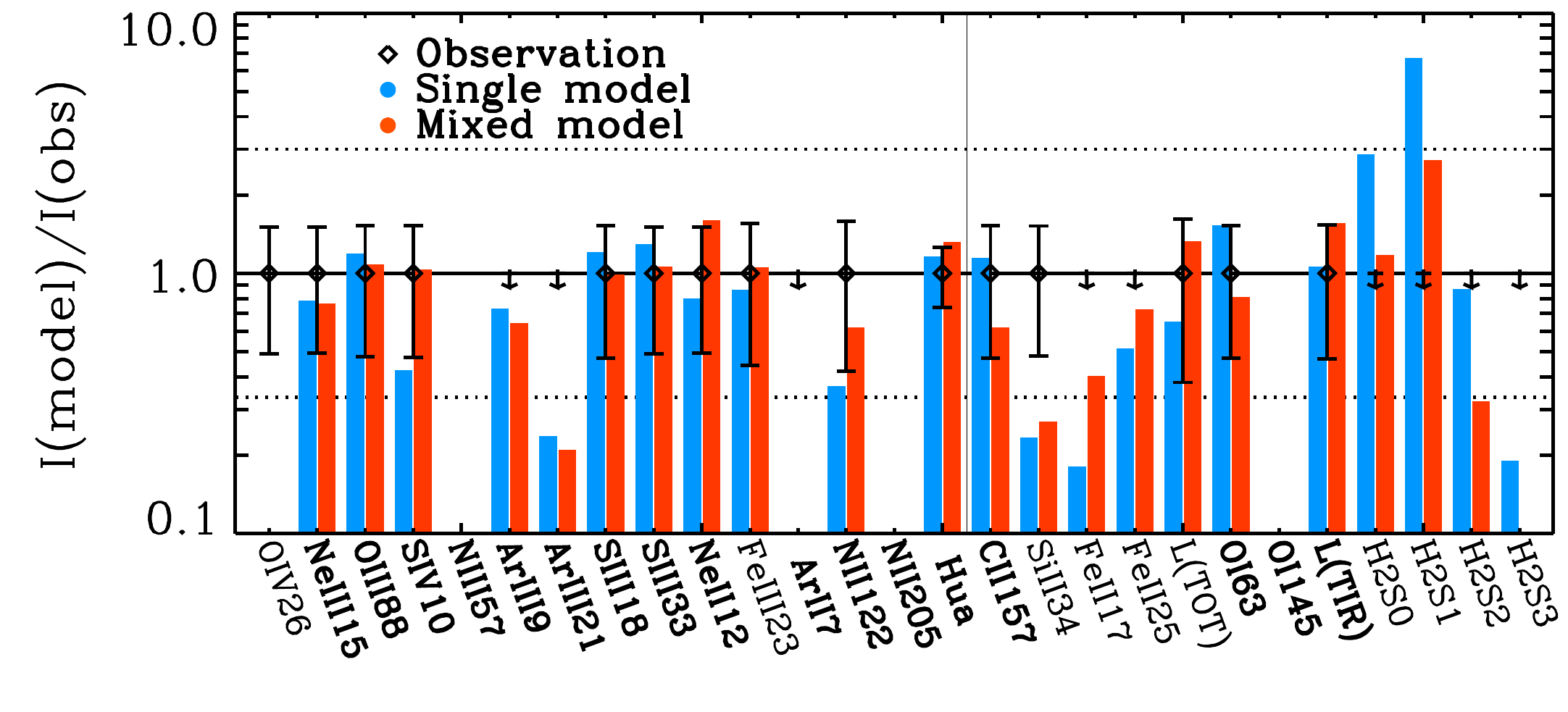} \hfill{}
(b)
\includegraphics[clip,width=6.2cm]{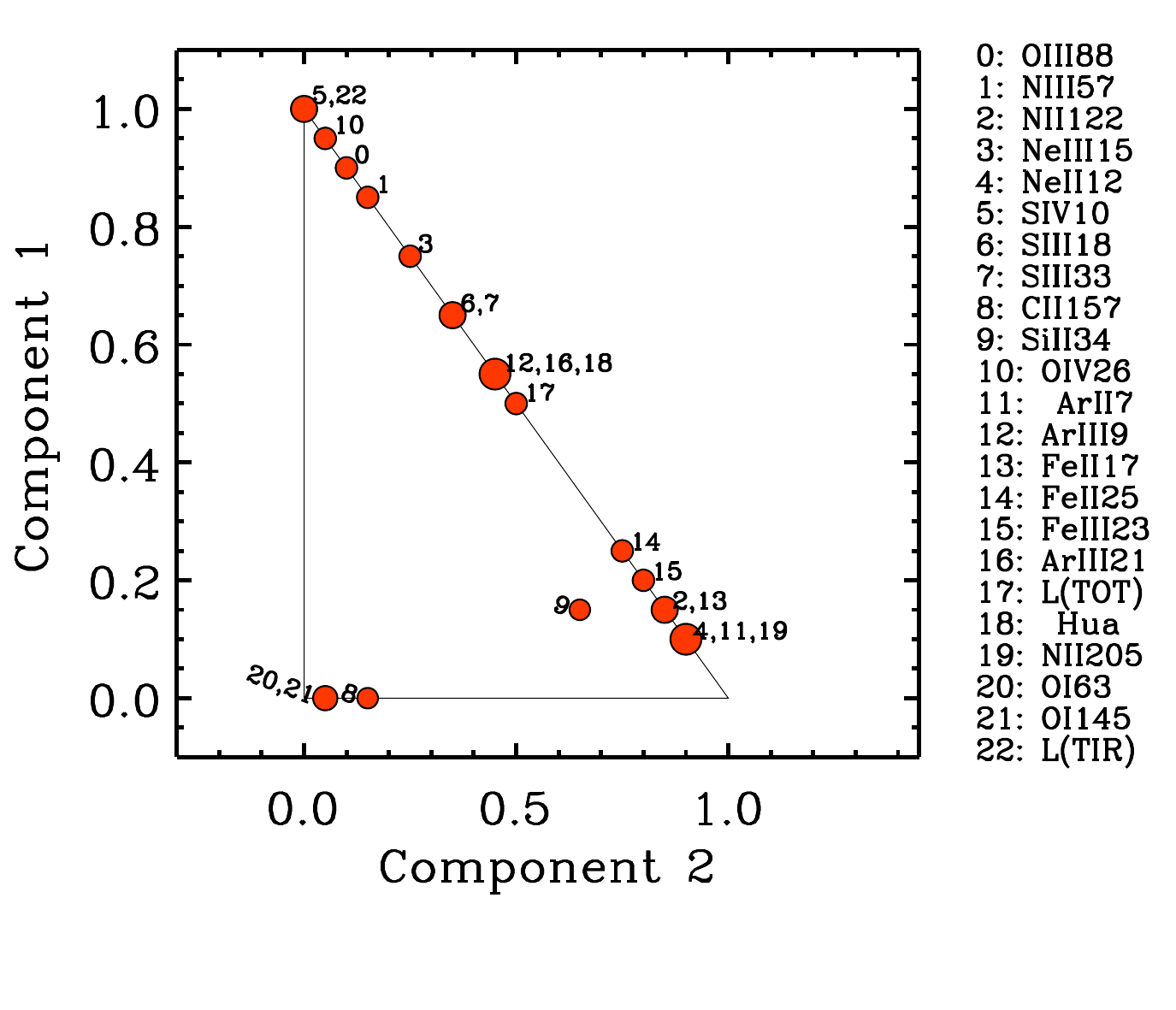}\\
(c)
\includegraphics[clip,width=8.8cm]{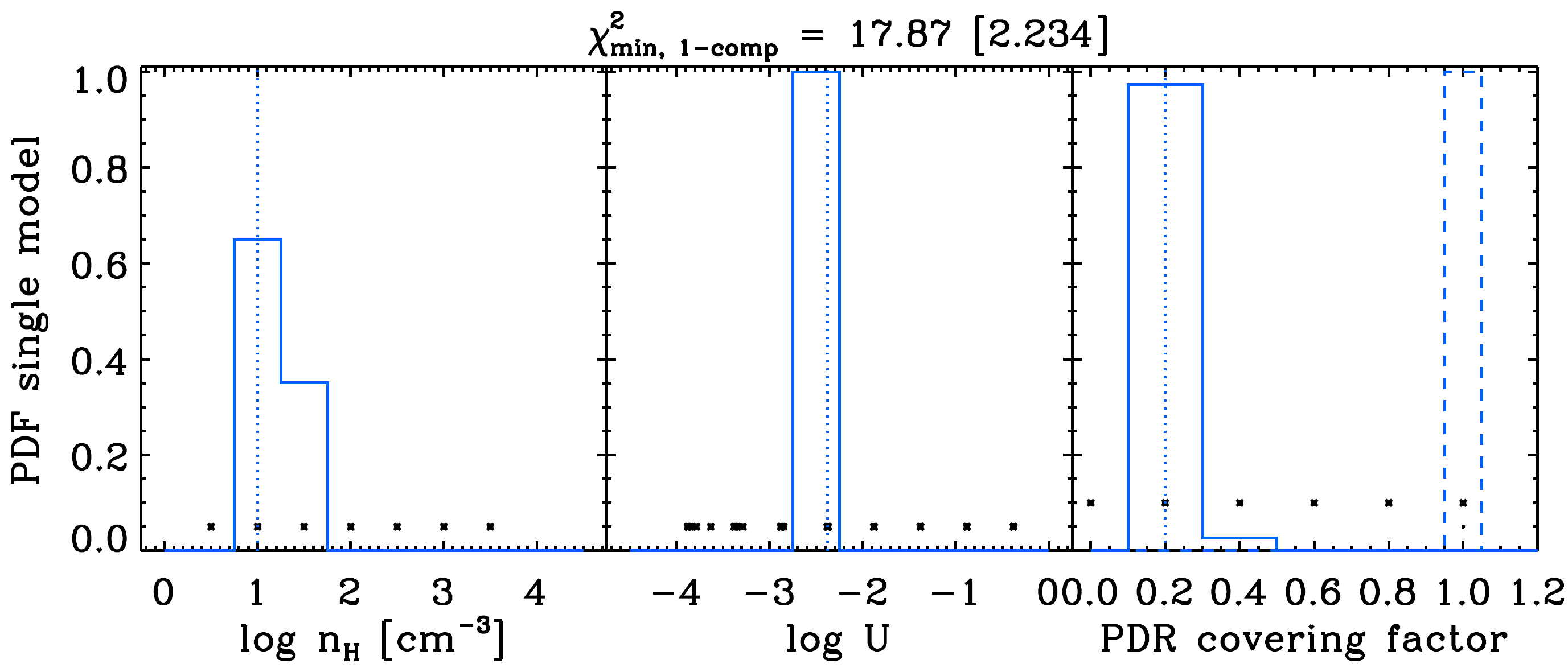} \hfill{}
\includegraphics[clip,width=8.8cm]{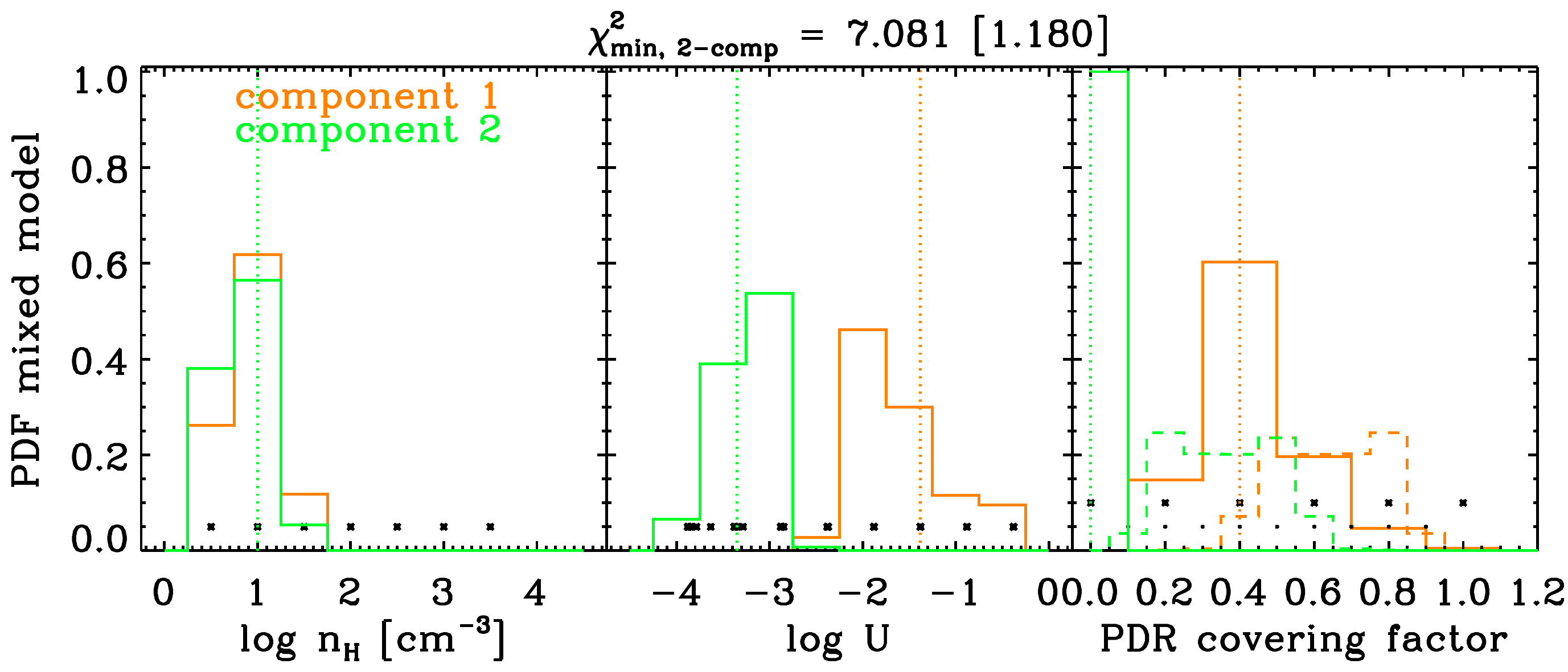}
\caption{ \textbf{Results for Mrk\,1450.}
See above for caption description.
}
\label{fig:bests-mrk1450}
\end{figure*}
\clearpage
 \begin{figure*}[thp]
\centering
(a)
\includegraphics[clip,width=11cm]{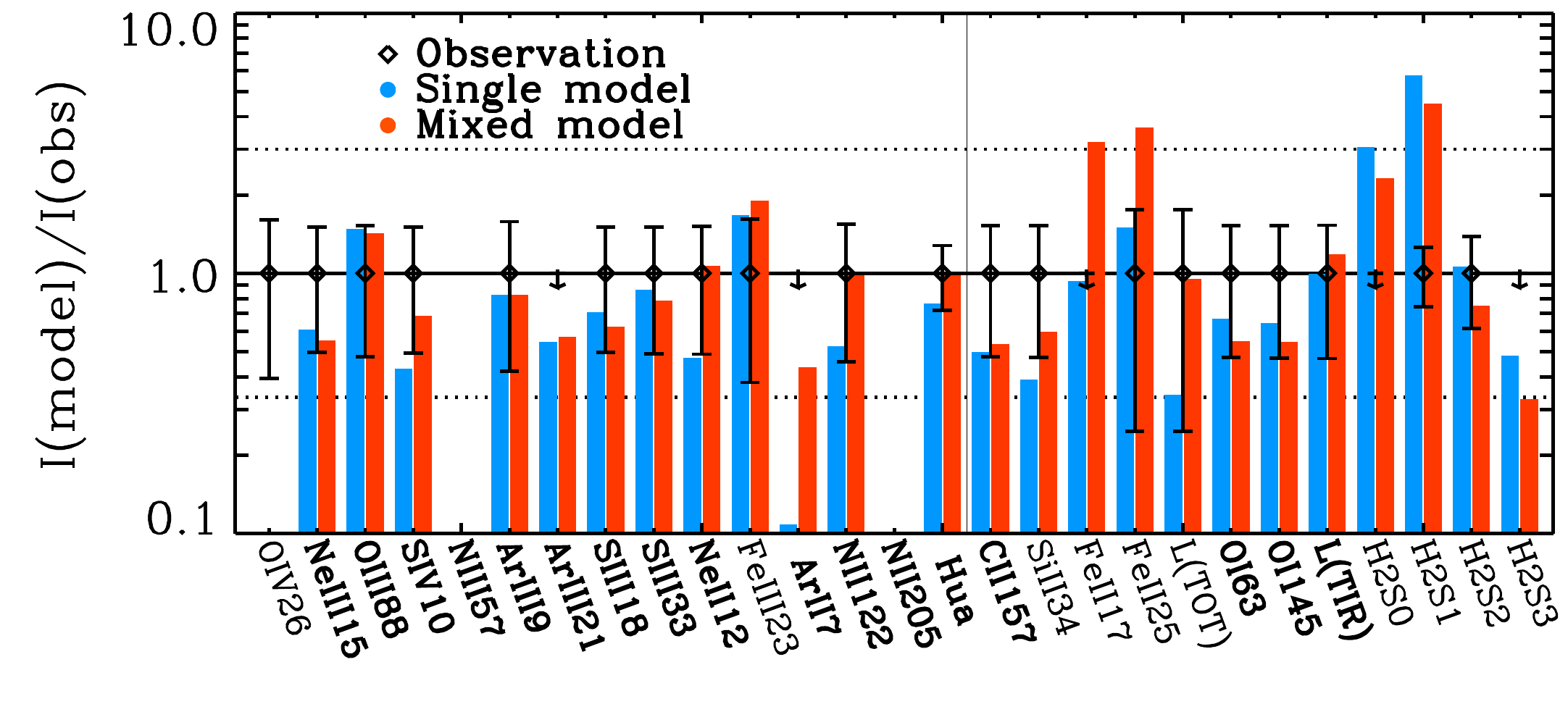} \hfill{}
(b)
\includegraphics[clip,width=6.2cm]{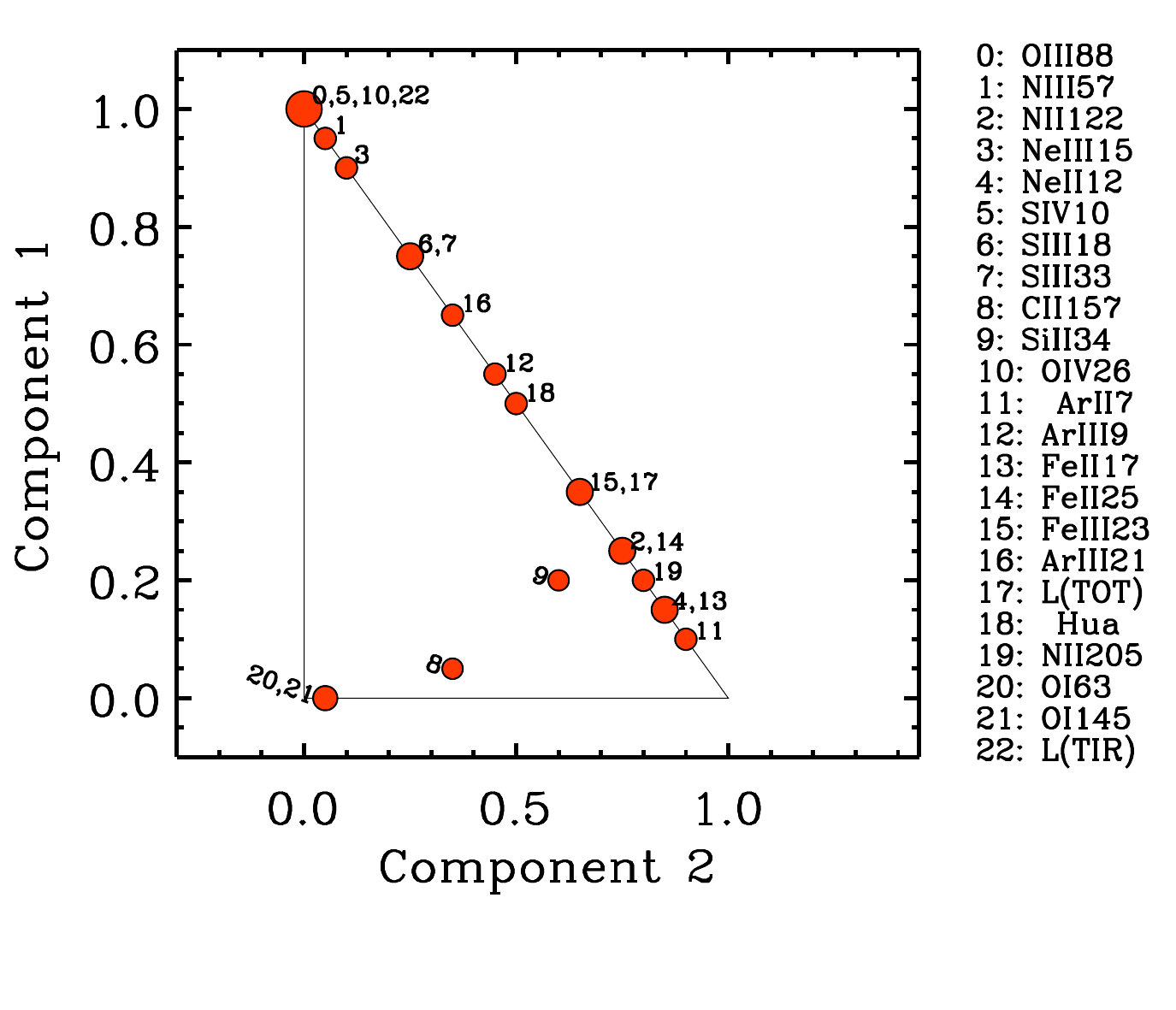}\\
(c)
\includegraphics[clip,width=8.8cm]{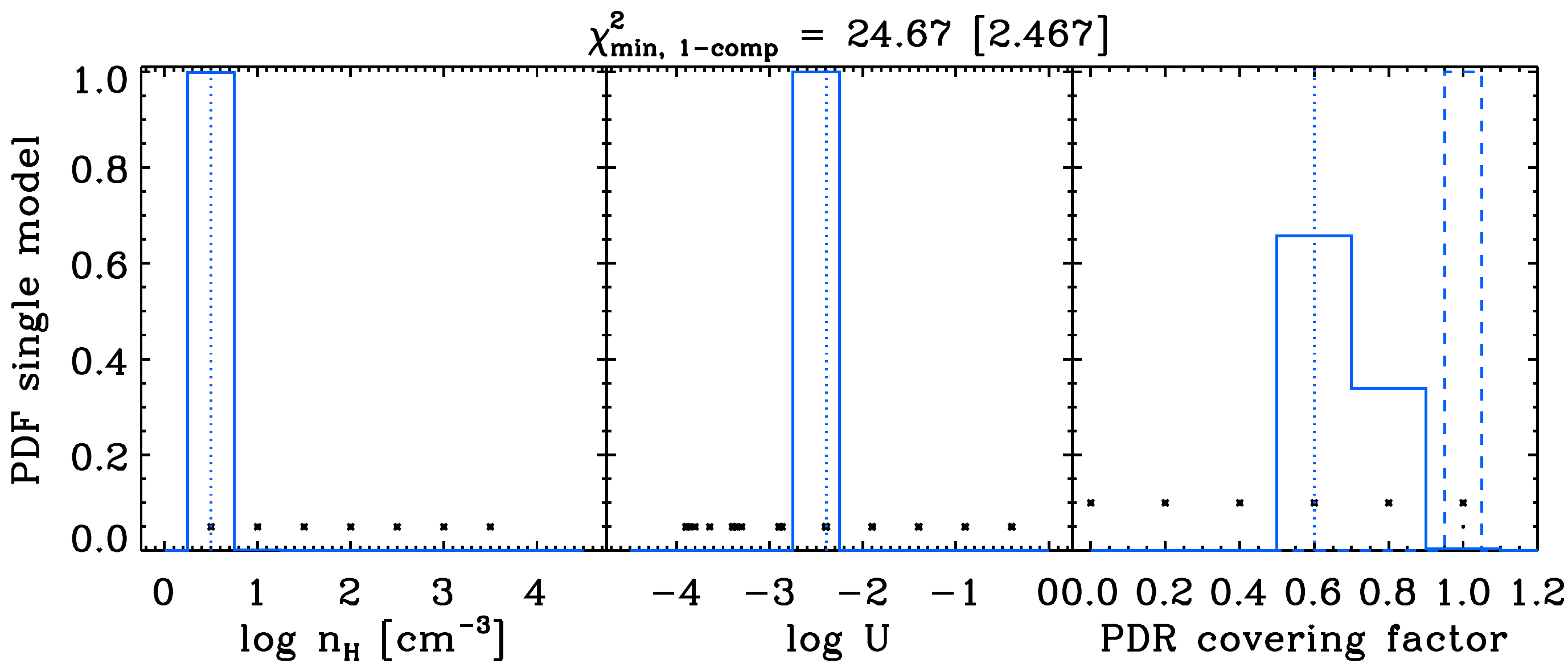} \hfill{}
\includegraphics[clip,width=8.8cm]{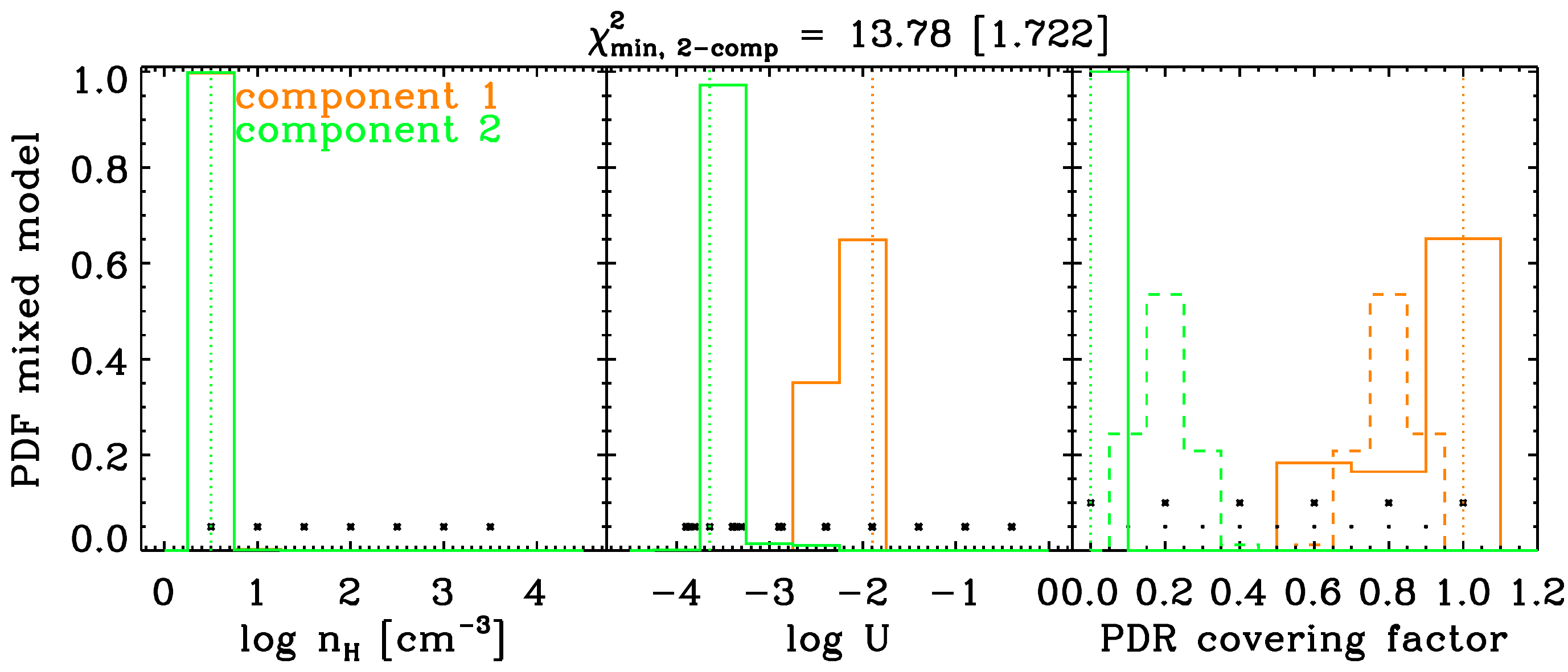}
\caption{ \textbf{Results for NGC\,1140.}
$(a)$~Comparison of the observed (losange) and predicted 
intensities for our best-fitting single (blue bar) and mixed (red bar)
\hii region+PDR models. Lines are sorted by decreasing energy.
The fitted lines have labels in boldface.
$(b)$~For mixed models: respective contributions from
component \#1 and component \#2 to the line prediction.
For PDR lines (\cii, \oi, and \silii), only the contribution from
the ionized gas is shown; the PDR contribution is
one minus the sum of the contributions from the plot.
Contributions are expected to be dominated by one
or the other component if their model parameters
are noticeably different.
$(c)$~Probability density functions of the model parameters
($n_{\rm H}$, $U$, $cov_{\rm PDR}$/scaling factor)
for single (left panel) and mixed (right panel) models.
The scaling factor corresponds to the proportion in which
we combine two models in the mixed model case. It is
equal to unity otherwise. It is shown with dashed lines in
the same panels as the PDR covering factor. Vertical dotted lines
show values of the best-fitting model. Black asterisks
show the range and step of values of the grid parameters.
}
\label{fig:bests-ngc1140}
\end{figure*}
 \begin{figure*}[thp]
\centering
(a)
\includegraphics[clip,width=11cm]{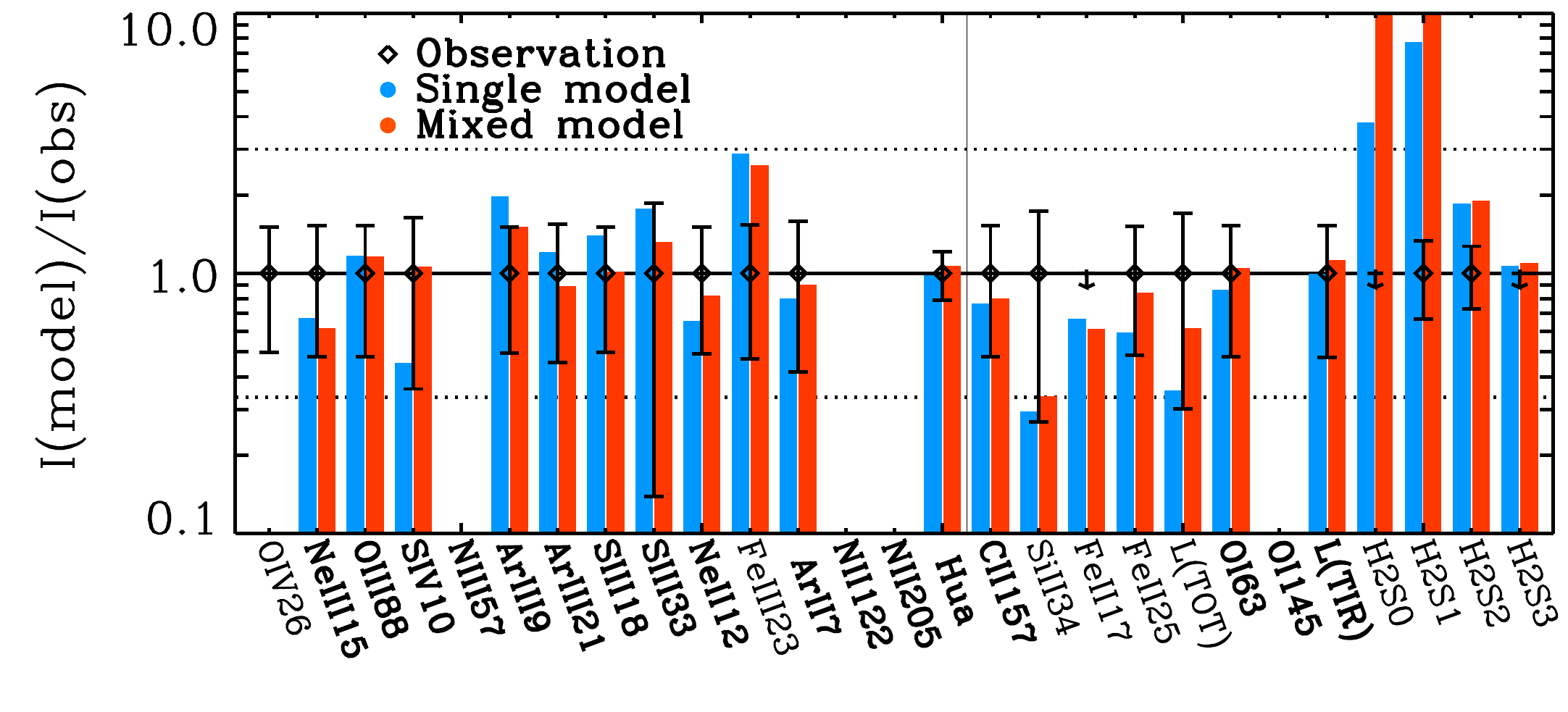} \hfill{}
(b)
\includegraphics[clip,width=6.2cm]{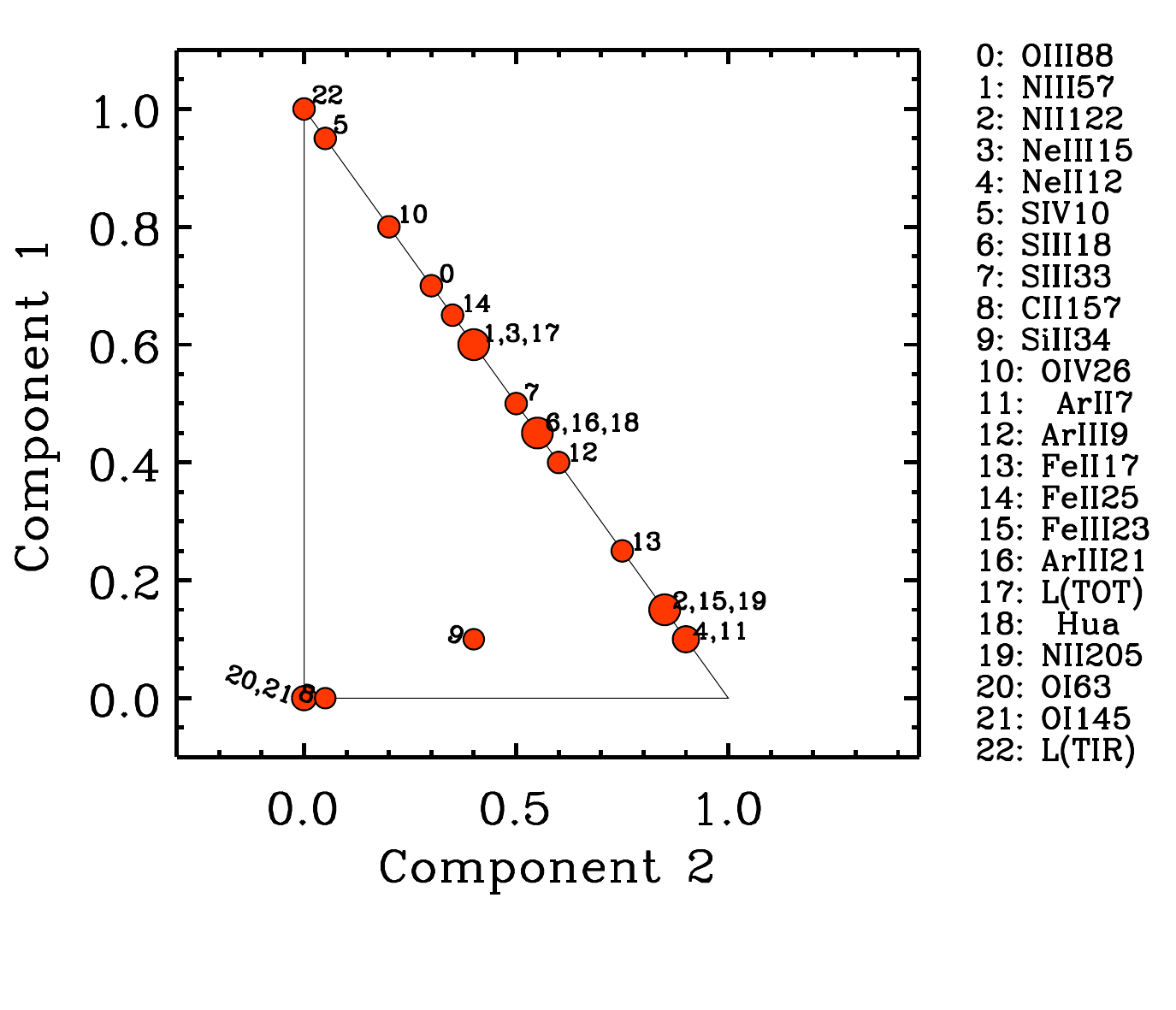}\\
(c)
\includegraphics[clip,width=8.8cm]{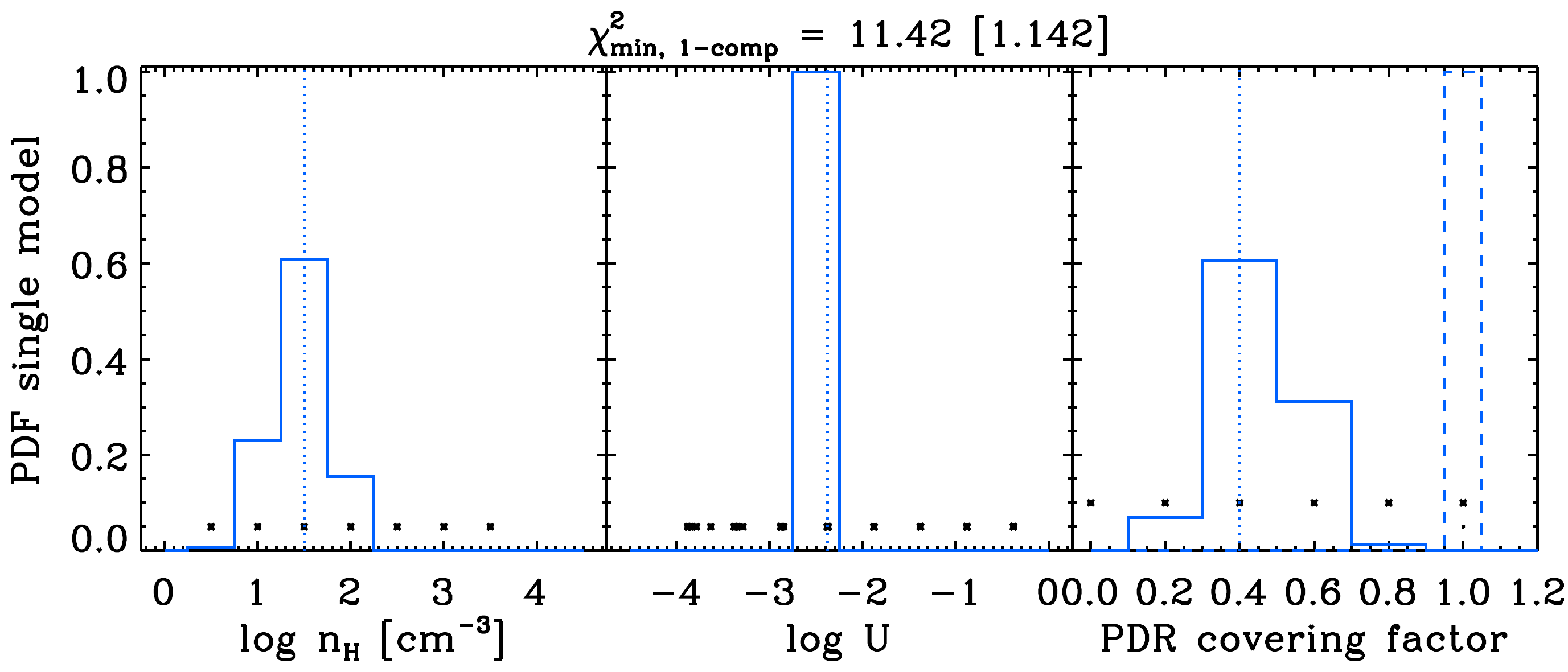} \hfill{}
\includegraphics[clip,width=8.8cm]{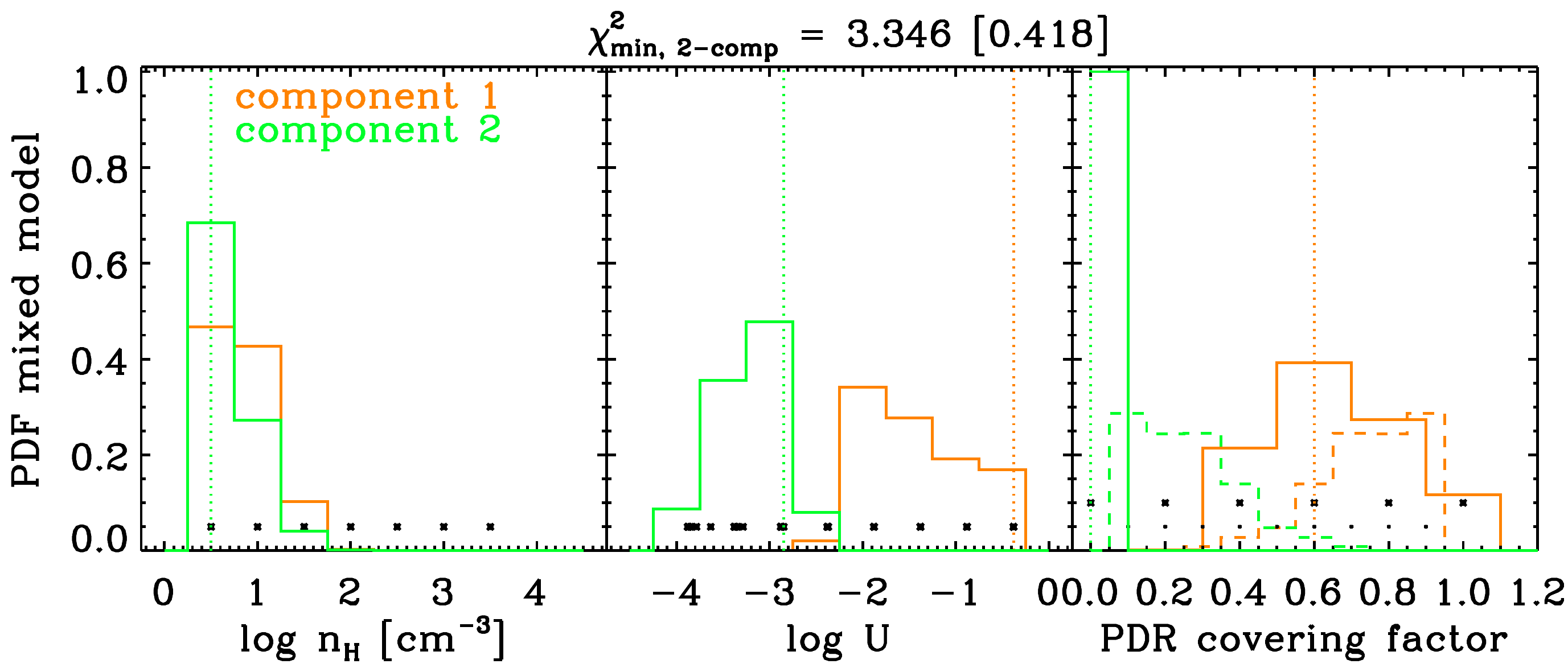}
\caption{ \textbf{Results for NGC\,1569.}
See above for caption description.
}
\label{fig:bests-ngc1569}
\end{figure*}
\clearpage
 \begin{figure*}[thp]
\centering
(a)
\includegraphics[clip,width=11cm]{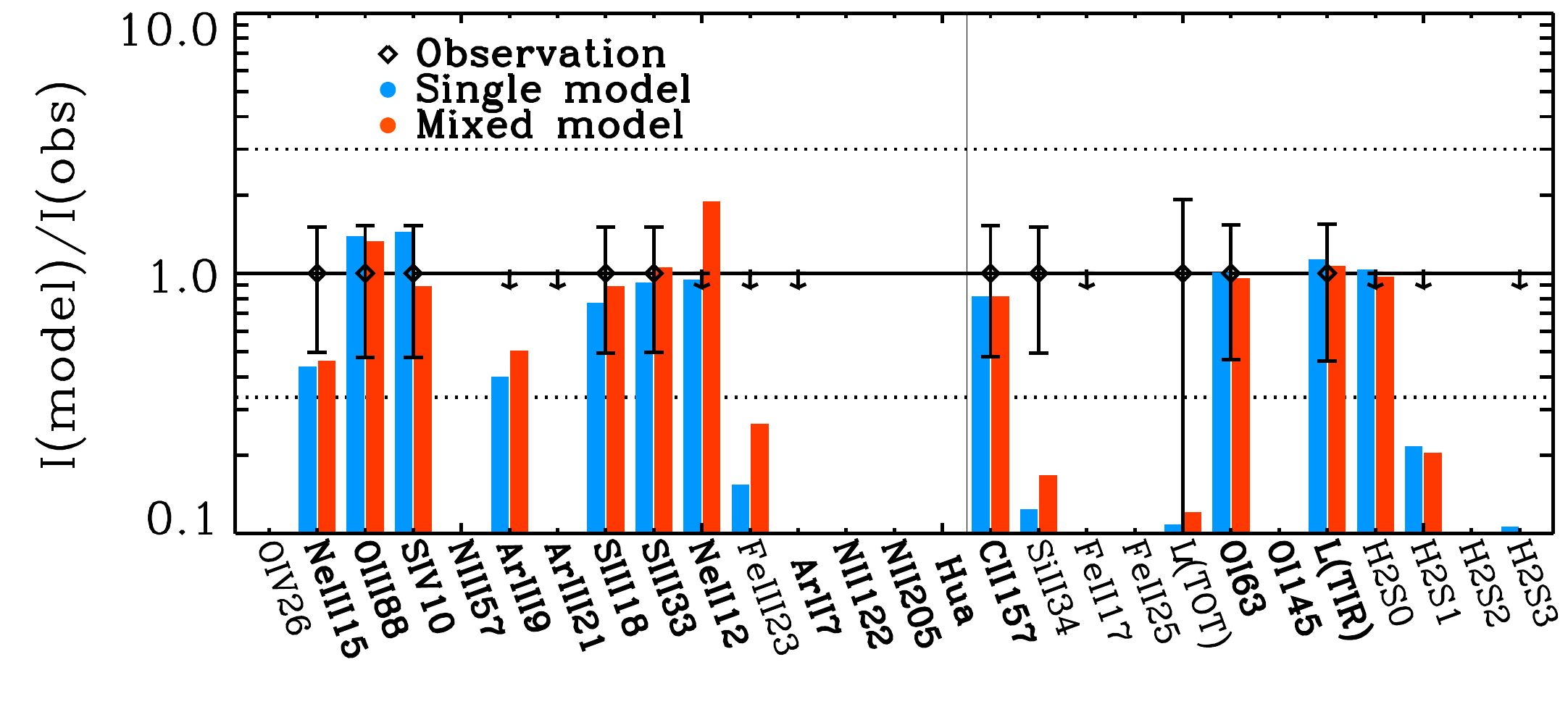} \hfill{}
(b)
\includegraphics[clip,width=6.2cm]{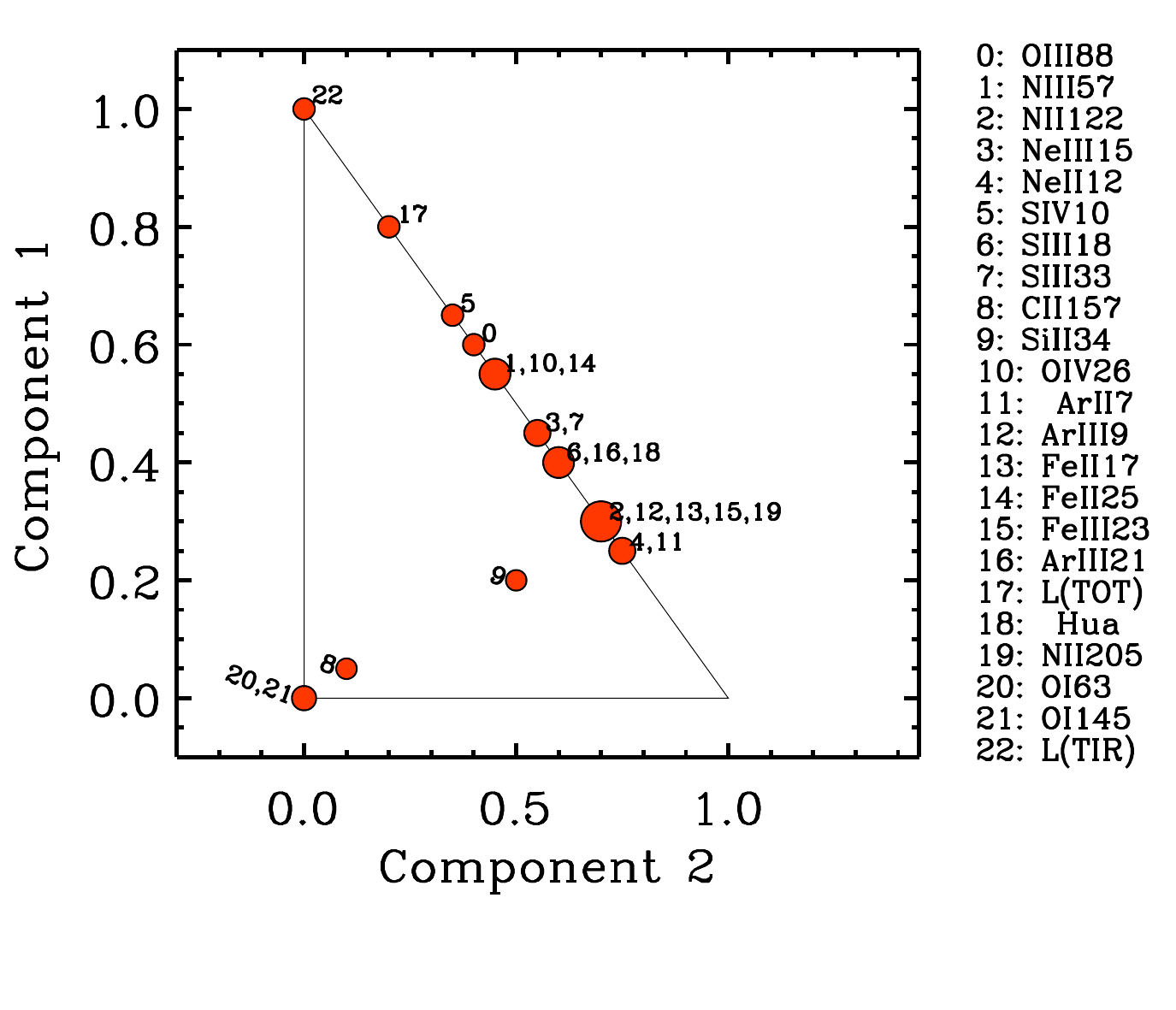}\\
(c)
\includegraphics[clip,width=8.8cm]{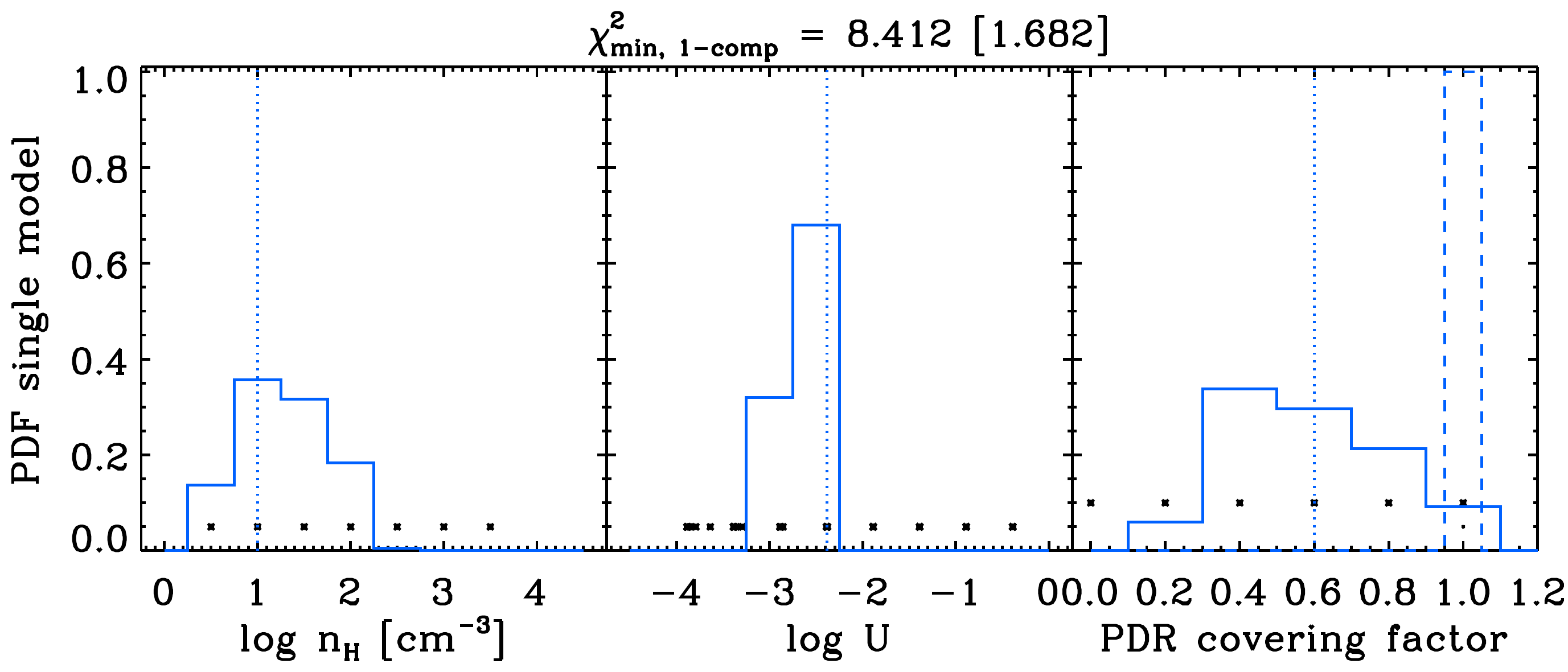} \hfill{}
\includegraphics[clip,width=8.8cm]{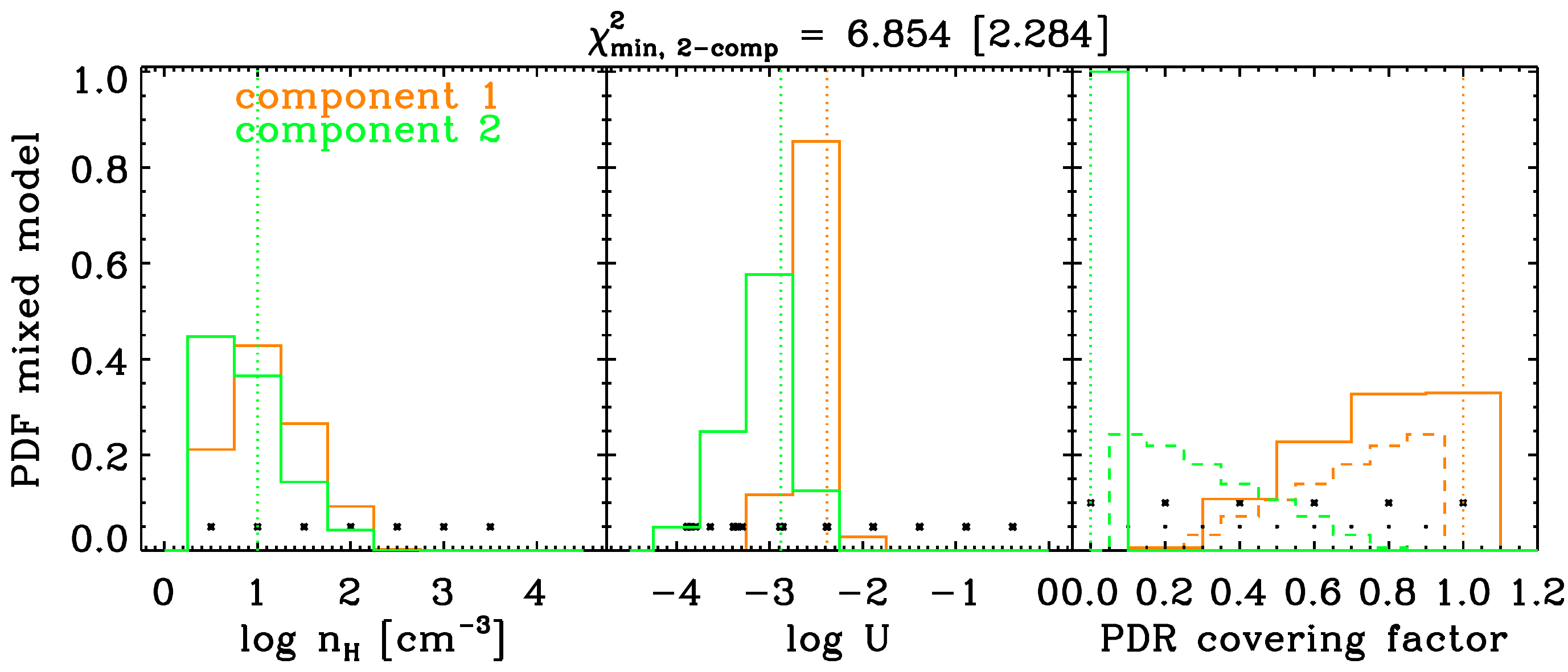}
\caption{ \textbf{Results for NGC\,1705.}
$(a)$~Comparison of the observed (losange) and predicted 
intensities for our best-fitting single (blue bar) and mixed (red bar)
\hii region+PDR models. Lines are sorted by decreasing energy.
The fitted lines have labels in boldface.
$(b)$~For mixed models: respective contributions from
component \#1 and component \#2 to the line prediction.
For PDR lines (\cii, \oi, and \silii), only the contribution from
the ionized gas is shown; the PDR contribution is
one minus the sum of the contributions from the plot.
Contributions are expected to be dominated by one
or the other component if their model parameters
are noticeably different.
$(c)$~Probability density functions of the model parameters
($n_{\rm H}$, $U$, $cov_{\rm PDR}$/scaling factor)
for single (left panel) and mixed (right panel) models.
The scaling factor corresponds to the proportion in which
we combine two models in the mixed model case. It is
equal to unity otherwise. It is shown with dashed lines in
the same panels as the PDR covering factor. Vertical dotted lines
show values of the best-fitting model. Black asterisks
show the range and step of values of the grid parameters.
}
\label{fig:bests-ngc1705}
\end{figure*}
 \begin{figure*}[thp]
\centering
(a)
\includegraphics[clip,width=11cm]{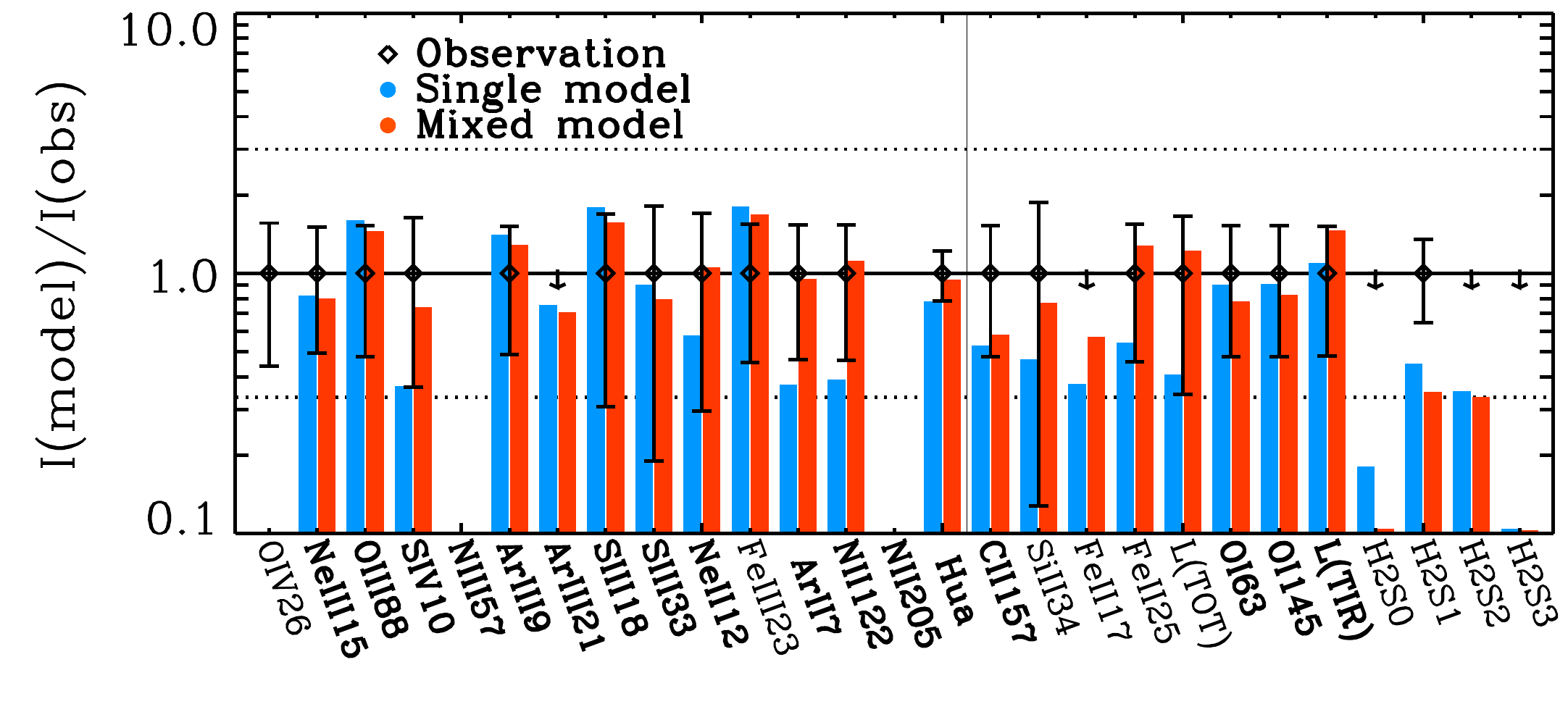} \hfill{}
(b)
\includegraphics[clip,width=6.2cm]{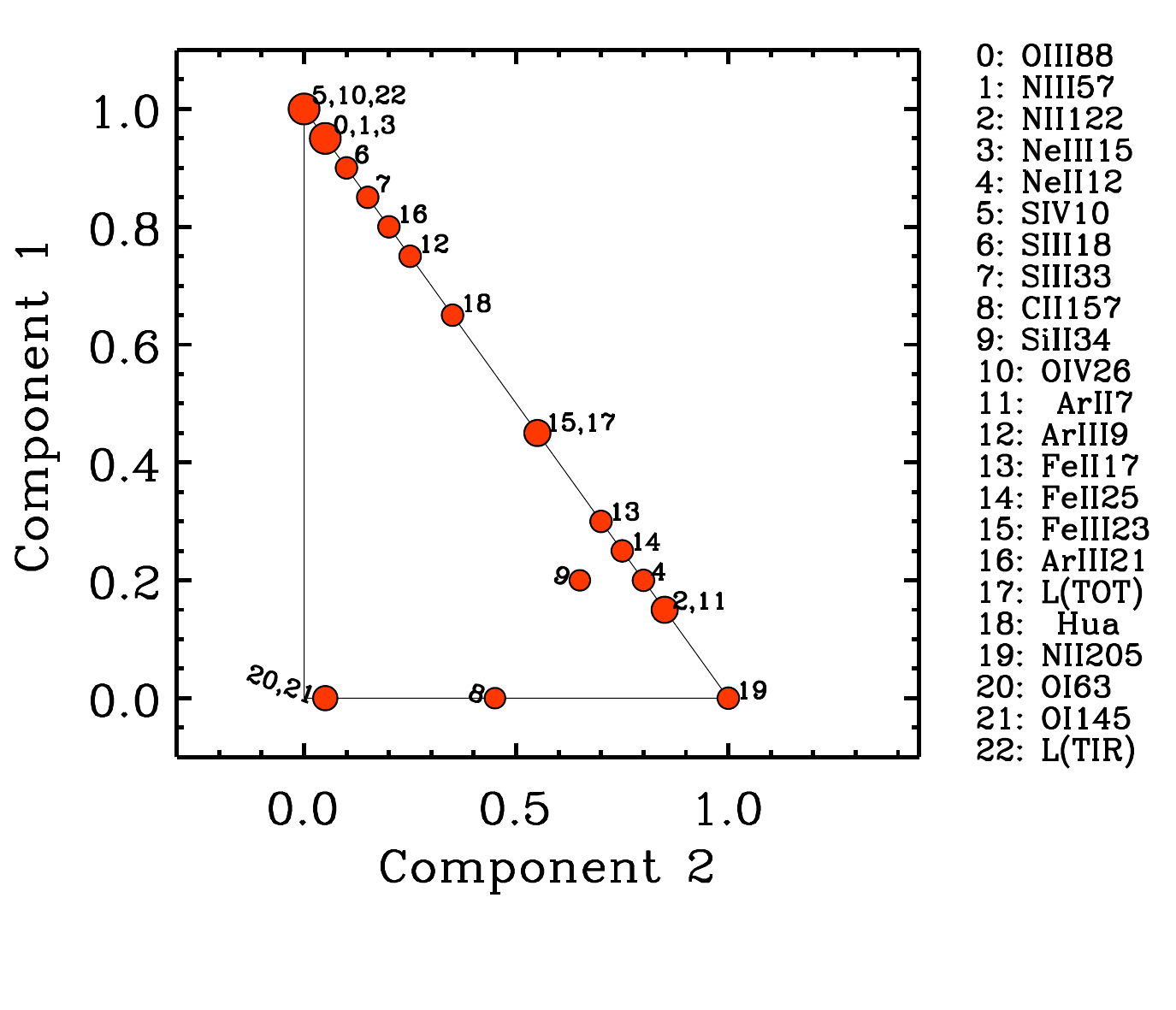}\\
(c)
\includegraphics[clip,width=8.8cm]{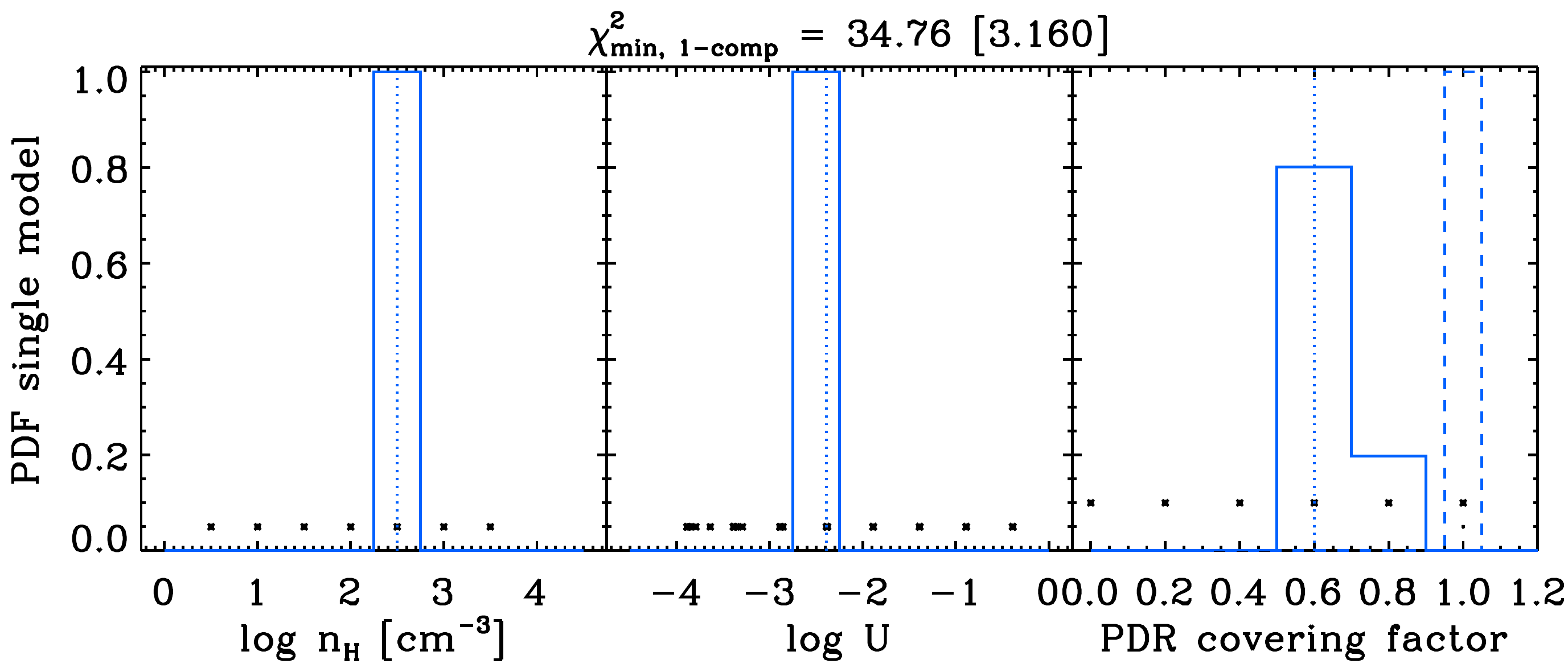} \hfill{}
\includegraphics[clip,width=8.8cm]{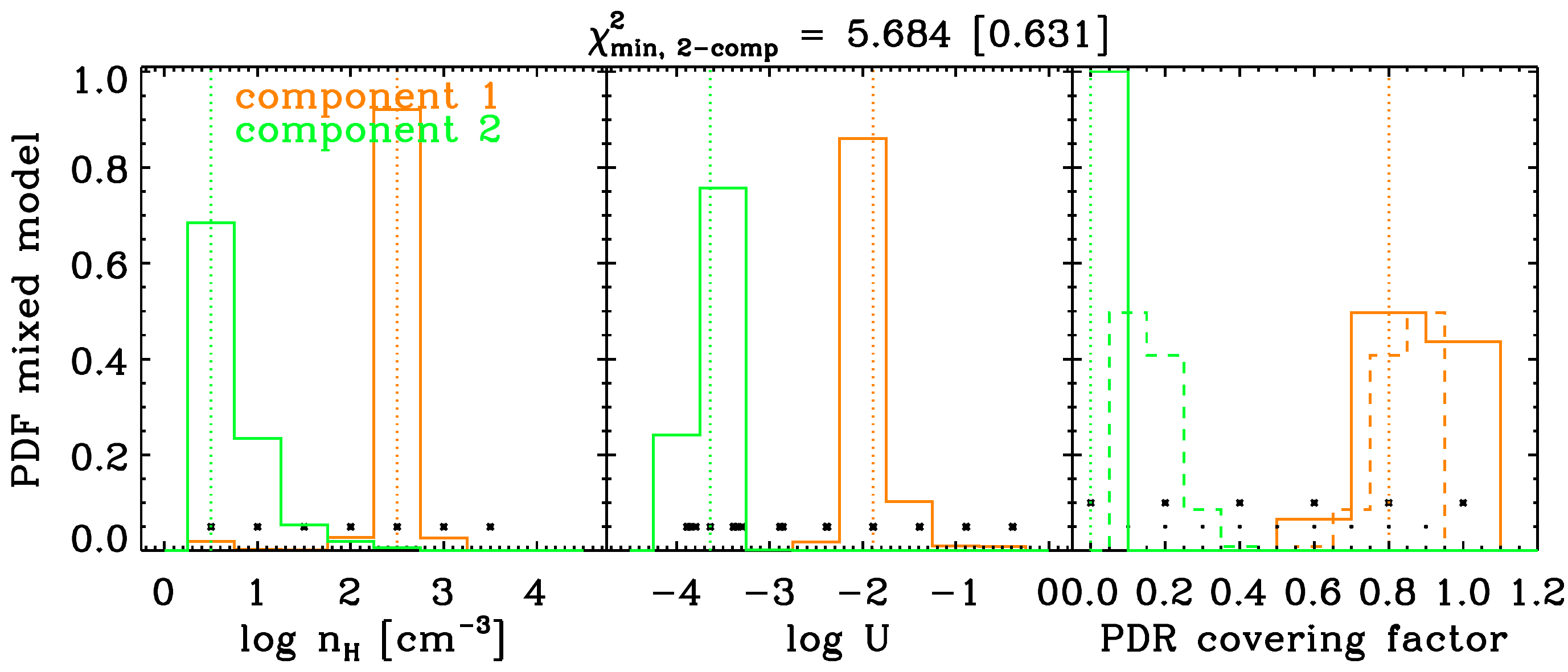}
\caption{ \textbf{Results for NGC\,5253.}
See above for caption description.
}
\label{fig:bests-ngc5253}
\end{figure*}
\clearpage
 \begin{figure*}[thp]
\centering
(a)
\includegraphics[clip,width=11cm]{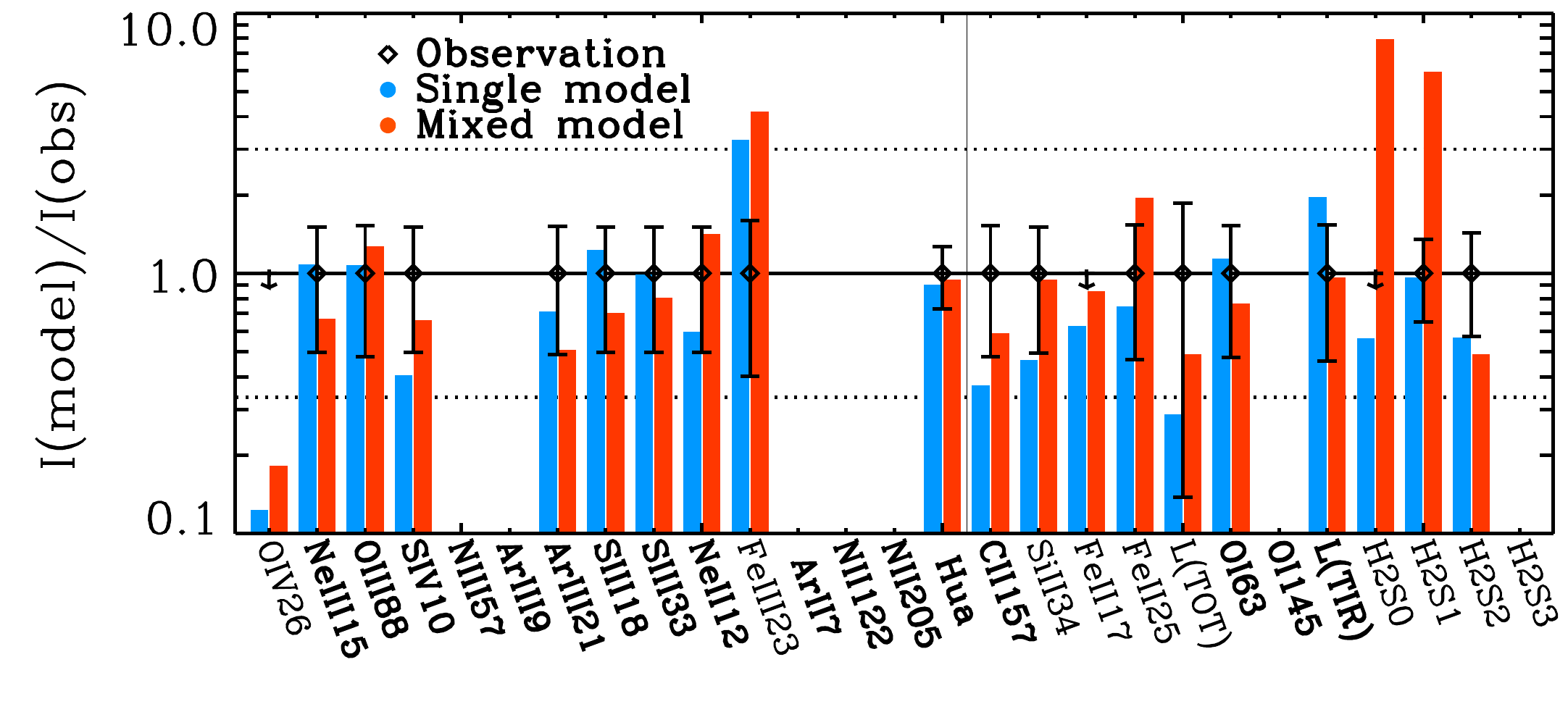} \hfill{}
(b)
\includegraphics[clip,width=6.2cm]{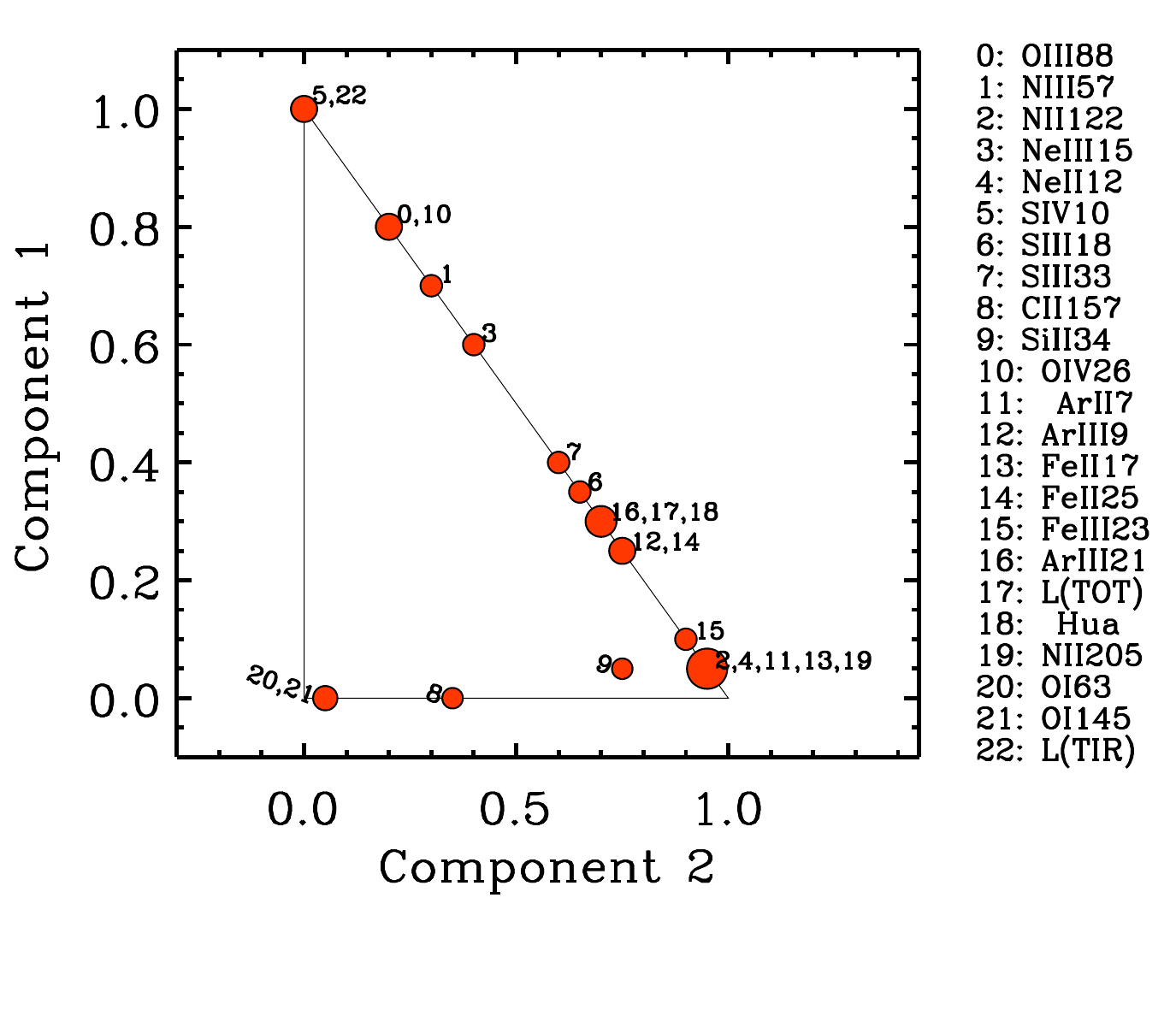}\\
(c)
\includegraphics[clip,width=8.8cm]{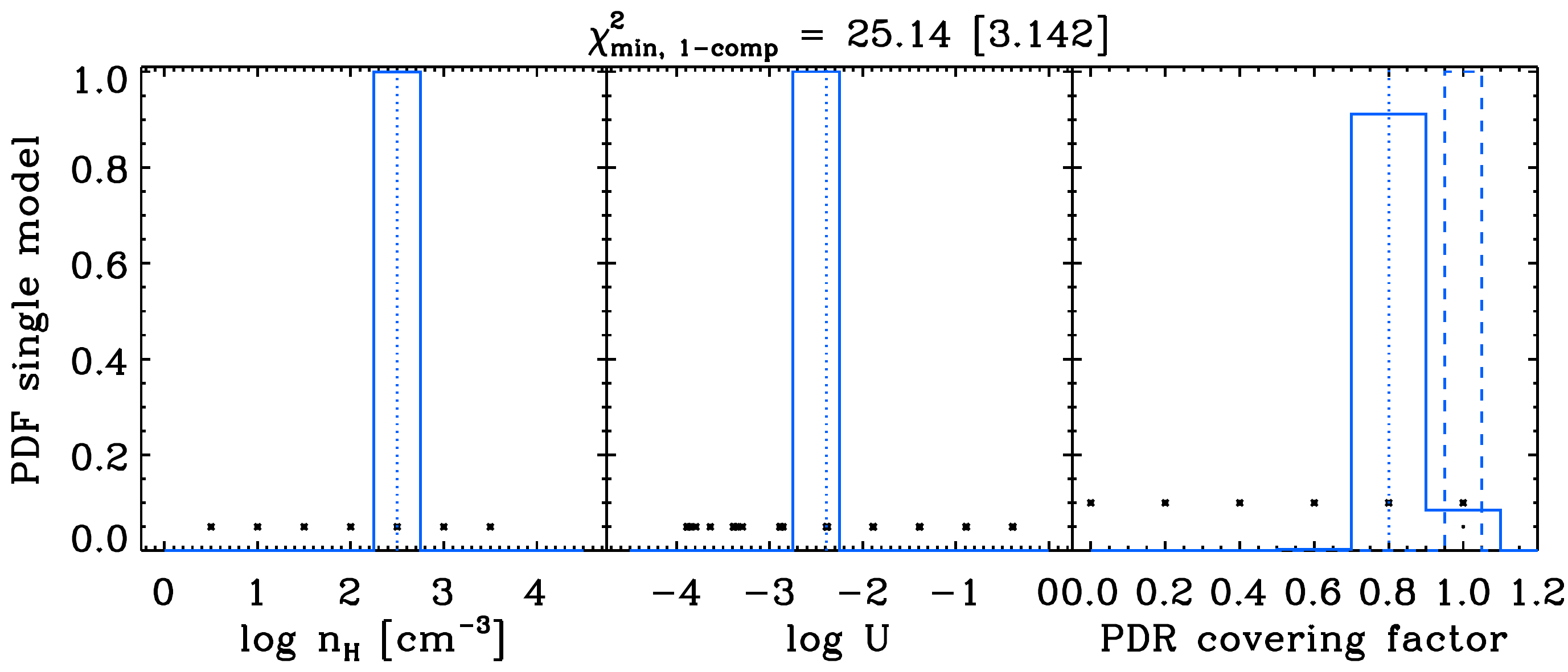} \hfill{}
\includegraphics[clip,width=8.8cm]{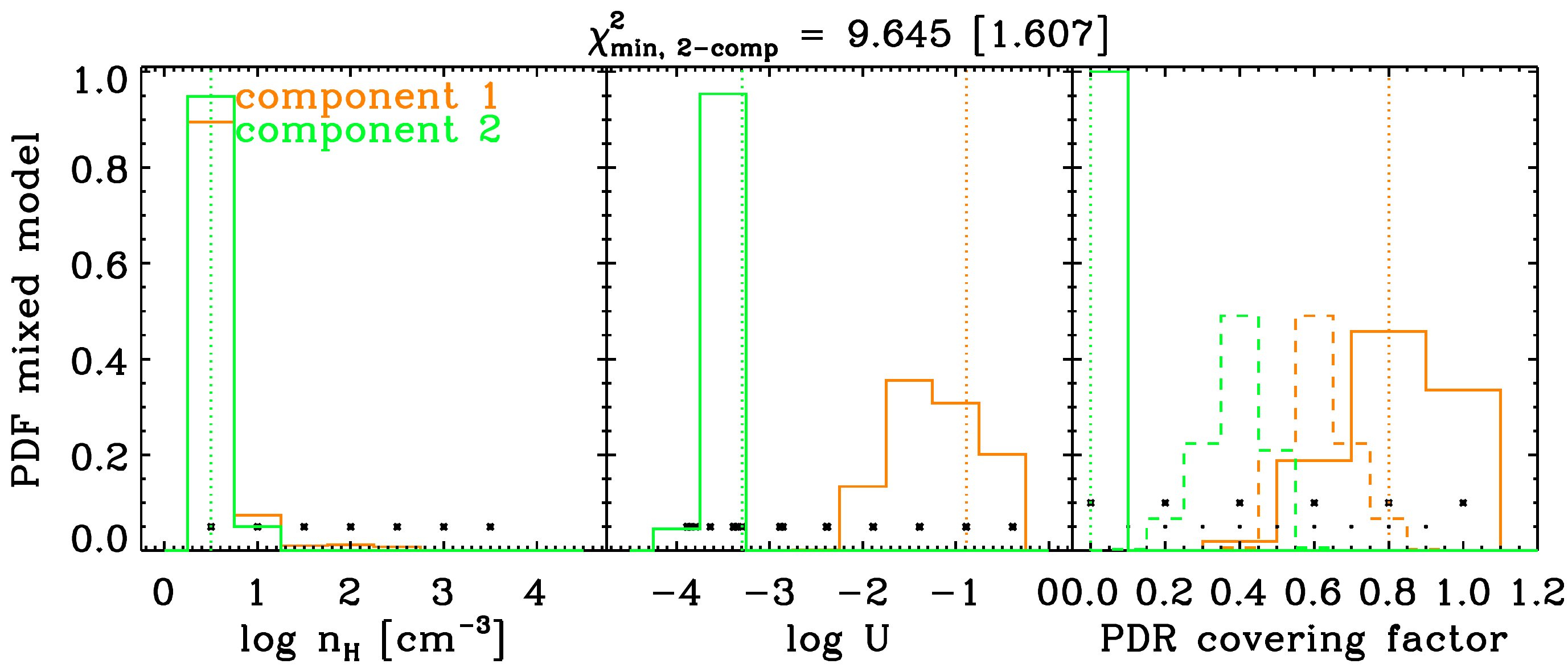}
\caption{ \textbf{Results for NGC\,625.}
$(a)$~Comparison of the observed (losange) and predicted 
intensities for our best-fitting single (blue bar) and mixed (red bar)
\hii region+PDR models. Lines are sorted by decreasing energy.
The fitted lines have labels in boldface.
$(b)$~For mixed models: respective contributions from
component \#1 and component \#2 to the line prediction.
For PDR lines (\cii, \oi, and \silii), only the contribution from
the ionized gas is shown; the PDR contribution is
one minus the sum of the contributions from the plot.
Contributions are expected to be dominated by one
or the other component if their model parameters
are noticeably different.
$(c)$~Probability density functions of the model parameters
($n_{\rm H}$, $U$, $cov_{\rm PDR}$/scaling factor)
for single (left panel) and mixed (right panel) models.
The scaling factor corresponds to the proportion in which
we combine two models in the mixed model case. It is
equal to unity otherwise. It is shown with dashed lines in
the same panels as the PDR covering factor. Vertical dotted lines
show values of the best-fitting model. Black asterisks
show the range and step of values of the grid parameters.
}
\label{fig:bests-ngc625}
\end{figure*}
 \begin{figure*}[thp]
\centering
(a)
\includegraphics[clip,width=11cm]{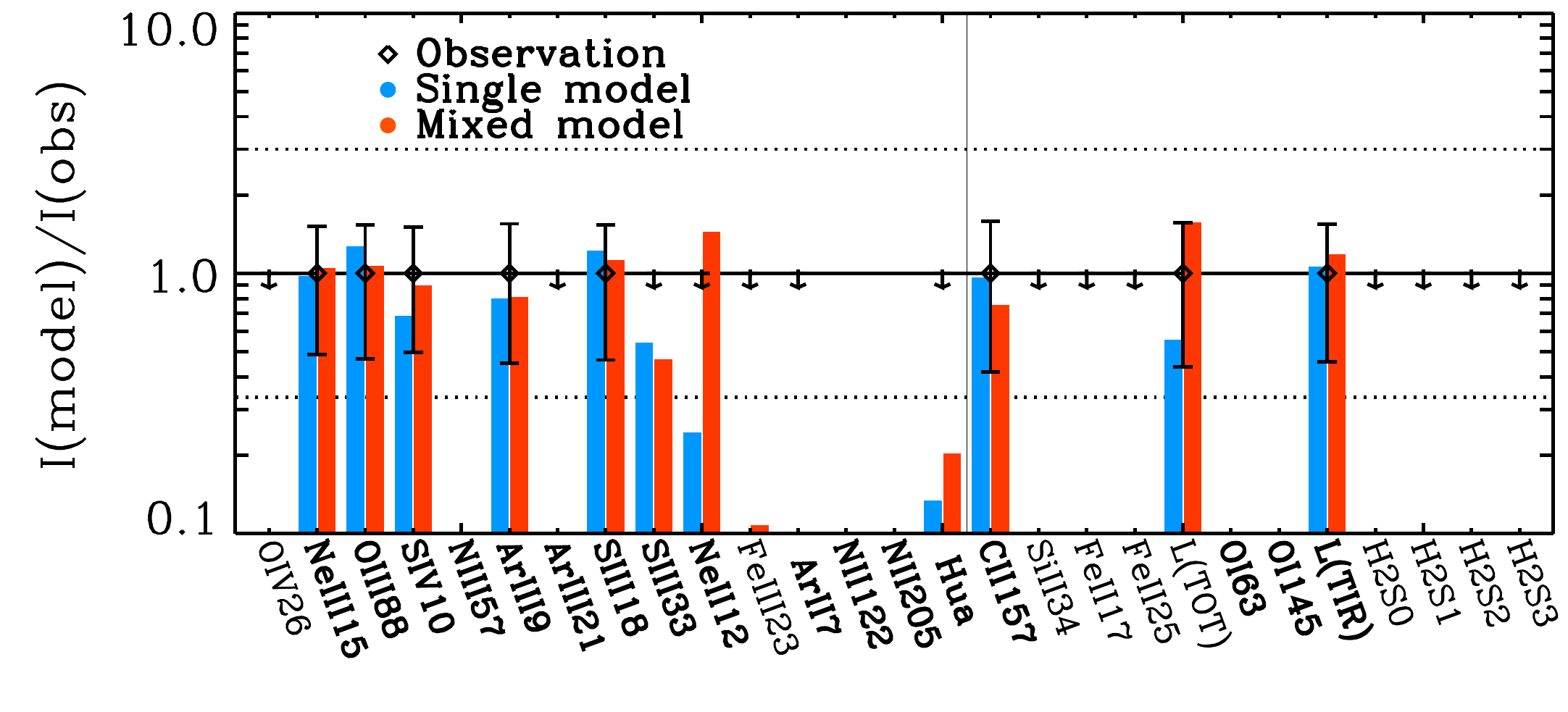} \hfill{}
(b)
\includegraphics[clip,width=6.2cm]{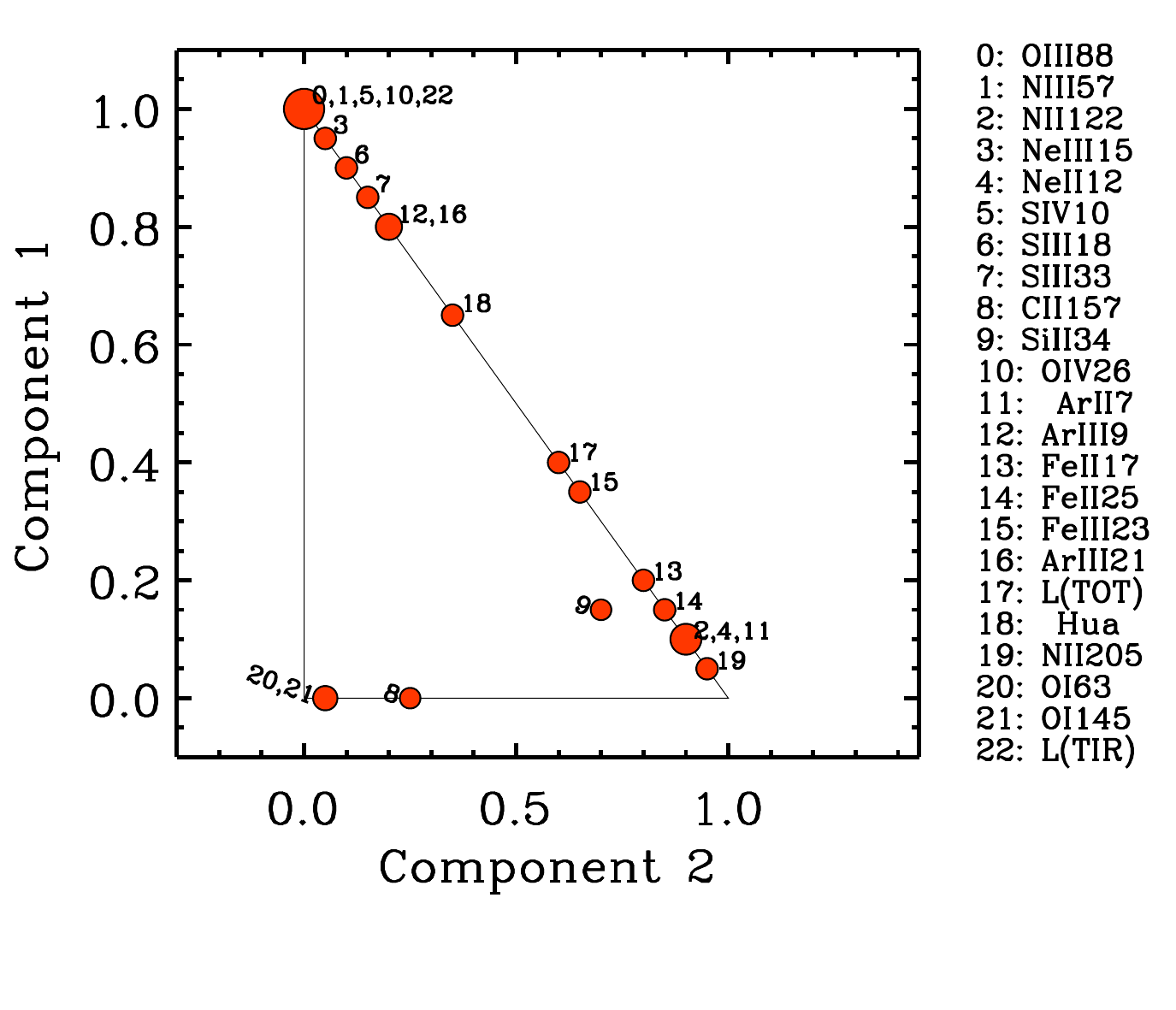}\\
(c)
\includegraphics[clip,width=8.8cm]{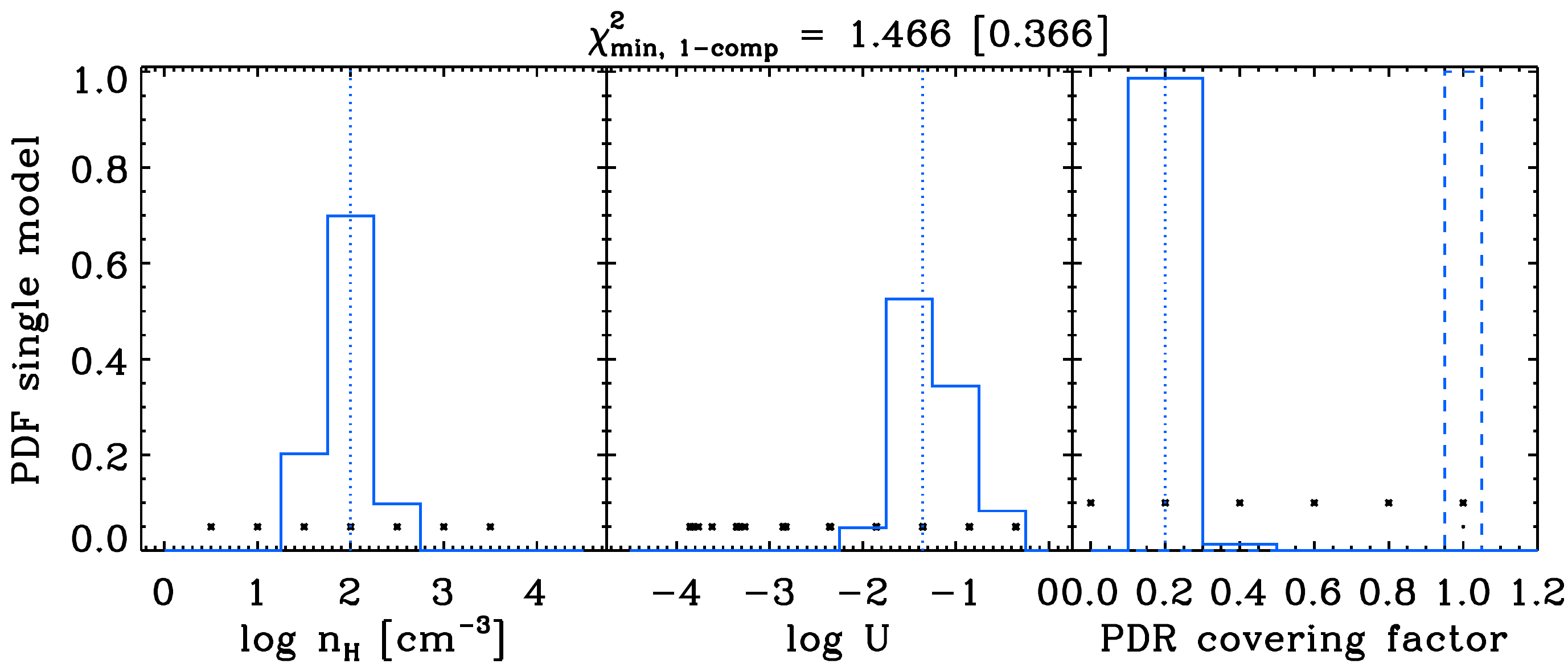} \hfill{}
\includegraphics[clip,width=8.8cm]{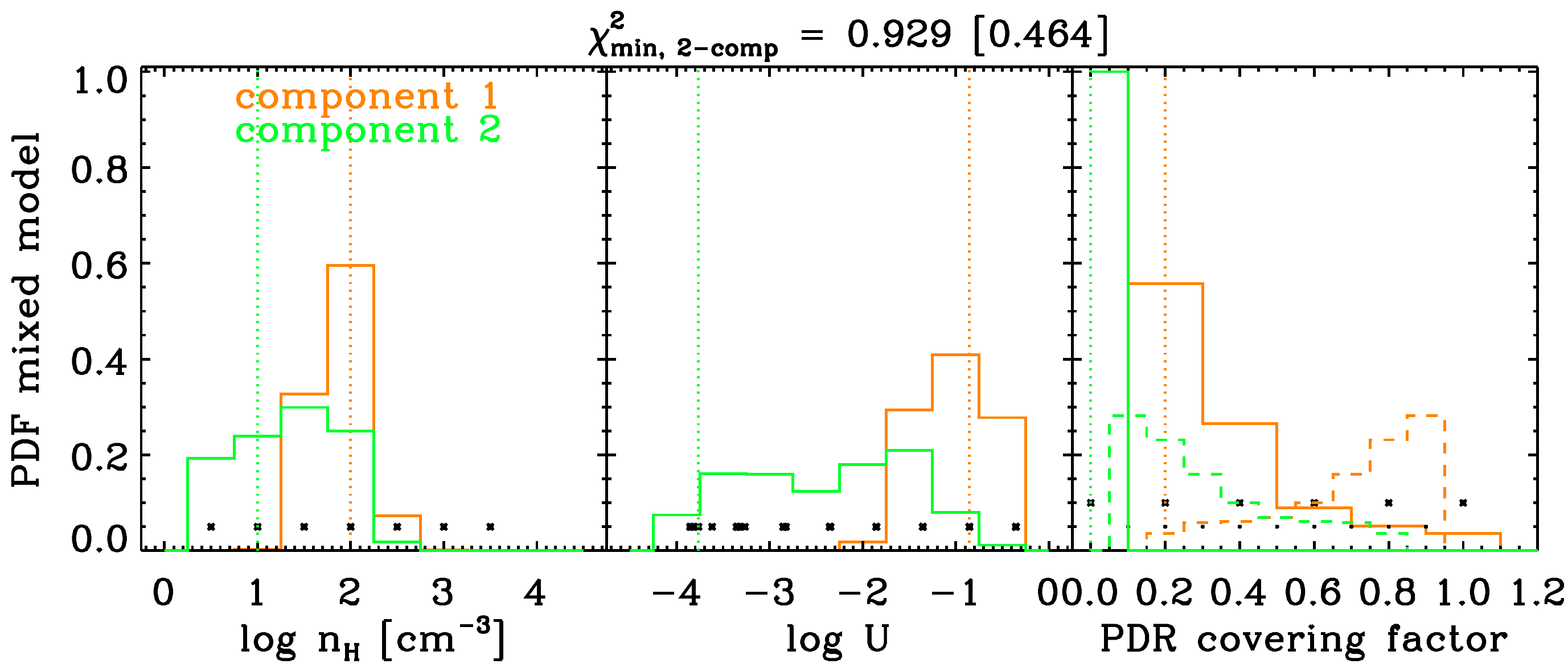}
\caption{ \textbf{Results for Pox\,186.}
See above for caption description.
}
\label{fig:bests-pox186}
\end{figure*}
\clearpage
 \begin{figure*}[thp]
\centering
(a)
\includegraphics[clip,width=11cm]{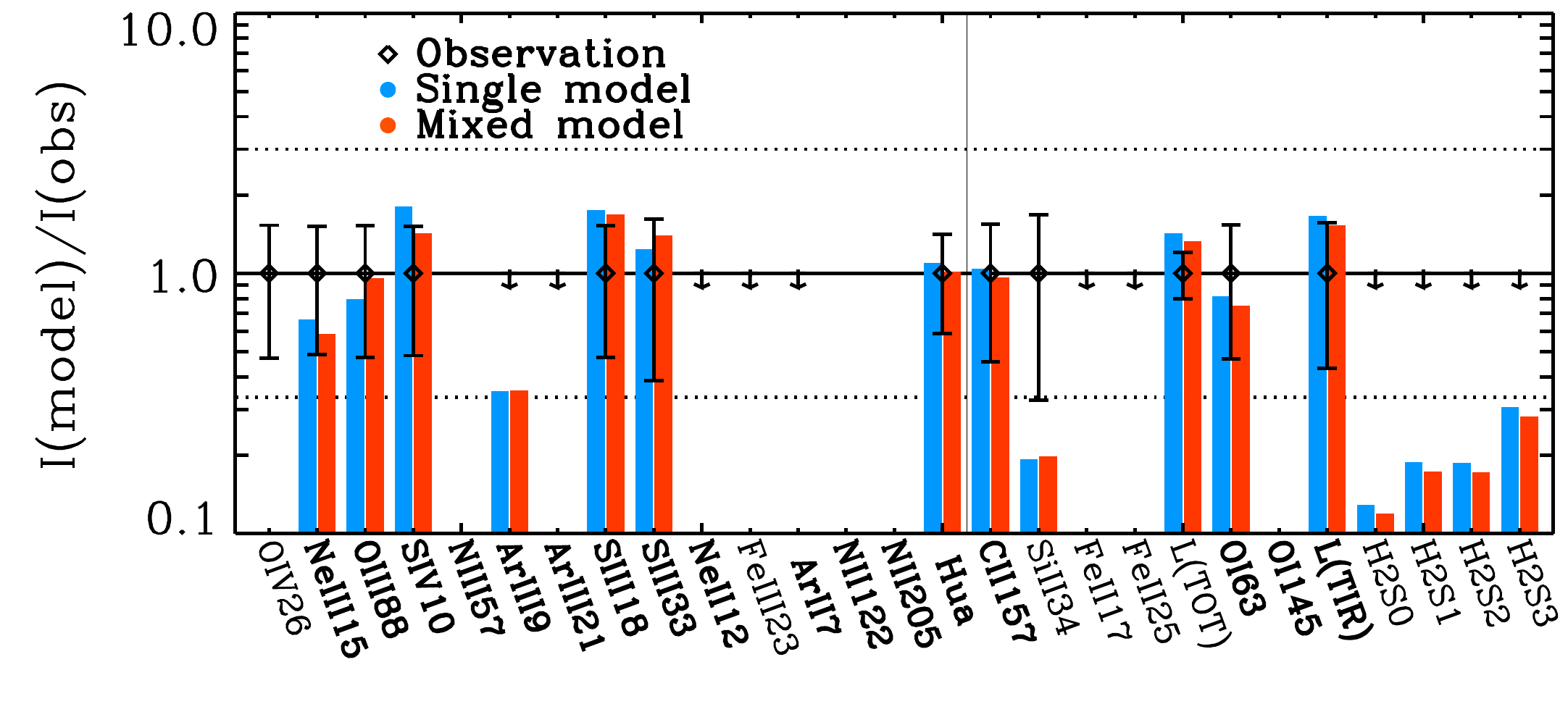} \hfill{}
(b)
\includegraphics[clip,width=6.2cm]{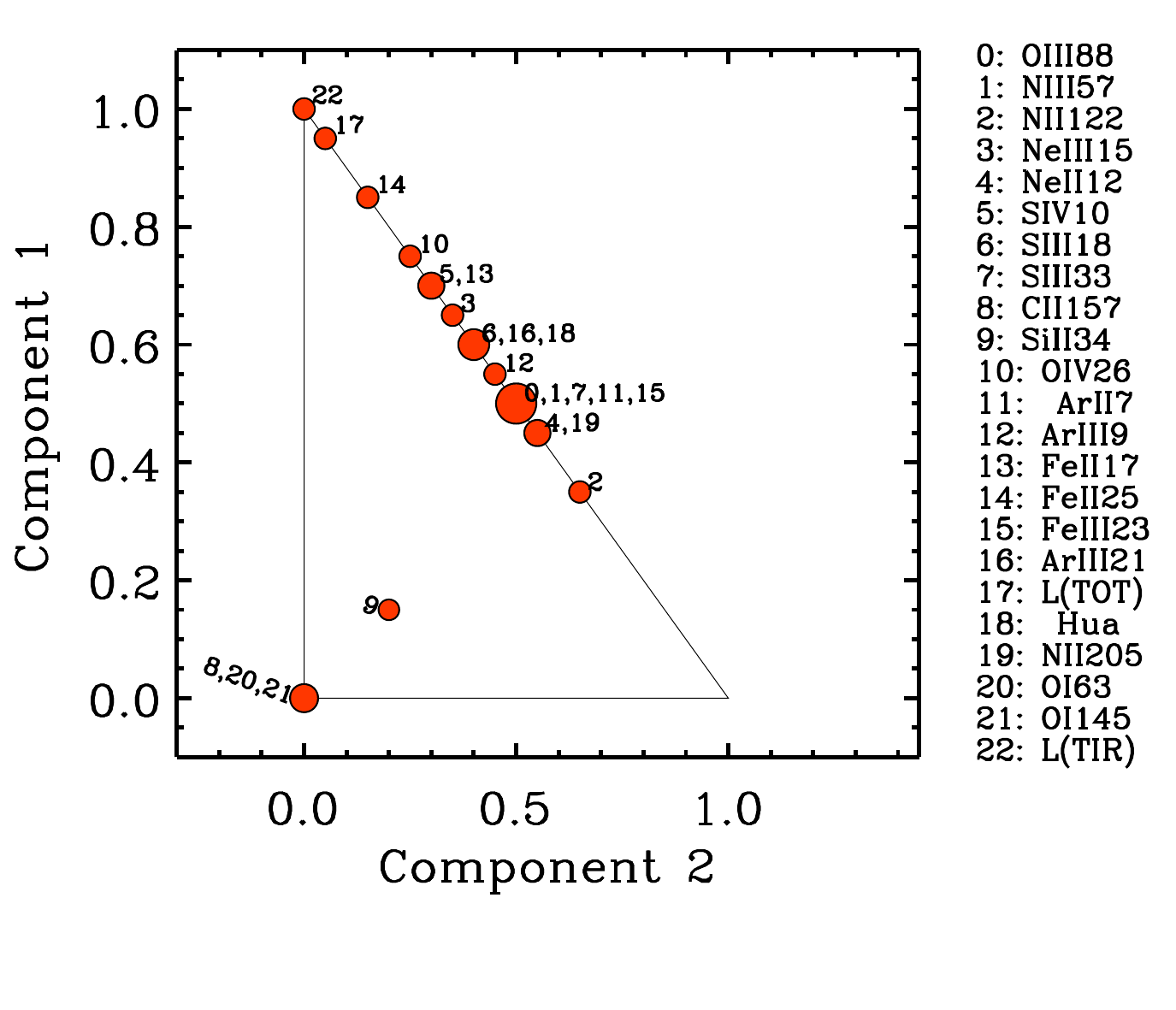}\\
(c)
\includegraphics[clip,width=8.8cm]{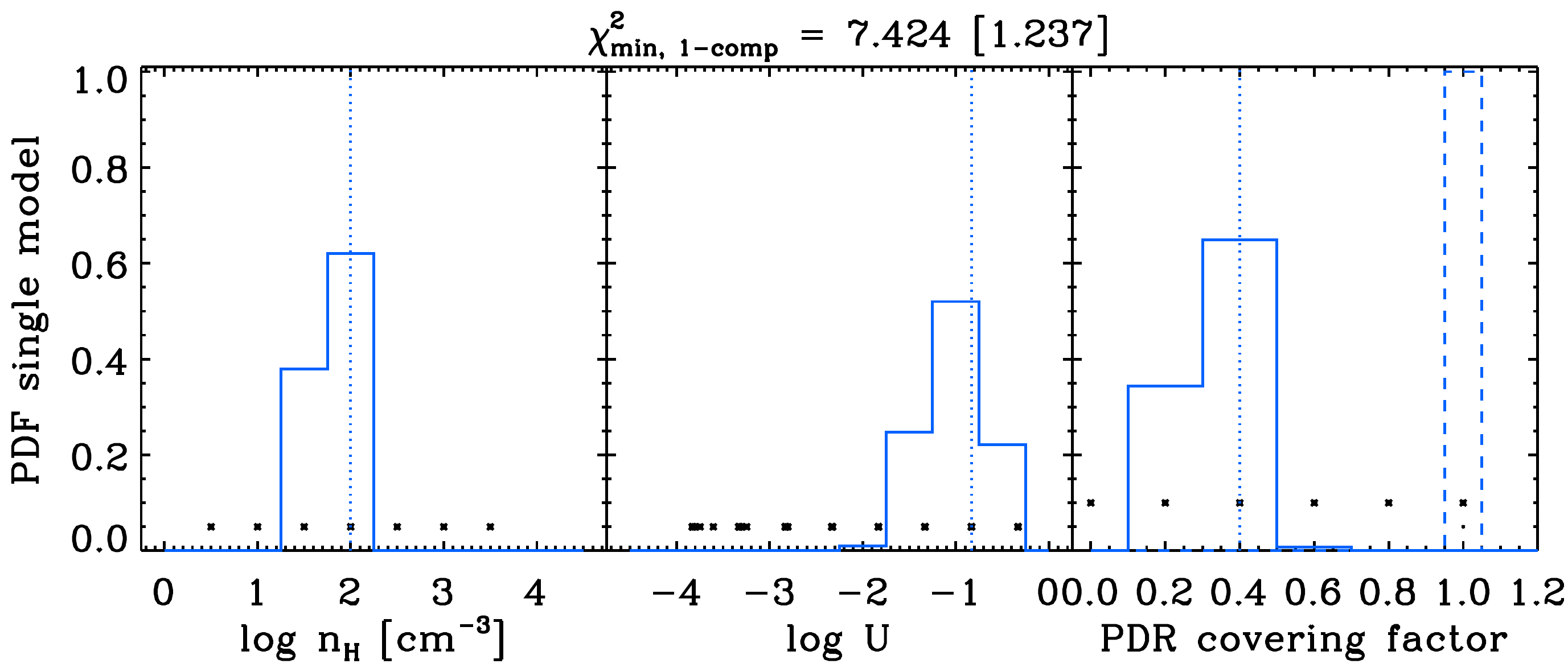} \hfill{}
\includegraphics[clip,width=8.8cm]{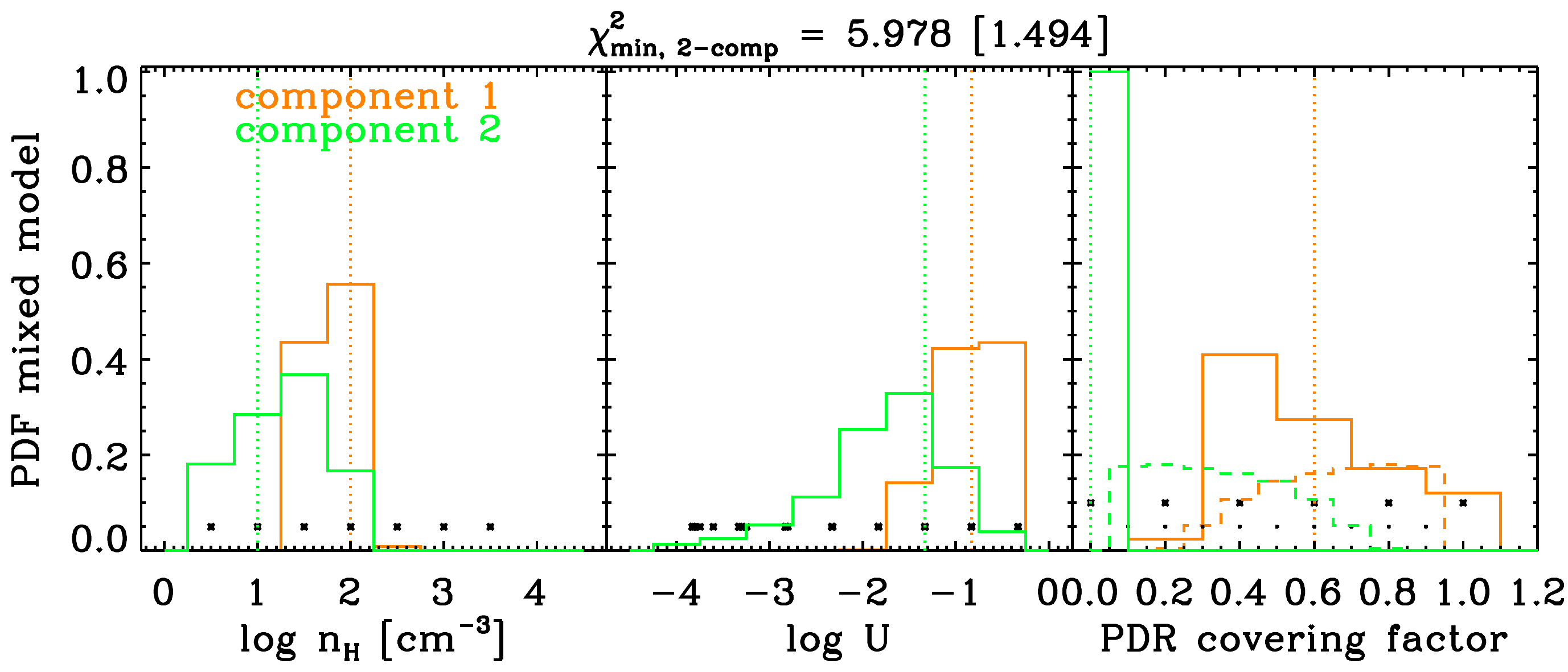}
\caption{ \textbf{Results for SBS\,0335.}
$(a)$~Comparison of the observed (losange) and predicted 
intensities for our best-fitting single (blue bar) and mixed (red bar)
\hii region+PDR models. Lines are sorted by decreasing energy.
The fitted lines have labels in boldface.
$(b)$~For mixed models: respective contributions from
component \#1 and component \#2 to the line prediction.
For PDR lines (\cii, \oi, and \silii), only the contribution from
the ionized gas is shown; the PDR contribution is
one minus the sum of the contributions from the plot.
Contributions are expected to be dominated by one
or the other component if their model parameters
are noticeably different.
$(c)$~Probability density functions of the model parameters
($n_{\rm H}$, $U$, $cov_{\rm PDR}$/scaling factor)
for single (left panel) and mixed (right panel) models.
The scaling factor corresponds to the proportion in which
we combine two models in the mixed model case. It is
equal to unity otherwise. It is shown with dashed lines in
the same panels as the PDR covering factor. Vertical dotted lines
show values of the best-fitting model. Black asterisks
show the range and step of values of the grid parameters.
}
\label{fig:bests-sbs0335}
\end{figure*}
 \begin{figure*}[thp]
\centering
(a)
\includegraphics[clip,width=11cm]{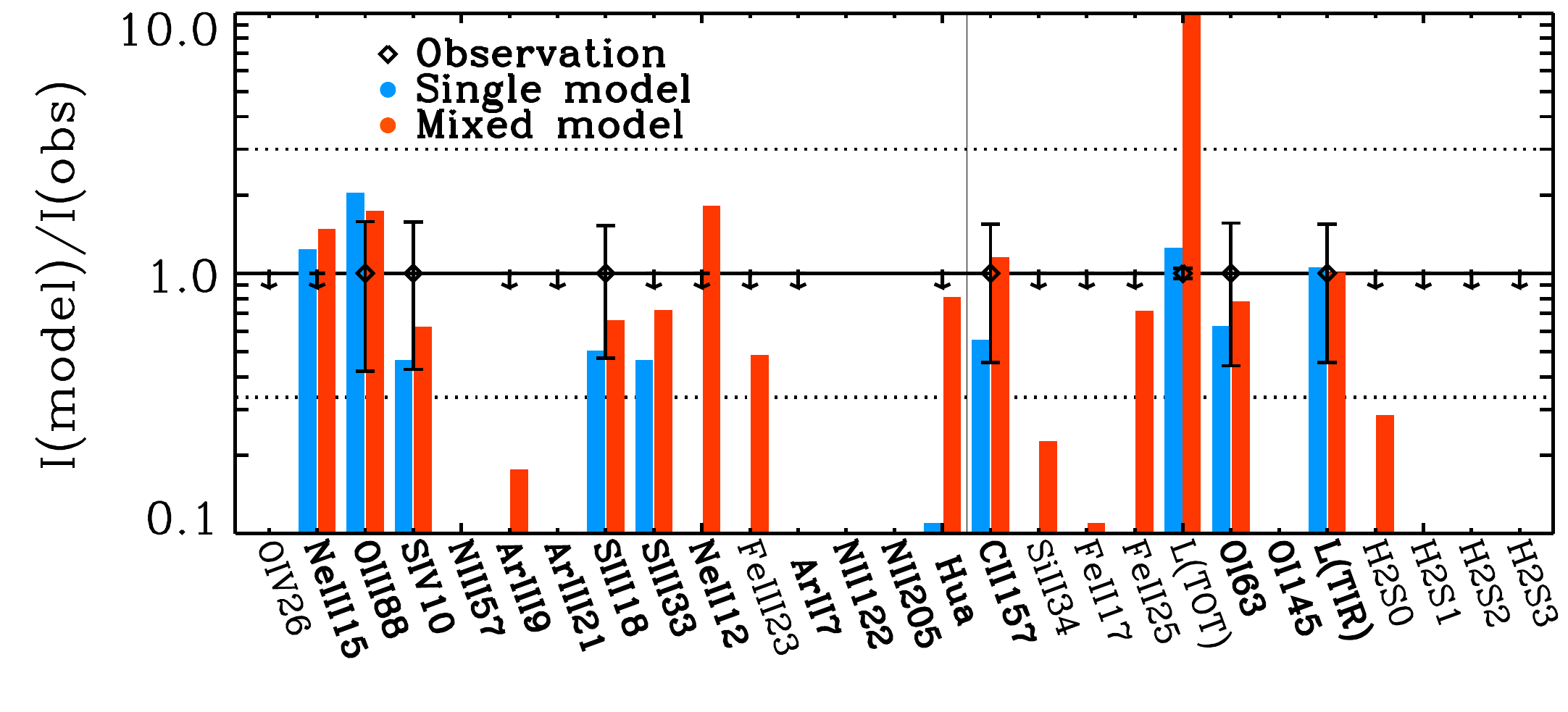} \hfill{}
(b)
\includegraphics[clip,width=6.2cm]{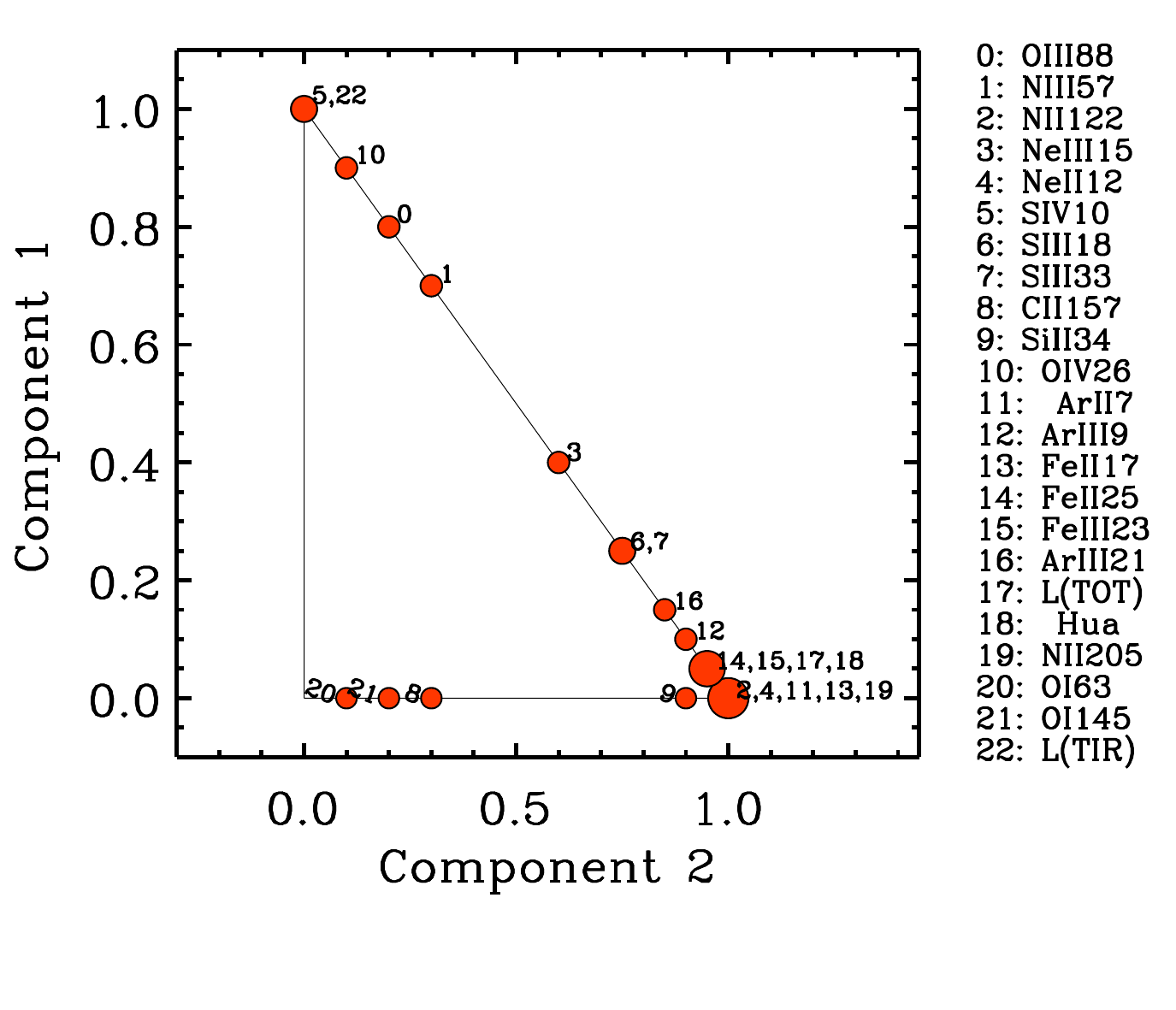}\\
(c)
\includegraphics[clip,width=8.8cm]{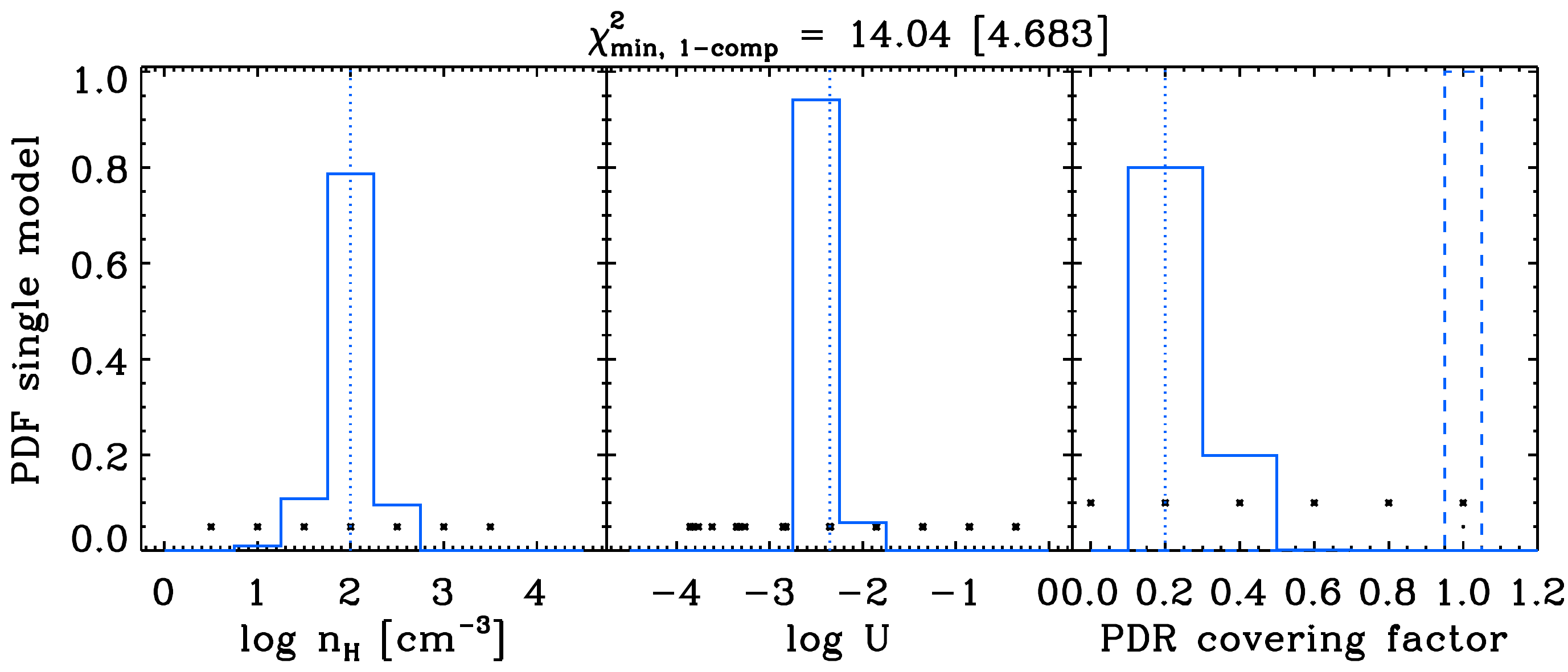} \hfill{}
\includegraphics[clip,width=8.8cm]{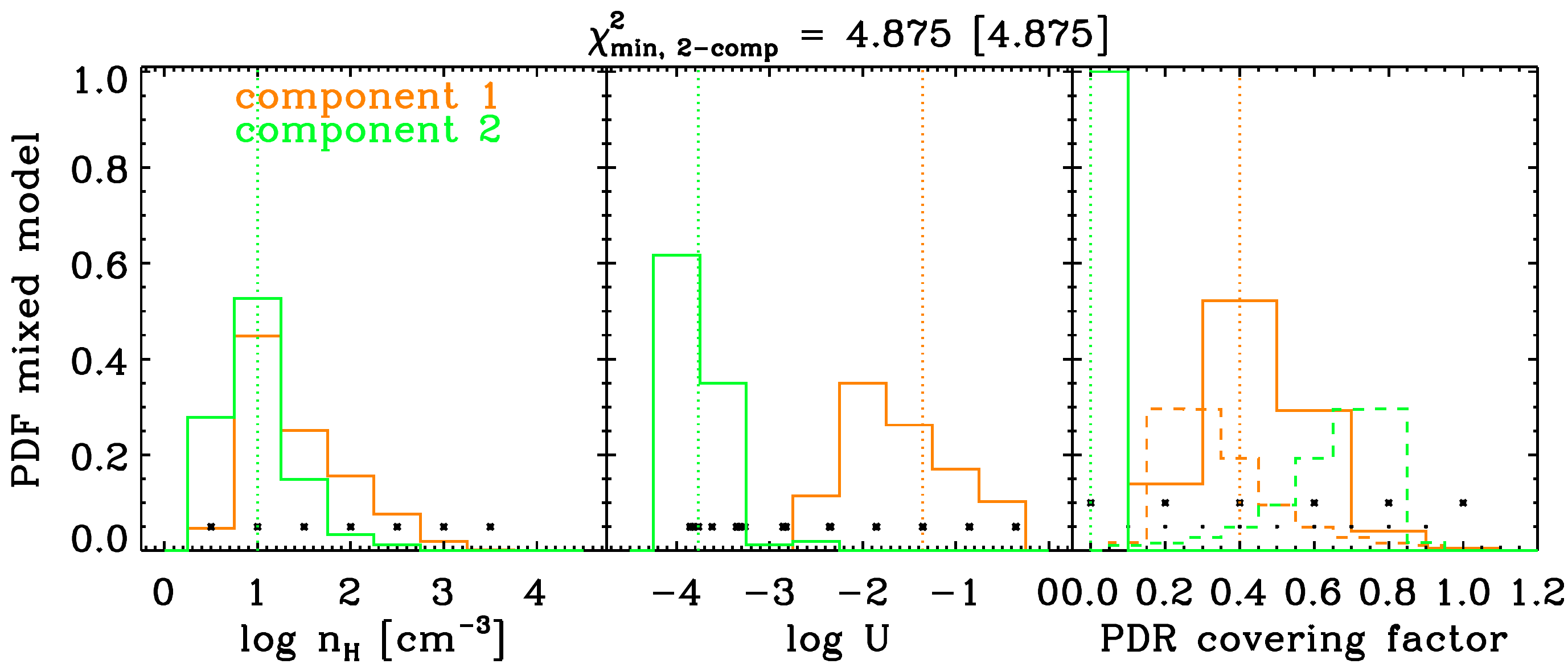}
\caption{ \textbf{Results for SBS\,1159.}
See above for caption description.
}
\label{fig:bests-sbs1159}
\end{figure*}
\clearpage
 \begin{figure*}[thp]
\centering
(a)
\includegraphics[clip,width=11cm]{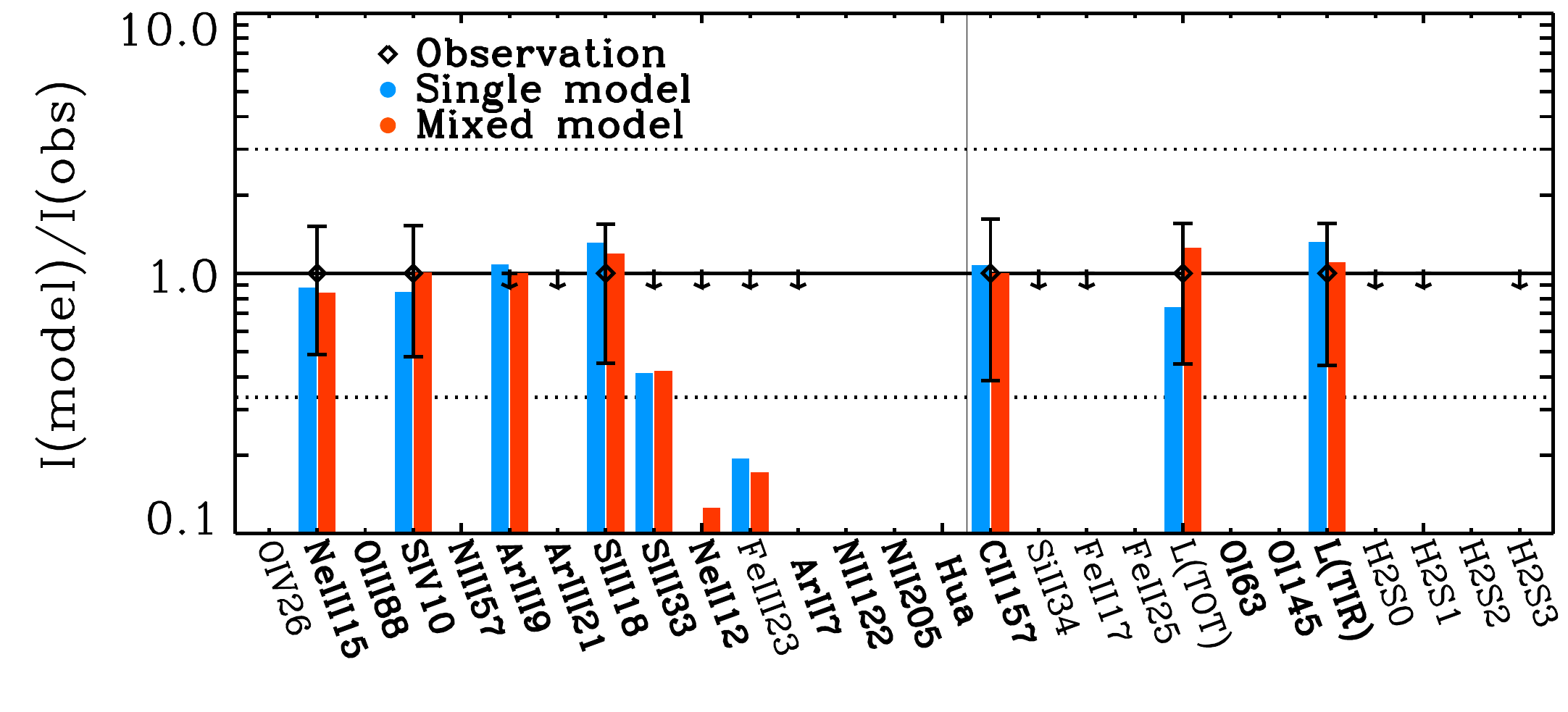} \hfill{}
(b)
\includegraphics[clip,width=6.2cm]{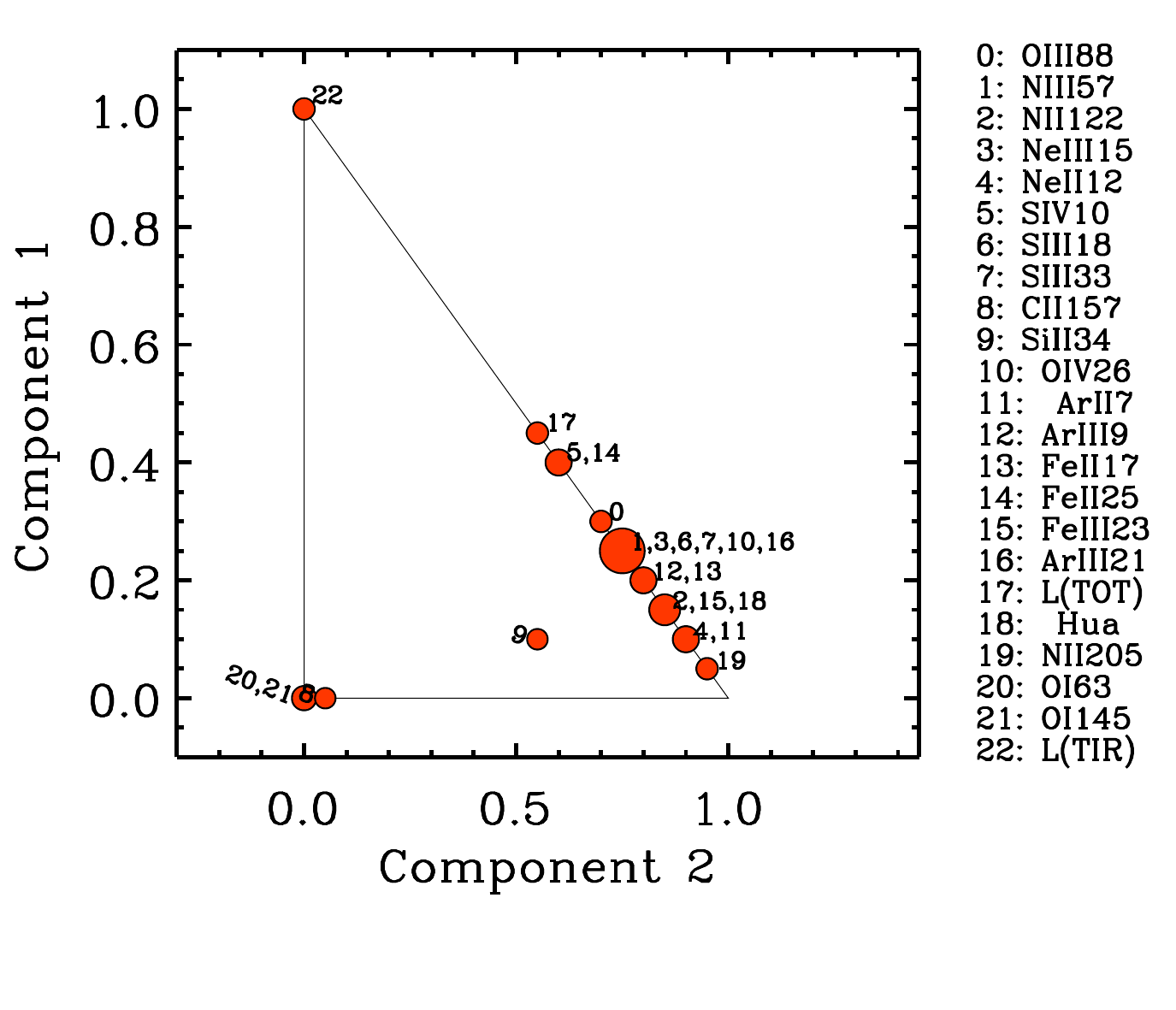}\\
(c)
\includegraphics[clip,width=8.8cm]{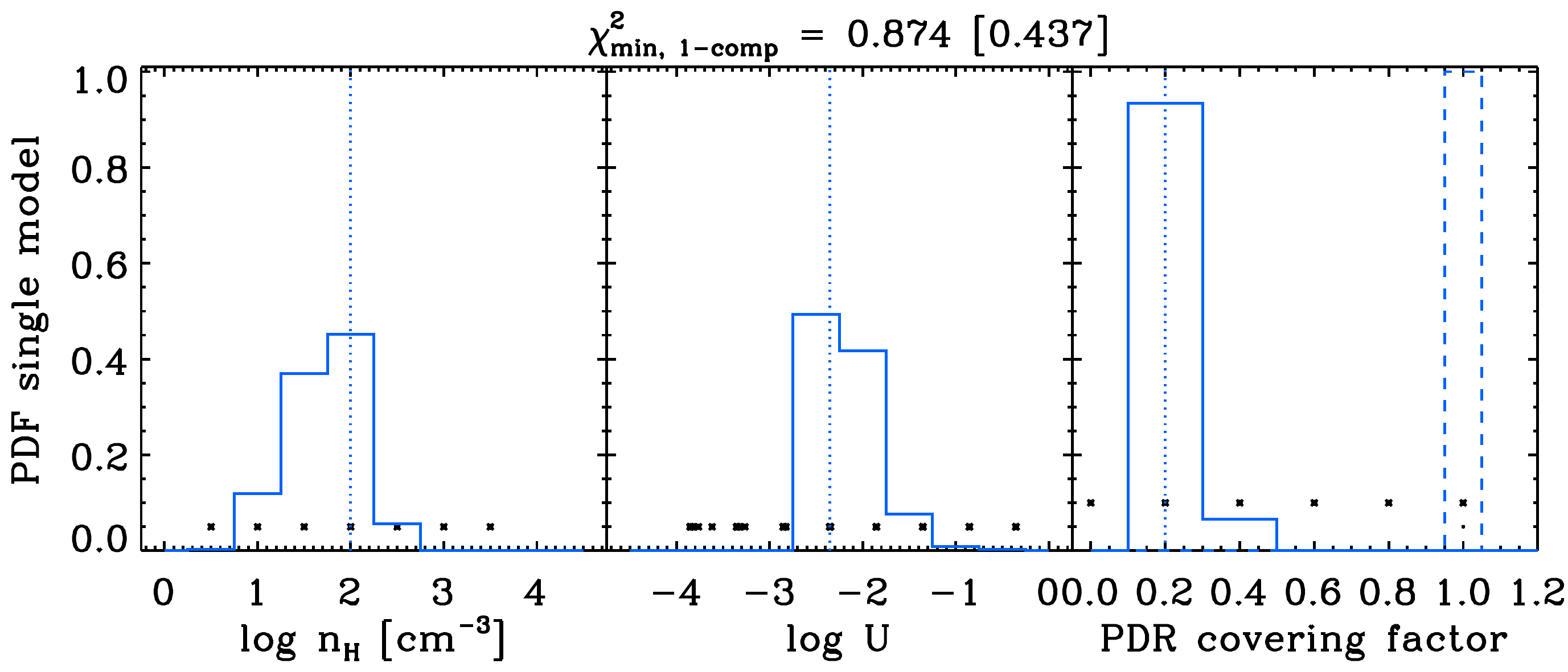} \hfill{}
\includegraphics[clip,width=8.8cm]{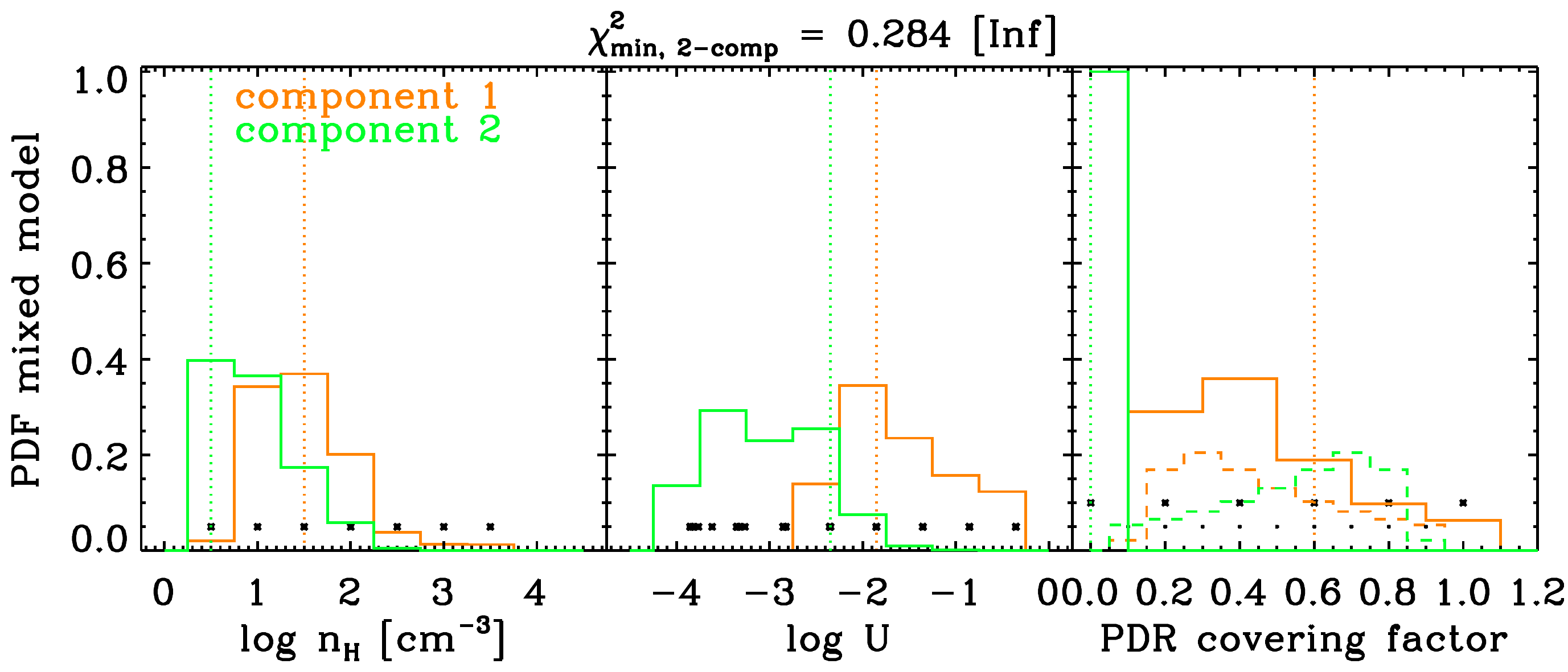}
\caption{ \textbf{Results for SBS\,1211.}
$(a)$~Comparison of the observed (losange) and predicted 
intensities for our best-fitting single (blue bar) and mixed (red bar)
\hii region+PDR models. Lines are sorted by decreasing energy.
The fitted lines have labels in boldface.
$(b)$~For mixed models: respective contributions from
component \#1 and component \#2 to the line prediction.
For PDR lines (\cii, \oi, and \silii), only the contribution from
the ionized gas is shown; the PDR contribution is
one minus the sum of the contributions from the plot.
Contributions are expected to be dominated by one
or the other component if their model parameters
are noticeably different.
$(c)$~Probability density functions of the model parameters
($n_{\rm H}$, $U$, $cov_{\rm PDR}$/scaling factor)
for single (left panel) and mixed (right panel) models.
The scaling factor corresponds to the proportion in which
we combine two models in the mixed model case. It is
equal to unity otherwise. It is shown with dashed lines in
the same panels as the PDR covering factor. Vertical dotted lines
show values of the best-fitting model. Black asterisks
show the range and step of values of the grid parameters.
}
\label{fig:bests-sbs1211}
\end{figure*}
 \begin{figure*}[thp]
\centering
(a)
\includegraphics[clip,width=11cm]{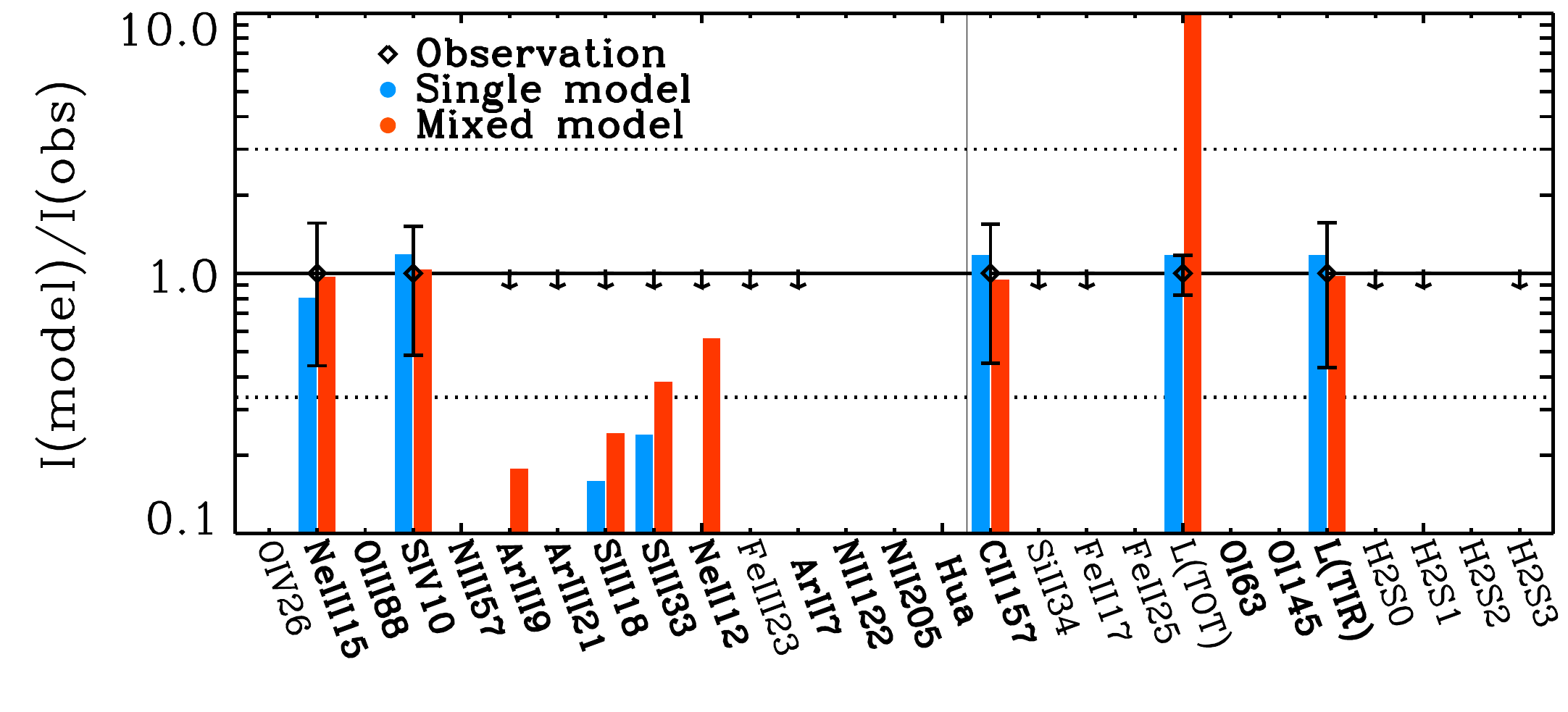} \hfill{}
(b)
\includegraphics[clip,width=6.2cm]{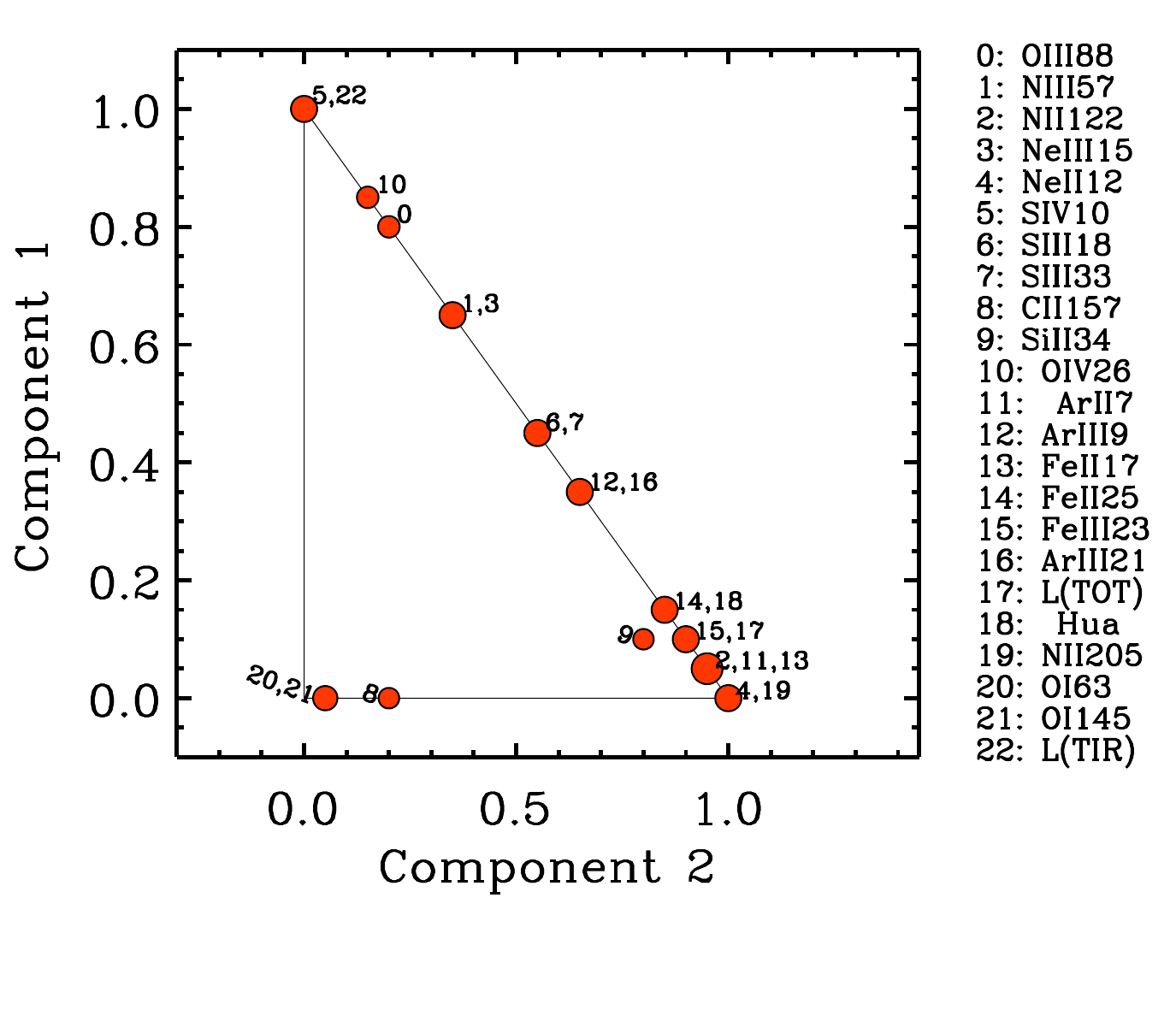}\\
(c)
\includegraphics[clip,width=8.8cm]{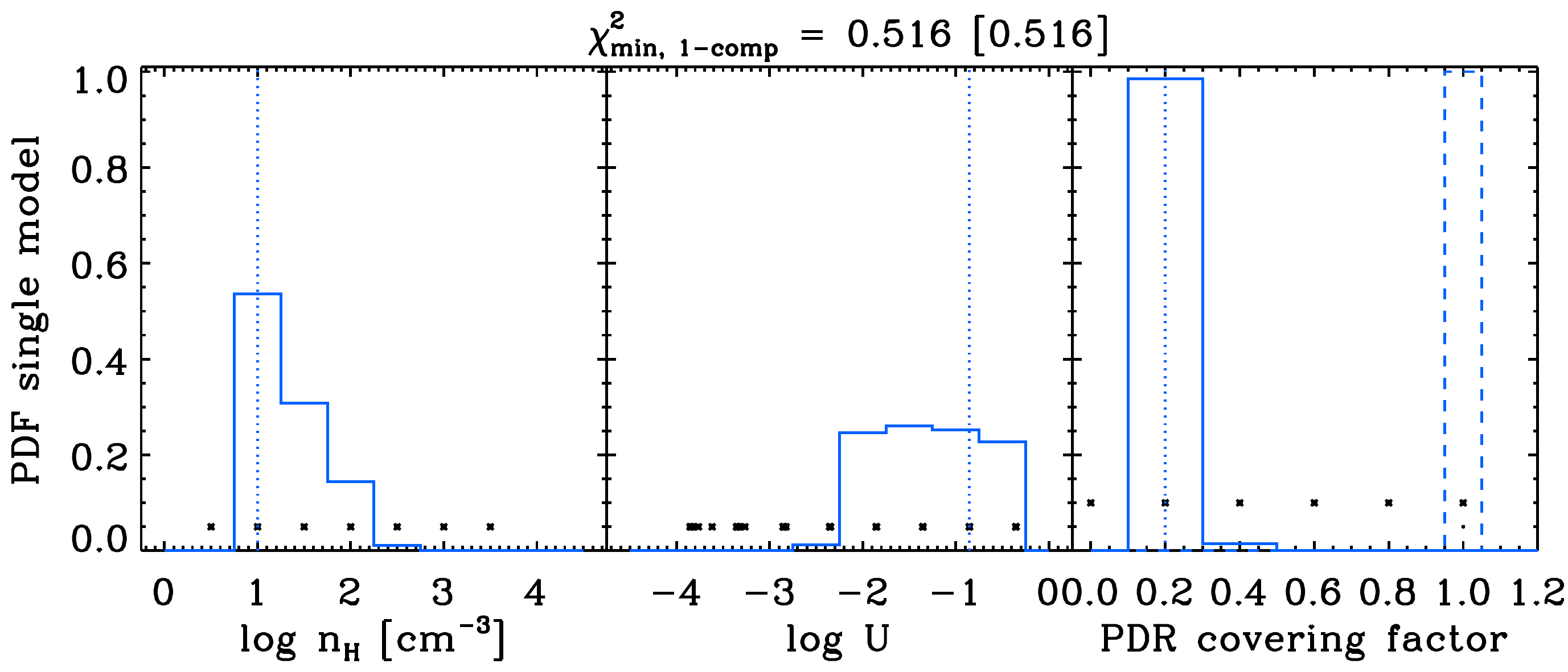} \hfill{}
\includegraphics[clip,width=8.8cm]{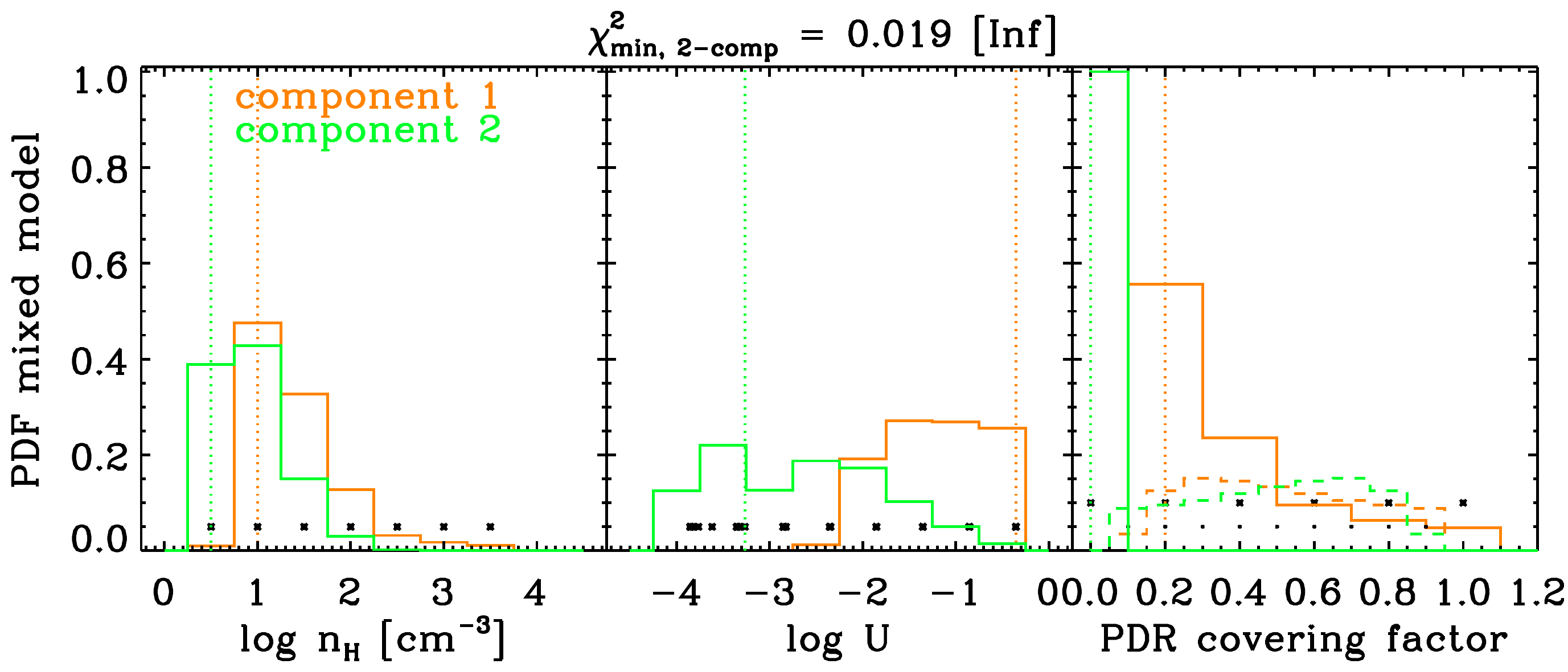}
\caption{ \textbf{Results for SBS\,1249.}
See above for caption description.
}
\label{fig:bests-sbs1249}
\end{figure*}
\clearpage
 \begin{figure*}[thp]
\centering
(a)
\includegraphics[clip,width=11cm]{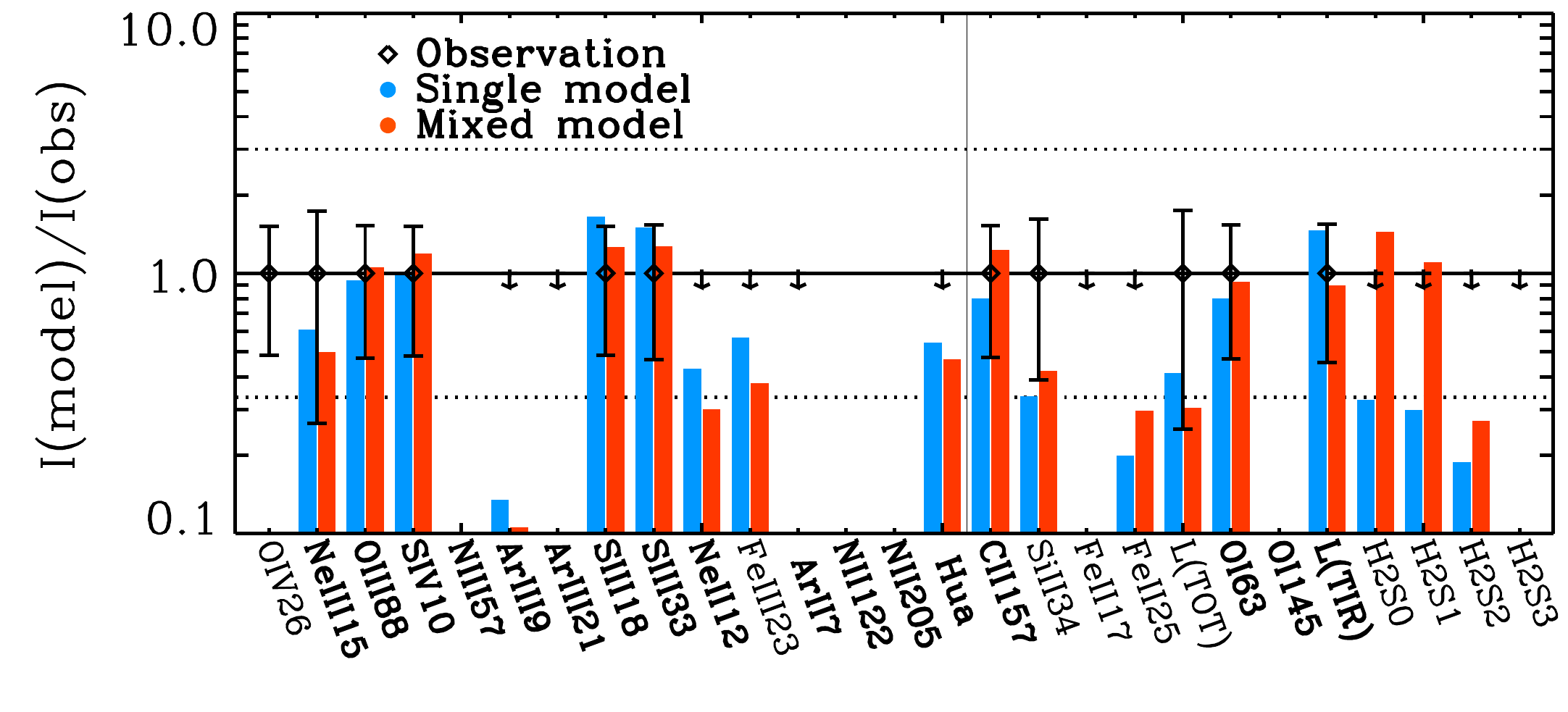} \hfill{}
(b)
\includegraphics[clip,width=6.2cm]{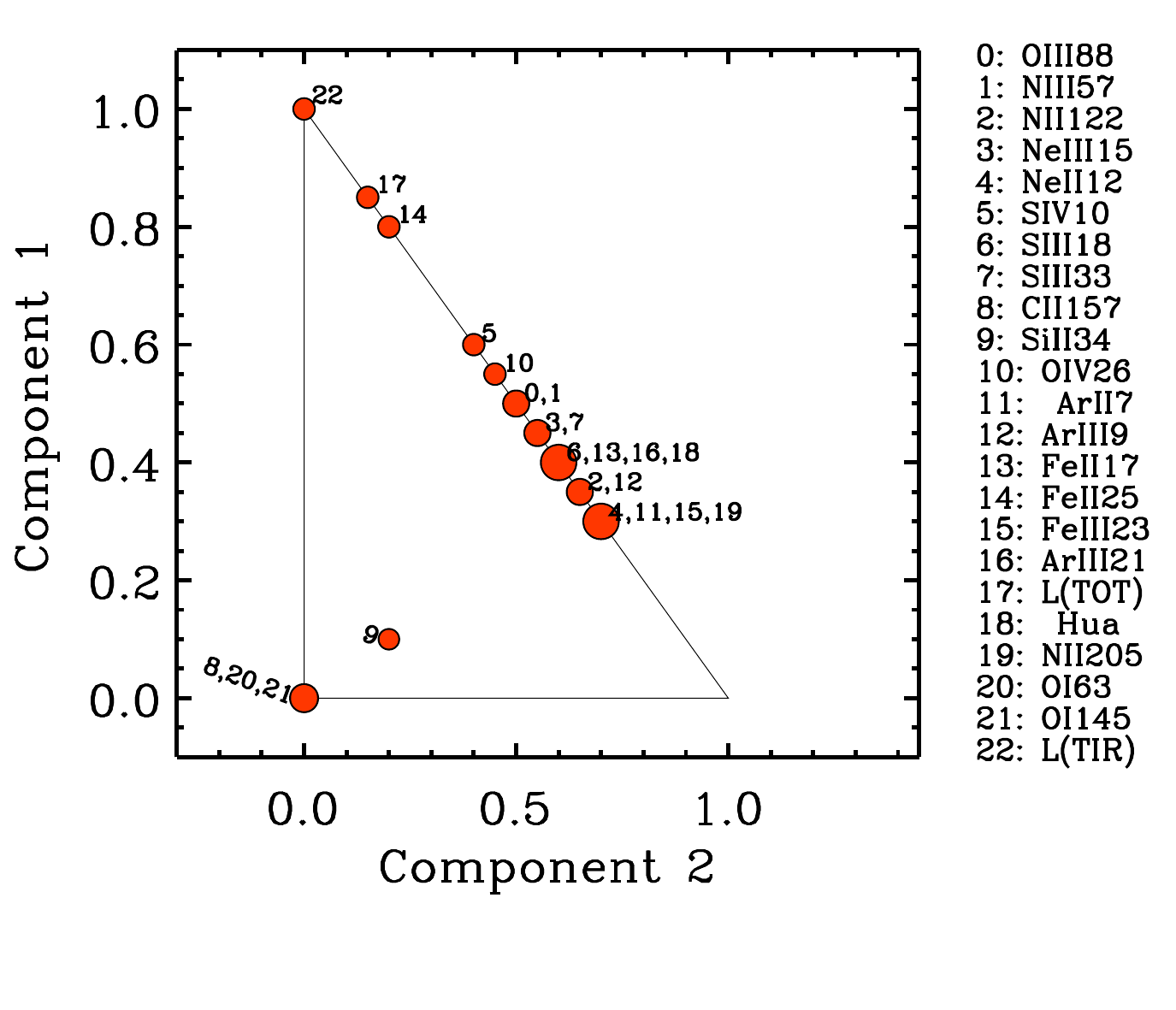}\\
(c)
\includegraphics[clip,width=8.8cm]{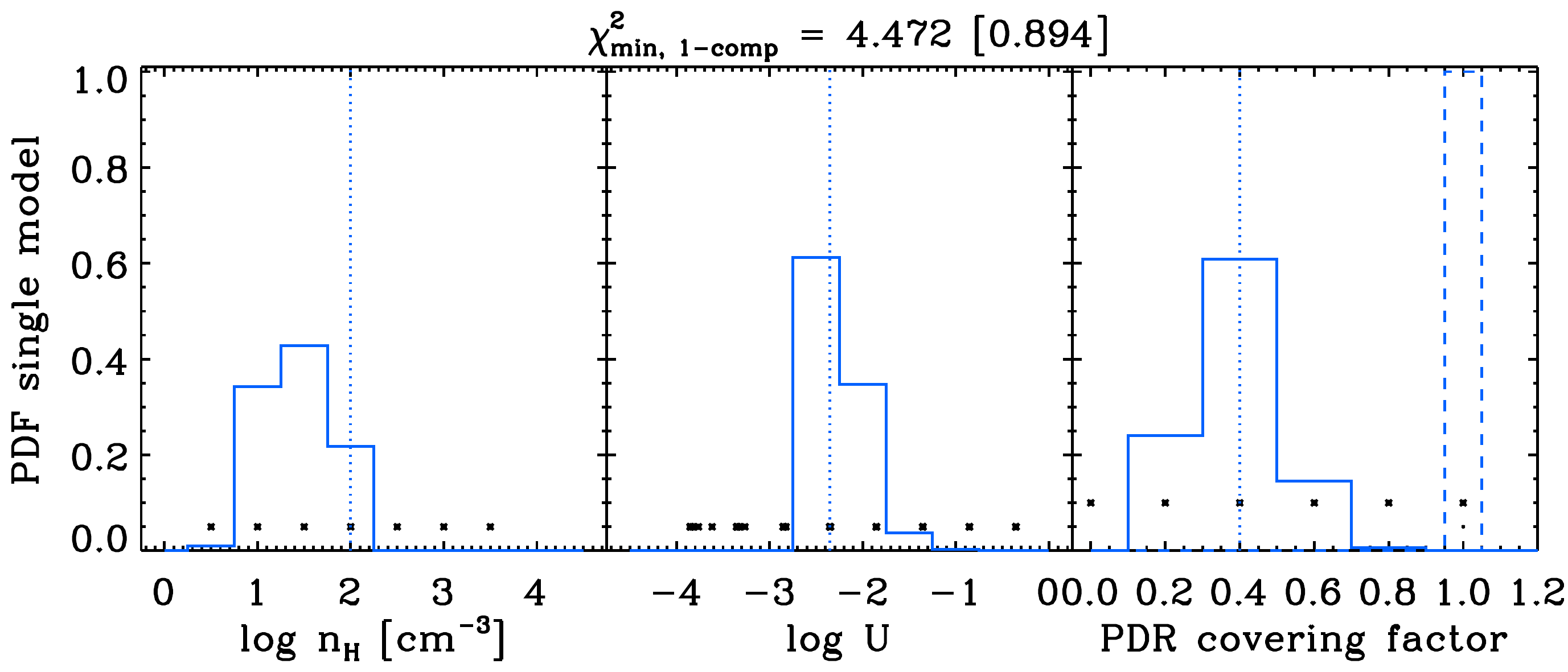} \hfill{}
\includegraphics[clip,width=8.8cm]{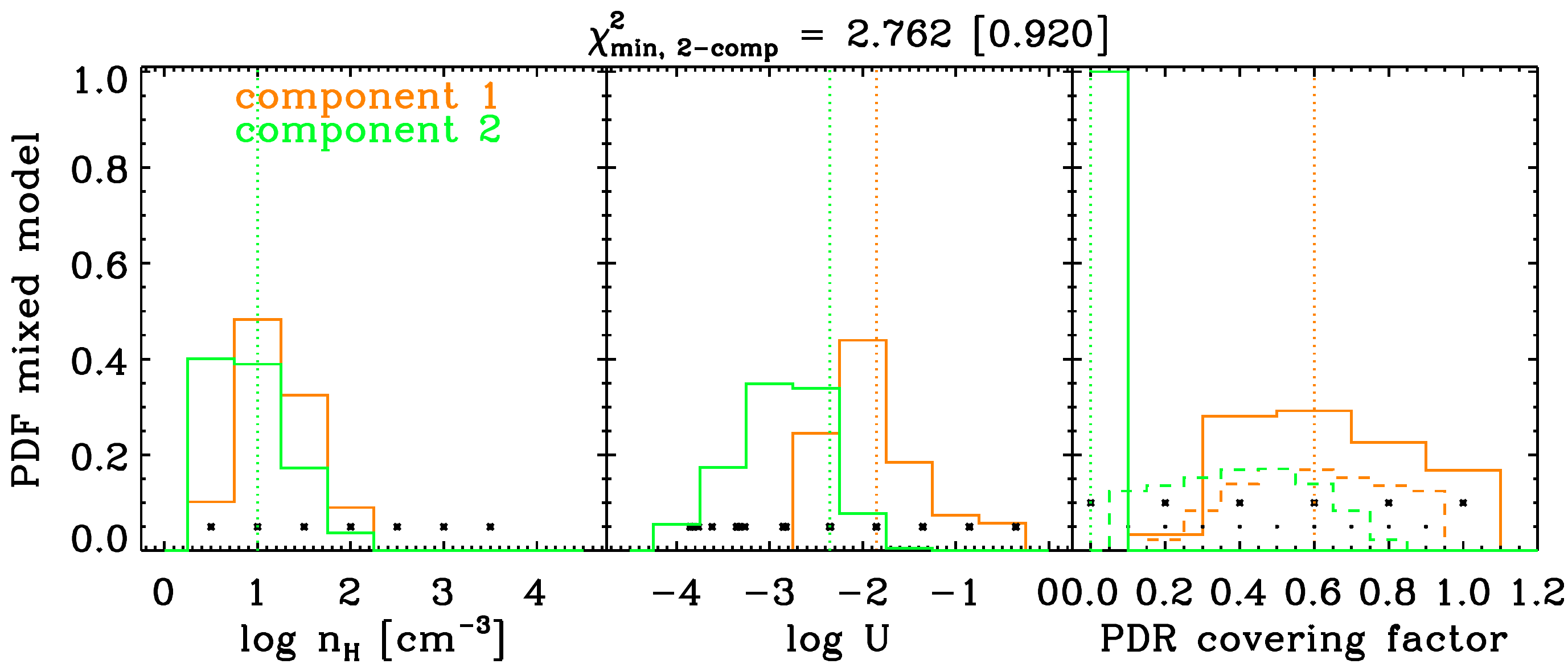}
\caption{ \textbf{Results for SBS\,1415.}
$(a)$~Comparison of the observed (losange) and predicted 
intensities for our best-fitting single (blue bar) and mixed (red bar)
\hii region+PDR models. Lines are sorted by decreasing energy.
The fitted lines have labels in boldface.
$(b)$~For mixed models: respective contributions from
component \#1 and component \#2 to the line prediction.
For PDR lines (\cii, \oi, and \silii), only the contribution from
the ionized gas is shown; the PDR contribution is
one minus the sum of the contributions from the plot.
Contributions are expected to be dominated by one
or the other component if their model parameters
are noticeably different.
$(c)$~Probability density functions of the model parameters
($n_{\rm H}$, $U$, $cov_{\rm PDR}$/scaling factor)
for single (left panel) and mixed (right panel) models.
The scaling factor corresponds to the proportion in which
we combine two models in the mixed model case. It is
equal to unity otherwise. It is shown with dashed lines in
the same panels as the PDR covering factor. Vertical dotted lines
show values of the best-fitting model. Black asterisks
show the range and step of values of the grid parameters.
}
\label{fig:bests-sbs1415}
\end{figure*}
 \begin{figure*}[thp]
\centering
(a)
\includegraphics[clip,width=11cm]{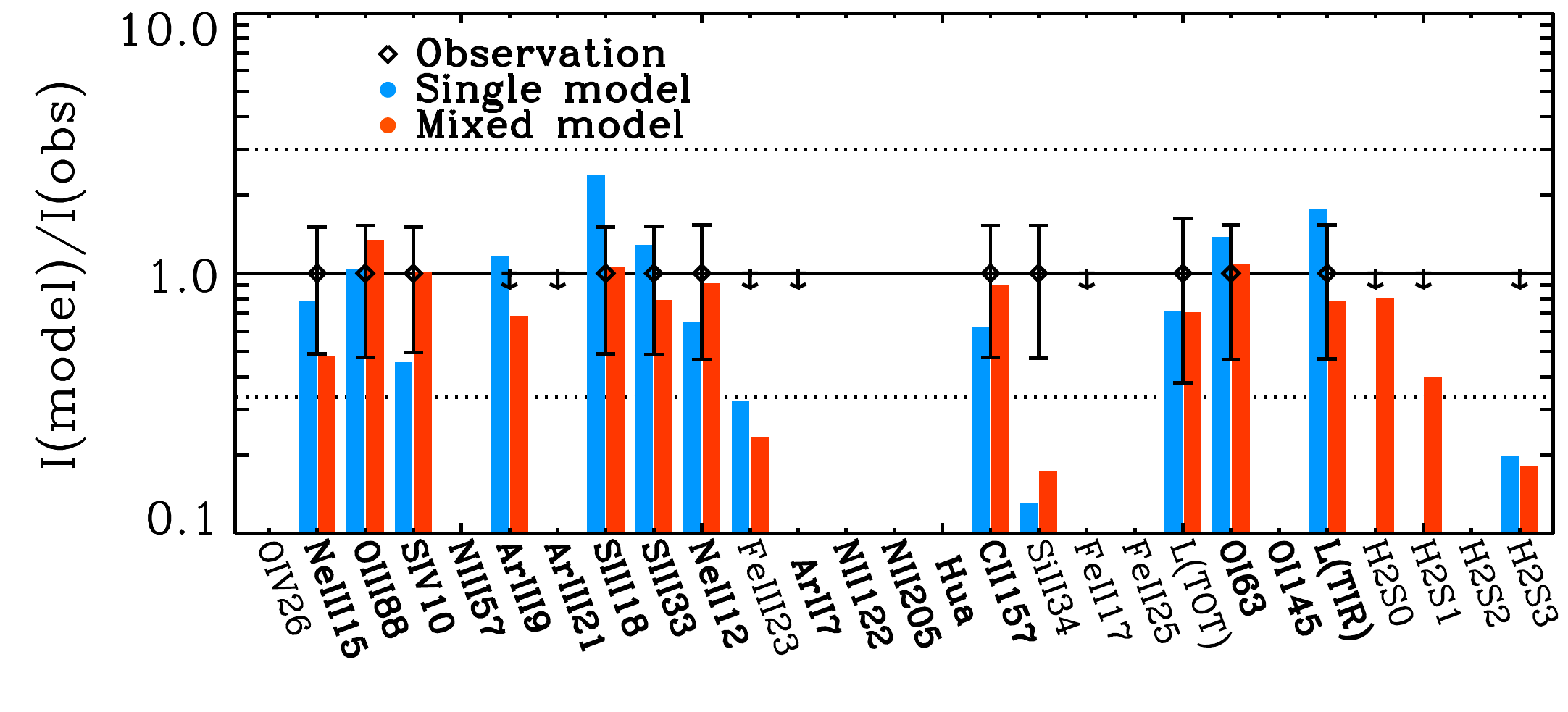} \hfill{}
(b)
\includegraphics[clip,width=6.2cm]{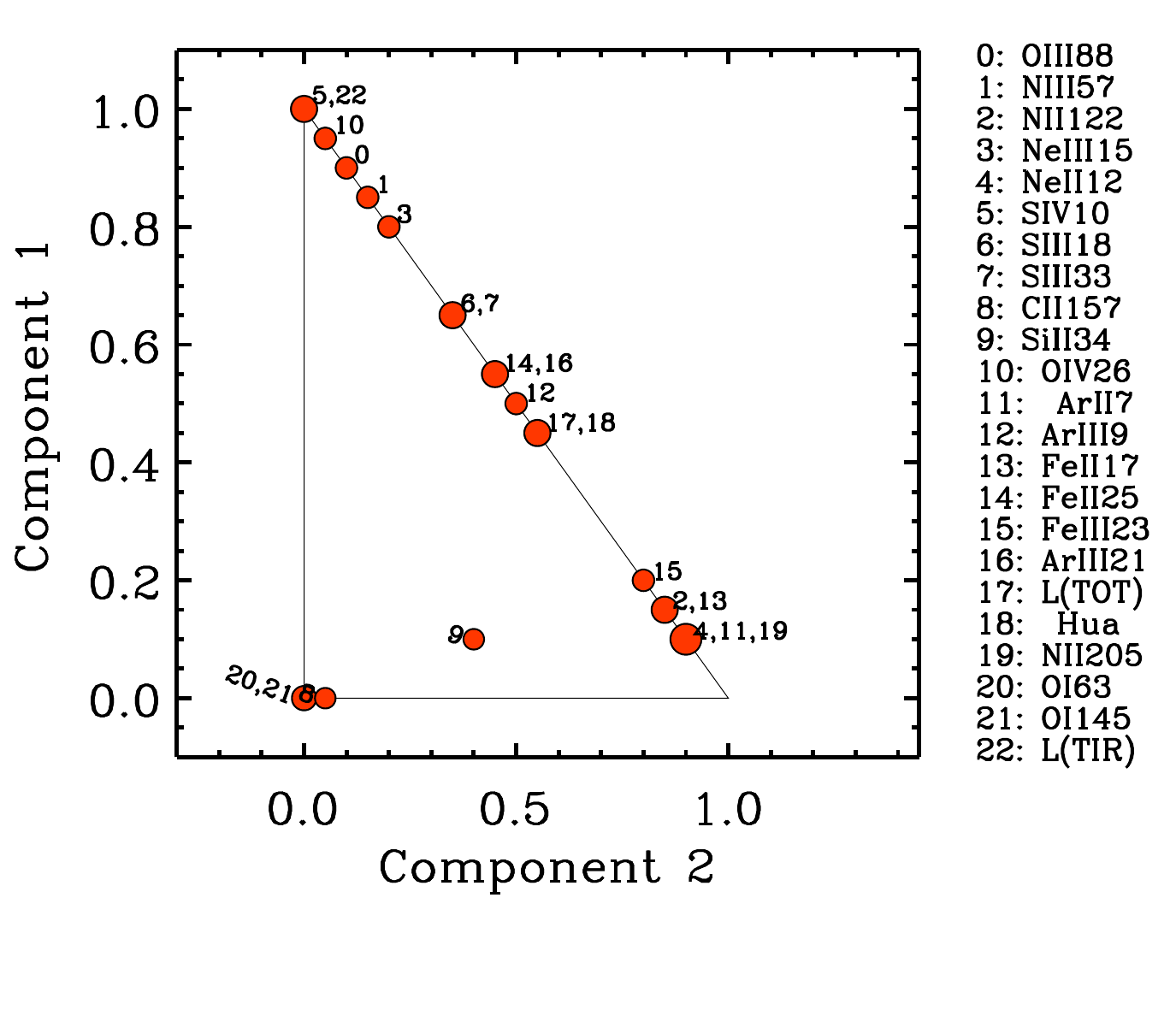}\\
(c)
\includegraphics[clip,width=8.8cm]{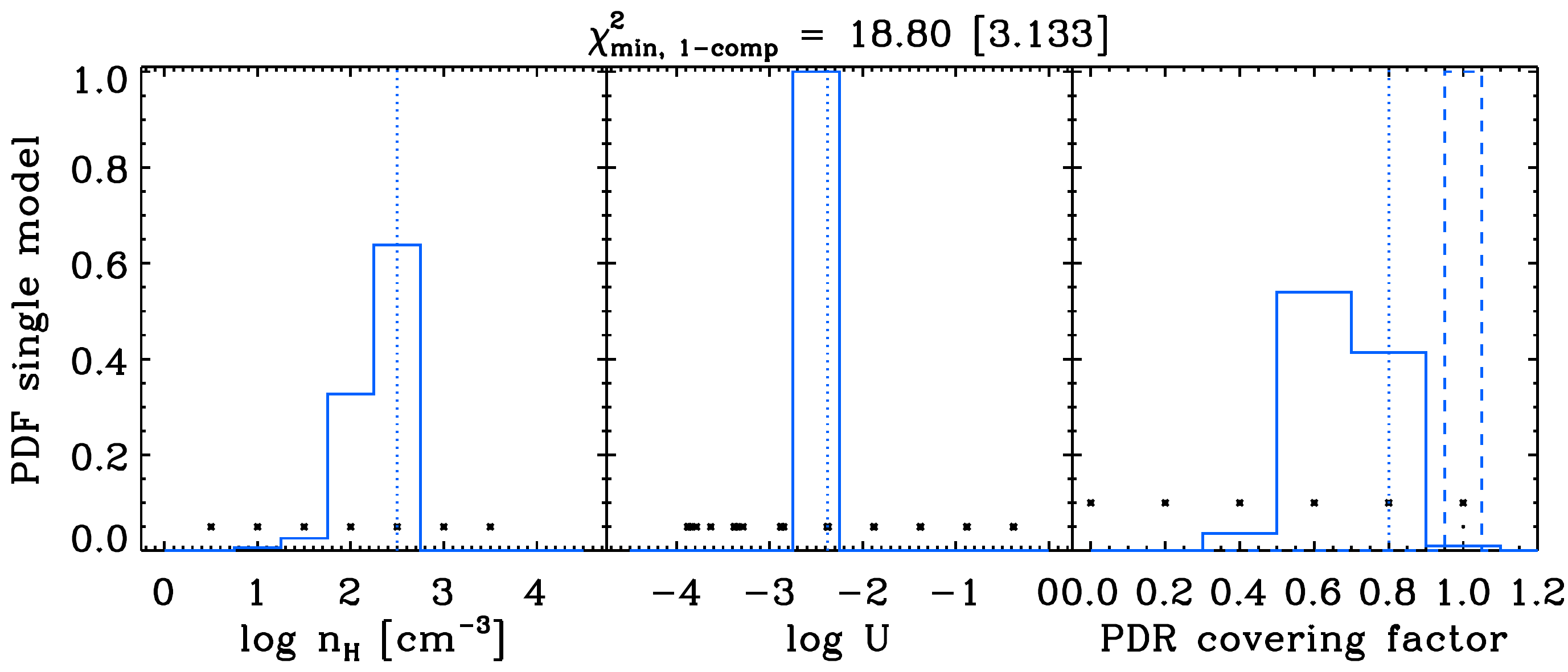} \hfill{}
\includegraphics[clip,width=8.8cm]{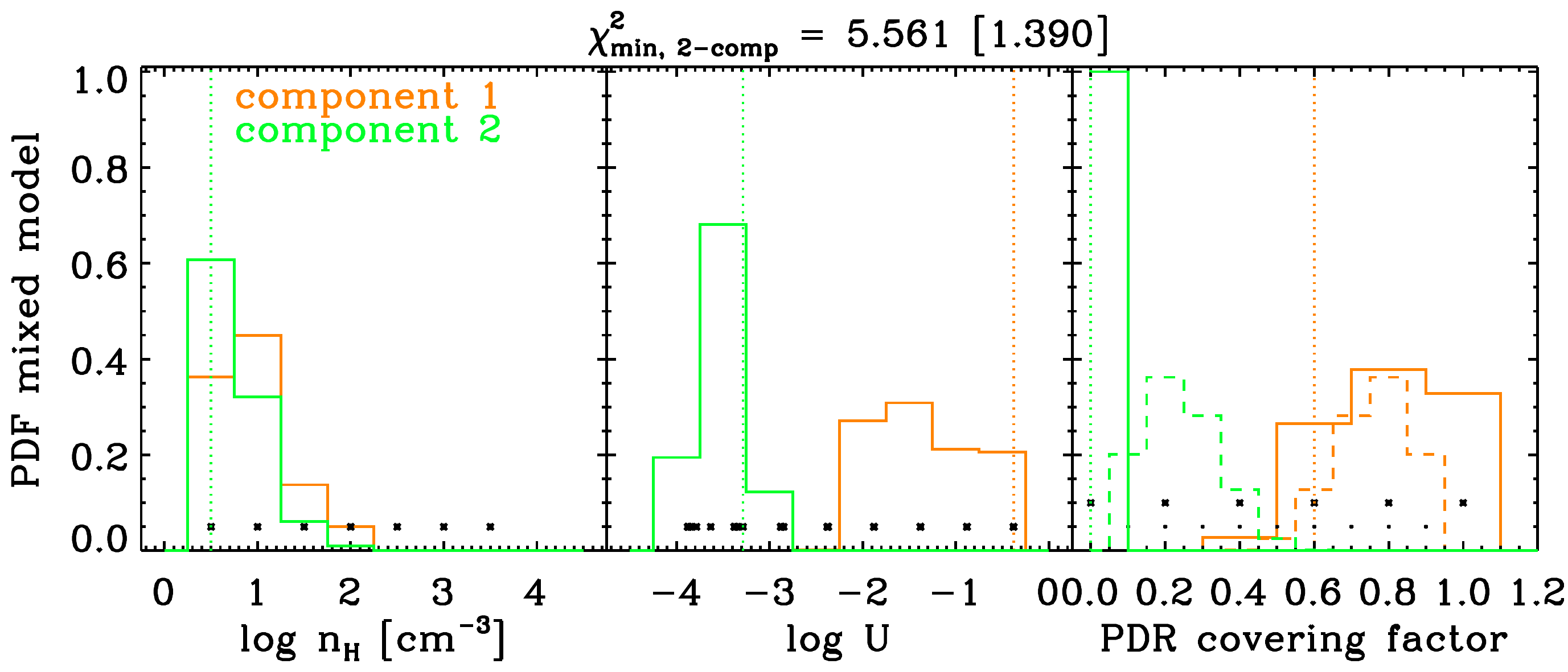}
\caption{ \textbf{Results for SBS\,1533.}
See above for caption description.
}
\label{fig:bests-sbs1533}
\end{figure*}
\clearpage
 \begin{figure*}[thp]
\centering
(a)
\includegraphics[clip,width=11cm]{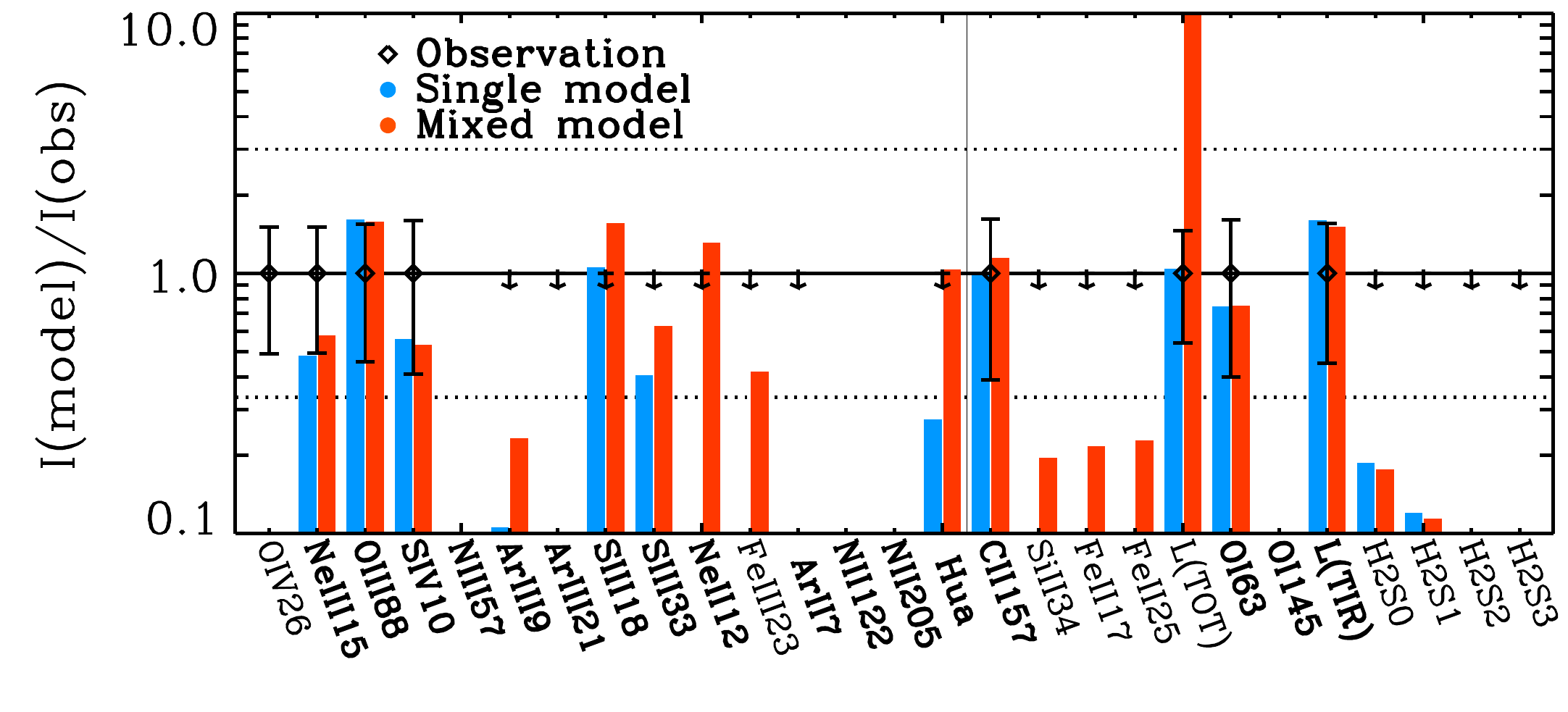} \hfill{}
(b)
\includegraphics[clip,width=6.2cm]{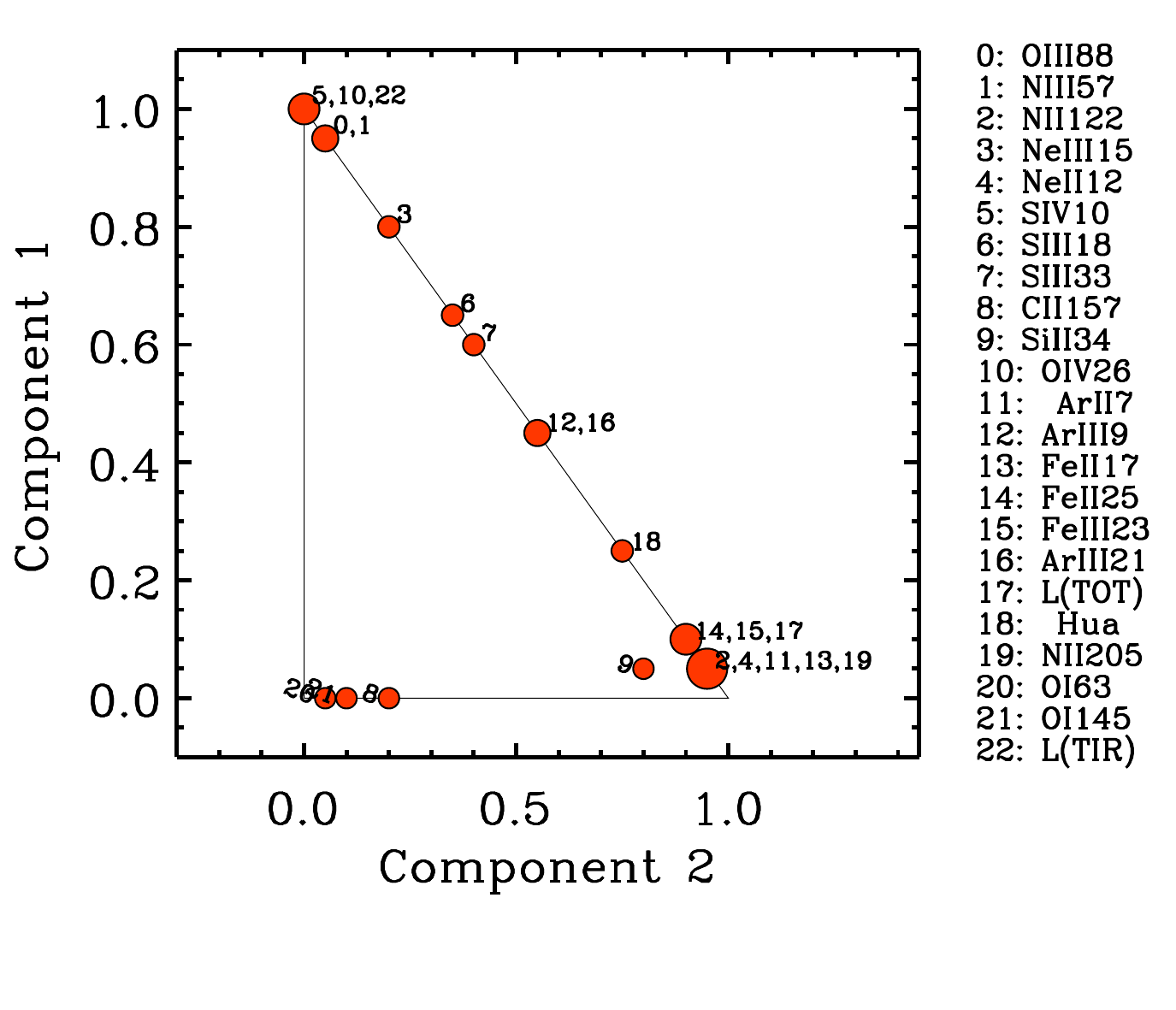}\\
(c)
\includegraphics[clip,width=8.8cm]{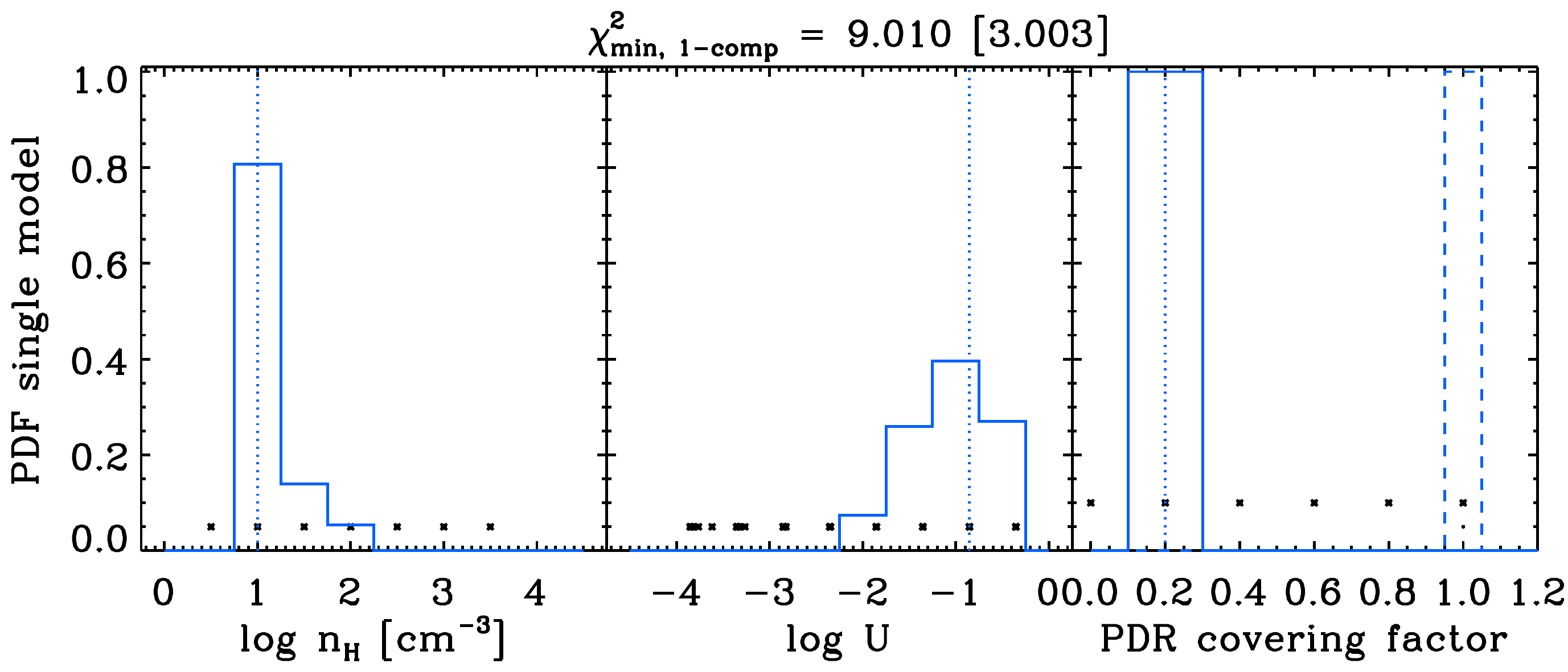} \hfill{}
\includegraphics[clip,width=8.8cm]{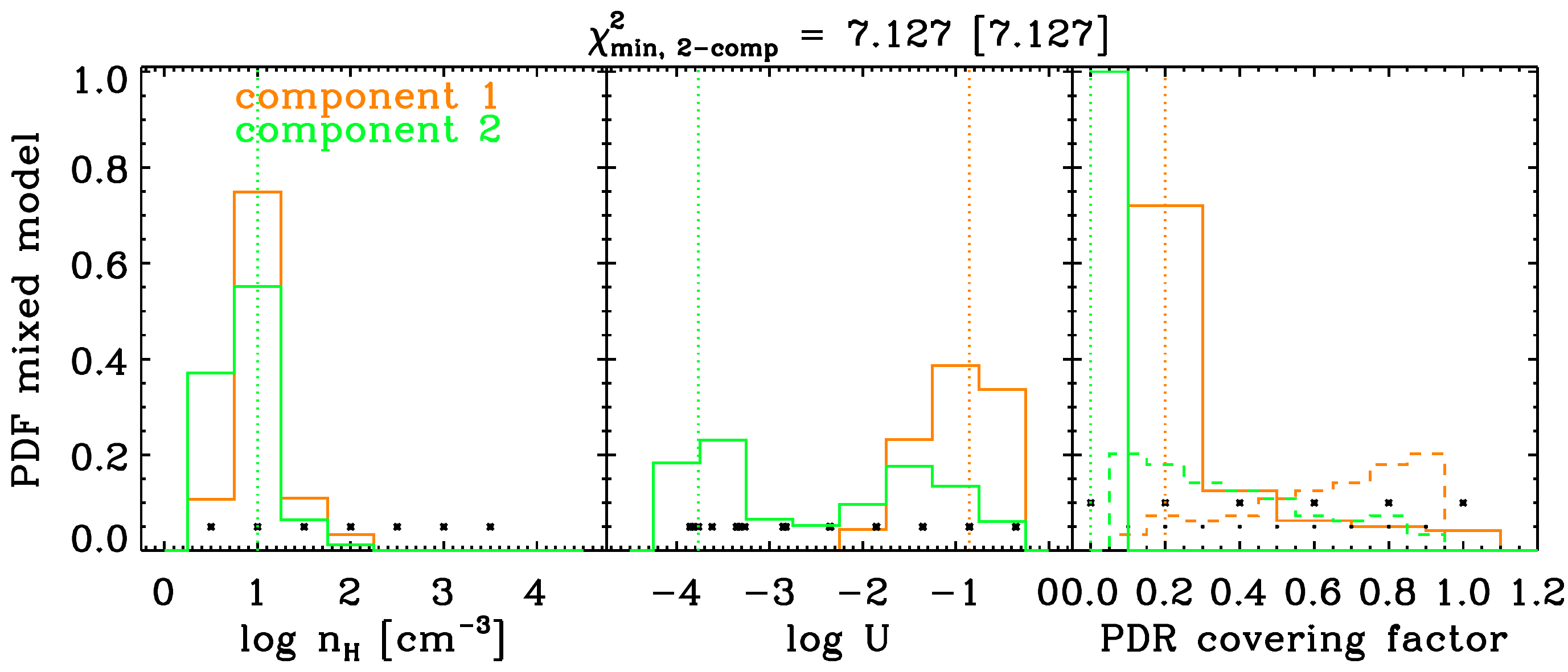}
\caption{ \textbf{Results for Tol\,1214.}
$(a)$~Comparison of the observed (losange) and predicted 
intensities for our best-fitting single (blue bar) and mixed (red bar)
\hii region+PDR models. Lines are sorted by decreasing energy.
The fitted lines have labels in boldface.
$(b)$~For mixed models: respective contributions from
component \#1 and component \#2 to the line prediction.
For PDR lines (\cii, \oi, and \silii), only the contribution from
the ionized gas is shown; the PDR contribution is
one minus the sum of the contributions from the plot.
Contributions are expected to be dominated by one
or the other component if their model parameters
are noticeably different.
$(c)$~Probability density functions of the model parameters
($n_{\rm H}$, $U$, $cov_{\rm PDR}$/scaling factor)
for single (left panel) and mixed (right panel) models.
The scaling factor corresponds to the proportion in which
we combine two models in the mixed model case. It is
equal to unity otherwise. It is shown with dashed lines in
the same panels as the PDR covering factor. Vertical dotted lines
show values of the best-fitting model. Black asterisks
show the range and step of values of the grid parameters.
}
\label{fig:bests-tol1214}
\end{figure*}
 \begin{figure*}[thp]
\centering
(a)
\includegraphics[clip,width=11cm]{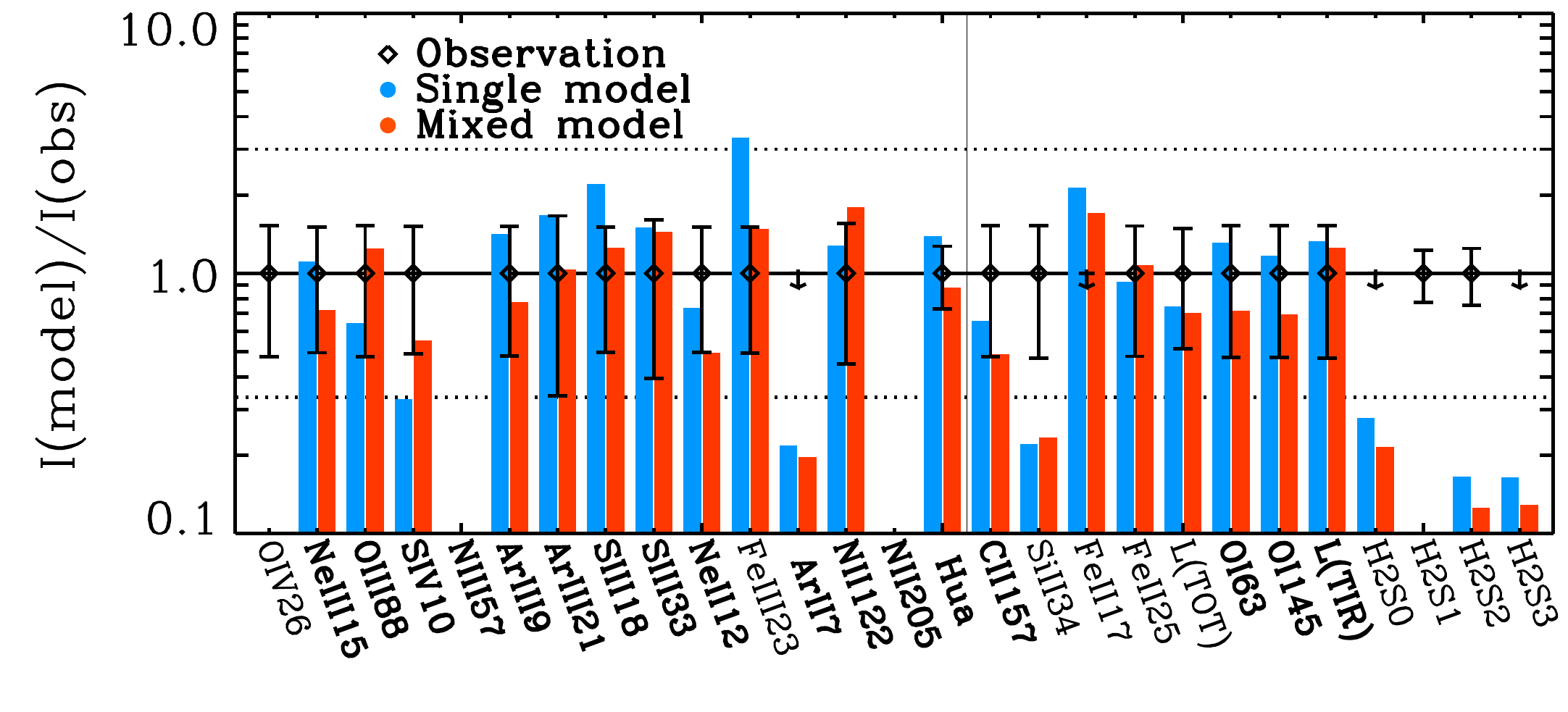} \hfill{}
(b)
\includegraphics[clip,width=6.2cm]{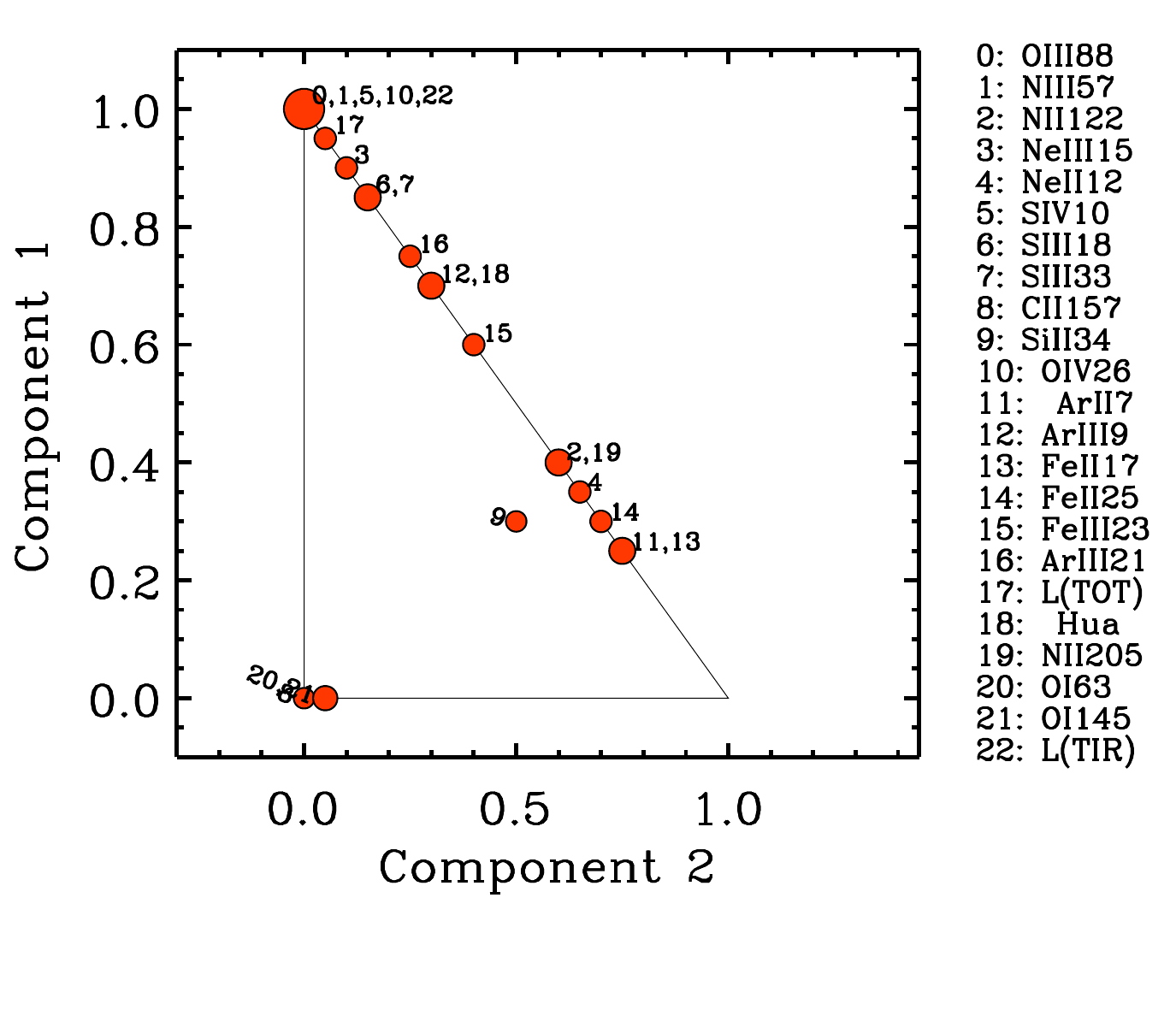}\\
(c)
\includegraphics[clip,width=8.8cm]{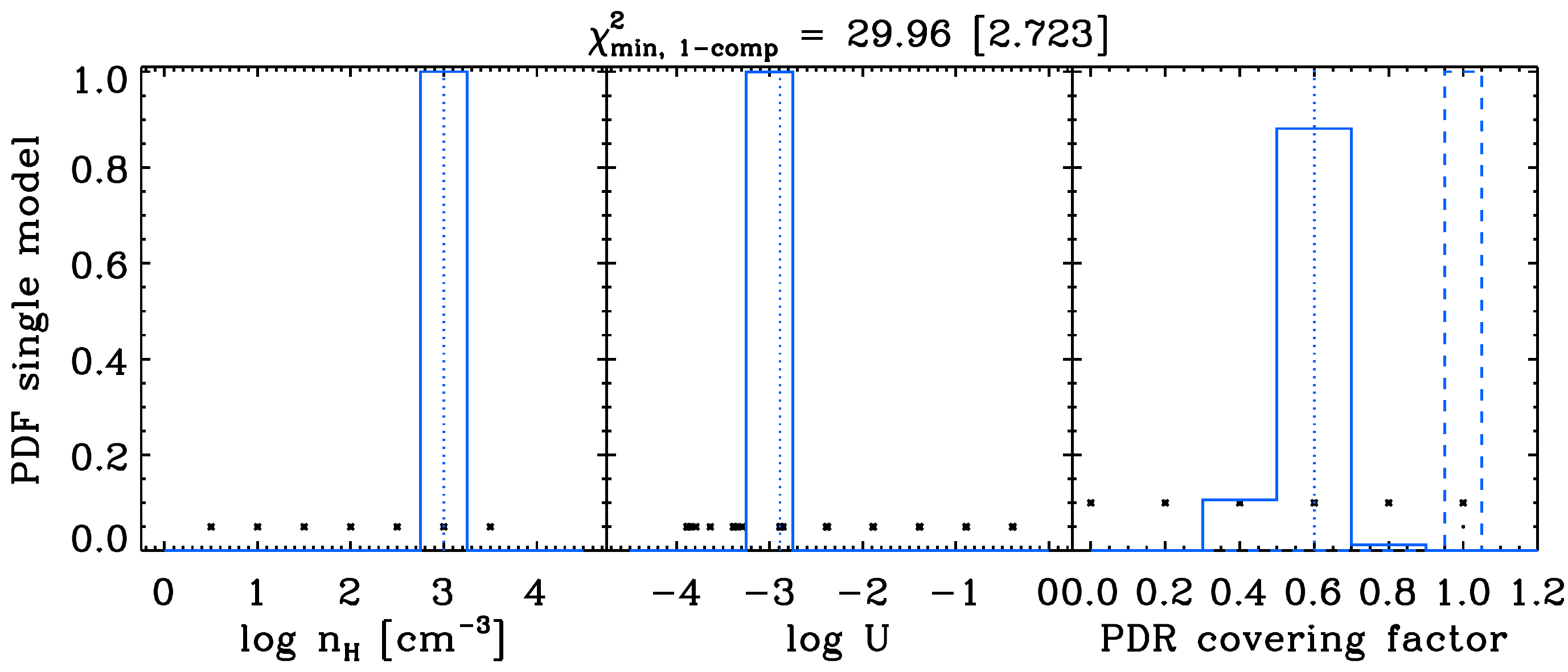} \hfill{}
\includegraphics[clip,width=8.8cm]{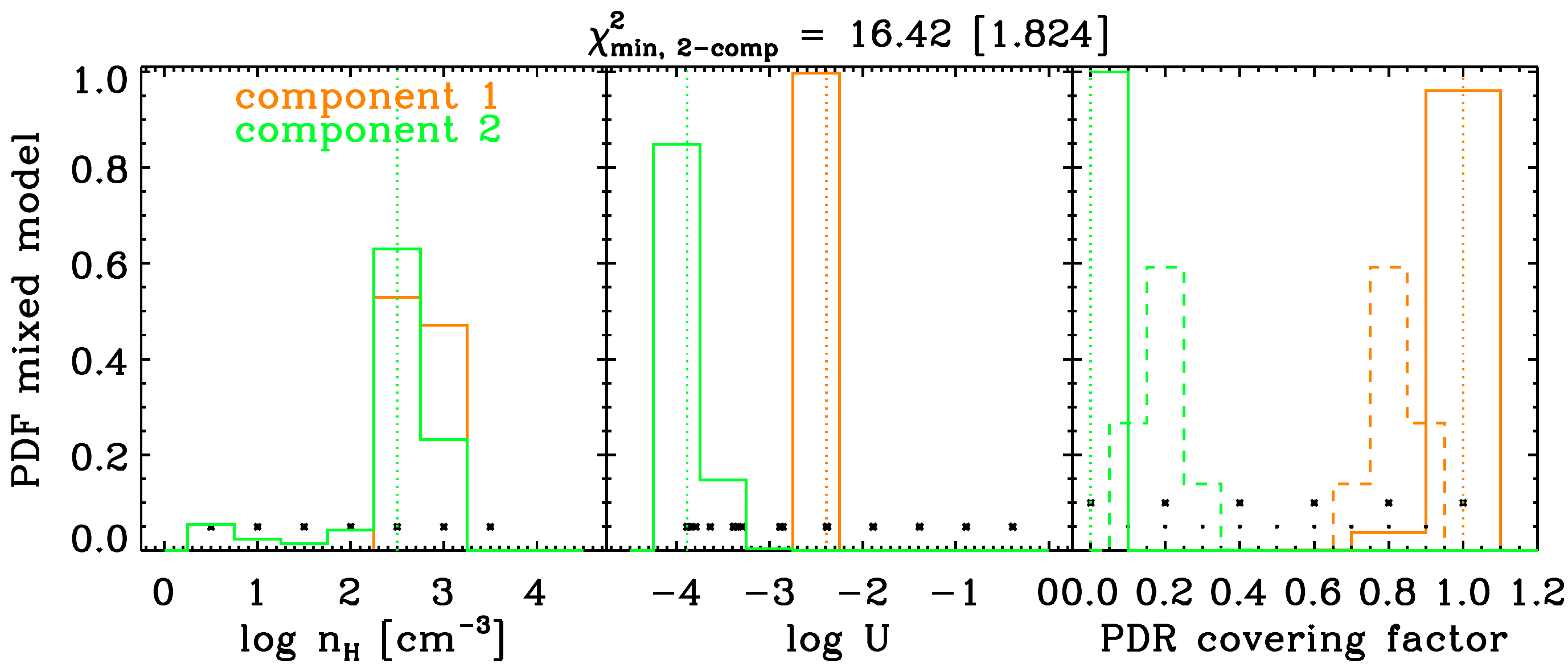}
\caption{ \textbf{Results for UM\,448.}
See above for caption description.
}
\label{fig:bests-um448}
\end{figure*}
\clearpage
 \begin{figure*}[thp]
\centering
(a)
\includegraphics[clip,width=11cm]{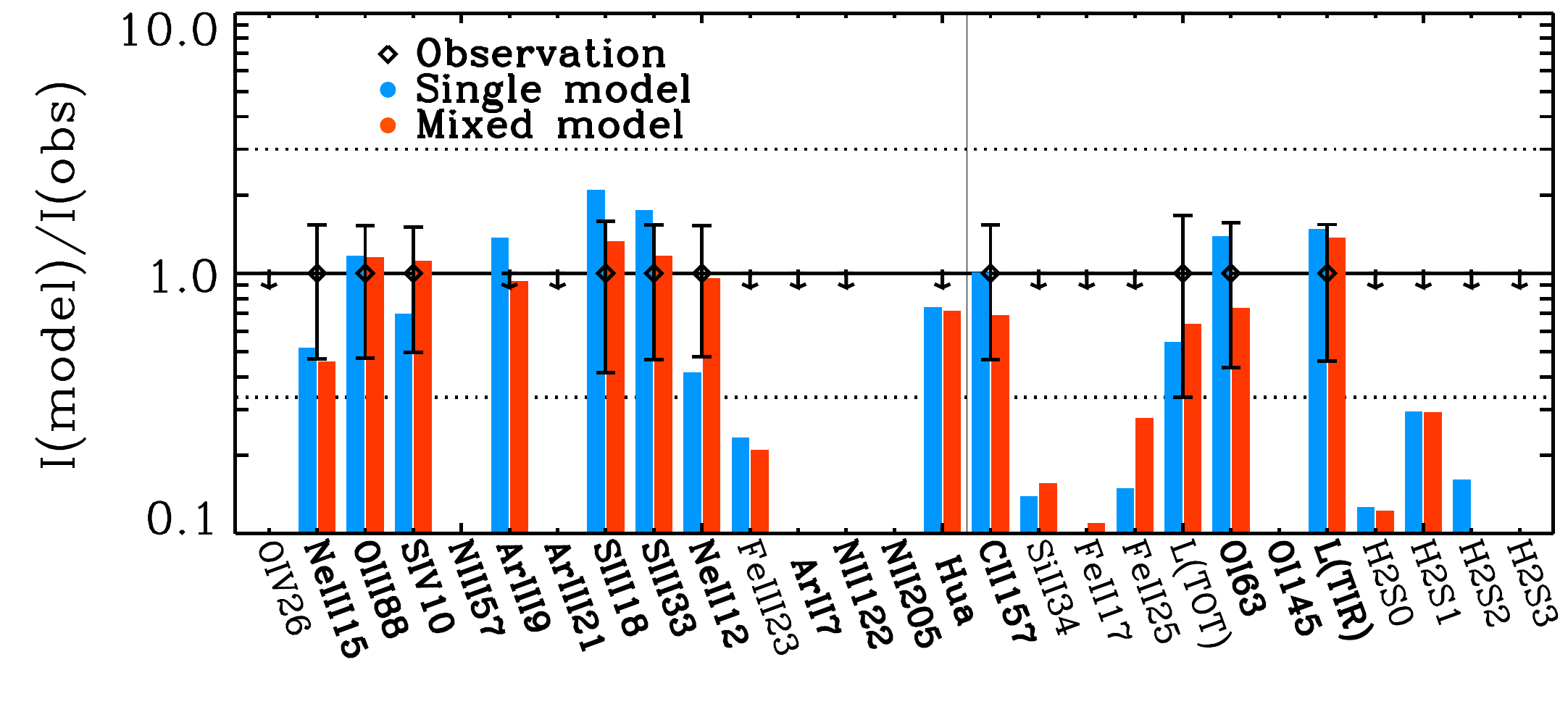} \hfill{}
(b)
\includegraphics[clip,width=6.2cm]{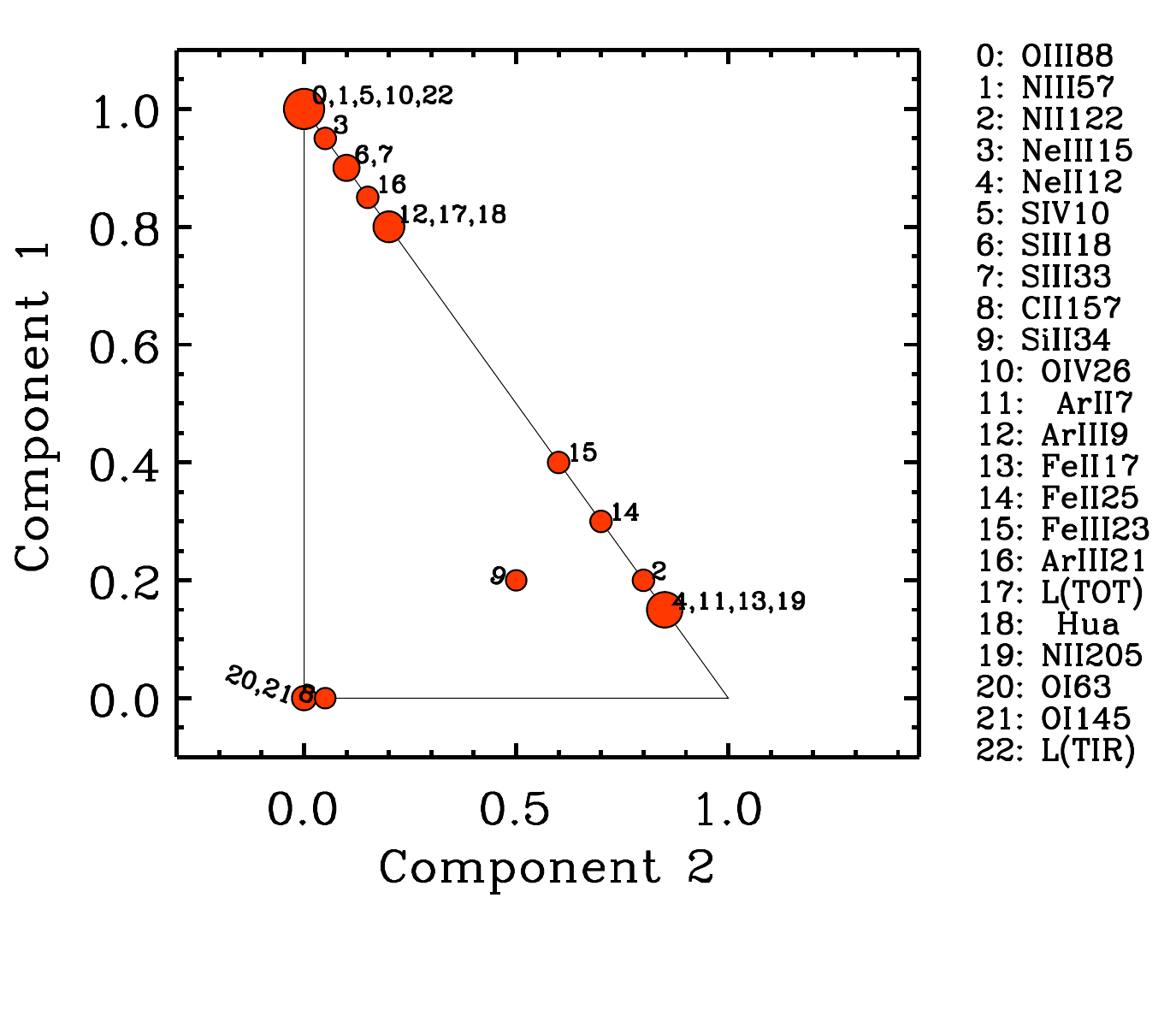}\\
(c)
\includegraphics[clip,width=8.8cm]{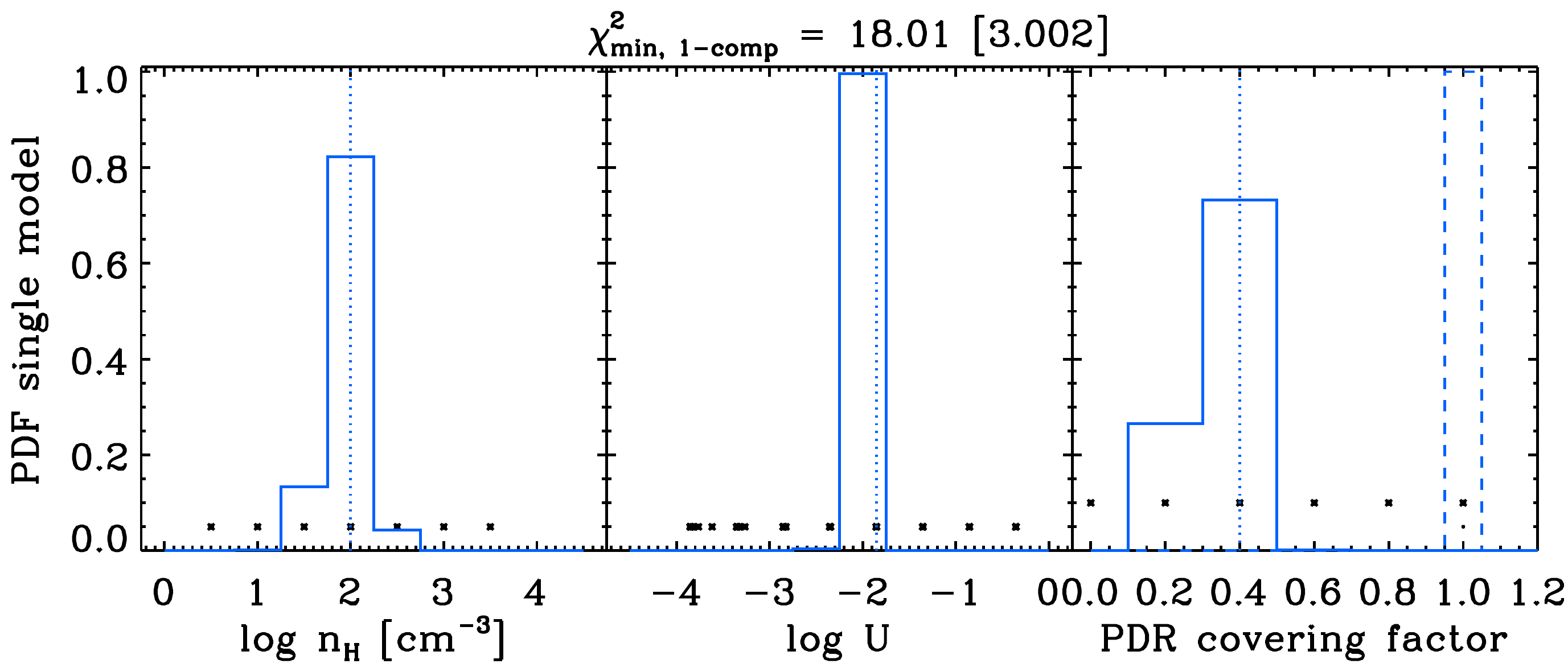} \hfill{}
\includegraphics[clip,width=8.8cm]{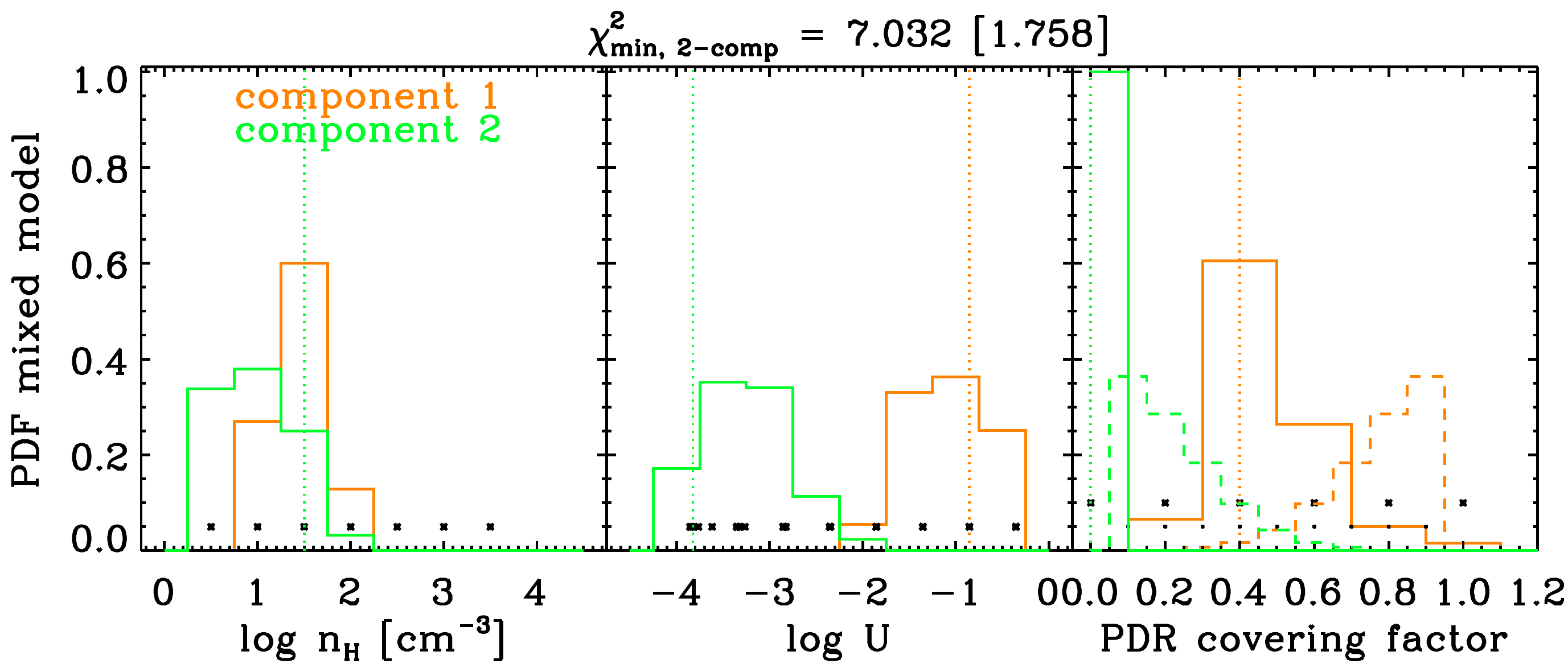}
\caption{ \textbf{Results for UM\,461.}
$(a)$~Comparison of the observed (losange) and predicted 
intensities for our best-fitting single (blue bar) and mixed (red bar)
\hii region+PDR models. Lines are sorted by decreasing energy.
The fitted lines have labels in boldface.
$(b)$~For mixed models: respective contributions from
component \#1 and component \#2 to the line prediction.
For PDR lines (\cii, \oi, and \silii), only the contribution from
the ionized gas is shown; the PDR contribution is
one minus the sum of the contributions from the plot.
Contributions are expected to be dominated by one
or the other component if their model parameters
are noticeably different.
$(c)$~Probability density functions of the model parameters
($n_{\rm H}$, $U$, $cov_{\rm PDR}$/scaling factor)
for single (left panel) and mixed (right panel) models.
The scaling factor corresponds to the proportion in which
we combine two models in the mixed model case. It is
equal to unity otherwise. It is shown with dashed lines in
the same panels as the PDR covering factor. Vertical dotted lines
show values of the best-fitting model. Black asterisks
show the range and step of values of the grid parameters.
}
\label{fig:bests-um461}
\end{figure*}
 \begin{figure*}[thp]
\centering
(a)
\includegraphics[clip,width=11cm]{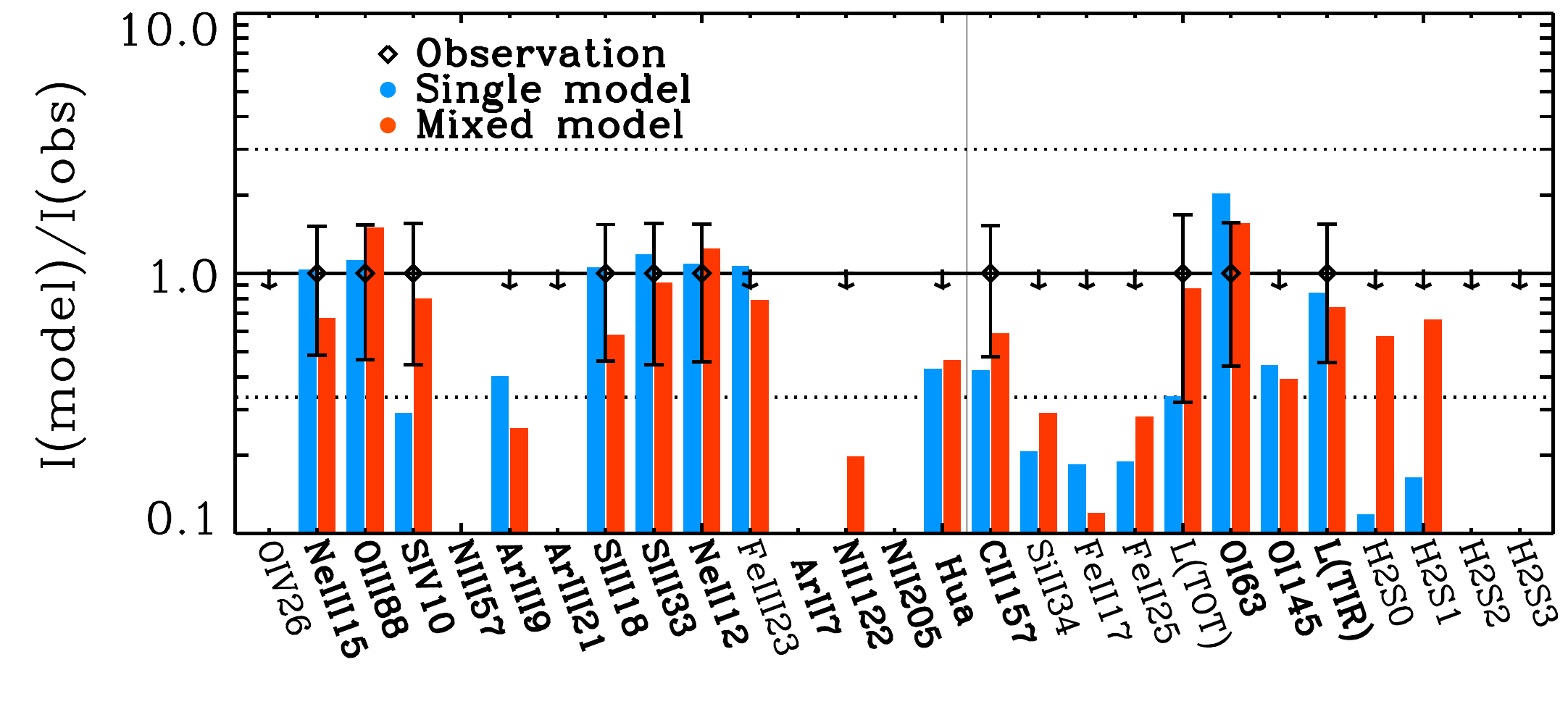} \hfill{}
(b)
\includegraphics[clip,width=6.2cm]{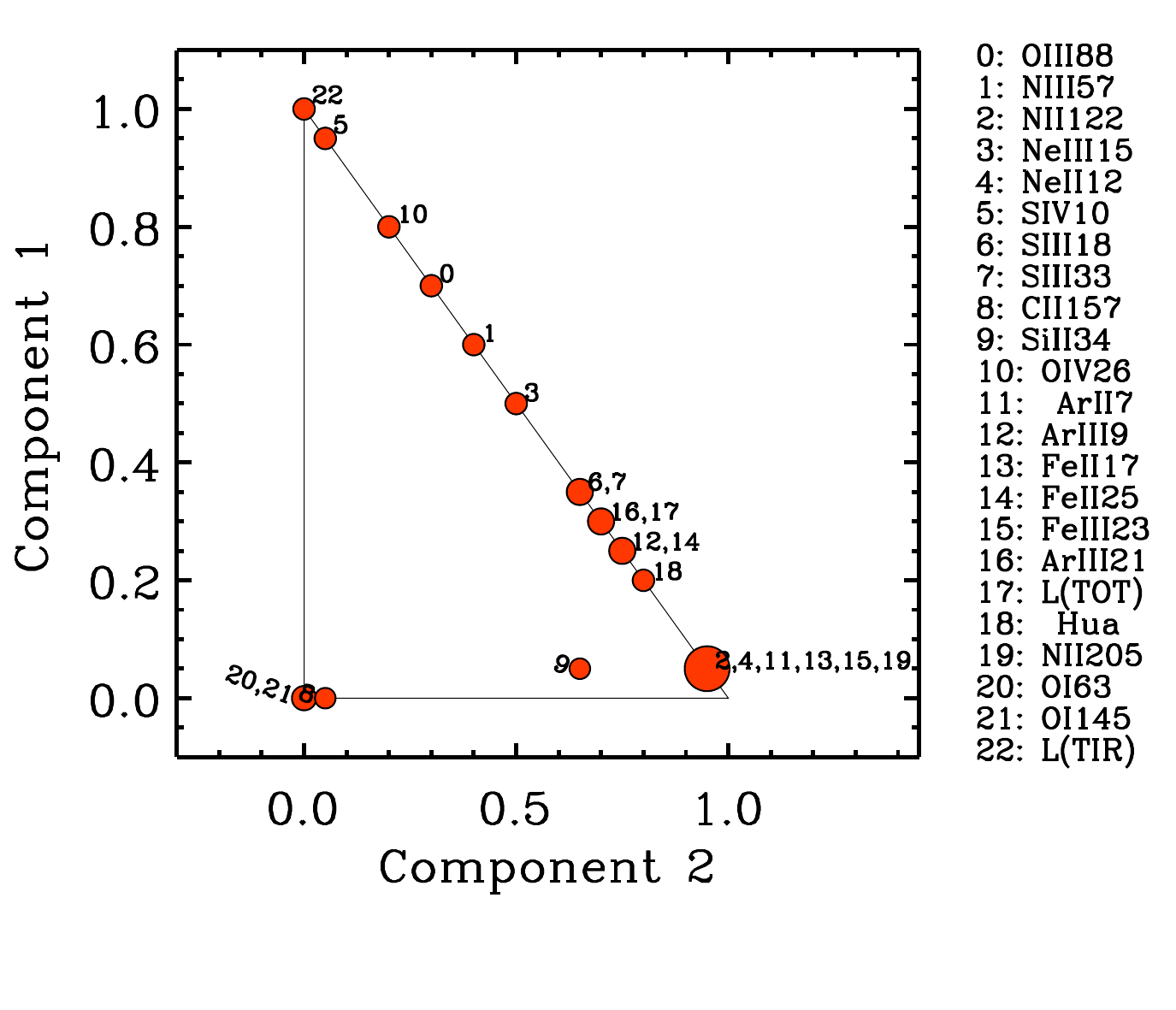}\\
(c)
\includegraphics[clip,width=8.8cm]{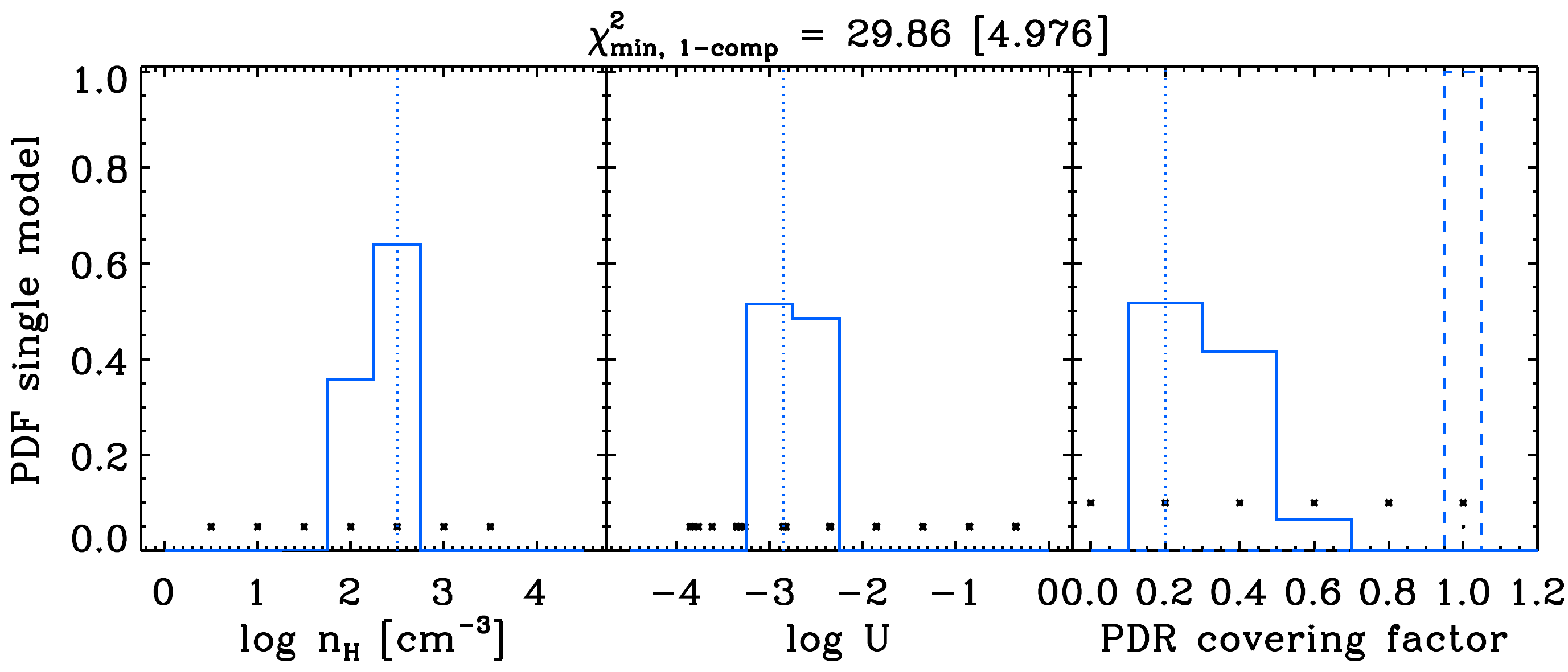} \hfill{}
\includegraphics[clip,width=8.8cm]{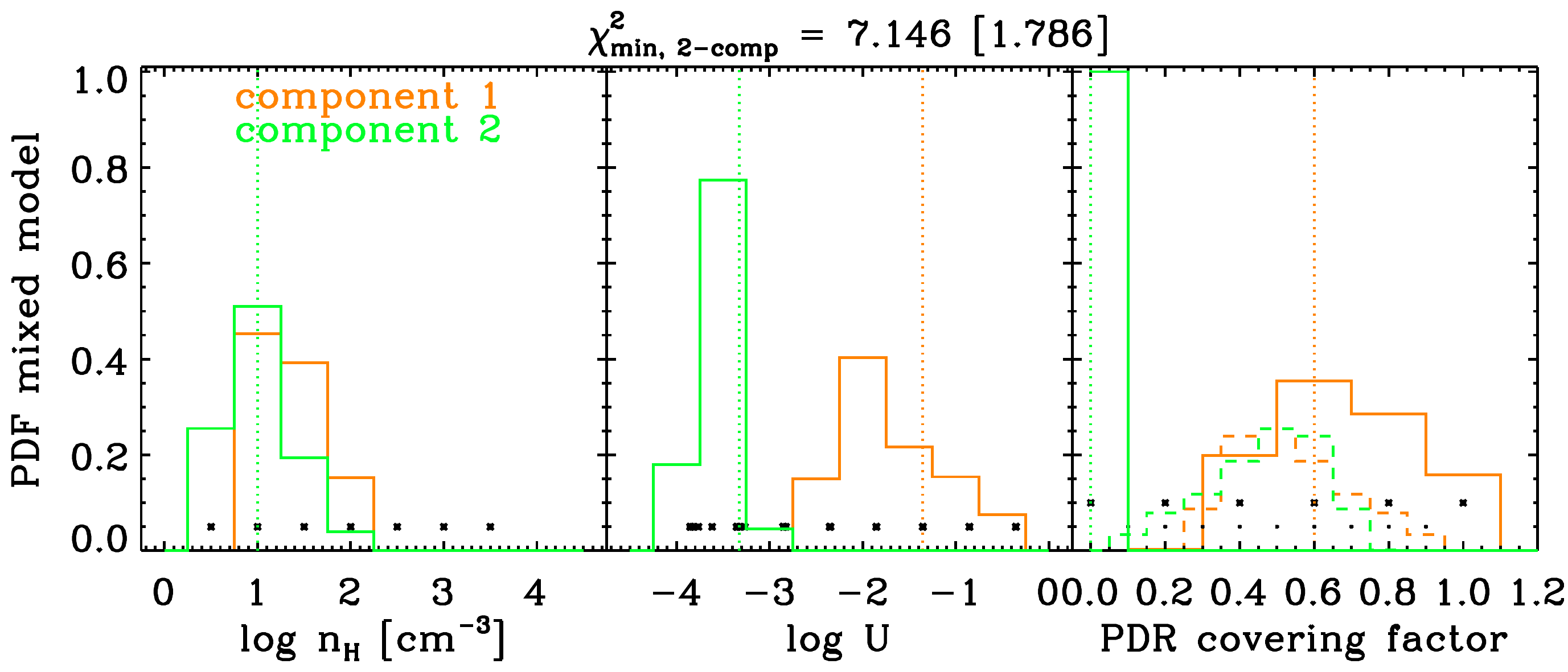}
\caption{ \textbf{Results for VII\,Zw\,403.}
See above for caption description.
}
\label{fig:bests-viizw403}
\end{figure*}
\clearpage
 \begin{figure*}[thp]
\centering
(a)
\includegraphics[clip,width=11cm]{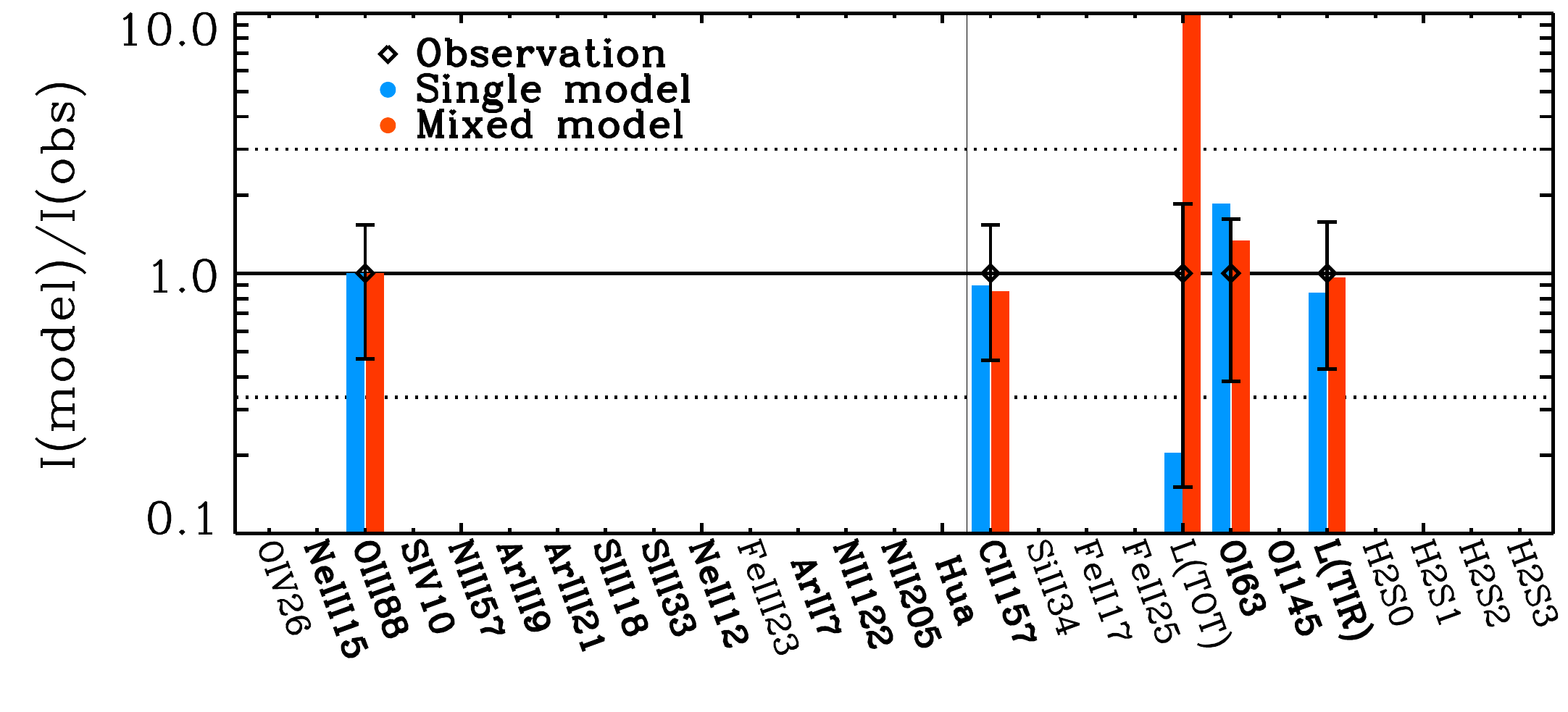} \hfill{}
(b)
\includegraphics[clip,width=6.2cm]{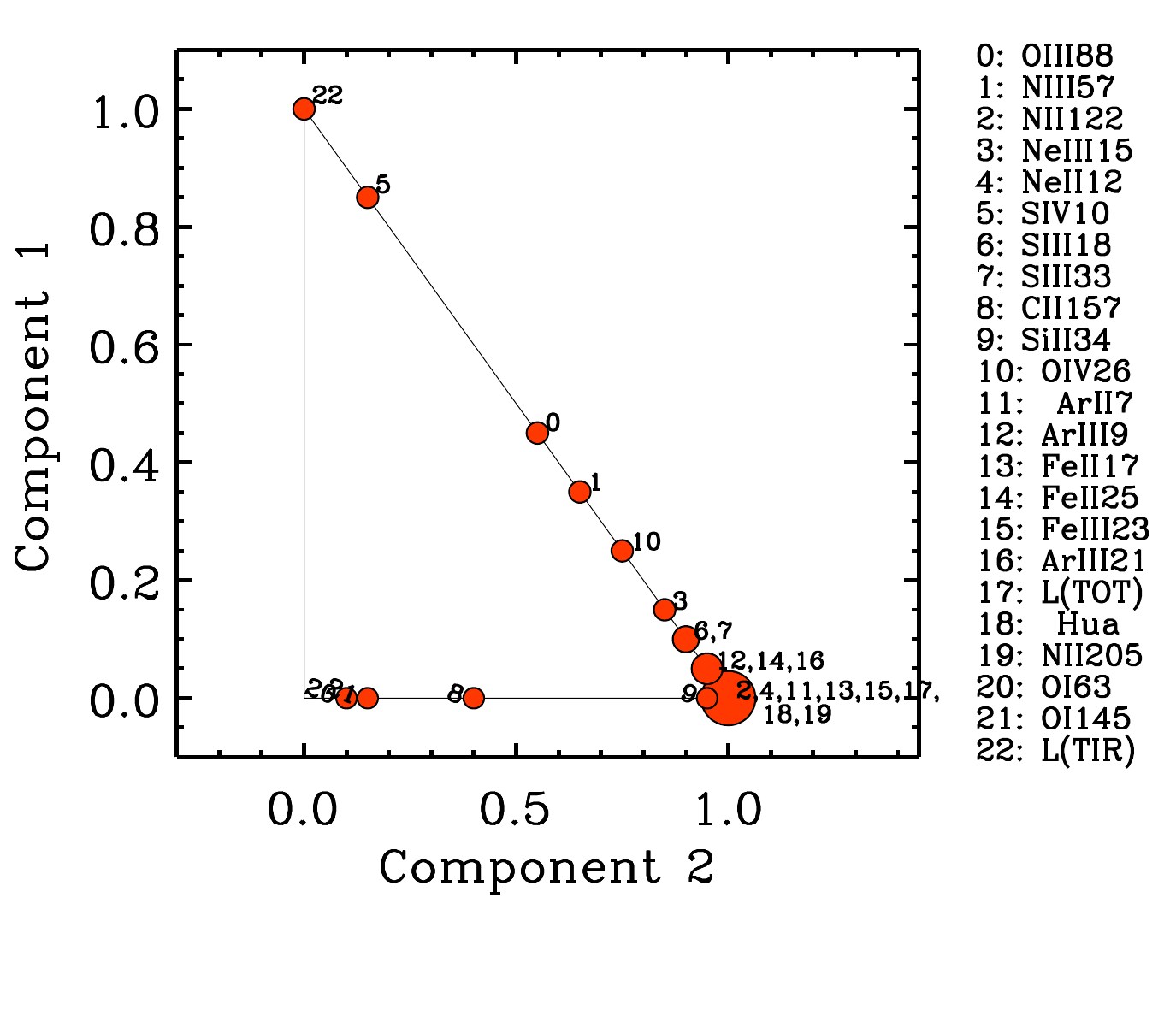}\\
(c)
\includegraphics[clip,width=8.8cm]{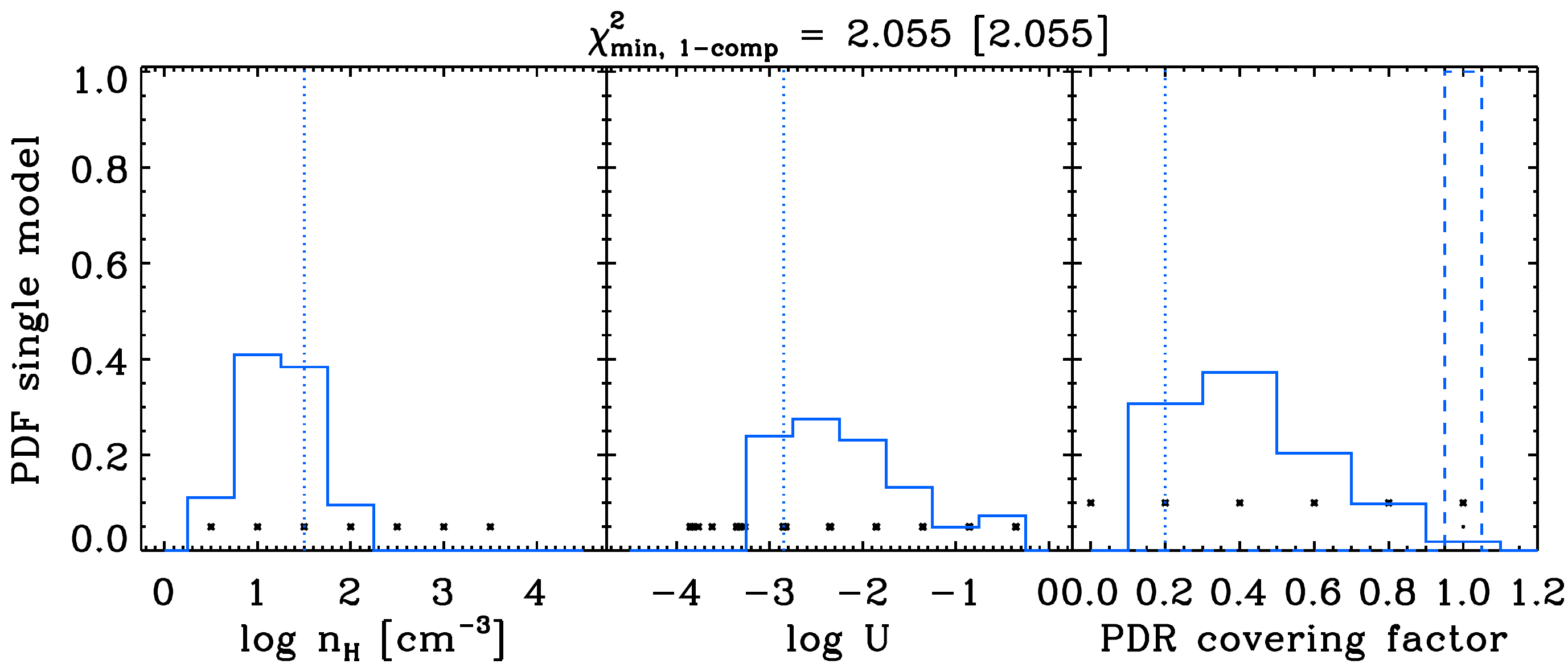} \hfill{}
\includegraphics[clip,width=8.8cm]{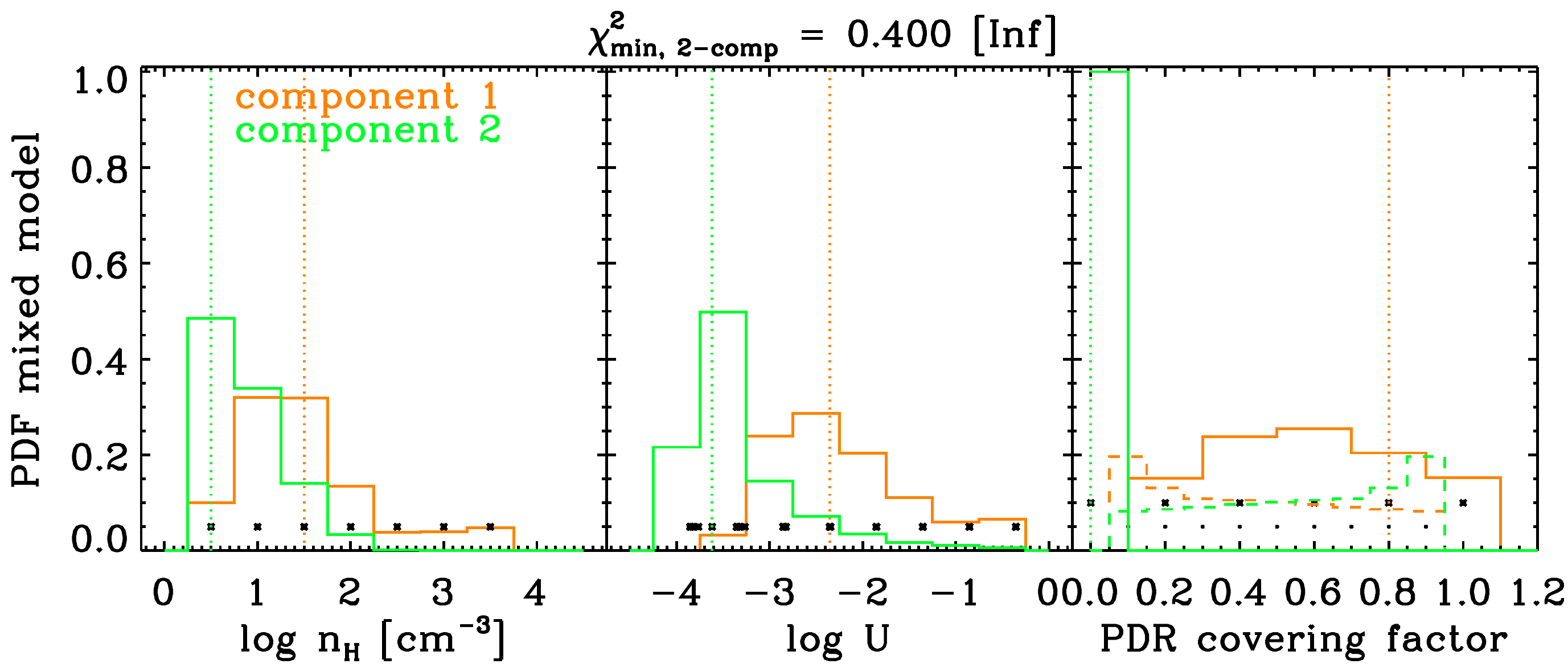}
\caption{ \textbf{Results for UGC\,4483.}
$(a)$~Comparison of the observed (losange) and predicted 
intensities for our best-fitting single (blue bar) and mixed (red bar)
\hii region+PDR models. Lines are sorted by decreasing energy.
The fitted lines have labels in boldface.
$(b)$~For mixed models: respective contributions from
component \#1 and component \#2 to the line prediction.
For PDR lines (\cii, \oi, and \silii), only the contribution from
the ionized gas is shown; the PDR contribution is
one minus the sum of the contributions from the plot.
Contributions are expected to be dominated by one
or the other component if their model parameters
are noticeably different.
$(c)$~Probability density functions of the model parameters
($n_{\rm H}$, $U$, $cov_{\rm PDR}$/scaling factor)
for single (left panel) and mixed (right panel) models.
The scaling factor corresponds to the proportion in which
we combine two models in the mixed model case. It is
equal to unity otherwise. It is shown with dashed lines in
the same panels as the PDR covering factor. Vertical dotted lines
show values of the best-fitting model. Black asterisks
show the range and step of values of the grid parameters.
}
\label{fig:bests-ugc4483}
\end{figure*}
 \begin{figure*}[thp]
\centering
(a)
\includegraphics[clip,width=11cm]{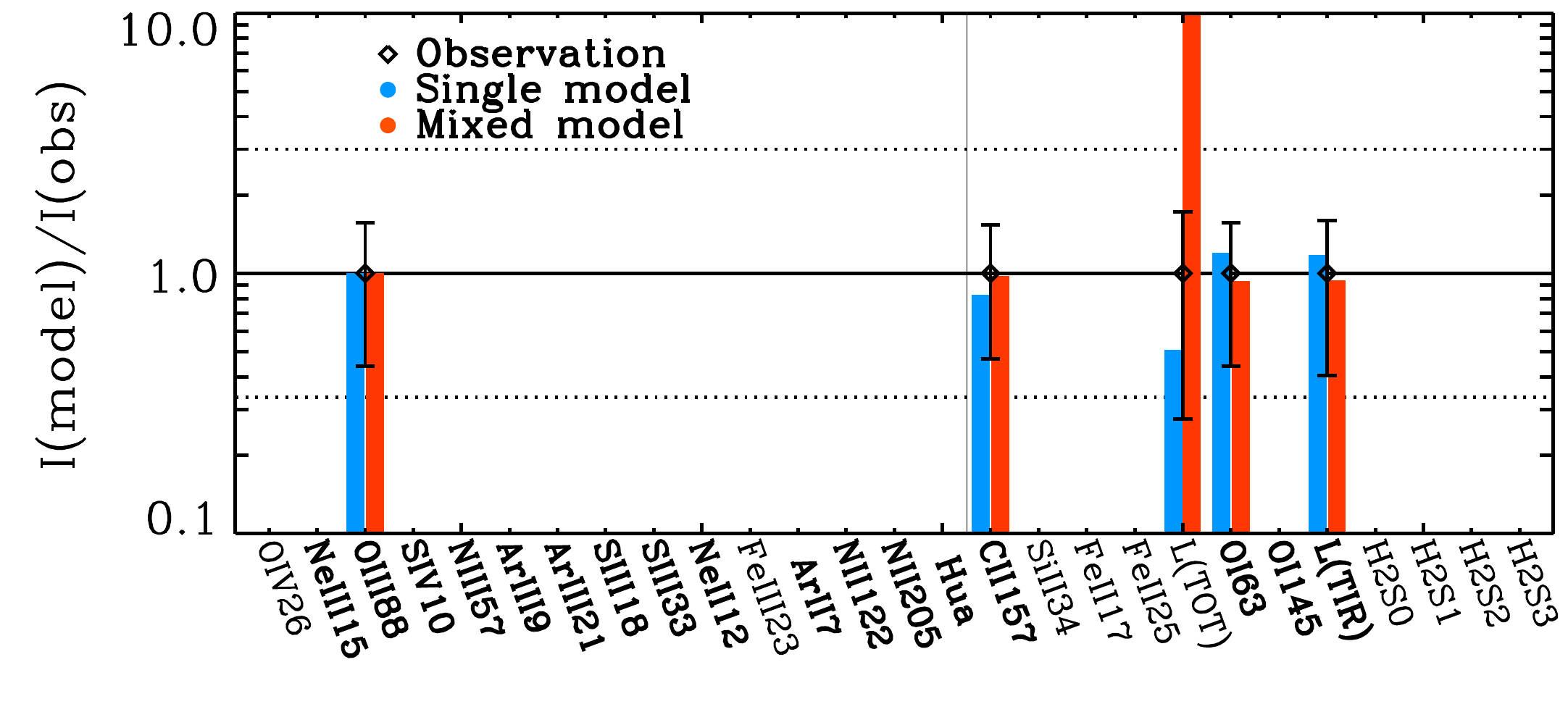} \hfill{}
(b)
\includegraphics[clip,width=6.2cm]{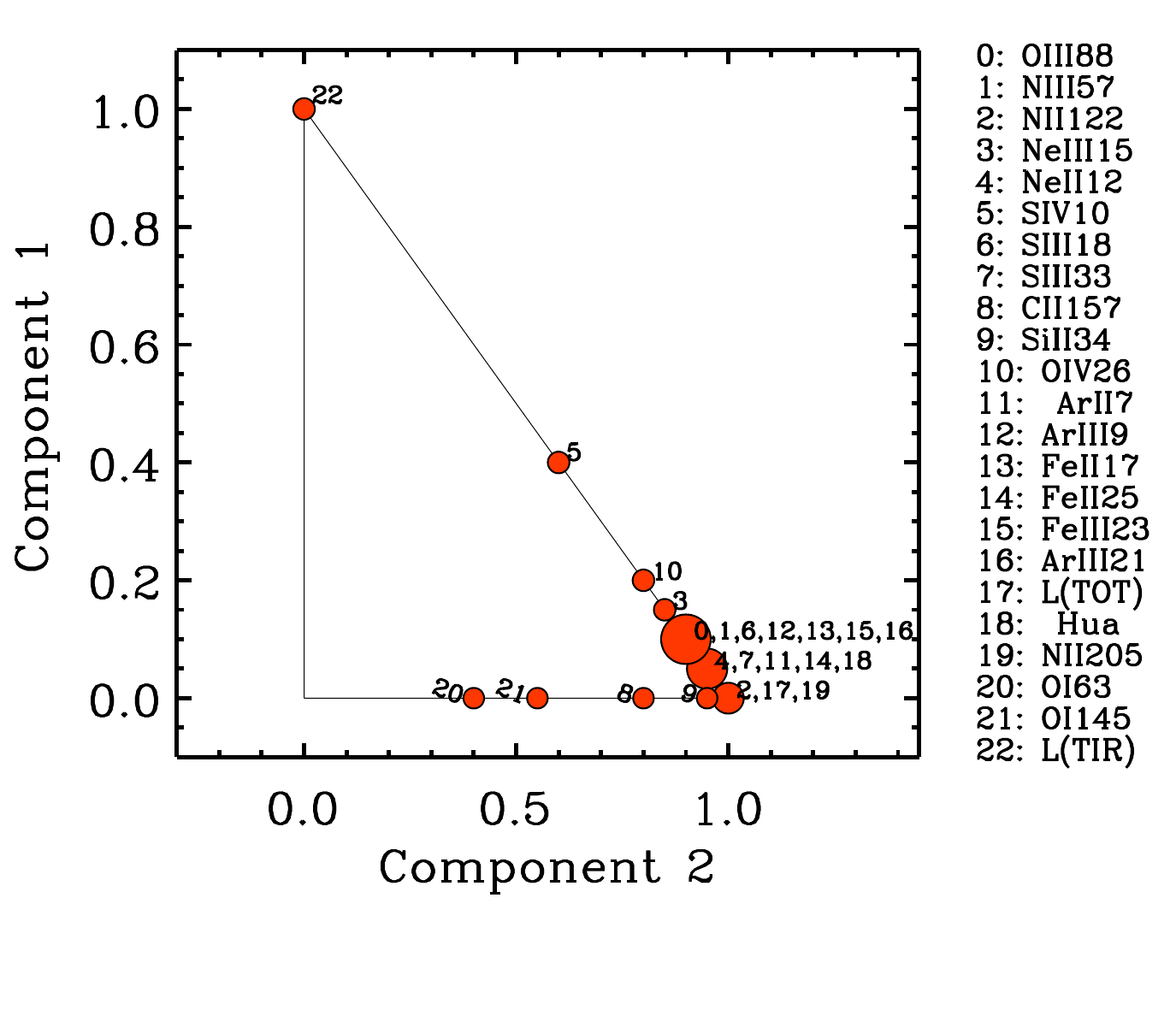}\\
(c)
\includegraphics[clip,width=8.8cm]{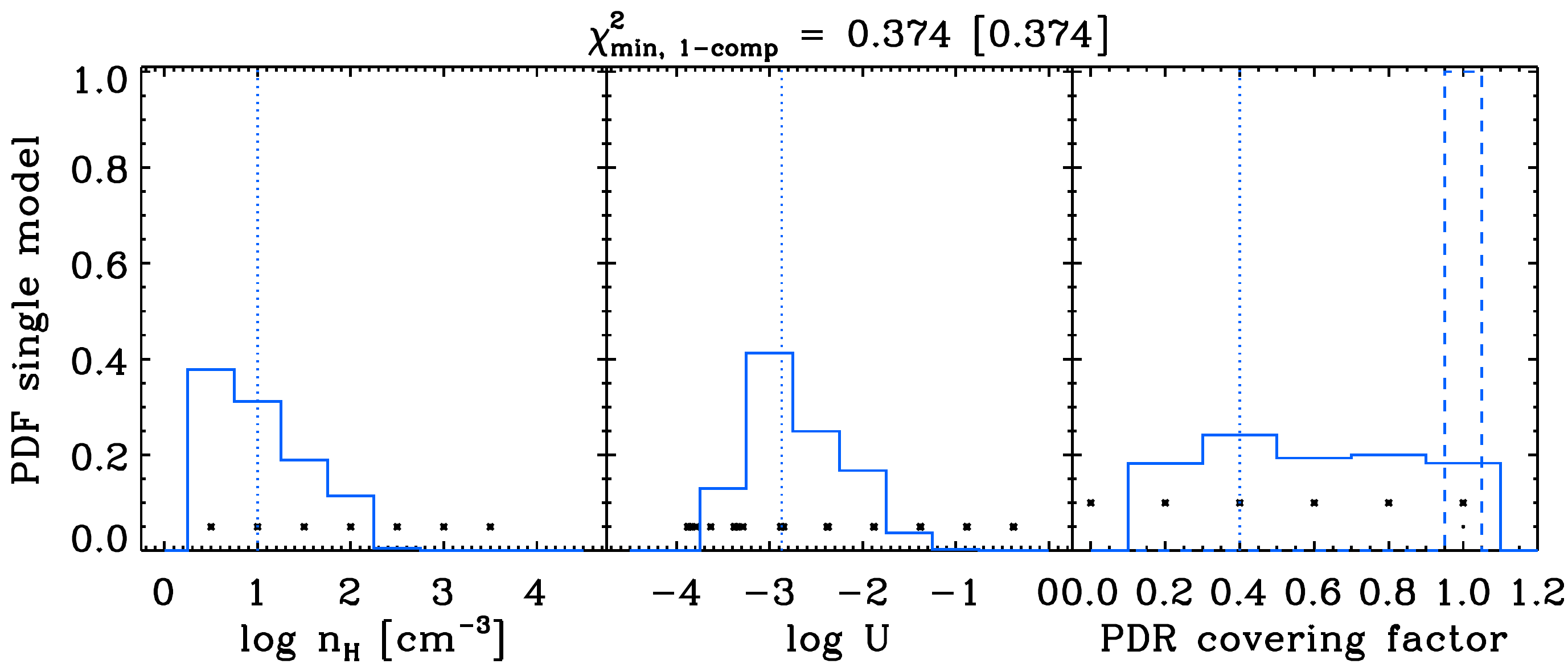} \hfill{}
\includegraphics[clip,width=8.8cm]{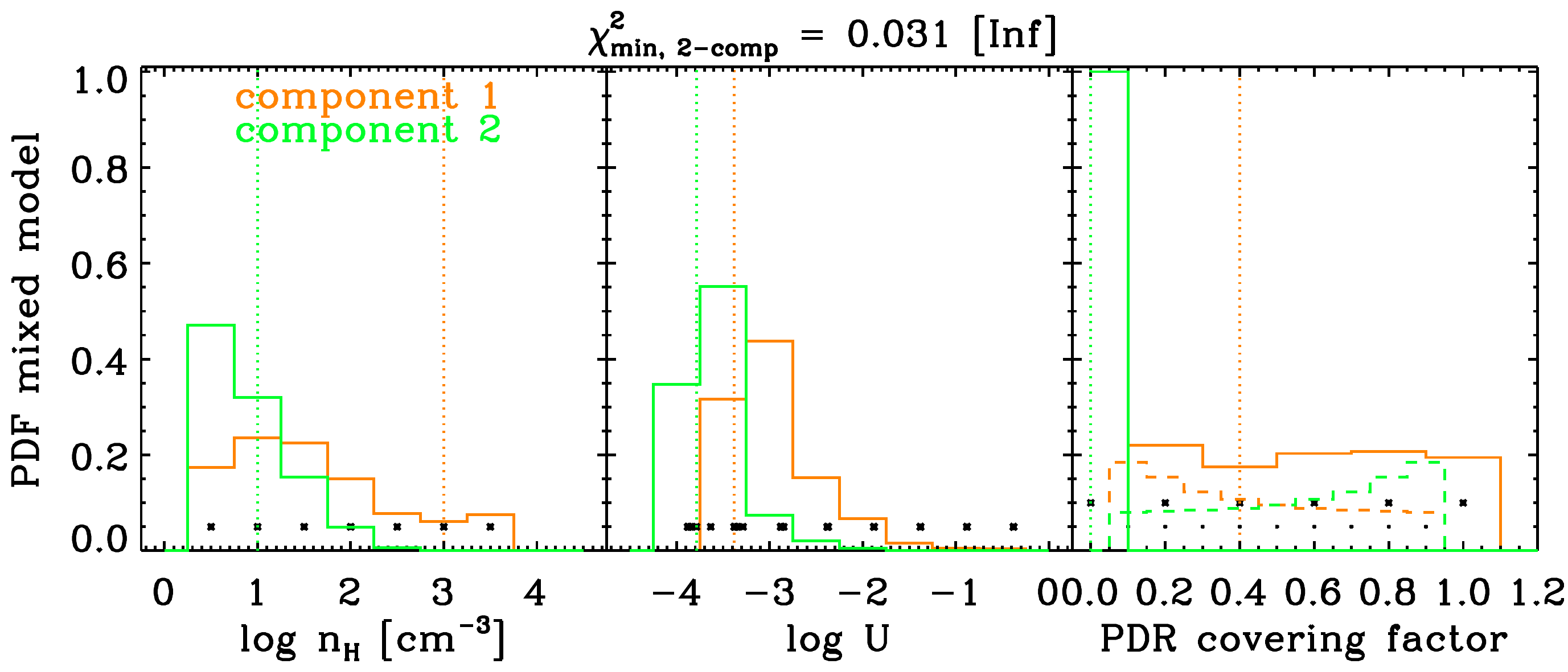}
\caption{ \textbf{Results for UM\,133.}
See above for caption description.
}
\label{fig:bests-um133}
\end{figure*}
\clearpage
 \begin{figure*}[thp]
\centering
(a)
\includegraphics[clip,width=11cm]{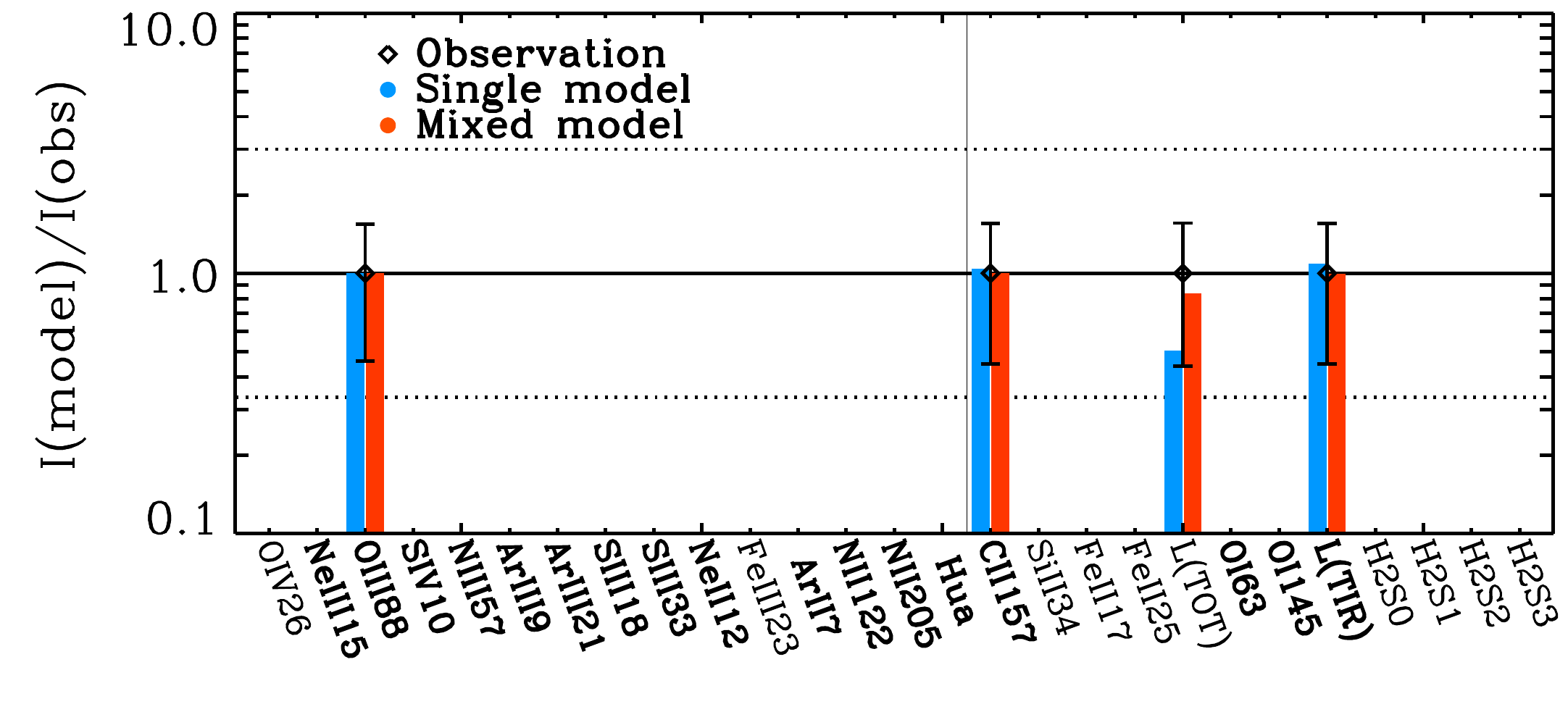} \hfill{}
(b)
\includegraphics[clip,width=6.2cm]{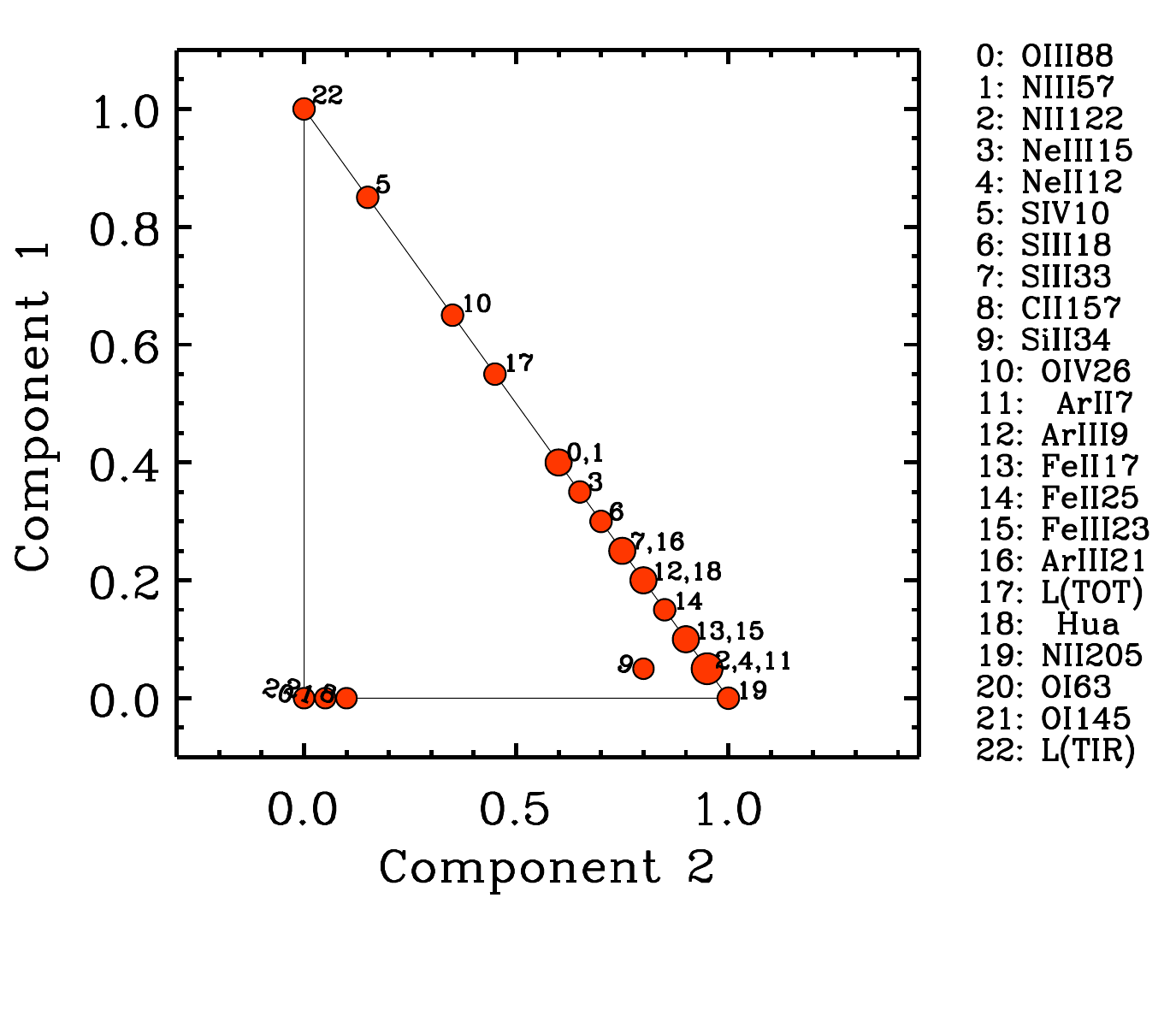}\\
(c)
\includegraphics[clip,width=8.8cm]{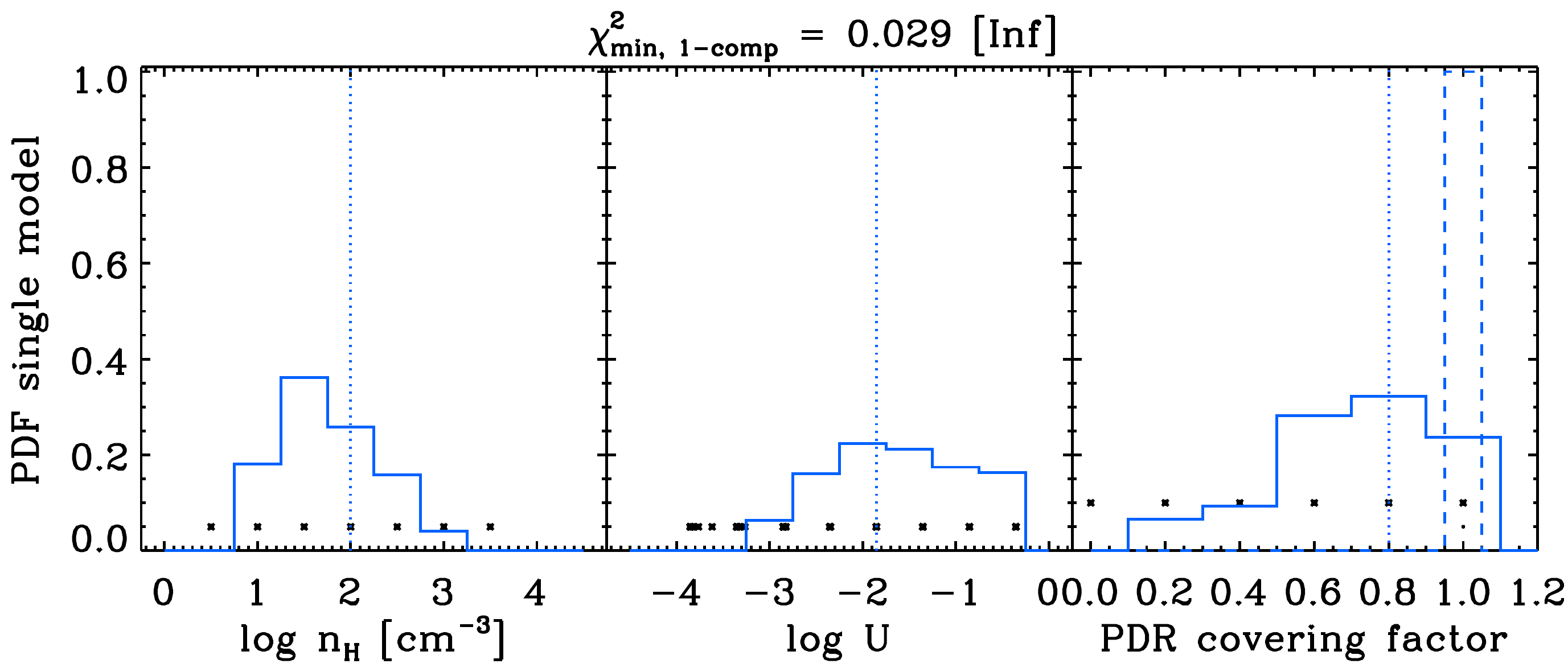} \hfill{}
\includegraphics[clip,width=8.8cm]{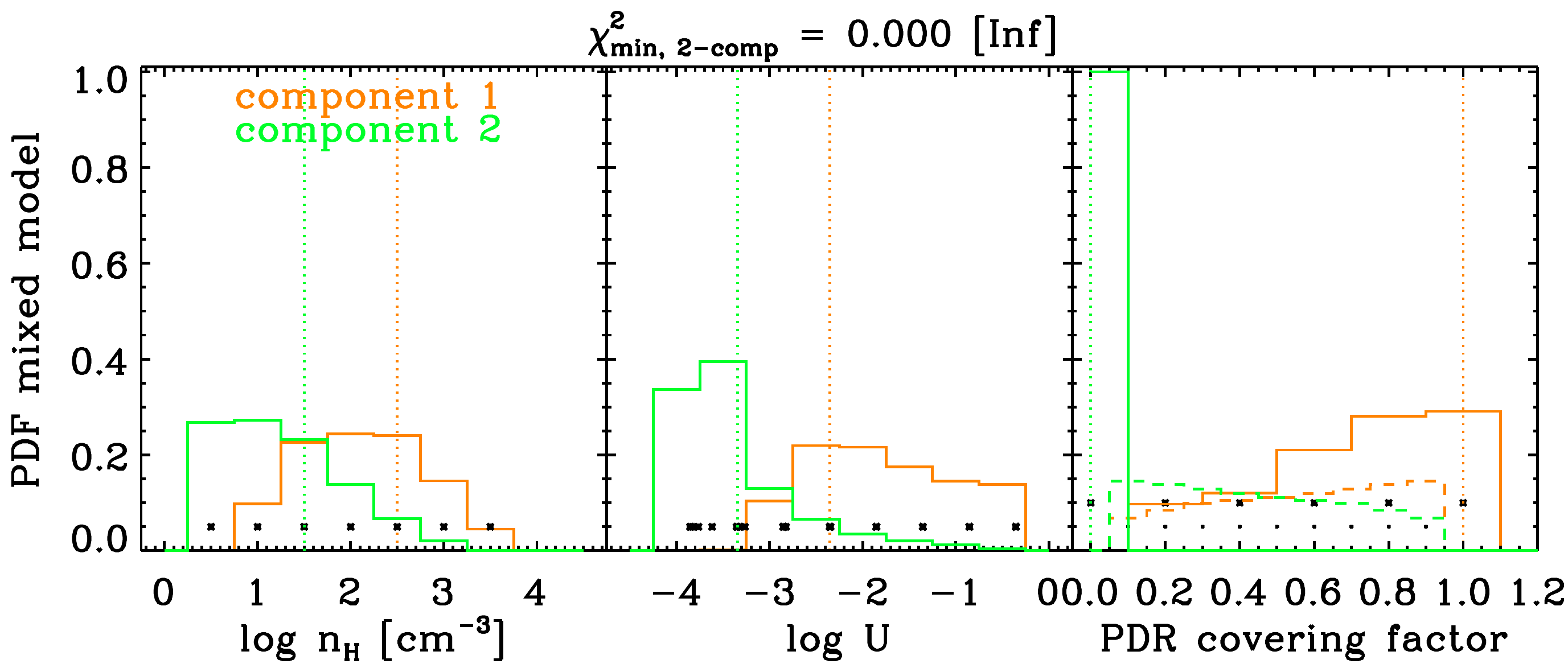}
\caption{ \textbf{Results for HS\,0017.}
$(a)$~Comparison of the observed (losange) and predicted 
intensities for our best-fitting single (blue bar) and mixed (red bar)
\hii region+PDR models. Lines are sorted by decreasing energy.
The fitted lines have labels in boldface.
$(b)$~For mixed models: respective contributions from
component \#1 and component \#2 to the line prediction.
For PDR lines (\cii, \oi, and \silii), only the contribution from
the ionized gas is shown; the PDR contribution is
one minus the sum of the contributions from the plot.
Contributions are expected to be dominated by one
or the other component if their model parameters
are noticeably different.
$(c)$~Probability density functions of the model parameters
($n_{\rm H}$, $U$, $cov_{\rm PDR}$/scaling factor)
for single (left panel) and mixed (right panel) models.
The scaling factor corresponds to the proportion in which
we combine two models in the mixed model case. It is
equal to unity otherwise. It is shown with dashed lines in
the same panels as the PDR covering factor. Vertical dotted lines
show values of the best-fitting model. Black asterisks
show the range and step of values of the grid parameters.
}
\label{fig:bests-hs0017}
\end{figure*}
 \begin{figure*}[thp]
\centering
(a)
\includegraphics[clip,width=11cm]{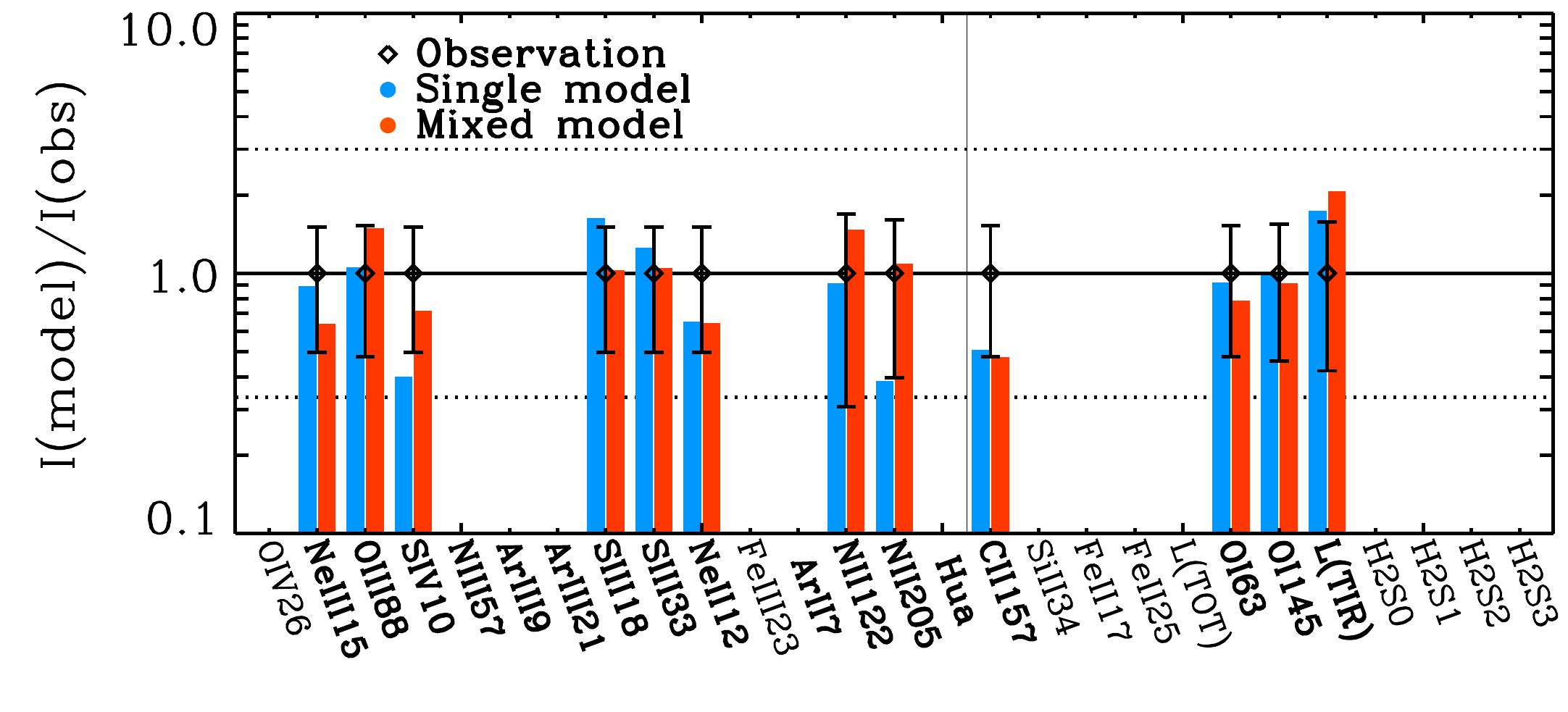} \hfill{}
(b)
\includegraphics[clip,width=6.2cm]{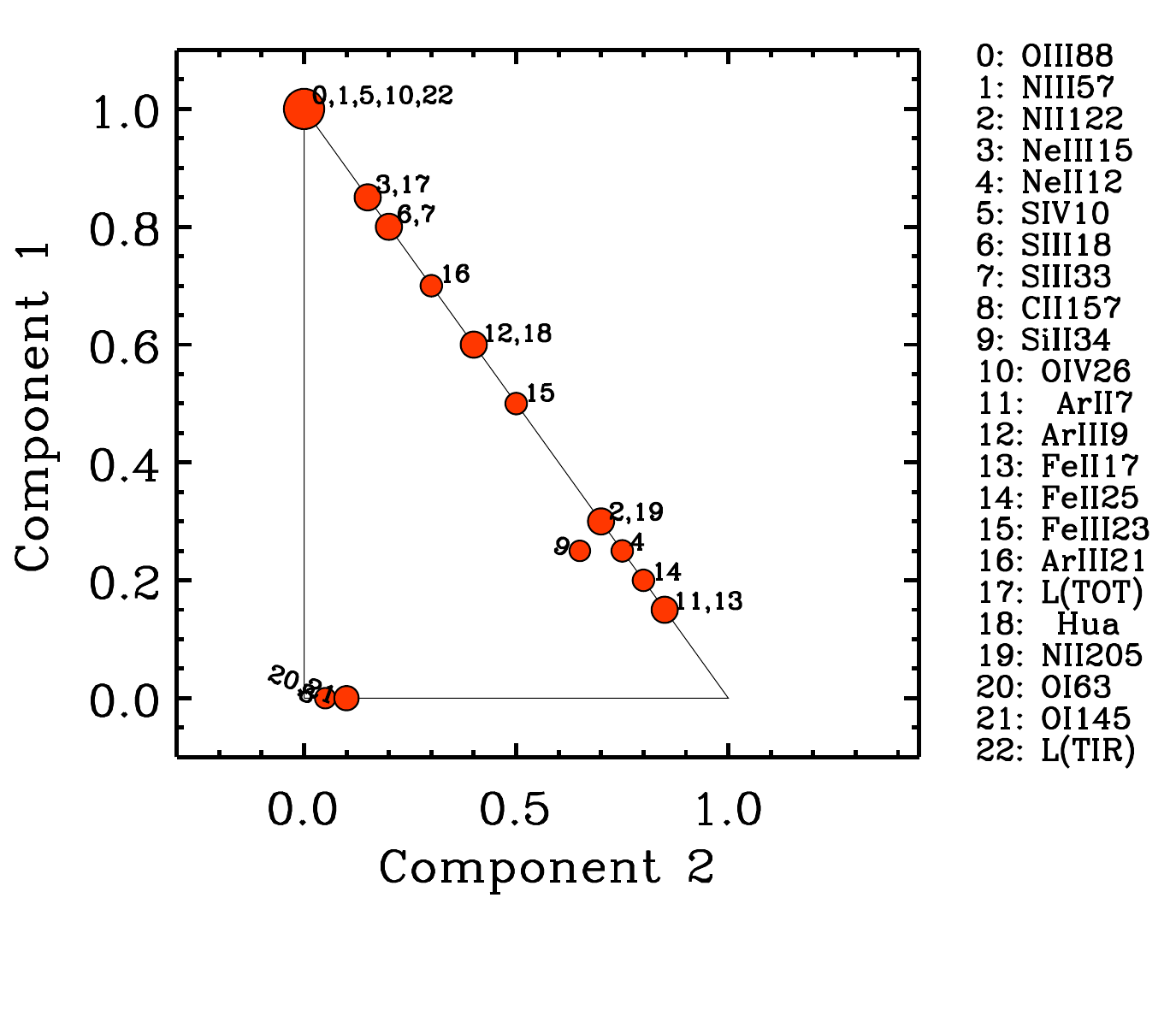}\\
(c)
\includegraphics[clip,width=8.8cm]{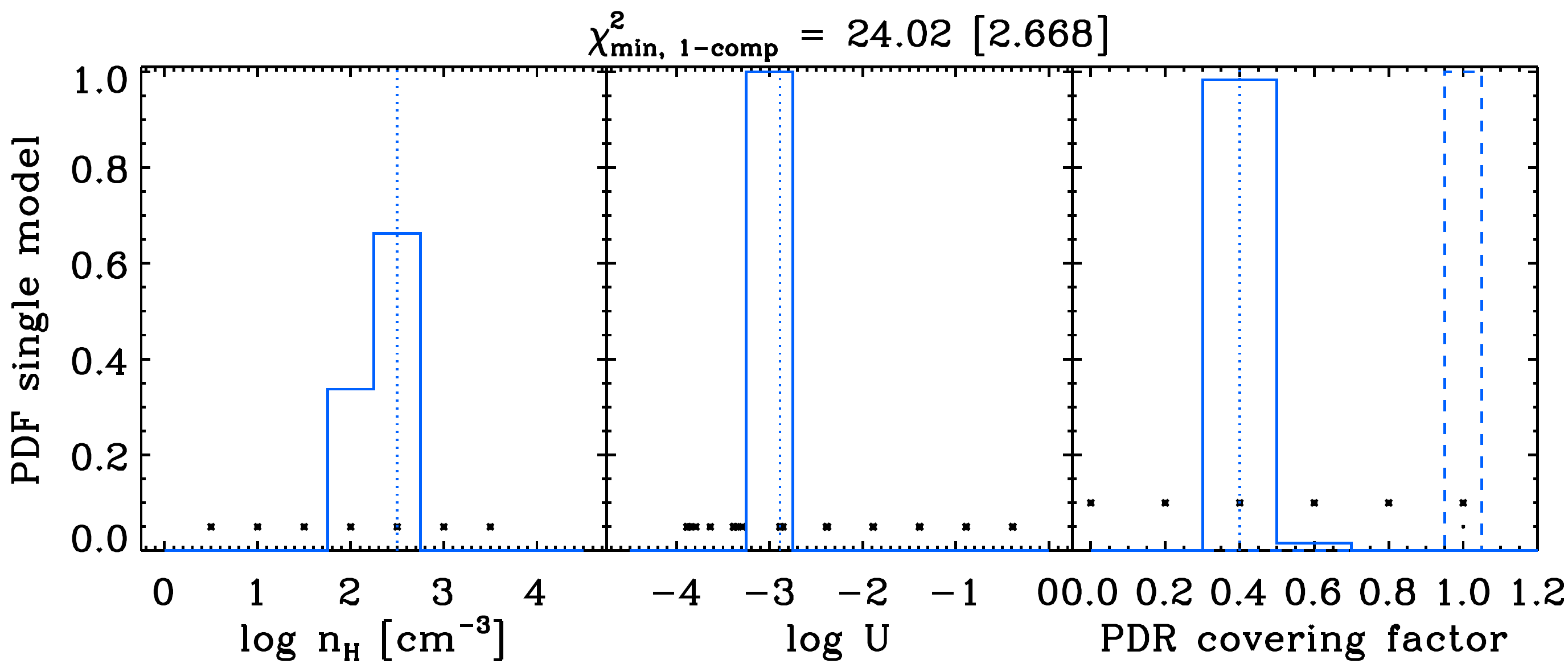} \hfill{}
\includegraphics[clip,width=8.8cm]{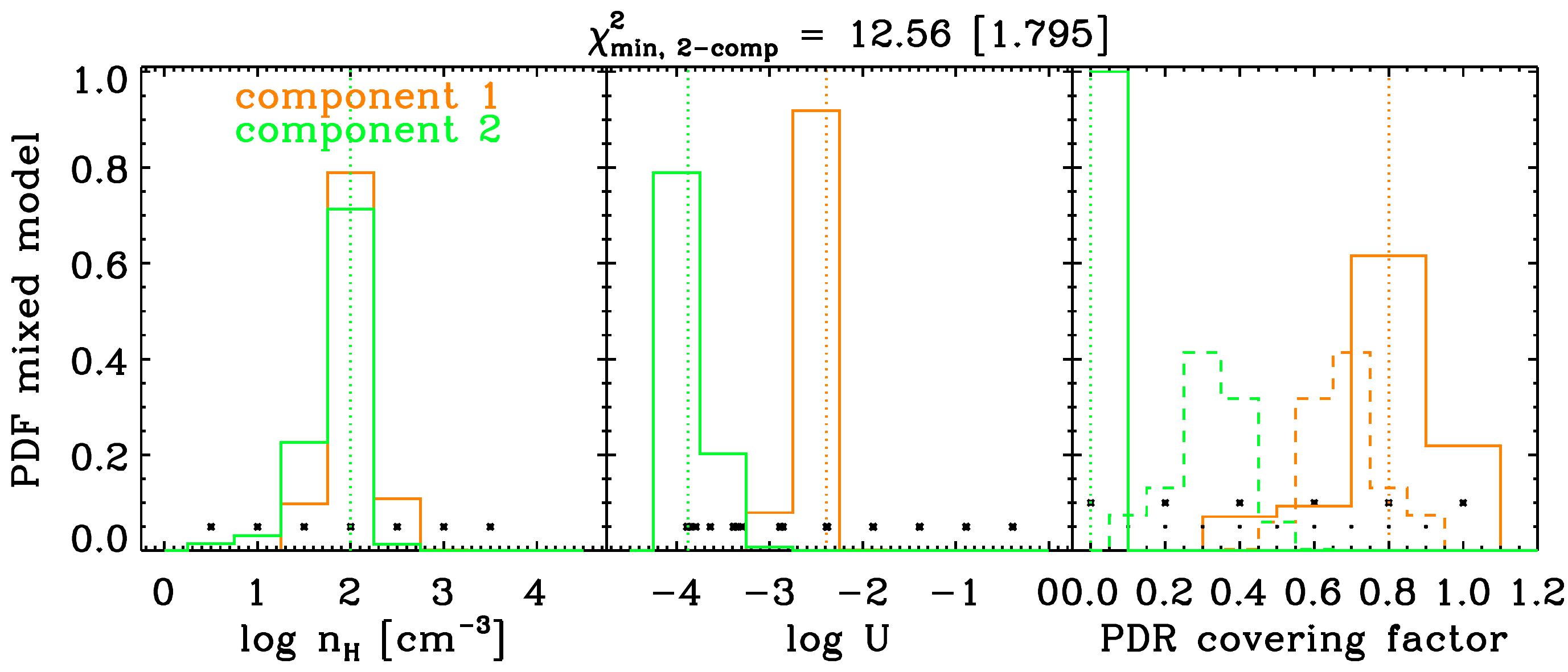}
\caption{ \textbf{Results for NGC\,4214-c.}
See above for caption description.
}
\label{fig:bests-ngc4214-c}
\end{figure*}
\clearpage
 \begin{figure*}[thp]
\centering
(a)
\includegraphics[clip,width=11cm]{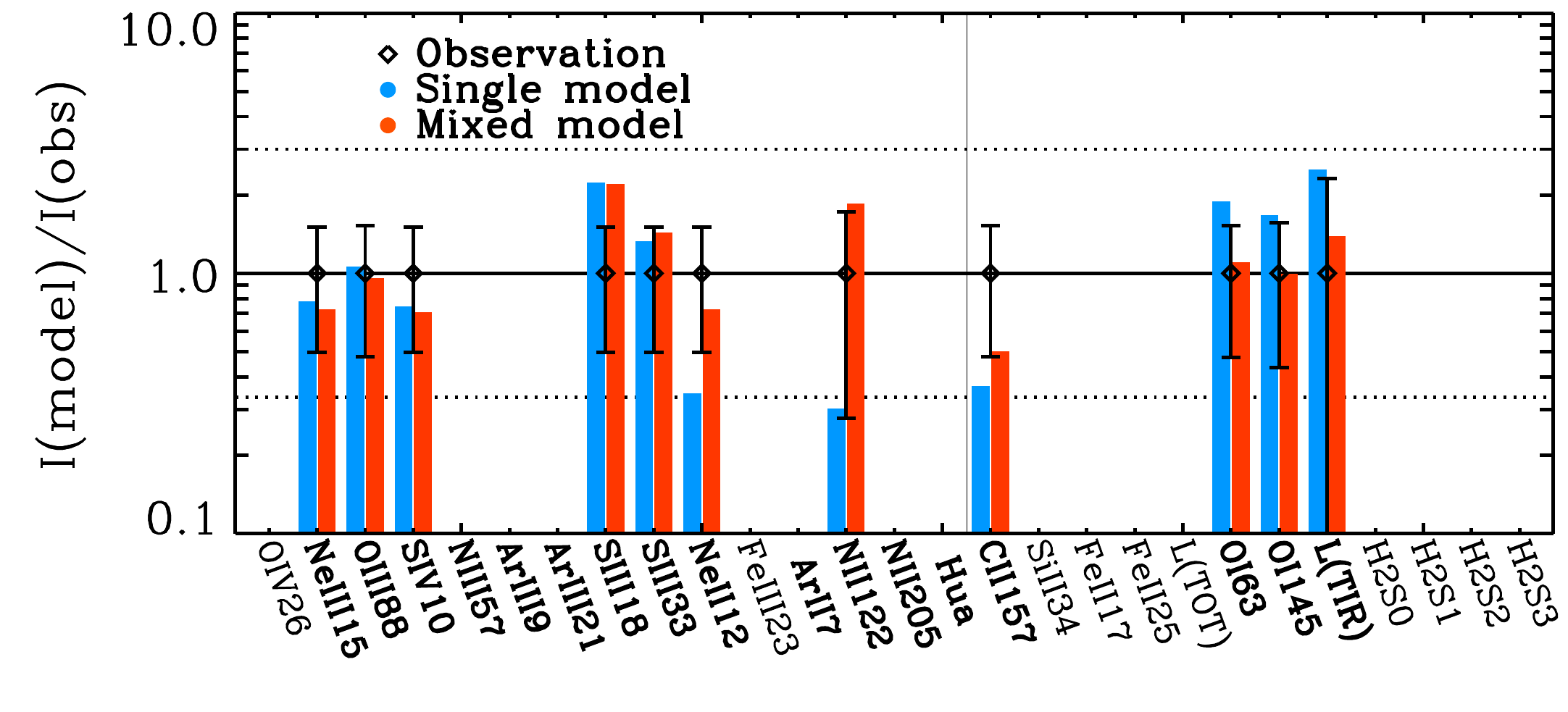} \hfill{}
(b)
\includegraphics[clip,width=6.2cm]{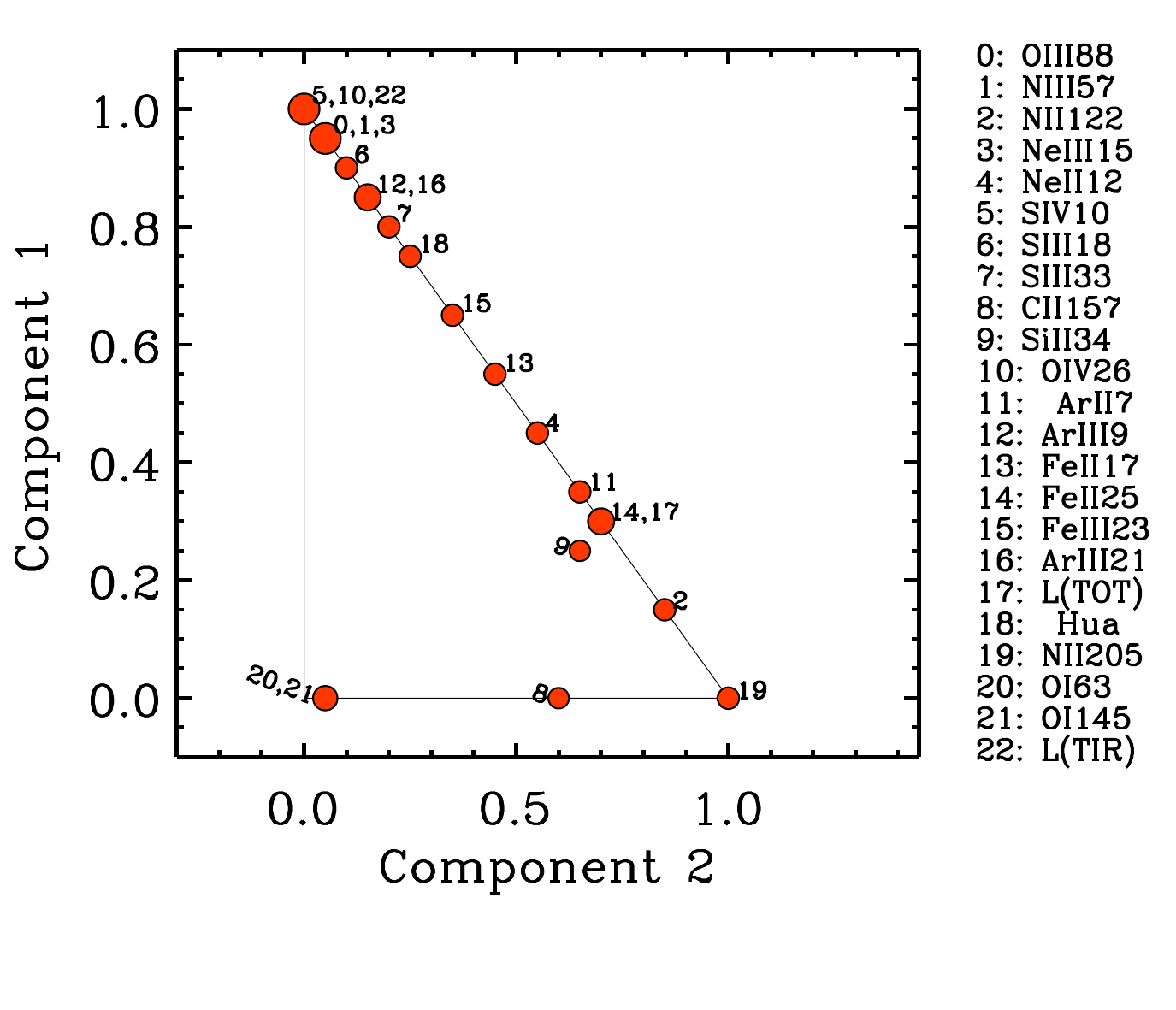}\\
(c)
\includegraphics[clip,width=8.8cm]{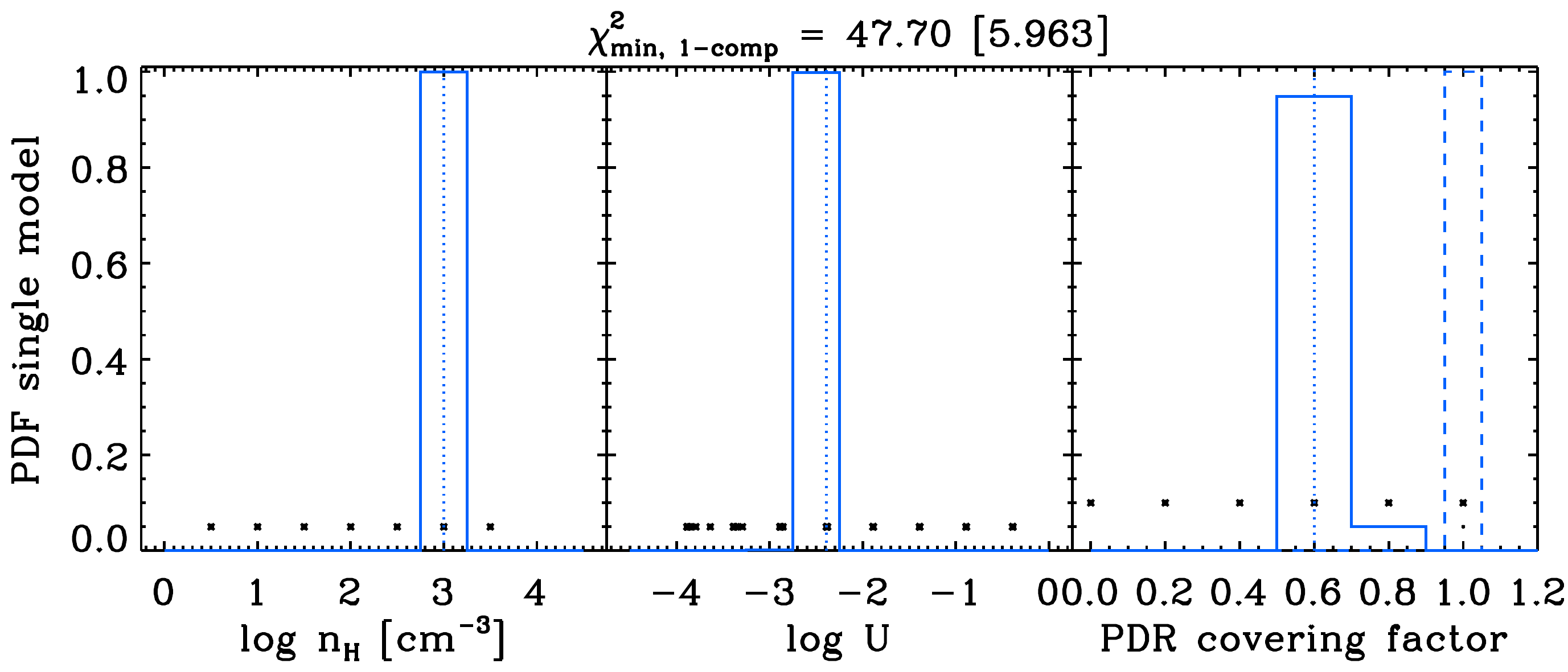} \hfill{}
\includegraphics[clip,width=8.8cm]{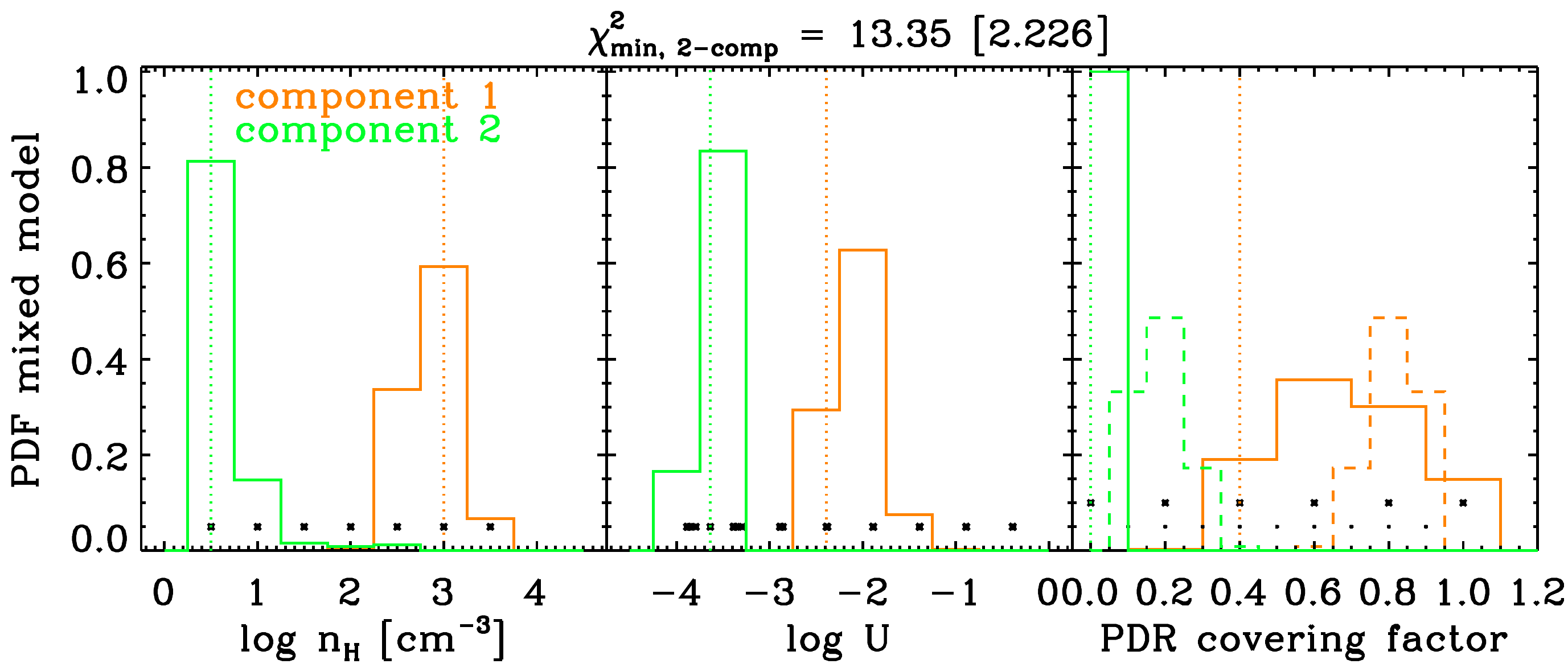}
\caption{ \textbf{Results for NGC\,4214-s.}
$(a)$~Comparison of the observed (losange) and predicted 
intensities for our best-fitting single (blue bar) and mixed (red bar)
\hii region+PDR models. Lines are sorted by decreasing energy.
The fitted lines have labels in boldface.
$(b)$~For mixed models: respective contributions from
component \#1 and component \#2 to the line prediction.
For PDR lines (\cii, \oi, and \silii), only the contribution from
the ionized gas is shown; the PDR contribution is
one minus the sum of the contributions from the plot.
Contributions are expected to be dominated by one
or the other component if their model parameters
are noticeably different.
$(c)$~Probability density functions of the model parameters
($n_{\rm H}$, $U$, $cov_{\rm PDR}$/scaling factor)
for single (left panel) and mixed (right panel) models.
The scaling factor corresponds to the proportion in which
we combine two models in the mixed model case. It is
equal to unity otherwise. It is shown with dashed lines in
the same panels as the PDR covering factor. Vertical dotted lines
show values of the best-fitting model. Black asterisks
show the range and step of values of the grid parameters.
}
\label{fig:bests-ngc4214-s}
\end{figure*}
%

\end{document}